\documentclass[10pt]{article}

\usepackage[lmargin=1in,rmargin=1in,tmargin=1in,bmargin=1in]{geometry}
\usepackage{amsthm,slashed,mathtools,amssymb}
\allowdisplaybreaks[1]
 
\usepackage[english]{babel}
\usepackage[T1]{fontenc}
\usepackage{lmodern}          

\usepackage{relsize}

\usepackage{microtype}      
\UseMicrotypeSet[protrusion]{basicmath}

\usepackage[shortlabels]{enumitem}
\usepackage[affil-it]{authblk}  

\usepackage[section]{placeins}
\usepackage{soul} 
\usepackage{csquotes}
\usepackage{bm,bbm}
\usepackage{mathrsfs, graphicx} 

\usepackage{tikz}
\usetikzlibrary{calc,intersections,backgrounds,patterns}
\usetikzlibrary{decorations.pathreplacing}

\usepackage{empheq}

\usepackage[%
bookmarks=true,
colorlinks=true,
linkcolor=black,
urlcolor=black,
citecolor=black,
plainpages=false,
pdfpagelabels,
final,hypertexnames=false]{hyperref}

\usepackage{aliascnt}
\usepackage[capitalise,nameinlink]{cleveref}

\usepackage[backend=biber,style=numeric,giveninits=true,doi=false,isbn=false,url=false,sorting=anyt,maxalphanames=10,maxbibnames=99,date=year]{biblatex}
\renewbibmacro{in:}{}
\DeclareMathOperator{\dvol}{dvol}

\newcommand{\f}{\frac}
\newcommand{\minusone}{{\text{-}1}}
\renewcommand{\div}{\operatorname{div}}
\renewcommand{\Re}{\operatorname{Re}}
\renewcommand{\Im}{\operatorname{Im}}
\DeclareMathOperator{\supp}{supp}
\DeclareMathOperator{\sgn}{sgn}
\DeclareMathOperator{\Ai}{Ai}
\DeclareMathOperator{\Bi}{Bi}

\newcommand{\ob}{\mathring{b}}
\newcommand{\oR}{\mathring{R}}
\newcommand{\oomega}{\mathring{\omega}}
\newcommand{\oV}{\mathring{V}}
\newcommand{\oE}{\mathring{e}}
\newcommand{\ov}{\mathring{v}}
\newcommand{\oF}{\mathring{F}}
\newcommand{\oy}{\mathring{y}}

\newcommand{\bigzero}{\mbox{\normalfont\large 0}}

\DeclareRobustCommand{\ind}{\mathop{\hbox{\Large$\mathbbm{1}$}}\nolimits}

\numberwithin{equation}{section}
\mathtoolsset{showonlyrefs=true}

\theoremstyle{plain}
\newtheorem{theorem}{Theorem} 
\crefname{theorem}{Theorem}{Theorems}
\newtheorem*{theorem*}{Theorem}

\newtheorem{proposition}{Proposition}[section]
\crefname{proposition}{Proposition}{Propositions}

\newaliascnt{lemma}{proposition}
\newtheorem{lemma}[lemma]{Lemma}
\aliascntresetthe{lemma}
\crefname{lemma}{Lemma}{Lemmas}

\newaliascnt{corollary}{proposition}
\newtheorem{corollary}[corollary]{Corollary}
\aliascntresetthe{corollary}
\crefname{corollary}{Corollary}{Corollaries}

\newaliascnt{conjecture}{proposition}
\newtheorem{conjecture}[conjecture]{Conjecture}
\aliascntresetthe{conjecture}
\crefname{conjecture}{Conjecture}{Conjectures}
\newtheorem*{conjecture*}{Conjecture}

\theoremstyle{definition}
\newaliascnt{definition}{proposition}
\newtheorem{definition}[definition]{Definition}
\aliascntresetthe{definition}
\crefname{definition}{Definition}{Definitions}

\newaliascnt{remark}{proposition}
\newtheorem{remark}[remark]{Remark}
\aliascntresetthe{remark}
\crefname{remark}{Remark}{Remarks}

\theoremstyle{plain}
\newtheorem{manualtheoreminner}{Theorem}
\newenvironment{manualtheorem}[1]{%
  \renewcommand\themanualtheoreminner{#1}%
  \manualtheoreminner
}{\endmanualtheoreminner}

\newcommand{\llangle}{\langle\mkern-4mu\langle}
\newcommand{\rrangle}{\rangle\mkern-4mu\rangle}

\makeatletter
\renewcommand\paragraph{\@startsection{paragraph}{4}{\z@}%
  {1.0ex plus .2ex minus .1ex}
  {-1em}
  {\normalfont\normalsize\itshape}}
\makeatother

\begin{document}

 \title{\makebox[\textwidth][c]{Weakly~turbulent~dynamics on~Schwarzschild--AdS~black~hole~spacetimes}}  
\author[1]{Christoph~Kehle\thanks{kehle@mit.edu}}
\author[2]{Georgios Moschidis\thanks{georgios.moschidis@epfl.ch}}
\affil[1]{\small  Massachusetts Institute of Technology, Department of Mathematics,

Building 2, 77 Massachusetts Avenue, Cambridge, MA 02139, United States of America \vskip.1pc \ 
}
\affil[2]{\small École Polytechnique Fédérale de Lausanne, Institute of Mathematics, 

	MA B1 595, CH-1015 Lausanne, Switzerland \vskip.1pc \  
	}

 \date{April 13, 2026}
\maketitle
\begin{abstract}
In the presence of confinement, small-data solutions to nonlinear dispersive equations can exhibit a gradual energy transfer from low to high frequencies, a mechanism driving the emergence of weakly turbulent dynamics.
 We show that such a forward energy transfer, manifested as arbitrary inflation of higher order Sobolev norms, occurs for small-data solutions of a quasilinear cubic wave equation on the Schwarzschild--AdS black hole exterior with Dirichlet conditions at infinity, for generic values of the mass parameter. 
This result is motivated by the question of nonlinear stability or instability of Schwarzschild--AdS as a solution to the Einstein vacuum equations, but the strategy of proof applies to a broader class of backgrounds exhibiting stable trapping of null geodesics. As an application, we obtain the analogous norm inflation statement on $\mathbb R \times \mathbb S^3_+$ for generic perturbations of the round metric on the hemisphere $\mathbb S^3_+$ preserving the trapping structure at the boundary. 
\end{abstract}

\thispagestyle{empty}

\newpage
 
\setcounter{tocdepth}{2}
\tableofcontents

 \newpage
\section{Introduction}	
   
The phenomenon of turbulence, characterized by the presence of an energy flux between different scales in systems far from thermodynamic equilibrium \cite{Z93,N11}, is ubiquitous in the world of nonlinear dispersive equations. In cases when the nonlinear interactions are appropriately weak, individual modes evolve almost linearly at short timescales, 
with the progressive exchange of energy among modes mediated by resonant interactions. 
The theory of weak (or wave) turbulence was developed in the 60s as a unifying framework to describe such 
processes in a broad class of systems modeled by Hamiltonian PDEs, from oceanic gravity waves to nonlinear optics systems \cite{H62,ZF67,Z68,ZLF92,MMcLT97,N11}.

The emergence of weakly turbulent dynamics in systems of finite energy typically involves a mechanism of confinement at the linear level: In bounded geometries (e.g.\ on compact manifolds, in cavities with reflecting boundaries) or in settings with strong trapping (e.g.\ spacetimes admitting stably trapped null geodesics, in the case of wave-type equations), linear perturbations persist as long-lived oscillatory modes. If certain resonance conditions are satisfied, these oscillatory modes allow weakly nonlinear interactions to accumulate over time, enabling the secular transfer of energy to progressively higher frequencies, a process known as a \emph{forward} (or direct) \emph{energy cascade}, signified by the inflation of higher order Sobolev norms of the solution. On the other hand, strong dispersion at the linear level obstructs a  direct cascade: rapid decay of linear perturbations prevents  nonlinear effects from accumulating and the nonlinear evolution remains close to the linear one.

The Einstein vacuum equations 
\begin{equation}\label{eq:EVE-Lambda}
\operatorname{Ric}[g]-\f12 R[g] g+\Lambda g=0
\end{equation}
is a prominent example of a geometric nonlinear system of PDEs of hyperbolic nature. As such, given a stationary spacetime solution with confining geometry, one would expect its nonlinear perturbations to exhibit weakly turbulent dynamics. In the case of a \emph{negative} cosmological constant $\Lambda$ (without loss of generality, we set $\Lambda=-3$ for the rest of the paper), the \emph{Schwarzschild--Anti-de Sitter} (AdS) exterior spacetime $\left( \mathcal M^{(M)}_{\mathrm{ext}}, g_M\right)$, where  
\begin{equation}\label{eq:metric-schwarzschild-ads}
g_M = -\Big(1-\frac{2M}{r} + r^2\Big)\,dt^2
      + \Big(1-\frac{2M}{r}+ r^2\Big)^{-1}\,dr^2
      + r^2\big(d\theta^2 + \sin^2\theta\, d\varphi^2\big)
\end{equation}
(with $M>0$), provides a one-parameter family of stationary solutions of \eqref{eq:EVE-Lambda} possessing both dispersive and confining geometric features. The function $f(r)=1-\frac{2M}{r}+r^2$ has a unique positive root $r_+$ corresponding to the event horizon, separating the spacetime into the black-hole region $r<r_+$ and the exterior region $\mathcal M^{(M)}_{\mathrm{ext}} = \mathbb R_t \times (r_+, +\infty)_r \times \mathbb S^2_{\theta, \varphi}$. The presence of the event horizon acts as a dispersive mechanism: waves that cross it are causally prohibited from returning to the exterior. At the same time, the exterior region admits stably trapped null geodesics between the photon sphere located at $\{ r=3M\}$ and the timelike conformal boundary at infinity $\mathcal I= \{ r=\infty\}$. Consequently, the question of whether the gravitational dynamics in a neighborhood of the Schwarzschild--AdS exterior spacetime  exhibit weak turbulence---a question directly related to the \emph{stability problem} of Schwarzschild--AdS, see \cref{sec:Stability Schw-AdS}---is one which remains open to debate \cite{DHS12,HS13,FR25,CrumpSantos25}.

In this work, we initiate the study of the dynamics of nonlinear models for the system \eqref{eq:EVE-Lambda} on the Schwarzschild--AdS exterior. We consider solutions $\phi: \mathcal M_{\mathrm{ext}}^{(M)}\to \mathbb C$ to the initial--boundary value problem of the following cubic quasilinear wave equation: 
	\begin{equation}\label{Initial Boundary Value Problem Phi}
		\begin{cases}
			\square_{g_M} \phi + 2 \phi - \mathcal N^{(3)}[\phi] =0,\\[5pt]
			(\phi, \partial_{t} \phi)|_{t=0} = (f_0, f_1), \\[5pt]
			r\phi|_{r=\infty} =0,
		\end{cases}
	\end{equation}
	where 
	\begin{equation}\label{eq:nonlinearity}
		\mathcal N^{(3)}[\phi] \doteq   \f1{r^6} \left(  |\partial_t \phi|^2 \phi + |\phi|^2 \partial_t^2 \phi \right) .
	\end{equation}
We will show that solutions to \eqref{Initial Boundary Value Problem Phi}, for an open and dense set of mass parameters $M>0$, exhibit inflation of their higher order Sobolev norms as a result of a forward energy transfer from low to high-frequency modes. 
Our choice of the  weight $1/r^6$ appearing in the definition of $\mathcal N^{(3)}$ is guided by the presence of a similar weight in the cubic terms in the formal reduction of \eqref{eq:EVE-Lambda} in generalized harmonic coordinates, see already our discussion in \cref{sec:motivation-nonlinearity}. Note also that smooth solutions to equation \eqref{Initial Boundary Value Problem Phi} have a conserved $H^1$-type energy given by  \eqref{Conserved energy intro}.

	\subsection{The main theorem}
Our main result, proved in \cref{sec:The proof of the main theorem}, is the following.

	\begin{theorem}[Weak turbulence on Schwarzschild--AdS]\label{thm:Main theorem}
		For any $\epsilon>0$ and any integer $s\ge 4$, there exists an open and dense set $\mathcal J_{\epsilon, s} \subseteq (0,+\infty)$ such that the following statement holds: For any black hole mass parameter $M\in \mathcal J_{\epsilon, s}$, there exists a smooth initial data set $(f_0, f_1)$ which is supported in the region $\{r>3M\}$ and satisfies
		\begin{equation*}
			\| f_0\|_{H^{s}} + \|f_1\|_{H^{s-1}} \le \epsilon,
		\end{equation*}
		as well as a time $T_1>0$ such that the corresponding solution  $\phi$ of the  initial--boundary value problem \eqref{Initial Boundary Value Problem Phi} exists for $0\le t^\ast \le T_1$ and satisfies
		\begin{equation}\label{Norm inflation}
			\|\phi |_{t^*=T_1}\|_{H^s} > \f1\epsilon.
		\end{equation} 
	\end{theorem}
In particular, the set
\[
\mathcal G \doteq \bigcap_{s\ge 4}\ \bigcap_{n\in\mathbb N}\ \mathcal J_{1/n,s},
\]
i.e.\ the set of Schwarzschild--AdS masses for which the statement of \cref{thm:Main theorem} holds for all $\epsilon>0$ and $s\geq 4$, is Baire-generic in $\mathbb R_{>0}$.
 The time function $t^\ast$ is a slight modification of $t$ which agrees with $t$ away from the event horizon, see  \cref{def:Modified time function}.  For the precise definition of the Sobolev norm $H^s$ along the hypersurfaces $\{t^*=\mathrm{\text{const}}\}$, we refer to \cref{sec:Sobolev norms}. For an outline of the proof of \cref{thm:Main theorem}, see \cref{sec:outline-proof-main-theorem}.

Let us make the following remarks:
\begin{remark}
The proof of \cref{thm:Main theorem} applies without any change to both the defocusing ``$+$'' and focusing ``$-$'' sign for the nonlinearity $\pm\mathcal N^{(3)}$; we merely fix the defocusing sign here for definiteness. 
\end{remark}
\begin{remark}
  The Schwarzschild--AdS family of metrics $(g_M)_{M> 0}$ sits within the broader 2-parameter family $g_{a,M}$ of Kerr--AdS spacetimes (with $a\in \mathbb R$ representing the angular momentum of the black hole). Our proof can be readily extended to apply to the case of slowly rotating Kerr--AdS black holes, yielding a weak turbulence result for an open and dense set of mass parameters $M$ for any fixed value of $a$; see \cref{rmk:Kerr}.
\end{remark}

Our proof of \cref{thm:Main theorem}  can in principle be applied to initial--boundary value problems of the form \eqref{Initial Boundary Value Problem Phi} defined on a much broader class of 1-parameter families of spacetimes, yielding a weak turbulence statement for an open and dense set of the corresponding parameters. Considering spherical coordinates $( y,\theta,\varphi)$ on the hemisphere $\mathbb S^3_+ =  [0,\pi/2)_y\times \mathbb  S_{\theta,\varphi}^2$, with the equator being  $\partial \mathbb S^3_+ = \{y=0\}$, the aforementioned class of spacetimes would include any $\mathbb S^2$-symmetric 1-parameter family of stationary Lorentzian  metrics on $\mathbb R\times \mathbb S^3_+$ (with parameter $M$), such that:
\begin{itemize}
\item Near $\{y=0\}$, the metric is a perturbation of  $g_{\mathbb R\times \mathbb S^3_+}=-dt^2+g_{\mathbb S^3_+}$ of size $\sim My^3$.
\item A global energy boundedness statement is available for the scalar wave equation via appropriate vector field multipliers.
\end{itemize}
 See also the discussion in \cref{sec:Proof of second theorem}. 
As a demonstration of this fact, we can deduce the following statement as an immediate corollary of our proof of \cref{thm:Main theorem}: Letting $(h^{M}_{\mathbb S^3_+})_{M\geq 0}$ be a smooth family of  $\mathbb S^2$-symmetric  metrics on $ \mathbb S^3_+$ of the form 
\[
h_{\mathbb S^3}^M= h_{\mathbb S^3_+} + y^3 \tilde h_M, 
\]
where $\tilde h_M$ is $\mathbb S^2$-symmetric (in particular, its components are functions only of $y$) and where $\tilde h_M|_{y=0}=2 M (d\theta^2 + \sin^2 \theta d\varphi^2)$, we have:

	\begin{theorem}[Weak turbulence on $\mathbb S^3_+$]
    \label{thm:Main theorem-S3}
		For any $\epsilon>0$ and any integer $s\ge 4$, there exists an open and dense set $\mathcal J_{\epsilon, s} \subseteq (0, +\infty)$ such that the following statement holds: For any parameter $M\in \mathcal J_{\epsilon, s}$, there exists a smooth initial data set $(f_0, f_1)$ on $(\mathbb S^3_+,h_{\mathbb S^3_+}^M)$ satisfying \begin{equation*}
			\| f_0\|_{H^{s}} + \|f_1\|_{H^{s-1}} \le \epsilon 
		\end{equation*} 
        for the initial--boundary value problem 
\begin{equation}\label{Equation for corollary problem}
		    \begin{cases}
		       - \partial_t^2 \phi +  \Delta_{h^M_{S^3_+}} \phi  =  y^6 \left(  |\partial_t \phi|^2 \phi + |\phi|^2 \partial_t^2 \phi \right)\\
                (\phi, \partial_t \phi)|_{t=0} = (f_0,f_1),\\
                \phi|_{\partial\mathbb S^3_+ }=0,
		    \end{cases}
            \end{equation}
		and a time $T_1>0$ such that a smooth solution  exists for $0\leq t\le T_1$ and satisfies
		\begin{equation}
			\|\phi |_{t=T_1}\|_{H^s} > \f1\epsilon.
		\end{equation}
	\end{theorem}

\noindent For an outline of how the proof of \cref{thm:Main theorem} extends (and, in fact, simplifies in many respects) to yield the proof of \cref{thm:Main theorem-S3}, see \cref{sec:Proof of second theorem}. 
	
\subsection{A test case for weak turbulence on Schwarzschild--AdS}\label{sec:motivation-nonlinearity}
The nonlinear initial--boundary value problem \eqref{Initial Boundary Value Problem Phi} can be considered as a testbed for the study of weakly turbulent phenomena in the dynamics of gravitational perturbations on the Schwarzschild--AdS exterior spacetime. From this perspective, let us briefly discuss the reasons motivating the particular choice of the nonlinearity $\mathcal N^{(3)}[\phi]$ in  \eqref{Initial Boundary Value Problem Phi}.

\subsubsection{A model for the vacuum Einstein equations} \label{sec:A toy model}
The well-posedness of the initial--boundary value problem for \eqref{eq:EVE-Lambda} in the asymptotically Anti-de\,Sitter setting was first considered by Friedrich \cite{Friedrich1995}, who established the existence and geometric uniqueness of solutions for a broad class of boundary conditions defined in terms of the conformal geometry of $\mathcal I$. This class included a unique Dirichlet-type condition, corresponding to the prescription of a conformally flat geometry on $\mathcal I$ (for the description of a geometric \emph{Neumann} boundary condition leading to a well-posed problem, see the recent work \cite{Sou25}).  His proof was based on the construction of a gauge tied to the conformal structure of the spacetimes under consideration.

For the corresponding problem expressed in the \emph{harmonic gauge}, well-posedness of the Dirichlet initial--boundary value problem was established by \cite{EncisoKamran2019} (based on the earlier work \cite{GrahamLee91} in the Riemannian setting). Denoting by $\tilde g = g_{\mathrm{AdS}}$ 
the reference AdS metric 
\begin{equation}\label{AdS metric Intro}
g_{\mathrm{AdS}} = -(1+r^2)dt^2+\f1{1+r^2}dr^2 + r^2\big(d\theta^2+\sin^2\theta d\varphi^2\big),
\end{equation}
 the harmonic gauge condition reads
\[
g^{\mu\nu}\big(\Gamma^\alpha_{\mu\nu} -\tilde\Gamma_{\mu\nu}^\alpha\big)=0.
\]
Under the above condition, the Einstein equations \eqref{eq:EVE-Lambda} (for $\Lambda=-3$) take the form
\begin{align}\label{Equations harmonic gauge intro}
g^{\alpha\beta}\partial_\alpha\partial_\beta g_{\mu\nu}  -6 g_{\mu\nu} & -2g^{\alpha\beta}g_{\gamma(\mu}\partial_{\nu)}\tilde\Gamma_{\alpha\beta}^\gamma \\
 +2g^{\alpha\beta}g_{\gamma\delta}\Gamma_{\mu\nu}^\gamma\tilde\Gamma_{\alpha\beta}^\delta& -2 g^{\alpha\beta}g_{\gamma\delta}\Gamma^\gamma_{\alpha\mu}\Gamma^\delta_{\beta\nu}-4g^{\alpha\beta}g_{\gamma(\mu}\Gamma^\delta_{\nu)\alpha}\Gamma^\gamma_{\beta\delta}=0.\nonumber
\end{align}
Denoting by $y=y(r)\sim \f1r$ the boundary-defining function of $\mathcal I=\{r=\infty\}$, regularity of the metric $g$ at the conformal boundary translates to regularity of the conformal metric \[\bar g = y^2 g\] in the $(t,y, \theta, \varphi)$ coordinates near $\mathcal I=\{y=0\}$; in this case,  \eqref{eq:EVE-Lambda} also implies that $\bar g_{\alpha\beta} (\bar g_{\mathrm{AdS}})^{\alpha\beta}|_{\mathcal I} = 4$ and $\bar g_{\alpha\beta} dy^{\beta} |_{\mathcal I} = (\bar g_{\mathrm{AdS}})_{\alpha\beta} dy^\beta|_{\mathcal I}$ (see  \cite{GrahamLee91}). The Dirichlet boundary condition for $g$ then takes the form
\begin{equation}\label{Boundary conditions}
\bar g|_{\mathcal I}  = \bar g_{\mathrm{AdS}}|_{\mathcal I}.
\end{equation}

In the case when the metric $g$ is a small perturbation of the Schwarzschild--AdS metric $g_M$, we can study the boundary value problem \eqref{Equations harmonic gauge intro}--\eqref{Boundary conditions} from a perturbative perspective: Setting 
\[
g = g_M+ h 
\]
and expanding each term in \eqref{Equations harmonic gauge intro} into a power series in $h$ (for instance, $g^{-1}  = y^2 \bar g_M-y^4(\bar g_M^{-1})^2 h+\dots $), we formally obtain:  
\begin{equation}\label{Perturbative system intro}
\begin{cases}
\square_{g_M} h+O(y) \bar\partial h+ 2(h-\mathrm{tr}_{g_M}h \cdot g_M+O(y)h) = \mathcal N^{(2)}_{\mathrm{Ein}}(h)+\mathcal N^{(3)}_{\mathrm{Ein}}(h)+\mathcal N^{(\ge 4)}_{\mathrm{Ein}}(h),
\\[5pt]
y^2 h|_{y=0}=0,
\end{cases}
\end{equation}
where $\square_{g_M}$ denotes the tensorial wave operator on Schwarzschild--AdS, $\bar\partial h$ denotes any linear combination of $\{\partial_t h, \partial_y h, \nabla_{\mathbb S^2}h\}$ with coefficients which are bounded up to $y=0$ (e.g.~formed from the components of $\bar g_M = y^2 g_M$) and:
\begin{itemize}\setlength\itemsep{5pt}
\item $\mathcal N^{(2)}_{\mathrm{Ein}}(h) = y^{4}\Big(h\cdot\bar\partial^2 h+\bar\partial h \cdot \bar\partial h+O(y^{-1}) h \cdot \bar\partial h\Big)$,
\item $\mathcal N^{(3)}_{\mathrm{Ein}}(h) = y^6\Big(h\cdot h\cdot\bar\partial^2 h+h\cdot \bar\partial h \cdot \bar\partial h+O(y^{-1}) h \cdot h\cdot \bar\partial h+O(y^{-2}) h\cdot h\cdot h$\Big),
\item $\mathcal N^{(\ge 4)}_{\mathrm{Ein}}(h)$ contains quasilinear terms of order $4$ or higher in $ h$.
\end{itemize}

Understanding whether a direct energy cascade appears in the evolution of solutions to \eqref{Perturbative system intro} requires a careful analysis of the resonant structure of the nonlinear terms. The quadratic terms $\mathcal N^{(2)}_{\mathrm{Ein}}$ satisfy the so-called \emph{weak null condition}, uncovered in \cite{LR2005}, which prohibits certain types of self-interactions in the solution. In the case of small data solutions to systems of nonlinear wave equations on $\mathbb R^{3+1}$, it is known (see \cite{S85,PS13}) that quadratic nonlinearities satisfying certain algebraic conditions (which include the classical null condition \cite{C86,K84}) can be ``transformed'' away in the context of their effect on the nonlinear dynamics.
 It is, therefore, not unreasonable to expect that a similar transformation might exist for the Einstein system, precluding the quadratic terms from contributing to the resonant system. On that basis, in our pursuit of a \emph{scalar} model for \eqref{Perturbative system intro}, we decided to drop the quadratic terms altogether,\footnote{Given a nonlinear equation of the form \eqref{Almost model cubic} with a quadratic (rather than cubic)  nonlinearity, our method of proof (and, crucially, our analysis of the error term in \cref{sec:Estimates error term}), would carry over without significant changes, as long as the nonlinearity had the property that the corresponding $3\times 3$ resonant system for the three dominant modes (see \cref{sec:Dominant modes}) exhibits norm inflation.} arriving at a quasilinear cubic nonlinear equation of the form
\begin{equation}\label{Almost model cubic}
\square_g \phi+2\phi = y^6 \Big( \phi\cdot \phi \cdot \bar\partial^2 \phi + \phi \cdot \bar\partial\phi \cdot \bar \partial\phi \Big).
\end{equation}
In the above, the conformally coupled value for the Klein--Gordon mass was chosen for simplicity; for that value, the Dirichlet boundary condition for $\phi$ reads
\[
r\phi|_{\mathcal I}=0.
\] 

In this work, we consider the nonlinearity $\mathcal N^{(3)}[\phi] = \displaystyle\f1{r^6} \Big(|\phi|^2\partial_t^2\phi +  |\partial_t \phi|^2 \phi\Big)$, which is a special case of the class of nonlinearities  appearing in \eqref{Almost model cubic}. The specific combination of terms in $\mathcal N^{(3)}[\phi]$ was motivated by the requirement that \eqref{Initial Boundary Value Problem Phi} should admit a conserved $H^1$-type energy for $\phi$; the absence of such a conservation law would inject ambiguity into the notion of an ``energy cascade''.\footnote{Once again, our strategy of proof would work for any other combinations of derivatives, as long as the resonant $3\times 3$ system exhibits norm inflation.} In the case of $\mathcal N^{(3)}[\phi]$, the conserved functional is
\begin{align}\label{Conserved energy intro}
\mathscr E[\phi](\tau) = \int_{t=\tau} \Bigg( & \big(1-\frac{2M}{r}+r^2\big)^{-1}|\partial_t\phi|^2+\big(1-\frac{2M}{r}+r^2\big)|\partial_r\phi|^2 \\
& \qquad+ \f1{r^2}|\nabla_{\mathbb S^2_{\theta,\varphi}}\phi|^2 -2|\phi|^2+\f1{r^6}|\phi|^2|\partial_t\phi|^2 \Bigg)\, r^2 \sin\theta dr d\theta d\varphi.  \nonumber
\end{align}
Note that, in the case of the Einstein system \eqref{eq:EVE-Lambda}, the conservation of the total (renormalized) ADM mass at $\mathcal I$ can be viewed as an $H^1$-type conservation law.

\subsubsection{Pure power nonlinearities: An obstruction to energy transfer}
\label{sec:obstruction-energy-transfer}

Attempting to set up a scalar model for the study of the weakly turbulent properties of the Einstein system \eqref{Perturbative system intro}, a natural first choice in place of \eqref{Initial Boundary Value Problem Phi} would be the simplest possible nonlinear wave equation with cubic nonlinearity, namely
\begin{equation}\label{eq:cubic-power-nonlinearity}
\square_{g_M}\phi  = \mathcal{N}^{(3)}_{\mathrm{pow}}[\phi] \doteq |\phi|^2\phi
\end{equation}
(or more generally $\mathcal{N}^{(p)}_{\mathrm{pow}}[\phi]= |\phi|^{p-1}\phi$ for $p \geq 3$). As it turns out, however, these nonlinearities have additional structure that does not allow the proof strategy for \cref{thm:Main theorem} to apply. 

While the quasilinear term $\mathcal{N}^{(3)}[\phi]$ in \eqref{eq:nonlinearity} and the semilinear term $\mathcal{N}^{(3)}_{\mathrm{pow}}[\phi] = |\phi|^2\phi$ both allow for the same set of temporal resonances (arithmetic progressions of frequencies) and the same reduction of the system to the system of resonant interactions of dominant modes (with all error terms controlled in exactly the same way), the corresponding resonant systems differ fundamentally in the structure of the \emph{interaction coefficients}.

 For a resonant triad of modes, the effective dynamics of the amplitudes are governed by a system of ODEs determined by these coefficients.
In the case of the semilinear power nonlinearity $\mathcal{N}^{(3)}_{\mathrm{pow}}[\phi]$, the symmetries of the interaction coefficients lead to a system that is linearly stable. Specifically, the energy flows between modes in an oscillatory manner rather than exhibiting the exponential growth required for a forward energy cascade on the relevant timescale. In contrast, the quasilinear structure of \eqref{eq:nonlinearity}, involving derivatives of $\phi$, breaks this symmetry. This results in interaction coefficients that allow for linear instability, creating a mechanism where energy is consistently pumped from low to high frequencies rather than merely exchanged. See already \cref{rem:intro-pure-power} for a more detailed discussion of this point.

\begin{remark}
As will become clear from the outline of the proof in \cref{sec:outline-proof-main-theorem}, the strategy of our proof carries over to the case when $\mathcal N^{(3)}[\phi]$ is replaced by any cubic nonlinearity almost verbatim, with the \emph{exception} of the analysis of the resonant $3\times 3$ system. The coefficients of that system (which can be viewed as spatial overlap integrals of the dominant mode eigenfunctions with weights determined by the nonlinearity) are sensitive to the precise choice of the nonlinear terms. For instance, dropping the $r^{-6}$ coefficient from $\mathcal N^{(3)}[\phi]$ results in a dominant system which does not exhibit norm inflation in the required timescale. See \cref{rem:intro-spatial-resonance}.
\end{remark}

\subsection{Weak turbulence and energy transfer in nonlinear dispersive  PDE}\label{sec:Weak Turbulence Intro}
 
The emergence of energy cascades in the evolution of weakly interacting nonlinear waves has been the object of extensive studies in the fields of water waves, plasma physics and nonlinear optics, beginning already in the '60s (see e.g.~\cite{H62,Zakharov65,Z68}). During that decade, it was heuristically understood that, for solutions to a wide class of Hamiltonian PDEs, the transfer of energy between different scales gives rise to dynamics resembling those associated to hydrodynamic \emph{turbulence}. In contrast to the dynamics of fluids, however, in the case of dispersive PDEs with weak nonlinearities, the energy flow is mediated by resonant nonlinear interactions among modes evolving in an approximately \emph{linear} fashion at short timescales, inspiring the name \emph{weak} or \emph{wave} turbulence for the theory describing the associated phenomena (see \cite{N11}). 

For a given nonlinear dispersive PDE, the length-scale $L$ associated with the spatial domain and the size of the initial data determine an effective timescale $T_{\mathrm{eff}}$ of interaction, at which the nonlinear nature of the equations dominates the dynamics. Using the scaling properties of the underlying equations (in the case of a homogeneous nonlinearity), one may transform the study of high-frequency interactions on domains with $L\sim 1$ to interactions of bounded frequencies in the large-box limit $L\rightarrow+\infty$.
In this limit, different asymptotic relations between the domain size $L$ and the timescale $T_{\mathrm{eff}}$ (or, equivalently, the size of the rescaled initial data) are expected to yield different asymptotic dynamics. In the case when the size of the initial data is not ``too'' small, the gaps between the time frequencies of the modes (these gaps vanish in the limit $L\rightarrow \infty$) are narrow compared to  $T^{-1}_{\mathrm{eff}}$ and the linear spectrum can be considered to be effectively continuous. In this regime of so-called \emph{kinetic} wave turbulence (see \cite{N11}), well-prepared \emph{randomized} initial data formally give rise to a nonlinear \emph{kinetic} equation governing the evolution of the (averaged) energy profile in phase space (\cite{GK63,ZF67,Zakharov65,Z68,BS66}, with a similar kinetic description appearing already in the study of phonons in anharmonic crystals by Peierls \cite{P29}). Stationary solutions of these kinetic models are characterized by energy spectra which qualitatively resemble those observed in the turbulent motion of a classical fluid (see \cite{ZF67,ZLF92}). On the other hand, in the case when the spectral gaps are large compared to $T^{-1}_{\mathrm{eff}}$ (i.e.~for sufficiently small initial data), the dynamics are dominated by exact resonances and the evolution is not expected to admit an effective kinetic description; this is sometimes referred to as the regime of \emph{discrete} wave turbulence (see \cite{N11,KNR08,K10} and references therein) and is the setting relevant to \cref{thm:Main theorem}.\footnote{The intermediate regime, in which the spectral gaps are at the same order as $T^{-1}_{\mathrm{eff}}$, is referred to as the regime of \emph{mesoscopic} turbulence \cite{N11,K10}.}
 We will not provide a representative survey of the (vast) literature on weak turbulence theory, but instead refer the reader to \cite{ZLF92,DNPZ92,MMcLT97,NNB01,N11}.

A rigorous derivation of the general picture painted above remains elusive to this day.  However, questions on the emergence of energy cascades in the dynamics of nonlinear dispersive PDEs have attracted considerable attention from the mathematical community in the past three decades, with Bourgain's expository work \cite{B00} being particularly influential in that direction. There, Bourgain posed the question of whether the cubic defocusing nonlinear Schr\"odinger equation 
\begin{equation}\label{NLS Intro}
i\partial_t u - \Delta u = -|u|^2 u
\end{equation}
on the torus $\mathbb T^d$ with $d\ge 2$ admits, for any Sobolev index $s>1$, solutions with $\|u[0]\|_{H^s}\ll 1$ and such that  $\|u[t]\|_{H^s}\xrightarrow{t\rightarrow+\infty}+\infty$, signifying an infinite energy cascade as $t\rightarrow +\infty$.
This question still remains open; however, in a major breakthrough,  Colliander, Keel, Staffilani,  Takaoka and Tao \cite{CKSTT10} constructed, for any $s>1$ and $\epsilon>0$, solutions to \eqref{NLS Intro} on $\mathbb{T}^2$ with $\|u[0]\|_{H^s} <\epsilon$ and such that $\|u[T]\|_{H^s}>\f1\epsilon$ at some time $T>0$.  This was the first proof of a direct energy transfer to arbitrarily large frequencies for small amplitude nonlinear waves in any setting and sparked a burst of activity on related questions: \cite{G14,HP15} obtained similar results for modifications of \eqref{NLS Intro}, while \cite{GK14} constructed solutions to \eqref{NLS Intro} undergoing norm inflation with better control of the timescale of growth (for data with small energy but relatively large initial $H^s$ norm); \cite{GG10}  obtained results analogous to \cite{CKSTT10} for the completely integrable cubic Szeg\"o equation in $1d$; \cite{P13} established the amplification of the $H^s$ norm of solutions to a $1d$ nonlinear wave equation by an arbitrary factor (with the $H^s$ norm of the solution, however, remaining $\ll 1$ on the whole time interval). In the case of a non-compact domain, \cite{ZPTV15} answered in the affirmative Bourgain's question when $\mathbb T^2$ is replaced by  $\mathbb R\times \mathbb T^2$; in that case, the slow decay at the linearized level due to the dispersive direction helped to isolate the resonant dynamics from the rest of the evolution, allowing for a global approximation of the evolution by the resonant dynamics (see also \cite{P11,BE22,GL24}). We should also mention Bourgain's earlier work \cite{B96} where, for a carefully spectrally designed nonlinear wave equation on the real line, he showed polynomial growth in time for higher order Sobolev norms.

\begin{remark}
The question proposed by Bourgain (as well as all the results mentioned above and the results of the current work) fall into the setting of discrete wave turbulence discussed earlier, which has been less extensively studied in the physics literature. On the other hand, in the regime of kinetic wave turbulence, the past few years have witnessed remarkable advances towards a rigorous justification of the wave kinetic theory: In the recent breakthrough \cite{DH23,DH24} (see also \cite{BGHS21,ST21}), Deng and Hani derived the wave kinetic equation as an effective statistical description (in the limit $L\rightarrow +\infty$) of the dynamics of the NLS equation \eqref{NLS Intro} on the torus $\mathbb T^d_L$, $d\ge 3$,  for the full range of scaling laws between the box dimension $L$ and the strength of the nonlinear interactions at which the theory of continuous wave turbulence is expected to be valid:  In terms of the timescale $T_{\mathrm{eff}}\sim L^{2\gamma}$, this range corresponds to  $\gamma\in (0,1]$. Note, however, that the initial data regime where the kinetic wave theory is expected to describe the dynamics of \eqref{NLS Intro} is strictly disjoint from that associated to Bourgain's question: Upon rescaling back to a torus of size $\sim 1$ (setting $u(x,t)\rightarrow L u(Lx, L^2t)$), the initial data associated to this range of scaling laws have size $\gtrsim 1$ and the interactions take place at timescales $\lesssim 1$.
\end{remark}

Our main result (namely \cref{thm:Main theorem}) should be viewed as the analog of the result of \cite{CKSTT10} in the setting of the quasilinear wave equation \eqref{Initial Boundary Value Problem Phi} on Schwarzschild--AdS spacetime; \cref{thm:Main theorem-S3} provides the analogous statement on  $\mathbb R\times \mathbb S^3_+$ equipped with the family of metrics $-dt^2+h^M_{\mathbb S^3_+}$. Some of the key differences between our setting and that of earlier works on (discrete) weakly turbulent dynamics can be summarized as follows:

\begin{enumerate}
		\item \emph{Quasilinear nature of the nonlinearity.} In contrast to the setting of previous works (with the exception of the very recent work \cite{LMMT26}), the nonlinear equation that we study, namely \eqref{Initial Boundary Value Problem Phi}, is  \emph{quasilinear} rather than semilinear. The quasilinear nature of \eqref{Initial Boundary Value Problem Phi} features prominently in the analysis of the error terms in \cref{sec:Estimates error term}.  Interestingly, our proof strategy does not provide Sobolev norm growth  in the case of semilinear power nonlinearities of the type $\mathcal N_{\mathrm{pow}} [\phi]= |\phi|^{p-1} \phi$, see \cref{sec:obstruction-energy-transfer} for a more detailed discussion.  

\item \emph{Equation of wave type.} Equation \eqref{Initial Boundary Value Problem Phi} is a $2^{nd}$ order wave-type equation rather than a Schr\"odinger-type one. As far as we know, this is the first higher order norm inflation result for such an equation in spatial dimension higher than $1$. In the case of a Schr\"odinger-type equation, 4-term (nontrivial) resonant interactions can only transfer energy from one scale to another one of comparable size; hence energy transferred from low to high frequencies has to gradually traverse the whole intermediate frequency spectrum. On the contrary, for a wave-type operator, 4-term resonant interactions can involve frequencies of arbitrary ratios (we exploit this fact in our selection of the resonant modes, see \cref{sec:Hierarchy of parameters}).\footnote{This non-local nature (in terms of scales) of the energy transfer associated to wave-type equations, as opposed to other dispersive PDEs, is a well-known fact, see already \cite{Zakharov65}.}

	\item \emph{Curved (non-compact) domain and stable trapping.} 
Another novel aspect of the setting of our work is that the underlying geometry (namely Schwarzschild--AdS spacetime) is curved, being, in particular, non-compact in the spatial directions; in the region where the interactions are localized the longest in time, namely near $\mathcal I$, the spatial geometry resembles the near-equatorial geometry of the round half sphere $\mathbb S^3_+$ (see \cref{sec:Proof of second theorem} for more details on the relation between the asymptotically AdS geometries and that of $\mathbb R\times \mathbb S^3_+$). The latter fact is reflected in the existence of stably trapped null geodesics near the conformal boundary $\mathcal I$. In view of the non-compactness in the spatial directions of the domain (and, in particular, the fact that linear solutions decay by gradually dispersing through the event horizon), we cannot use normal mode linear solutions to obtain an ansatz for a solution to \eqref{Initial Boundary Value Problem Phi}. However, in view of the stable trapping near $\mathcal I$, we can construct well-localized \emph{quasi-modes} instead. 

Unlike previous works (which considered a flat background), the curvature of the underlying geometry makes keeping track of the spatial interactions of those quasi-modes more challenging. For instance, in the case of the flat torus $\mathbb T^d$, where the spatial profile of linear modes for the wave (or Schr\"odinger) equation is given by $E_k = e^{i k \cdot x}$ ($k\in \mathbb Z^d$, the orthogonal projection $\mathbb P_k(E_{k_1} \bar E_{k_2} E_{k_3})$ onto $E_k$ is nontrivial only when $k=k_1-k_2+k_3$; there is no similar statement in the case of a curved geometry. Instead, for the evaluation of such projections of quasimodes in our setting, we have to rely on careful estimates on the large frequency asymptotics of the corresponding eigenfunctions.  Moreover, the spatial profiles of the linear quasimodes that remain localized near $\mathcal I$ do not form a complete basis in $L^2$; therefore, any decomposition of a solution to \eqref{Initial Boundary Value Problem Phi} into a sum of such modes with slowly evolving amplitudes (such as \eqref{Ansatz phi Intro}) must necessarily involve a non-trivial error term, and controlling that term requires taking into consideration the global geometry of the spacetime.

	\item  \emph{Supercriticality.} The nonlinear wave equation  \eqref{Initial Boundary Value Problem Phi} has the same scaling properties as the Einstein equations and, in particular, is supercritical with respect to the (conserved) $H^1$ energy \eqref{Conserved energy intro}. Therefore, the existence of the solution up until the time $t=T_1$ cannot be obtained as in the case of \eqref{NLS Intro} on $\mathbb T^2$. In order to obtain this long-time existence result, we have to carefully exploit a certain quantitative non-resonant structure in the spectrum as well as energy estimates associated to a novel \emph{helical vector field} adapted to our construction. Note that establishing that all lower order norms remain small in this time interval (as part of the proof of the existence of a solution) is crucial in distinguishing an energy cascade statement from a blow-up scenario, in which lower order norms are also inflated.

\end{enumerate}

We should remark that, as in the case of \cite{CKSTT10},  \cref{thm:Main theorem} does not provide information about the behavior of Sobolev norms beyond the time $t=T_1$. In particular, in both settings, Bourgain's question remains open. Unlike \cite{ZPTV15} where, on $\mathbb R\times \mathbb T^2$, solutions to the linear equation decay at a $t^{-\f12}$ rate, in our setting the dispersion due to the event horizon is too weak (leading to a merely logarithmic decay rate at the linearized level); thus, controlling the difference between the resonant dynamics and the true solution for longer timescale is a much more challenging problem. 
In the energy-supercritical setting of \eqref{Initial Boundary Value Problem Phi} studied here, there is also the possibility of a  finite-time singularity formation in the evolution of smooth initial data. This motivates the following  question concerning the ultimate fate of the solutions constructed in \cref{thm:Main theorem}:
\begin{quote}
Do the solutions constructed in \cref{thm:Main theorem} exist for all time or do they form a singularity at some finite time $T_{\mathrm{blowup}}>T_1$? If the former is true, do higher order Sobolev norms blow up asymptotically as $t^\ast\to\infty$?
\end{quote}

 \subsection{The stability problem for the Schwarzschild--AdS exterior spacetime}\label{sec:Stability Schw-AdS}
The question of whether the Einstein vacuum equations \eqref{eq:EVE-Lambda} (which our scalar model \eqref{Initial Boundary Value Problem Phi} aims to simulate) exhibit weakly turbulent dynamics on the exterior of a Schwarzschild--AdS black hole is intimately connected to  the problem of nonlinear stability of those spacetimes as solutions to \eqref{eq:EVE-Lambda}. Let us briefly review the current status of this problem and explain how our result fits into this context.

\label{sec:intro-stability}
\subsubsection{Instability of Anti-de\,Sitter spacetime}
The maximally symmetric solution of the vacuum Einstein equations \eqref{eq:EVE-Lambda} with negative cosmological constant $\Lambda=-3$ is \emph{Anti-de\,Sitter} spacetime $(\mathbb R^{3+1}, g_{\mathrm{AdS}})$, with $g_{\mathrm{AdS}}$ given by \eqref{AdS metric Intro} (which is formally obtained from  \eqref{eq:metric-schwarzschild-ads} for $M=0$). This spacetime is conformal to the half cylinder $(\mathbb R\times \mathbb S^3_+, -dt^2+g_{\mathbb S^3_+})$. Therefore, imposing \emph{reflecting} boundary conditions at conformal infinity $\mathcal I = \mathbb R \times \mathbb S^3_+$ (see \cref{sec:A toy model} for a brief discussion about such boundary conditions) introduces confinement in the evolution of perturbations of $g_{AdS}$, with the nonlinear nature of \eqref{eq:EVE-Lambda} raising the prospect of an energy cascade emerging in this setting.

In 2006, Dafermos--Holzegel \cite{DH06} proposed the following conjecture:
\begin{quote}
\textbf{The AdS instability conjecture.} There exist arbitrarily small perturbations of Anti-de\,Sitter spacetime solving \eqref{eq:EVE-Lambda} with reflecting boundary conditions at $\mathcal I$ which, after sufficiently long time, lead to the emergence of a trapped surface (signifying, in particular, the creation of a black hole). Thus, AdS spacetime is nonlinearly unstable.
\end{quote}
The picture proposed by the above conjecture can be viewed as a geometric statement of an energy cascade, since, in light of Thorne's hoop conjecture \cite{T73}, the formation of a trapped surface is expected to be possible only when sufficient ``energy'' is concentrated at a small enough scale.

A plethora of numerical and heuristic works have been devoted to the study of the conjecture in the class of \emph{spherically symmetric} solutions of the Einstein equations coupled to matter fields \cite{DHS12Instability,BLL12,DHS12,DFLY15,MR13,BBGLL14,GMLL15,DY15}. This flurry of activity was initiated by the seminal work of Bizon--Rostworowski \cite{BR11}, where the authors studied spherically symmetric perturbations of AdS for the Einstein--Klein-Gordon system
\begin{equation}\label{Einstein--scalar field Intro}
\begin{cases}
\mathrm{Ric}_{\mu\nu}-\f12\mathrm{R} g_{\mu\nu}-3g_{\mu\nu}=8\pi \big(\partial_\mu\phi \partial_\nu\phi-\f12 \partial^\alpha\phi\partial_\alpha\phi g_{\mu\nu}\big),\\[5pt]
\square_g \phi+\alpha\phi=0,
\end{cases}
\end{equation}
 with $\alpha=0$ and with Dirichlet conditions at $\mathcal I$. Noting that the spectrum of the linear operator is a rescaling of $\mathbb Z$ (and hence contains infinitely long arithmetic progressions of temporal frequencies), they conjectured that the instability is driven by a mechanism of resonant interactions of spherically symmetric modes, leading to weakly turbulent dynamics, at least at the early stages of the instability. That mechanism was further explored in \cite{CEV14,CEV15,BMR15}. At the same time, the existence of special perturbations of AdS leading to quasiperiodic solutions (not collapsing, in particular, into black holes) has been investigated both numerically and heuristically \cite{MR13,BBGLL14,GMLL15,DY15}. A rigorous construction of periodic small amplitude solutions for nonlinear models of \eqref{Einstein--scalar field Intro} was carried out in \cite{CS24,CS25}, with their stability at exponential timescales being explored in \cite{CS23}. For numerical and heuristic works outside spherical symmetry, see \cite{HS15,DS16,R17}.

While the AdS instability conjecture remains open in the case of the vacuum equations, it has been shown to hold in the case of Einstein--null dust and Einstein--massless Vlasov systems in spherical symmetry, at least within classes of initial data of relatively low regularity \cite{Mos20,Mos23}.  However, a rigorous construction of perturbations which are small in a high regularity class and exhibit the weakly turbulent behavior predicted by Bizon--Rostworowski in any matter model is still open.

	\subsubsection{The case of Schwarzschild--AdS: Quasimodes and resonant interactions}\label{sec:Schwarzschild--AdS}
	\label{sec:stability-problem-sads}
Determining the ultimate fate of the black hole spacetimes postulated by the AdS instability conjecture---deciding, in particular, whether these solutions asymptotically settle down to \emph{stationary} black holes---appears to be at present a highly intractable problem. Understanding the dynamics of these spacetimes would involve at first understanding the global nonlinear stability properties of stationary asymptotically AdS black holes, such as the Schwarzschild--AdS exterior spacetime $(\mathcal M_{\mathrm{ext}}^{(M)},g_M)$.

In the case of  the \emph{spherically symmetric} Einstein--Klein-Gordon system \eqref{Einstein--scalar field Intro} with Dirichlet boundary conditions at $\mathcal I$, it was shown by Holzegel--Smulevici \cite{HS13-stab} that perturbations of Schwarzschild--AdS decay to $0$ at an exponential rate (under the technical restriction $\alpha\ge 2$ in \eqref{Einstein--scalar field Intro}). Such a strong decay rate is consistent with the fact that radial null geodesics (along which high-frequency spherically symmetric perturbations are expected to propagate) fall through the black hole horizon after at most one reflection off conformal infinity $\mathcal I$. 

Outside spherical symmetry, however, the stability problem becomes more delicate, owing to the presence  on $(\mathcal M_{\mathrm{ext}}^{(M)},g_M)$ of \emph{stably trapped} null geodesics that are reflected infinitely many times off $\mathcal I$ but never approach the horizon. These null geodesics have the capacity of supporting long-lived perturbations, as reflected on the existence of solutions $\phi$ to  the \emph{linear} Klein--Gordon equation with Dirichlet boundary conditions
\begin{equation}\label{Linear wave equation Introduction}
\begin{cases}
\square_{g_M} \phi +2\phi =0,\\
r\phi|_{r=\infty}=0
\end{cases}
\end{equation}
which remain localized near $r=\infty$ for a time interval which is exponentially long compared to the size of their frequency support. More precisely, Holzegel--Smulevici \cite{HS14} constructed a family of  \emph{quasimodes} of the form
\begin{equation}\label{Separated form quasimode Intro}
\mathring{\phi}_{(n,\ell,m)} = e^{\pm i \omega_{n,\ell} t} \frac{R_{n,\ell}(r)}{r}  Y_{\ell,m}(\theta,\varphi), \text{ for }  n\in \mathbb N, m \in \mathbb Z, \ell \in \mathbb N_{\geq |m|}
\end{equation}
(with $Y_{\ell,m}$ being a spherical harmonic of order $(\ell,m)$ on $\mathbb S^2$), i.e.~approximate solutions of \eqref{Linear wave equation Introduction} satisfying the following properties in the $\ell\gg 1$ regime (setting $k=(n,\ell,m)$ for simplicity): 
\begin{enumerate}
\item $\omega_{n,\ell} \in \mathbb R_{\ge 0}$ (hence $\mathring\phi_k$ is periodic in time) with $\omega_{\ell,n}\rightarrow +\infty $ as $\ell\rightarrow \infty$,
\item $\mathring\phi_{k}$ solves \eqref{Linear wave equation Introduction} with an $O\big(e^{-c_0 \ell}\|\mathring \phi_{k}\|_{L^\infty}\big)$ error term on the right-hand side,
\item $\mathring\phi_k$ is localized near $r=\infty$ in the sense that it satisfies in the region away from  $r=\infty$ the quantitative ``smallness'' bound
\begin{equation}\label{Smallness bound phi k intro}
 \big\|\partial \mathring \phi_k \big\|^2_{L^\infty(\{r\ll \ell^{\f12}\})}  \lesssim e^{-c_1 \ell} \mathcal E[\mathring \phi_k](0).
\end{equation}
\end{enumerate}
Note that Property 3 above can be viewed as an $O(e^{-c_1\ell})$-approximate Dirichlet condition for $\mathring{\phi}_k$ at $r=r_0$ for any fixed $r_0\ll \ell^{\f12}$; thus, the frequency $\omega_{n,\ell}$ satisfies
\begin{equation}\label{Approximate Sturm Liouville eigenvalues}
\omega_{n,\ell} = \omega_{n,\ell}^{(\mathrm{SL})}+O(e^{-c_1\ell}),
\end{equation}
where $\omega_{n,\ell}^{(\mathrm{SL})}$ is the Sturm--Liouville eigenvalue associated to the boundary value problem for $R_{n,\ell}$ implied by \eqref{Linear wave equation Introduction} after imposing an exact Dirichlet boundary condition at $r=r_0$. For fixed $\ell$, this is a discrete family of eigenvalues parametrized by the overtone $n$, so that $\omega_{n,\ell}$ is increasing in $n$; the corresponding Sturm--Liouville eigenfunction is localized at $r\gtrsim \ell^{\f12}$ (and thus is close to the radial profile of a quasimode with the properties above) if and only if $n\ll \ell$, in which case one also has
$\omega_{n,\ell}^{(\mathrm{SL})}\sim \ell$ as $\ell\rightarrow +\infty$
(see \cref{sec:Modes near boundary Intro} for more details). 

It  follows via a standard energy estimate that, for each quasimode $\mathring\phi_k$ as above with $\ell\gg 1$ and for $0\le t \lesssim e^{c' \ell}$, $c'<c_0$, the exact solution $\phi_k$ of \eqref{Linear wave equation Introduction} with initial data $(\phi_k, \partial_t \phi_k)|_{t=0} = (\mathring\phi_k, \partial_t \mathring\phi_k)|_{t=0}$ remains $O\big(e^{-c'' \ell}\|\mathring \phi_k\|_{L^\infty}\big)$-close to $\mathring\phi_k$ for some $0<c''<c_0-c'$. 
Consequently, the existence of such long-lived quasimodes implies a slow \emph{logarithmic} lower bound for the decay rate of generic solutions to  \eqref{Linear wave equation Introduction}: As shown in \cite{HS14}, a simple optimization argument in the frequency $\ell$  using the properties 1--3 above and the approximation of $\phi_k$ by $\mathring\phi_k$ for $t\lesssim  \exp{c \ell}$ implies that, for any $s\ge 1$:
\[
\sup\left\{\limsup_{\tau \rightarrow +\infty} \left((\log(\tau))^{s-1}\f{\|\partial \phi|_{t^*=\tau}\|_{L^{2\hphantom{-1}}}}{\|\partial \phi|_{t^*=0}\|_{H^{s-1}} }\right) \quad \Big| \quad \phi \text{ is a smooth solution of \eqref{Linear wave equation Introduction}}\, \right\}>0.
\]
A logarithmic \emph{upper} bound for the decay of solutions to  \eqref{Linear wave equation Introduction}, namely
\[
\|\partial \phi|_{t^*=\tau}\|_{H^1} \lesssim (\log(\tau))^{1-s} \|\partial \phi|_{t^*=0}\|_{H^{s-1}},
\]
was established in \cite{HS13}, together with the more refined angular frequency-projected bound
\begin{equation}\label{eq:exponential-decay-angular-mode}
\|\mathbb P_{\ell} \phi|_{t^*=\tau}\| \lesssim  \exp\left(-c_1 e^{-c_2\ell} \tau \right) \|\partial \phi|_{t^*=0}\|_{L^2}
\end{equation}
(for some constants $c_1, c_2>0$), thus completely characterizing the long time asymptotics of generic solutions to    \eqref{Linear wave equation Introduction} on Schwarzschild--AdS. These results were generalized to the full system of the linearized Einstein vacuum equations in the recent \cite{GH24I,GH24III}. 

\begin{remark}
An alternative approach to establish a logarithmic lower bound for the decay rate of generic solutions to \eqref{Linear wave equation Introduction} is by studying, in place of quasimodes, families of \emph{quasinormal} modes, namely functions of the form \eqref{Separated form quasimode Intro} which are \emph{exact} solutions of \eqref{Linear wave equation Introduction} with infalling boundary conditions on the horizon and with a radial profile concentrated in the region $r\gg 1$. In this case, the quasinormal mode frequency is necessarily complex valued, reflecting the dispersion through the event horizon. Quasinormal modes on Schwarzschild--AdS have been constructed by Gannot \cite{Gannot14}; their frequencies satisfy \[
\left|\mathrm{Im}(\omega_{n,\ell})\right|\lesssim e^{-c\ell} \quad \text{and} \quad\left|\mathrm{Re}(\omega_{n,\ell})-\omega_{n,\ell}^{(\mathrm{SL})}\right|\lesssim e^{-c\ell}
\]
 for some $c>0$, i.e.~they are exponentially close to the (real) frequencies of the quasimodes.
\end{remark}

The slow decay rate of generic solutions $\phi$ to the linear problem \eqref{Linear wave equation Introduction} on Schwarzschild--AdS (and, more generally, on the Kerr--AdS black hole exteriors with parameters satisfying the Hawking--Reall bound) and, in particular, the exponentially long (in terms of their frequency) lifespan of high-frequency linear quasimodes, led Holzegel and Smulevici \cite{HS13} to propose the following nonlinear instability conjecture:

\begin{quote}
\textbf{Instability of Kerr--AdS conjecture} (Section 1.4 of \cite{HS13}). The family of Kerr--AdS spacetimes is nonlinearly unstable for generic perturbations.
\end{quote}

Any attempt of proving (or disproving) the above conjecture should necessarily involve an investigation of the resonant nature of the nonlinear interactions of the long-lived linear quasimodes. Fixing, for definiteness, a model nonlinear equation with a nonlinearity of \emph{cubic} type (such as \eqref{Initial Boundary Value Problem Phi}, for instance), a quartet of quasimodes will be said to be in temporal resonance if their combined interaction through the nonlinearity is not out of phase, i.e.
\begin{equation}\label{eq:resonance-condition-intro}
		\pm	\omega_{n_1,\ell_1}  \pm	\omega_{n_2,\ell_2}  \pm	\omega_{n_3,\ell_3}  \pm	\omega_{n_4,\ell_4}  \approx 0
	\end{equation}
(with the size of the allowed error on the right hand side being smaller than the inverse of the effective timescale of interaction). The existence of quartets of quasimodes of arbitrarily large frequencies which are temporally (but also spatially!) resonant opens the gate for a nonlinear transfer of energy between the corresponding frequency scales and can thus be a first indication of nonlinear instability (though in itself is not a guarantee of an instability, since for energy transfer to occur the resonant system must in addition exhibit an appropriate instability; see \cref{rem:intro-pure-power}).

The study of the resonant properties of the spectrum of \eqref{Linear wave equation Introduction} on a broad class of stationary and asymptotically AdS spacetimes (including Schwarzschild--AdS)  was initiated in \cite{DHS12}. There, the authors derived the following asymptotic expression for the long-lived-quasimode frequencies $\omega_{n,\ell}$ (more precisely, the real parts of the quasinormal mode frequencies discussed above) on Schwarzschild--AdS:
\begin{equation}\label{Asymptotic expansion HDS}
\omega_{n,\ell} = \omega^{\mathrm{(AdS)}}_{n,\ell}+ \f{C(n,M)}{\ell^{\f12}} +O_n(\ell^{-1}),
\end{equation}
where $\omega^{\mathrm{(AdS)}}_{n,\ell} = 2n+\ell$ are the frequencies of the corresponding \emph{normal} modes on pure AdS spacetime and the constant $C(n,M)$ depends only on the overtone $n$ and the Schwarzschild--AdS mass $M$ and is non-zero when $M>0$. A similar formula was also deduced for more general asymptotically AdS backgrounds. As a consequence of \eqref{Asymptotic expansion HDS}, for  the lowest lying modes, namely the ones satisfying $n\lesssim 1$ (these are the ones localized closest to conformal infinity and having the longest lifespan), the cubic resonant condition \eqref{eq:resonance-condition-intro} can never be satisfied if $\ell \gg 1$ (and the same is true for the resonant condition associated to a nonlinearity of any fixed homogeneous order); these quasimodes are quantitatively \emph{non-resonant}. The absence of the resonant energy transfer mechanism for the class of lowest-lying quasimodes, as well as the fact that such linear quasimodes have an exponentially long but nevertheless finite timescale of nonlinear interaction, motivated \cite{DHS12} to suggest that, contrary to the conjecture of \cite{HS13} stated earlier, the Kerr--AdS family might in fact be nonlinearly \emph{stable}.

	The main result of the present paper (namely \cref{thm:Main theorem}) points in the opposite direction: Our proof of an instability for the nonlinear equation \eqref{Initial Boundary Value Problem Phi}  is built upon the  observation that, once all modes with radial overtone $n\gg 1$ are taken into account, then a resonant structure emerges in the linear spectrum, at least for a generic set of mass parameters $M>0$. This is a consequence, in particular, of the refinement \eqref{Taylor expansion omega intro} of the formula \eqref{Asymptotic expansion HDS}. See \cref{sec:outline-proof-main-theorem} below for more details.

  \subsubsection{Weakly turbulent instability for the Einstein vacuum equations}
Based on \cref{thm:Main theorem} and the discussion above (as well as the fact that \eqref{Initial Boundary Value Problem Phi} was selected as a model for the Einstein vacuum equations, see \cref{sec:A toy model}), it is natural to conjecture that an analogous instability of weakly
turbulent nature should occur for the full Einstein vacuum equations on the Schwarzschild--AdS
exterior with reflecting boundary conditions at $\mathcal I$. In particular, a ``maximalistic'' version of this expectation could be formulated as follows (refining the statement of the instability conjecture of \cite{HS13}):

\begin{conjecture}[Nonlinear instability of Schwarzschild--AdS]\label{con:Trapped surface}
For any $s\in \mathbb N$,  there exists an open and dense set $J^{\mathrm{EVE}}_s \subset (0,+\infty)$ of
black hole masses $M$ with the following property: There exists a one-parameter family of asymptotically AdS initial data $(\gamma_\epsilon,K_\epsilon)$ which converge in the $H^s$ topology to the Schwarzschild--AdS initial data $(\gamma_0, K_0)$ of mass $M$ such that their evolution under the vacuum Einstein equations with reflecting boundary conditions at $\mathcal I$ leads to the creation of a trapped surface in the region
far from the background event horizon.
\end{conjecture}

In light of the hoop conjecture, the emergence of a trapped surface should be interpreted as an indication of the growth of the scale invariant size of the solution. In the case of the model problem \eqref{Initial Boundary Value Problem Phi} studied in \cref{thm:Main theorem}, such a statement would correspond to the growth of the $H^{\f32}$ norm of a solution $\phi$ emanating from data which are initially small in $H^s$, $s\gg 1$. At present, such a result seems to be out of reach for the techniques developed in this work. A weaker conjecture, which is more in line with the statement of  \cref{thm:Main theorem}, would be the following:

\begin{conjecture}[Norm inflation for the Einstein vacuum equations on Schwarzschild--AdS] \label{conj:weak-turbulence-eve}
There exists an integer $s_0 \in \mathbb N$ such that for every $s \geq s_0$ the following holds.
There exists an open and dense set $J^{\mathrm{EVE}}_s \subset (0,+\infty)$ of
black hole masses with the following property: for every $M \in J^{\mathrm{EVE}}_s$ and every 
$0<\epsilon\ll1$, there exist smooth, asymptotically AdS initial data $(\gamma_\epsilon, K_\epsilon)$
for the Einstein vacuum equations \eqref{eq:EVE-Lambda} on a Cauchy hypersurface $\Sigma$, with reflecting
boundary conditions at $\mathcal I$, such that
\[
\|(\gamma_\epsilon,K_\epsilon) - (\gamma_0,K_0)\|_{H^s(\Sigma)\times H^{s-1}(\Sigma)} \leq \epsilon,
\]
where $(\gamma_0,K_0)$ denote the Schwarzschild--AdS data of mass $M$ and $H^s$ is a ``geometrically'' defined norm, and with the following
properties:

\begin{enumerate}
\item In wave coordinates, the corresponding maximal Cauchy development $(\mathcal M,g)$ of $(\gamma_\epsilon,K_\epsilon)$ exists at
      least up to some time $T_1>0$.
\item On the time interval $[0,T_1]$, the metric remains $C^2$-close to that of Schwarzschild--AdS but exhibits higher order norm inflation in the
      sense that the deviation from Schwarzschild--AdS obeys
      \[
      \sup_{0\leq t\leq T_1}
      \|(g(t) - g_M, K(t) - K_M)\|_{H^s(\Sigma_t)\times H^{s-1}(\Sigma_t)}
      \;\geq\; \f1\epsilon.
      \]
\end{enumerate}
\end{conjecture}

\begin{remark}
In the numerical work \cite{FR25} on scalar perturbations of very slowly rotating Kerr--AdS, the authors also observed a nonlinear instability associated to mode interactions. The perturbations in \cite{FR25} are however of a different nature and are not  ``small'' in the sense of \cref{thm:Main theorem}. In particular, our weak turbulence mechanism in \cref{thm:Main theorem} occurs at much later timescales than those instabilities observed in \cite{FR25}.
\end{remark}
\begin{remark}
Another class of spacetimes exhibiting stable trapping are ultracompact neutron stars: horizonless solutions possessing a photon sphere. In \cite{K16} it was shown that linear waves on such backgrounds decay only inverse-logarithmically. Recent numerical studies \cite{BCPS25,RYCA25} further indicate that this slow decay indeed gives rise to an instability of turbulent nature. We expect that the mechanism and results of the present paper likewise apply to quasilinear wave equations on such ultracompact objects; see also the recent review \cite{P25} and references therein. 
\end{remark}

\begin{remark}
In \cite{BCPS25}, the authors study numerically the cubic nonlinear wave equations on spacetimes with stable trapping and propose the following scenario for the late-time dynamics of ``generic'' gravitational perturbations: While higher order Sobolev norms are inflated as a result of an energy cascade, the energy of the perturbation should spread out in a homogeneous fashion among the stably trapped modes. As a consequence of this scenario, energy should not accumulate in a region of small diameter, and hence (in view of the hoop conjecture) trapped surface formation in that region should not be possible for such ``generic'' perturbations. However, the proof of \cref{thm:Main theorem} seems to be at odds with this picture: In the time interval $t\in[0, T_1]$ where the energy transfer takes place, the solution $\phi$ remains localized mostly at spherical harmonics with $\ell=|m|$, and the same is true for an \emph{open} set of initial data around the given ones. This is a consequence of the fact that the vast majority of the frequencies in the linear spectrum are quantitatively non-resonant with respect to the initially excited frequency triad, and hence homogenization of the solution cannot be achieved at the same timescale at which energy is transferred through the resonant interactions. As a result, we believe that, in the statement of \cref{con:Trapped surface}, the space of initial data leading to the creation of a trapped surface in the far away region should contain an open neighborhood of the stated one-parameter family.\footnote{Note, however, that, in view of the hoop conjecture,  exciting the $\ell=|m|$ modes, as we do in the proof of \cref{thm:Main theorem}, might not be enough to create a trapped surface, since for such modes the energy is spread out along the equator and is thus not contained in a set of small spatial diameter.}
\end{remark}

    \subsubsection{Stability of Kerr--AdS for analytic perturbations}

In our main result \cref{thm:Main theorem}, we measure regularity in Sobolev norms and show growth of
these norms as a consequence of a forward energy cascade. A natural question is whether there exists a regularity threshold beyond which such an energy transfer ceases to exist; a natural class of data to search for such a threshold is that of \emph{real-analytic initial data}, as suggested already by the late time dynamics of analytic solutions to the linear Klein--Gordon equation \eqref{Linear wave equation Introduction} on Schwarzschild--AdS.

Assume that the initial data $(f_0,f_1)$ for the linear boundary value problem \eqref{Linear wave equation Introduction} are real analytic and satisfy, for some radius of analyticity
$\rho>0$,
\begin{equation}\label{Real analytic norm}
\|(f_0,f_1)\|^2_{\rho}
\;\doteq\;
\|f_0\|^2_{L^2}
+ \sum_{s\geq 1} \frac{\rho^{2s}}{(2s)!}
\Bigl(\|f_0\|^2_{\dot H^s} + \|f_1\|^2_{\dot H^{s-1}}\Bigr)
< \infty.
\end{equation}
Then, in view of the estimate \eqref{eq:exponential-decay-angular-mode} of
Holzegel--Smulevici~\cite{HS13}, one obtains for the corresponding solution $\phi$ to
\eqref{Linear wave equation Introduction} after decomposing in spherical harmonics and applying a simple interpolation argument that:
\begin{equation*}
\|\phi|_{t^*=\tau}\|_{H^1}
\;\lesssim_{\varepsilon}\;
\|(f_0,f_1)\|_\rho\,(1+\tau)^{-\rho/c_2 + \varepsilon}
\end{equation*}
for every $\varepsilon>0$ and all $\tau \ge 0$, where $c_2>0$ is the constant from
\eqref{eq:exponential-decay-angular-mode} and the constant implicit in $\lesssim_\varepsilon$ depends on $\varepsilon$ and $\|(f_0,f_1)\|^2_{\rho}$. In particular, by taking $\rho$ large, one obtains arbitrarily
fast inverse-polynomial decay for real analytic data. 

Switching our attention to the nonlinear setting, the above observation would imply that, in the analytic category, high-frequency
linear modes are damped sufficiently rapidly for the nonlinear interactions not to have sufficient time to alter the dynamics. Consequently, analytic
data with sufficiently large radius of analyticity may avoid a forward energy cascade and instead yield global-in-time decay without weak turbulence for \eqref{Initial Boundary Value Problem Phi}. This picture is consistent with the numerical results of \cite{CrumpSantos25} for gravitational perturbations of Schwarzschild--AdS in 5D, where small smooth data that are analytic in the angular variables exhibit late-time decay and no evidence of a turbulent cascade on the time scales probed.

Motivated by the above discussion, we propose the following stability statement:

\begin{conjecture}[Stability of Kerr--AdS for analytic perturbations]\label{conj:stability-analytic}
There exists a radius of analyticity $\rho_0>0$ such that the following holds. For every Kerr--AdS parameters $(M,a)$ satisfying the Hawking--Reall bound (with $M>0$) and every
$0<\epsilon\ll 1$, there exists $\delta=\delta(M,a,\epsilon,\rho_0)>0$ with the following property: For any
real-analytic, asymptotically Kerr--AdS initial data $(\gamma_0,K_0)$ for \eqref{eq:EVE-Lambda} satisfying
\[
\|(\gamma_0,K_0) - (\gamma_{M,a},K_{M,a})\|_{\rho} \leq \delta
\]
for some $\rho\ge \rho_0$, where $\|\cdot\|_{\rho}$ is a geometrically defined real analytic norm of the form \eqref{Real analytic norm} and $(\gamma_{M,a},K_{M,a})$ are the Kerr--AdS initial data, we have:

\begin{enumerate}
\item The corresponding maximal Cauchy development $(\mathcal M,g)$ of $(\gamma_0,K_0)$ with reflecting boundary conditions at $\mathcal I$ has complete conformal infinity and the black hole exterior region is $C^2$ close to the background Kerr--AdS exterior.
\item There exist Kerr--AdS parameters $(M',a')$ with $|(M',a')-(M,a)|<\epsilon$ such that, in an appropriate gauge, the black hole exterior of $(\mathcal M,g)$ asymptotically settles down
      to the Kerr--AdS exterior with parameters $(M',a')$ at an inverse–polynomial rate:
      \[
      \|(g(t)-g_{M',a'},K(t)-K_{M',a'})\|_{H^s(\Sigma_t)\times H^{s-1}(\Sigma_t)}
      \;\lesssim_s\; \epsilon \,(1+t)^{-p}
      \]
for some $p>1$ and any $s>0$.
\end{enumerate}
\end{conjecture}

\subsection{Outline of the proof of \texorpdfstring{\cref{thm:Main theorem}}{Theorem 1}}
\label{sec:outline-proof-main-theorem}

In this section, we will lay out a detailed description of the main ideas involved in the proof of \cref{thm:Main theorem}.  
\subsubsection{The linear wave equation on Schwarzschild--AdS: From trapped quasimodes to normal modes with an inner mirror}
\label{sec:Modes near boundary Intro}

As was mentioned earlier (see \cref{sec:Schwarzschild--AdS} above),
 the stably trapped (broken) null geodesics of Schwarzschild--AdS exterior spacetime $(\mathcal M^{(M)}_\mathrm{ext}, g_M)$ give rise to exponentially long-lived quasimodes $\mathring\phi_k$, i.e.~approximate solutions of the Dirichlet boundary value problem for the linear Klein--Gordon equation \eqref{Linear wave equation Introduction} with the properties 1--3 listed below \eqref{Linear wave equation Introduction}.
Our proof of \cref{thm:Main theorem} will be based on the study of the dynamics of the ``nonlinear interactions'' of the  quasimodes $\mathring\phi_k$. For this reason, let us revisit the construction of the linear quasimodes in more detail.

 The smallness condition \eqref{Smallness bound phi k intro}  implies that the quasimodes $\mathring \phi_k$ nearly vanish in the region $r\lesssim 1$. As a result, it is reasonable to attempt to approximate  $\mathring\phi_k$ by exact time-periodic solutions of \eqref{Linear wave equation Introduction} satisfying in addition a Dirichlet condition at a radius $r=r_\mathrm{mirror}\lesssim 1$, namely solving
\begin{equation}\label{Linear wave equation with mirror Introduction}
\begin{cases}
\square_{g_M} \phi +2\phi =0,\\
r\phi|_{r=\infty}=0, \quad \phi|_{r=r_{\mathrm{mirror}}} =0.
\end{cases}
\end{equation}
Fixing a smooth cut-off function $\chi:(2M, +\infty)\rightarrow [0,1]$ such that
\begin{equation}\label{Cut off function intro}
\chi(r) = 
\begin{cases}
0, \quad r\le r_{\mathrm{mirror}},\\
1, \quad r\ge 2r_{\mathrm{mirror}},
\end{cases}
\end{equation}
it will then follow that, provided $\phi_k$ satisfies an appropriate smallness estimate in the region $\left\{r\ll L_k^{\f12}\right\}$, the function 
\begin{equation}\label{quasimode construction intro}
\mathring\phi_k(t,r,\theta, \varphi) \doteq \chi(r) \phi_k(t,r,\theta, \varphi)
\end{equation}
is a \emph{quasimode}. 
In order to simplify our analysis, we will choose $r_\mathrm{mirror}>3M$, i.e.~that the mirror radius is larger than the radius of the photon sphere.

Unlike \eqref{Linear wave equation Introduction} (every solution of which eventually decays due to dispersion through the event horizon, see \cite{HS13}), the boundary value problem \eqref{Linear wave equation with mirror Introduction} admits \emph{normal mode} solutions, i.e.~time periodic solutions of the form
\[
\phi_k (t,r,\theta, \varphi) = \f{a_k(t)}{r} E_k(r,\theta, \varphi),
\]
where $k=(n_k, \ell_k, m_k) \in \mathbb N^* \times \mathbb N_{\ge|m_k|} \times \mathbb Z$ (or $(n, \ell, m)$ for short) is the \emph{frequency parameter}. The functions appearing in the above expression depend on the frequency parameter as follows:
\begin{itemize}
\item The functions $E_k$ separate as
\begin{equation}\label{Eigenfunctions Intro}
E_k(r,\theta,\varphi) = R_{n, \ell}(r) Y_{\ell, m}(\theta, \varphi),
\end{equation}
where $Y_{\ell,m}$ is the spherical harmonic of order $(\ell, m)$ and $R_{n,\ell}$ is the $n$-th eigenfunction of the
 Sturm--Liouville problem (formulated with respect to the variable $y=y(r) \sim \f1r$ defined by \eqref{The y coordinate}):
\begin{equation}\label{Radial model boundary value problem introduction}
\begin{cases}
-\f{d^2}{dy^2} R+ V_\ell\left(r(y)\right) R={\omega}^2 R,\\
R|_{y=0} = R|_{y=y_\mathrm{mirror}}=0,
\end{cases}
\end{equation}
with
\begin{equation}\label{Radial potential introduction}
V_\ell\left(r(y)\right) = \Big(1 + \f{1}{r(y)^2}-\f{2M}{r(y)^3} \Big)\Big(\ell(\ell+1)+\f{2M}{r(y)}\Big).
\end{equation}
The Sturm--Liouville eigenvalue problem \eqref{Radial model boundary value problem introduction} determines, for each $\ell\in \mathbb N$, a discrete set of real frequencies 
\[
\pm \omega =\pm \omega_{(n, \ell)}, \quad n\in \mathbb N,
\]
depending smoothly on the mass parameter $M$. The eigenfrequency/eigenfunction pair $(\omega_{n,\ell}, R_{n,\ell})$ is parametrized by $n$ so that $\omega_{(n, \ell)}$ is increasing in $n$; in this case, $n$ is known as the \emph{radial overtone} parameter and $\omega_{n,\ell}$ can be shown to satisfy the asymptotic estimate $\omega_{(n,\ell)}-\ell \sim n$; see  \cref{lem:Crude Weyl law}.  The functions $E_k$ (appropriately normalized) form an orthonormal basis of $L^2\big([0, y_\mathrm{mirror}]\times \mathbb S^2,  dy d\mathrm{\text{vol}}_{\mathbb S^2}\big)$.

\begin{figure}
\centering
\begin{tikzpicture}[
    scale=2.3, 
    font=\small,
    declare function={
        V(\x) = (1 + 1/(\x^2) - 2*0.2/(\x^3))*(1+0.2/(3*\x));
        X(\x) = 2.5*sqrt(\x); 
    },
       x node/.style={below, text height=1.5ex, text depth=1ex}
]
 
\def\rplus{0.355}
\def\rmirror{0.8}
\def\rmax{6}
\def\omegaSq{1.4}
\def\rtp{1.45} 
 
\fill[black!35] 
  ({X(\rmirror)}, \omegaSq) -- 
  ({X(\rmirror)}, {V(\rmirror)}) --
  plot[domain=\rmirror:\rtp, smooth, variable=\x] ({X(\x)}, {V(\x)}) --
  ({X(\rtp)}, \omegaSq) -- cycle;

\fill[black!20] 
  ({X(\rtp)}, \omegaSq) -- 
  plot[domain=\rtp:\rmax, smooth, variable=\x] ({X(\x)}, {V(\x)}) --
  ({X(\rmax)}, \omegaSq) -- cycle;

\draw ({X(\rplus)},0) -- ({X(\rmax)},0) node[pos=0.8, below] {$r$};
\draw[->] ({X(\rplus)},0) -- ({X(\rplus)},2.5) node[above] {$V_\ell$};
 
\node[x node] at ({X(\rplus)},-0.05) {$r=r_+$};
\node[x node] at ({X(\rmirror)},-0.05) {$r_{\mathrm{mirror}}$};
\node[x node] at ({X(\rtp)},-0.05) {$r(y_\mathrm{c})$};
\node[x node] at ({X(\rmax)},-0.05) {$r=\infty$};

\draw[thick] ({X(\rplus)},\omegaSq) -- ({X(\rmax)},\omegaSq) node[pos=0.65, above] {$\omega^2$};
 
\draw[thick] ({X(\rmirror)},0) -- ({X(\rmirror)},1.935);
\draw[thin] ({X(\rmax)},0) -- ({X(\rmax)},1.935);
 
\draw[dashed, thick, domain=\rplus:\rmirror, smooth, variable=\x]
  plot ({X(\x)}, {V(\x)});

\draw[thick, domain=\rmirror:\rmax, smooth, variable=\x]
  plot ({X(\x)}, {V(\x)});

\draw[dashed,thick,black] ({X(\rtp)},\omegaSq) -- ({X(\rtp)},0);
\filldraw ({X(\rtp)},\omegaSq) circle (1pt); 

\draw[decorate,decoration={brace,amplitude=5pt,mirror},yshift=-4mm] 
  ({X(\rmirror)},0) -- ({X(\rtp)},0) 
  node[midway,below=5pt,align=center] {Classically \\ forbidden\\ region};

\draw[decorate,decoration={brace,amplitude=5pt,mirror},yshift=-4mm] 
  ({X(\rtp)},0) -- ({X(\rmax)},0) 
  node[midway,below=5pt] {Classically allowed region};

\end{tikzpicture}
\caption{The graph of the potential $V_\ell(r)$, as defined by \eqref{Radial potential introduction}, is strictly decreasing in $r$ for $r\ge r_{\mathrm{mirror}}$. Thus, when the Sturm--Liouville eigenvalue $\omega^2=\omega^2_{(n,\ell)}$ lies in the interval $\big(V_\ell|_{r=\infty}, V_\ell|_{r=r_{\mathrm{mirror}}})$ (which is true, for instance, when $n\ll \ell$), the corresponding mode solution is localized in the classically allowed region (where $V_\ell \le \omega^2$), with exponentially small tails in the classically forbidden region   (where $V_\ell \geq\omega^2$). These regions are separated by the unique turning point $r(y_c)$ determined by $V_\ell(r) = \omega^2$.}
\label{fig:potential}
\end{figure}
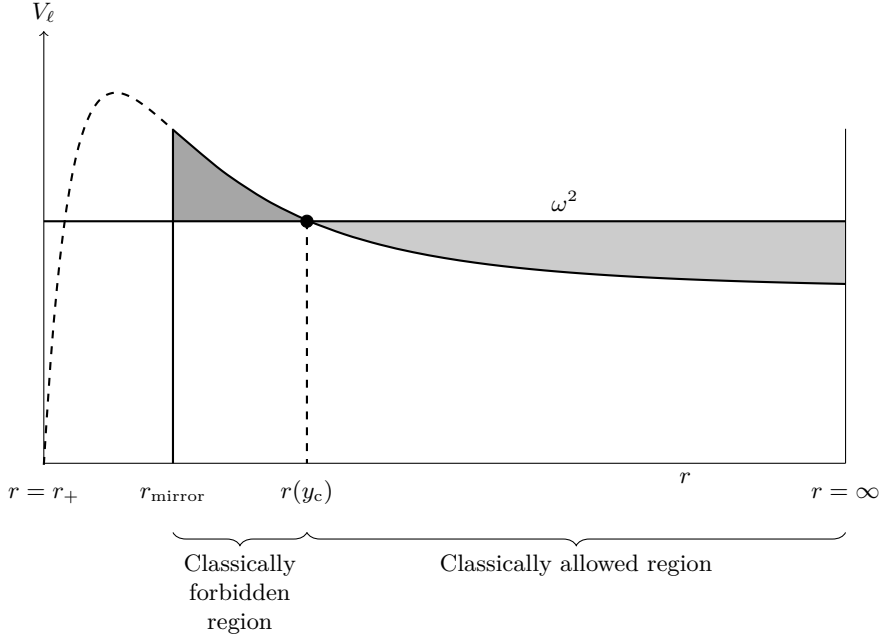

\item The amplitude functions $a_k(t)$ solve 
\[
\ddot a_k + \omega_k^2 a_k =0,
\]
where $\omega_k = \omega_{(n,\ell)}$. Thus, $a_k(t) = A_k e^{i\omega_k t}+B_k e^{-i\omega_k t}$ for some constants $A_k, B_k \in \mathbb C$.
\end{itemize}

The potential $V_\ell$ has a local minimum at $\{y=0\}=\{r=\infty\}$; this corresponds to the stable trapping phenomenon at $r=\infty$ that we described earlier. Our choice $r_\mathrm{mirror}>3M$ implies that $y=0$ is in fact a global minimum (since $V_\ell$ is strictly increasing for $y\in \big(0, y(3M)\big)$). As a result, in the case when the frequency $\omega_k$ satisfies 
\begin{equation}\label{Intro condition for localization}
\omega_k^2 < V_\ell(r_{\mathrm{mirror}})
\end{equation}
(this is always true, for instance, when $n\ll \ell$), there is a unique  \emph{turning point} $y=y_c \in (0, y_\mathrm{mirror})$ for \eqref{Radial model boundary value problem introduction}: This is defined as the root of $\omega_k^2-V_{\ell}\left(r(y)\right)$ in $(0,y_\mathrm{mirror})$ (its value has to satisfy $y_c \sim \sqrt{\f{n}{\ell}}$), and the behavior of $R_{n,\ell}(y)$ changes across $y=y_c$ as follows: 
\begin{itemize}
\item In the region  $y\in [0, y_c]$ (the so-called \emph{classically allowed} region), the function $R_{n,\ell}$ is oscillatory, undergoing $\sim n$ oscillations.
\item In the region $y\in [y_c, y_\mathrm{mirror}]$ (the so-called \emph{classically forbidden} region),  the function $R_{n,\ell}$ is exponentially decaying in $y$. Moreover, in the case when $n\lesssim \ell^{1-\delta}$ for some $\delta>0$, the following estimate holds for some absolute constant $c>0$ and for any $\delta'>0$:
\begin{equation}\label{Exponentially small bound introduction}
|R_{n,\ell}(y)|+|R^\prime_{n,\ell}(y)| \lesssim_{\delta,\delta'} e^{-c \ell} \quad \text{for }\, y\in[ y_c+\delta', y_\mathrm{mirror}].
\end{equation}
\end{itemize}
This is illustrated in \cref{fig:potential}. For more precise estimates on the behavior of $R_{n,\ell}$ and the asymptotics of the eigenvalues $\omega_k$, see \cref{subsection: Estimates R}. In particular, 
 the exponential smallness \eqref{Exponentially small bound introduction} of $\phi_k$ in the region $r\lesssim 1$ implies that, for any frequency parameter $k=(n,\ell, m)$ such that $n\le \ell^{1-\delta}$, the function $\mathring\phi_k$ defined  by \eqref{quasimode construction intro}
is a \emph{quasimode} with the properties 1--3 described below \eqref{Linear wave equation Introduction}. 
 
\subsubsection{The nonlinear ansatz and the mode system}
\label{sec:nonlinear-ansatz-intro}
Turning our attention to the nonlinear equation \eqref{Initial Boundary Value Problem Phi}, our aim is to study the long-time dynamics of solutions $\phi$ arising from the initial data induced at $t=0$ by a sum of small amplitude quasimodes of \eqref{Linear wave equation Introduction}.  For the rest of this section, we will fix a small parameter $\epsilon>0$ (which will be used to measure the size of the initial data), a Sobolev index $s\geq 4$, as well as a small absolute constant $\delta_0\in (0,1)$ and a large frequency scale $L\gg_{\epsilon, s} 1$. 

We will define 
\begin{equation}\label{nonlinearity Intro}
\mathcal N[\phi_1, \phi_2,\phi_3] \doteq \f1{r^6} \Big(\partial_t \phi_1 \partial_t \bar\phi_2 \phi_3 + \phi_1 \bar\phi_2 \partial_t^2 \phi_3\Big),
\end{equation}
so that the nonlinearity in \eqref{Initial Boundary Value Problem Phi} becomes:
\[
\mathcal N^{(3)}[\phi]=\mathcal N[\phi,\phi,\phi].
\]
We will use the following ansatz for the solution $\phi$ for \eqref{Initial Boundary Value Problem Phi}:
\begin{equation}\label{Ansatz phi Intro}
\phi(t,r,\theta, \varphi) = \chi(r) \sum_{k\in \mathcal K_\mathrm{D}} \tilde \phi_k(t,r,\theta, \varphi) + \chi(r)\sum_{k\in \mathcal K_\mathrm{ND}} \tilde \phi_k(t,r,\theta, \varphi) + \psi(t,r,\theta,\varphi),
\end{equation}
where, in the above:
\begin{itemize}
\item The functions $\tilde\phi_k(t,r,\theta, \varphi)$, which  we will refer to as the \emph{mode solutions} (through a slight abuse of terminology), are of the form
\[
\tilde\phi_k(t,r,\theta, \varphi) = \f{a_k(t)}{r} E_k (r,\theta, \varphi),
\]
where $k=(n_k, \ell_k, m_k)\in \mathbb N \times \mathbb N_{\ge |m|}\times \mathbb Z$ is a frequency parameter and $E_k (r,\theta, \varphi)$ is the Dirichlet eigenfunction \eqref{Eigenfunctions Intro}.  

\item The frequency parameter sets $\mathcal K_{\mathrm{D}}$ and $\mathcal K_{\mathrm{ND}}$  will be called the \emph{dominant} and \emph{non-dominant} sets of parameters, respectively. Their union $\mathcal K =\mathcal K_{\mathrm{D}} \cup \mathcal K_{\mathrm{ND}}$, which we will refer to as the set of \emph{permissible} frequency parameters, consists of all $(n,\ell, m)\in \mathbb N^*\times \mathbb N_{\ge |m|} \times \mathbb Z$ satisfying  $n \le \ell^{1-\delta_0}$ for a small constant $\delta_0>0$.  The dominant set $\mathcal K_{\mathrm{D}}$ will be chosen to consist of three parameters $k_\minusone$, $k_0$ and $k_1$, with corresponding angular frequencies $\ell_k$ of size $\sim L$, satisfying $\frac{\ell_{k_\minusone}}{\ell_{k_0}}, \frac{\ell_{k_{1}}}{\ell_{k_0}}\gg 1$. The corresponding \emph{dominant} modes $\left\{\tilde\phi_{k}\right\}_{k\in \mathcal K_{\mathrm{D}}}$ will serve as the leading terms in the expansion \eqref{Ansatz phi Intro}, while the rest of the terms will remain relatively smaller in size; see the remarks below.

\item The modes $\left\{\tilde\phi_k\right\}_{k\in \mathcal K}$ are chosen to solve the following system of boundary value problems:
\begin{equation}\label{Boundary value problem tilde phi intro}
\begin{cases}
\displaystyle\square_g \tilde\phi_k + 2\tilde\phi_k = \f{1}{r\big(1-\f{2M}r+r^2\big)}\mathlarger{\sum}_{k_1, k_2, k_3\in \mathcal K_{\mathrm{D}}} \mathbb P_k \Big(r\big(1-\f{2M}r+r^2\big) \mathcal N[\tilde\phi_{k_1}, \tilde\phi_{k_2}, \tilde\phi_{k_3}]\Big), \\[5pt]
\tilde\phi_k|_{y=y_{mirror}}=0, \quad r\tilde\phi_k|_{y=0}=0,
\end{cases}
\end{equation} 
where, $\mathbb P_k$ denotes the $L^2(dy\dvol_{\mathbb S^2})$ projection on $E_k$. A direct calculation shows that the amplitudes $a_k(t)$ solve the following \emph{mode system} of nonlinear ODEs:

\begin{equation}\label{Mode system Intro}
\f{d^2 a_k}{dt^2} +{\omega}_k^2 a_k = \sum_{k_1, k_2, k_3 \in \mathcal K_{\mathrm{D}}}\Bigg(\langle E_{k_1} \bar E_{k_2} E_{k_3} \bar E_k \rangle  \Big( \f{d a_{k_1}}{dt} \f{d \bar a_{k_2}}{dt} a_{k_3} + a_{k_1} \bar a_{k_2} \f{d^2 a_{k_3}}{dt^2} \Big) \Bigg),
\end{equation}
where $\langle \cdot \rangle$ denotes a spatial averaging operator, see already \eqref{Definition average}. 

\medskip
\begin{remark}
Note that the nonlinear term in the right-hand side of \eqref{Boundary value problem tilde phi intro} has been chosen to include only contributions from the dominant modes.
\end{remark}
\medskip

\item The function $\psi(t,r,\theta,\varphi)$ will be referred to as the \emph{error term} and satisfies a quasilinear initial--boundary value system with a source term:
\begin{equation}\label{IVP Psi Intro}
\begin{cases}
\square_g \psi + 2 \psi +\widetilde{\mathcal N}[\chi\tilde\phi; \psi] = - \mathcal F[\tilde \phi],\\[5pt]
(\psi, \partial_{t} \psi)|_{t=0} = (0,0), \\
r\psi|_{r=\infty} =0,
\end{cases}
\end{equation}
where
\begin{equation}\label{Approximate solution intro}
\tilde\phi = \sum_{k\in \mathcal K_{D}} \tilde \phi_k + \sum_{k\in \mathcal K_{\mathrm{ND}}} \tilde \phi_k,
\end{equation}
\[
\widetilde{\mathcal N}[\chi\tilde\phi; \psi] \doteq - \left(\mathcal N[\chi\tilde\phi+\psi, \chi\tilde\phi+\psi, \chi\tilde\phi+\psi]-\mathcal N[\chi\tilde\phi, \chi\tilde\phi, \chi\tilde\phi]\right)
\]
and the source term $\mathcal F[\tilde \phi]$ measures the failure of the mode sum $\chi \tilde\phi$ to solve equation \eqref{Initial Boundary Value Problem Phi}:
\begin{equation}\label{Source Psi Intro}
\mathcal F[\tilde \phi] \doteq   \square_g (\chi \tilde\phi) + 2 \chi\tilde\phi - \mathcal N[\chi\tilde\phi, \chi\tilde\phi, \chi\tilde\phi].
\end{equation}
Note that \eqref{IVP Psi Intro} ensures that $\phi = \chi \tilde \phi + \psi$ satisfies \eqref{Initial Boundary Value Problem Phi}.
\end{itemize}
See \cref{sec:Ansatz} for more details.

Given the above ansatz for $\phi$, the proof of  \cref{thm:Main theorem} consists of the following steps:
\begin{enumerate}
\item We first identify a dense set of Schwarzschild--AdS mass parameters $M\in(0,+\infty)$ for which  the \emph{linear} boundary value problem \eqref{Linear wave equation with mirror Introduction} enjoys the following spectral property: Its normal modes $\phi_k$ corresponding to the (carefully chosen) dominant frequency parameters $\mathcal K_{\mathrm{D}}=\left\{k_\minusone, k_0, k_{1}\right\}$ satisfy a set of spatial and temporal resonant conditions, while at the same time being resonantly ``separated'' from the modes corresponding to the non-dominant parameters $\mathcal K_{\mathrm{ND}}$.

\begin{remark}
Note that \cref{thm:Main theorem} asks for an \emph{open} and \emph{dense} set of mass parameters $M$ for which norm inflation holds. However, openness follows readily using a soft Cauchy stability statement; see \cref{sec:Completion of the proof of Main theorem}.
\end{remark}

\item The system of ODEs \eqref{Mode system Intro} for the dominant mode amplitudes $\left\{a_{k_\minusone}, a_{k_0}, a_{k_{1}}\right\}$ decouples from those for the rest of the modes. The resonant conditions satisfied by the dominant frequencies then allow us to approximate the dynamics of $\left\{a_{k_\minusone}, a_{k_0}, a_{k_{1}}\right\}$ by those described by a \emph{resonant} $3\times 3$ system of ODEs. The structure of this resonant system (which is sensitive to the precise form of the coefficients $\langle E_{k_1} \bar E_{k_2} E_{k_3} \bar E_{k} \rangle $ in \eqref{Mode system Intro}, which are in turn dependent on the form of the nonlinearity in $\eqref{Initial Boundary Value Problem Phi}$ and our precise choice of the frequency parameters $\mathcal K_{\mathrm{D}}$) allows us to establish that a non-trivial energy cascade from the low-frequency mode $\tilde\phi_{k_0}$ to the high-frequency pair $\tilde\phi_{k_\minusone}, \tilde\phi_{k_{1}}$ takes place at a timescale $T\sim L^{2s+1}$. In particular, at that timescale, the sum $\sum_{k\in \mathcal K_{\mathrm{D}}}\tilde\phi_k$ exhibits norm inflation in $H^s$.

\item Exploiting the quantitative resonant ``separation'' between the dominant and non-dominant modes, we show that the non-dominant amplitudes $\{a_k\}_{k\in \mathcal K_{\mathrm{ND}}}$ remain small for $t\in [0,T]$.

\item Finally, we show that the error term $\psi$ remains small in $H^s$  for $t\in [0,T]$ via an energy estimate based on a delicate commutation scheme.
\end{enumerate}

In the remainder of this section, we will highlight the technical challenges arising in each of the steps above. 

\medskip
\begin{remark}\label{rmk:Smallness of source term} Our precise choice of the range of frequency parameters appearing in the ansatz \eqref{Ansatz phi Intro} and the mode system \eqref{Mode system Intro} is motivated by our need to address certain difficulties regarding the control of the subdominant terms in \eqref{Ansatz phi Intro}:
\begin{itemize}
\item We ultimately show that the non-dominant modes $\left\{\tilde\phi_k\right\}_{k\in \mathcal K_{\mathrm{ND}}}$ and the error term $\psi$ in \eqref{Ansatz phi Intro} remain smaller compared to the dominant mode sum 
\[
\tilde\phi_D = \left\{\tilde\phi_k\right\}_{k\in \mathcal K_{\mathrm{D}}},
\]
thus one could view 
\begin{equation}\label{Combined error term Intro}
\psi' = \chi\sum_{k\in \mathcal K_{\mathrm{ND}}}\tilde\phi_k+\psi
\end{equation}
 as describing a combined error term in our approximation of the dynamics of $\phi$ by the dominant mode system. Note that $\psi'$ satisfies an initial boundary value problem similar to \eqref{IVP Psi Intro} for $\psi$, sourced by $\tilde\phi_D$ in place of $\tilde\phi$, i.e. 
\begin{equation}\label{IVP Psi' Intro}
\begin{cases}
\square_g \psi' + 2 \psi' +\widetilde{\mathcal N}[\chi\tilde\phi_D; \psi'] = - \mathcal F[\tilde \phi_D],\\[5pt]
(\psi', \partial_{t} \psi')|_{t=0} = (0,0), \\
r\psi'|_{r=\infty} =0.
\end{cases}
\end{equation}
It would be, therefore, natural to attempt at first to control $\psi'$ directly, instead of decomposing it further as in \eqref{Combined error term Intro}. Unfortunately, this is not possible via energy estimates that do not take into account the additional structure of the source term
\[
\mathcal F[\tilde \phi_D] = \square_g (\chi \tilde\phi_D) + 2 \chi\tilde\phi_D - \mathcal N[\chi\tilde\phi_D, \chi\tilde\phi_D, \chi\tilde\phi_D]
\] 
(measuring the failure of the dominant mode sum $\chi \cdot \tilde\phi_D$ to solve the original equation). This term is of size 
\[
\left\| r^3 \mathcal F[\tilde \phi_D] \right\|_{L^1_t\left([0,T]\right) L^2_{y,\theta,\varphi}} \sim L^{\f32-s} \, \gg \, L^{1-s} \sim \left\|\partial (r\tilde\phi_D)\right\|_{L^\infty_t L^2_{y,\theta,\varphi}}
\]
(with the weighted norm on the left-hand side being the one naturally arising for the source term in the standard $\partial_t$-energy estimate on an asymptotically AdS background) and, thus, energy estimates alone do not suffice to show that $\psi'$ remains smaller compared to $\tilde\phi_D$. Instead, our approach consists of considering separately the ``first order'' contributions of the nonlinear interactions of $\tilde\phi_D$ to the rest of the permissible modes (described by the system \eqref{Mode system Intro} for the non-dominant modes $\{\tilde\phi_K\}_{k\in \mathcal K_{\mathrm{ND}}}$) and then subtracting this contribution from the combined error term, leading us to consider the ``refined'' error term $\psi = \psi' - \chi\sum_{k\in \mathcal K_{\mathrm{ND}}}\tilde\phi_k$. The resulting right hand side for the equation for $\psi$, namely \eqref{Source Psi Intro}, now satisfies
\[
\left\| r^3 \mathcal F[\tilde \phi_D] \right\|_{L^1_t\left([0,T]\right) L^2_{y,\theta,\varphi}} \lesssim L^{-2-s}
\]
(see \cref{lem:Bounds for F tilde phi combined}) and, thus, controlling $\psi$ via energy estimates becomes feasible. 

\item The set of permissible frequency parameters $\mathcal K = \left\{(n,\ell,m): \, n\le \ell^{1-\delta_0}\right\}$ contains parameters for which the corresponding Dirichlet eigenfunction $E_k$ is exponentially small (in terms of the frequency $\ell$) near the mirror at $r=r_{\mathrm{mirror}}$; see \eqref{Exponentially small bound introduction}. As a result, the corresponding normal mode solutions $\phi_k$ of the boundary value problem \eqref{Linear wave equation with mirror Introduction} yield quasimodes $\chi \cdot \phi_k$ for  \eqref{Linear wave equation Introduction} with exponentially small error term. This is no longer true for modes with $n\gtrsim \ell$. 

\item The restriction to considering only modes with $n\le \ell^{1-\delta_0}$ places a lower bound on the smallness of the source term $\mathcal F[\tilde\phi]$ appearing in equation \eqref{IVP Psi Intro} for the error term $\psi$: Expanding $\mathcal F[\tilde\phi]$ in the $L^2$ basis formed by the eigenfunctions $E_k$, one can easily check (see the proof of \cref{lem:Bounds for F tilde phi combined}) that its $L^2_{y,\theta,\varphi}$-norm contains a contribution from terms of the form 
\[
\mathbb P_{k;k_1,k_2,k_3} \doteq \langle E_k \overline{E_{k_1}} E_{k_2} \overline{E_{k_3}}\rangle,
\]
 where $k\notin \mathcal K$ and $k_1,k_2,k_3\in \mathcal K_{\mathrm{D}}$. The above term can be thought of as a projection of the ``low-frequency'' product $ E_{k_1} \overline{E_{k_2}} E_{k_3}$ to the ``high frequency'' eigenfunction $E_k$. The integral, which is non trivial only in the case when the angular frequency $\ell_k$ is of size $\sim L$, decays to $0$ as the radial overtone $n_k\rightarrow +\infty$ at a polynomial rate:
\[
 n_k^{-C_0} \lesssim L^{2}\left| \mathbb P_{k;k_1,k_2, k_3}\right| \lesssim  n_k^{-5}
\]
for some absolute constant $C_0\ge 5$.\footnote{In particular, the right-hand side of the bound \eqref{Polynomial bound projection radial modes} cannot be upgraded to a polynomial function of $n_k$ of arbitrarily large power.} Since $\min \left\{n_k\, | \, k\notin \mathcal K, \,  \ell_k\sim L\right\} \sim L^{1-O(\delta_0)}$, this implies that the source term $\mathcal F[\tilde\phi]$ is only polynomially small in terms of $L$. Peculiarly, this fact seems to be connected to the dimension of our background: In the case of Schwarzschild--AdS spacetimes of \emph{even} space dimension, the term $\mathbb P_{k;k_1, k_2, k_3}$ is in fact exponentially small in $n$ and, consequently, $\mathcal F[\tilde\phi]$ is exponentially small in terms of $L$; in that case, it seems plausible that our proof of smallness for the error terms could simplify substantially. 

\end{itemize}
\end{remark}

\subsubsection{The resonant properties of the linear spectrum and the selection of the dominant parameters}
\label{sec:Spectrum intro}

The construction of the dominant set of frequency parameters $\mathcal K_{\mathrm{D}}$ at the given frequency scale $L\gg 1$ relies on a delicate analysis of the dependence of the temporal frequencies $\omega_{n,\ell}$ on the background mass parameter $M$.  Our main tool is the following asymptotic expansion, which we derive in \cref{sec:Spectral analysis radial} using techniques from perturbative Sturm--Liouville theory:
\begin{equation}\label{Taylor expansion omega intro}
\omega_{n,\ell} = (2n+\ell) - f_3(n) \ell^{-1/2} M - \frac 12 h(n)\ell^{-1} M^2 + O(\ell^{-1-5\delta_0}),
\end{equation}
valid under the assumption that $n\ll \ell^{\delta_0}$, with explicit coefficients $f_3(n), h(n)$ expressed as integrals of products of Hermite functions (see \cref{sec:Integrals Hermite}). Note that the above is a refinement of \eqref{Asymptotic expansion HDS}.

Using the above expansion, we show that, for a $L^{-5\delta_0}$-\emph{dense} set of mass parameters $M>0$, there exists a parameter triad $\mathcal K_{\mathrm{D}} = \{k_0, k_1, k_\minusone\}$, $k_i=(n_i,\ell_i,m_i)$, $i=0,1,\minusone$, satisfying the spatial and temporal resonance conditions
\begin{equation} \label{eq:resonant-frequency-relation-intro}
{m}_{k_1}-2{m}_{k_0}+{m}_{k_\minusone} = 0 \quad \text{and} \quad {\omega}_{k_1}-2{\omega}_{k_0}-{\omega}_{k_\minusone}=0,
\end{equation}
where \begin{equation}
    \label{eq:dominant-modes-choice}
  \frac{\omega_{k_0}}{\omega_{k_1} }, \frac{\omega_{k_0}}{\omega_{k_\minusone}} \ll \frac{\omega_{k_\minusone}}{\omega_{k_{1}}}\sim 1 \quad \text{and} \quad |m_{k_i}| = \ell_{k_i} \sim L.
\end{equation}
At the same time, the non-dominant frequency parameters $k\in \mathcal K_{\mathrm{ND}}= \mathcal K \setminus \mathcal K_{\mathrm{D}}$ are resonantly separated from the dominant ones in the following sense:
\begin{equation}\label{Non resonant condition time frequencies intro}
\min_{j_1, j_2, j_3 \in \mathcal K_{\mathrm{D}}}  \left[ \Big| {\omega}_k - \big| \varepsilon_{j_1}{\omega}_{j_1} - \varepsilon_{j_2}{\omega}_{j_2} + \varepsilon_{j_3}{\omega}_{j_3}\big| \Big| + |m_k - m_{j_1} + m_{j_2} - m_{j_3}| \right]  \gtrsim L^{-\f12},
\end{equation}
where the signs $\varepsilon_k \in \{\pm1\}$ for $k\in \mathcal K_{\mathrm{D}}$ are given by 
\begin{equation}\label{Signs intro}
 \varepsilon_{k_\minusone} = -1,   \varepsilon_{k_0} =+1,   \varepsilon_{k_1} = +1.
\end{equation}
The specification of the precise form of the dominant frequency parameters and the derivation of the above statement is carried out in  \cref{sec:Ansatz}; see \cref{prop:Resonance conditions}.

Let us note at this point that the set of non-dominant frequency parameters $\mathcal K_{\mathrm{ND}}$ is actually further split into two disjoint sets: The \emph{almost resonant} parameters $\mathcal K_{\mathrm{AR}}$, consisting of those parameters $k =(n,\ell,m)\in \mathcal K\setminus \mathcal K_{\mathrm{D}}$ that would satisfy 
\[
 {\omega}_k - \big| \varepsilon_{j_1}{\omega}_{j_1} - \varepsilon_{j_2}{\omega}_{j_2} + \varepsilon_{j_3}{\omega}_{j_3}\big| =0 \quad \text{and} \quad  m_k - m_{j_1} + m_{j_2} + m_{j_3} =0 \quad \text{for some }\, j_1, j_2, j_3\in \mathcal K_{\mathrm{D}}
\]
in the case $M=0$ (i.e.~on pure AdS spacetime), and the \emph{non-resonant} parameters $\mathcal K_{\mathrm{NR}} = \mathcal K \setminus (\mathcal K_{\mathrm{D}} \cup \mathcal K_{\mathrm{AR}})$. 
Showing that the almost resonant modes satisfy the non-resonant condition \eqref{Non resonant condition time frequencies intro} for the values of $M>0$ under consideration relies heavily on the delicate asymptotics of the coefficients $f(n)$ and $h(n)$ in the expansion \eqref{Taylor expansion omega intro} for $\omega_{n,\ell}$. On the other hand, the same lower bound for modes with $k\in \mathcal K_{\mathrm{NR}}$ follows almost trivially from the corresponding bound when $M=0$. See \cref{sec:Resonant structure} for more details.

\subsubsection{Dynamics of the dominant modes and the slowly oscillating approximation} \label{sec:dominant-modes-intro}

The mode system of ODEs \eqref{Mode system Intro} for the amplitudes $a_k$, $k\in \mathcal K$, yields a decoupled $3\times 3$ nonlinear ODE system for the dominant mode amplitudes $a_k$, $k\in \mathcal K_{\mathrm{D}} = \left\{k_\minusone, k_0, k_{1}\right\}$.  The analysis of this system occupies  
\cref{sec:Dominant modes}.

 In order to separate the fast linear oscillations from the genuinely nonlinear dynamics, we introduce the renormalized, slowly varying amplitudes
\[
\tilde b_k(t) \doteq L^{s} a_k(t) e^{i\varepsilon_k \omega_k t},\qquad k\in \mathcal K_{\mathrm{D}},
\]
where the signs $\varepsilon_k\in\{\pm 1\}$ are given by \eqref{Signs intro}. In terms of the variables $\tilde b_k$ and the rescaled time variable
\[
\tilde t \doteq L^{-1-2s} t,\qquad \tilde\omega_k \doteq L^{-1}\varepsilon_k\omega_k,
\]
the system \eqref{Mode system Intro} for $k\in \mathcal K_{\mathrm{D}}$ becomes a $3\times 3$ second-order ODE system of the form (using the shorthand notation $\{\minusone, 0,1\}$ for indices associated to the dominant parameters $\{k_\minusone, k_0,k_{1}\}$, respectively):
\begin{align}\label{Dominant mode system Intro}
L^{-2s -2}    \frac{d^2}{d \tilde t^2} \tilde b_k  - 2i \tilde {\omega}_k \frac{d}{d\tilde t} \tilde b_k =& \sum_{j_1,j_2,j_3 \in \{\minusone, 0, 1 \}} e^{i L^{2s +2 } (  \tilde {\omega}_k - \tilde {\omega}_{j_1} + \tilde {\omega}_{j_2} - \tilde {\omega}_{j_3} )\tilde t} \llangle \bar E_k E_{j_1} \bar E_{j_2} E_{j_3}   \rrangle \\ 
\nonumber
       &\quad  \times \bigg[ \left( \tilde {\omega}_{j_1}\tilde  {\omega}_{j_2} - \tilde {\omega}_{j_3}^2 \right)\tilde b_{j_1} \overline{ \tilde b_{j_2}} \tilde b_{j_3}   \\
 \nonumber
 &\quad\qquad + L^{-2s -2} \left( - i\tilde  {\omega}_{j_1} \tilde  b_{j_1} \frac{d}{d\tilde t}{\overline{\tilde b_{j_2}}}  \tilde b_{j_3} +  i \tilde  {\omega}_{j_2}   \frac{d}{d\tilde t}{\tilde b_{j_1} }{\overline{\tilde b_{j_2}}} \tilde  b_{j_3}   - 2i \tilde  {\omega}_{j_3} \tilde b_{j_1} {\overline{ \tilde b_{j_2}}} \frac{d}{d\tilde t}{\tilde{b}_{j_3}}  \right) \\ 
&\quad\qquad    +L^{-4s -4} \left( \frac{d}{d\tilde t}{\tilde b_{j_1}}  \frac{d}{d\tilde t}{\overline{\tilde b_{j_2}}} \tilde b_{j_3}  + \tilde b_{j_1} \overline{\tilde b_{j_2}} \frac{d^2}{d\tilde t^2}{\tilde b_{j_3}}  \right) \bigg], \nonumber
\end{align}
 whose coefficients are given by rescaled spectral overlap integrals $\llangle \overline{E_k}E_{j_1}\overline{E_{j_2}}E_{j_3}\rrangle$. Note that the right-hand side contains terms that are either suppressed by negative powers of $L$ or carry rapidly oscillating phases of the form
$\exp\left(iL^{2s+2}(\tilde\omega_k-\tilde\omega_{j_1}+\tilde\omega_{j_2}-\tilde\omega_{j_3})\tilde t\right)$.

The temporal resonance relation \eqref{eq:resonant-frequency-relation-intro}  implies that there is a distinguished class of terms $k,j_1,j_2,j_3\in \mathcal K_{\mathrm{D}}$ for which the phase $\tilde\omega_k-\tilde\omega_{j_1}+\tilde\omega_{j_2}-\tilde\omega_{j_3}$ in the exponential above vanishes, whereas all remaining combinations oscillate in $\tilde t$ with frequency $\gtrsim 1$. These oscillating contributions average out on the time-scale $\tilde t\in[0,\tilde T]$. At the same time, the higher order terms with extra powers of $L^{-2s-2}$ can be considered to be perturbative. Motivated by this, we define the slowly oscillating approximation of the dominant system by formally discarding all terms in \eqref{eq:full-system-for-tildeak 2} whose coefficients either decay with $L$ or involve non-zero oscillatory phases; the resulting limit system is a first-order autonomous $3\times 3$ ODE system for slowly varying amplitudes  $b^{\mathrm{sl}}_j(\tilde t)$, $j\in\{-1,0,1\}$:
\begin{equation}\label{Slowly oscillating system Intro}
\begin{cases}
     - 2 i {\tilde {\omega}}_{\minusone} \frac{d}{d\tilde t} b^{\mathrm{sl}}_\minusone = \! \Big( \!\displaystyle\sum_{j=-1}^1 \! \llangle |E_1|^2 |E_{j}|^2    \rrangle \left( \tilde{\omega}_{\minusone} \tilde {\omega}_j - \tilde {\omega}_\minusone^2 \right) |b^{\mathrm{sl}}_{j}|^2\Big)  b^{\mathrm{sl}}_{\minusone}
+\! \Big(\llangle E_{-1} \bar E_0^2  E_{1}   \rrangle \left( \tilde {\omega}_{0} \tilde  {\omega}_{1} - \tilde {\omega}_{0}^2 \right) (b^{\mathrm{sl}}_{0})^2\Big)\overline{ b^{\mathrm{sl}}_{1}}, \\[5pt]
      - 2 i {\tilde {\omega}}_{0} \frac{d}{d\tilde t} b^{\mathrm{sl}}_0 =  \! \Big(\! \displaystyle\sum_{j=-1}^1 \! \llangle |E_0|^2 |E_{j}|^2    \rrangle \left(  \tilde{\omega}_0 \tilde {\omega}_j - \tilde {\omega}_0^2 \right) |b^{\mathrm{sl}}_{j}|^2\Big)  b^{\mathrm{sl}}_{0}
+\! \Big(\llangle E_{-1} \bar E_0^2  E_{1}   \rrangle \left( (\tilde {\omega}_{1}+\tilde{\omega}_{\minusone}) \tilde  {\omega}_{0} - \tilde {\omega}_{\minusone}^2-{\omega}_1^2 \right) b^{\mathrm{sl}}_{\minusone} b^{\mathrm{sl}}_{1} \Big)\overline{ b^{\mathrm{sl}}_{0}}, \\[5pt]
      - 2 i {\tilde {\omega}}_{1} \frac{d}{d\tilde t} b^{\mathrm{sl}}_1 = \! \Big( \!\displaystyle\sum_{j=-1}^1 \! \llangle |E_1|^2 |E_{j}|^2    \rrangle \left( \tilde{\omega}_1 \tilde {\omega}_j - \tilde {\omega}_1^2 \right) |b^{\mathrm{sl}}_{j}|^2\Big)  b_{1}^{\mathrm{sl}}
+\! \Big(\llangle E_{-1} \bar E_0^2  E_{1}   \rrangle \left( \tilde {\omega}_{0} \tilde  {\omega}_{\minusone} - \tilde {\omega}_{0}^2 \right) (b^{\mathrm{sl}}_{0})^2\Big)\overline{ b_{\minusone}^{\mathrm{sl}}}.
\end{cases}
\end{equation}
Note that the  coefficients of the system above are independent of $L$.

In \cref{sec:slowly-oscillating-approximation}, we show that the slowly oscillating approximation is quantitatively accurate on the time-scale of interest. More precisely, \cref{prop:Slowly oscillating approximation} establishes that, for any fixed $\tilde T>0$, if $L$ is chosen sufficiently large, then the solution $  \tilde b_k(\tilde t)$, $k\in\{-1,0,1\}$,  of the full renormalized system \eqref{Dominant mode system Intro} arising from suitable initial data (which are ``consistent'' with the slowly oscillating ansatz) exists on $\tilde t\in[0,\tilde T]$ and remains $O(L^{-2s-2})$-close to the solution $b^{\mathrm{sl}}_k(\tilde t)$, $k\in\{-1,0,1\}$, of the slowly oscillating system \eqref{Slowly oscillating system Intro}.

The dynamics of this approximating system is studied in \cref{sec:Analysis of the approximating system}, which contains the technical heart of \cref{sec:Dominant modes}.
 As a first step, we linearize \eqref{Slowly oscillating system} around the constant solution in which only the low-frequency dominant mode is present, i.e.~$b^{\mathrm{sl}}_{-1}=b^{\mathrm{sl}}_{1}=0$, $b^{\mathrm{sl}}_{0}\equiv \varepsilon$. Arranging the linearization into a vector $\mathring b^{\mathrm{sl}}$, see \eqref{Vector for linearized system}, one finds that its evolution is governed by the linear system
\begin{equation}\label{eq:linearized-system-intro}
\frac{d}{d\tilde t} \ob^{\mathrm{sl}}= \frac{\varepsilon^2}{2}\mathbb M  \cdot \ob^{\mathrm{sl}},
\end{equation}
where $\mathbb M$ is the $6\times 6$ real matrix in \eqref{eq:matrix-M}, and the evolution of the high-frequency modes $(\mathring b^{\mathrm{sl}}_{\minusone},\mathring b^{\mathrm{sl}}_{1})$ is governed by the $4\times 4$ real submatrix $\widetilde{\mathbb M}$ in \eqref{The reduced M matrix}. In particular, the spectral properties of $\widetilde{\mathbb M}$ encode the linear stability properties for the resonant triad of dominant modes. Using detailed bounds on the spectral coefficients derived in \cref{sec:Estimates spectral coefficients}, in particular the strict inequality \eqref{Spectral projection ratio}, \cref{lem:The matrix spectrum} shows that $\widetilde{\mathbb M}$ possesses two pairs of eigenvalues $\pm\sigma_1,\pm\sigma_2$ with
\[ \Re\sigma_1 = \Re\sigma_2 \sim 1,\qquad \Im\sigma_1 = -\Im\sigma_2  \gg 1,
\]
so that the linearized system possesses an exponentially growing mode associated with an eigenvector of $\widetilde{\mathbb M}$. This growing mode is the precise mechanism by which the low-frequency mode transfers energy into the high-frequency modes. 
 Building on this, by choosing the initial data for $b_j^{\mathrm{sl}}$  along the unstable eigenvector of $\widetilde{\mathbb M}$ and controlling the nonlinear error, we show in \cref{prop:Growth slowly oscillating system} that one can arrange the associated renormalized $H^s$-energy
\[
\sum_{j\in\{-1,0,1\}} \tilde\omega_j^{2s}\,\bigl|b_j^{\mathrm{sl}}(\tilde t)\bigr|^2
\]
being amplified by an arbitrary factor $C_\mathrm{amp}$ at time $\tilde T$. 

Translating back to the original variables $a_k(t)$ and physical time $t$ yields a definite $H^s$-norm inflation statement for the triad of dominant modes $\{\tilde\phi_k\}_{k\in\mathcal K_D}$; see \cref{cor:The 3 times 3 system}.

\begin{remark}[The resonant system for pure power nonlinearities]
\label{rem:intro-pure-power}
The turbulent properties of the resonant system are sensitive to the precise form of the nonlinearity in the right-hand side of \eqref{eq:nonlinearity}. If one were to replace $\mathcal{N}^{(3)}[\phi]$ by the semilinear
power nonlinearity
\[
\mathcal{N}^{(p)}_{\mathrm{pow}}[\phi] = |\phi|^{p-1}\phi, \qquad p \geq 3
\]
and investigate the associated dynamics by using the same ansatz that we employ here, the steps that we have laid out so far can be repeated mutatis mutandis: Since the spectrum of the linear operator remains the same, one would obtain (for the same values of the Schwarzschild--AdS mass) the same conditions on the triad of dominant parameters  $\mathcal K_{\mathrm{D}}$ (satisfying the same  resonance relation \eqref{eq:resonance-condition-intro}), as well as a $3\times 3$ slowly oscillating system for the dominant mode amplitudes analogous 
to \eqref{Slowly oscillating system Intro}. Linearizing around the configuration where only the low-frequency mode is present (see \eqref{eq:linearized-system-intro}) leads 
again to a $4 \times 4$ real submatrix $\tilde{\mathbb M}^{\mathrm{pow}}$ governing the high-frequency components of 
$\mathring b^{\mathrm{sl}}$. A direct computation shows that 
 $\tilde{\mathbb M}^{\mathrm{pow}}$  is real skew-symmetric, hence all eigenvalues are purely imaginary and the triad exhibits only oscillatory energy exchange, with no exponentially growing mode. 

In the cubic case $p=3$, one can show that the above stability statement extends beyond the linear regime: The analog of the resonant system \eqref{Slowly oscillating system Intro} now becomes:
\begin{equation}\label{Slowly oscillating system cubic Intro}
\begin{cases}
     - 2 i {\tilde {\omega}}_{\minusone} \frac{d}{d\tilde t} b^{\mathrm{sl}}_\minusone =  \Big( \displaystyle\sum_{j=-1}^1\llangle |E_1|^2 |E_{j}|^2    \rrangle |b^{\mathrm{sl}}_{j}|^2\Big)  b^{\mathrm{sl}}_{\minusone}
+ \llangle E_{-1} \bar E_0^2  E_{1}   \rrangle (b^{\mathrm{sl}}_{0})^2 \overline{ b^{\mathrm{sl}}_{1}}, \\[5pt]
      - 2 i {\tilde {\omega}}_{0} \frac{d}{d\tilde t} b^{\mathrm{sl}}_0 =  \Big( \displaystyle\sum_{j=-1}^1\llangle |E_0|^2 |E_{j}|^2    \rrangle |b^{\mathrm{sl}}_{j}|^2\Big)  b^{\mathrm{sl}}_{0}
+ \llangle E_{-1} \bar E_0^2  E_{1}   \rrangle b^{\mathrm{sl}}_{\minusone} b^{\mathrm{sl}}_{1} \overline{ b^{\mathrm{sl}}_{0}}, \\[5pt]
      - 2 i {\tilde {\omega}}_{1} \frac{d}{d\tilde t} b^{\mathrm{sl}}_1 =  \Big( \displaystyle\sum_{j=-1}^1\llangle |E_1|^2 |E_{j}|^2    \rrangle  |b^{\mathrm{sl}}_{j}|^2\Big)  b_{1}^{\mathrm{sl}}
+ \llangle E_{-1} \bar E_0^2  E_{1}   \rrangle  (b^{\mathrm{sl}}_{0})^2\overline{ b_{\minusone}^{\mathrm{sl}}}.
\end{cases}
\end{equation}
Notice that, for the above system, the following positive definite expression is conserved: 
\[
\mathcal E^{(cubic)} \doteq (-\tilde\omega_\minusone)|b^{\mathrm{sl}}_\minusone|^2 + \tilde\omega_{1}|b^{\mathrm{sl}}_{1}|^2
\]
(recall that $\tilde\omega_\minusone = L^{-1}\varepsilon_\minusone \omega_\minusone <0$). The conservation of $\mathcal E$ precludes the transfer of energy from the low frequency mode $b^{(\mathrm{sl})}_0$ to the high frequency modes  $\{b^{(\mathrm{sl})}_{\minusone}, b^{(\mathrm{sl})}_{1}\}$. Let us note that, in the case of the system \eqref{Slowly oscillating system Intro} studied in this work, there is an analogous conserved quadratic form, namely 
\[
\mathcal E = \frac{(-\tilde\omega_\minusone)}{\tilde\omega_0 \tilde\omega_1-\tilde\omega_0^2}|b^{\mathrm{sl}}_\minusone|^2 - \f{\tilde\omega_{1}}{\tilde\omega_0 (-\tilde\omega_\minusone)+\tilde\omega^2_0}|b^{\mathrm{sl}}_{1}|^2,
\]
which is indefinite and therefore does not lead to a similar stability statement.
\end{remark}

\begin{remark}[Spatial resonances and the $r^{-6}$ weight]
\label{rem:intro-spatial-resonance}
  Noting that the eigenvalues of the matrix $\widetilde{\mathbb M}$ in \eqref{eq:linearized-system-intro} depend
crucially on the values of the spatial overlap integrals entering the spectral coefficients of \cref{sec:Estimates spectral coefficients} , it is
the key inequality \eqref{Spectral projection ratio} which implies (via \cref{lem:The matrix spectrum}) that
$\widetilde{\mathbb M}$ has eigenvalues with strictly positive real part, thus showing the instability of the dominant mode system. One can check that, in the model where one drops completely the $r^{-6}$ weight from the right hand side of \eqref{eq:nonlinearity}, the spectral coefficients (which can be computed in a similar fashion) now yield a reduced matrix $\widetilde{\mathbb M}$ with purely imaginary spectrum, hence the resonant system in that case is (asymptotically) stable at the linearized level (despite the resonant structure of the spectrum being identical to our case).
\end{remark}

\subsubsection{Control of the non-dominant modes}
Switching our attention to the non-dominant modes $\{\tilde\phi_k\}_{k\in\mathcal K_{\mathrm{ND}}}$, the mode system  \eqref{Mode system Intro} implies that each non-dominant amplitude $a_k(t)$ solves a (decoupled) \emph{linear} ODE in time with a forcing term given by the ($E_k$-projection of) the nonlinear interactions of the dominant modes:
\begin{equation}\label{eq:ND-mode-ODE-intro}
\frac{d^2 a_k}{dt^2} + \omega_k^2 a_k
 = \sum_{j_1,j_2,j_3\in \mathcal K_{\mathrm{D}}}
e^{i(-\varepsilon_{j_1}\omega_{j_1}+\varepsilon_{j_2}\omega_{j_2}-\varepsilon_{j_3}\omega_{j_3})t}
\langle\bar{E_k}E_{j_1}\bar{E_{j_2}}E_{j_3}\rangle\,
\mathcal  B_{j_1,j_2,j_3}(t),
\end{equation}
where the coefficients $\mathcal  B_{j_1,j_2,j_3}(t)$ are explicit cubic expressions in the slowly oscillating amplitudes
$b_j(t)$ and their time derivatives (cf.~\eqref{ODE system non dominant modes}). We initialize the above system by $\left(a_k(0), \frac{da_k}{dt}(0)\right)=(0,0)$ for all $k\in \mathcal K_{\mathrm{ND}}$.

The analysis of the dynamics of \eqref{eq:ND-mode-ODE-intro} is carried out in \cref{sec:Non dominant modes} and is based on two crucial inputs: The first one is the \emph{resonant separation} of the non-dominant frequencies from the dominant ones (recall our condition \eqref{Non resonant condition time frequencies intro}). This allows one to integrate by parts in time in the Duhamel representation of $a_k$, gaining a factor of order $\lesssim L^{-1/2}$ from the oscillatory phase. Secondly, we show that the spectral projection integrals
\[
P_k \coloneqq \max_{j_1,j_2,j_3\in \mathcal K_{\mathrm{D}}}
L^2\langle\bar{E_k}E_{j_1}\bar{E_{j_2}}E_{j_3}\rangle,
\]
vanish unless the angular momentum $\ell_k$ lies in the window $[L,3(\lambda+2)L]$ and $m_k=m_{j1}-m_{j_2}+m_{j_3}$. In addition, one has strong decay in the overtone index:
\[
|P_k|\lesssim n_k^{-5}
\]
 (see  \cref{lem:Bounds P k}). Thus, only a very sparse subset of non-dominant modes is actually excited, and high overtones are strongly suppressed.

Combining these structural facts with the uniform bounds for the dominant amplitudes $b_j, j\in \mathcal K_\mathrm{D}$ obtained previously, we derive mode-wise estimates for $a_k$ and its time derivatives.
 After translating these bounds back to  $\tilde\phi_k$ and summing over $k\in \mathcal K\setminus \mathcal K_{\mathrm{D}}$, one obtains the following global $H^s$ estimate from \cref{prop:Total bound non-dominant modes}:
\[
\sup_{t\in[0,T_1]}
\sum_{k\in \mathcal K\setminus \mathcal K_{\mathrm{D}}}
\;\sum_{p_1+p_2+p_3\leq s}
\bigl\|\partial_t^{p_1}\partial_y^{p_2}\nabla_{S^2}^{p_3}(r\tilde\phi_k)\bigr\|_{L^2(dy\,d\mathrm{vol}_{S^2})}
\;\leq\; C_p\,L^{-2s+\frac12}.
\]
In particular, the combined contribution of all non-dominant modes to the $H^s$-size of the solution remains much smaller than that of the dominant triad on the time-scale $[0,T_1]$.

\subsubsection{Control of the error term and the helical commutation vector field}

The last step in the proof of \cref{thm:Main theorem} is bounding the error term $\psi$ in the decomposition
\begin{equation*}
\phi = \chi \tilde\phi + \psi
\end{equation*}
(see \eqref{Ansatz phi Intro}), 
where $\tilde\phi$ is the approximate solution \eqref{Approximate solution intro} built from the modes $\{ \tilde\phi_k\}_{k\in \mathcal K}$. Recall that $\psi$ solves  \eqref{IVP Psi Intro}, which can be reexpressed as:
\begin{equation}
    \label{eq:psi-equation-intro}
\begin{cases}
\square_g \psi + 2 \psi - r^{-6} |\chi\tilde\phi+\psi|^2 \partial_{t^*}^2\psi - \chi^2  \mathcal N^{(1)}[\tilde \phi;\psi]   - \chi \mathcal N^{(2)}[\tilde\phi; \psi] -\mathcal N^{(3)}[\psi] = - \mathcal F[\tilde \phi],\\[5pt]
(\psi,\partial_t \psi)|_{t=0}=(0,0),\\[5pt]
r\psi|_{r=+\infty}=0,
\end{cases}
\end{equation}
where $\mathcal N^{(1)}$, $\mathcal N^{(2)}$ and $\mathcal N^{(3)}$ denote the terms which are, respectively, linear, quadratic and cubic in $\psi$ and contain derivatives of $\psi$ of order $\le 1$. Our aim is to show that, on the time interval $[0,T_1]$, the error term $\psi$ remains smaller (in the $H^s$ norm) compared to $\tilde \phi$; this is carried out in \cref{sec:Estimates error term}. 

Let us focus on the \emph{linearized} version of \eqref{eq:psi-equation-intro}:
\begin{equation}
    \label{eq:psi-equation-intro-linear}
\begin{cases}
\square_g \psi + 2 \psi - r^{-6} \chi^2 \Big(|\tilde\phi|^2 \cdot \partial_{t^*}^2\psi  + 2 \tilde\phi \Re\{\partial_{t^*} \tilde\phi \cdot \partial_{t^*} \bar\psi\}+   |\partial_{t^*} \tilde\phi|^2\cdot  \psi  + 2\partial_{t^*}^2 \tilde\phi \, \Re\{\tilde\phi \cdot \bar \psi\} \Big) = - \mathcal F[\tilde \phi],\\[5pt]
(\psi,\partial_t \psi)|_{t=0}=(0,0),\\[5pt]
r\psi|_{r=+\infty}=0
\end{cases}
\end{equation}
(our estimates for \eqref{eq:psi-equation-intro-linear} readily extending to the nonlinear problem \eqref{eq:psi-equation-intro} via a standard bootstrap argument utilizing Sobolev-type estimates to control lower order norms in $\psi$, see e.g.~\cref{lem:Sobolev estimates driving norms}). 
Ignoring, for a moment, the near horizon region (where the red-shift vector field \eqref{Red shift vector field} can be used to obtain non-degenerate energy estimates as first shown in \cite{DR09}), attempting to establish an energy estimate for \eqref{eq:psi-equation-intro-linear} using solely the background Killing vector field $T=\partial_t$ as a multiplier and commutator, i.e.~using only energy norms of the form
\[
\mathcal E^{(\partial_t)}_{\kappa}[\psi](\tau) = \left\| \partial(L^{-\kappa}\partial_t^\kappa r \psi)|_{t=\tau}\right\|^2_{L^2_{y,\theta,\varphi}},
\]
 is doomed to fail; we will now briefly explain why. Noting that the approximate solution $\tilde\phi$ (which appears both in the source term and the coefficients of \eqref{eq:psi-equation-intro-linear}) satisfies with respect to all reasonable Sobolev norms
\[
\| \partial_t^\kappa \tilde\phi\| \sim L^\kappa \|\tilde\phi\|,
\]
it is natural to define the higher order norms by commuting with $L^{-1}\partial_t$, since the higher order terms $L^{-\kappa}\partial_t^\kappa\psi$ would then satisfy equations with coefficients and source term obeying similar bounds as in the $\kappa=0$ case. Using $\partial_t \psi$ as a multiplier for \eqref{eq:psi-equation-intro-linear} and integrating by parts (again, ignoring the near horizon region), one readily arrives at the following energy inequality:
\begin{align}\label{Model false energy estimate intro}
\left\| \partial(r \psi)|_{t=\tau}\right\|^2_{L^2_{y,\theta,\varphi}} \lesssim &\left\| r^3 \mathcal F[\tilde \phi] \right\|^2_{L^1_t L^2_{y,\theta,\varphi}} \\ &
\quad + \int_{0\le t\le \tau}  r^{-6} \Big(|r\tilde\phi| |\partial_t(r\tilde\phi)| | \partial_t(r\psi)|^2 +   \big(|\partial_t(r\tilde\phi)|^2+|r\tilde\phi||\partial_t^2(r\tilde\phi)| \big)|r\psi|| \partial_t(r\psi)|\Big) \, dt dy d\sigma_{\theta,\varphi}.\nonumber
\end{align}
The dominant modes (and, hence, the approximate solution $\tilde\phi$) satisfy the estimate
\begin{equation}\label{Lp norm model solution Intro}
\|r^{-3}\partial^j(r\tilde \phi_k)\|_{L^\infty_t L^p_{y,\theta,\varphi}} \sim \f{L^{j-\f12-\f1p}}{ T_1^{\f12}} \quad \text{for }\, j\in \mathbb N, \, p\in [1,+\infty],
\end{equation}
implying that the term $ r^{-6} |r\tilde\phi| |\partial_t(r\tilde\phi)| | \partial_t(r\psi)|^2$  on the right hand side can be readily absorbed (using Gr\"onwall's inequality) into the left hand side. However, the lower order term $ r^{-6}|\partial_t(r\tilde\phi)|^2 |r\psi|| \partial_t(r\psi)|$ cannot be effectively controlled using the same norms, since, for instance
\begin{align*}
\int_{t=s} r^{-6}|\partial_t(r\tilde\phi)|^2 |r\psi|| \partial_t(r\psi)| \,  dy d\sigma_{\theta,\varphi} 
& 
\lesssim \left\|r^{-6} |\partial_t (r\tilde\phi)|^2\right\|_{L^3_{y,\theta,\varphi}} \|r \psi\|_{L^6_{y,\theta,\varphi}} \|\partial_t (r\psi)\|_{L^2_{y,\theta,\varphi}}\\
&  
\lesssim
\f{L^{\f23}}{T_1}  \|\partial (r\psi)\|^{2}_{L^2_{y,\theta,\varphi}}
\end{align*}
(any other attempt to control this expression using H\"older and Sobolev-type inequalities, i.e.~without exploiting additional cancellations in the integrals involved in the energy estimate, will result in at least the same polynomial loss in $L$).\footnote{Even assuming that $\mathcal E^{(\partial_t)}_{1}[\psi]$ can be bounded in the same way as $\mathcal E^{(\partial_t)}_{0}[\psi]$, higher order Sobolev norms do not mitigate this loss.} As a result, an application of Gr\"onwall's inequality in \eqref{Model false energy estimate intro} would merely yield a prohibitively large bound of the form
\[
\left\| \partial(r \psi)\right\|_{L^\infty_t L^2_{y,\theta,\varphi}} \lesssim e^{c L^{\f23}} \left\| r^3 \mathcal F[\tilde \phi] \right\|_{L^1_t L^2_{y,\theta,\varphi}}
\]
(recall that $|\mathcal F[\tilde\phi]|$ is only polynomially small in $L$; see also the last paragraph of \cref{rmk:Smallness of source term}). What is missing in the $\partial_t$-energy norm is the control of the localization properties of $\psi$ near the asymptotic boundary $\mathcal I$.\footnote{Assuming, for instance, that $\psi$ is localized in the region $|y|,|\theta-\f\pi2|\lesssim L^{-\f12}$, as $\tilde\phi$ does, one could use the Poincare bound $\|r\psi\|_{L^2}\lesssim L^{-\f12}\|\partial(r\psi)\|_{L^2}$ to absorb the corresponding $L$ loss; unfortunately, it is not even clear if $\psi$ is so strongly localized.}

Our argument for obtaining energy estimates for $\psi$ relies, instead, on using commutators which are adapted both to the geometry of the Schwarzschild--AdS exterior and to the symmetry properties of the approximate solution $\tilde\phi$. The  \emph{helical} Killing field
\begin{equation*}
V \doteq (1+L^{-1})\partial_{t} + \partial_\varphi
\end{equation*}
 is timelike for $r>2M$, extends smoothly to the conformal boundary $\mathcal I=\{r=+\infty\}$, and becomes null there as $L\to\infty$. From the point of view of the conformal compactification, $V$ generates a ``helical'' flow tangent to the family of null geodesics along which the high-frequency modes of $\tilde\phi$ are concentrated. A key ingredient is that, for $k\in \mathcal K_{\mathrm{D}}$, the eigenfunctions $E_k$ are localized near the equatorial null geodesics at $\mathcal I$, and the corresponding modes satisfy
\[
V\bigl(e^{-i\varepsilon_k\omega_k t^*} E_k\bigr)
= O(1) \cdot e^{-i\varepsilon_k\omega_k t^*} E_k,
\]
so that differentiation along $V$ only incurs a frequency of size $O(1)$ (rather than $O(L)$ as for $\partial_{t^*}$).\footnote{This improvement only holds for the first differentiation of $\tilde\phi$ in $V$; any additional differentiation will incur a growth in the corresponding estimates by a factor of $L$.} In other words, $V$ is almost tangent to the stably trapped rays at infinity and almost commutes with the building blocks of $\tilde\phi$. As a result, $V$ is expected to commute ``well'' with deformations of the scalar wave operator with coefficients which are functions of $\tilde\phi$  (such as the linear operator on the left hand side of \eqref{eq:psi-equation-intro-linear}).

The energy norm used in \cref{sec:Estimates error term} to control $\psi$ is that provided by the $\partial_t$-energy flux after one commutation with $V$, i.e.~$\mathcal E_0^{(\partial_t)}[V\psi]$ (with higher order norms obtained by further commuting with $L^{-1}\partial_t$), with suitable modifications near the horizon using the red-shift vector field (for the sake of simplicity of the exposition, we will ignore that region for the rest of this section). See \cref{def:Norms} and \cref{lem:Energy estimates Psi}. Near the conformal boundary at $y=0$, this norm controls (up to $\mathcal F[\tilde\phi]$ terms, which arise when using equation \eqref{eq:psi-equation-intro-linear} to exchange certain combinations of $2^{nd}$ order derivatives of $\psi$ in the expression for $\mathcal E_0^{(\partial_t)}[V\psi]$):
\begin{align*}
\mathcal E[\psi](\tau) \doteq \mathcal E^{(\partial_t)}_0[V\psi] \gtrsim \int_{\{t=\tau\}\cap\{y\lesssim 1\}} &  \Bigg[|\bar\partial^2(r\psi)  |^2+\big(w(y, \theta)\big)^2|\partial\bar\partial (r\psi)|^2+\big( w(y, \theta)\big)^4|\partial^2(r\psi)|^2  \\
& \hphantom{ \Bigg[|\bar\partial^2(rf)  |^2+\ }\qquad
+  |\partial V(r\psi)|^2+ |\partial (r\psi)|^2 + \frac{1}{y^2}|r\psi|^2\Bigg]\, \sin\theta dy d\theta d\varphi,
\end{align*}
where, in the above:
\begin{itemize}
\item $\bar\partial$ denotes any derivative in $\{\partial_y, \partial_{\theta}\}$, while $\partial$ denotes any derivative in $\{\partial_t, \f{1}{\sin(\theta)}\partial_{\varphi},\partial_y, \partial_{\theta}\}$,
\item The weight function $w(y,\theta) \sim |y|+|\theta-\f\pi2|+L^{-\f12}$ measures the distance from the equator $\{y=0\}\cap\{\theta=\f\pi2\}$ of  the conformal boundary $\mathcal I \simeq \mathbb R \times \mathbb S^2$.
\end{itemize}
In particular, $\mathcal E[\psi](\tau)$ is a \emph{coercive} norm, which degenerates as $L\rightarrow +\infty$ in the region of the phase space where $\tilde\phi$ is (microlocally) supported.

In order to close an energy estimate for \eqref{eq:psi-equation-intro-linear} with respect to the norm $\mathcal E[\psi]$, we commute \eqref{eq:psi-equation-intro-linear} and multiply the resulting expression with $\partial_t V\psi$. Upon integrating by parts in the region $\{0\le t\le \tau\}$, we obtain the following schematic bound (analogous to \eqref{Model false energy estimate intro}): 
\begin{align}\label{Model V energy estimate intro}
\mathcal E[\psi](\tau)  \lesssim &\left\| r^3 \mathcal F[\tilde \phi] \right\|^2_{L^1_t L^2_{y,\theta,\varphi}} \\ &
\quad +\Bigg|  \int_{0\le t\le \tau}  r^{-6} \Big(r\tilde\phi\cdot \partial_t(r\tilde\phi) \cdot ( \partial_tV(r\psi))^2
+V\big(r\tilde\phi\cdot \partial_t(r\tilde\phi)\big) \cdot  \partial_t(r\psi)\cdot \partial_tV(r\psi)\nonumber \\
& \hphantom{\quad +\Bigg|  \int_{0\le t\le \tau}  r^{-6} \Big(}
+ V\big[(\partial_t(r\tilde\phi))^2 + r\tilde\phi\cdot \partial_t^2(r\tilde\phi)\big]\cdot r\psi\cdot  \partial_tV(r\psi)
\\
& \hphantom{\quad +\Bigg|  \int_{0\le t\le \tau}  r^{-6} \Big(}
+\big((\partial_t(r\tilde\phi))^2 + r\tilde\phi\cdot \partial_t^2(r\tilde\phi)\big) \cdot V(r\psi)\cdot  \partial_tV(r\psi)\Big) \, dt dy d\sigma_{\theta,\varphi}\Bigg|.\nonumber
\end{align}
Using bounds similar to  \eqref{Lp norm model solution Intro} for the model solution $\tilde\phi$, as well as a number of (borderline!) Hardy-type inequalities controlling lower order norms of $\psi$ in terms of $\mathcal E[\psi]$  (based on \cref{lem: WKB asymptotics Airy} in the Appendix), the first three terms in the  integral in the right hand side of \eqref{Model V energy estimate intro} can be estimated by terms of the form
\begin{equation}\label{Log loss bound Intro}
\f{(\log L)^{\f12}}{T_1} \int_0^\tau \mathcal E[\psi](t)\, dt.
\end{equation}
The $(\log L)^{\f12}$ loss seems to be necessary for the Hardy-type inequalities used; however, since the corresponding Gr\"onwall factor $e^{C (\log L)^{\f12}}$ is subpolynomial in $L$, these terms can be absorbed into the left hand side of \eqref{Model V energy estimate intro}.

For the last term in the integral in the right hand side of \eqref{Model V energy estimate intro}, we have to take advantage of the oscillations appearing in the $\tilde\phi$ factors, since estimating this expression in the same way as the first three terms would require $V(r\psi)$ to satisfy the same bounds (in terms of $\mathcal E[\psi]$) as $r\psi$, which is incompatible with the Hardy-type inequalities we establish. To this end, we schematically decompose 
\[
\partial_{t}^2\tilde\phi\cdot\tilde\phi + \partial_{t}\tilde\phi \cdot  \partial_{t}\tilde\phi= L^{-1}\partial_{t^*}\Phi^\sharp + L^{-1}\Phi^\star,
\]
where $\Phi^\star$ satisfies improved estimates, while $\Phi^\sharp$ satisfies similar estimates as $\partial_t \tilde\phi \cdot \partial_t \tilde\phi$; effectively, this relation allows us to ``pull out'' an $L^{-1}\partial_t$ derivative out of the coefficient of $V(r\psi) \cdot \partial_t V(r\psi)$ (the reason that this is possible is that the precise combination of products of the dominant modes appearing in $\partial_{t}^2\tilde\phi\cdot\tilde\phi + \partial_{t}\tilde\phi \cdot  \partial_{t}\tilde\phi$ has a non-trivial oscillatory phase of frequency $\gtrsim L$).  We then implement a delicate integration-by-parts scheme (see, in particular, the scheme employed in estimating \eqref{def B k 1} and \eqref{def B k 2}), in order to end up with an expression that can be estimated by terms of the form \eqref{Log loss bound Intro}. This allows us to finally obtain from \eqref{Model V energy estimate intro}:
\[
\mathcal E[\psi](\tau)  \lesssim \left\| r^3 \mathcal F[\tilde \phi] \right\|^2_{L^1_t L^2_{y,\theta,\varphi}} + \frac{(\log L)^{\f12}}{T_1} \int_0^\tau \mathcal E[\psi](t)\, dt,
\]
which, via an application of Gr\"onwall's lemma, yields the bound
\[
\sup_{\tau\in [0,T_1]}\mathcal E[\psi](\tau) \lesssim L^{\delta_0}  \left\| r^3 \mathcal F[\tilde \phi] \right\|^2_{L^1_t L^2_{y,\theta,\varphi}} \ll \mathcal E[\tilde\phi](0).
\]
  For more details on closing the estimates for the energy norm $\mathcal E[\psi]$, see \cref{sec:A priori energy bounds}.

Applying the same arguments to the higher order energy norms obtained after further commuting with $L^{-1}\partial_t$, utilizing also appropriate bootstrap bounds in terms of $L^\infty$-type norms for $\psi$ (see \cref{subsec: Bootstrap norms}) to deal with the nonlinear terms, we finally obtain that, for $\psi$ solving \eqref{eq:psi-equation-intro}, we have
\[
\|\psi\|_{L^\infty_t([0,T_1]) H^s} \ll \|\tilde\phi\|_{L^\infty_t([0,T_1]) H^s},
\]
i.e.~the error term $\psi$ remains negligible compared to $\tilde\phi$ throughout the whole time interval $[0,T_1]$.

\subsubsection{Completion of the proof of \texorpdfstring{\cref{thm:Main theorem}}{Theorem 1}} 
\label{sec:Completion of the proof of Main theorem}
In \cref{sec:The proof of the main theorem}, we assemble the above ingredients in order to conclude the statement of \cref{thm:Main theorem}. For any frequency scale $L\gg 1$ and any Schwarzschild--AdS mass $M>0$ for which the spectrum of the corresponding linear operator possesses a frequency triad (with the properties outlined in \cref{sec:Hierarchy of parameters}) satisfying the resonance conditions \eqref{eq:resonant-frequency-relation-intro} and \eqref{Non resonant condition time frequencies intro}, the analysis of \cref{sec:Dominant modes,sec:Non dominant modes,sec:Estimates error term} applies and yields the growth of the $H^s$-energy of the solution given by the ansatz \eqref{Ansatz phi Intro} on the time interval $[0,T_1]$. In \cref{sec:Ansatz}, we show that these values of the Schwarzschild--AdS mass are $L^{\delta_0}$-dense in $(0,+\infty)$. In particular, the statement of \cref{thm:Main theorem} holds for a dense set of masses $M>0$.

 The openness of the same set follows from a soft \emph{Cauchy stability} argument in the mass parameter: For fixed initial data and fixed $T_1$, the initial-boundary value problem \eqref{Initial Boundary Value Problem Phi} depends continuously on the mass $M$ in the space $C^0([0,T_1], H^s(\Sigma^*_{t^*}) \times H^{s-1}(\Sigma^\ast_{t^\ast}))$; thus, if for some $M_0$, the conclusion of \cref{thm:Main theorem} holds, then it also holds for all $M$ sufficiently close to $M_0$. Consequently, the conclusion of \cref{thm:Main theorem} holds for an open and dense set of mass parameters $M>0$.

\subsection{Norm inflation on the hemisphere: The proof of \texorpdfstring{\cref{thm:Main theorem-S3}}{Theorem~2}}\label{sec:Proof of second theorem}

In this section, we will outline how the proof of \cref{thm:Main theorem} can be extended to provide a norm inflation statement for nonlinear equations of the type \eqref{Initial Boundary Value Problem Phi} on a broader class of spacetimes exhibiting stable trapping; a special application of this extension will be  \cref{thm:Main theorem-S3}.
First, let us see how to recast \cref{thm:Main theorem} as a statement on the spatially compactified domain $\mathbb R\times \mathbb S^3_+$: By modifying the coordinate function $y=y(r)$ (by the relation \eqref{The y coordinate}) in the near horizon region so that it becomes a smooth function up to $r=r_H$, and conformally rescaling the Schwarzschild--AdS metric $g_M$ by $\tilde g_M = \sin^2 y \cdot g_M$, we can readily conformally identify  $(\mathcal M_{\mathrm{ext}}^{(M)},g_M)$ with the spacetime $\left(\mathbb R\times (\mathbb S^3_+\setminus \mathcal B), \tilde g_M\right)$ ($\mathcal B$ being a closed ball in $\mathbb S^3$ centered at the north pole), with the metric $\tilde g_M$ satisfying near the boundary $\{y=0\} = \mathbb R \times \partial \mathbb S^3_+$:
\begin{equation}\label{Asymptotics metric in y Intro}
\tilde g_M =  -dt^2 + g_{\mathbb S^3_+} +M y^3 h,
\end{equation}
where $g_{\mathbb S^3_+} = dy^2 + \cos^2(y) \big(d\theta^2 +\sin^2\theta d\varphi^2\big)$ is the round metric on $\mathbb S^3_+$ and
\[
h =  3\big(1+O(y) \big) dt^2  -3\big( 1 +O(y) \big) dy^2 + \big( 1 +O(y) \big) \big(d\theta^2 + \sin^2\theta\, d\varphi^2 \big).
\]
In particular, the metric $\tilde g_M$ is smooth up to the boundary $\mathcal I = \{y=0\}$ and, since $\tilde g_M$ is merely a perturbation of the cylinder metric $ -dt^2 + g_{\mathbb S^3_+}$ of size $O(y^3)$ in a neighborhood of $\{y=0\}$, the hypersurface $\mathcal I$ remains totally geodesic, spanned by null geodesics which are \emph{stably trapped}.
The conformally coupled wave operator $\square_{g_M}+2$ on $\left(\mathcal M^{(M)}_{\mathrm{ext}}, g_M\right)$ is related to the corresponding operator for $\tilde g_M$ by the relation (setting $\phi' = \f{\phi}{\sin y}$):
\[
\square_{g_M} \phi+2\phi = \sin^3 y \left(\square_{g_{\tilde M}}\phi' + \big(1+O(My)\big)\phi'\right)
\]
(see also \cref{sec:Expressions for wave equation} for a more convenient expression in terms of the variable $r\phi$).
Through the above conformal identification, one can restate \cref{thm:Main theorem} in the compactified picture, i.e.~as a weak turbulence result for solutions on the one parameter family of Lorentzian spacetimes  $\left(\mathbb R\times (\mathbb S^3_+\setminus B), \tilde g_M\right)$  of the following initial--boundary value problem (equivalent to \eqref{Initial Boundary Value Problem Phi}):
        \begin{equation}\label{Initial Boundary Value Problem Phi Rescaled}
                \begin{cases}
                        \square_{g_{\tilde M}}\phi' + \big(1+O(My)\big)\phi' = \big(y^6+O(y^7)\big)\left(|\partial_t \phi'|^2 \phi'+|\phi'|^2 \partial_t^2 \phi'\right) ,\\[5pt]
                        (\phi', \partial_{t} \phi')|_{t=0} = (f'_0, f'_1), \\
                        \phi'|_{y=0} =0.
                \end{cases}
        \end{equation}
From this perspective, our proof of \cref{thm:Main theorem} can be immediately generalized to apply to an initial-boundary value problem of the form \eqref{Initial Boundary Value Problem Phi Rescaled} with the left hand side being replaced by the wave operator of any other $\mathbb S^2$-symmetric 1-parameter family of stationary Lorentzian  metrics on $\mathbb R\times \mathbb S^3_+$ (parametrized by $M>0$) having the following properties:\footnote{Let us also note that our result also immediately generalizes to cover the case when additional smooth (up to $y=0$) lower order linear terms are included in \eqref{Initial Boundary Value Problem Phi Rescaled}.
}
\begin{itemize}
\item Near the boundary $\{y=0\}$, the metric is a perturbation of $g_{\mathbb R\times \mathbb S^3_+}$ of size $\sim My^3$
\item A global energy boundedness statement is available for the corresponding scalar wave equation via vector field multipliers which, near the boundary, resemble a timelike Killing vector field (not necessarily $\partial_t$, see \cref{rmk:Kerr}).
\end{itemize}
An example of such a family of spacetimes is $\left(\mathbb R\times \mathbb S^3_+, -dt^2 +h^M_{\mathbb S^3_+}\right)$, i.e.~the ones appearing in the statement of \cref{thm:Main theorem-S3} (since this is the Lorentzian metric giving rise to the wave operator $-\partial_t^2 + \Delta_{h^M_{\mathbb S^3_+}}$ in \eqref{Equation for corollary problem}).
\medskip
\noindent \emph{Outline of the proof of \cref{thm:Main theorem-S3}.} The proof of  \cref{thm:Main theorem-S3} follows by repeating exactly the same steps as in the proof of \cref{thm:Main theorem}, with only minor modifications, stemming from the absence of an event horizon for $\left(\mathbb R\times \mathbb S^3_+, -dt^2 +h^M_{\mathbb S^3_+}\right)$. In particular, in this case there is no need to place an inner mirror at some finite radius in order to obtain a system admitting normal mode solutions, as we did in \eqref{Linear wave equation with mirror Introduction} to model the stably trapped quasinormal modes for \eqref{Linear wave equation Introduction}. Instead, the boundary value problem
\[
\begin{cases}
-\partial_t^2 \phi +\Delta_{h^M_{\mathbb S^3_+}}\phi =0,\\
\phi|_{y=0} =0
\end{cases}
\]
admits a decomposition into a complete system of normal modes: The Dirichlet eigenfunctions $\tilde E_k :\mathbb S^3_+\rightarrow \mathbb \mathbb C $ of $\Delta_{h^M_{\mathbb S^3_+}}$,
\begin{equation}\label{Dirichlet eigenfunction Theorem 2}
\begin{cases}
\Delta_{h^M_{\mathbb S^3_+}} \tilde E_k +\omega^2_{k} \tilde E_k=0,\\[5pt]
\tilde E_k|_{y=0}=0,
\end{cases}
\end{equation}
form a complete orthonormal basis of $L^2\left(\mathrm{dvol}_{h^M_{\mathbb S^3}}\right)$.
In view of the $\mathbb S^2$-symmetry of $h^M_{\mathbb S^3_+}$, the eigenvalue problem \eqref{Dirichlet eigenfunction Theorem 2}  separates (as in the case of \eqref{Eigenfunctions Intro}): Denoting by $(\rho,\theta,\varphi)$ the standard polar coordinates on $\mathbb S^3_+$ and defining the function $y=y(\rho)$ so that it satisfies
\[
y(\rho) =\f\pi2-\rho \,\text{ for }\, \rho\ge \f\pi4, \quad y(0)=\frac{3\pi}{8}, \quad y'<0,
\]
(so, in particular, $y=0$ on $\partial \mathbb S^3$), we have
\begin{equation}\label{Eigenfunctions theorem 2}
\tilde E_k(y,\theta,\varphi) = \f{1}{\cos(y)}R_{n,\ell}(y) Y_{\ell,m}(\theta, \varphi)
\end{equation}
with $k=(n,\ell,m)\in \mathbb N\times \mathbb N_{\ge |m|}\times \mathbb Z$.
The problem \eqref{Dirichlet eigenfunction Theorem 2} then becomes a radial Sturm--Liouville problem for $R_{n,\ell}$ analogous to \eqref{Radial model boundary value problem introduction}:
\begin{equation}\label{Radial model boundary value problem Theorem 2 introduction}
\begin{cases}
-\f{d^2}{dy^2} R_{n,\ell}(y) + O(y^2) \f{d}{dy} R_{n,\ell}(y)+ \tilde V_\ell(y) R_{n,\ell}(y)={\omega_{n,\ell}}^2 R_{n,\ell}(y),\\
R_{n,\ell}|_{y=0} = 0,
\end{cases}
\end{equation}
where $\omega_k=\omega_{n,\ell} $ and
\begin{equation}\label{Tilde V intro}
\tilde V_\ell(y) = \Big(1 +y^2-2My^3 +O(y^4) \Big)\Big(\ell(\ell+1)+O(1)\Big).
\end{equation}
In particular, to top order in $\ell$ and up to terms of order $y^3$, $\tilde V_\ell$ coincides with $V_\ell$. As a result, the potential $\tilde V_\ell$ has a strict local minimum at $y=0$ when $\ell \gg 1$. We will fix the discrete parameter $n$ (at least in the regime $n\ll \ell$) by the condition that, for $1\le n \ll \ell$, the sequence of eigenvalues $\omega_{n,\ell}$ is increasing in $n$ and $\omega_{1,\ell}$ is the minimum eigenvalue such that $\omega^2_{1,\ell}>\tilde V_\ell(0)$.\footnote{Unlike the case of the potential , in this case it is not guaranteed that $\tilde V_\ell$ has a \emph{global} minimum at $y=0$; hence, there can exist modes at a lower energy level, which are necessarily supported away from $y=0$ when $\ell \gg 1$.} As a result, for $n\ll \ell$, the corresponding eigenfunctions $R_{n,\ell}$ will be localized in the region $\{y\lesssim n^{\f12}\ell^{-\f12}\}$ and $n$ will be the radial overtone (measuring the number of oscillations of $R_{n,\ell}$). These will be the class of modes that will be of interest to us and (as we will explain in a moment) their spectral properties are (to top order in the frequency parameters) the same as for the corresponding modes for \eqref{Radial model boundary value problem introduction}.
In analogy with \eqref{Ansatz phi Intro}, let us employ the following ansatz for the proof of \cref{thm:Main theorem-S3}:
\begin{equation}\label{Ansatz phi Theorem 2 Intro}
\phi(t,y,\theta, \varphi) =  \sum_{k\in \mathcal K_\mathrm{D}} \tilde \phi_k(t,y,\theta, \varphi) + \sum_{k\in \mathcal K_\mathrm{ND}} \tilde \phi_k(t,y,\theta, \varphi) + \psi(t,y,\theta,\varphi),
\end{equation}
where the mode solutions $\tilde\phi_k$ are defined similarly as before by
\[
\tilde\phi_k(t,r,\theta, \varphi) = a_k(t) \tilde E_k (y,\theta, \varphi),
\]
while the amplitudes $a_k$ are defined (as before) as solutions of the mode system \eqref{Mode system Intro} (with $\tilde E_k$ in place of $E_k$ in the corresponding coefficients) and the sets of frequency parameters $\mathcal K_{\mathrm{D}}$ and $\mathcal K_{\mathrm{ND}}$ are defined as in \cref{sec:Spectrum intro}. Notice that, compared to \eqref{Ansatz phi Intro}, there is no need to multiply the normal modes with a cut-off function in the near region, since we have disposed of the mirror for this problem.\footnote{For the same reason, in this setting we could choose $\mathcal K_{\mathrm{D}}\cup \mathcal K_{\mathrm{ND}}= \mathbb N \times \mathbb N_{\ge |m|}\times \mathbb Z$, i.e.~the permissible set $\mathcal K$ to contain all the frequency parameters.}
Given the above ansatz and a frequency scale $L\gg 1$, the proof of  \cref{thm:Main theorem-S3} follows by repeating exactly the same steps as for the proof of  \cref{thm:Main theorem}:
\begin{itemize}
\item In view of the fact that the radial potential $\tilde V_\ell(y)$ in  \eqref{Radial model boundary value problem Theorem 2 introduction} coincides with $V_\ell(r(y))$ in \eqref{Radial model boundary value problem introduction} to top order in $\ell(\ell+1)$, up to (and including) coefficients of size $O(My^3)\ell(\ell+1)$, the two boundary value problems enjoy the same spectral properties (to top two orders in the frequency parameter $\ell$) in the phase space regime where $n\ll \ell$ (see our comment on the normalization of the parameter $n$ below \eqref{Tilde V intro}). In particular, the crucial expansion \eqref{Taylor expansion omega intro} for $\omega_{n,\ell}$ also holds in this case for $n\lesssim \ell^{\delta_0}$, and the corresponding eigenfunctions $E_k$ satisfy the same estimates as in the case of \eqref{Eigenfunctions Intro}. As a result, the spectral coefficients $\langle \tilde E_{k_1} \bar{ \tilde E}_{k_2} \tilde E_{k_3} \bar{\tilde E}_k \rangle $ for $k_1, k_2, k_3\in \mathcal K_D$ and $k\in \mathcal K$ in the mode system are the same (to top two orders in $L$) as for the system \eqref{Mode system Intro}.
\item The validity of \eqref{Taylor expansion omega intro} implies that, by selecting the dominant frequency parameters $\mathcal K_D$ exactly as in the case of \cref{thm:Main theorem}, we can ensure that the dominant modes $\{\tilde\phi_k\}_{k\in \mathcal K_D}$ satisfy the same resonant conditions as the ones described in \cref{sec:Spectrum intro} for the same values of the parameter $M>0$.  
\item The $3\times 3$ decoupled dominant system for $\{\tilde\phi_k\}_{k\in \mathcal K_D}$ has the same coefficients (to top two orders in $L$) as the corresponding system for \cref{thm:Main theorem}, therefore the results of \cref{sec:dominant-modes-intro} carry over without any change.
\item Similarly, the system of non-dominant modes satisfies exactly the same estimates as for the corresponding system for \cref{thm:Main theorem}.
\item The analysis of the error term $\psi$ can be performed in exactly the same way as for \cref{thm:Main theorem}, with some simplifications coming from the fact that, due to the absence of an event horizon in this case, there is no need to introduce the red-shift vector field $Y$, while the helical vector field $V$ remains timelike on the whole of $\mathbb R\times \mathbb S^3_+$ (therefore, one only needs to commute the equation for $\psi$ with $V$ and $L^{-1}\partial_t$ and use the energy estimate provided by the $\partial_t$ multiplier).
\end{itemize}
\begin{remark}\label{rmk:Kerr}
In a similar fashion, our proof of \cref{thm:Main theorem} can also be applied to the case of a Kerr--AdS exterior  spacetime $(\mathcal M^{(a,M)}_{\mathrm{ext}}, g_{a,M})$ with a fixed small value of $a$. Even though these spacetimes are not spherically symmetric, the wave operator $\square_{g_{a,M}}$ separates, and, for modes with $\ell=|m|\gg 1$,  the corresponding radial ODE has the same asymptotic structure near infinity as in our case, thus allowing one to repeat the same estimates as the ones we employ for the proof of \cref{thm:Main theorem}. In that case, the proof of energy estimates for the error term $\psi$ follows by using the Hawking vector field rather than the $\partial_t$ vector field.
\end{remark}

\subsection*{Acknowledgements}
The authors would like to thank Mihalis Dafermos, Gustav Holzegel, Igor Rodnianski and Jacques Smulevici for many insightful conversations. C.K.~also thanks Gigliola Staffilani for helpful discussions. 

\section{Preliminaries}

\subsection{The geometry of the Schwarzschild--AdS exterior spacetime}

The Schwarzschild--AdS metric $g_M$ of mass parameter $M>0$ takes the following form in the standard polar coordinates on $\mathcal M^{(M)}_\mathrm{ext} = \mathbb R_t \times (r_+, +\infty)_r \times \mathbb S^2_{\theta, \varphi}$:
\[
g_M = -\Big(1-\f{2M}{r}+r^2\Big) dt^2 + \Big(1-\f{2M}{r}+r^2\Big)^{-1} dr^2 + r^2 \big(d\theta^2 + \sin^2\theta d\varphi^2 \big),
\]
where $r_+>0$ is the unique (real) solution of 
\[
1-\f{2M}{r_+}+r^2_+=0,
\]
i.e.
\[
r_+(M) = \sqrt[3]{M+\sqrt{M^2+\f1{27}}} - \sqrt[3]{\sqrt{M^2+\f1{27}}-M}.
\]
Note that $(\mathcal M^{(M)}_\mathrm{ext}, g_M)$ is \emph{spherically symmetric} and \emph{static}, with corresponding Killing vector field $T = \partial_t$.

\begin{remark} Throughout this work, we will frequently drop the index $M$ from the exterior domain $\mathcal M^{(M)}_\mathrm{ext}$ and metric $g_M$, denoting them simply as $\mathcal M_\mathrm{ext}$ and  $g$. 
\end{remark} 

\subsubsection{The asymptotically AdS region}
The region $\{r\gg 1\} \subset \mathcal M_\mathrm{ext}$ is known as the \emph{asymptotically AdS} region of $(\mathcal M_\mathrm{ext}, g_M)$. Let us introduce the function $y=y(r)$ defined by
\begin{equation}\label{The y coordinate}
y(r) \doteq \int_r^{+\infty} \f{1}{1-\f{2M}{\bar r}+\bar r^2} \, d\bar r.
\end{equation}
Note that $y|_{r=r_+}=+\infty$ and $y|_{r=+\infty}=0$; in particular, near $r=+\infty$,
\[
y(r) = \f1r + O_M(r^{-3}).
\]
We also note that for $r(y,M)$ defined as the inverse of $y(r,M)$, we have
\begin{equation}\label{eq:AsymptoticspartialMr(y,M)}
\frac{\partial r(y,M)}{\partial M} =  \left(1-\frac{2M}{r} + r^2\right) \int_{r}^{+\infty} \frac{2}{\bar r (1-\frac{2M}{\bar r}+\bar r^2)^2} \, d\bar r = O(y^2)\quad \text{and} \quad \frac{\partial^2 r(y,M)}{\partial M^2} = O(y^5) 
\end{equation}
as $ y\rightarrow 0.$
It can be readily verified that the conformal metric $\tilde g_M \doteq y^2 \cdot g_M$ can be smoothly extended across $\{y=0\}$; this can be seen by noting that $g_M$ takes the following form in the $(t,y,\theta,\varphi)$ coordinate system:
\begin{equation}\label{Asymptotics metric in y}
g_M =  \f{1}{\sin^2y}\Bigg(  - \big(1+ O(My^3) \big) dt^2 + \big( 1 + O(M y^3) \big) dy^2 + \big( \cos^2(y) + O(M y^3) \big) \big(d\theta^2 + \sin^2\theta\, d\varphi^2 \big)  \Bigg).
\end{equation}
The conformal boundary
\begin{equation}\label{Definition conformal boundary}
\mathcal I \doteq \{y=0\}
\end{equation}
is frequently called the \emph{conformal infinity} of $(\mathcal M_\mathrm{ext}, g_M)$, and has the conformal structure of a timelike hypersurface diffeomorphic to $\mathbb R \times \mathbb S^2$. Note also that $y$ acts as a boundary defining function for $\mathcal I$. 

\begin{remark} Through the coordinate system $(t^*, y, \theta, \varphi)$, the region $\{0<y\lesssim 1\}$ of $\mathcal M_\mathrm{ext}$ can be identified with a time-invariant one-sided neighborhood of the equator $\mathbb R \times \partial \mathbb S^3_+ \simeq \mathbb R \times \mathbb S^2$ in the half-cylinder $\mathbb R \times \mathbb S^3_+$ via the map $\mathcal F: (t^*, y, \theta, \varphi)\rightarrow (\bar t, \bar r, \bar\theta, \bar\varphi)= (t^*, \f\pi2-y, \theta,\varphi)$, where $(\bar r, \bar\theta, \bar\varphi)$ denotes the standard polar coordinate system on $\mathbb S^3$ centered at the north pole (with $\bar r=\f\pi2$ corresponding to $\partial\mathbb S^3_+$). This map identifies $\mathcal I$ with $\mathbb R \times \partial \mathbb S^3_+$ and is asymptotically \emph{conformal} near $\mathcal I$ with respect to the standard metric 
\[
g_E \doteq -d\bar t^2 + g_{\mathbb S^3} = -d\bar t^2 + d\bar r^2 + \sin^2 \bar r \big( d\bar\theta^2 + \sin^2 \bar\theta d\bar\varphi^2\big)
\] 
on $\mathbb R \times \mathbb S^3_+$: The relation \eqref{Asymptotics metric in y} implies that
\[
\tilde g_M \doteq \sin^2y \cdot g_M = \mathcal F^* g_E + O(M y^3).
\]  
In particular, with respect to the conformal metric $\tilde g_M$ on $\mathcal M_\mathrm{ext}$, the boundary $\mathcal I$ is \emph{totally geodesic} (as is the boundary $\mathbb R\times \partial \mathbb S^3_+$ of $(\mathbb R\times \mathbb S^3_+, g_E)$). This fact gives rise to stably trapped dynamics for solutions to wave-type equations on $(\mathcal M_\mathrm{ext}, g_M)$ with reflecting boundary conditions on $\mathcal I$.
\end{remark}

\subsubsection{The event horizon.} The spacetime $(\mathcal M_\mathrm{ext}, g_M)$ can be isometrically embedded into a larger smooth spacetime, which extends $\mathcal M_\mathrm{ext}$ beyond $\{r=r_+\}$. The ``largest'' such extension $(\mathcal M, g_M)$ is known as the \emph{maximally extended} Schwarzschild--AdS spacetime: this is a black hole spacetime, with $\mathcal M_\mathrm{ext}$ being one connected component of the black hole exterior. 

In this paper, we will study functions and tensors on $(\mathcal M_\mathrm{ext}, g_M)$ which are smooth up to the boundary $\partial \mathcal M_\mathrm{ext}$ of $\mathcal M_\mathrm{ext}$ inside $\mathcal M$. In order to address questions of regularity near $\partial \mathcal M_\mathrm{ext}$, we will have to work in a coordinate system on $\mathcal M_\mathrm{ext}$ which extends smoothly beyond $\partial \mathcal M_\mathrm{ext}$. An example of such a coordinate system is the double null system $(U,V, \theta, \varphi)$ with 
\[
U=-e^{-(t+y)}, \quad V=e^{t-y} 
\]
(with $y=y(r)$ defined by \eqref{The y coordinate}). In these coordinates, $g_M$ takes the form
\[
g_M =  \f{\Big(1-\f{2M}{r(U,V)}+\big(r(U,V)\big)^2\Big)}{U \cdot V} dU dV + \big(r(U,V)\big)^2 \big(d\theta^2 + \sin^2\theta d\varphi^2 \big).
\]
Note that the coefficients of $g_M$ in the above coordinates can be smoothly extended across $\{U=0\}$ and $\{V=0\}$; in this extension, $\mathcal M_\mathrm{ext}$ can be identified with (a connected component of) the domain $\{U<0, V>0\}$, while $\partial \mathcal M_\mathrm{ext}$ (on which $r=r_+$) can be written as the union of two smooth null hypersurfaces (with boundary)
\[
\partial \mathcal M_\mathrm{ext} = \mathcal H^+ \cup \mathcal H^-.
\]
In the above, $\mathcal H^+ = \{U=0\} \cap \partial \mathcal M_\mathrm{ext}$ and $\mathcal H^-= \{V=0\} \cap \partial \mathcal M_\mathrm{ext}$ called the \emph{future} and \emph{past event horizons}, respectively. The $2$-sphere $\mathcal H^+\cap \mathcal H^-$ is known as the \emph{bifurcation sphere}. We will also denote with $\overline{\mathcal M}_\mathrm{ext}$ the topological closure of $\mathcal M_\mathrm{ext}$ in this extended spacetime, i.e.
\[
\overline{\mathcal M}_\mathrm{ext} \doteq \mathcal M_\mathrm{ext}\cup \mathcal H^+ \cup \mathcal H^-.
\]

\subsection{The \texorpdfstring{$t^*$}{t*} coordinate on \texorpdfstring{$\mathcal M_\mathrm{ext}$}{M\_ext}}\label{sec:Coordinate systems}
It will be convenient for us to work in coordinate systems that are well adapted to the AdS asymptotics in the region $\{r\gg 1\}\subset \mathcal M_\mathrm{ext}$ and at the same time can be extended smoothly across the future event horizon $\mathcal H^+$. In particular, we will mainly work in the coordinate systems $(t^*, r, \theta, \varphi)$ and $(t^*, y, \theta, \varphi)$, where $t^*:\mathcal M_\mathrm{ext} \rightarrow \mathbb R$ is a modified time function that we will now proceed to define.

Let us fix a smooth function $h:[r_+, +\infty) \rightarrow [0,1]$ satisfying
\begin{equation}\label{First property h}
h(r) = 1 \quad \text{for } r_+ < r \le r_+ + \delta_1 \quad \text{and} \quad h(r) \equiv 0 \quad \text{for } r\ge r_++2\delta_1 
\end{equation}
for some 
\[
0<\delta_1(M) <3M-r_+,
\]
as well as
\begin{equation}\label{Second property h}
\inf_{r\in (r_+,+\infty)} \Bigg[ \Big(1-\f{2M}{r}+r^2\Big)^{-1}\big(1-h^2(r)\big)+\Big(2-\big(1-\f{2M}{r}+r^2\big)\Big) h^2(r)  \Bigg] >0
\end{equation}
(note that this is possible provided $\delta_1$ is chosen sufficiently small in terms of $M$). 

\begin{definition}\label{def:Modified time function}
We will define the modified time function $t^*:\mathcal M_\mathrm{ext}\rightarrow \mathbb R$ by the relation
\[
t^* \doteq t  - \int^{+\infty}_r  \Big[ \Big(1-\f{2M}{\bar r}+\bar r^2\Big)^{-1}-1\Big]\cdot h(\bar r) \, d\bar r.
\]
For any $\tau\in \mathbb R$, we will denote by $\Sigma^*_\tau$ the level set $\{t^*=\tau\}$.
\end{definition}
Note that $t^*:\mathcal M_\mathrm{ext}\rightarrow \mathbb R$ can be extended smoothly across $\mathcal H^+$ and satisfies the following properties:
\begin{itemize}
\item The hypersurfaces $\Sigma^*_\tau$ intersect $\mathcal H^+$ transversally and are \emph{smooth} and \emph{spacelike} uniformly up to $\mathcal H^+$ (the latter following from the condition \eqref{Second property h} on $h$).
\item In the region $\{r\ge 3M\}$, we have $\Sigma^*_\tau = \{t=\tau\}$.
\end{itemize}

The coordinate system $(t^*, r, \theta, \varphi)$ can be smoothly extended across $\mathcal H^+$ (but not across $\mathcal H^-$); in these coordinates, the Schwarzschild--AdS metric takes the form
\begin{align}\label{Expression metric t star coordinate}
g = -  \Big(1-\frac{2M}{r}+r^2\Big) (dt^*)^2 +2\Big(\frac{2M}{r}-r^2\Big) h(r) dt^* dr+ \tilde h(r) dr^2 
 + r^2 \big( d\theta^2 + \sin^2\theta d\varphi^2\big),
\end{align}
where
\[
 \tilde h(r) \doteq \Big(1-\f{2M}{r}+r^2\Big)^{-1}\big(1-h^2(r)\big)+\Big(2-\big(1-\f{2M}{r}+r^2\big)\Big) h^2(r) >0.
\]
We can then readily compute that
\[
g_{t^* t^*} g_{rr} - g_{t^*r}^2 = -1
\]
and
\begin{equation}\label{Inverse metric components}
g^{t^* t^*} =  - g_{rr}, \quad g^{rr} = -g_{t^* t^*}, \quad g^{t^* r} = g_{t^* r}.
\end{equation}
Note that, along $\mathcal H^+$, we have
\[
g_{t^* t^*}|_{r=r_+}=0, \quad g_{t^*r}|_{r=r_+} =1 \quad \text{and} \quad g_{rr}|_{r=r_+}=2>0.
\]

\begin{remark}
In the region $\{r\ge 3M\}$, the coordinate systems $(t^*, r, \theta, \varphi)$ and $(t, r, \theta, \varphi)$ coincide. 
When working near the conformal boundary $\mathcal I = \{r=+\infty\}$, we will sometimes switch to the coordinate system $(t^*, y, \theta, \varphi)$, where $y=y(r)$ is the function defined by \eqref{The y coordinate}. Recall that in the asymptotically AdS region, the Schwarzschild--AdS metric $g$ has the asymptotic form \eqref{Asymptotics metric in y}. It can be, thus, readily verified, that the volume form $\dvol_g$ and the induced volume form $\dvol_{\Sigma^*_\tau}$ have the asymptotic forms
\begin{align*}
\dvol_g & = \frac{1}{y^4} \big(1+O(y^2)\big) \, \sin\theta d\theta d\varphi dy dt^*\\
\dvol_{\Sigma^*_\tau} & = \frac{1}{y^3} \big(1+O(y^2)\big)\, \sin\theta  d\theta d\varphi dy.
\end{align*}
\end{remark}

\begin{remark}
Note that in both the coordinate systems $(t^*, r, \theta, \varphi)$ and $(t^*, y, \theta, \varphi)$, the coordinate vector field $\partial_{t^*}$ coincides with the Killing vector field $T$ (which is timelike away from $\mathcal H^+$ and null along $\mathcal H^+$).    
\end{remark}

 \subsection{The reflecting sphere radius \texorpdfstring{$r_\mathrm{mirror}$}{rmirror}}
We will introduce the parameters $y_\mathrm{mirror}$ and $r_\mathrm{mirror}$ defined as follows:
\begin{definition}\label{def:Mirror radius}
For any $M_0>0$, we will define $y_\mathrm{mirror}=y_\mathrm{mirror}(M_0)\in (0, \f\pi4]$ by the relation
\begin{equation}\label{Definition y mirror}
y_\mathrm{mirror} \doteq \min\big\{  y_*(M_0) , \, \f\pi4\big\},
\end{equation}
where 
\begin{align*}
y_*(M_0) \doteq \inf_{\bar M\in (0,M_0]} \Bigg\{ & \f{r_+(\bar M)}{2(1+3r_+^2(\bar M))}\log\Big(\f{9\bar M^2+ 3\bar M r_+(\bar M)  + r_+^2(\bar M)+1}{\big(3\bar M-r_+(\bar M)\big)^2}\Big) \\ & + \f{3r_+^2(\bar M)+2}{2(3r_+^2(\bar M)+1)\sqrt{3r_+^2(\bar M)+4}}\Big( \pi - 2\arctan\big(\f{6\bar M+r_+(\bar M)}{\sqrt{3r_+^2(\bar M)+4}}\big)\Big) \Bigg \}
\end{align*}
and where $r_+(\bar M)\in (0, 2\bar M)$ is the real root of the cubic equation $r^3+r-2\bar M=0$.

For any Schwarzschild--AdS mass parameter $M\in [0,M_0]$, we will define $r_\mathrm{mirror} = r_\mathrm{mirror}(M;M_0)$ by the relation
\begin{equation}\label{Definition mirror value}
r_\mathrm{mirror} \doteq r(y_\mathrm{mirror})
\end{equation}
(with $r(y)$ interpreted as the inverse of the function $y(r)$ defined by \eqref{The y coordinate}). 
\end{definition}

\begin{remark}
 Using the fact that
\begin{align*}
y(r) = \int_r^{+\infty} \f{1}{1+\bar r^2-\f{2M}{\bar r}} \, d \bar r & = \int_r^{+\infty} \f{\bar r}{(\bar r-r_+)(\bar r^2+r_+ \bar r + r_+^2+1)} \, d \bar r \\
& = \f{r_+}{2(1+3r_+^2)}\log\Big(\f{r^2+ r_+ r + r_+^2+1}{(r-r_+)^2}\Big) \\ & \hphantom{=}+ \f{3r_+^2+2}{2(3r_+^2+1)\sqrt{3r_+^2+4}}\Big( \pi - 2\arctan\big(\f{2r+r_+}{\sqrt{3r_+^2+4}}\big)\Big),
\end{align*}
we can readily see that 
\[
 r_\mathrm{mirror}\ge  \max\{3M_0,1\},
\] 
i.e.~the region $\{r\ge r_\mathrm{mirror}\}$ does not include the photon sphere of $(\mathcal M_\mathrm{ext}, g_M)$ for any $M\in (0, M_0]$. In particular, the function $1+\f{1}{r^2}-\f{2M}{r^3}$ is strictly decreasing in $r$ for $r\ge r_\mathrm{mirror}$.

We will later use the timelike hypersurface $\{r=r_\mathrm{mirror}\}$ to place reflecting boundary conditions in an auxiliary construction involved in the proof of \cref{thm:Main theorem}; see  \cref{sec:Linear wave equation mirror}. It will be important for us that, for fixed $M_0$, the value $y_\mathrm{mirror}$ is constant with respect to the mass parameter $M\in [0, M_0]$. 
\end{remark}

\subsection{The conformally coupled wave equation}
The inhomogeneous \emph{conformally coupled} wave equation on the Schwarzschild--AdS exterior spacetime $(\mathcal M_\mathrm{ext}, g)$ takes the form
\begin{equation}\label{Conformally coupled wave equation}
\square_g f + 2 f = F,
\end{equation}
 where, as usual, $\square_g = \f{1}{\sqrt{-\det g}} \partial_{\alpha}\big(\sqrt{-\det g} g^{\alpha\beta}   \partial_\beta\big)$ denotes the wave operator associated to $g$. In view of the presence of the timelike conformal boundary $\mathcal I$, the appropriate framework to study solutions to \eqref{Conformally coupled wave equation} is provided by the \emph{initial--boundary value problem}.

\subsubsection{The initial--boundary value problem}
\label{sec:Expressions for wave equation}
 In the region $0<y\le y_\mathrm{mirror}$ (where $t^*=t$), the conformal wave operator $\square_g +2$ takes the following form:
\begin{equation}\label{Wave operator}
r^3 \big(\square_g f + 2 f\big) =  -\f{r^2}{1-\frac{2M}{r}+r^2} \partial_{t^*}^2(rf) + \f{r^2}{1-\frac{2M}{r}+r^2}  \partial_y^2(rf) +  \Delta_{\mathbb S^2}(rf)-\frac{2M}{r} (rf).
\end{equation}
Therefore, near $y=0$, the conformal wave operator behaves asymptotically as follows:
\begin{equation*}
\f1{y^3} \big(\square_g f + 2 f\big) = \big(1+O(y^2)\big)\Big(-\partial_{t^*}^2(rf) + \partial_y^2 (rf) \Big) + \big(1+O(y^2)\big) \Delta_{\mathbb S^2}(rf) + O(y) (rf).
\end{equation*}
 In particular, equation \eqref{Conformally coupled wave equation}, when expressed in the $(t^*, y, \theta, \varphi)$ coordinate system, turns into a wave equation for $r f$ with \emph{bounded} (in fact, smooth) coefficients uniformly up to $\mathcal I = \{y=0\}$(provided the source term $\f1y F$ is bounded and smooth up to $y=0$). Thus, any choice of boundary conditions for $r f$ at $y=0$ in the $(t^*, y, \theta, \varphi)$ coordinate system that is known to yield a well-posed initial-boundary value problem on $(\mathbb R \times \mathbb S^3_+, g_E)$ also leads to a well-posed initial-boundary value problem for  \eqref{Wave operator} on $(\mathcal M_\mathrm{ext}, g)$; for instance, Dirichlet boundary conditions correspond to prescribing $rf|_{y=0}$, while Neumann conditions correspond to prescribing $\partial_y (rf)|_{y=0}$. 
 
 In this work, we will study the initial--boundary value problem for \eqref{Conformally coupled wave equation} with \emph{homogeneous Dirichlet} (hence, reflecting) boundary conditions on $\mathcal I$, with initial data prescribed along $\Sigma^*_0$:
 \begin{equation}\label{Initial boundary value problem inhomogeneous wave equation}
 \begin{cases}
 \square_g f +2f = F,\\
 (f, \partial_{t^*} f) |_{t^*=0} = (f_0, f_1),\\
 rf|_{\mathcal I}=0.
 \end{cases}
 \end{equation}

\begin{remark}The precise value of the Klein--Gordon mass in \eqref{Conformally coupled wave equation} was crucial in order to obtain a smooth equation in the conformally compactified picture. In the case of the more general massive wave equation
 \begin{equation}\label{General Klein Gordon equation}
 \square_g f - \mu f = F,
 \end{equation}
 the corresponding expression for \eqref{General Klein Gordon equation} in the $(t^*, y, \theta, \varphi)$ \emph{does not} yield an equation with smooth coefficients for any quantity of the form $r^\alpha f$ (unless $\mu=-2$ and $\alpha=1$). Thus, in this more general setting, identifying a set of boundary conditions giving rise to a well-posed initial--boundary value problem is a highly non-trivial task. For masses above the Breitenlohner--Freedman bound $\mu \ge -\tfrac{9}{4}$, well-posed
initial--boundary value problems can be formulated by working with weighted energy norms and ``twisted derivatives''
adapted to the asymptotics at $\mathcal I$; see, for instance, \cite{W13}.
 \end{remark}
 
 \subsubsection{The energy-momentum tensor and vector field multipliers}\label{sec:Energy momentum}
The energy-momentum tensor associated to \eqref{Conformally coupled wave equation} is defined by
\begin{equation}\label{Energy momentum tensor}
T_{\mu\nu}[f] \doteq \Re\big\{\partial_\mu f \partial_\nu \bar f \big\} - \f12 \big(\partial^\alpha f \partial_\alpha \bar f - 2|f|^2\big)g_{\mu\nu}
\end{equation}
and satisfies
\[
\nabla^\mu T_{\mu\nu}[f] = \Re\big\{ F \cdot \partial_\nu \bar f\big\}.
\]

For any smooth vector field $X$ on $\mathcal M_\mathrm{ext}$, we will define the \emph{energy current} associated to $X$ as the 1-form defined by
\[
J^{(X)}_\mu[f] \doteq T_{\mu\nu}[f] X^\nu.
\]
Note that $J^{(X)}[f]$ satisfies the divergence identity
\begin{equation}\label{Basic energy identity div}
\div J^{(X)}[f] = T_{\mu\nu}[f](\pi^{(X)})^{\mu\nu} + \Re\big\{ F \cdot X \bar f\big\},
\end{equation}
where
\begin{equation}\label{Deformation tensor}
\pi^{(X)}_{\mu\nu} \doteq \f12 \mathcal L_X g_{\mu\nu} = \f12 \Big( \nabla_\mu X_\nu + \nabla_\nu X_\mu\Big).
\end{equation}
Note, in particular, that if $X$ is a Killing vector field and $F=0$, then $J^{(X)}[f]$ is conserved (i.e.~it is divergence free).

For any vector field $X$ on $\mathcal M_\mathrm{ext}$, we will refer to $\int_{t=\tau} J^{(X)}_\mu[f] \, \hat n^\mu \, \mathrm{\text{dvol}}_{\Sigma^*}$ as the \emph{energy flux} of $f$ through $\Sigma^*_\tau$ associated to $X$; here,  $\hat n_{\Sigma^*}$ denotes the future directed unit normal to the hypersurfaces $\Sigma^*$, i.e.
\[
\hat n_{\Sigma^*} \doteq -(-g^{t^*t^*})^{-\f12} \Big(g^{t^*t^*}\partial_{t^*}+g^{t^* r}\partial_r \Big).
\]

\begin{remark} We will mainly use the identity \eqref{Basic energy identity div} (with an appropriate choice of a vector field $X$, see \cref{subsec:special-vector-fields} right below) to establish energy estimates for solutions to the initial--boundary value problem \eqref{IVP Psi} in  \cref{sec:Estimates error term}.
\end{remark}

\subsection{Special vector fields on the Schwarzschild--AdS exterior}
\label{subsec:special-vector-fields}

In \cref{sec:Estimates error term}, we will establish a number of a priori energy estimates for the initial--boundary value problem \eqref{IVP Psi} using three distinguished vector fields on $\mathcal M_\mathrm{ext}$ as \emph{multipliers} and \emph{commutators}: The Killing vector field $T$ (i.e.~the vector field $\partial_{t^*}$ in the $(t^*, r, \theta, \varphi)$ coordinate system), the red-shift vector field $Y$ and the helical vector field $V$. 
We will now proceed to define the vector fields $Y$ and $V$ and describe their basic properties in more detail. We will also reexpress the conformal wave operator $\square_g+2$ using the above vector fields.

\subsubsection{The red-shift vector field \texorpdfstring{$Y$}{Y}}\label{sec:The red shift vector field}

Let us fix a $\delta_2>0$ sufficiently small in terms of the geometry of $(\mathcal M_\mathrm{ext}, g_M)$. Following \cite{DR09}, we will define the vector field $Y$ as follows.
\begin{definition}\label{def:Red shift vector field}
With respect to the $(t^*, r, \theta, \varphi)$ coordinate system on $\mathcal M_\mathrm{ext}$, the red-shift vector field $Y$ is fixed by the relation:
\begin{equation}\label{Red shift vector field}
Y = \zeta(r)\big(\partial_{t^*} - \eta \partial_r) + \partial_{t^*},
\end{equation}
where $\eta \in (0, 1]$ is chosen so that
\[
\max_{r\in [r_+, r_+ +\delta_2]} \Big(g_{t^* t^*} -2\eta g_{t^* r} + \eta^2 g_{rr}\Big) <0
\]
 (note that this is possible for small enough $\delta_2=\delta_2(M)$ since, at $r=r_+$, we have $g_{t^*t^*}=0$, $g_{t^* r} =1$)  and  the function $\zeta:[r_+, +\infty)\rightarrow [0,2]$ satisfies
\[
\zeta(r_+)=1, \quad \zeta(r) =0 \quad \text{for } r\ge r_+ +\delta_2 
\]
and
\[
\f{d\zeta}{dr}(r_+) = C_1 \gg 1
\]
for some constant $C_1$ depending only on $\eta$ and the geometry of $(\mathcal M_\mathrm{ext}, g_{M})$ near $\mathcal H^+$.
\end{definition}

 Note that $Y$ satisfies the following properties:
\begin{enumerate}
\item $Y$ is $t^*$-invariant, i.e.~satisfies
\[
[Y, \partial_{t^*}]=0
\]
and is \emph{uniformly timelike} on $\Big(\mathcal M_\mathrm{ext}\cup \mathcal H^+\Big)\cap \{t^*\ge 0\}$.

\item $Y=\partial_{t^*}$ for $r\ge r_+ +\delta_2$.

\item The deformation tensor $\pi^{(Y)}_{\mu\nu}\doteq \f12 \big( \nabla_\mu Y_\nu + \nabla_\nu Y_\mu\big)$ of $Y$ vanishes on $r\ge r_++\delta_2$ and satisfies
\[
\pi^{(Y)}_{\mu\nu} = -\f12 \eta\zeta(r) \partial_r g_{\mu\nu} + \zeta'(r) \nabla_{(\mu} r  \cdot \big(\partial_{t^*}-\eta\partial_r\big)_{\nu)}.
\]
 In particular, provided $C_1$ has been chosen large enough, there exist $\delta_3 \in (0, \delta_2)$ and $C_2, C_3 >0$ such that, for any $f\in C^1(\mathcal M_\mathrm{ext})$:
\begin{equation}\label{Red shift positivity}
T^{\mu\nu}[f] \pi^{(Y)}_{\mu\nu} \ge C_2 \big(|\partial_{t^*} f|^2+|\partial_r f|^2+\|\nabla_{\mathbb S^2} f\|^2_{\mathbb S^2}\big) - C_3 |f|^2 \quad \text{on} \quad \{r_+ \le r \le r_++\delta_3\},
\end{equation}
where $T_{\mu\nu}[f] $ is the energy momentum tensor defined by \eqref{Energy momentum tensor}.
\end{enumerate}

\subsubsection{The helical vector field \texorpdfstring{$V$}{V}}
Let $L\gg 1$ be a large parameter (that will be introduced in \cref{def:Dominant mode parameters}). We will define the vector field $V$ as follows:

\begin{definition}\label{def:Helical vector field}
With respect to the $(t^*, r, \theta, \varphi)$ coordinate system on $\mathcal M_\mathrm{ext}$, we will fix $V$ by the relation
\begin{equation}\label{Helical vector field}
V \doteq \big(1+ L^{-1}\big)\partial_{t^*} + \partial_\varphi.
\end{equation}
\end{definition}

Note that $V$ satisfies the following properties:
\begin{itemize}
\item $V$ is a Killing vector field for $(\mathcal M_\mathrm{ext}, g)$, 
\item It is timelike in the region $r>2M$ (but ceases to be timelike near $\mathcal H^+$), 
\item In the conformally compactified picture (i.e.~in the $(t^*, y, \theta, \varphi)$ coordinate system), $V$ extends smoothly on $\{y=0\}$ and becomes null on  $\{y=0\}\cap\{ \theta = \frac{\pi}{2} \}$ as $L\rightarrow +\infty$. 
\end{itemize}

\begin{remark} The properties listed above imply that $V$ has ``good'' commutation properties with respect to both the conformal wave operator $ \square_g +2$ and operators with coefficients which are functions of wave-packets of frequency $\lesssim L$ concentrated around null geodesics on $\mathcal I$. For this reason, $V$ will be crucial (in the role of a commutation vector field) in deriving the energy estimates of  \cref{sec:Estimates error term}.
\end{remark}
\subsubsection{Some useful expressions for \texorpdfstring{$\square_g+2$}{Boxg+2}}\label{sec:More expressions for wave operator}
We will now proceed to list a few expressions for the conformally coupled wave operator involving the vector fields $V$ and $Y$. These expressions will be useful in \cref{sec:Estimates error term} when establishing higher order energy estimates for \eqref{Conformally coupled wave equation} using $V$ and $Y$ as commutation vector fields.

By rearranging the terms in the expression \eqref{Wave operator} for  $\square_g+2$ in the region $\{r\ge 3M\}$ and using the expression \eqref{Helical vector field}  for the helical vector field $V$, we can readily infer that, for any sufficiently regular function $f:\mathcal M_\mathrm{ext}\rightarrow \mathbb C$:
\begin{equation}\label{Wave operator V}
 r\Big(1-\frac{2M}{r}+r^2 \Big)\big(\square_g f+2f\big) = -\frac{1}{1+L^{-1}} \partial_{t^*}V(rf) + \frac{1}{(1+L^{-1})^2} \partial_{\varphi}V(rf) 
+  \mathfrak L(rf) ,
\end{equation}
where $\mathfrak L$ is a $2^{nd}$ order operator given by the expression
\begin{align}\label{Elliptic operator}
\mathfrak L (rf) \doteq \partial_y^2 & (rf) + \Big(1+\frac{1}{r^2}  -\frac{2M}{r^3}\Big)\frac{1}{\sin\theta}\partial_\theta\Big(\sin\theta\partial_\theta (rf)\Big) +  \Big( \frac{1+\frac{1}{r^2}-\frac{2M}{r^3}}{ \sin^2\theta}- \frac{1}{(1+L^{-1})^2} \Big)\partial_{\varphi}^2(rf) \\
&  -\frac{2M}{r^3}\Big(1-\frac{2M}{r}+r^2\Big)(rf). \nonumber
\end{align}
Note that $\mathfrak L$ is elliptic (which is a consequence of the fact that $V$ is timelike in the region $\{r> 2M\}$).

\begin{remark}
Note that, as $y\rightarrow 0$, the coefficients of the $\partial_y^2$ and $\partial_\theta^2$ terms in \eqref{Elliptic operator} approach a positive constant, while the coefficient of $\partial_{\varphi}^2$ behaves like $\sim \frac{1}{\sin^2\theta}\Big(L^{-1}+\frac{1}{r^2}+|\theta-\frac{\pi}{2}|^2\Big)$, i.e.~it is of size $\sim L^{-1}\ll 1$ near $\{y=0\}\cap\{\theta=\frac{\pi}{2}\}$.
The above formula will be useful for us  in \cref{sec:Estimates error term}, when deriving energy estimates for solutions to \eqref{Conformally coupled wave equation} using $V$ as a commutator.
\end{remark}

Similarly, using the fact that in the $(t^*, r, \theta, \varphi)$ coordinate system we have 
\begin{align*}
\square_g f + 2 f & =
 - \tilde h(r) \partial_{t^*}^2 f + \f{1}{r^2} \partial_r \Big( r^2 \big(1-\f{2M}{r}+r^2\big) \partial_r f\Big) + 2\big(\f{2M}{r}-r^2\big)h(r)\partial_r \partial_{t^*} f\\
 & \hphantom{=} + \f{1}{r^2} \Delta_{\mathbb S^2} f  + \partial_r\Big( \big(\f{2M}{r}-r^2\big)h(r) \Big)\partial_{t^*} f+2 f
\end{align*}
(where $h$, $\tilde h$ are the functions appearing in the expression \eqref{Expression metric t star coordinate} of $g$) and combining the above with the expression \eqref{Red shift vector field} for the red-shift vector field $Y$, we obtain the following identity:
\begin{align}\label{Two expression wave elliptic}
r^3 \big(\square_g f + 2f\big) =  & A(r) Y^2(rf) + B(r) \partial_{r} Y (r f) + C^{\alpha}(r) \partial_{\alpha} (r f) \\
&  + \mathcal L_* (rf) + \big(D(r) -\f{2M}{r} \big) r f,   \nonumber
\end{align}
where:
\begin{itemize}
\item The function $A:[r_+, +\infty)\rightarrow \mathbb R$ is smooth and satisfies
\[
A(r) = -\f{r^2}{1-\f{2M}r+r^2} \quad \text{for } r\ge 3M,
\]
\item The functions $B, C^\alpha, D:[r_+, +\infty)\rightarrow \mathbb R$ are smooth and satisfy
\[
B(r), C^\alpha(r), D(r) = 0 \quad \text{for } r\ge r_\mathrm{mirror},
\]
\item $\mathcal L_*$ is a $2^{nd}$ order \emph{elliptic} operator in the spatial variables, which satisfies 
\[
\mathcal L_* =  \f{r^2}{1-\f{2M}{r}+r^2} \partial_y^2 + \Delta_{\mathbb S^2}  \quad \text{for } r\ge r_\mathrm{mirror}
\]
and
\[
\mathcal L_*  = \big(\f\eta2+O(r-r_+)\big) \partial_r^2 +  \Delta_{\mathbb S^2}  \quad \text{as} \quad r\rightarrow r_+,
\]
where $\eta\in (0,1)$ is the constant appearing in \eqref{Red shift vector field}.
\end{itemize}

\begin{remark}
Note that the uniform ellipticity of $\mathcal L_*$ up to $r=r_+$ is a direct consequence of the fact that $Y$ is uniformly timelike.
\end{remark} 
 
\subsection{Shorthand notation on derivatives}\label{sec:Derivative notation}
Throughout this work (but especially in \cref{sec:Estimates error term}), we will use the shorthand notation $\partial$ in two different ways, depending on whether we are working in the region away from the event horizon or not:
\begin{itemize}
\item In the region $\{0<y\lesssim 1\}$, we will adopt the following notation:
\begin{itemize}
\item We will use $\partial$ to denote any derivative of the form $\partial_{t^*}$, $\partial_y$ or $\nabla_{\mathbb S^2}$.
\item We will denote with $\bar\partial$ any derivative of the form $\partial_y, \partial_{\theta}$. 
\end{itemize}
In the same spirit, we will denote for any sufficiently regular function $f$:
\begin{align}\label{Shorthand square derivatives}
|\partial f|^2 \, & \doteq |\partial_{t^*} f|^2 + |\partial_{y} f|^2+\|\nabla_{\mathbb S^2} f\|^2_{g_{\mathbb S^2}} ,\\
|\bar\partial f|^2 \, & \doteq  |\partial_{y} f|^2+|\partial_{\theta} f|^2,   \nonumber\\
|\partial^2 f|^2 \, & \doteq |\partial \partial_{t^*} f|^2+ |\partial\partial_{y} f|^2+\|\nabla^2_{\mathbb S^2} f\|^2_{g_{\mathbb S^2}},  \nonumber\\
|\bar\partial^2 f|^2 \, & \doteq |\bar\partial\partial_{y} f|^2+|\bar\partial\partial_{\theta} f|^2,   \nonumber\\
|\partial \bar\partial f|^2 \, & = |\bar\partial^2 f|^2+|\partial_{t^*} \partial_y f|^2+\f{1}{\sin^2\theta} |\partial_{\varphi} \partial_y f|^2+ |\partial_{t^*} \partial_\theta f|^2 + \f{1}{\sin^2\theta}|(\nabla_{\mathbb S^2}^2)_{\theta\varphi} f|^2  \nonumber
\end{align}
and similarly for higher order derivatives of $f$.
\item In the near horizon region $\{r_+ \le r \lesssim 1\} = \{y\gtrsim 1\}$, we will use $\partial$ to denote any derivative of the form $\partial_{t^*}$, $\partial_r$ or $\nabla_{\mathbb S^2}$. We will also denote 
\begin{align}\label{Shorthand square derivatives near horizon}
|\partial f|^2 \, & \doteq |\partial_{t^*} f|^2 + |\partial_{r} f|^2+\|\nabla_{\mathbb S^2} f\|^2_{g_{\mathbb S^2}},\\
|\partial^2 f|^2 \, & \doteq |\partial \partial_{t^*} f|^2+ |\partial\partial_{r} f|^2+\|\nabla^2_{\mathbb S^2} f\|^2_{g_{\mathbb S^2}} \nonumber
\end{align}
and similarly for higher order derivatives.

\end{itemize}

\subsection{Sobolev norms along \texorpdfstring{$\Sigma^*_\tau$}{Sigma*\_tau}}
\label{sec:Sobolev norms}
Let $f: \mathcal M_{\mathrm{ext}} \rightarrow \mathbb C$ be a smooth function.
\begin{definition}\label{def:Integer Sobolev norm}
For any $s\in \mathbb N$, we will define the Sobolev norm $\|f\|_{H^s(\Sigma^*_\tau)}$ of $f$ by the relation
\begin{equation}\label{Integer Sobolev norm}
\|f\|_{H^s(\Sigma^*_\tau)}^2 \doteq \int_{\Sigma^*_\tau} \sum_{|\alpha|=0}^s \big\|\partial_{t^*}^{\alpha_1} \big((1+r^2)\partial_r\big)^{\alpha_2} \nabla_{\mathbb S^2}^{\alpha_3} (rf)\big\|^2_{\mathbb S^2} \, \f1{r^3} \dvol_{\Sigma^*_\tau}
\end{equation}
(where the coordinate derivative $\partial_\mu \nabla_{\mathbb S^2}^k$ should be understood as a Lie derivative $\mathcal L_{\partial_\mu} \nabla_{\mathbb S^2}^k$ when $k\ge 1$).  
\end{definition}

\begin{remark} One can extend the above definition to any $s\in [0,+\infty)$ by (complex) interpolation: For any $s\in (0, +\infty)\setminus \mathbb N$, we will define the space $H^s(\Sigma^*_\tau)$ to be the complex interpolation space $[H^{\lfloor s\rfloor}(\Sigma^*_\tau), H^{\lfloor s\rfloor+1}(\Sigma^*_\tau)]_{\bar\theta}$ with $\bar\theta = s-\lfloor s \rfloor$.
\end{remark}

\begin{remark} Note that for a function $f$ supported in the region $\{y\lesssim 1\}$, \eqref{Integer Sobolev norm} yields (for $s\in \mathbb N$):
\[
\|f\|_{H^s(\Sigma^*_\tau)}^2 \sim \int_{\Sigma^*_\tau} \sum_{j=0}^s |\partial^j (rf)|^2 \, \sin\theta dy d\theta d\varphi 
\]
(see \cref{sec:Derivative notation} for the $\partial$ notation). In particular, for any $s\ge 0$, $\|f\|_{H^s}$ is equivalent, in this case, to the corresponding Sobolev (Bessel-potential for non-integer $s$) norm of $rf$ when viewed as a function on $\mathbb S^3_+$.
\end{remark}
 
We will sometimes need to consider $L^p$-type norms along more general hypersurfaces $\Sigma \subset \mathcal M_\mathrm{ext}$, with respect to a measure $d\mu$ which does not necessarily carry geometric significance. In this case, we will use the notation 
\[
\|f\|_{L^p(\Sigma;d\mu)} \doteq \Big( \int_{\Sigma} |f|^p \, d\mu\Big)^{\f1p}.
\]
In order to simplify our notations, we will sometimes drop $\Sigma$ from the corresponding subscripts when it is clear from the context to which hypersurface we are referring to.

\subsection{The Hermite functions}
In this work, we will make extensive use of the Hermite functions $\psi_n:[0,+\infty)\rightarrow \mathbb R$ for $n \equiv 1 \mod 2$, defined by
\[
\psi_n(x) \doteq H_n(x) e^{-\f{x^2}2},
\]
where $H_n$ is the $n$-th Hermite polynomial, defined by
\begin{equation}\label{Rodrigues formula again}
H_n (x) = (-1)^n e^{x^2} \Big( \frac{d^n}{dx^n} e^{-x^2}\Big).
\end{equation}
Note that $\psi_n$ satisfies
\[
\begin{cases}
\f{d^2 \psi_n}{dx^2} + \big(2n+1-x^2\big) \psi_n =0,\\[5pt]
\psi_n(0)=0, \quad \f{d \psi_n}{dx}(0) = (-1)^{\f{n-1}2}2\f{\Gamma(n+1)}{\Gamma(\f{n+1}2)}.
\end{cases}
\]
We will also set for $n\in \mathbb N^*$
\begin{equation}\label{Normalized Hermite function}
e_n(x) = c_{n}\psi_{2n-1}(x) , \text{ with } c_{n}^2 = \frac{ 1}{\sqrt \pi}  \cdot \frac{1}{2^{2n-2} \Gamma(2n)  }.
\end{equation}
Note that the functions $e_n$ satisfy  
\begin{equation}
 e_n^{\prime\prime} +\big(4n-1-x^2\big) e_n =0.
\end{equation}
Moreover, the functions $\{e_n\}_{n \in \mathbb N^*}$ form an \emph{orthonormal} basis of $L^2\big([0,+\infty)\big)$.

The Hermite functions will appear as high frequency approximations of the radial eigenfunctions associated to the initial-boundary value problem \eqref{Initial boundary value problem inhomogeneous wave equation} on $(\mathcal M_\mathrm{ext}, g_M)$. Thus, estimates for the functions $\psi_n$ and their products as $n\rightarrow +\infty$ will play a crucial role in our analysis of the spectral properties of  \eqref{Initial boundary value problem inhomogeneous wave equation}; those estimates are stated and proved in \cref{sec:Integrals Hermite,sec:WKB asymptotics} of the Appendix.

\subsection{Notations on constants and inequalities}
We will adopt the standard convention regarding the inequality symbols $\lesssim$, $\sim$ and $O(\cdot)$: For any real valued functions $f_1, f_2>0$, the expression $f_1 \lesssim f_2$ will be used to denote that there exists a constant $C>0$ such that $f_1 \le C f_2$, while $f_1 \sim f_2$ and $f_1=O(f_2)$ will be used to denote, respectively, that $f_2 \lesssim f_1 \lesssim f_2$ and $|f_1| \lesssim f_2$ (in the latter case, we will allow $f_1$ to be complex valued). 

Throughout this work, it will be crucial to keep track of the dependence of the constants implicit in the $\lesssim, \sim, O(\cdot)$ notation on the various parameters introduced in the proof of \cref{thm:Main theorem} (such as $\epsilon$, $s$, $N$, $\lambda$ and $L$). It will always be stated clearly (usually at the beginning of the proof of each lemma, proposition, etc.) on which parameters those constants are allowed to depend. If no such specification is provided, it will be assumed that those constants depend only on the geometry of $(\mathcal M_\mathrm{ext}, g_M)$. With respect to the latter, we will assume that the dependence of all constants on the Schwarzschild--AdS mass parameter $M\in (0,+\infty)$ (the only parameter determining the geometry of $(\mathcal M_\mathrm{ext}, g_M)$) is \textbf{uniform over any given pre-compact subset} of $(0,+\infty)$ (but might ``degenerate'' in the limit as $M$ approaches $0$ or $+\infty$), unless otherwise stated. In \cref{sec:Linear wave equation mirror} to \cref{sec:Non dominant modes}, the dependence of our constants on $M$ will be uniform on pre-compact subsets of $[0,+\infty)$ (i.e.~they will not degenerate in the limit $M\rightarrow 0$).

\section{The linear wave equation with an inner mirror}\label{sec:Linear wave equation mirror}
Let $M_0>0$ be a given constant and $M$ be a Schwarzschild--AdS mass parameter in the range $[\tfrac 12 M_0,M_0]$. Our construction of a solution $\phi$ for the initial--boundary value problem \eqref{Initial Boundary Value Problem Phi} with the properties required by \cref{thm:Main theorem} will utilize an ansatz based on the \emph{normal mode} solutions of the linear problem with an \emph{additional} mirror placed at $\{r=r_\mathrm{mirror}\}$ (where $r_\mathrm{mirror}=r_\mathrm{mirror}(M;M_0)$ was defined by \eqref{Definition mirror value}), namely:
\begin{equation}\label{Initial boundary value problem mirror}
\begin{cases}
\square_g \phi + 2\phi =0 ,\\
(\phi, \partial_{t}\phi)|_{t=0}=( f_0, f_1),\\
\phi|_{r=r_\mathrm{mirror}}=0, ~ r\phi|_{r=\infty}=0.
\end{cases}
\end{equation}
Recall that, in the region $\{r\ge r_\mathrm{mirror}\}\subset \{r\ge 3M\}$, the coordinate system $(t,r,\theta,\varphi)$ coincides with $(t^*, r, \theta, \varphi)$. In this section, we will study the properties of those high-frequency normal mode solutions to \eqref{Initial boundary value problem mirror} that are ``trapped'' near $r=+\infty$ (and, therefore, have small amplitude in the region $r\sim 1$).

\subsection{Separation of variables and parametrization of the normal modes}
The boundary value problem \eqref{Initial boundary value problem mirror} is separable. It formally admits solutions of the form
\[
\phi(t,r,\theta, \varphi) = \f{a(t)}{r} R(r) Y(\theta,\varphi)
\]
where:
\begin{itemize}
\item $Y:\mathbb S^2 \rightarrow \mathbb C$ satisfies for some $\ell\in \mathbb N$
\[
\Delta_{\mathbb S^2} Y + \ell(\ell+1) Y=0.
\]
\item $R(r)$ satisfies the radial boundary value problem:
\[
\begin{cases}
\f{d}{dr}\Big(g^{rr}\f{d}{dr} R\Big) + \Big( (-g^{tt}){\omega}^2-\f{\ell(\ell+1)}{r^2} +   \big(2-\f1r \f{d g^{rr}}{dr}\big)\Big) R =0,\\
R(r_\mathrm{mirror})=0, ~R(\infty)=0,
\end{cases}
\]
where $g^{rr}= (-g^{tt})^{-1}=1-\f{2M}{r}+r^2$. With respect to the $y$ variable (defined by \eqref{The y coordinate}), the above boundary value problem becomes: 
\begin{equation}\label{Radial model boundary value problem}
\begin{cases}
-\f{d^2}{dy^2} R+ V_\ell(y) R={\omega}^2 R,\\
R|_{y=0} = R|_{y=y_\mathrm{mirror}}=0,
\end{cases}
\end{equation}
where
\begin{equation}\label{Radial potential}
V_\ell(y) = \Big(1 + \f{1}{r(y)^2}-\f{2M}{r(y)^3} \Big)\Big(\ell(\ell+1)+\f{2M}{r(y)}\Big).
\end{equation}
Note that
\begin{equation}\label{Monotonicity radial potential}
\f{d}{dy} V_{\ell}(y) >0 \quad \text{for} \quad y\in (0,y_\mathrm{mirror}),
\end{equation}
since $r_\mathrm{mirror}>3M$.

\item The function $a(t)$ solves the ODE
\begin{equation*}
\f{d^2 a}{dt^2} + {\omega}^2 a =0.
\end{equation*}
\end{itemize}

A solution of \eqref{Initial boundary value problem mirror} of the above form for ${\omega}\in \mathbb R$ will be called a \emph{normal mode} solution. 

Note that the problem \eqref{Radial model boundary value problem} is a regular Sturm--Liouville problem. Traditional Sturm--Liouville theory and the fact that $V_{\ell}>0$ tell us that, for each fixed $\ell\in \mathbb N$, the eigenvalue problem \eqref{Radial model boundary value problem} has a discrete set of \emph{real and non-negative} eigenvalues ${\omega}^2_{n,\ell}$ corresponding to \emph{real} eigenfunctions $R_{n,\ell}$, parametrized by $n\in \mathbb N^*$ (the so-called \emph{overtone}). In particular, the functions 
 satisfy
 \begin{equation}\label{Radial boundary value problem}
\begin{cases}
-\f{d^2}{dy^2} R_{n,\ell} + V_\ell(y) R_{n,\ell}={\omega}^2_{n,\ell}R_{n,\ell},\\
R_{n,\ell}|_{y=0} = R_{n, \ell}|_{y=y_\mathrm{mirror}}=0,
\end{cases}
\end{equation}
 and form an orthonormal basis for $L^2([0,y_\mathrm{mirror}],dy)$, satisfying in particular
\begin{equation}\label{Normalization radial eigenfunctions}
\int_{0}^{y_\mathrm{mirror}} R_{n_1, \ell}(y) R_{n_2, \ell}(y) \, dy = \boldsymbol{\delta}_{n_1 n_2}.
\end{equation}

\begin{remark}
     We will adopt the sign convention that 
\[
{\omega}_{n,\ell}>0.
\]
We will also frequently denote ${\omega}_{n,\ell}$ by ${\omega}_k$. For the eigenfunctions $R_{n,\ell}$, we will adopt the sign convention that $\sgn(\f{d R_{n,\ell}}{dy}(0)) = \sgn(e'_n(0))$ as defined in \eqref{Normalized Hermite function}.
\end{remark}

\begin{remark}\label{remark:Smooth dependence eigenvalues}
  For a fixed $M_0>0$, the Sturm--Liouville problem \eqref{Radial boundary value problem} depends on the mass parameter $M\in (0,M_0]$ in a \emph{smooth} way, in the sense that the potential $V_{\ell}$ (viewed as a function of $y\in[0,y_\mathrm{mirror}]$ and $M\in (0,+\infty)$) depends \emph{smoothly} on the parameter $M$, and can be smoothly extended at $M=0$ (recall that $y_\mathrm{mirror}= y_\mathrm{mirror}(M_0)$ defined by \eqref{Definition y mirror} is \textbf{independent} of $M\in [0,M_0]$).  Moreover, the map $n\rightarrow \omega_{n,\ell}$ is strictly increasing (as is always the case with non-singular Sturm--Liouville problems).
As a result, it follows by standard results from Sturm--Liouville theory that the eigenvalue $\omega_{n,\ell}$ is a \emph{smooth} function of $M\in [0,M_0]$.
\end{remark}

We will parametrize the normal mode solutions of  \eqref{Initial boundary value problem mirror} as follows: For any triplet $k=(n, \ell, m)$, where $n\in \mathbb N^*$, $\ell \in \mathbb N_{\ge |m|}$ and $m\in \mathbb Z$, we will define
\begin{equation}\label{Normal mode}
\phi_k(t,r,\theta,\varphi) = \f{a_k(t)}{r} R_{n,\ell}(r) Y_{\ell,m}(\theta,\varphi),
\end{equation}
where:
\begin{itemize}
\item $Y_{\ell,m}:\mathbb S^2\rightarrow \mathbb C$ denotes the complex valued \emph{spherical harmonic} of order $(\ell, m)$, i.e.
\begin{equation}\label{Expression spherical harmonic}
Y_{\ell,m}(\theta, \varphi)=(-1)^m \sqrt{\f{2\ell+1}{4\pi}\f{\Gamma(\ell-m+1)}{\Gamma(\ell+m+1)}} P^m_\ell(\cos\theta) e^{im\varphi},
\end{equation}
where $P^m_\ell(\cdot)$ is the associated Legendre polynomial of order $(m;\ell)$.

\item $R_{n,\ell}(r)$ is the $n$'th eigenfunction of the Sturm--Liouville problem \eqref{Radial boundary value problem}.

\begin{remark} From now on, we will only consider $R_{n, \ell}$ as a \textbf{function of $y$} through the relation for $r=r(y)$ obtained by inverting \eqref{The y coordinate}.
\end{remark}

\item The function $a_k(t)$ solves the ODE
\begin{equation}\label{Time ODE a}
\f{d^2 a_k}{dt^2} + {\omega}_{n,\ell}^2 a_k =0
\end{equation}
where ${\omega}_{n,\ell}^2 $ is the Sturm--Liouville eigenvalue associated to $R_{n,\ell}$.
Note that the amplitude $a_k(t)$ can be simply represented as
\[
a_k(t) = a_{k,+} e^{i{\omega}_{n,\ell} t} + a_{k,-} e^{-i{\omega}_{n,\ell} t},
\]
where $a_{k,+}$ and $a_{k,-}$ are constants.
\end{itemize}

For any $k = (n, \ell, m) \in \mathbb N^* \times \mathbb N_{\ge |m|} \times \mathbb Z$, we will define the function $E_k:[0,y_\mathrm{mirror}]\times \mathbb S^2\rightarrow \mathbb C$ by
\begin{equation}\label{Spatial eigenfunction}
E_k(y,\theta,\varphi) \doteq R_{n,\ell}(y) Y_{\ell,m}(\theta,\varphi).
\end{equation}
Note that the $E_k$'s form an orthonormal basis for $L^2\big([0,y_\mathrm{mirror}]\times \mathbb S^2; dy \, \dvol_{\mathbb S^2}\big)$ and the normal mode \eqref{Normal mode} takes the simplified form
\[
\phi_k (t,r,\theta,\varphi) = \f{a_k(t)}{r} E_k(r,\theta,\varphi).
\]
We will denote with $\mathbb P_k$ the projection on the eigenspace spanned by $E_k$, i.e.:
\begin{equation}\label{Projection k}
\mathbb P_k f \doteq \Big(\int_{0}^{y_\mathrm{mirror}} \int_{\mathbb S^2} f \cdot \bar E_k \sin\theta d\theta d\varphi dy\Big) \cdot E_k.
\end{equation}
We will also denote with $\langle \cdot \rangle$ the following averaging operator:
\begin{equation}\label{Definition average}
\langle f \rangle \doteq \int_{0}^{y_\mathrm{mirror}} \int_{\mathbb S^2} f \cdot\frac{1}{\big(r(y)\big)^6} \Big(1+\f1{\big(r(y)\big)^2}-\f{2M}{\big(r(y)\big)^3}\Big) \sin\theta d\theta d\varphi dy.
\end{equation}

The rest of this section will be devoted to the study of the properties of ${\omega}_k$ and $E_k$ in the regime $n\ll \ell$.

\subsection{Estimates for the radial eigenfunctions \texorpdfstring{$R_{n,\ell}$}{R\_{n,l}}}\label{subsection: Estimates R}
In this section, we will establish a number of approximation formulas for the radial eigenfunctions $R_{n,\ell}$ of the Sturm--Liouville problem \eqref{Radial boundary value problem} in terms of the renormalized Hermite functions $e_n$ in the low-lying regime $n\lesssim \ell^\delta$, $\delta\ll 1$. We will also establish a number of crude estimates for $R_{n,\ell}$ and the frequencies $\omega_{n,\ell}$ in the case when $n$ is not necessarily small with respect to $\ell$.

We will need the following general Agmon-type estimates for solutions to a $2^{nd}$ order ODE with a potential that changes sign:

\begin{lemma}\label{lem: Agmon type estimates}
Let us consider the following inhomogeneous initial value problem on the interval $[-1,x_0]$, $x_0>0$:
\begin{equation}\label{Model problem for Carleman estimates}
\begin{cases}
u'' +  W u =F,\\
u(-1)=u'(-1)=0,
\end{cases}
\end{equation}
where $F\in C^0([-1,x_0])$ and the potential $W \in C^1\big([-1,x_0]\big)$ is real valued and satisfies the following conditions:
\begin{enumerate}
\item Sign change: 
\[
W(x)\ge 0 \quad \text{for} \quad x<0 \quad \text{and} \quad W(x)< 0 \quad \text{for}\quad  x>0.
\]
\item In the case when $x_0\ge 1$:
\[
-W(x) \ge 1 \quad \text{for} \quad x\ge 1.
\]
\item Monotonicity:
\[
W'(x) \le 0 \quad \text{for} \quad x\in [-1,0] 
\]
and
\begin{equation}
\label{Monotonicity forbidden region}
-W'(x) \le \kappa  -W(x) \quad \text{for} \quad x\in [0,x_0]
\end{equation}
for some constant $\kappa \ge 1$
\end{enumerate}
Then, the  solution $u:[-1,x_0]\rightarrow \mathbb R$ of \eqref{Model problem for Carleman estimates} satisfies the following pointwise bound in the classical region $x\in [-1,0]$:
\begin{equation}\label{Bound pointwise classical}
\sup_{x\in [-1, 0]} \Big( (u')^2 + \big(W +1\big)u^2 \Big) \le 8\int_{-1}^{0} F^2(s) \, ds
\end{equation}
and the following integrated decay estimate in the forbidden region $x\in [0,x_0]$:
\begin{align}\label{Bound pointwise forbidden}
\int_0^{x_0} \Big[\f{1}{1-W}  (u'')^2 + \f{(1-W)\sqrt{-W}}{\kappa+(-W)^{\f32}} & (u')^2   +(1-W) u^2 \Big]\,e^{\int_0^x \sqrt{-W(s)}\, ds}  dx \\
\le & 10^5 \Bigg( \int_0^{x_0}  \f1{1-W} F^2 \, e^{\int_0^x \sqrt{-W(s)}\, ds} dx +\int_{-1}^{0} F^2(x) \, dx   \nonumber \\
&\hphantom{10^4 \Bigg( +}
 + \Big(\big|u' (x_0) u(x_0)\big|+ \sqrt{-W(x_0)} u^2(x_0)\Big) e^{\int_0^{x_0} \sqrt{-W(s)}\, ds}\Bigg),   \nonumber
\end{align}
where $\kappa\ge 1$ is the constant appearing in \eqref{Monotonicity forbidden region}.
\end{lemma}

\begin{proof}
We will start by performing an energy-type estimate for \eqref{Model problem for Carleman estimates} in the classical region $x\in [-1,0]$: Multiplying \eqref{Model problem for Carleman estimates} with $u'(x)$ and integrating by parts over $[-1,x]$ for any $x\in (-1,0]$, we obtain:
\[
\big(u'(x)\big)^2 + W(x) \big(u(x)\big)^2 = \int_{-1}^x W'(s)  u^2(s) \, ds + 2\int_{-1}^x F(s) u'(s) \, ds 
\stackrel{W'|_{[-1,0]}\le 0}{\le} 2\int_{-1}^x F(s) u'(s) \, ds.
\]
Therefore, applying a Cauchy--Schwarz inequality for the right-hand side (together with the trivial inequality $ u^2(x) \le (x+1) \int_{-1}^{x} (u')^2 \, ds$ in the left-hand side), we infer that
\begin{equation}\label{Estimate classical region}
\sup_{x\in [-1, 0]} \Big( (u')^2 + \big(W +1\big)u^2 \Big) \le 8\int_{-1}^{0} F^2(s) \, ds.
\end{equation}
In particular,  \eqref{Bound pointwise classical} holds.

In order to estimate the solution $u$ in the forbidden region $x\in [0, x_0]$, we will apply an Agmon-type estimate.  Let us set for $x\ge 0$
\[
w(x) \doteq \f12 \int_0^x \sqrt{-W(\bar z)} \, d\bar z
\]
and note that
\begin{equation}\label{Positivity W h difference}
-W - (w')^2 \ge \f12 (-W) \ge  0.
\end{equation}

Defining the function
\begin{equation}\label{Substitution u v}
v(x) \doteq e^{w(x)} u(x),
\end{equation}
note that $v$ solves for $x\ge 0$ the ODE
\begin{equation}\label{Elliptic ODE v}
v'' - 2 w' v' + \Big((w')^2 -w'' + W\Big) v = e^w F.
\end{equation}
Multiplying \eqref{Elliptic ODE v} with $v$ and integrating by parts over $[0,x]$ for any $0\le x\le x_0$, we infer that
\[
\int_0^{x} \Big[(v')^2  + \Big( (-W) -(w')^2 \Big) v^2 \Big]\, dz - \Big[ v' v(z) - w' v^2(z)\Big]\Big|_{z=0}^{x} = - \int_0^{x} e^w F v dz.
\]
Using a Cauchy--Schwarz inequality for the right-hand side above, together with the trivial estimate (recalling that $-W\ge 1$ for $x\ge 1$)
\[
\int_0^{x} \f12\big(1 - W\big)v^2 \, dz \le 4\int_0^{x} \Big( (v')^2 -\f12 W v^2 \Big) \, dz
\]
and the lower bound \eqref{Positivity W h difference}, we deduce that
\begin{align*}
\int_0^{x} \Big[(v')^2  +(1-W) v^2 \Big]\, dz \le 2^4 & \int_0^{x} e^{2w} \f{1}{1-W} \cdot F^2 \, dz \\
& + 2^4 \Big( \Big|v' v(0) - w' v^2(0) \Big|+ \Big|v' v(x) - w' v^2(x) \Big|\Big).
\end{align*}
Using the expression \eqref{Substitution u v} to obtain from the above an estimate for $u= v e^{-w}$, and noting that 
\[
(v')^2 = e^{2w}\Big((u')^2 - \f14\big((-W) +\f{-W'}{\sqrt{-W}}\big)u^2 \Big) +\big(e^{2w} w' u^2 \big)'
\]
and
\[
w(0)=w'(0)=0 \quad \text{and} \quad -W'(x) \le \kappa -W(x),
\]
 we infer that
\begin{align}\label{First Carleman estimate}
\int_0^{x} \Big[\f{(1-W)\sqrt{-W}}{\kappa-W+(-W)^{\f32}}(u')^2  +(1-W) u^2 \Big]\,e^{2w}  dz &\le 2^8   \int_0^{x}  \f1{1-W} \cdot F^2 \, e^{2w} dz \\
& + 2^8 \Big( \big|u'(0) u(0) \big|+ \big(\big|u' (x) u(x)\big|+ \sqrt{-W(x)} u^2(x)\big) e^{2w(x)} \Big). \nonumber 
\end{align}
Using equation \eqref{Model problem for Carleman estimates} to express $u''$ in terms of $u$ and $F$, we have
\[
\int_0^{x}\f1{1-W} (u'')^2 \,e^{2w}  dz \le 2 \int_0^{x} (1-W) u^2  \,e^{2w}  dz + 2 \int_0^{x}  \f1{1-W} \cdot F^2 \, e^{2w} dz.
\]
Therefore, the bound \eqref{First Carleman estimate} yields (using also the trivial estimate $\kappa+(-W)^{\f32} \ge \f12\big(\kappa-W+(-W)^{\f32}\big)$)
\begin{align}\label{Second Carleman estimate}
\int_0^{x} \Big[\f1{1-W}  (u'')^2 + \f{(1-W)\sqrt{-W}}{\kappa+(-W)^{\f32}} (u')^2 &  +(1-W) u^2 \Big]\,e^{2w}  dz\\
   \le & 2^{12}  \int_0^{x} \f1{1-W} \cdot F^2 \, e^{2w} dz \\
& + 2^{12} \Big( \big|u'(0) u(0) \big|+ \big(\big|u' (x) u(x)\big|+ \sqrt{-W(x)} u^2(x)\big) e^{2w(x)} \Big). \nonumber 
\end{align}
Using the bound \eqref{Estimate classical region} to control the term $|u'(0)u(0)|$ in the right-hand side above, 
we finally obtain \eqref{Bound pointwise forbidden}.

\end{proof}

The following result will provide the main approximation estimates for the radial eigenfunctions $R_{n,\ell}$ that will be used throughout this work:

\begin{lemma}\label{lem:eigenfunctions-estimate-m>0}
Let $M_0>0$ be a given constant. For any Schwarzschild--AdS mass parameter $M\in [0,M_0]$ and any $(n,\ell)\in \mathbb N^*$ such that $\ell$ is sufficiently large in terms of $M_0$ and satisfies
\[
n \le \ell^{\delta} \quad \text{for some given} \quad 0<\delta\ll 1,
\]
let $R_{n,\ell}:[0,y_\mathrm{mirror}]\rightarrow \mathbb R$ be the radial eigenfunctions defined by  \eqref{Radial boundary value problem} and normalized according to \eqref{Normalization radial eigenfunctions}. 
Then, the functions $R_{n,\ell}(r)$ satisfy the approximation formula
\begin{align}\label{eq:bound-on-Rnl-M}
R_{n,\ell}(r(y)) = \ell^{\f14} e_n \big(\ell^{\f12}y\big) + O\big(\ell^{-\f14+5\delta}\big),\\
\frac{d}{dy} R_{n,\ell}(r(y)) = \ell^{\f34} e_n' \big( \ell^{\f12}y\big) + O\big(\ell^{\f14+5\delta}\big),\label{eq:bound-on-Rnl-prime-M}
\end{align}
where $e_n(y)$ are the normalized Hermite functions \eqref{Normalized Hermite function} and the constants implicit in the $O(\cdot)$ notation depend only on $\delta, M_0$ (and are, therefore, \textbf{independent} of $n, \ell$).
 Moreover, the following smallness bound holds for $R_{n,\ell}$ outside the classical region $0\le y\lesssim n^{\f12}\ell^{-\f12}$:
\begin{equation}\label{eq:expo-decay}
\big|R_{n,\ell}(y)\big| + \f1{n^{\f12} \ell^{\f12}} \Big|\f{d}{dy} R_{n,\ell}(y)\Big| \lesssim e^{-\f12 n^{\f14}\ell^{\f34}\Big(y-\sqrt{\f{4n-1}{\ell}}\Big)^{\f32}}\ell^{\f14} \quad \text{for} \quad y\ge \sqrt{\f{4n-1}{\ell}} .
\end{equation}
The corresponding Sturm--Liouville eigenvalue $\omega_{n, \ell}$ satisfies
\begin{equation}\label{Approximation estimate frequency}
\Big|\omega_{n,\ell}-(2n+\ell) \Big| \lesssim \ell^{-\f12+2\delta}
\end{equation}
(where, again the constant implicit in the $\lesssim$ notation depend only on $\delta, M_0$ and \textbf{not} on $n, \ell$).
\end{lemma}

\begin{proof}
The proof of \cref{lem:eigenfunctions-estimate-m>0} will consist of two consecutive bootstrap arguments to deduce the approximation estimate \eqref{Approximation estimate frequency} (the bounds \eqref{eq:bound-on-Rnl-M}--\eqref{eq:expo-decay} then following readily by applying \cref{lem: Agmon type estimates}): At first, we establish estimate \eqref{Approximation estimate frequency} for the eigenvalues of an auxiliary Sturm--Liouville problem, corresponding to the construction of normal mode solutions to the conformally coupled wave equation on \emph{pure} AdS spacetime (i.e.~with $M=0$) with a reflecting mirror placed at some finite radius around $r=0$; we will use a continuity argument with respect to the mirror radius. In the second step, we will obtain \eqref{Approximation estimate frequency} via a continuity argument on the Schwarzschild--AdS mass parameter $M$.

\begin{remark}
For this proof, we will assume that the constants implicit in the $O(\cdot)$ and $\lesssim$ notation are independent of $n, \ell$ (but are allowed to depend on $M_0, \delta$).
\end{remark}

\noindent \textbf{The model problem: The case $M=0$.} Let us consider the following auxiliary Sturm--Liouville problem:
\begin{equation}\label{Sturm Liouville problem zero mass}
\begin{cases}
-\f{d^2}{dy^2} \oR_{n,\ell}+ \f{\ell(\ell+1)}{\cos^2(y)} \oR_{n, \ell}={\oomega_{n,\ell}}^2 \oR,\\
\oR_{n,\ell}|_{y=0} = \oR_{n,\ell}|_{y=y_{\mathrm{D}}}=0,
\end{cases}
\end{equation}
for some $y_{\mathrm{D}} \in (0, \f{\pi}{2}]$. Here, $\oomega^2_{n,\ell}$ denotes the $n$-th eigenvalue of the above problem (i.e.~$n$ is the overtone parameter), and $n,\ell$ are as in the statement of \cref{lem:eigenfunctions-estimate-m>0}. We will think of $\oomega_{n,\ell}$ and $\oR_{n,\ell}$ as dependent on the parameter $y_{\mathrm{D}} \in (0, \f{\pi}{2}]$. Note that the Sturm--Liouville potential in \eqref{Sturm Liouville problem zero mass} is formally obtained from the corresponding term in \eqref{Radial boundary value problem} after setting $M=0$. We will also assume that $\oR_{n,\ell}$ is $L^2$-normalized, i.e.
\begin{equation}\label{L 2 normalization oR}
\int_0^{y_{\mathrm{D}}} \oR_{n,\ell}^2(y) \, dy =1.
\end{equation}

\begin{remark} Note that, when $y_{\mathrm{D}}=\f{\pi}{2}$, we have
\begin{equation}\label{Eigenvalue zero mass no boundary}
\oomega_{n,\ell}\left(y_{\mathrm{D}}=\f\pi2\right) = 2n +\ell.
\end{equation}
In this case, 
\[
f(y, \theta, \varphi) = \f1{\tan(\f\pi2-y)} \oR_{n,\ell}(y) Y_{\ell,m}(\theta, \varphi)
\] 
is a standard spherical harmonic on $(\mathbb S^3_+, g_{\mathbb S^3})$ vanishing on $\partial \mathbb S^3_+ = \{y=0\}$ (with the center of symmetry for the half-sphere lying at $y=\f\pi2$).
\end{remark}

We will assume that $\ell$ is sufficiently large in terms of $M_0$ so that
\begin{equation}\label{Smallness condition turning point}
\ell^{-\f12+3\delta}\ll 1.
\end{equation}
We will define the following bootstrap set of parameters $y_{\mathrm{D}}$:
\begin{equation}\label{Bootstrap set zero mass}
\mathscr B_0 \doteq \Big\{ y_{\mathrm{D}} \in [y_\mathrm{mirror}(M_0),\f\pi2]:\, \text{ For all } \, \bar y \in [y_{\mathrm{D}}, \f\pi2] \, \text{ we have } \, \big|\oomega_{n,\ell}(\bar y)-(2n+\ell)\big| \le \ell^{-1}.  \Big\}
\end{equation}

Our aim is to show that
\[
\mathscr B_0 = [y_\mathrm{mirror}, \f\pi2].
\]
We will achieve this via the method of continuity; note that $\mathscr B_0$ is manifestly closed (since $\omega_{n,\ell}(y_{\mathrm{D}})$ depends continuously on $y_{\mathrm{D}}$) and non-empty, since $\f\pi2 \in \mathscr B_0$ in view of \eqref{Eigenvalue zero mass no boundary}. In order to show that $\mathscr B_0$ is also an open subset of   $[y_\mathrm{mirror}, \f\pi2]$, it suffices to show that, for any $y_{\mathrm{D}}\in \mathscr B_0$, the following improved bound holds for $\omega_{n,\ell}(y_{\mathrm{D}})$:
\begin{equation}\label{First improved bound to show zero mass}
\big|\oomega_{n,\ell}(y_{\mathrm{D}})-(2n+\ell)\big| \le \f12 \ell^{-1}.
\end{equation}

\medskip
\noindent \emph{Estimates for $\oR_{n,\ell}$ using the bootstrap assumption.} Let $y_{\mathrm{D}} \in \mathscr B_0$. Note that this implies, in particular, that
\begin{equation}\label{Bound definition bootstrap set zero mass}
\big|\oomega_{n,\ell}(y_{\mathrm{D}})-(2n+\ell)\big| \le \ell^{-1}.
\end{equation}
The boundary value problem \eqref{Sturm Liouville problem zero mass} takes the form
\begin{equation}
\begin{cases}
\f{d^2}{dy^2} \oR_{n,\ell}+\ell \oV_{n,\ell} \oR_{n, \ell}=0,\\
\oR_{n,\ell}|_{y=0} = \oR_{n,\ell}|_{y=y_{\mathrm{D}}}=0,
\end{cases}
\end{equation}
where
\[
\oV_{n,\ell}(y) \doteq  \f{\oomega_{n,\ell}^2}{\ell} - \f{\ell+1}{\cos^2(y)}.
\]
Note that, in view of \eqref{Bound definition bootstrap set zero mass} and our assumption that $n\le \ell^{\delta}$, we have
\begin{equation} \label{o V}
\oV_{n,\ell}(y) = 4n-1-\ell y^2+O(y^2 + \ell y^4) +O(\ell^{-1}).
\end{equation}
Let us also consider the $\ell$-rescaled, $L^2$ normalized Hermite functions 
\[
\oE_{n,\ell}(y) \doteq \ell^{\f14} e_n(\ell^{\f12} y)
\]
(where $e_n$ is fixed as in \eqref{Normalized Hermite function}). Note that $\oE_{n,\ell}$ satisfies
\[
\begin{cases}
\f{d^2}{dy^2} \oE_{n,\ell} + \ell \Big(4n-1 - \ell y^2\Big) \oE_{n,\ell}=0,\\
\oE_{n,\ell}(0)=0, \quad \f{d}{dy}\oE_{n,\ell}(0) \neq 0.
\end{cases}
\]
We will show that $\oE_{n, \ell}$ approximates $\oR_{n, \ell}$ well; to this end, let us set
\[
A \doteq \f{\f{d \oR_{n,\ell}}{dy}(0)}{\f{d \oE_{n,\ell}}{dy}(0)}
\]
(we will later show that $A=1+O(\ell^{-1+})$) and 
\[
\ov_{n, \ell}(y) = \oR_{n,\ell}(y)- A \oE_{n,\ell}(y).
\]
Note that $\ov_{n,\ell}$ solves
\begin{equation}\label{y equation ov}
\begin{cases}
\f{d^2 \ov_{n,\ell}}{dy^2} + \ell \oV_{n,\ell} \ov_{n,\ell} =\ell \oF_{n,\ell},
\\
\ov_{n,\ell}(0) = \f{d\ov_{n,\ell}}{dy}(0)=0,
\end{cases}
\end{equation}
where
\[
\oF_{n,\ell}(y) = A \cdot \Big( 4n-1-\ell y^2 -  \oV_{n,\ell}(y)\Big) \oE_{n,\ell}(y).
\]

Let us denote with $\oy_*$ the root of the potential $\oV_{n,\ell}$, i.e.
\[
\oy_* = \arccos\Big(\sqrt{\f{\ell(\ell+1)}{\oomega_{n,\ell}^2}}\Big)
=  \arccos\Big(1-\f12 \f{4n-1}{\ell} +O(\ell^{-2+3\delta})\Big) = \sqrt{\f{4n-1+O(\ell^{-1+3\delta})}\ell}.
\] 
Note that, as a consequence of our assumption \eqref{Smallness condition turning point}, we have $\oy_* \in [0,y_{\mathrm{D}}]$.
Let us define the new coordinate
\[
z \doteq \f{y}{\oy_*} \text{ and set } z_{\mathrm{D}} \doteq z(y_{\mathrm{D}}).
\]
 Note that \eqref{Smallness condition turning point} implies that
\[
z_{\mathrm{D}} \ge \ell^{\delta} \gg 1.
\]

We will denote differentiation with respect to $z$ by $^\prime$. With respect to $z$, \eqref{y equation ov} takes the form
\begin{equation}\label{z equation ov}
\begin{cases}
\ov_{n,\ell}'' + (\oy_*^2\ell) \oV_{n,\ell} \ov_{n,\ell} =(\oy_*^2\ell) \oF_{n,\ell},
\\
\ov_{n,\ell}(0) = \ov_{n,\ell}'(0)=0.
\end{cases}
\end{equation}
Note also that 
\[
\oy_*^2\ell=4n-1+O(\ell^{-1+3\delta})
\]
 and that, reexpressed as a function of $z$, the potential  $\oV_{n,\ell}$ satisfies
\[
\oV_{n,\ell} = (4n-1) \big( 1-z^2\big) + O\big(\ell^{-1+3\delta}(1+z^4)\big),
\]
\[
\oV_{n,\ell}>0 \quad \text{for} \quad z<1, \quad \oV_{n,\ell}<-(z-1) \quad \text{for} \quad z>1,
\]
as well as 
\[
\oV_{n,\ell}' \le 0 \quad \text{for} \quad z\in [0,1]
\]
and
\[
0 \le -(\oy_*^2\ell) \oV_{n,\ell}' \lesssim n^2 +(-  (\oy_*^2\ell) \oV_{n,\ell}) \quad \text{for} \quad z\in [1,z_{\mathrm{D}}].
\]
Therefore, the conditions of \cref{lem: Agmon type estimates} are satisfied for \eqref{z equation ov} with $\kappa\sim n^2$; we thus obtain the following estimates for $\ov_{n,\ell}$:
\begin{itemize}
\item In the region $z\in [0,1]$:
\begin{equation}\label{Energy estimate ov}
\sup_{z\in [0,1]} \Big( \big(\ov_{n,\ell}'\big)^2 + n^2 \ov_{n,\ell}^2\Big)   \lesssim  n^2  \int_0^1 \oF_{n,\ell}^2 \, dz.
\end{equation}
\item In the region $z\in [1,z_{\mathrm{D}}]$:
\begin{align*}
\int_1^{z_{\mathrm{D}}} \Big[\f{1}{(n^2 z^2)} \big(\ov_{n,\ell}''\big)^2 & +  \f{\sqrt{z^2-1}}{z}\big(\ov_{n,\ell}'\big)^2 + n^2 z^2 \ov_{n,\ell}^2\Big]  e^{2n \big(z \sqrt{z^2-1}+O(z-1)\big)} \, dz \\
\lesssim & n^2 \int_{0}^1 \oF_{n,\ell}^2 \, dz + \int_1^{z_{\mathrm{D}}}\f{1}{1+z^2} \oF_{n,\ell}^2  e^{2n \big(z \sqrt{z^2-1}+O(z-1)\big)} \, dz \\
& + \Big( \Big| \ov_{n,\ell}'(z_{\mathrm{D}}) \ov_{n,\ell}(z_{\mathrm{D}})\Big| + n \cdot z_{\mathrm{D}}\cdot  \ov_{n,\ell}^2(z_{\mathrm{D}})\Big) e^{2n \big(z_{\mathrm{D}} \sqrt{z^2_{\mathrm{D}}-1}+O(z_{\mathrm{D}})\big)}.
\end{align*}
Using the trace estimate (recalling that $z_{\mathrm{D}}\gg 1$)
\[
\f{1}{n z_{\mathrm{D}}} \big(\ov_{n,\ell}'(z_{\mathrm{D}})\big)^2 e^{2n \big(z_{\mathrm{D}} \sqrt{z^2_{\mathrm{D}}-1}+O(z_{\mathrm{D}})\big)} \lesssim  \int_{z_{\mathrm{D}}-1}^{z_{\mathrm{D}}} \Big[ \f{1}{n^2 z^2}\big(\ov_{n,\ell}''\big)^2 + \big(\ov_{n,\ell}'(z_{\mathrm{D}})\big)^2 \Big] e^{2n \big(z \sqrt{z^2-1}+O(z)\big)} \, dz,
\]
to control the term $\ov_{n,\ell}'(z_{\mathrm{D}})$ in the right-hand side above, we obtain:
\begin{align}\label{Integrated Agmon estimate for ov}
\int_1^{z_{\mathrm{D}}} \Big[\f{1}{(n^2 z^2)} \big(\ov_{n,\ell}''\big)^2 & +  \f{\sqrt{z^2-1}}{z}\big(\ov_{n,\ell}'\big)^2 + n^2 z^2 \ov_{n,\ell}^2\Big]   e^{2n \big(z \sqrt{z^2-1}+O(z-1)\big)} \, dz \\
\lesssim & n^2 \int_{0}^1 \oF_{n,\ell}^2 \, dz + \int_1^{z_{\mathrm{D}}}\f{1}{1+z^2} \oF_{n,\ell}^2  e^{2n \big(z \sqrt{z^2-1}+O(z-1)\big)} \, dz 
\nonumber \\
& \hphantom{+++}
+ n \cdot z_{\mathrm{D}}\cdot  \ov_{n,\ell}^2(z_{\mathrm{D}}) e^{2n \big(z_{\mathrm{D}} \sqrt{z^2_{\mathrm{D}}-1}+O(z_{\mathrm{D}})\big)}.   \nonumber
\end{align}
\end{itemize}

Using the asymptotic properties of Hermite functions \cite[Chapter~12]{Olver97}, we can readily estimate
\begin{equation}\label{First bounds normalized Hermite function}
|e_n(x)|+\f1{\sqrt n} \Big|\f{d e_n}{dx}(x)\Big|\lesssim
\begin{cases}
1, \quad  x\le \sqrt{4n-1},\\
e^{-(\f23-\delta) (4n-1)^{\f14} x^{\f32} }, \quad  x\ge \sqrt{4n-1}.
\end{cases}
\end{equation}
Therefore, since
\[
\oF_{n,\ell} = A  \cdot O\big(\ell^{-1+3\delta}(1+z^4)\big) \cdot \ell^{\f14} e_n(\sqrt{4n-1+O(\ell^{-1+3\delta})} \cdot z),
\]
we can readily bound
\[
\int_0^1 \oF_{n,\ell}^2 \, dz \lesssim A^2 \ell^{-\f32+6\delta}
\]
and
\[
\int_1^{z_{\mathrm{D}}}\f{1}{1+z^2} \oF_{n,\ell}^2  e^{2n \big(z \sqrt{z^2-1}+O(z-1)\big)} \, dz \lesssim  A^2 \ell^{-\f32+6\delta}.
\]
Moreover, in view of the boundary condition $\oR_{n,\ell}(y_{\mathrm{D}})=0$, we have
\[
|\ov_{n,\ell}(z_{\mathrm{D}})| = A |\oE_{n,\ell}(z_{\mathrm{D}})| \lesssim A \ell^{\f14} e^{-(\f23-2\delta)(4n-1)^{\f34} z_{\mathrm{D}}^{\f32}}. 
\]
Using the above bounds to control the terms in the right-hand sides of \eqref{Energy estimate ov} and  \eqref{Integrated Agmon estimate for ov}, we obtain
\begin{equation}\label{Energy estimate ov again}
\sup_{z\in [0,1]} \Big( \big(\ov_{n,\ell}'\big)^2 + n^2 \ov_{n,\ell}^2\Big)   \lesssim  A^2 \ell^{-\f32+8\delta}
\end{equation}
and
\begin{equation}\label{Integrated Agmon estimate for ov again}
\int_1^{z_{\mathrm{D}}} \Big[\f{1}{(n^2 z^2)} \big(\ov_{n,\ell}''\big)^2 +  \f{\sqrt{z^2-1}}{z}\big(\ov_{n,\ell}'\big)^2 + n^2 z^2 \ov_{n,\ell}^2\Big]   e^{2n \big(z \sqrt{z^2-1}+O(z-1)\big)} \, dz  \lesssim A^2 \ell^{-\f32+8\delta}.
\end{equation}
The above estimate also gives the pointwise bound
\begin{equation}\label{Pointwise Agmon estimate ov}
\sup_{z\in [1,z_{\mathrm{D}}]} \Big[\Big( \f1n \big(\ov_{n,\ell}'\big)^2 + n \ov_{n,\ell}^2\Big)e^{(2-\delta)n (z-1)^{\f32} }\Big]   \lesssim  A^2 \ell^{-\f32+8\delta}.
\end{equation}
Switching back to the $y$ coordinates, we therefore obtain from \eqref{Energy estimate ov again} and \eqref{Pointwise Agmon estimate ov}:
\begin{align}\label{Bounds ov y}
\sup_{y\in [0,\oy_*]} \Big( \f1{\ell n} \big(\f{d \ov_{n,\ell}}{dy}\big)^2 +  \ov_{n,\ell}^2\Big)   & \lesssim  A^2 \ell^{-\f32+8\delta},\\
\sup_{y\in [\oy_*, y_{\mathrm{D}}]} \Big[\Big( \f1{\ell n} \big(\f{d \ov_{n,\ell}}{dy}\big)^2 + \ov_{n,\ell}^2\Big)e^{(2-\delta)n^{\f14}\ell^{\f34} (y-\oy_*)^{\f32}} \Big]  &  \lesssim  A^2 \ell^{-\f32+8\delta}. \nonumber 
\end{align}

Recall that $\oR_{n,\ell}$ satisfies the $L^2$-normalization condition $\int_0^{y_{\mathrm{D}}} \oR^2_{n,\ell} \,dy=1$, while $e_n$ satisfies
\[
\int_0^{y_{\mathrm{D}}} \ell^{\f12} e_{n}^2(\ell^{\f12} y) \, dy= 1 - O(e^{-\ell^{\delta}}).
\]
Therefore, since $\ov_{n,\ell}(y) = \oR_{n,\ell}(y) - A \ell^{\f14} e_N(\ell^{\f12}y)$, we can estimate
\[
\big|1-A\cdot (1 - O(e^{-\ell^{\delta}}))\big| \lesssim \Big(\int_0^{y_{\mathrm{D}}}\ov_{n,\ell}^2(y) \, dy \Big)^{\f12} \lesssim A \ell^{-1+5\delta},
\]
from which we infer that
\begin{equation}\label{Bound A close to 1}
\big|1-A\big| \lesssim \ell^{-1+5\delta}.
\end{equation}
Returning to \eqref{Bounds ov y} and using the fact that $\ov_{n,\ell}(y) = \oR_{n,\ell}(y) - A \ell^{\f14} e_n (\ell^{\f12}y)$ and $e_n$ satisfies the bounds \eqref{First bounds normalized Hermite function}, we obtain, for any $y\in [0,y_{\mathrm{D}}]$:
\begin{align}\label{Bounds oR in y}
\oR_{n,\ell}(y) = \ell^{\f14} e_n \big( \ell^{\f12}y\big) + O\big(\ell^{-\f34+5\delta}\big),\\
\frac{d}{dy} \oR_{n,\ell}(y) = \ell^{\f34} e_n' \big( \ell^{\f12}y\big) + O\big(\ell^{-\f14+5\delta}\big)
\end{align}
and
\begin{equation}\label{Exponential decay oR in y}
\big|\oR_{n,\ell}(y)\big| + \f1{n^{\f12} \ell^{\f12}} \Big|\f{d}{dy} \oR_{n,\ell}(y)\Big| \lesssim e^{-\f12 n^{\f14}\ell^{\f34}\Big(y-\sqrt{\f{4n-1}{\ell}}\Big)^{\f32}}\ell^{\f14}  \quad \text{for} \quad y\ge \sqrt{\f{4n-1}{\ell}}.
\end{equation}

\medskip
\noindent \emph{Closing the bootstrap for $\oomega_{n,\ell}$.} It follows from the standard perturbation theory for Sturm--Liouville problems \cite{Zettl05} that the function $y_{\mathrm{D}} \mapsto \oomega_{n,\ell}^2(y_{\mathrm{D}})$ satisfies
\[
\f{d}{dy_{\mathrm{D}}} \oomega^2_{n,\ell} = -\Big(\f{d}{dy} \oR_{n,\ell}(y_{\mathrm{D}})\Big)^2.
\]
In view of the bound \eqref{Exponential decay oR in y} (and the fact that $y_{\mathrm{D}} \ge y_\mathrm{mirror} \gg \ell^{-\f12+3\delta}$), we therefore have
\[
\Big|\f{d}{dy_{\mathrm{D}}} \oomega^2_{n,\ell}\Big| \lesssim e^{-\ell^\delta}.
\]
Using the fact that, if $y_{\mathrm{D}}\in \mathscr B_0$, then any $\bar y_{\mathrm{D}} \in [y_{\mathrm{D}}, \f\pi2]$ also belongs to $\mathscr B_0$, we obtain by integrating the above estimate  (and using that $\oomega_{n,\ell}(y_{\mathrm{D}}=\f\pi2)=2n+\ell$)
\begin{equation}\label{Improved bound oomega exponential}
\Big| \oomega_{n,\ell} - (2n+\ell)\Big| \lesssim e^{-\ell^{\f12\delta}},
\end{equation}
from which we deduce the improvement \eqref{First improved bound to show zero mass}. Therefore, $\mathscr B_0 = [y_\mathrm{mirror}, \f\pi2]$ and \eqref{Improved bound oomega exponential} holds for all $y_{\mathrm{D}} \in [y_\mathrm{mirror}, \f\pi2]$.

\medskip

\noindent \textbf{The case $M>0$: Proof of \eqref{eq:bound-on-Rnl-M}--\eqref{Approximation estimate frequency}.} We will now return to the Sturm--Liouville problem \eqref{Radial boundary value problem} for the functions $R_{n,\ell}$ when $M\in [0,M_0]$, namely
\begin{equation}\label{Sturm Liouville problem non zero mass}
\begin{cases}
-\f{d^2}{dy^2} R_{n,\ell}+ V_{\ell} R_{n, \ell}=\omega_{n,\ell}^2 R,\\
R_{n,\ell}|_{y=0} = R_{n,\ell}|_{y=y_\mathrm{mirror}}=0,
\end{cases}
\end{equation}
where  $V_\ell$ is given by \eqref{Radial potential} and $R_{n,\ell}$ satisfies the $L^2$ normalization condition  \eqref{Normalization radial eigenfunctions}. We will think of $R_{n,\ell}$ and $\omega_{n,\ell}$ as functions of the mass parameter $M$. Recall that $y_\mathrm{mirror}$ is independent of $M\in [0,M_0]$.

We will define the following bootstrap set of mass parameters $M$:
\[
\mathscr M_0 \doteq \Big\{ M \in [0,M_0]: \quad \text{For all }\, \bar M\in [0,M], \, \text{ we have } \, \big|\omega_{n,\ell}(\bar M) -(2n+\ell)\big| \le 2 f_3(n) M_0 \ell^{-\f12} + 2\ell^{-1+\delta},   \Big\}
\]
where $f_3(n)$ is defined by
\[
f_3(n) \doteq \int_0^\infty x^3 \big(e_n(x)\big)^2 \, dx.
\]
Note that, as a consequence of  \cref{lem:Asymptotic-functions-f-and-h} in the Appendix, we have the following asymptotic formula for $f_3$:
\begin{equation}\label{Asymptotic formula f3 again}
f_3(n) = \f{32}{3\pi}n^{\f32} - \f4\pi n^{\f12}+O(n^{-\f12}) \quad \text{as} \quad n\rightarrow +\infty
\end{equation}
and so we can estimate $f_3(n)\lesssim n^{\f32}$ for all $n\in \mathbb N^*$.

The set $\mathscr M_0\subset [0,M_0]$ is closed and non-empty (since $0\in \mathscr M_0$), as a consequence of the fact that $\omega_{n,\ell}(M=0)=\oomega_{n,\ell}(y_{\mathrm{D}}=y_\mathrm{mirror}) = (2n+\ell)+O(e^{-\ell^{\f12\delta}})$; see \eqref{Improved bound oomega exponential} and  the remark below \eqref{Normalization radial eigenfunctions} on the regularity of $\omega_{n,\ell}$ as a function of $M\in [0,+\infty)$). Our aim is to show that it is also an open subset of $[0,M_0]$, which will imply that $\mathscr M_0 = [0,M_0]$. For this, it will suffice to show for any $M\in \mathscr M_0$ the improved estimate
\begin{equation}\label{Improvement bootstrap M non zero}
\big|\omega_{n,\ell}( M) -(2n+\ell)\big| \le f_3(n) M_0 \ell^{-\f12} + O(\ell^{-1})
\end{equation}

We will argue similarly as in the case of the bootstrap set $\mathscr B_0$. Let $M\in \mathscr M_0$, and let us consider the solution $\oR_{n,\ell}$ of the model Sturm--Liouville problem \eqref{Sturm Liouville problem zero mass} corresponding to 
\[
y_{\mathrm{D}} =y_\mathrm{mirror}(M_0).
\]
Let us define the constant $B$ by
\[
B = \f{\f{d R_{n,\ell}}{dy}(0)}{\f{\oR_{n,\ell}}{dy}(0)}
\]
(we will later show that $B=1+O(\ell^{-\f32+})$) and let us set
\[
v_{n,\ell}(y) = R_{n,\ell}(y)-B \cdot \oR_{n,\ell}(y).
\]
Note that $v_{n,\ell}$ solves
\begin{equation}\label{y equation v}
\begin{cases}
\f{d^2 v_{n,\ell}}{dy^2} + \ell \tilde V_{n,\ell} \cdot v_{n,\ell} =\ell F_{n,\ell},
\\
v_{n,\ell}(0) = \f{dv_{n,\ell}}{dy}(0)=0,
\end{cases}
\end{equation}
where
\begin{align*}
\tilde V_{n, \ell}(y) = &   \f{\omega^2_{n,\ell}}{\ell} - \Big(1 + \f{1}{r(y)^2}-\f{2M}{r(y)^3} \Big)\Big(\ell+1+\f{2M}{\ell r(y)}\Big)\\ 
= & 4n-1 -(\ell+1)y^2 +O(n^{\f32} \ell^{-\f12}) + O(\ell^{-1} y+\ell y^3),
\end{align*}
\[
F_{n,\ell}(y) = B \cdot \Bigg(  \f{\omega^2_{n,\ell}-\oomega^2_{n,\ell}}{\ell}  - \Big(1 + \f{1}{r(y)^2}-\f{2M}{r(y)^3} \Big)\Big(\ell+1+\f{2M}{\ell r(y)}\Big) + \f{\ell+1}{\cos^2 y}\Bigg) \oR_{n,\ell}(y)
\]
and $v_{n,\ell}$ satisfies, in addition (in view of the boundary conditions on $R_{n,\ell}$, $\oR_{n,\ell}$):
\begin{equation}\label{Other boundary v}
v_{n,\ell}|_{y=y_\mathrm{mirror}}=0.
\end{equation}
Note that
 there exists a $y_* =y_*(M)\in (0,y_\mathrm{mirror})$ satisfying 
\[
y_* = \sqrt{\f{4n-1+O(n^{\f32}\ell^{-\f12})}{\ell}}
\]
such that
\[
\tilde V_{n,\ell} (y) \ge 0 \quad \text{for }\, y\le y_*, \quad \tilde V_{n,\ell}(y) <0 \quad \text{for }\, y>y_*.
\]
Defining (as before) the rescaled coordinate 
\[
z = \f{y}{y_*},
\]
we note that $\ell y_*^2 = 4n-1+O(n^{\f32}\ell^{-\f12})$ and that $\tilde V_{n,\ell}$ (expressed as a function of $z$) satisfies 
and
\[
\f{d}{dz} \tilde V_{n,\ell} (z) \le 0 \quad \text{for }\, z\le 1, \quad  - \f{d}{dz} \tilde V_{n,\ell} (z)  \le 1 + (-\tilde V_{n,\ell}(z)) \quad \text{for }\, z>1.
\]
Therefore, applying \cref{lem: Agmon type estimates} for \eqref{y equation v} (reexpressed in terms of the $z$ coordinate), using the relation \eqref{Other boundary v} to treat the boundary terms at $z=z_\mathrm{mirror} \doteq z(y_\mathrm{mirror})$ and the relation
\[
 F_{n,\ell}(y) = B \cdot O(n^{\f32} \ell^{-\f12}) \cdot \oR_{n,\ell}(y)
\]
(in conjunction with the bounds \eqref{Bounds oR in y}--\eqref{Exponential decay oR in y} for $\oR_{n,\ell}$ and the fact that $n\le \ell^\delta$), we obtain (by arguing exactly as for the derivation of \eqref{Bounds ov y}):
\begin{align}\label{Bounds v y}
\sup_{y\in [0,y_*]} \Big( \f1{\ell n} \big(\f{d v_{n,\ell}}{dy}\big)^2 +  v_{n,\ell}^2\Big)   & \lesssim  B^2 \ell^{-1+8\delta},\\
\sup_{y\in [y_*, y_\mathrm{mirror}]} \Big[\Big( \f1{\ell n} \big(\f{d v_{n,\ell}}{dy}\big)^2 + v_{n,\ell}^2\Big)e^{(2-\delta)n^{\f14}\ell^{\f34} (y-y_*)^{\f32}} \Big]  &  \lesssim  B^2 \ell^{-1+8\delta}. \nonumber 
\end{align}
Using the fact that $\oR_{n,\ell}$ and $R_{n,\ell}$ are $L^2$ normalized and that $\int_0^{y_\mathrm{mirror}} v_{n,\ell}^2 \, dy\lesssim \ell^{-\f32+9\delta}$, arguing exactly as for the derivation of \eqref{Bound A close to 1} we obtain
\[
\big|1-B\big| \lesssim \ell^{-\f34+5\delta}.
\]
Returning to \eqref{Bounds v y} and using the fact that $v_{n,\ell} = R_{n,\ell}-B \cdot \oR_{n,\ell}$ and that $\oR_{n,\ell}$ satisfies \eqref{Bounds oR in y}--\eqref{Exponential decay oR in y}, we infer that
\begin{align}\label{Bounds R in y}
R_{n,\ell}(y) = \ell^{\f14} e_n \big( \ell^{\f12}y\big) + O\big(\ell^{-\f12+5\delta}\big),\\
\frac{d}{dy} R_{n,\ell}(y) = \ell^{\f34} e_n' \big( \ell^{\f12}y\big) + O\big(\ell^{\f14+5\delta}\big)
\end{align}
and
\begin{equation}\label{Exponential decay R in y}
\big|R_{n,\ell}(y)\big| + \f1{n^{\f12} \ell^{\f12}} \Big|\f{d}{dy} R_{n,\ell}(y)\Big| \lesssim e^{-\f12 n^{\f14}\ell^{\f34}\Big(y-\sqrt{\f{4n-1}{\ell}}\Big)^{\f32}} \ell^{\f14} \quad \text{for} \quad y\ge \sqrt{\f{4n-1}{\ell}}.
\end{equation}

\medskip
\noindent \emph{Closing the bootstrap for $\omega_{n,\ell}$.} It follows from the standard perturbation theory for eigenvalue problems that the derivative of the eigenvalue $\omega^2_{n,\ell}$ with respect to the parameter $M$ satisfies
\begin{align*}
\Big|\f{d}{dM}\omega^2_{n,\ell}(M)\Big| = & \Big|\int_0^{y_\mathrm{mirror}} \f{\partial V_{\ell}}{\partial M}R_{n,\ell}^2(y) \, dy\Big| \\
\leq  & 
 \Big|\int_0^{y_\mathrm{mirror}} \Big(-\f{2}{r(y)^3} \big(\ell(\ell+1)+\f{2M}{r(y)}\big) + \f{2}{r(y)}\big(1+\f{1}{r(y)^2}-\f{2M}{r(y)^3} \Big)R_{n,\ell}^2(y) \, dy\Big|\\
   & +
 \Big|\int_0^{y_\mathrm{mirror}}\frac{\partial V_\ell(r,M)}{\partial r} \frac{\partial r(y,M)}{\partial M}R_{n,\ell}^2(y) \, dy\Big|
\\
\le&   2 \big(\ell(\ell+1)+O(1)\big) \Big|\int_0^{y_\mathrm{mirror}} \big(y^3+O(y^4)\big) R_{n,\ell}^2(y) \, dy\Big| \\
 & +  O(1) \Big|\int_0^{y_\mathrm{mirror}} \Big(\f{1}{r(y)}+\f{1}{r(y)^3}\Big) R_{n,\ell}^2(y) \, dy\Big|.
\end{align*}
Using the approximation estimates \eqref{Bounds R in y}--\eqref{Exponential decay R in y} for $R_{n,\ell}$  and the fact that $y_\mathrm{mirror} \ge \ell^{-\f12+3\delta}$, we infer that
\[
\Big|\f{d}{dM}\omega^2_{n,\ell}(M)\Big| \le 2\big(\ell^{\f12}+O(\ell^{-\f12})\big)\int_0^{+\infty} (\ell^{\f12}y) ^3 \ell^{\f12} e_n^2(\ell^{\f12} y) \, dy + O(1) = 2f_3(n) \ell^{\f12} +O(1),
\]
from which we obtain (using the fact that $\omega_{n,\ell}=2n+\ell+O(\ell^{-\f12+2\delta}$):
\[
\Big|\f{d}{dM}\omega_{n,\ell}(M)\Big| \le f_3(n) \ell^{-\f12}+O(\ell^{-1}).
\]
Integrating the above over $M\in (0,M_0]$ and using the fact that $\omega_{n,\ell}(M=0) = \oomega_{n,\ell}(y_{\mathrm{D}} = y_\mathrm{mirror})=2n+\ell + O(e^{-\ell^{\f12\delta}})$ (see \eqref{Improved bound oomega exponential}  and \cref{remark:Smooth dependence eigenvalues}), we obtain \eqref{Improvement bootstrap M non zero}. Therefore, $\mathscr M_0 = [0,M_0]$ and the bounds \eqref{Improvement bootstrap M non zero} and \eqref{Bounds R in y}--\eqref{Exponential decay R in y} hold for all $M\in [0,M_0]$.

The bound \eqref{Approximation estimate frequency} now follows directly from \eqref{Improvement bootstrap M non zero}, while \eqref{eq:bound-on-Rnl-M}--\eqref{eq:bound-on-Rnl-prime-M} and \eqref{eq:expo-decay} follow from \eqref{Bounds R in y} and \eqref{Exponential decay R in y}, respectively.
\end{proof}

We will also make use of the following crude estimate,  obtained using standard results in Sturm--Liouville theory, regarding the asymptotic behavior of the frequencies $\omega_{n,\ell}$ when $n$ does not necessarily lie in the low-lying regime:

\begin{lemma}\label{lem:Crude Weyl law}
Let $M_0>0$ be a given constant. For any Schwarzschild--AdS mass parameter $M\in [0,M_0]$ and any $(n,\ell)\in (\mathbb N^*)^2$ such that $\ell$ is sufficiently large in terms of $M_0$, the eigenvalues of the radial Sturm--Liouville problem \eqref{Radial boundary value problem} satisfy
\begin{equation}\label{Crude bounds frequency}
\f1C n \le \omega_{n,\ell}-\ell \le C n,
\end{equation}
where $C>0$ depends only on $M_0$.
\end{lemma}

\begin{proof}
For the proof of this lemma, we will assume that all the constants implicit in the $O(\cdot)$ notation are allowed to depend on $M_0$ (but are independent of $n,\ell$). We will fix a $\delta\in (0,1)$ small enough in terms of $M_0$, and we will assume, without loss of generality, that
\[
n\ge \ell^{\delta}
\]
(since, in the case when $n<\ell^\delta$, \eqref{Crude bounds frequency} follows immediately from \eqref{Approximation estimate frequency}).

In order to establish \eqref{Crude bounds frequency}, we will make use of the standard \emph{Pr\"ufer} substitution \cite{P26}: We will define the functions $\mathfrak R: [0,y_\mathrm{mirror}]\rightarrow (0,+\infty)$ and $ \Theta:[0,y_\mathrm{mirror}]\rightarrow \mathbb R$ by the relations
\[
R_{n,\ell} = (-1)^{n-1} \mathfrak R \cdot \sin\Theta, \quad \f{d R_{n,\ell}}{dy} = (-1)^{n-1} \mathfrak R \cdot \cos\Theta.
\]
with 
\[
\Theta(0)=0
\]
(note that this choice is consistent with the boundary condition $R_{n,\ell}(0)=0$ and the sign convention for $R_{n,\ell}$). Let us also set
\[
Q(y) \doteq \omega^2_{n,\ell}- V_\ell(y).
\]
Then the equation
\begin{equation}\label{Radial equation rewritten}
\f{d^2 R_{n,\ell}}{dy^2} + Q(y) R_{n,\ell} =0
\end{equation}
is formally equivalent to
\[
\f{d \mathfrak R}{dy} = \f12 (1-Q) \cdot \mathfrak R \cdot \sin(2\Theta), \quad \f{d \Theta}{dy} = \cos^2\Theta + Q \cdot \sin^2 \Theta.
\]
Note, in particular, the following facts: 
\begin{itemize}
\item $\Theta$ satisfies a first order equation which is decoupled from $\mathfrak R$.
\item The points where $R_{n,\ell}$ vanishes correspond precisely to those points where $\Theta = m \pi$ for some $m\in \mathbb Z$.
\item At any point $y_* \in [0,y_\mathrm{mirror}]$ such that $\Theta(y_*) \in \mathbb Z \pi$, we have
\[
\f{d \Theta }{dy}(y_*) >0.
\] 
Therefore, for any $m\in \mathbb N$, if $y_m \in [0,y_\mathrm{mirror}]$ is such that
\begin{equation}\label{Definition y m}
\Theta(y_m) = m \pi,
\end{equation}
then
\[
\Theta(y) 
\begin{cases}
< m\pi \quad \text{for } y<y_m,\\
>m\pi \quad \text{for } y>y_m.
\end{cases}
\]
In particular, there is at most one such $y_m$.
\item If $Q(y)\le 0$ on an interval $y\in [a,b]$, then $R_{n,\ell}$ is convex where $R_{n,\ell}>0$ and concave where $R_{n,\ell}<0$ and thus it can only vanish at most once in $[a,b]$ since $R_{n,\ell}\not\equiv 0$.
\end{itemize}

According to standard Sturm--Liouville theory, $R_{n,\ell}$ vanishes exactly $n+1$ times on $[0,y_\mathrm{mirror}]$ (counting also the boundary points $0,y_\mathrm{mirror}$). As a result, 
\[
\Theta(y_\mathrm{mirror}) = n \pi.
\]
For any $m\in \{0, \ldots, n\}$, we will define $y_m$ to be the unique point in $[0,y_\mathrm{mirror}]$ such that \eqref{Definition y m} holds (noting that $y_0=0$, $y_{n}=y_\mathrm{mirror}$). Note that the sequence of points $\{y_m\}_{m=0}^n$ is strictly increasing in $m$.

The monotonicity property \eqref{Monotonicity radial potential} for $V_{\ell}(y)$ in the region $y\in (0, y_\mathrm{mirror})$ implies that
\[
V_\ell(y_\mathrm{mirror}) \ge V_{\ell}(y) \ge V_\ell (0) = \ell (\ell+1) \quad \text{for} \quad y\in [0,y_\mathrm{mirror}]
\]
and
\begin{equation}\label{Monotonicity Q}
\f{d}{dy}Q(y) < 0 \quad \text{for} \quad y\in (0,y_\mathrm{mirror}).
\end{equation}
Note also that, in view of the fact that $R_{n,\ell}(y_\mathrm{mirror})=0$ and $\f{d}{dy}R_{n,\ell}(y_\mathrm{mirror}) <0 $, equation \eqref{Radial equation rewritten} implies that if, for some $\bar y\in [0, y_\mathrm{mirror})$ we have
\[
Q(y) \le 0 \quad \text{for} \quad y\in [\bar y, y_\mathrm{mirror}],
\]
then $\f{d^2}{dy^2} R_{n,\ell} \ge 0$ on $[\bar y, y_\mathrm{mirror}]$ and, therefore
\[
R_{n,\ell}(\bar y)>0.
\]
From the above, we infer that
\[
\omega_{n,\ell} > \min_{y\in [0, y_\mathrm{mirror}]} V_{\ell} = \ell(\ell+1)
\]
(since, otherwise, we would have $Q(y)\le 0$ on the whole interval $ [0, y_\mathrm{mirror}]$, which would imply that $R_{n,\ell}(0)>0$, a contradiction). Moreover, the monotonicity bound \eqref{Monotonicity Q} for $Q$ implies that, for any $n\in \mathbb N^*$, there exists a unique $y_\mathrm{tp}^{(n,\ell)} \in (0, y_\mathrm{mirror}]$ with the property that
\[
Q(y) > 0 \, \text{ for }\, y<y_\mathrm{tp}^{(n,\ell)} \quad \text{and} \quad
Q(y) < 0 \, \text{ for }\, y>y_\mathrm{tp}^{(n,\ell)}.
\]
Note that
\[
y_\mathrm{tp}^{(n,\ell)} < y_\mathrm{mirror} \, \text{ if } \, \omega_{n,\ell}^2 < V_\ell(y_\mathrm{mirror}) \quad \text{and} \quad y_\mathrm{tp}^{(n,\ell)} = y_\mathrm{mirror} \, \text{ if } \, \omega_{n,\ell}^2 \ge  V_\ell(y_\mathrm{mirror}).
\]
Moreover, in the case when $y_\mathrm{tp}^{(n,\ell)} < y_\mathrm{mirror}$ (in which case $Q<0$ for $y\in (y_\mathrm{tp}^{(n,\ell)}, y_\mathrm{mirror}]$), our previous discussion implies that
\begin{equation}\label{No root R beyond turning point}
R_{n,\ell}(y)  >0 \quad \text{for} \quad y \in [y_\mathrm{tp}^{(n,\ell)}, y_\mathrm{mirror}).
\end{equation}
As a consequence of \eqref{No root R beyond turning point}, we have
\begin{equation}\label{Upper bound maximum root}
y_{n-1} < y_\mathrm{tp}^{(n,\ell)},
\end{equation}
while the fact that $Q>0$ for $y\in [0, y_\mathrm{tp}^{(n,\ell)})$ implies that:
\[
\min_{y\in [0, y_{n-1}]} Q(y) >0.
\]

The equation for $\f{d \Theta}{dy}$ and the fact that $Q$ is decreasing and non-negative on $ [0, y_\mathrm{tp}^{(n,\ell)}]$ implies that, on any interval $[a,b]\subset [0, y_\mathrm{tp}^{(n,\ell)}]$, we have
\begin{equation}\label{Inequality Theta}
\cos^2\Theta + Q(b) \sin^2\Theta \le \f{d\Theta}{dy} \le \cos^2\Theta + Q(a) \sin^2\Theta.
\end{equation}
Integrating the above over $[y_m, y_{m+1}]$ for any $m\in \{0, \ldots, n-2\}$, we obtain:
\begin{equation}\label{Estimate successive roots}
\f{\pi}{\sqrt{Q(y_{m})}} \le y_{m+1} - y_{m} \le \f{\pi}{\sqrt{Q(y_{m+1})}}.
\end{equation}
More generally, \eqref{Inequality Theta} implies that on any interval $[a,b]\subset [0, y_\mathrm{tp}^{(n,\ell)}]$ on which $Q(b) > \Big(\f{\pi}{b-a}\Big)^2$, $\Theta(y)$ must attain a value which is an integer multiple of $\pi$ (and, hence, $[a,b]$ must contain some $y_m$, $m\in \{0, \ldots, n-1\}$).

We will now consider two cases, depending on whether $Q(y)$ is uniformly positive on $(0, y_\mathrm{mirror})$ or not; we will make use of the fact that we assumed that $n\ge \ell^\delta$, which implies, in view of the monotonicity of $\omega_{n,\ell}$ in $n$ and the bound \eqref{Approximation estimate frequency} for $\omega_{\ell^\delta, \ell}$:
\begin{equation}\label{Lower bound omega from monotonicity}
\omega_{n,\ell}-\ell \ge \omega_{\ell^\delta, \ell} -\ell \gtrsim \ell^\delta.
\end{equation}
\begin{enumerate}
\item Let us first consider the case when 
\begin{equation}\label{First case positivity Q}
\omega^2_{n,\ell} >  V_\ell(y_\mathrm{mirror}) + \f12 \Big( \omega^2_{n,\ell}-\ell(\ell+1)\Big),
\end{equation}
in which case $y_{pt}^{(n,\ell)}=y_\mathrm{mirror}$ and 
\[
Q(y) > \f12 \Big( \omega^2_{n,\ell}-\ell(\ell+1)\Big) >0 \quad \text{for all} \quad y\in [0, y_\mathrm{mirror}].
\]
Since $Q$ is decreasing in $y$ and $Q(0) = \omega^2_{n,\ell}-\ell(\ell+1)$, we infer that, in this case,
\begin{equation}\label{Estimate positive Q}
Q(y) \sim \omega^2_{n,\ell}-\ell(\ell+1) \quad \text{for all} \quad y\in [0, y_\mathrm{mirror}].
\end{equation}
In view of the lower bound \eqref{Lower bound omega from monotonicity} for $\omega_{n,\ell}$, we deduce that
\[
Q(y) \gtrsim \ell^{1+\delta}.
\]
The discussion below \eqref{Estimate successive roots} then implies that at least one of the $y_m$'s, $m\in \{0, \ldots, n-1\}$ lies in the interval $[\f12 y_\mathrm{mirror}, y_\mathrm{mirror}]$. Since $y_m$ is increasing in $m$, we infer that
\[
\f12 y_\mathrm{mirror} \le y_{n-1} < y_\mathrm{mirror}.
\]
Therefore,
\[
\sum_{m=0}^{n-2} (y_{m+1}-y_m) = y_{n-1} \sim 1.
\]
Using the estimate \eqref{Estimate successive roots} for $y_{m+1}-y_m$ together with the estimate \eqref{Estimate positive Q} for $Q(y)$, we obtain:
\[
\f{n-1}{\sqrt{\omega^2_{n,\ell}-\ell(\ell+1)}} \sim 1,
\]
i.e.
\begin{equation}\label{Asymptotic expression omega uniform positivity}
\omega^2_{n,\ell}-\ell^2 \sim n^2 +\ell.
\end{equation}

Our assumption \eqref{First case positivity Q} and the explicit expression \eqref{Radial potential} for $V_\ell$ imply that
\[
\f12 \omega^2_{n,\ell} > V_\ell(y_\mathrm{mirror})-\ell(\ell+1) \gtrsim \ell^2.
\]
In view of the expression \eqref{Asymptotic expression omega uniform positivity}, the above bound yields
\[
n\gtrsim \ell.
\]
Thus, returning to \eqref{Asymptotic expression omega uniform positivity}, we obtain
\begin{equation}\label{Asymptotic formula omega 1}
\omega_{n,\ell}-\ell \sim n.
\end{equation}

\item Let us now consider the case when
\[
\omega^2_{n,\ell} \le  V_\ell(y_\mathrm{mirror}) + \f12 \Big( \omega^2_{n,\ell}-\ell(\ell+1)\Big). 
\]
Note that when $\omega^2_{n,\ell} \le V_\ell(y_\mathrm{mirror}) $, we have $Q(y_\mathrm{tp}^{(n,\ell)})=0$, while when $\omega^2_{n,\ell}>V_\ell(y_\mathrm{mirror})$ we have $y_\mathrm{tp}^{(n,\ell)}=y_\mathrm{mirror}$ and $Q(y_\mathrm{tp}^{(n,\ell)})=\omega^2_{n,\ell}-V_\ell(y_\mathrm{mirror})$. In either case, the following bound holds:
\begin{equation}\label{Bound Q turning point}
0\le Q(y_\mathrm{tp}^{(n,\ell)}) \le \f12 \Big( \omega^2_{n,\ell}-\ell(\ell+1)\Big).
\end{equation}

 The explicit expression \eqref{Radial potential} of $V_\ell(y)$ implies that, for $y\in [0, y_\mathrm{mirror}]$, we have
\[
\f{d}{dy}Q(y) = - \f{d}{dy}V_\ell (y) \sim -\ell^2 y,
\]
which, upon integration from $y=y_\mathrm{tp}^{(n,\ell)}$, yields
\begin{equation}\label{Estimate Q turning point}
Q(y)-Q(y_\mathrm{tp}^{(n,\ell)}) \sim \ell^2 \cdot \big( (y_\mathrm{tp}^{(n,\ell)})^2-y^2 \big) \quad \text{for} \quad y\in [0, y_\mathrm{tp}^{(n,\ell)}].
\end{equation}
Since $Q(0)=\omega^2_{n,\ell}-\ell(\ell+1)$, we readily obtain in this case (using the bound  \eqref{Bound Q turning point} for $Q(y_\mathrm{tp}^{(n,\ell)})$) that
\[
y_\mathrm{tp}^{(n,\ell)} \sim \f{\sqrt{\omega^2_{n,\ell}-\ell(\ell+1)}}{\ell}.
\]
Note that, since $y_\mathrm{tp}^{(n,\ell)} \le y_\mathrm{mirror} \lesssim 1$, a consequence of the above estimates is that
\begin{equation}\label{Upper bound omega reduced case}
\omega^2_{n,\ell}  \lesssim \ell^2.
\end{equation}
In view of the lower bound \eqref{Lower bound omega from monotonicity} for $\omega_{n,\ell}-\ell$, we infer that
\begin{equation}\label{Turning point estimate}
y_\mathrm{tp}^{(n,\ell)} \gtrsim \ell^{-\f12+\f12 \delta}.
\end{equation}

On the interval $[\f14 y_\mathrm{tp}^{(n,\ell)}, \f12 y_\mathrm{tp}^{(n,\ell)}]$, we have in view of \eqref{Estimate Q turning point}:
\[
\min_{y \in [\f14 y_\mathrm{tp}^{(n,\ell)}, \f12 y_\mathrm{tp}^{(n,\ell)}]} Q(y) \gtrsim \ell^2 (y_\mathrm{tp}^{(n,\ell)})^2,
\]
which implies (since $\ell^\delta \gg 1$) that
\[
\min_{y \in [\f14 y_\mathrm{tp}^{(n,\ell)}, \f12 y_\mathrm{tp}^{(n,\ell)}]} Q(y) \gg \f{1}{(y_\mathrm{tp}^{(n,\ell)} )^2}.
\]
As a consequence of our discussion below \eqref{Estimate successive roots}, we therefore infer that $[\f14 y_\mathrm{tp}^{(n,\ell)}, \f12 y_\mathrm{tp}^{(n,\ell)}]$ contains some $y_m$, $m\in \{0, \ldots, n-1\}$ and, thus
\[
\f14 y_\mathrm{tp}^{(n,\ell)} \le y_{n-1} \le y_\mathrm{tp}^{(n,\ell)}.
\]

Using the fact that 
\[
\sum_{m=0}^{n-2}(y_{m+1}-y_m) = y_{n-1} \sim y_\mathrm{tp}^{(n,\ell)}
\]
together with the estimate \eqref{Estimate successive roots} for $y_{m+1}-y_m$ and the fact that 
\[
 Q(0) \Big(1-\f{y^2}{(y_\mathrm{tp}^{(n,\ell)})^2}\Big) \lesssim Q(y) \le  Q(0),
\]
an application of the pigeonhole principle readily implies that
\[
\f{1}{\sqrt{Q(0)}} \sim \f{y_\mathrm{tp}^{(n,\ell)}}{n}.
\]
In view of the estimate \eqref{Turning point estimate} for $y_\mathrm{tp}^{(n,\ell)}$ and the fact that $Q(0) = \omega^2_{n,\ell}-\ell(\ell+1)$, we therefore obtain
\begin{equation}\label{First asymptotic formula omega}
\omega_{n,\ell}^2-\ell^2 \sim n \ell.
\end{equation}
The above asymptotic formula implies, in view of \eqref{Upper bound omega reduced case}, that
\[
n\lesssim \ell,
\]
and therefore we infer from \eqref{First asymptotic formula omega} that
\begin{equation}\label{Asymptotic formula omega 2}
\omega_{n,\ell} - \ell \sim n.
\end{equation}
\end{enumerate}
Combining \eqref{Asymptotic formula omega 1} and \eqref{Asymptotic formula omega 2}, we obtain \eqref{Crude bounds frequency}.

\end{proof}

The following result provides an exponentially small bound for $R_{n,\ell}$ in the region $\{\f12 y_\mathrm{mirror} \le y \le y_\mathrm{mirror}\}$ when $n$ lies in the intermediate regime $n\lesssim \ell^{1-\delta}$:

\begin{lemma}
Let $M_0>0$ be a given constant. For any Schwarzschild--AdS mass parameter $M\in [0,M_0]$ and any $0<\delta\ll 1$, let $(n,\ell)\in (\mathbb N^*)^2$ be a pair of parameters such that $\ell\gg 1$ is sufficiently large in terms of $M_0, \delta$ and $n$ satisfies
\[
n\le \ell^{1-\delta}.
\]
Let us also set
\[
y_\mathrm{cr} \doteq \sqrt{\f{n}{\ell}}.
\]
Then, for any $y_*$ in the interval
\[
 \ell^{\f14\delta} y_\mathrm{cr} \le y_* \le \f12 y_\mathrm{mirror}, 
\]
we have
\begin{equation}\label{Exponentially small bound R general}
\sup_{y \in [y_* , y_\mathrm{mirror}]} \Big( \big| R_{n,\ell}(y)\big| + \Big| \f{d}{dy} R_{n,\ell}(y)\Big| \Big) \le e^{-c y_*^2 \ell,}
\end{equation}
where the constant $c>0$ depends only on $M_0, \delta$.
\end{lemma}

\begin{proof}
In this proof, we will assume that the constants implicit in the $O(\cdot)$ and $\lesssim $ notation are allowed to depend on $M_0, \delta$ (but not on $n, \ell$).

Using \cref{lem:Crude Weyl law} for $\omega_{n,\ell}$ and our assumption that $n\lesssim \ell^{1-\delta}$ and $y_* \gtrsim n^{\f12}\ell^{\f14\delta-\f12}$, we can readily estimate that:
\[
 V_{\ell}(y) -  \omega_{n,\ell}^2 \gtrsim \ell^2 y_*^2 \quad \text{for all} \quad y \in [\f12 y_*, y_\mathrm{mirror}].
\]
Therefore, multiplying equation
\begin{equation}\label{R equation once more}
\f{d^2}{dy^2}R_{n,\ell} - \Big( V_\ell - \omega^2_{n,\ell}\Big) R_{n,\ell}=0
\end{equation}
with $e^{c_* y_*\ell \cdot (y-\f12 y_*)} R_{n,\ell}(y)$ for some $c_*>0$ small enough in terms of $M_0, \delta$ (to be determined more precisely later)  and integrating by parts over $y \in [\f12 y_*, y_\mathrm{mirror}]$, we obtain:
\begin{multline*}
  \int_{\f12 y_*}^{y_\mathrm{mirror}} \Big( \big( \f{d R_{n,\ell}}{dy}\big)^2 + \Big( V_\ell - \omega^2_{n,\ell} - \f12 c_*^2 y_*^2 \ell^2\Big) R^2_{n,\ell}\Big) e^{c_* y_* \ell \cdot (y-\f12 y_*)}\, dy \\ = \Big[ \Big( - R_{n,\ell} \f{d R_{n,\ell}}{dy} + \f12 c_* y_* \ell R_{n,\ell}^2
  \Big) e^{c_* y_* \ell \cdot (y-\f12 y_*)}\Big]_{y=\f12 y_*}^{y_\mathrm{mirror}}.
\end{multline*}
Assuming that $c_*$ is small enough so that 
\[
V_{\ell}(y) -  \omega_{n,\ell}^2 \gtrsim 2 c_*^2 y_*^2 \ell^2 \quad \text{for all} \quad y \in [\f12 y_*, y_\mathrm{mirror}].
\]
and using the fact that $R_{n,\ell}(y_\mathrm{mirror})=0$ (and $y_* \lesssim 1$), we therefore obtain:
\[
\int_{\f12 y_*}^{y_\mathrm{mirror}} \Big( \big( \f{d R_{n,\ell}}{dy}\big)^2 + \ell^2 y_*^2 R^2_{n,\ell}\Big) e^{c_* y_*\ell \cdot (y-\f12 y_*)}\, dy \lesssim \ell^{-1} \Big(  \f{d R_{n,\ell}}{dy} \Big)^2|_{y=\f12 y_*} + \ell R_{n,\ell}^2 |_{y=\f12 y_*}.
\]
Using the fact that $R_{n,\ell}$  satisfies the $2^{nd}$ order ODE  \eqref{R equation once more}, the above bound immediately yields:
\begin{equation}\label{Crude Carleman estimate}
\int_{\f12 y_*}^{y_\mathrm{mirror}} \Big(\f1{\ell^2 y_*^2}\big( \f{d^2 R_{n,\ell}}{dy^2}\big)^2 +  \big( \f{d R_{n,\ell}}{dy}\big)^2 + \ell^2 y_*^2  R^2_{n,\ell}\Big) e^{c_* y_* \ell \cdot (y-\f12 y_*)}\, dy \lesssim \ell^{-1} \Big(  \f{d R_{n,\ell}}{dy} \Big)^2|_{y=\f12 y_*} + \ell R_{n,\ell}^2|_{y=\f12 y_*}.
\end{equation}

We will now make use of the fact that $R_{n,\ell}$ is $L^2$-normalized and satisfies \eqref{R equation once more} to infer that
\[
\int_0^{y_\mathrm{mirror}} \Big( \big( \f{d^2 R_{n,\ell}}{dy^2}\big)^2 + \ell^4 R_{n,\ell}^2 \Big) \, dy \lesssim \ell^4
\]
and, as a result:
\[
\sup_{y\in [0,y_\mathrm{mirror}]} \Big( \Big| \f{d R_{n,\ell}}{dy}\Big| + \ell \big| R_{n,\ell}\big| \Big) \lesssim \ell^{2}. 
\]
Therefore, using the above bound to estimate the right-hand side of \eqref{Crude Carleman estimate}, we obtain:
\begin{equation}\label{Crude Carleman estimate again}
\int_{\f12 y_*}^{y_\mathrm{mirror}} \Big(\f1{\ell^2 y_*^2}\big( \f{d^2 R_{n,\ell}}{dy^2}\big)^2 +  \big( \f{d R_{n,\ell}}{dy}\big)^2 + \ell^2 y_*^2  R^2_{n,\ell}\Big) e^{c_* y_* \ell \cdot (y-\f12 y_*)}\, dy \lesssim \ell^3.
\end{equation}
The trivial Sobolev-type bound
\begin{align*}
\sup_{y\in [\f12 y_*, y_\mathrm{mirror}]} \Bigg( & \Big(\f{d}{dy}\big( e^{\f12c_* y_* \ell \cdot (y-\f12 y_*)} R_{n,\ell}\big)\Big)^2 + \big( e^{\f12c_* y_*\ell \cdot (y-\f12 y_*)} R_{n,\ell}\big)^2 \Bigg)\\
&  \lesssim 
\int_{\f12 y_*}^{y_\mathrm{mirror}} \Big(\f{d^2 (e^{\f12 c_* y_* \ell \cdot (y-\f12 y_*)} R_{n,\ell})}{dy^2}\big)^2 +  \big( \f{d (e^{\f12 c_* y_* \ell \cdot (y-\f12 y_*)} R_{n,\ell})}{dy}\big)^2 + e^{c_* y_* \ell \cdot (y-\f12 y_*)}  R^2_{n,\ell}\Big) \, dy \\
&  \lesssim 
\int_{\f12 y_*}^{y_\mathrm{mirror}} \Big(\big(\f{d^2 R_{n,\ell}}{dy^2}\big)^2 + \ell^2 y_*^2 \big( \f{d  R_{n,\ell}}{dy}\big)^2 + \ell^4 y_*^4  R^2_{n,\ell}\Big) e^{c_* y_* \ell \cdot (y-\f12 y_*)} \, dy
\end{align*}
then implies that
\[
\sup_{y\in [\f12 y_*, y_\mathrm{mirror}]} \Bigg[ e^{c_*y_*\ell \cdot (y-\f12 y_*)} \Bigg( \Big(\f{d}{dy} R_{n,\ell}\Big)^2 + \big( R_{n,\ell}\big)^2 \Bigg)\Bigg]  \lesssim \ell^7 y_*^2,
\]
from which \eqref{Exponentially small bound R general} follows readily by choosing a $c=c(c_*)$ such that
\[
\max_{y_*\le y \le y_\mathrm{mirror}} \big(\ell^7 y_*^2 e^{-c_* y_* \ell \cdot (y-\f12 y_*)} \big) = \ell^7 y_*^2 e^{-\f12 c_* y_*^2 \ell} \le e^{-2c y_*^2 \ell}
\]
(this is possible in view of the fact that $\ell^{\f14\delta-\f12} \lesssim y_* \lesssim 1$).

\end{proof}

The following lemma will provide a crude ``non-stationary phase'' estimate for products of radial eigenfunctions $R_{n,\ell}$.

\begin{lemma}\label{lem:Crude non stationary phase}
Let $M_0>0$ and $0< \delta \ll 1$ be given constants. Let $(n, \ell)$ and $(n_i, \ell_i)$, $i=1,2,3$, be pairs of positive integers satisfying the following conditions:
\begin{itemize}
\item The angular momenta $\ell$ and $\ell_i$ are sufficiently large in terms of $M_0, \delta$ and are comparable in size, in the sense that there exist $\lambda_-, \lambda_+ \in (0,+\infty)$ such that
\begin{equation}\label{Comparable ells}
\lambda_- \le \f{\ell_i}{\ell} \le \lambda_+ \quad \text{for all} \quad i=1,2,3
\end{equation}
and 
\[
\lambda_+ \le \ell^\delta.
\]
\item The overtones $n_i$, $i=1,2,3$, satisfy
\[
n_i \le \ell_i^{\delta}.
\]
\item The overtone $n$ is large in terms of $M_0$ and $\delta$ and satisfies
\begin{equation}\label{Largeness of overtone n}
\f{n}{\ell} \ge n^{\delta} \max_{i=1,2,3} \f{n_i}{\ell_i}.
\end{equation}
\end{itemize}
Then, for any Schwarzschild--AdS mass parameter $M\in [0,M_0]$ and any function $V\in C^{10} \big([0,y_\mathrm{mirror}]\big)$, the radial eigenfunctions $R_{n,\ell}$ and $R_{n_i,\ell_i}$, $i=1,2,3$, satisfy
\begin{align}\label{Estimate oscillatory integral in Rs}
\int_0^{y_\mathrm{mirror}} y^6 \, V(y) R_{n,\ell}(y)  & R_{n_1,\ell_1}(y) R_{n_2,\ell_2}(y) R_{n_3,\ell_3}(y) \, dy \\
& \le C \| V\|_{C^{10}([0,y_\mathrm{mirror}]} \Big(\max_{i=1,2,3} \f{n_i}{\ell_i}\Big)^3 \cdot  \f{\max_{i=1,2,3} n_i^5}{n^5}  \ell^{\f12}, \nonumber
\end{align}
where the constant $C>0$ depends only on $M_0$, $\delta$ and the constants $\lambda_\pm$. 
\end{lemma}

\begin{remark} The decay rate of the right-hand side of \eqref{Estimate oscillatory integral in Rs} in terms of the high frequency overtone $n$ (namely $n^{-5}$) is not optimal. However, the proof shows that the optimal decay rate depends on the regularity of the function $y^{6+3}\cdot V(y)=y^9 \cdot V(y)$ at $y=0$ when $V$ is extended as an \emph{even} function across $y=0$, with the decay rate being superpolynomial only when $V$ extends as a $C^\infty$ function.\footnote{Extending $V$ as an even function and $R_{n,\ell}$, $R_{n_i,\ell_i}$ as odd functions across $y=0$, estimating the integral \eqref{Estimate oscillatory integral in Rs} can be done via a simple non-stationary phase-type argument.} In this paper, we will only consider the case when $V(y)=1+\f{1}{r(y)^2}-\f{2M}{r(y)^3}$; in this case, extending $V$ evenly, the function $y^9 \cdot V(y)$ is merely of class $C^{12}$ (and not $C^{13}$) at $y=0$ unless $M=0$. 
\end{remark}

\begin{proof}
In this proof, we will assume that all constants implicit in the $\sim$, $\lesssim$ and $O(\cdot)$ notation might depend on $\lambda_\pm$, $M_0$ and $\delta$ (but are independent of the parameters $(n,\ell)$ and $(n_i,\ell_i)$). With this convention, we have
\[
\ell_i \sim \ell \quad \text{for all} \quad i=1,2,3.
\]

Let us set
\[
y_\mathrm{cr} \doteq \max_{i=1,2,3} \sqrt{\f{4n_i-1}{\ell_i}},
\]
fix a smooth cut-off function $\chi_c: [0,+\infty)\rightarrow [0,1]$ such that $\chi_c\equiv 1$ on $[0,1]$ and $\chi_c\equiv 0$ on $[2,+\infty)$ and let us split
\begin{align*}
 \int_0^{y_\mathrm{mirror}}  & y^6 V(y) R_{n,\ell}(y)  R_{n_1,\ell_1}(y) R_{n_2,\ell_2}(y) R_{n_3,\ell_3}(y) \, dy \\
 = &   \int_0^{y_\mathrm{mirror}}  \chi_c\big(n^{-\f14\delta}\f{y}{y_\mathrm{cr}}\big) y^6 V(y) R_{n,\ell}(y)  R_{n_1,\ell_1}(y) R_{n_2,\ell_2}(y) R_{n_3,\ell_3}(y) \, dy \\
 &+  \int_0^{y_\mathrm{mirror}} \Big(1-\chi_c\big(n^{-\f14\delta}\f{y}{y_\mathrm{cr}}\big) \Big) y^6 V(y) R_{n,\ell}(y)  R_{n_1,\ell_1}(y) R_{n_2,\ell_2}(y) R_{n_3,\ell_3}(y) \, dy.
\end{align*}
\begin{itemize}
\item In the region $\{y\gtrsim n^{\f14\delta} y_\mathrm{cr}\}$, where $1-\chi_c\big(n^{-\f14\delta}\f{y}{y_\mathrm{cr}}\big)$ is supported, the functions $R_{n_i, \ell_i}$, $i=1,2,3$, satisfy the exponential decay estimate \eqref{eq:expo-decay}, from which we obtain in particular that
\[
\sup_{y\gtrsim n^{\f14\delta} y_\mathrm{cr}} \big| R_{n_i,\ell_i}(y)\big| \lesssim e^{-n^{\f14\delta}}\ell_i^{\f14}.
\]
Therefore, using also the fact that $R_{n,\ell}$ and the $R_{n_i, \ell_i}$'s are $L^2$-normalized (and the $\ell$ and $\ell_i$'s satisfy \eqref{Comparable ells}), we can estimate:
\begin{equation} 
 \int_0^{y_\mathrm{mirror}}  \Big(1-\chi_c\big(n^{-\f14\delta}\f{y}{y_\mathrm{cr}}\big) \Big)y^6 V(y) R_{n,\ell}(y)  R_{n_1,\ell_1}(y) R_{n_2,\ell_2}(y) R_{n_3,\ell_3}(y) \, dy \lesssim \|V\|_{L^\infty} e^{-2 n^{\f14\delta}} \ell^{\f12}   \label{Estimate in forbidden region oscillations}
\end{equation}

\item In the region $\{y\lesssim n^{\f14\delta} y_\mathrm{cr}\}$, where $\chi_c\big(n^{-\f14\delta}\f{y}{y_\mathrm{cr}}\big)$ is supported, we have as a consequence of our assumption \eqref{Largeness of overtone n} on $n$ (and bounding $\omega_{n,\ell}$ from below using \cref{lem:Crude Weyl law}):
\begin{equation}\label{Lower bound coefficient of high frequency term}
 \omega^2_{n,\ell} - V_{\ell}(y)\sim \ell n \quad \text{for all} \quad y\lesssim n^{\f14\delta} y_\mathrm{cr}.
\end{equation}
Using the equation \eqref{Radial boundary value problem} for $R_{n,\ell}$ to write
\[
R_{n,\ell} = \Big(\f{1}{V_\ell-\omega^2_{n,\ell}}\f{d^2}{dy^2}\Big)^q R_{n,\ell}
\]
for any $q\in \mathbb N$, we can express (choosing, in particular, $q=5$):\footnote{In general, for a weight function of the form $y^{2A} V(y)$, $A\in \mathbb N$, we would have chosen $q=A+2$.}
\begin{align*}
 \int_0^{y_\mathrm{mirror}} &  \chi_c\big(n^{-\f14\delta}\f{y}{y_\mathrm{cr}}\big) y^6 \cdot V(y) R_{n,\ell}(y)  R_{n_1,\ell_1}(y) R_{n_2,\ell_2}(y) R_{n_3,\ell_3}(y) \, dy\\
 & = \int_0^{y_\mathrm{mirror}}  \chi_c\big(n^{-\f14\delta}\f{y}{y_\mathrm{cr}}\big) y^6 \cdot V(y)\Big(\f{-1}{\omega^2_{n,\ell}-V_\ell}\f{d^2}{dy^2}\Big)^5 R_{n,\ell}(y) R_{n_1,\ell_1}(y) R_{n_2,\ell_2}(y) R_{n_3,\ell_3}(y) \, dy.
 \end{align*}
 Integrating by parts with respect to $\f{d}{dy}$ $10$ times and using the boundary conditions
 \[
 R_{n,\ell}(0) = R_{n_i,\ell_i}(0)=0, \quad \f{d^{2}}{dy^{2}}R_{n_i,\ell_i}(0)=0 
 \]
(the latter one following directly from the equation \eqref{Radial boundary value problem}) and 
\[
\frac{d^i}{dy^i} \big(y^6 \cdot V(y) \big)\big|_{y=0} \quad \text{for }\, i=0, \ldots, 5
\]
 we can thus readily estimate
 \begin{align*}
\int_0^{y_\mathrm{mirror}} &  \chi_c\big( n^{-\f14\delta}\f{y}{y_\mathrm{cr}}\big) y^6 V(y) R_{n,\ell}(y)  R_{n_1,\ell_1}(y) R_{n_2,\ell_2}(y) R_{n_3,\ell_3}(y) \, dy\\
 & \lesssim \| V\|_{C^{10}}  \sum_{ \begin{subarray}{l} 0 \le a_1 + b_1 + \dots + b_5\\+ j_1+ j_2  + j_3 \le 10 \end{subarray} }  \int_0^{y_\mathrm{mirror}} R_{n,\ell}(y)  \cdot \Big|\f{d^{a_1}}{dy^{a_1}}\big( y^6 \cdot \chi_c\big(n^{-\f14\delta}\f{y}{y_\mathrm{cr}}\big)\big)\Big|\cdot  \prod_{i=1}^5 \Big|\f{d^{b_i}}{dy^{b_i}}\f{1}{\omega^2_{n,\ell}-V_\ell}\Big| \\
 & \hphantom{\lesssim \|y^6 \cdot V\|_{C^{10}}  \int_0^{y_\mathrm{mirror}}}
 \times \Big| \f{d^{j_1}}{dy^{j_1}} R_{n_1,\ell_1}(y)\Big|  \cdot \Big| \f{d^{j_2}}{dy^{j_2}} R_{n_2,\ell_2}(y)\Big|  \cdot \Big| \f{d^{j_3}}{dy^{j_3}} R_{n_3,\ell_3}(y)\Big| \, dy.
 \end{align*}
Using the facts that
\begin{itemize}
\item  $R_{n,\ell}$ is $L^2$ normalized, 
\item $\omega^2_{n,\ell}-V_\ell$ satisfies the bound  \eqref{Lower bound coefficient of high frequency term} 
\item The functions  $R_{n_i,\ell_i}$ satisfy  for any $j\in \mathbb N$
\[
 \Big| \f{d^{j}}{dy^{j}} R_{n_i,\ell_i}(y)\Big| \lesssim_j \ell_i^{\f{j}2+\f14} n_i^{\f{j}2}
\]
and, for any  $j\in \mathbb N$ and $A\ge 0$
\[
\int_0^{y_\mathrm{mirror}} \Big| y^A \f{d^{j}}{dy^{j}} R_{n_i,\ell_i}(y)\Big|^2 \, dy \lesssim_{A,j} \ell_i^{j-\f{A}2} n_i^{j+\f{A}2},
\]
which both follow from the approximation formulas \eqref{eq:bound-on-Rnl-M}--\eqref{eq:bound-on-Rnl-prime-M} and the fact that higher order derivatives of $R_{n_i,\ell_i}$ can be computed in terms of lower order terms using equation  \eqref{Radial boundary value problem},
\end{itemize}
we infer that
\begin{align}\label{Estimate in classical region oscillations}
\int_0^{y_\mathrm{mirror}}  \chi_c\big(n^{-\f14\delta}\f{y}{y_\mathrm{cr}}\big)  & y^6 V(y) R_{n,\ell}(y)  R_{n_1,\ell_1}(y) R_{n_2,\ell_2}(y) R_{n_3,\ell_3}(y) \, dy \\
& \lesssim \|V\|_{C^{10}} \ell^{\f12}  \Big(\max_{i=1,2,3} \f{n_i}{\ell_i}\Big)^3 \cdot \Big( \f{\max_{i=1,2,3} (n_i^5)}{n^5}\Big).   \nonumber 
\end{align}

\end{itemize}
Combining \eqref{Estimate in forbidden region oscillations} and \eqref{Estimate in classical region oscillations} (and noting that $e^{-n^{\f14\delta}} \ll  \f{\max_{i=1,2,3} (n_i^5)}{n^5}$), we finally obtain \eqref{Estimate oscillatory integral in Rs}.

\end{proof}

\subsection{An expansion formula for the frequencies \texorpdfstring{$\omega_{n,\ell}$}{omega n,l}}
\label{sec:Spectral analysis radial}

In this section, we will establish the following Taylor expansion formula for $\omega_{n,\ell}$ as a function of the mass parameter $M$:

\begin{proposition}\label{prop:Taylor expansion omega}
Let $M_0>0$ and let $\ell$ be sufficiently large in terms of $M_0$ and satisfy 
\[
n\le \ell^{\delta}
\]
 for some given $0< \delta \ll 1$. Then, for any Schwarzschild--AdS mass parameter $M\in [0,M_0]$, the eigenvalue $\omega_{n,\ell}$ of the radial Sturm--Liouville problem \eqref{Radial boundary value problem} satisfies
\begin{equation}\label{Taylor expansion omega}
\omega_{n,\ell} = \omega_{n,\ell}^{\mathrm{(AdS)}} -\f{f_3(n)}{\ell^{\f12}} M - \f{h(n)}{\ell} \f{M^2}2 + O(\ell^{-1-5\delta}),
\end{equation}
where 
\[
\omega_{n,\ell}^{\mathrm{(AdS)}} \doteq 2n+\ell,
\]
\[
f_3(n) = \int_0^{+\infty} x^3 e_n^2(x) \, dx, \quad h(n) = \sum_{ n'\in \mathbb N^* \setminus \{n\}} \f{\Big| \int_0^{+\infty} x^3 e_n (x) e_{n'}(x)\, dx\Big|^2}{n'-n}
\]
and the constants implicit in the $O(\cdot)$ notation in \eqref{Taylor expansion omega} depend only on $\delta, M_0$ (see also \cref{lem:Asymptotic-functions-f-and-h} for the asymptotics of $f_3(n), h(n)$ for $n\gg 1$).
\end{proposition}

\begin{proof}
In the proof of \cref{prop:Taylor expansion omega}, we will always assume that the constants implicit in the $\lesssim, O(\cdot)$ notation depend only on $M_0, \delta$ (and are therefore independent of $n,\ell$).

Let us first consider a general 1-parameter family (with respect to the parameter $s\in [s_0,s_1]$) of Sturm--Liouville problems:
\begin{equation}\label{General Sturm--Liouville problem}
\begin{cases}
-\f{d^2 X_n}{dz^2} + V[s] X_n = \lambda_n X_n,\\
X_n|_{z=0}=X_n|_{z=1}=0,
\end{cases}
\end{equation}
with $V[s] \in C^\infty([0,1])$ and with the eigenfunctions $X_n$ being $L^2$-normalized:
\[
\int_0^1 X_n^2(z) \, dz = 1.
\]
 The family \eqref{General Sturm--Liouville problem} is said to be smooth if the map $s\rightarrow V[s]$ is smooth. In that case, the standard Sturm--Liouville theory  \cite{Zettl05}  implies that the eigenvalues $\lambda_n$ are distinct and that the eigenfunctions $X_n$ and the eigenvalues $\lambda_n$ are smooth functions of $s$ and satisfy (denoting by $\dot~$ derivatives with respect to $s$ and by $\langle \cdot, \cdot \rangle$ the standard $L^2\big([0,1]\big)$ inner product)
 \begin{equation}\label{First variation formula spectrum}
 \dot\lambda_n = \langle X_n , \dot V X_n\rangle, \quad \dot X_n  = -\sum_{n'\neq n} \f{\langle X_{n'}, \dot V X_{n}\rangle}{\lambda_{n'}-\lambda_n} X_{n'}
 \end{equation}
and
\begin{equation}\label{Second variation formula spectrum}
\ddot \lambda_n = \langle X_n , \ddot V X_n\rangle + 2 \langle \dot X_n , \dot V X_n\rangle.
\end{equation}

Let us now focus on the family of Sturm--Liouville problems \eqref{Radial boundary value problem} for $M\in [0,M_0]$. This family is of the form \eqref{General Sturm--Liouville problem}, with $V[M]= V_\ell$ and $\lambda_n = \omega^2_{n,\ell}$. 
Recall also that this is a smooth family with respect to $M\in [0,M_0]$ (see the remark below \eqref{Normalization radial eigenfunctions}) and the eigenvalues $\omega^2_{n,\ell}$ are distinct for $n\le \ell^{\delta}$ (as a consequence of \eqref{Approximation estimate frequency} and the fact that $\omega_{n,\ell}^2$ is increasing in $n$ by its definition). Therefore, the spectral variation formulas \eqref{First variation formula spectrum}--\eqref{Second variation formula spectrum} imply that, for any $M\in [0,M_0]$:
\begin{equation}\label{First variation formula omega squared}
\f{d}{dM} \omega_{n,\ell}^2 
= \int_0^{y_\mathrm{mirror}} \f{\partial V_{\ell}}{\partial M} \cdot R_{n,\ell}^2(y) \, dy
\end{equation}
and
\begin{equation}\label{Second variation formula omega squared}
\f{d^2}{dM^2}\omega_{n,\ell}^2 =  \int_0^{y_\mathrm{mirror}} \f{\partial^2 V_{\ell}}{\partial M^2} \cdot R_{n,\ell}^2(y) \, dy
- 2  \sum_{n'\neq n} \f{\Big( \int_0^{y_\mathrm{mirror}} \f{\partial V_\ell}{\partial M} \cdot R_{n',\ell}(y) R_{n,\ell}(y) \, dy \Big)^2}{\omega^2_{n',\ell}-\omega^2_{n,\ell}}.
\end{equation}

The expansion formula \eqref{Taylor expansion omega} will follow from the standard Taylor expansion for $\omega_{n,\ell}(M)$ around $M=0$, namely
\begin{equation}\label{General Taylor expansion formula omega}
\omega_{n,\ell}(M) = \omega_{n,\ell}(0) + \Big(\f{d}{dM}\omega_{n,\ell}(0)\Big) \cdot M + \int_0^M \f{d^2}{dM^2}\omega_{n,\ell}(\bar M) \cdot (M-\bar M) \, d\bar M.
\end{equation}
To this end, we will need to estimate $\f{d}{dM}\omega_{n,\ell}\big|_{M=0}$ and $\f{d^2}{d M^2} \omega_{n,\ell}\big|_{M\in [0,M_0]}$.

At $M=0$, we can readily compute (using also \eqref{eq:AsymptoticspartialMr(y,M)}) that
\begin{align*}
\f{\partial V_\ell(r(y,M),M)}{\partial M}\Big|_{M=0} & = -\f{2\ell(\ell+1)}{r(y)^3} + \f{2}{r(y)}\Big(1+\f{1}{r(y)^2} \Big) + \frac{\partial V_\ell(r,M)}{\partial r} \frac{\partial r(y,M)}{\partial M} \Big|_{M=0}  \\ & = \ell(\ell+1)\big(-2 y^3+ O(y^{5}) + \ell^{-2} O(y)\big).
\end{align*}
Moreover, in the language of the proof of \cref{lem:eigenfunctions-estimate-m>0}, we have
\[
\omega_{n,\ell}|_{M=0} = \oomega_{n,\ell}|_{y_{\mathrm{D}}=y_\mathrm{mirror}} \quad \text{and} \quad R_{n,\ell}|_{M=0} = \oR_{n, \ell}|_{y_{\mathrm{D}}=y_\mathrm{mirror}}.
\]
Therefore, in view of the estimates \eqref{Improved bound oomega exponential} for $\oomega_{n,\ell}$ and  \eqref{Bounds oR in y}--\eqref{Exponential decay oR in y} for $\oR_{n,\ell}$, we have:
\begin{equation}\label{Omega zero mass}
\omega_{n,\ell}|_{M=0} = 2n+\ell +O\big(e^{-\ell^{\f12\delta}}\big)
\end{equation}
and
\begin{equation}\label{R n l zero mass}
R_{n,\ell}|_{M=0}(y) = \ell^{\f14} e_n \big( \ell^{\f12}y\big) + O\big(\ell^{-\f34+5\delta}\big)  \text{ and }   \big|R_{n,\ell}|_{M=0}(y)\big| \lesssim e^{-\f12 n^{\f14}\ell^{\f34}\Big(y-\sqrt{\f{4n-1}{\ell}}\Big)^{\f32}}\ell^{\f14}  \text{ for }  y\ge \sqrt{\f{4n-1}{\ell}}.
\end{equation}
As a result, from \eqref{First variation formula omega squared} we compute:
\begin{align}\label{First variation omega zero mass}
\f{d}{dM}\omega_{n,\ell}\big|_{M=0} &  = \f{1}{2\big(\ell+2n+O(e^{-\ell^{\f12\delta}})\big)}\Bigg(-\int_0^{y_\mathrm{mirror}} 2\ell(\ell+1)y^3 \ell^{\f12} e_n^2(\ell^{\f12} y) \, dy 
+ O(\ell^{-\f12+6\delta})\Bigg) \\
& = -\ell^{-\f12} \int_0^{+\infty} x^3 e_n^2(x) \, dx + O(\ell^{-\f32+7\delta})  \nonumber \\
& = -\ell^{-\f12} f_3(n) + O(\ell^{-\f32+7\delta})  \nonumber.
\end{align}
Arguing similarly but using the weaker bounds \eqref{eq:bound-on-Rnl-M}--\eqref{Approximation estimate frequency} for $R_{n,\ell}$ and $\omega_{n,\ell}$, (instead of \eqref{Omega zero mass}--\eqref{R n l zero mass}) and the fact that 
\begin{align*}
\f{\partial V_\ell}{\partial M} & = -\f{2\ell(\ell+1)+\f{4M}{r(y)}}{r(y)^3}+ \f{2}{r(y)}\Big(1+\f{1}{r(y)^2}-\f{2M}{r(y)^3} \Big) + \frac{\partial V_\ell(r,M)}{\partial r} \frac{\partial r(y,M)}{\partial M} \\
    & = -2\ell(\ell+1)\Big(y^3 +O(y^5)+O(\ell^{-2} y)\Big)\ \quad \text{for all } \, M\in [0,M_0],
\end{align*}
 we obtain for any $M\in [0,M_0]$:
\begin{equation}\label{First variation omega general}
\f{d}{dM}\omega_{n,\ell}\big|_{M\in [0,M_0]} = -\ell^{-\f12} f_3(n) + O(\ell^{-1+7\delta}).
\end{equation}

We can readily compute for any $M\in [0,M_0]$ that
\[
\f{\partial^2 V_\ell}{\partial M^2} = \f{-8}{r(y)^4} + 2\frac{\partial^2 V_\ell(r,M)}{\partial M \partial r} \frac{\partial r}{\partial M} + \frac{\partial V_\ell(r,M)}{\partial r} \frac{\partial^2 r}{\partial M^2}+ \frac{\partial^2 V_\ell}{\partial r^2} \left(\frac{\partial r}{\partial M}\right)^2 = O(y^4) +  O(\ell^2 y^6).
\]
Therefore, in view of the bounds \eqref{eq:bound-on-Rnl-M}--\eqref{eq:expo-decay} for $R_{n,\ell}$, we have
\begin{equation}\label{Term d 2 V}
\int_0^{y_\mathrm{mirror}} \f{\partial^2 V_{\ell}}{\partial M^2} \cdot R_{n,\ell}^2(y) \, dy = O(\ell^{-1+3\delta}).
\end{equation}

In order to estimate the term 
\[
 \sum_{n'\neq n} \f{\Big( \int_0^{y_\mathrm{mirror}} \f{\partial V_\ell}{\partial M} \cdot R_{n',\ell}(y) R_{n,\ell}(y) \, dy \Big)^2}{\omega^2_{n',\ell}-\omega^2_{n,\ell}},
\]
we will split the sum into the parts corresponding to $n'\le \ell^{10\delta}$ and $n'>\ell^{10\delta}$ and we will argue as follows:
\begin{itemize}
\item \textbf{The case $n'\le \ell^{10\delta}$:} By applying \cref{lem:eigenfunctions-estimate-m>0} (for $10\delta$ in place of $\delta$), we obtain for any $n'\le \ell^{10\delta}$
\[
\omega_{n',\ell}^2-\omega_{n,\ell}^2 =4\big( \ell + n + n' +O(\ell^{-\f12+20\delta}) \big)\big(n'-n+ O(\ell^{-\f12+20\delta})\big) = 4\ell(n'-n)+O(\ell^{\f12+20\delta})
\]
and
\begin{align*}
\int_0^{y_\mathrm{mirror}} \f{\partial V_\ell}{\partial M} \cdot R_{n',\ell}(y) R_{n,\ell}(y) \, dy 
& = -2 \int_0^{y_\mathrm{mirror}} \ell^2 y^3 \cdot \ell^{\f12} e_n (\ell^{\f12} y) e_{n'}(\ell^{\f12}y) \, dy +O(\ell^{100\delta}) \\
& = - 2 \ell^{\f12} \int_0^{+\infty} x^3 e_n(x) e_{n'}(x) \, dx +O(\ell^{100\delta}).
\end{align*}
Therefore, 
\begin{equation}\label{Low n' sum}
 \sum_{\substack{n'\neq n,\\n'\le \ell^{10\delta}}} \f{\Big( \int_0^{y_\mathrm{mirror}} \f{\partial V_\ell}{\partial M} \cdot R_{n',\ell}(y) R_{n,\ell}(y) \, dy \Big)^2}{\omega^2_{n',\ell}-\omega^2_{n,\ell}}
  = \sum_{\substack{n'\neq n,\\n'\le \ell^{10\delta}}} \f{\Big(\int_0^{+\infty} x^3 e_n(x) e_{n'}(x) \, dx\Big)^2}{n'-n} + O(\ell^{-\f12+100\delta}).
\end{equation}

\item \textbf{The case $n' > \ell^{10\delta}$:} In this case, we will argue similarly as for the proof of \cref{lem:Crude non stationary phase}. Using the fact that $\omega_{n',\ell}$ satisfies the lower bound provided by \cref{lem:Crude Weyl law}, we can readily estimate for any $n'>\ell^{10\delta}$:
\[
\omega^2_{n',\ell}-\omega^2_{n,\ell} = (\omega_{n',\ell}+\omega_{n,\ell}) (\omega_{n',\ell} - \omega_{n,\ell})\gtrsim \ell n'.
\] 

Let us fix a smooth cut-off function $\chi_c: [0,+\infty)\rightarrow [0,1]$ such that $\chi_c\equiv 1$ on $[0,1]$ and $\chi_c\equiv 0$ on $[2,+\infty)$ and let us split
\begin{align*}
 \int_0^{y_\mathrm{mirror}} \f{\partial V_\ell}{\partial M} \cdot R_{n',\ell}(y) R_{n,\ell}(y) \, dy = & \int_0^{y_\mathrm{mirror}}  \chi_c\big((n')^{-\delta}\ell^{\f12-\f34\delta}y\big) \f{\partial V_\ell}{\partial M} \cdot R_{n',\ell}(y) R_{n,\ell}(y) \, dy \\
 &+  \int_0^{y_\mathrm{mirror}} \Big(1-\chi_c\big((n')^{-\delta}\ell^{\f12-\f34\delta}y\big) \Big)\f{\partial V_\ell}{\partial M} \cdot R_{n',\ell}(y) R_{n,\ell}(y) \, dy.
\end{align*}
In view of the fact that $R_{n',\ell}$ is $L^2$ normalized and $R_{n,\ell}$ satisfies the exponential decay estimate \eqref{eq:expo-decay} in the region $\big\{y\gtrsim (n')^{\delta}\ell^{-\f12+\f34\delta} \big\} \subset \big\{y\gtrsim n^{\f12}(n')^{\delta}\ell^{-\f12+\f14\delta} \big\}$ (i.e.~in the region where $1-\chi_c\big((n')^{-\delta}\ell^{\f12-\f34\delta}y\big)$ is supported), we have
\begin{equation}\label{Exponential decay mixed product R}
\Bigg| \int_0^{y_\mathrm{mirror}} \Big(1-\chi_c\big((n')^{-\delta}\ell^{\f12-\f34\delta}y\big) \Big)\f{\partial V_\ell}{\partial M} \cdot R_{n',\ell}(y) R_{n,\ell}(y) \, dy \Bigg| \lesssim e^{-\f18\ell^{\f14\delta}(n')^{\delta}}.
\end{equation}
On the other hand, in the region $\big\{ y \lesssim (n')^{\delta}\ell^{-\f12+\f34\delta}\big\}$ (where $\chi_c\big((n')^{-\delta}\ell^{\f12-\f34\delta}y\big)$ is supported), we have (in view of \cref{lem:Crude Weyl law}):
\begin{equation}\label{Lower bound potential classical region}
\omega^2_{n',\ell}-V_{\ell}(y) \ge (\ell+c n')^2-V_{\ell}(y) = (\ell+c n')^2 - \Big(\ell(\ell+1)+O(y)\Big)\Big(1+y^2 +O(y^3)\Big) \gtrsim \ell n'.
\end{equation}
Using the equation \eqref{Radial boundary value problem} for $R_{n',\ell}$ to write
\[
R_{n',\ell} = -\f{1}{\omega^2_{n',\ell}-V_\ell}\f{d^2}{dy^2} R_{n',\ell},
\]
we obtain after integrating by parts:
\begin{align*}
\int_0^{y_\mathrm{mirror}} &  \chi_c\big(\ell^{\f12+\f34\delta}y\big) \f{\partial V_\ell}{\partial M} \cdot R_{n',\ell}(y) R_{n,\ell}(y) \, dy
\\
= & \int_0^{y_\mathrm{mirror}} \chi_c\big((n')^{-\delta}\ell^{\f12+\f34\delta}y\big) \f{\partial V_\ell}{\partial M} \cdot  \Big(-\f{1}{\omega^2_{n',\ell}-V_\ell}\f{d^2}{dy^2}\Big) R_{n',\ell} \cdot  R_{n,\ell} \, dy \\
= &\int_0^{y_\mathrm{mirror}}  \Big(-\f{d^2}{dy^2}\big(\f{1}{\omega^2_{n',\ell}-V_\ell}\chi_c\big((n')^{-\delta}\ell^{\f12+\f34\delta}y\big) \f{\partial V_\ell}{\partial M}  R_{n,\ell}\big)\Big) \cdot R_{n',\ell}  \, dy \\
& \hphantom{\sum} + 
 \Bigg[ \Big(\f{d}{dy}\big(-\f1{\omega^2_{n',\ell}-V_\ell}  \f{\partial V_\ell}{\partial M}  R_{n,\ell}\big)\Big)\cdot R_{n',\ell} 
 -  \f{\partial V_\ell}{\partial M}  R_{n,\ell} \cdot \Big(-\f{1}{\omega^2_{n',\ell}-V_\ell}\f{d}{dy}\Big)  R_{n',\ell}\Bigg] \Bigg|_{y=0}.
\end{align*}
Using the boundary conditions $R_{n',\ell}|_{y=0}= R_{n,\ell}|_{y=0}=0$ to deduce that the boundary terms above at $y=0$ vanish, we infer that:
\begin{align*}
\Bigg| \int_0^{y_\mathrm{mirror}}   &\chi_c\big((n')^{-\delta}\ell^{\f12+\f34\delta}y\big)   \f{\partial V_\ell}{\partial M} \cdot R_{n',\ell}(y) R_{n,\ell}(y) \, dy \Bigg| 
 \\ & \lesssim \Bigg( \int_0^{y_\mathrm{mirror}}  \Big|-\f{d^2}{dy^2}\Big(\f{1}{\omega^2_{n',\ell}-V_\ell}\chi_c\big((n')^{-\delta}\ell^{\f12+\f34\delta}y\big) \f{\partial V_\ell}{\partial M}  R_{n,\ell}\Big)\Big|^2  \, dy  \Bigg)^{\f12} 
\Bigg( \int_0^{y_\mathrm{mirror}} R_{n',\ell}^2 \, dy \Bigg)^{\f12} \\
& \lesssim \Bigg( 
\int_0^{y_\mathrm{mirror}}  \Big(\Big|\f{d^2}{dy^2}\big(\f{1}{\omega^2_{n',\ell}-V_\ell}\chi_c\big((n')^{-\delta}\ell^{\f12+\f34\delta}y\big) \f{\partial V_\ell}{\partial M}\big)  R_{n,\ell}\Big|^2 
\\
&\hphantom{\lesssim \Bigg( \int_0^{y_\mathrm{mirror}} }
+ \Big|\f{d}{dy}\big(\f{1}{\omega^2_{n',\ell}-V_\ell}\chi_c\big((n')^{-\delta}\ell^{\f12+\f34\delta}y\big) \f{\partial V_\ell}{\partial M}\big) \cdot \f{d}{dy} R_{n,\ell}\Big|^2 
\\
& \hphantom{\lesssim \Bigg( \int_0^{y_\mathrm{mirror}} }
+  \Big|\f{1}{\omega^2_{n',\ell}-V_\ell}\chi_c\big((n')^{-\delta}\ell^{\f12+\f34\delta}y\big) \f{\partial V_\ell}{\partial M} \cdot \f{d^2}{dy^2} R_{n,\ell}\Big|^2 
\Big)\, dy 
\Bigg)^{\f12}
\lesssim \f{\ell^{\f12+4\delta}}{(n')^{1-6\delta}}
\end{align*}
where, in passing to the last line above, we made use of the following facts:
\begin{itemize}
\item The function $R_{n',\ell}$ is $L^2$-normalized,
\item The lower bound \eqref{Lower bound potential classical region} for $\omega^2_{n',\ell}-V_\ell$,
\item The fact that $\f{d}{dy}V_\ell = O(\ell^2 y+1)$, $\f{d^2}{dy^2}V_\ell = O(\ell^2)$,
\item The fact that  $\big(y\f{d}{dy}\big)^j\f{\partial V_\ell}{\partial M}=O(\ell^2 y^3)$ for $j=0,1,2$,
\item The  estimates \eqref{eq:bound-on-Rnl-M}--\eqref{eq:expo-decay} for $R_{n,\ell}$ and $\f{d}{dy}R_{n,\ell}$,
\item The fact that $\f{d^2}{dy^2} R_{n,\ell} = (V_\ell-\omega^2_{n,\ell}) R_{n,\ell}$.
\end{itemize}
Collecting the above bounds for $n'>\ell^{10\delta}$, we infer that:

\begin{equation}\label{High n' sum}
\sum_{n'> \ell^{10\delta}} \f{\Big( \int_0^{y_\mathrm{mirror}} \f{\partial V_\ell}{\partial M} \cdot R_{n',\ell}(y) R_{n,\ell}(y) \, dy \Big)^2}{\omega^2_{n',\ell}-\omega^2_{n,\ell}}
 \lesssim \sum_{n'> \ell^{10\delta}} \f{\Big( \f{\ell^{\f12+4\delta}}{(n')^{1-6\delta}} +e^{-\f18\ell^{\f14\delta}(n')^{\delta}} \Big)^{2}}{\ell n'} \lesssim \ell^{-10\delta}
\end{equation}

\end{itemize}

Combining \eqref{Low n' sum} and \eqref{High n' sum}, we deduce that
\[
\sum_{n'\neq n} \f{\Big( \int_0^{y_\mathrm{mirror}} \f{\partial V_\ell}{\partial M} \cdot R_{n',\ell}(y) R_{n,\ell}(y) \, dy \Big)^2}{\omega^2_{n',\ell}-\omega^2_{n,\ell}} = \sum_{\substack{n'\neq n\\ n'\le \ell^{10\delta}}} \f{\Big(  \int_0^{+\infty} x^3 e_n(x) e_{n'}(x) \, dx \Big)^2}{n'-n} +O(\ell^{-10\delta}).
\]
Arguing similarly as for the derivation of \eqref{High n' sum} (namely, by writing $e_{n'}(x) = -\f{1}{4n'-1-x^2} \f{d}{dx^2} e_{n'}(x)$, integrating by parts, and using the fact that $e_{n}(x) \lesssim e^{-\f12 n^{\f14}(x-\sqrt{4n-1})^{\f32}}$ for $x\ge \sqrt{4n-1}$), we can readily estimate
\[
\int_0^{+\infty} x^3 e_{n}(x) e_{n'}(x) \, dx =O\big(\f{n^2}{(n')^{\f12-10\delta}}\big),
\]
from which we infer that
\[
\sum_{n'> \ell^{10\delta}} \f{\Big(  \int_0^{+\infty} x^3 e_n(x) e_{n'}(x) \, dx \Big)^2}{n'-n}  \lesssim O(\ell^{-5\delta}).
\]
Therefore, 
\begin{equation}\label{Sum of projections for second order term}
\sum_{n'\neq n} \f{\Big( \int_0^{y_\mathrm{mirror}} \f{\partial V_\ell}{\partial M} \cdot R_{n',\ell}(y) R_{n,\ell}(y) \, dy \Big)^2}{\omega^2_{n',\ell}-\omega^2_{n,\ell}} = \sum_{n'\neq n} \f{\Big(  \int_0^{+\infty} x^3 e_n(x) e_{n'}(x) \, dx \Big)^2}{n'-n} +O(\ell^{-5\delta}).
\end{equation}

Returning to the formula \eqref{Second variation formula omega squared} for $\f{d^2}{dM^2}\omega^2_{n,\ell}$ and using \eqref{Term d 2 V}, \eqref{Sum of projections for second order term}, and the fact that $\omega_{n,\ell}$ and $\f{d}{dM}\omega_{n,\ell}$ satisfy \eqref{Approximation estimate frequency} and \eqref{First variation omega general}, respectively, we obtain for any $M\in [0,M_0]$:
\begin{equation}\label{Second variation formula omega}
\f{d^2}{dM^2}\omega_{n,\ell} = \f{1}{2\omega_{n,\ell}} \f{d^2}{dM^2}\omega^2_{n,\ell} - \f1{\omega_{n,\ell}} \Big(\f{d}{dM}\omega_{n,\ell}\Big)^2 = -  \f1\ell \sum_{n'\neq n} \f{\Big(  \int_0^{+\infty} x^3 e_n(x) e_{n'}(x) \, dx \Big)^2}{n'-n} +O(\ell^{-1-5\delta}).
\end{equation}

Using the expression \eqref{Omega zero mass} for $\omega_{n,\ell}(0)$, \eqref{First variation omega zero mass} for $\f{d}{dM}\omega_{n,\ell}(0)$ and \eqref{Second variation formula omega} for $\f{d^2}{dM^2}\omega_{n,\ell}|_{M\in [0,M_0]}$, the Taylor expansion formula \eqref{General Taylor expansion formula omega} finally yields \eqref{Taylor expansion omega}.

\end{proof}

\section{Construction of the solution \texorpdfstring{$\phi$}{phi}}\label{sec:Ansatz}
In this section, we will introduce the ansatz that we will use for the function $\phi$ appearing in the statement of \cref{thm:Main theorem}. This ansatz (see \eqref{The ansatz} below) will be built using the normal mode solutions of the initial--boundary value problem \eqref{Initial boundary value problem mirror} with an inner mirror; it will consist of a triad of \emph{dominant} modes with frequency parameters satisfying a set of resonance conditions, a larger set of \emph{non-dominant} modes which are resonantly separated from the dominant ones, plus an additional error term. Enforcing those (non-)resonance conditions will necessitate choosing the Schwarzschild--AdS mass parameter $M$ to lie in a carefully designed  dense set $\mathcal S_{\epsilon,s}\subset (0,+\infty)$. The analysis of the dynamics of the modes and the error term will constitute the bulk of this paper and will occupy  \cref{sec:Dominant modes,sec:Non dominant modes,sec:Estimates error term}.

\begin{remark} For the rest of this section, we will assume that we are given a constant $M_0>0$ and that the mass parameter $M$ takes values in $(0, M_0]$. Given such $M_0>0$ and $M\in (0, M_0]$, the mirror parameters $r_\mathrm{mirror}$ and $y_\mathrm{mirror}$ will be fixed according to \cref{def:Mirror radius}.
\end{remark}

\subsection{The hierarchy of parameters} \label{sec:Hierarchy of parameters}
Let $0<\delta_0 \ll 1$ be a given constant. We will define the set of ``permissible'' mode parameters $\mathcal K$ by
\begin{equation}\label{Mode parameters}
\mathcal K \doteq  \Big\{ (n, \ell, m) \in \mathbb N^* \times \mathbb N_{\ge |m|}\times \mathbb Z: \, n \le \ell^{1-\delta_0}   \Big\}.
\end{equation}
Note that, for any $M_0>0$ and for each triad $k=(n, \ell, m)\in \mathcal K$ and any Schwarzschild--AdS mass parameter $M\in (0, M_0]$, the corresponding eigenfunction $E_k$ defined by  \eqref{Spatial eigenfunction} is concentrated near $y=0$ provided $\ell$ is sufficiently large in terms of $M_0$.

Among the frequency parameters contained in $\mathcal K$, we will single out a triplet $k_\minusone, k_0, k_1$ which we will refer to as the set of \emph{dominant} mode parameters, and will be defined as follows:

\begin{definition}\label{def:Dominant mode parameters}
Let $M_0>0$ and $0<\delta_0 \ll 1$ be given constants and let $N, L \in \mathbb N^*$ and $\lambda \in \lceil \f1L\rceil \mathbb N^*$ be parameters which are large in terms of $M_0$ and satisfy $N\le L^{\delta_0}$. We will define the frequency triads $k_\minusone=(n_\minusone,\ell_\minusone,m_\minusone)$, $k_0=(n_0,\ell_0,m_0)$ and $k_1=(n_1,\ell_1,m_1)$ by the relations:
\begin{empheq}[box=\fbox]{alignat=3}
n_\minusone &= N, \quad & \ell_\minusone &= \lambda L, \quad  & m_\minusone &= - \ell_\minusone\\
n_0 &= 1, \quad & \ell_0 &= L, \quad & m_0 &= \ell_0\\
n_1 &= N+2, \quad & \ell_1 &= (\lambda + 2)L, \quad  & m_1 &= \ell_1.
\end{empheq} 
We will also define
\begin{equation}\label{Associated signs}
\boxed{
\varepsilon_{k_\minusone} = -1,  \quad \varepsilon_{k_0} =+1, \quad  \varepsilon_{k_1} = +1.
}
\end{equation}
\end{definition}

\begin{remark} We will later fix $\lambda$ and $L$ in a way that $N$ and $\lambda$ are large in terms of the parameters $s$ and $\epsilon$ of \cref{thm:Main theorem}, while $L$ will be large compared to $N$ and $\lambda$; see also \cref{def:Relation lambda N} below. Schematically,
\[
1 \ll_{s,\epsilon} N\sim \lambda \ll_{s,\epsilon, N,\lambda} L. 
\]
Note that $k_0$ and $k_\pm$ satisfy the resonance conditions
\begin{equation}\label{Resonant condition AdS}
m_1-2m_0+m_\minusone=0 \quad \text{and} \quad \varepsilon_{k_\minusone}{\omega}^{\mathrm{(AdS)}}_{k_\minusone} - 2 \varepsilon_{k_0}{\omega}^{\mathrm{(AdS)}}_{k_0} +\varepsilon_{k_1} {\omega}^{\mathrm{(AdS)}}_{k_1} =0,
\end{equation}
where
\[
{\omega}^{\mathrm{(AdS)}}_{(n,\ell,m)} \doteq 2n+\ell.
\]
\end{remark}

Having defined $k_0$ and $k_{\pm1}$ as in \cref{def:Dominant mode parameters}, we will split the set $\mathcal K$ of permissible mode parameters as follows:

\begin{definition}\label{def:Splitting of parameter set}
The set $\mathcal K$ can be expressed as the disjoint union of the subsets $\mathcal K_{\mathrm{D}}$, $\mathcal K_{\mathrm{AR}}$ and $\mathcal K_{\mathrm{NR}}$, where:
\begin{itemize}
\item The \emph{dominant} subset $\mathcal{K}_{\mathrm{D}}$ is defined as  
\[
\mathcal K_{\mathrm{D}} \doteq \{ k_\minusone, k_0, k_1\}.
\]
\item The \emph{almost resonant} subset $\mathcal K_{\mathrm{AR}}$ is defined as
\begin{align*}
\mathcal K_{\mathrm{AR}} \doteq \Big\{  k & =(n,\ell,m)\in \mathcal K \setminus \mathcal K_{\mathrm{D}}: ~  \text{ There exist } j_1, j_2, j_3 \in \mathcal K_{\mathrm{D}} \text{ such that } \\
& m=m_{j_1}-m_{j_2}+m_{j_3} 
 \text{ \textbf{and} }{\omega}_k^{\mathrm{(AdS)}} = \Big|\varepsilon_{j_1}{\omega}_{j_1}^{\mathrm{(AdS)}}-\varepsilon_{j_2}{\omega}_{j_2}^{\mathrm{(AdS)}}+\varepsilon_{j_3}{\omega}_{j_3}^{\mathrm{(AdS)}}\Big| 
\Big\},
\end{align*}
where the signs $\varepsilon_{j_i}\in \{-1, +1\}$ are defined by \eqref{Associated signs}.
\item The \emph{purely non-resonant} subset $\mathcal K_{\mathrm{NR}}$ is defined as
\begin{align*}
\mathcal K_{\mathrm{NR}} \doteq \Big\{  k & =(n,\ell,m)\in \mathcal K \setminus \mathcal K_{\mathrm{D}}: ~  \text{ For all } j_1, j_2, j_3 \in \mathcal K_{\mathrm{D}}: \\
& m\neq m_{j_1}-m_{j_2}+m_{j_3} 
 \text{ \textbf{or} }{\omega}_k^{\mathrm{(AdS)}} \neq \Big| \varepsilon_{j_1}{\omega}_{j_1}^{\mathrm{(AdS)}}-\varepsilon_{j_2}{\omega}_{j_2}^{\mathrm{(AdS)}}+\varepsilon_{j_3}{\omega}_{j_3}^{\mathrm{(AdS)}}\Big|
\Big\}.
\end{align*}
\end{itemize}
\end{definition}

\subsection{The mode-based ansatz for \texorpdfstring{$\phi$}{phi}}\label{sec:The mode based ansatz}
For any constant $M_0>0$ and any Schwarzschild--AdS mass parameter $M\in (0, M_0]$, let us fix a smooth cut-off function $\chi:[r_+, +\infty) \rightarrow [0,1]$ such that
\[
\chi(r) =0 \, \text{ for } \, r\le  r_\mathrm{mirror}+\f12 \, \text{ and } \, \chi(r)= 1 \, \text{ for } \, r\ge  r_\mathrm{mirror}+1.
\]
We will assume that $\chi$ has been fixed so that, for any $k\in \mathbb N$, $\|\chi\|_{C^k}$ is uniformly bounded in $M\in (0, M_0]$.

\begin{definition}\label{def:Ansatz}
For any $M_0>0$ and any  Schwarzschild--AdS mass parameter $M\in (0, M_0]$, we will use the following ansatz for the solution $\phi$ to \eqref{Initial Boundary Value Problem Phi} appearing in the statement of \cref{thm:Main theorem}:
\begin{equation}\label{The ansatz}
\phi(t^*, r, \theta,\varphi) = \chi(r) \tilde\phi(t^*, r, \theta, \varphi) + \psi(t^*, r, \theta, \varphi),
\end{equation}
where 
\begin{itemize}
\item $\tilde\phi$ is supported in the region $\{r\ge r_\mathrm{mirror}\}$ and has the form
\[
\tilde\phi = \sum_{k \in \mathcal K} \tilde\phi_k,
\]
where the functions $\tilde\phi_k$ are of the form
\[
\tilde\phi_k(t,y,\theta, \varphi) = \f{a_k(t)}{r} E_k(y, \theta, \varphi),
\]
(with $E_k$ being the spatial eigenfunction \eqref{Spatial eigenfunction}) and satisfy the following system of boundary value problems:
\begin{equation}\label{Boundary value problem tilde phi}
\begin{cases}
\square_g \tilde\phi_k + 2\tilde\phi_k = \f{1}{r\big(1-\f{2M}r+r^2\big)}\sum_{k_1, k_2, k_3\in \mathcal K_{\mathrm{D}}} \mathbb P_k \Big(r\big(1-\f{2M}r+r^2\big) \mathcal N[\tilde\phi_{k_1}, \tilde\phi_{k_2}, \tilde\phi_{k_3}]\Big), \\
\tilde\phi_k|_{y=y_\mathrm{mirror}}=0, \quad r\tilde\phi_k|_{y=0}=0.
\end{cases}
\end{equation} 
In the above, $\mathcal N[\cdot, \cdot, \cdot]$ was defined by \eqref{nonlinearity Intro}, while $\mathbb P_k$ denotes the $L^2(dy\dvol_{\mathbb S^2})$ spectral projection on $E_k$ (defined by \eqref{Projection k}). Note that the right-hand side of \eqref{Boundary value problem tilde phi} involves the nonlinear interactions only of those $\tilde\phi_k$ corresponding to the \emph{dominant} mode parameters $k\in \mathcal K_{\mathrm{D}}$.
\item The function $\psi$ satisfies initially
\begin{equation}\label{Initial data for psi}
\psi|_{t^*=0}=\partial_{t^*}\psi|_{t^*=0}=0.
\end{equation}
\end{itemize}

We will refer to $\tilde\phi$ as the \textbf{approximate solution} of the problem \eqref{Initial Boundary Value Problem Phi} (in contrast to $\phi$ being the true solution), while $\psi$ will be called the \textbf{error term}.
\end{definition}
Let us make the following remarks regarding the above ansatz:
\begin{itemize}
\item In view of our assumption that the functions $\tilde\phi_k$ satisfy the system of boundary value problems \eqref{Boundary value problem tilde phi}, the fact that the true solution $\phi$ satisfies the initial--boundary value problem \eqref{Initial Boundary Value Problem Phi} is \textbf{formally equivalent} to the statement that\footnote{Note that the finite speed of propagation for equation \eqref{Initial Boundary Value Problem Phi} implies that $\phi=0$ in the region $\{t\ge 0\}\cap\{t^*\le 0\}$, hence one can exchange the initial hypersurface $\{t^*=0\}$ with $\{t=0\}$ in \eqref{Same initial data phi and tilde phi}.}
\begin{equation}\label{Same initial data phi and tilde phi}
 (\chi\tilde\phi, \chi\partial_{t^*} \tilde \phi)|_{t^*=0} = (f_0, f_1)
\end{equation}
and that the error term $\psi$ satisfies the initial--boundary value problem
\begin{equation}\label{IVP Psi again}
\begin{cases}
\square_g \psi + 2 \psi - r^{-6} |\chi\tilde\phi+\psi|^2 \partial_{t^*}^2\psi - \chi^2  \mathcal N^{(1)}[\tilde \phi;\psi]   - \chi \mathcal N^{(2)}[\tilde\phi; \psi] -\mathcal N^{(3)}[\psi] = - \mathcal F[\tilde \phi],\\[5pt]
(\psi, \partial_{t^*} \psi)|_{t^*=0} = (0,0), \\
r\psi|_{r=\infty} =0.
\end{cases}
\end{equation}
where
\[
\mathcal F[\tilde \phi] \doteq   \square_g (\chi \tilde\phi) + 2 \chi\tilde\phi - \mathcal N[\chi\tilde\phi, \chi\tilde\phi, \chi\tilde\phi]
\]
and $\mathcal N^{(1)}[\tilde\phi; \psi]$, $\mathcal N^{(2)}[\tilde\phi; \psi]$ and $\mathcal N^{(3)}[\psi]$ denote the lower-order linear, quadratic and cubic terms in $\psi$, respectively, i.e.
\begin{align*}
\mathcal N^{(1)}[\tilde\phi; \psi] \doteq & \,  r^{-6} \left( 2\partial_{t^*}^2 \tilde\phi \, \Re\{\tilde\phi \bar \psi\}  +  |\partial_{t^*} \tilde\phi|^2 \psi + 2 \tilde\phi \Re\{\partial_{t^*} \tilde\phi \partial_{t^*} \bar\psi\} \right) \\[5pt]
	\mathcal N^{(2)}[\tilde\phi; \psi] \doteq &\,  r^{-6} \left(  \partial_{t^*}^2 \tilde\phi |\psi|^2 + \tilde\phi |\partial_{t^*} \psi|^2 + 2\Re\{\partial_{t^*} \tilde\phi \partial_{t^*} \bar \psi\} \psi \right) \\[5pt]
\mathcal N^{(3)}[\psi] \doteq& \,  r^{-6} \psi |\partial_{t^*} \psi|^2.
\end{align*}
\item The system of boundary value problems \eqref{Boundary value problem tilde phi} for the $\tilde\phi_k$'s is formally equivalent to the following system of ODEs for the coefficients $a_k$:
\begin{equation}\label{System of ODEs ak}
\f{d^2 a_k}{dt^2} +{\omega}_k^2 a_k = \sum_{k_1, k_2, k_3 \in \mathcal K_{\mathrm{D}}}\Bigg(\langle E_{k_1} \bar E_{k_2} E_{k_3} \bar E_k \rangle  \Big( \f{d a_{k_1}}{dt} \f{d \bar a_{k_2}}{dt} a_{k_3} + a_{k_1} \bar a_{k_2} \f{d^2 a_{k_3}}{dt^2} \Big) \Bigg),
\end{equation}
where the averaging operator $\langle \cdot \rangle$ is defined by \eqref{Definition average}. 

Note that, in the regime when the $a_k$'s are small (in an appropriate sense), the system \eqref{System of ODEs ak} can be thought of as a perturbation of the linear ODE \eqref{Time ODE a} satisfied by the amplitude of a normal mode solution of the linear boundary value problem \eqref{Initial boundary value problem mirror}. In particular, in this case, the function $\tilde\phi_k$ can be thought of as a mode solution of \eqref{Initial boundary value problem mirror} multiplied by a slowly-varying in time amplitude. For this reason, through a slight abuse of terminology, we will refer to the functions $\tilde\phi_k$ as \emph{mode solutions}; the functions $\{\tilde \phi_k\}_{k\in \mathcal K_{\mathrm{D}}}$ will be called the \emph{dominant} modes, while $\{\tilde \phi_k\}_{k\in \mathcal K \setminus \mathcal K_{\mathrm{D}}}$ will constitute the set of \emph{non-dominant} modes. By a further abuse of terminology, we will frequently also refer to the amplitudes $a_k$ simply as modes.

\item The system \eqref{System of ODEs ak} for the $a_k$'s decouples in the following sense: The dominant modes satisfy a $3\times 3$ ODE system (which will be analyzed in \cref{sec:Dominant modes}), while the non-dominant modes satisfy a decoupled system of inhomogeneous linear ODEs, with a source term that depends only on the dominant modes (for the analysis of this system, see \cref{sec:Non dominant modes}).

\item Our choice of the terms dominant modes, non-dominant modes etc.~is based on the expected dynamic behavior of these functions: In the next sections, we will ensure (via an appropriate choice of the parameters $N, \lambda, L$, a perturbation of the mass parameter $M$ and a careful choice of the corresponding initial data for the $\tilde\phi_k$'s) that the dominant modes $\{\tilde \phi_k\}_{k\in \mathcal K_{\mathrm{D}}}$ contribute the most to the $H^s$ size of the true solution $\phi$, while $\{\tilde \phi_k\}_{k\in \mathcal K \setminus \mathcal K_{\mathrm{D}}}$ and $\psi$ remain relatively small.
\end{itemize}

\subsection{The resonant structure of the system of equations for \texorpdfstring{$\tilde\phi_k$}{tilde-phi k}}\label{sec:Resonant structure}

In this section, we will study the structure of the set of linear frequencies ${\omega}_k$ associated to the modes $\tilde\phi_k$ using the spectral estimates of \cref{sec:Spectral analysis radial}. In particular, we will examine the resonant structures appearing under careful choices of the parameters $\lambda, N, L$ in \cref{def:Dominant mode parameters} and the Schwarzschild--AdS mass parameter $M>0$.

We will start by fixing the relation between $\lambda$ and $N$ in \cref{def:Dominant mode parameters}  as follows:

\begin{definition}\label{def:Relation lambda N}
Let $M_0>0$ be a given constant and let  $M\in [\tfrac 12 M_0, M_0]$  be a Schwarzschild--AdS mass parameter. Assume that $N,\lambda, L$ and $\{k_i\}_{i=-1,0,+1}$ are as in \cref{def:Dominant mode parameters} and such that, moreover,  $N \gg_{M_0} 1$ and $L$ is sufficiently large in terms of $N$. The relation between  $N\in \mathbb N^*$ and $\lambda \in \f{1}{L} \mathbb N^*$ will be fixed by the requirement that
\begin{equation*}
\lambda = \f1L \lceil L \cdot \lambda_*\rceil,
\end{equation*}
where $\lambda_*$ is defined as the unique solution to the equation
\begin{equation}\label{eq:C_* again}
\f{f_3(N)}{\lambda_*^{\f12}}+2f_3(1)-\f{f_3(N+2)}{(\lambda_*+2)^{\f12}} = \f1{2 L^{\f12}} \Big(\frac{h(N+2)}{\lambda_*+2} - 2h(1)  - \frac{h(N)}{\lambda_*}\Big) M 
\end{equation}
in the interval $(0,N)$, where
\[
f_3(n) \doteq \int_0^\infty x^3 \big(e_n(x)\big)^2 \, dx.
\]
Note that 
\[
f_3(1) = \f{4}{\sqrt\pi}
\]
and
\begin{equation}\label{Asymptotic formula f3}
f_3(n) = \f{32}{3\pi}n^{\f32} - \f4\pi n^{\f12}+\f{5}{12\pi}n^{-\f12}+O(n^{-\f32}) \quad \text{as} \quad n\rightarrow +\infty
\end{equation}
(see \cref{lem:Asymptotic-functions-f-and-h} in the Appendix for the proof of the above relations). See already \eqref{Estimate N over lambda} below for an asymptotic formula for $\lambda$ for $N\gg 1$.
\end{definition}

The condition \eqref{eq:C_* again} between $\lambda$ and $N$ ensures that we can achieve the additional resonance condition
\begin{equation}\label{Resonant condition Schwarzschild AdS}
 \varepsilon_{k_\minusone}{\omega}_{k_\minusone} - 2 \varepsilon_{k_0}{\omega}_{k_0} +\varepsilon_{k_1} {\omega}_{k_1} =0
\end{equation}
with respect to the time frequencies associated to the Schwarzschild--AdS wave operator via a \emph{small} (i.e.~of size $\lesssim L^{-5\delta_0}$) perturbation of the Schwarzschild--AdS mass parameter; see \cref{prop:Resonance conditions}.

Applying the asymptotic formula \eqref{Asymptotic formula f3} for $f_3(N)$ and $f_3(N+2)$ and using the ansatz 
\begin{equation} \label{eq:ansatz-lambda}
    \lambda_\ast = \beta^{-2} N + \gamma + O(N^{-1}), 
\end{equation}
the relation \eqref{eq:C_* again} yields (provided $N\gg 1$ and the parameter $L$ is sufficiently large in terms of $N$): 
\[
\frac{32}{3 \pi } [\beta^3 - 3 \beta + \frac{3}{16} \pi f_3 (1) ] + \frac{4}{\pi\beta N } (\beta^2-1) (4 \beta^2(1+\gamma) -3)= O(\lambda_\ast^{-\f12} N^{-\f32}+\lambda_\ast^{-\f32} N^{-\f12})
\]
from which we infer that $\beta=1.441448\dots$ is the unique solution to 
\begin{equation}\label{Cubic equation beta}
\beta^3 - 3 \beta +\f{3\sqrt\pi}{4}=0, \quad \beta>1
\end{equation}
and $\gamma = -0.6390369\dots $ satisfies 
\begin{equation}
\gamma =\frac{3}{4} \beta^{-2} -1.
\end{equation}

Note that, since $\lambda = \lambda_* + O(L^{-1})$, we have
\begin{align}\label{Estimate N over lambda}
\f{N}{\lambda} & = \beta^2 -\f{\beta^4\gamma}{N}+  O(N^{-2})\\
& = 2.0777\ldots + O(N^{-1}). \nonumber
\end{align}
The following relation will also be useful to us in   \cref{sec:Estimates spectral coefficients}:
\begin{equation}\label{Estimate difference N over lambda}
\f{N}{\lambda}-\f{N+2}{\lambda+2} = \f2{\lambda}\big( \beta^2-1 \big)+O(\lambda^{-2}).
\end{equation}

The main result of this section is the following \cref{prop:Resonance conditions} which we will prove at the end of the section as a consequence of \cref{lem:existence-of-barM}, \cref{lem:decomposition-of-almost-resonant}, \cref{lem:lower-bound-on-AR} and \cref{lem:estim-pure-nonres} below.
\begin{proposition}\label{prop:Resonance conditions}
Let $M_0>0$ be a given positive constant and let $M'\in[\tfrac{1}{2}M_0,M_0]$ be a given Schwarzschild--AdS mass parameter. Let also $L\gg N \gg 1$ be positive integers such that $N$ is sufficiently large in terms of $M_0$, while $L$ is sufficiently large in terms of $N$ and $M_0$. Then, there exists a $\bar M \in \mathbb R$ satisfying
\begin{equation}
\label{eq:estimate-on-Mbar}
    |\bar M | \lesssim_N L^{-5\delta_0}
\end{equation}
(with the implicit constant above depending only on $N$ and $M_0$, while $\delta_0\ll 1$ is as in \cref{def:Dominant mode parameters}), such that, if we define $\lambda\in \f1L \mathbb N^*$ according to \cref{def:Relation lambda N} in terms of $N$, $L$, $M'$ and we set
\begin{equation*}
  M = M' +\bar M,  
\end{equation*}
the following statement is true: On the Schwarzschild--AdS exterior spacetime with mass parameter $M$, 
the dominant frequency triplets $k_0$ and $k_{\pm1}$ (expressed in terms of $N, \lambda, L$ as in \cref{def:Dominant mode parameters}) satisfy the resonant condition
\begin{equation}\label{Resonant condition time frequencies}
{\omega}_{k_1}-2{\omega}_{k_0}-{\omega}_{k_\minusone}=0.
\end{equation}
At the same time, the non-dominant frequency triplets $k\in \mathcal K \setminus \mathcal K_{\mathrm{D}}$ are resonantly separated from the dominant frequencies in the following sense:
\begin{equation}\label{Non resonant condition time frequencies}
\min_{j_1, j_2, j_3 \in \mathcal K_{\mathrm{D}}}  \left[ \Big| {\omega}_k - \big| \varepsilon_{j_1}{\omega}_{j_1} - \varepsilon_{j_2}{\omega}_{j_2} + \varepsilon_{j_3}{\omega}_{j_3}\big| \Big| + |m_k - m_{j_1} + m_{j_2} - m_{j_3}| \right]  \gtrsim_N L^{-\f12},
\end{equation}
where the signs $\varepsilon_k \in \{\pm1\}$ for $k\in \mathcal K_{\mathrm{D}}$ were fixed in \cref{def:Dominant mode parameters} and the implicit constant above depends only on $N$ and $M_0$.
\end{proposition}

Let us introduce the following subset of Schwarzschild--AdS mass parameters $M$ that will play a central role in the analysis carried out in the next sections (and the definition of which is motivated precisely by the statement of \cref{prop:Resonance conditions}):

\begin{definition}\label{def:Dense set of mass parameters}
For any $M_0>0$ and any integers $1\ll_{M_0} N \ll_{M_0} L^{\delta_0}$, we will define the set $\mathcal C_{N, L;M_0} \subset (0,+\infty)$ of special Schwarzschild--AdS parameters as follows: A parameter $M\in (0,M_0]$ will belong to the set $\mathcal C_{N,L;M_0}$ if and only if the Schwarzschild--AdS spacetime $(\mathcal M_\mathrm{ext}^{(M)}, g_M)$ satisfies the following condition: If we define $\lambda \in \lceil \f1L\rceil \mathbb N^*$ in terms of $N,L,M$ according to   \cref{def:Relation lambda N}, the dominant frequency triplets $k_0$ and $k_{\pm1}$ (expressed in terms of $N, \lambda, L$ as in   \cref{def:Dominant mode parameters}) satisfy the resonant condition
\eqref{Resonant condition time frequencies}, while the non-dominant frequency triplets $k\in \mathcal K \setminus \mathcal K_{\mathrm{D}}$ satisfy the non-resonance condition \eqref{Non resonant condition time frequencies}. 

\end{definition}

As a corollary of \cref{prop:Resonance conditions}, the set $\mathcal C_{N, L;M_0}$ is $L^{-\delta_0}$-\textbf{dense} in $[\f12 M_0,M_0]$ provided $N$ is large enough in terms of $M_0$ and $L$ is large enough in terms of $N, M_0$.

\medskip
We will now establish a number of lemmas that will constitute the heart of the proof of  \cref{prop:Resonance conditions}. 

\begin{remark} For the rest of this section, we will assume that the constants implicit in the $O(\cdot)$ and $\lesssim$ notation might depend on $s$ and $M_0$, but are independent of $M', N, \lambda, L$; the notation $O_N(\cdot)$ and $\lesssim_N$ will imply that the implicit constants are, in addition, allowed to depend on $N$. We will also use the simplified notation $\omega_{\pm 1}$ and $\omega_0$ in place of $\omega_{k_{\pm}}$ and $\omega_0$ (and similarly for other expressions with indices $k\in \mathcal K_{\mathrm{D}}=\{k_-, k_0, k_+\}$). 
\end{remark}

\begin{lemma}\label{lem:existence-of-barM}
Let $M_0>0$, $M'\in [\f12 M_0, M_0]$ and $1\ll N \ll L$ be as in \cref{prop:Resonance conditions}. Then, there exists a $\bar M$ satisfying \eqref{eq:estimate-on-Mbar} such that,  if we define $\lambda\in \f1L \mathbb N^*$ according to \cref{def:Relation lambda N} in terms of $N$, $L$ and $M'$, and we set
\[
M = M'+\bar M,
\]
then the dominant frequencies $\{\omega_k\}_{k\in \mathcal K_{\mathrm{D}}}$ defined with respect to the Schwarzschild--AdS mass parameter $M$ and the frequency parameters $N,\lambda, L$ satisfy the resonance condition \eqref{Resonant condition time frequencies}. 
\begin{proof}
In view of the Taylor-type expansion \eqref{Taylor expansion omega} for the frequencies $\{\omega_k\}_{k\in \mathcal K_{\mathrm{D}}}$ and the expression of the frequency parameters $\ell_k$ in terms of $\lambda, L$ (see \cref{def:Dominant mode parameters}), the resonance condition \eqref{Resonant condition time frequencies} becomes
\begin{equation*}
 \Big(\frac{f_3(n_\minusone)}{l_\minusone^{\f12}} +2\frac{f_3(n_0)}{l_0^{\f12}}-\frac{f_3(n_1)}{l_1^{\f12}}\Big) (M'+\bar M) + \Big(\frac{h(n_\minusone)}{l_\minusone} + 2\frac{h(n_0)}{l_0}  - \frac{h(n_1)}{l_1}\Big) \f{(M'+\bar M)^2}{2} +  O(L^{-1 - 5 \delta_0} ) =0,
\end{equation*}
which can be reexpressed as
\begin{equation}\label{Reexpressing the resonant condition}
\bar M = \frac{2L^{\f12}\Big(\frac{f_3(N+2)}{(\lambda_* + 2)^{\f12}} -2f_3(1)-\frac{f_3(N)}{\lambda_*^{\f12}}\Big)+\Big(\frac{h(N+2)}{\lambda_*+2} - 2h(1)  - \frac{h(N)}{\lambda_*})\Big) M' +  O(L^{-5\delta_0})\cdot \f{1}{M'+\bar M}}
{\frac{h(N)}{\lambda_*} + 2h(1)  - \frac{h(N+2)}{\lambda_*+2}} 
\end{equation}
The proof of \cref{lem:existence-of-barM} will follow if we can show that there exists a $\bar M =O_N(L^{-5\delta_0})$ such that  \eqref{Reexpressing the resonant condition} holds.

In view of the fact that $h(n) = h_1 n^2 +h_2n + O(1)$ with $h_1 = -2 h_2 = 30-\f{512}{3\pi^2}$ (see \cref{lem:Asymptotic-functions-f-and-h}) and $\lambda_\ast = \beta^{-2} N + \gamma +O(N^{-1})$ (see \eqref{eq:ansatz-lambda}), we can calculate: 
\begin{align*}
\frac{h(N)}{\lambda_*} + 2h(1)  - \frac{h(N+2)}{\lambda_*+2}=2\beta^2 h_1 \left[  \beta^2-2 + \frac{h(1)}{h_1} \beta^{-2} \right] + O(N^{-1}) .
\end{align*}
Since $h(1)\ge \f{123}{5\pi}>0$ (see \cref{lem:Asymptotic-functions-f-and-h}), $h_1>0$  and $\beta^2 >2$, we obtain that the denominator in \eqref{Reexpressing the resonant condition} is uniformly bounded from below independently of $N, L$ (provided $N\gg 1$), i.e.
\[
\frac{h(N)}{\lambda_*} + 2h(1)  - \frac{h(N+2)}{\lambda_*+2} \gtrsim 1.
\]
Moreover, using the fact that $\lambda = \lambda_* +O(\f 1L)$, the defining property \eqref{eq:C_* again} of $\lambda_\ast$ implies that\footnote{Recall that, in the statement of \cref{prop:Resonance conditions}, we defined $\lambda_\ast$ so that it satisfies \eqref{eq:C_* again} with $M'$ in place of $M$.}
\[
2L^{\f12}\Big(\frac{f_3(N+2)}{(\lambda_* + 2)^{\f12}} -2f_3(1)-\frac{f_3(N)}{\lambda_*^{\f12}}\Big)+\Big(\frac{h(N+2)}{\lambda_*+2} - 2h(1)  - \frac{h(N)}{\lambda_*}\Big) M' = O_N(L^{-\f12}).
\]
Thus, returning to \eqref{Reexpressing the resonant condition}, we infer that
\[
\bar M = O_N(L^{-\f12}) + O(L^{-5\delta_0})\cdot \f{1}{M'+\bar M}.
\]
As a result, the resonance condition \eqref{Resonant condition time frequencies} is achieved for a $\bar M$ satisfying
\begin{equation}
    |\bar M|\lesssim L^{-5 \delta_0},
\end{equation}
which concludes the proof of this lemma.
\end{proof}
\end{lemma}

\begin{lemma}\label{lem:decomposition-of-almost-resonant}

For all frequency triplets $k=(n_k, \ell_k, m_k)$ in the almost resonant set $K_{\mathrm{AR}}$ (see \cref{def:Splitting of parameter set}), the pair $(\omega_k^{\mathrm{(AdS)}}, m_k) = (2n_k+\ell_k, m_k)$ satisfies either
\begin{equation}\label{eq:case-1-in-KAR} 
\omega_k^{\mathrm{(AdS)}} = \omega_j^{\mathrm{(AdS)}}\quad \text{\textbf{and}} \quad   m_k = m_j\quad  \text{for some }\,  j\in \{\minusone,1\}
\end{equation}
  or 
  \begin{equation}\label{eq:case-2-in-KAR} 
  \omega_k^{\mathrm{(AdS)}} = \big|a \omega_0^{\mathrm{(AdS)}} + b \omega_1^{\mathrm{(AdS)}}\big|\quad  \text{\textbf{and}}\quad  m_k  = a m_0 + b m_1 \quad  \text{for some }\,  (a,b) \in \big\{ (-1,2),\, (-2,3),\, (3, -2),\, (4, -3)\big\}.
        \end{equation}
Moreover,  in both cases we have that 
\begin{equation}\label{Upper bound n k Kar}
n_k\leq 3 N+4 \quad \text{and} \quad   \ell_k \ge L.
\end{equation}

    \begin{proof}
    Recall that the set $\mathcal K_{\mathrm{AR}}$ is defined as those $k=(n_k, \ell_k, m_k)\in \mathcal K \setminus \mathcal K_{\mathrm{D}}$ for which there exists $j_1, j_2, j_3 \in \{\minusone, 0, +1\}$ such that
 \begin{equation}\label{M relation for Kar}
 m_k=m_{j_1}-m_{j_2}+m_{j_3} 
 \end{equation}
 and
 \begin{equation}\label{Omega relation for Kar}
 \omega_k^{\mathrm{(AdS)}} = \Big|\varepsilon_{j_1}\omega_{j_1}^{\mathrm{(AdS)}}-\varepsilon_{j_2}{\omega}_{j_2}^{\mathrm{(AdS)}}+\varepsilon_{j_3}{\omega}_{j_3}^{\mathrm{(AdS)}}\Big|,
 \end{equation}
 where
 \[
 \omega^{\mathrm{(AdS)}}_k \doteq 2n_k + \ell_k.
 \]
The above relations are invariant under the symmetry $j_1\leftrightarrow j_3$. Thus,  we can assume without loss of generality that $j_1 \le j_3$. We will show that, for each triplet in the set $\big\{(j_1, j_2, j_3)\in \{\minusone, 0, +1\}^3:\, j_1\le j_3\big\}$ (there are $18$ elements in this set), if there is a $k\in \mathcal K \setminus \mathcal K_{\mathrm{D}}$ satisfying both \eqref{M relation for Kar} and \eqref{Omega relation for Kar}, then $k$ satisfies either \eqref{eq:case-1-in-KAR} or \eqref{eq:case-2-in-KAR}; in addition, each such $k$ satisfies \eqref{Upper bound n k Kar}.

    ~

    \textbf{Case I: $j_1= j_2$ or $j_3 = j_2$.} In this case, \eqref{M relation for Kar} and \eqref{Omega relation for Kar} imply that
    \begin{equation}\label{Equal omega and m frequencies}
    \omega_k^{\mathrm{(AdS)}} = \omega_j^{\mathrm{(AdS)}} \quad \text{and}\quad m_k = m_j \quad \text{for some }\, j\in \{\minusone, 0 , +1\}.
    \end{equation}
 If $j=0$, the above relation becomes $2n_k + \ell_k = 2n_0 + \ell_0 = 2+\ell_0$ and $m_k=m_0$.  Since  $l_k \geq m_k = m_0 = l_0$,  we obtain $n_k \leq 1 \Rightarrow n_k=1$. Therefore, $l_k = l_0$. This, however, means that $(n_k, \ell_k, m_k)=(n_0, \ell_0, m_0)$, i.e. that $k=k_0 \in \mathcal K_{\mathrm{D}}$, which contradicts our assumption that $k\in \mathcal K\setminus \mathcal K_{\mathrm{D}}$.  Thus, in this case, we must have  $j\in \{\minusone, 1\}$. The upper bound \eqref{Upper bound n k Kar} for $n_k$ then  follows from  \eqref{Equal omega and m frequencies}:
 \begin{equation}\label{Upper bound nk strong}
 2 n_k = 2n_j + \ell_j - \ell_k = 2n_j + |m_j| - \ell_k = 2n_j + |m_k| - \ell_k   \leq 2n_j\leq 2 (N+2),
 \end{equation}
  where we used that $\ell_j = |m_j|$ for $j \in \{0,\minusone,1\} $ and $\ell_k\ge |m_k|$ for all $k\in \mathcal K$; the lower bound for $\ell_k$ in \eqref{Upper bound n k Kar} follows from the fact that $\ell_k \ge |m_j| \ge L$.
    
    ~
    
    \textbf{Case II: $j_1 \neq j_2$ and $j_3 \neq j_2$.} There are $9$ triplets in $\big\{(j_1, j_2, j_3)\in \{\minusone, 0, +1\}^3:\, j_1\le j_3\big\}$ satisfying $j_1 \neq j_2$ and $j_3 \neq j_2$:
    \begin{align*}
      (j_1,j_2,j_3) \in   \Big\{&  (\minusone,0,\minusone),  (\minusone,0,1), (\minusone,1,0), (\minusone,1,\minusone),\\  
      & \quad
      (0,\minusone,0), (0,1,0), (0,\minusone,1), (1,\minusone,1), (1,0,1) \Big\}.
    \end{align*}
    However, using the relation \eqref{Resonant condition AdS} for the dominant frequency parameters $j\in \mathcal K_{\mathrm{D}}$, in the cases $(\minusone,0,1)$, $(0, \minusone,0) $ and $(0,1,0)$, the relations \eqref{M relation for Kar}--\eqref{Omega relation for Kar}  give $(\omega_k^{\mathrm{(AdS)}},m_k)= (\omega_0^{\mathrm{(AdS)}},m_0)$, $(\omega_k^{\mathrm{(AdS)}},m_k)= (\omega_1^{\mathrm{(AdS)}},m_1)$ and  $(\omega_k^{\mathrm{(AdS)}},m_k)= (\omega_\minusone^{\mathrm{(AdS)}},m_\minusone)$, respectively. Thus, we end back in the setting of Case I; arguing as before, we infer that, in these cases, \eqref{eq:case-1-in-KAR} and \eqref{Upper bound nk strong} hold.

 Using the condition \eqref{Resonant condition AdS} to express $\omega_\minusone$ in terms of $\omega_0, \omega_{1}$ and $m_\minusone$ in terms of $m_0, m_{1}$,  an explicit computation shows that, in each of the remaining 6 cases for $(j_1, j_2, j_3)$, the relations \eqref{M relation for Kar}--\eqref{Omega relation for Kar} reduce to 
        \begin{equation}\label{Almost final relation for Kar}
            \begin{cases}
                \omega_k^{\mathrm{(AdS)}} = \big| a \omega_0^{\mathrm{(AdS)}} + b \omega_1^{\mathrm{(AdS)}} \big| \quad \text{and} \\
                m_k = a m_0 + b m_1,
            \end{cases}
        \end{equation}
 with $(a,b) \in \big\{ (3,-2), (4,-3), (-1,2),(-2,3)\big\}$. In particular, \eqref{Almost final relation for Kar} holds and, in each of the four cases, we have (since $m_1= (\lambda+2) L \gg L = m_0$):
 \[
 \ell_k \ge |m_k| \ge |b|m_1-|a| m_0 \ge L.
 \]
  Moreover, using the fact that  $m_0 = \ell_0$ and $m_1 = \ell_1$, we compute from \eqref{Almost final relation for Kar} that
 \begin{multline*}
     2n_k =  \big|a ( 2n_0 + \ell_0) + b (2n_1 + \ell_1)\big| - \ell_k = \big|2 a n_0 + 2 bn_1 + a m_0 + b m_1\big| - \ell_k \\  = \big|2 a n_0 + 2 bn_1  + m_k\big| -\ell_k\le \big|2 a n_0 + 2 bn_1\big|+|m_k|-\ell_k.
 \end{multline*}
 Since $|m_k| \leq l_k$, $n_0=1$ and $n_1=N+2\gg 1$, we infer that
 \[
 n_k \le |bN+a+2b|.
 \]
Thus,  \eqref{Upper bound n k Kar} holds.
    \end{proof}
\end{lemma}

\begin{lemma}\label{lem:lower-bound-on-AR}
Let $M_0>0$, $M'\in [\f12 M_0, M_0]$ and $1\ll N \ll L$ be as in \cref{prop:Resonance conditions} and  $\bar M$, $\lambda$ be as in \cref{lem:existence-of-barM}. We will set 
\[
M=M'+\bar M.
\]
Then, for all $k\in \mathcal K_{\mathrm{AR}}$, the   non-resonant condition \eqref{Non resonant condition time frequencies} holds for the frequencies $\omega_k$ and $\{\omega_j\}_{j\in \mathcal K_{\mathrm{D}}}$ associated to the Schwarzschild--AdS mass parameter $M$ and the frequency parameters $N,\lambda, L$.
   \end{lemma}

\begin{proof}
Let $k=(n_k, \ell_k, m_k) \in \mathcal K_{\mathrm{AR}}$. In view of the upper bound \eqref{Upper bound n k Kar} for $n_k$, we can apply \cref{prop:Taylor expansion omega} to calculate $\omega_k$. In particular, using the Taylor-type expansion formula \eqref{Taylor expansion omega}  for $\omega_k$ and $\{\omega_i\}_{i\in \mathcal K_{\mathrm{D}}}$, we can readily see that the minimum in the expression
\[
\min_{j_1, j_2, j_3 \in \mathcal K_{\mathrm{D}}}  \left[ \Big| {\omega}_k - \big| \varepsilon_{j_1}{\omega}_{j_1} - \varepsilon_{j_2}{\omega}_{j_2} + \varepsilon_{j_3}{\omega}_{j_3}\big| \Big| + |m_k - m_{j_1} + m_{j_2} - m_{j_3}| \right] 
\]
is achieved when both the azimuthal frequencies $m$ and the AdS time frequencies $\omega^{\mathrm{(AdS)}}$ are resonant, namely for $(j_1, j_2, j_3)\in \{\minusone, 0, +1\}^3$ such that
\begin{equation}\label{Resonant condition AdS frequencies and m}
m_k = m_{j_1}-m_{j_2}+m_{j_3} \quad \text{and} \quad  {\omega}^{\mathrm{(AdS)}}_k = \big| \varepsilon_{j_1}{\omega}^{\mathrm{(AdS)}}_{j_1} - \varepsilon_{j_2}{\omega}^{\mathrm{(AdS)}}_{j_2} + \varepsilon_{j_3}{\omega}^{\mathrm{(AdS)}}_{j_3}   \big|.
\end{equation}
In view of \cref{lem:decomposition-of-almost-resonant}, either \eqref{eq:case-1-in-KAR} or \eqref{eq:case-2-in-KAR} hold; we will consider these two cases separately.  

\textbf{ Case I.} Suppose that \eqref{eq:case-1-in-KAR} holds, i.e.~$2n_k+\ell_k=2n_j+\ell_j$ and $m_k=m_j$ for some $j\in \{\minusone, +1\}$. Since $\ell_k \geq |m_k|=|m_j|=\ell_j$, we infer in this case that 
\[
1\leq n_k < n_j
\] 
(the last inequality being strict, since $n_k=n_j$ would imply in this case that $k=(n_k, \ell_k, m_k)=(n_j, \ell_j, m_j)=k_j \in \mathcal K_{\mathrm{D}}$, which contradicts our assumption that $k\in \mathcal K_{\mathrm{AR}}$). Without loss of generality, we will assume that  $j=\minusone$ (the case $j=1$ being completely analogous). In this case, we have
\[
m_k = m_{\minusone} = -\lambda L,  \quad \ell_k = \ell_{\minusone} +2n_{\minusone}-2n_k = \lambda L +2(N-n_k)
\]
and
\[
1\le n_k < N.
\]
Note that, in this case, \eqref{Resonant condition AdS frequencies and m} is achieved only if $j_1 = \minusone, j_2=j_3$ or  $j_3=\minusone, j_1=j_2$ or $j_1=j_3=0, j_2=1$. In each of those cases, we infer that
\[
\min_{j_1, j_2, j_3 \in \mathcal K_{\mathrm{D}}}  \left[ \Big| {\omega}_k - \big| \varepsilon_{j_1}{\omega}_{j_1} - \varepsilon_{j_2}{\omega}_{j_2} + \varepsilon_{j_3}{\omega}_{j_3}\big| \Big| + |m_k - m_{j_1} + m_{j_2} - m_{j_3}| \right] \ge \big|\omega_k -\omega_\minusone  \big|.
\]
Using, again, the Taylor expansion formula \eqref{Taylor expansion omega}, we compute
\begin{equation*}
    \omega_k - \omega_{\minusone} = -\left( \frac{f_3(n_k)}{l_k^{\frac 12}} - \frac{f_3(n_{\minusone})}{l_{\minusone}^{\frac 12}}\right) M + O_N(L^{-1})
\end{equation*}
and thus,
\begin{align*}
    |\omega_k - \omega_{\minusone} | &\gtrsim   M_0 \left| \frac{f_3(n_k)}{(l_{\minusone} + 2(N-n_k))^{\frac 12}} - \frac{f_3(N)}{l_{\minusone}^{\frac 12}}\right|
 - O_N(L^{-1})\\ 
 &\gtrsim \frac{M_0}{l_{\minusone}^{\frac 12}}|f_3(n_k) - f_3(N)| - O_N(L^{-1}) \\
 & \gtrsim_N L^{-\frac 12},
\end{align*}
where, in the last step above, we used that $|f_{3}(n_k) - f_3(N) |\gtrsim 1$ for $n_k < N$ (which follows from the asymptotic expression \eqref{Asymptoics f3} for $f_3(n)$ in \cref{lem:Asymptotic-functions-f-and-h}). In particular,  \eqref{Non resonant condition time frequencies} holds in this case.

\textbf{Case II.} Suppose, now, that \eqref{eq:case-2-in-KAR} holds, i.e.
\begin{equation}\label{eq:case-2-in-KAR again} 
  \omega_k^{\mathrm{(AdS)}} = \big|a \omega_0^{\mathrm{(AdS)}} + b \omega_1^{\mathrm{(AdS)}}\big|\quad  \text{\textbf{and}}\quad  m_k  = a m_0 + b m_1 \quad  \text{for some }\,  (a,b) \in \big\{ (-1,2),\, (-2,3),\, (3, -2),\, (4, -3)\big\}.
        \end{equation} 
In this case, \eqref{Resonant condition AdS frequencies and m} is only attained when the triplet $(j_1, j_2, j_3) \in \{\minusone, 0, +1\}^3$ satisfies
\begin{align*}
\big|\varepsilon_{j_1}\omega^{\mathrm{(AdS)}}_{j_1}-\varepsilon_{j_2}\omega^{\mathrm{(AdS)}}_{j_2}+\varepsilon_{j_3}\omega^{\mathrm{(AdS)}}_{j_3}\big| =  \big|a \omega_0^{\mathrm{(AdS)}} + b \omega_1^{\mathrm{(AdS)}}\big|\quad  \text{\textbf{and}}\quad  m_{j_1}-m_{j_2}+m_{j_3}  = a m_0 + b m_1.
\end{align*}
In this case, we trivially have
\[
\big|\varepsilon_{j_1}\omega_{j_1}-\varepsilon_{j_2}\omega_{j_2}+\varepsilon_{j_3}\omega_{j_3}\big| =  \big|a \omega_0 + b \omega_1\big|
\]
Therefore, similarly as in \textbf{Case I}, \eqref{Non resonant condition time frequencies} will follow if we show that
\begin{equation}\label{Lower bound omega difference to show}
\Big|\omega_k - \big|a\omega_0+b\omega_1\big|\Big| \gtrsim_N L^{-\f12}.
\end{equation}

In view of the upper bound \eqref{Upper bound n k Kar} for $n_k$, we can apply the Taylor expansion formula \eqref{Taylor expansion omega} for $\omega_k$; thus, we compute (using the fact that $|a\omega_0+b\omega_1|$ equals $+(a\omega_0+b\omega_1)$ when $b>0$ and $-(a\omega_0+b\omega_1)$ when $b<0$, since $\omega_1 \gg \omega_0$ as a consequence of \eqref{Taylor expansion omega}):
\begin{align}\label{Almost final bound for omega difference}
   \omega_k - |a \omega_0 + b \omega_{1}| =& - \left( \frac{f_3(n_k)}{\ell_k^{\frac 12}} -\Big| a \frac{f_3(n_0)}{\ell_0^{\frac 12}} + b \frac{f_3(n_1)}{\ell_1^{\frac 12}}\Big| \right) M + O_N(L^{-1}).
\end{align}
Recall that $\ell_0 = L, \ell_1 = ( \lambda +2 )L$ and, as a consequence of  \eqref{eq:case-2-in-KAR} and \eqref{Upper bound n k Kar}:
\[
\ell_k = L \big| a +b(\lambda + 2)\big| + O(N).
\]
Recall also that $\lambda = \frac 1 L \lceil L \cdot \lambda_*\rceil= \lambda_*  + O(L^{-1}) $ (with $\lambda_\ast=\lambda_\ast(N)$ fixed by \cref{def:Relation lambda N}).
Thus, \eqref{Almost final bound for omega difference} yields
\begin{align*}
 \frac{ \big| \omega_k - |a \omega_0 + b \omega_{1}| \big|}{L^{-\frac 12}M} =& A_{N}(n_k) + O_N(L^{-\frac 12}),
\end{align*}
where
\[
A_{N} (n) \doteq \Bigg| \frac{f_3(n)}{|a+b(\lambda_\ast +2)|^{\frac 12}}  - \Big| a f_3(1) + b \frac{f_3(N+2)}{(\lambda_\ast +2)^{\frac 12}}\Big|\Bigg|.
\]
Thus, \eqref{Lower bound omega difference to show} will follow if we show that
\begin{equation}\label{Lower bound to show AN}
\inf_{n\in \mathbb N^*} A_{N} (n) >0.
\end{equation}

Recall that $f_3(n) = \frac{ 32}{3\pi} n^{\frac 32} - \frac{4}{\pi} n^{\frac 12} + O(n^{-\frac 12})$ (see \eqref{Asymptoics f3}) and $\lambda_\ast = \beta^{-2} N + \gamma + O(N^{-1})$ (see \eqref{eq:ansatz-lambda}). Moreover, since $\lambda_\ast\gg 1$ and $\f{f_3(N+2)}{\lambda_\ast+2)^{\f12}} \sim N \gg 1$, we have
\[
|a+b(\lambda_\ast+2)| = \sgn(b) a + |b|(\lambda_\ast+2) \quad \text{and} \quad
\Big| a f_3(1) + b \frac{f_3(N+2)}{(\lambda_\ast +2)^{\frac 12}}\Big| = \sgn(b) a f_3(1) + |b| \frac{f_3(N+2)}{(\lambda_\ast +2)^{\frac 12}} ,
\]
where
\[
\sgn(b) \doteq \f{b}{|b|}.
\]
Therefore,
\begin{align*}
    &|a+b(\lambda_\ast +2)|^{-\frac 12} = \frac{1}{ \sqrt{|b| \beta^{-2}}}N^{-\f12}   -  \frac{\sgn(b)a + (\gamma +2)|b|}{2 (|b| \beta^{-2})^{\frac 32}} N^{-\f32} + O(N^{-\frac 52}),\\
    &(\lambda_\ast +2)^{-\frac 12} = \frac{1}{ \sqrt{  \beta^{-2}}}N^{-\f12}   -  \frac{ \gamma +2}{2 ( \beta^{-2} )^{\frac 32}}N^{-\f32} + O(N^{-\frac 52}),\\
    & f_3(N+2) =   \frac{32}{3\pi}   N^{\frac 32} + \frac{28}{ \pi}N^{\frac 12} + O(N^{-\frac 12}).
\end{align*}  
In view also the fact that $f_3(1)=\f4{\sqrt\pi}$, we calculate:
\begin{align}
A_{N} (n)  &= \Bigg| 
\big(\frac{ 32}{3\pi} n^{\frac 32} - \frac{4}{\pi} n^{\frac 12}\Big)\Big(\frac{1}{ \sqrt{|b| \beta^{-2}}}N^{-\f12}  -  \frac{\sgn(b)a + (\gamma +2)|b|}{2 (|b| \beta^{-2})^{\frac 32}} N^{-\f32}\Big)  \nonumber  \\
&\hphantom{= \Bigg| 
\big(}
- \f{4\sgn(b)a}{\sqrt\pi} - |b| \Big(\frac{32}{3\pi}   N^{\frac 32} + \frac{28}{ \pi}N^{\frac 12}\Big) \Big(\frac{1}{ \sqrt{  \beta^{-2}}}N^{-\f12}   -  \frac{ \gamma +2}{2 ( \beta^{-2} )^{\frac 32}}N^{-\f32}\Big)+O(n^{\f32} N^{-\f52}+N^{-1}) 
\Bigg| 
 \nonumber  \\
& = \Bigg|
\f{32\beta|b|}{3\pi}\Big(\f{n^{\f32}}{|b|^{\f32}N^{\f32}}-1\Big) N
 \nonumber \\
&\hphantom{=\Bigg|}
-\Big(\f{16\beta^3}{3\pi}\cdot \big(\sgn(b) a + (\gamma+2)|b|\big)\f{n^{\f32}}{|b|^{\f32} N^{\f32}}
+\f{4\beta}{\pi } \f{n^{\f12}}{|b|^{\f12}N^{\f12}}
-\f{4|b| \beta \big(4\beta^2(\gamma+2)-21 \big)-12\sgn(b) a\sqrt{\pi}}{3\pi}\Big) \nonumber \\ &
+O(n^{\f32} N^{-\f52}+N^{-1})    \Bigg|.   \label{Expression asymptotic A n}
\end{align}
From the above expression, it follows readily that, in the case when $\big|\f{n}{|b|N}-1\big|\gg \f1N$, the first term in the right-hand side dominates, yielding $A_N(n)\gtrsim 1$. Thus, in order to establish \eqref{Lower bound to show AN}, it remains to examine the behavior of $A_N(n)$ for $n$ satisfying $\big|\f{n}{|b|N}-1\big|\lesssim \f1N$, i.e.~the case when
\begin{equation}\label{Ansatz n b}
n=|b| N +\kappa
\end{equation}
for some $\kappa \in \mathbb Z$ satisfying $\kappa=O(1)$ (i.e.~being uniformly bounded independently of $N$). Using the ansatz \eqref{Ansatz n b} for $n$ in \eqref{Expression asymptotic A n}, noting that, in this case,
\[
\f{n^{\f32}}{|b|^{\f32}N^{\f32}} = 1+ \f32 \f{\kappa}{|b|N} +O(N^{-2}), \quad \f{n^{\f12}}{|b|^{\f12}N^{\f12}} = 1+ \f12 \f{\kappa}{|b|N} +O(N^{-2}),
\]
we infer that: 
\begin{align*}
A_N \big(|b|N+\kappa\big) & = \Bigg| \f{16\beta }{\pi}\kappa - \f{16\beta^3}{3\pi}\cdot \big(\sgn(b) a + (\gamma+2)|b|\big) -\f{4\beta}{\pi} \\
& \hphantom{= \Bigg|}+ \f{4|b| \beta \big(4\beta^2(\gamma+2)-21 \big)-12\sgn(b) a\sqrt{\pi}}{3\pi} +O(N^{-1})\Bigg|
\\
& = \f{4\beta}{\pi}\Bigg| 4\kappa -\f{4\sgn(b)a (\beta^3+\f{3\sqrt\pi}{4})+3\beta +21|b|\beta }{3\beta}+O(N^{-1})\Bigg|.
\end{align*}
(note that the $\gamma+2$ terms cancel out).
Using the fact that
\[
\beta^3 - 3\beta+\f{3\sqrt\pi}{4}=0,
\]
the above expression becomes:
\[
A_N \big(|b|N+\kappa\big)  = \f{4\beta}{\pi} \Big|
4\kappa 
 -7|b|-4\sgn(b)a-1
 +O(N^{-1})\Big|.
 \]
 In each of the four cases $(a,b) = (3,-2), (4,-3),(-1,2),(-2,3)$, the integer $7|b|+4\sgn(b)a+1$ is \textbf{not} a multiple of $4$; therefore,
 \[
 \big|4\kappa  -7|b|-4\sgn(b)a-1\big|\ge 1 \quad \text{for all }\, \kappa\in \mathbb Z
 \]
 and, therefore, $A_N\big(|b|N+\kappa\big)\gtrsim 1$ for all integers $\kappa=O(1)$. Thus, the proof of \eqref{Lower bound to show AN} (and, therefore, \eqref{Lower bound omega difference to show}) is complete.

\end{proof}

\begin{lemma}\label{lem:estim-pure-nonres}
Let $M_0>0$, $M'\in [\f12 M_0, M_0]$ and $1\ll N \ll L$ be as in  \cref{prop:Resonance conditions} and  $\bar M$, $\lambda$ be as in    \cref{lem:existence-of-barM}. We will set 
\[
M=M'+\bar M.
\]
Then, for all purely non-resonant frequency triplet $k\in \mathcal K_{\mathrm{NR}}$ and for all $j_1,j_2,j_3 \in \{ \minusone,0,1\}$, at least one of the following two conditions hold:
\begin{itemize}
    \item $m\neq m_{j_1}-m_{j_2}+m_{j_3} $
    \item $\Big| {\omega}_k - \big| \varepsilon_{j_1}{\omega}_{j_1} - \varepsilon_{j_2}{\omega}_{j_2} + \varepsilon_{j_3}{\omega}_{j_3}\big|  \Big|\geq \frac 12$
    \end{itemize}
  In the above, the $\omega_j$'s are the mode frequencies associated to the Schwarzschild--AdS mass parameter $M$ and the frequency parameters $N,\lambda, L$.
\begin{proof} 
Let $k\in \mathcal K_{\mathrm{NR}}$ and $j_1, j_2, j_3\in \{\minusone, 0, +1\}$.
It suffices to consider the case in which 
\begin{equation}\label{Condition for resonant ms}
m_k= m_{j_1}-m_{j_2}+m_{j_3}.
\end{equation}
 In this case, the definition of the set $\mathcal K_{\mathrm{NR}}$ (see \cref{def:Splitting of parameter set}) implies that the corresponding AdS frequencies $\omega^{\mathrm{(AdS)}}_i = \ell_i+2n_i$ satisfy
\[
\Bigg|{\omega}_k^{\mathrm{(AdS)}} - \Big| \varepsilon_{j_1}{\omega}_{j_1}^{\mathrm{(AdS)}}-\varepsilon_{j_2}{\omega}_{j_2}^{\mathrm{(AdS)}}+\varepsilon_{j_3}{\omega}_{j_3}^{\mathrm{(AdS)}}\Big|\Bigg| \ge 1.
\]
Note that, since $m_j = \varepsilon_j \ell_j$ (and therefore  $\varepsilon_j\omega^{\mathrm{(AdS)}}_j = m_j +2\varepsilon_j n_j$) for all $j\in \{\minusone, 0, +1\}$, the above inequality yields
\begin{equation}\label{Lower bound from completely non resonant condition}
\Bigg|2n_k +\ell_k - \Big|m_k + 2\big( \varepsilon_{j_1}n_{j_1}-\varepsilon_{j_2}n_{j_2}+\varepsilon_{j_3}n_{j_3}\big)\Big|\Bigg| \ge 1.
\end{equation}
We have to show that 
\[
\Big| {\omega}_k - \big| \varepsilon_{j_1}{\omega}_{j_1} - \varepsilon_{j_2}{\omega}_{j_2} + \varepsilon_{j_3}{\omega}_{j_3}\big|  \Big|\geq \frac 12.
\]
In view of  the Taylor expansion formula \eqref{Taylor expansion omega} for the dominant frequencies $\omega_j$, $j\in \{\minusone, 0, +1\}$, it suffices to show that
\[
\Big| {\omega}_k - \big| \varepsilon_{j_1}{\omega}^{\mathrm{(AdS)}}_{j_1} - \varepsilon_{j_2}{\omega}^{\mathrm{(AdS)}}_{j_2} + \varepsilon_{j_3}{\omega}^{\mathrm{(AdS)}}_{j_3}\big|  \Big|\geq \frac34
\]
or, equivalently,
\begin{equation}\label{Bound to show for completely non resonant}
\Big| {\omega}_k - \big| m_k + 2\big( \varepsilon_{j_1}n_{j_1}-\varepsilon_{j_2}n_{j_2}+\varepsilon_{j_3}n_{j_3}\big)\big|  \Big|\geq \frac34.
\end{equation}

We will consider two cases, depending on the size of the overtone $n_k$. Let us note that the condition \eqref{Condition for resonant ms} implies (in view of the fact that $\min_{j_1, j_2, j_3\in \{\minusone, 0, +1\}}\big| m_{j_1}-m_{j_2}+m_{j_3}\big| = L$):
\[
\ell_k \ge |m_k| \ge L.
\]

~

\textbf{Case I: $n_k \le L^{\delta_0}$.} In this case, we can apply the Taylor expansion formula \eqref{Taylor expansion omega} for $\omega_k$ and obtain
\[
\omega_k = 2n_k+\ell_k +O(L^{-\f12}).
\]
In this case, the bound \eqref{Bound to show for completely non resonant} follows directly from \eqref{Lower bound from completely non resonant condition}.

~

\textbf{Case II: $n_k \ge L^{\delta_0}$.} In this case, applying \cref{lem:Crude Weyl law}, we get
\[
\omega_k \ge \ell_k + L^{\f12\delta_0}.
\]
Therefore, using the trivial bound $\ell_k \ge |m_k|$, we have
\[
{\omega}_k - \big| m_k + 2\big( \varepsilon_{j_1}n_{j_1}-\varepsilon_{j_2}n_{j_2}+\varepsilon_{j_3}n_{j_3}\big)\big| \ge L^{\f12\delta_0} + \ell_k -|m_k|-2(n_{j_1}+n_{j_2}++n_{j_3}) \ge L^{\f12\delta_0} -6(N+2) \gg 1
\]
(provided $L$ is sufficiently large in terms of $N$). Thus, \eqref{Bound to show for completely non resonant} holds.
\end{proof}
\end{lemma}

 \begin{proof}[Proof of \cref{prop:Resonance conditions}] The estimate \eqref{eq:estimate-on-Mbar}  on $\bar M$  and the resonance condition \eqref{Resonant condition time frequencies}  follow from \cref{lem:existence-of-barM}. 
The non-resonant condition \eqref{Non resonant condition time frequencies} follows from \cref{lem:lower-bound-on-AR} and \cref{lem:estim-pure-nonres}  for $k \in \mathcal K_{\mathrm{AR}}$ and $k \in \mathcal K_{\mathrm{NR}}$, respectively. This concludes the proof of \cref{prop:Resonance conditions}. 
\end{proof}

\section{The system of dominant modes}\label{sec:Dominant modes}

In this section, we will study the evolution in time of the dominant modes $\{\tilde\phi_k\}_{k\in \mathcal K_{\mathrm{D}}}$. Recall that the system of equations satisfied by $\{\tilde \phi_k\}_{k\in \mathcal K_{\mathrm{D}}}$  decouples from the rest of the  equations for $\{\tilde \phi_{k}\}_{k\in \mathcal K \setminus \mathcal K_{\mathrm{D}}}$ (see \eqref{Boundary value problem tilde phi}), with the corresponding amplitudes $\{a_k(t)\}_{k\in \mathcal K_{\mathrm{D}}}$ solving the following $3\times 3$ $2^{nd}$ order ODE system:
\begin{equation}\label{System of ODEs ak again}
\f{d^2 a_k}{dt^2} +{\omega}_k^2 a_k = \sum_{k_1, k_2, k_3 \in \mathcal K_{\mathrm{D}}}\Bigg(  \langle \bar E_k E_{k_1} \bar E_{k_2} E_{k_3} \rangle  \Big( \f{d a_{k_1}}{dt} \f{d \bar a_{k_2}}{dt} a_{k_3} + a_{k_1} \bar a_{k_2} \f{d^2 a_{k_3}}{dt^2} \Big) \Bigg)
\end{equation}
(recall that the averaging operator $\langle \cdot \rangle$ is defined by \eqref{Definition average}).

For the rest of this section, it will be convenient for us to use the shorthand notation
\[
{\omega}_0 \doteq {\omega}_{k_0}, \quad {\omega}_{\pm1} \doteq {\omega}_{k_{\pm1}}
\]
and
\[
a_0 \doteq a_{k_0}, \quad a_{\pm1} \doteq a_{k_{\pm1}}
\]
(and similarly for any other instance of a variable with index $k\in \mathcal K_{\mathrm{D}}$). 
We will also use the following ansatz for $\{a_k(t)\}_{k\in \mathcal K_{\mathrm{D}}}$:
\begin{equation}\label{B k variables}
a_k(t) = b_k(t) e^{-i \varepsilon_k {\omega}_k t},
\end{equation}
(where the signs $\varepsilon_k\in \{\pm1\}$ were fixed in \eqref{Associated signs}). Solving for the $b_k$'s, we obtain:
\[
b_{\minusone} = a_\minusone e^{-i {\omega}_\minusone t}, \quad b_0 = a_0 e^{i{\omega}_0 t}, \quad b_1 = a_1 e^{i {\omega}_1 t}.
\]
The system  of equations \eqref{System of ODEs ak again} recast in terms of the $b_k$ variables takes the following form:
\begin{align} \nonumber
     \f{d^2 b_k}{dt^2}  - 2i \varepsilon_k {\omega}_{k} \f{d b_k}{dt} = \sum_{k_1,k_2,k_3 \in \mathcal K_{\mathrm{D}} } & e^{i (\varepsilon_k {\omega}_k - \varepsilon_{k_1}{\omega}_{k_1} + \varepsilon_{k_2}{\omega}_{k_2} - \varepsilon_{k_3}{\omega}_{k_3})t  } \langle \bar E_k E_{k_1} \bar E_{k_2} E_{k_3} \rangle \\  \nonumber
     \cdot &\big[ \left( \varepsilon_{k_1}\varepsilon_{k_2}{\omega}_{k_1} {\omega}_{k_2} - {\omega}_{k_3}^2 \right)b_{k_1} \overline{ b_{k_2}} b_{k_3}   \\  \nonumber & - i \varepsilon_{k_1}{\omega}_{k_1}  b_{k_1} \f{d{\overline{b_{k_2}}}}{dt}  b_{k_3}   +  i \varepsilon_{k_2}{\omega}_{k_2}   \f{d b_{k_1}}{dt} {\overline{b_{k_2}}}  b_{k_3}   - 2i \varepsilon_{k_3}{\omega}_{k_3} b_{k_1} {\overline{b_{k_2}}} \f{d b_{k_3}}{dt} \\ & +    \f{d b_{k_1}}{dt}  \f{d \overline{b_{k_2}}}{dt} b_{k_3} + b_{k_1}\overline{b_{k_2}} \f{d^2 b_{k_3}}{dt^2}  \big]. \label{eq:ode-system-for-ak}
\end{align}

We will renormalize the system \eqref{eq:ode-system-for-ak} in accordance with the scaling properties (with respect to the frequency parameter $L$) of the initial data for $\{b_k\}_{k\in \mathcal K_{\mathrm{D}}}$, the frequencies ${\omega}_k$ and the expected timescale of the evolution. In particular, we will set
\begin{empheq}[box=\fbox]{equation}
    \tilde b_k \doteq L^{s} b_k, \quad  \tilde t \doteq L^{-1-2s} t,\quad   \tilde {\omega}_k \doteq L^{-1} \varepsilon_k {\omega}_k, \quad   \llangle \bar E_k E_{j_1} \bar E_{j_2} E_{j_3}\rrangle \doteq L^{2} \langle \bar E_k E_{j_1} \bar E_{j_2} E_{j_3}   \rangle.    \label{Renormalization}
\end{empheq}

\begin{remark}
     Note that we have incorporated the sign $\varepsilon_k$ in the definition of $\tilde\omega_k$, so that $\tilde\omega_\minusone<0$ and $\tilde\omega_0, \tilde\omega_1>0$.
    \end{remark}

The renormalized quantities $\big\{ \tilde b_{\minusone},  \tilde b_{0}, \tilde b_{1}\big\}$ satisfy the following system of ODEs:
\begin{align} \nonumber
L^{-2s -2}    \frac{d^2}{d \tilde t^2} \tilde b_k  - 2i \tilde {\omega}_k \frac{d}{d\tilde t} \tilde b_k =& \sum_{j_1,j_2,j_3 \in \{\minusone, 0, 1 \}} e^{i L^{2s +2 } (  \tilde {\omega}_k - \tilde {\omega}_{j_1} + \tilde {\omega}_{j_2} - \tilde {\omega}_{j_3} )\tilde t} \llangle \bar E_k E_{j_1} \bar E_{j_2} E_{j_3}   \rrangle \\ 
\nonumber
       &\quad  \times \bigg[ \left( \tilde {\omega}_{j_1}\tilde  {\omega}_{j_2} - \tilde {\omega}_{j_3}^2 \right)\tilde b_{j_1} \overline{ \tilde b_{j_2}} \tilde b_{j_3}   \\
 \nonumber
 &\quad\qquad + L^{-2s -2} \left( - i\tilde  {\omega}_{j_1} \tilde  b_{j_1} \frac{d}{d\tilde t}{\overline{\tilde b_{j_2}}}  \tilde b_{j_3} +  i \tilde  {\omega}_{j_2}   \frac{d}{d\tilde t}{\tilde b_{j_1} }{\overline{\tilde b_{j_2}}} \tilde  b_{j_3}   - 2i \tilde  {\omega}_{j_3} b_{j_1} {\overline{ \tilde b_{j_2}}} \frac{d}{d\tilde t}{\tilde{b}_{j_3}}  \right) \\ 
&\quad\qquad    +L^{-4s -4} \left( \frac{d}{d\tilde t}{\tilde b_{j_1}}  \frac{d}{d\tilde t}{\overline{\tilde b_{j_2}}} \tilde b_{j_3}  + \tilde b_{j_1} \overline{\tilde b_{j_2}} \frac{d^2}{d\tilde t^2}{\tilde b_{j_3}}  \right) \bigg]. \label{eq:full-system-for-tildeak 2}
\end{align}

The main result of this section is the following:
\begin{proposition}\label{prop:The 3 times 3 system}
For any $C_{\mathrm{amp}}>1$, $s\ge 4$,  $M_0>0$ and $0< \epsilon \le 1$, let us assume that $N$ is sufficiently large in terms of $C_{\mathrm{amp}}, s, M_0$ and that $L$ is sufficiently large in terms of $N, \epsilon, C_{\mathrm{amp}}, s, M_0$. Let also $M$ be a Schwarzschild--AdS mass parameter in the set $\mathcal C_{N,L;M_0}$ (where the special set of mass parameters $\mathcal C_{N,L;M_0}$ was introduced in \cref{def:Dense set of mass parameters}) and assume that $\lambda$ is fixed in terms of $N, L, M$ as in \cref{def:Relation lambda N}.

In this case, the following statement is true: There exist a $\tilde T = \tilde T(C_{\mathrm{amp}}, \epsilon, \lambda, N, s, M_0)>0$ (in particular, $\tilde T$ is independent of $L$) and a pair $\big\{\big( \tilde B^{(0)}_{j}, \tilde B^{(1)}_j\big) \big\}_{j\in \mathcal K_{\mathrm{D}}}$ of initial data for the system \eqref{eq:full-system-for-tildeak 2} satisfying
\begin{equation} \label{eq:initial-data-for-ak 2}
       \tilde B^{(0)}_{j} = \begin{cases}
            \epsilon c_\minusone \lambda^{-s} & \text{ if } j=\minusone, \\
              \epsilon  & \text{ if } j=0, \\  
              \epsilon c_1 \lambda^{-s} & \text{ if } j=1. 
        \end{cases} 
\end{equation}
for some $c_{\pm 1} \in \mathbb C$ with
 \begin{equation*} 
|c_\minusone|^2 + |c_1|^2 =1 
\end{equation*}
and
\begin{equation}\label{Initial derivative tilde a bound}
\big|\tilde B^{(1)}_j \big| \le C \quad \text{for all} \quad j\in \{\minusone, 0, 1\}
\end{equation}
(for some $C>0$ depending on $\lambda, N, s, M_0$ but \textbf{not} on $L$), such that the maximal solution $\{\tilde b_k(\tilde t)\}_{k\in \{\minusone, 0, 1\}}$ of \eqref{eq:full-system-for-tildeak 2} with initial data 
\begin{equation}\label{Initial data b tilde dominant}
\big( \tilde b_k (0), \frac{d \tilde b_k}{d\tilde t}(0) \big) = \big( \tilde B_0, \tilde B_1\big)
\end{equation}
satisfies the following properties:
\begin{enumerate}
\item The $\tilde b_k(\tilde t)$'s exist on the whole interval $\tilde t\in [0, \tilde T]$ and satisfy the bounds:
\begin{equation}\label{Bound C 1 norm tilde a}
\sum_{j\in \{\minusone, 0, 1\}} \Big(\big\| \tilde b_j \|_{L^\infty([0,\tilde T])} + \big\| \frac{d}{d\tilde t}\tilde b_j  \big\|_{L^\infty([0,\tilde T])}\Big) \le C
\end{equation}
and, for any integer $q \ge 2$,
\begin{equation}\label{Bound C q norm tilde a}
\sum_{j\in \{\minusone, 0, 1\}} \big\| \frac{d^q}{d\tilde t^q}\tilde b_j \|_{L^\infty([0,\tilde T])} \le C_q L^{(2s+2)(q-1)},
\end{equation}
where the constants $C,C_q$ above depend on $C_{\mathrm{amp}}, \epsilon, \lambda, N, s, M_0$ but \textbf{not} on $L$.
\item At $\tilde t = \tilde T$, we have the following amplification estimate:
\begin{equation}\label{Amplification lower bound renormalized system}
\frac{\displaystyle\sum_{j\in \{\minusone, 0, 1\}} \tilde{\omega}_j^{2s} \big| \tilde b_j (\tilde T)\big|^2}{\displaystyle\sum_{j\in \{\minusone, 0, 1\}} \tilde{\omega}_j^{2s} \big| \tilde b_j (0)\big|^2} \ge C_{\mathrm{amp}}.
\end{equation}
\end{enumerate}
\end{proposition}

The proof of  \cref{prop:The 3 times 3 system} will be given at the end of \cref{sec:Dominant modes} and will be obtained in two steps: First, we will show that, by choosing the initial data $\big(\tilde b_k(0), \frac{d\tilde b_k}{d\tilde t}(0)\big)$ appropriately, the solution to the corresponding initial value problem for \eqref{eq:full-system-for-tildeak 2} can be well approximated by the solution to a $1^{st}$ order $3\times 3$ system which arises as the \textbf{slowly oscillating} limit of \eqref{eq:full-system-for-tildeak 2}, i.e.~by formally setting $L=+\infty$ and dropping terms which oscillate in time with frequency $\gtrsim L^{2s+2}$; the resulting system of equations (namely \eqref{Slowly oscillating system}) has coefficients which do not depend on the parameter $L$. See already \cref{prop:Slowly oscillating approximation}. We will then show that the solutions to the $1^{st}$-order system \eqref{Slowly oscillating system} arising from the initial data set \eqref{eq:initial-data-for-ak 2} exhibits the amplification property \eqref{Amplification lower bound renormalized system}; see \cref{prop:Growth slowly oscillating system}. In order to obtain the aforementioned results, the hierarchy of inequalities
\[
C_{\mathrm{amp}} \ll \lambda \ll L
\]
will be crucial. In order to establish the growth of the solutions to the slowly oscillating system, we will need, in addition, very precise estimates on the value of the spectral coefficients $\llangle \bar E_k E_{j_1} \bar E_{j_2} E_{j_3}\rrangle$ appearing in \eqref{eq:full-system-for-tildeak 2}; this will be achieved in \cref{sec:Estimates spectral coefficients}.

\begin{remark} In the statement of   \cref{prop:The 3 times 3 system}, the parameters $\epsilon$ and $C_{\mathrm{amp}}$ can be chosen independently of each other (and the frequency parameters $N,\lambda$ need only be large in terms of $C_{\mathrm{amp}}$). However, we will later be interested in the special case when $C_{\mathrm{amp}}$ is a function of $\epsilon$ (in particular, $C_{\mathrm{amp}}\sim \f1{\epsilon^4}$; see   \cref{sec:The proof of the main theorem}). 
\end{remark}

As a corollary of \cref{prop:The 3 times 3 system}, we will obtain the following estimates for the dominant modes $\{\tilde \phi_k\}_{k\in \mathcal K_{\mathrm{D}}}$ in the original (unrenormalized) $(t,y,\theta,\varphi)$ coordinates:
\begin{corollary}\label{cor:The 3 times 3 system}
For any $C_{\mathrm{amp}}>1$, $s\in \mathbb N_{\ge 4}$, $M_0>0$ and $0< \epsilon \le 1$, let $N$ be sufficiently large in terms of $C_{\mathrm{amp}}, s, M_0$ and let $L$ be sufficiently large in terms of $N, \epsilon, C_{\mathrm{amp}}, s, M_0$. Let also $M$ be a Schwarzschild--AdS mass parameter in the set $\mathcal C_{N,L;M_0}$ and assume that $\lambda$ is fixed in terms of $N, L, M$ as in \cref{def:Relation lambda N}. Let also $\tilde T = \tilde T(C_{\mathrm{amp}}, \epsilon, \lambda, N , s, M_0)$ and $\big\{\big(  \tilde B^{(0)}_j, \tilde B^{(1)}_j \big)\big\}_{j\in \mathcal K_{\mathrm{D}}}$ be as in the statement of \cref{prop:The 3 times 3 system}.

Then, the solutions $\{\tilde\phi_j\}_{j\in \mathcal K_{\mathrm{D}}}$ of the boundary value problem  \eqref{Boundary value problem tilde phi} with initial data\footnote{Note that, with respect to the representation $r \tilde\phi_j(t,y,\theta, \varphi) = b_j(t) e^{-i\varepsilon_j \omega_j t} E_j(y, \theta, \varphi)$ of $\phi_j$, $j\in \mathcal K_{\mathrm{D}}$, after setting $\tilde t = L^{-1-2s}t$ and $\tilde b_j = L^{s} b_j$, the choice \eqref{Initial data dominant modes final} of initial data corresponds precisely to the choice \eqref{Initial data b tilde dominant}
for $\big( \tilde b_k (0), \frac{d \tilde b_k}{d\tilde t}(0) \big)$.}  
\begin{align}\label{Initial data dominant modes final}
 \tilde\phi_j(0,y, \theta, \varphi) = & L^{-s} \tilde B^{(j)}_0 \cdot \f{1}{r(y)}E_j(y, \theta, \varphi) ,
 \\
  \partial_{t^*} \tilde\phi_j(0, y, \theta, \varphi)  = &  L^{-s} \big(-i\varepsilon_j \omega_j \tilde B^{(j)}_0 + L^{-2s-1}\tilde B^{(j)}_1\big) \cdot \f{1}{r(y)}E_j (y, \theta, \varphi) ,\nonumber 
\end{align}
(where $E_j$ are the spatial eigenfunctions \eqref{Spatial eigenfunction})  satisfy the following bounds for a constant $C>0$ which is \textbf{independent} of $C_{\mathrm{amp}}, \epsilon, N, \lambda$ and $L$ (but which is allowed to depend on $M_0$ and $s$):
\begin{equation}\label{Smallness initial norm dominant}
\Big\|\chi \cdot  \sum_{j\in \mathcal K_{\mathrm{D}}} \sum_{s_1+s_2+s_3=0}^s \partial_t^{s_1} \partial_y^{s_2} \nabla_{\mathbb S^2}^{s_3} (r\tilde\phi_j)\Big|_{t=0} \Big\|^2_{L^2(\sin\theta dy d\theta d\varphi)} \le C \epsilon^2
\end{equation}
and, for $T_1 \doteq L^{2s+1}\tilde T$:
\begin{equation}\label{Largeness final norm dominant}
\Big\|\chi \cdot  \sum_{j\in \mathcal K_{\mathrm{D}}} \sum_{s_1+s_2+s_3=0}^s \partial_t^{s_1} \partial_y^{s_2} \nabla_{\mathbb S^2}^{s_3} (r\tilde\phi_j)\Big|_{t=T_1} \Big\|^2_{L^2(\sin\theta dy d\theta d\varphi)} \ge \f1{C} C_{\mathrm{amp}} \epsilon^2.
\end{equation}
\end{corollary}

The proof of \cref{cor:The 3 times 3 system}  will be given in \cref{sec:Proof of corollary dominant modes}.

\subsection{Estimates for the spectral coefficients}\label{sec:Estimates spectral coefficients}

In this section, we will establish a number of estimates for the spectral projection coefficients $\llangle \bar E_k E_{j_1} \bar E_{j_2} E_{j_3}\rrangle$ appearing in the system \eqref{eq:full-system-for-tildeak 2}. 

\begin{proposition}\label{prop:Estimates for spectral projections}
Let $N, \lambda, L$ be as in the statement of \cref{prop:The 3 times 3 system}. Then, the spectral projections $\llangle |E_1|^2 |E_0|^2\rrangle$, $\llangle |E_\minusone|^2 |E_0|^2\rrangle$ and $\llangle E_\minusone \bar E_0^2 E_1\rrangle$ are real-valued and satisfy
\begin{equation}
\llangle |E_1|^2 |E_0|^2\rrangle, \llangle |E_\minusone|^2 |E_0|^2\rrangle,\big|\llangle E_\minusone \bar E_0^2 E_1\rrangle\big| \sim 1,
\end{equation}
\begin{equation}
\llangle |E_1|^2 |E_0|^2\rrangle = \llangle |E_\minusone|^2 |E_0|^2\rrangle +O(\f1\lambda), \quad \llangle E_\minusone \bar E_0^2 E_1\rrangle = \llangle |E_\minusone|^2 |E_0|^2\rrangle +O(\f1\lambda)
\end{equation}
(where the constants implicit in the $\sim$ and $O(\cdot)$ notation above are \textbf{independent} of $N,\lambda, L$), as well as the difference bound
\begin{equation}\label{Crucial spectral estimate}
0.96 < \lambda \f{\Big|\llangle |E_1|^2 |E_0|^2\rrangle - \llangle |E_\minusone|^2 |E_0|^2\rrangle \Big|}{2\Big|\llangle E_\minusone \bar E_0^2 E_1\rrangle\Big|} < 0.97.
\end{equation}
\end{proposition}

\begin{remark}
    The upper bound in the estimate \eqref{Crucial spectral estimate} is crucial for our proof of \cref{thm:Main theorem}: The fact that the spectral ratio \eqref{Crucial spectral estimate} is strictly less than $1$ makes the slowly oscillating system \eqref{Slowly oscillating system} linearly unstable around its trivial solution (see the proof of \cref{prop:Growth slowly oscillating system} and, in particular, \cref{lem:The matrix spectrum}). Had this constant been larger than $1$, the system \eqref{Slowly oscillating system} would be linearly orbitally stable, and it would not be a priori clear whether the system of dominant modes (and hence the actual solution $\phi$) even exhibits turbulence at all. The fact that this ratio turns out to be less than $1$ with our precise choice of the spectral parameters $(k_\minusone, k_0, k_1)$ in \cref{sec:Hierarchy of parameters} appears to us to be a mere coincidence; there are alternative choices for the parameters  $(k_\minusone, k_0, k_1)$  for which the set of eigenvalues $\{\omega_k\}_{k\in \mathcal K}$ has the same resonance properties that we use, but for which the ratio \eqref{Crucial spectral estimate} is larger than $1$.
\end{remark}

\begin{proof}[Proof of \cref{prop:Estimates for spectral projections}]
In this proof, we will assume that the constants implicit in the $\lesssim_N$, $O_N(\cdot)$ notation are independent of $L$ (but are allowed to depend on $N$, $\lambda$). 

For any $j_1, j_2, j_3, j_4 \in \mathcal K_{\mathrm{D}}$, our definition \eqref{Renormalization} of $\llangle E_{j_1} \bar E_{j_2} E_{j_3} \bar E_{j_4} \rrangle$ and the definition \eqref{Spatial eigenfunction} of $E_k$ imply that
\begin{align*}
\llangle E_{j_1} \bar E_{j_2} E_{j_3} \bar E_{j_4} \rrangle = L^2 \Bigg(&\int_0^{y_\mathrm{mirror}} \! \! \! \!   R_{n_{j_1},\ell_{j_1}}(y) R_{n_{j_2},\ell_{j_2}}(y) R_{n_{j_3},\ell_{j_3}}(y) R_{n_{j_4},\ell_{j_4}}(y) (r(y))^{-6}\Big(1+\f{1}{r(y)^2} -\f{2M}{r(y)^3}\Big)\, dy \Bigg)\\
& \times   \Bigg(\int_0^{2\pi}\int_0^\pi Y_{\ell_{j_1},m_{j_1}}(\theta, \varphi) \bar Y_{\ell_{j_2},m_{j_2}}(\theta, \varphi) Y_{\ell_{j_3},m_{j_3}}(\theta, \varphi) \bar Y_{\ell_{j_4},m_{j_4}}(\theta, \varphi) \, \sin\theta d\theta d\varphi \Bigg).
\end{align*}
According to \cref{lem:eigenfunctions-estimate-m>0}, we have for any $(n, \ell)$ associated to a frequency triad $k\in \mathcal K_{\mathrm{D}}$:
\[
R_{n,\ell}(y) = \ell^{\f14} e_n(\ell^{\f12} y) + O(\ell^{-\f14+5\delta_0}).
\]
Combining the above bound with the exponential decay of $R_{n,\ell}(y)$ in the region $y\gtrsim n^{\f12} \ell^{-\f12}$ (see the bound \eqref{eq:expo-decay}), we therefore obtain:
\begin{align}\label{Separation expression average}
\llangle E_{j_1} \bar E_{j_2} E_{j_3} \bar E_{j_4} \rrangle = \big(\ell_{j_1} \ell_{j_2}\ell_{j_3}\ell_{_4}\big)^{\f14} L^{\f 52} \langle  e_{n_{j_1}} & (\ell^{\f12}_{j_1}\cdot) e_{n_{j_2}}(\ell^{\f12}_{j_2}\cdot) e_{n_{j_3}}(\ell^{\f12}_{j_3}\cdot) e_{n_{j_4}}(\ell^{\f12}_{j_4} \cdot) y^6 \rangle_y\\
& \times L^{-\f12} \langle Y_{\ell_{j_1},m_{j_1}} \bar Y_{\ell_{j_2},m_{j_2}} Y_{\ell_{j_3},m_{j_3}} \bar Y_{\ell_{j_4},m_{j_4}}\rangle_{\mathbb S^2} + O(L^{-\f12+5\delta_0}), \nonumber 
\end{align}
where 
\[
\langle f \rangle_y \doteq \int_0^{+\infty} f(y) \, dy
\]
and
\[
\langle f \rangle _{\mathbb S^2} \doteq \int_0^{2\pi} \int_0^\pi f(\theta,\varphi) \, \sin\theta d\theta d\varphi.
\]
The statements of \cref{prop:Estimates for spectral projections} then follow by combining \cref{lem:Spherical harmonics asymptotics} below for the asymptotic expression for the integral of spherical harmonics appearing in \eqref{Separation expression average} with  \cref{cor:Radial estimates spectral coefficients} at the end of \cref{sec:Estimates spectral coefficients} for the integral in \eqref{Separation expression average} involving the normalized Hermite functions.
\end{proof}

\begin{lemma}\label{lem:Spherical harmonics asymptotics}
In the regime $1\leq a,b, c \ll L$, the following asymptotic formula holds
\begin{equation}\label{eq:spherical-harmonics-asymp}
L^{-\f12}\langle Y_{L,L} \bar Y_{aL,aL} Y_{bL, bL} \bar Y_{cL, cL} \rangle_{\mathbb S^2}= \ind_{\{1-a+b-c=0\}} \cdot \left(\f1{\sqrt 2 \pi^{\f32}} \frac{ ( abc )^{\frac 14}}{(1 + a+b+c)^{\frac 12}} + O_{a,b,c}(L^{-1})\right), 
\end{equation}
where 
\[
\ind_{\{m=0\}} =
\begin{cases}
1, \quad m=0,\\
0, \quad m\neq 0
\end{cases}
\]
and the constant implicit in the $O_{a,b,c}(\cdot)$ notation depends on $a,b,c$ but is independent of $L$. As a result (in view of the fact that $Y_{\ell, -\ell} = (-1)^\ell \bar{Y}_{\ell, \ell}$), 
\[
L^{-\f12} \big\langle |Y_{L,L}|^2 |Y_{\lambda L,\lambda L}|^2 \big\rangle_{\mathbb S^2} = \f{1}{2 \pi^{\f32}}  +O(\lambda^{-1}) = -L^{-\f12}\big\langle Y_{\lambda L, -\lambda L} |Y_{\lambda L, \lambda L}|^2 Y_{(\lambda+2)L, (\lambda+2)L}\big\rangle_{\mathbb S^2} + O(\lambda^{-1})
\]
and
\[
L^{-\f12} \Big(\big\langle |Y_{L,L}|^2 |Y_{\lambda L,\lambda L}|^2 \big\rangle_{\mathbb S^2} - \big\langle |Y_{L,L}|^2 |Y_{(\lambda+2) L,(\lambda+2) L}|^2 \big\rangle_{\mathbb S^2} \Big)= O(\lambda^{-2}),
\]
where the constants implicit in the $O(\cdot)$ notation above are independent of $N, \lambda, L$.
\end{lemma}

\begin{proof}
    For $m=\ell$, the spherical harmonic $Y_{\ell,\ell}(\theta, \varphi)$ takes the form:
    \[
Y_{\ell,\ell}(\theta, \varphi)= \f{\gamma_\ell}{\sqrt{2\pi}} (\sin\theta)^\ell  e^{i\ell \varphi},
\]
    where
    \[
    \gamma_\ell \doteq \left( \frac{ \Gamma(  \ell+\frac 32)}{\sqrt\pi \Gamma(\ell + 1)})\right)^{\frac 12 }.
    \]
     Thus, for any $\ell_1, \ell_2, \ell_3, \ell_4 \in \mathbb N^*$:
    \begin{align}
     \int_0^{2\pi} \int_0^\pi Y_{\ell_1, \ell_1 }(\theta, \varphi) \bar Y_{\ell_2, \ell_2 }(\theta, \varphi)  Y_{\ell_3, \ell_3 }(\theta, \varphi) \bar Y_{\ell_4, \ell_4 }(\theta, \varphi) \, \sin \theta d \theta d\varphi 
     &= \f{1}{4\pi^2} \frac{\gamma_{\ell_1} \gamma_{\ell_2} \gamma_{\ell_3} \gamma_{\ell_4} }{ \gamma^2_{\frac{\ell_1 + \ell_2 + \ell_3 + \ell_4}{2}}} \int_0^{2\pi} e^{i(\ell_1-\ell_2+\ell_3-\ell_4)\varphi} d\varphi    \nonumber  \\
     & = \f{\ind_{\{\ell_1-\ell_2+\ell_3-\ell_4=0\}}}{2\pi} \frac{\gamma_{\ell_1} \gamma_{\ell_2} \gamma_{\ell_3} \gamma_{\ell_4} }{ \gamma^2_{\frac{\ell_1 + \ell_2 + \ell_3 + \ell_4}{2}}}. \label{Explicit expression product spherical harmonics}
\end{align}
    Using Stirling's approximation for the asymptotics of the Gamma function, we obtain \begin{equation*} 
        \gamma_\ell  =\ell^{\frac 14} \left(  \frac{1}{\pi^{\frac 14}} + \frac{3}{16 \pi^{\frac 14} \ell} + O(\ell^{-2}) \right).
    \end{equation*}
After substituting the above asymptotics  in the right-hand side of \eqref{Explicit expression product spherical harmonics} and setting $\ell_1=L, \ell_2=aL, \ell_3 = bL, \ell_4 = cL$, we obtain \eqref{eq:spherical-harmonics-asymp}.
\end{proof}

The rest of the section is dedicated to proving \cref{cor:Radial estimates spectral coefficients} at the end of this subsection.
To this end, let us define the \emph{radial} spectral projection coefficients $\mathbb I_{-1,1,0,0}$, $\mathbb I_{-1,-1,0,0}$ and $\mathbb I_{1,1,0,0}$ by the relations
\begin{equation}\label{Mixed spectral projection}
\mathbb I_{-1,1,0,0} \doteq \int_0^{+\infty} \lambda^{\f14}e_{n_\minusone}(\lambda^{\f12}x) \cdot (\lambda+2)^{\f14} e_{n_1}((\lambda+2)^{\f12} x) \cdot e_{n_0}^2(x) \cdot x^6 \, dx
\end{equation}
and
\begin{equation}\label{Pure spectral projections}
\mathbb I_{-1,-1,0,0} \doteq
\int_0^{+\infty} \lambda^{\f12}e^2_{n_\minusone}(\lambda^{\f12}x) \cdot e_{n_0}^2(x) \cdot x^6 \, dx
,
\quad
\mathbb I_{1,1,0,0} \doteq 
\int_0^{+\infty}  (\lambda+2)^{\f12} e^2_{n_1}((\lambda+2)^{\f12} x) \cdot e_{n_0}^2(x)\cdot x^6  \, dx.
\end{equation}
We begin by computing $\mathbb I_{-1,1,0,0}$.
\begin{lemma}\label{lem:Mixed spectral projection estimate}
Let $N, \lambda, L$ be as in the statement of \cref{prop:The 3 times 3 system}. The mixed spectral projection coefficient $\mathbb I_{-1,1,0,0}$ defined by \eqref{Mixed spectral projection} satisfies (in the regime $N\gg 1$):
\begin{equation}\label{Estimate mixed spectral projection radial}
   \mathbb I_{-1,1,0,0}= \frac{ 8 \beta^8 e^{-2\beta^2}}{\pi^{\frac 12}}(14 - 8 \beta^2 + \beta^4) +O(N^{-1}) , 
\end{equation}
where the constants implicit in the $O(\cdot)$ notation are independent of $N, \lambda,L$ and  with $\beta=1.441448\dots$ defined as the solution to \eqref{Cubic equation beta} with $\beta>1$:
\begin{equation} \frac{ 8 \beta^8 e^{-2\beta^2}}{\pi^{\frac 12}}(14 - 8 \beta^2 + \beta^4)  = 2.235287 \dots.\end{equation}
\end{lemma}

\begin{proof}
For this proof, we will assume that the constants implicit in the $\lesssim$, $\sim$ and $O(\cdot)$ notation are independent of $N, \lambda, L$. Moreover, we will use the shorthand notation 
\[
\lambda' =\lambda+2.
\]
We recall from \eqref{Normalized Hermite function} that 
\begin{equation*}
e_{n_0}(x) = \f{2}{\pi^{\f14}} x e^{-\f{x^2}2}.
\end{equation*}
and then compute 
\begin{align*}
\mathbb I_{-1,1,0,0} \doteq \, & \int_0^{+\infty} \lambda^{\f14}e_{n_\minusone}(\lambda^{\f12}x) \cdot (\lambda')^{\f14} e_{n_1}((\lambda')^{\f12} x) \cdot e_{n_0}^2(x) x^6  \, dx \\
= &  \lambda^{\frac 14} (\lambda')^{-\frac 72+\f14}
\int_0^{+\infty} e_{n_\minusone} (\lambda^{\frac 12} (\lambda')^{-\frac 12} x) e_{n_1}(x) e^2_{n_0}( (\lambda ')^{-\frac 12}x) x^6 dx 
\\
= &  \frac{4 c_{n_\minusone} c_{n_1}}{\pi^{\frac 12}} \lambda^{\frac 14} (\lambda')^{-\frac 92+\f14}
\int_0^{+\infty} H_{2N-1} (\lambda^{\frac 12} (\lambda')^{-\frac 12} x) H_{2N +3} (x)  x^8 e^{-x^2 [ (\lambda')^{-1} + \frac 12 + \frac 12 \lambda  (\lambda')^{-1} ]}  dx 
\\
= &  \frac{4 c_{n_\minusone} c_{n_1}}{\pi^{\frac 12}} \lambda^{\frac 14} (\lambda')^{-\frac 92+\f14}
\int_0^{+\infty} H_{2N-1} (\lambda^{\frac 12} (\lambda')^{-\frac 12} x) H_{2N +3} (x)  x^8 e^{-x^2}  dx 
\\
= &  \frac{4 c_{n_\minusone} c_{n_1}}{\pi^{\frac 12}} \lambda^{-\frac{15}{4}} (\lambda')^{-\frac{1}{4}}\int_0^{+\infty}  (\lambda^{\frac 12} (\lambda ')^{-\frac 12} x)^8 H_{2N-1} (\lambda^{\frac 12} (\lambda')^{-\frac 12} x) H_{2N +3} (x)  e^{-x^2}  dx 
\\  = &  \frac{4 c_{n_\minusone} c_{n_1}}{\pi^{\frac 12}} \lambda^{-\frac{15}{4}} (\lambda')^{-\frac{1}{4}} \sum_{j=4}^8 p^{j}_{2N-1}  \int_0^{+\infty} H_{2N-1+j} (\lambda^{\frac 12} (\lambda')^{-\frac 12} x) H_{2N +3} (x)  e^{-x^2}  dx,
\end{align*}
where, in the last step, we used \cref{lem:iteration-hermite} (the coefficients $p_{2N-1}^j$ are given by \eqref{Coefficients in final recurrence}) and dropped the terms coming from $j< 4$ in view of the orthogonality properties of the Hermite functions.\footnote{In particular, note that $\int_0^{+\infty} p(x) H_{2N+3}(x) e^{-x^2} \, dx =0$ for any odd polynomial $p(x)$ of order $\le 2N+1$.} We now use the scaling property of the Hermite polynomials  (see \cite[p.416, 18.18.13]{NIST-10}): For any $\gamma>0$, we have:
\begin{equation}\label{eq:scaling-hermite}
    H_n(\gamma x) = \sum_{i=0}^{\lfloor \frac n2 \rfloor} \gamma^{n-2i} (\gamma^2 -1)^i \binom{n}{2i} \frac{(2i)!}{i!} H_{n-2i}(x).
\end{equation}
Upon setting $\gamma \doteq \lambda^{\frac 12} (\lambda')^{-\frac 12} $, we therefore compute (using, again the orthogonality properties of Hermite polynomials and the fact that $p_n^j=0$ for $j$ odd):
\begin{align*}
   \sum_{j=4}^8 p^{j}_{2N-1}  &  \int_0^{+\infty} H_{2N-1+j} (\gamma  x) H_{2N +3} (x)  e^{-x^2}  dx  \\ &  = 
  p^{4}_{2N-1}   \int_0^{+\infty} H_{2N+3} (\gamma x) H_{2N +3} (x)  e^{-x^2}  dx \\
    & \hphantom{=} +   p^{6}_{2N-1} \int_0^{+\infty} H_{2N+5} (\gamma x) H_{2N +3} (x)  e^{-x^2}  dx \\
    &  \hphantom{=} + p^{8}_{2N-1}   \int_0^{+\infty} H_{2N+7} (\gamma x) H_{2N +3} (x)  e^{-x^2}  dx \\
    & = c_{n_1}^{-2}  \gamma^{2N+3}  \left(  p_{2N-1}^4 + p_{2N-1}^6  (\gamma^2 -1)(2N-1)(2N-2)+ p_{2N-1}^8 \frac{1}{2} (\gamma^2-1)^2 \frac{(2N-1)!}{(2N-5)! }\right).
\end{align*}
Note that  \begin{equation*}
\gamma^{2N+3} = \left(\frac{ \lambda}{\lambda +2} \right)^{N+\frac 32}  = \left( 1- \frac{2\beta^2}{N+\f32} + O(N^{-2})\right)^N = e^{-2\beta^2} (1+O(N^{-1})).\end{equation*}
Moreover, $\lambda = \beta^{-2} N +O(1) $,  $\lambda' =\lambda+2 = \beta^{-2} N +O(1)$ and $\gamma^2 -1 = -2 \beta^2 N^{-1} + O(N^{-2})$.  We can thus calculate:
\begin{equation*}
     p_{2N-1}^4 + p_{2N-1}^6  (\gamma^2 -1)(2N-1)(2N-2)+ \frac{1}{2}p_{2N-1}^8  (\gamma^2-1)^2 \frac{(2N-1)!}{(2N-5)! } = \frac{1}{8} ( 14-8 \beta^2 + \beta^4) N^2 +O(N). 
\end{equation*}
Combining the above, we obtain:
\begin{align*}
    \mathbb I_{-1,1,0,0} &= \frac{4 c_{n_\minusone} }{\pi^{\frac 12} c_{n_1} } \beta^8 N^{-4}e^{-2\beta^2}\left( \frac 18 (14 - 8 \beta^2 + \beta^4) N^2 \right) (1+O(N^{-1})) \\
    &= \frac{\beta^8 e^{-2\beta^2}}{2\pi^{\frac 12}}(14 - 8 \beta^2 + \beta^4) \frac{ c_{n_\minusone} }{N^2 c_{n_1} } (1+O(N^{-1})) \\
& =\frac{ 8 \beta^8 e^{-2\beta^2}}{\pi^{\frac 12}}(14 - 8 \beta^2 + \beta^4)  (1+O(N^{-1})) ,
\end{align*}
where we used that  $\frac{ c_{n_\minusone} }{N^2 c_{n_1} }   = 16  +O(N^{-1})$. 
\end{proof}

Before we estimate the difference of the pure spectral coefficients we will prove the following auxiliary lemma. To state our result, we  will make use of  the modified Bessel function of the first kind (see, e.g. \cite[Chapter~2, \S10]{Olver97}), defined as 
\begin{equation*}
     I_k (x) = \sum_{i=0}^\infty \frac{1}{i! (i+k)!} \left( \frac{x}{2}\right)^{2i+k}.
\end{equation*}
Note that $\frac{d}{dx} I_0(x) = I_1(x)$ and $\frac{d}{dx} I_1 (x) = I_0(x) - x^{-1} I_1(x)$.
\begin{lemma}\label{lem:weighted-hermite-functions-integral}
Let $\alpha_0>0$ be a given constant.  Then, for any  $n\in\mathbb N$ and any $|\alpha|\leq \alpha_0$, the following relations hold: 
\begin{equation}\label{Simple weighted-projections}
    \frac{1}{\sqrt \pi 2^{n-1} n!} \int_{0}^{\infty} e^{- ( 1 + \frac{\alpha}{n})  x^2} H^2_{n}(x) dx = e^{-\alpha} I_0(\alpha) -\frac{e^{-\alpha } \alpha  (I_0(\alpha )-I_1(\alpha ))}{2 n} + O(n^{-2})
\end{equation}
and, more generally, for any $k\in \mathbb N$:
\begin{equation}
     \frac{1}{\sqrt \pi 2^{n-1} n!} \frac{1}{(-n)^k} \int_{0}^{\infty} x^{2k} e^{- ( 1 + \frac{\alpha}{n})  x^2} H^2_{n}(x) dx =  \frac{d^k}{d\alpha^k}\left( e^{-\alpha} I_0(\alpha) -\frac{e^{-\alpha } \alpha  ( I_0(\alpha )-I_1(\alpha ))}{2 n} \right) + O(n^{-2}) \label{eq:weighted-projections}
\end{equation}
In the above, the constants implicit in the $O(\cdot)$ notation may depend on $\alpha_0$ and on $k$, but they are \textbf{independent} of $n$ and $\alpha \in [-\alpha_0, \alpha_0]$.

As a special case, we have
\begin{align*}
    \frac{1}{\sqrt \pi 2^{n-1} n!} \frac{1}{n^4} \int_{0}^{\infty} x^{8} e^{-( 1 + \frac{\alpha}{n})  x^2} H^2_{n}(x) dx = & L_{8,0}(\alpha) + L_{8,1}(\alpha) n^{-1} +O(n^{-2}),
\end{align*}
where
\begin{align*}
    L_{8,0}(\alpha)& \doteq  \frac{\alpha  \left(8 \alpha ^2+4 \alpha +3\right) I_0(\alpha )-2 \left(4 \alpha  \left(\alpha ^2+\alpha +1\right)+3\right) I_1(\alpha )}{\alpha ^3 e^\alpha } ,
\\  L_{8,1}(\alpha)&\doteq  \frac{(1+ 20 \alpha - 16 \alpha^2) I_0(\alpha )-    (2 + 5 \alpha + 12 \alpha^2 - 16 \alpha^3) I_1(\alpha )}{ \alpha^2 e^{\alpha }}  .
\end{align*}

\end{lemma}
\begin{proof}
Let us set 
\[
A =  1 + \frac{\alpha}{n}.
\]
We can then compute using the scaling identity \eqref{eq:scaling-hermite} and the orthogonality properties of Hermite polynomials:
\begin{align}
    \int_{0}^{\infty} e^{- A  x^2} H^2_{n}(x) dx \nonumber
& =\frac{1}{\sqrt  A } \int_0^\infty e^{-x^2} H_n^2\left(\frac{x}{\sqrt  A}\right) dx \\
 & = \frac{1}{\sqrt  A } \int_0^\infty e^{-x^2} \left[\sum_{i=0}^{\lfloor n/2 \rfloor }  \left( \frac{1}{A}\right)^{n-2i}  \left(\frac{1}{A}-1\right)^i \frac{n!}{(n-2i)! i! } H_{n-2i}(x) \right]^2 dx   \nonumber\\
    & =\frac 12 \frac{\sqrt \pi}{\sqrt A } \sum_{i=0}^{\lfloor n/2 \rfloor } 
 \left( \frac{1}{{A}}\right)^{n-2i}  \left(\frac{1}{A}-1\right)^{2i} \frac{(n!)^2}{(n-2i)! (i!)^2 } 2^{n-2i}     \nonumber\\
     & = \frac 12 \frac{\sqrt \pi}{ A^n \sqrt A } 2^n n! \sum_{i=0}^{\lfloor n/2 \rfloor } \frac{2^{-2i}}{(i!)^2} \frac{n!}{(n-2i)!} (A-1)^{2i}    \nonumber\\ 
& = \frac 12 \frac{\sqrt \pi}{ A^n \sqrt A } 2^n n! \sum_{i=0}^{\infty } \frac{1}{(i!)^2} \left( \frac{\alpha}{2}\right)^{2i} \frac{n! }{(n-2i)!n^{2i}} \ind_{\{ i \leq \lfloor n/2 \rfloor \} } . \label{Sum in terms of alpha}
    \end{align}
  
Let us  split   
    \begin{equation*}\frac{n! }{(n-2i)!n^{2i}} \ind_{\{ i \leq \lfloor n/2 \rfloor \} } =   \frac{n! }{(n-2i)!n^{2i}} \ind_{\{ i \leq \lfloor n/4 \rfloor \} }  +  \frac{n! }{(n-2i)!n^{2i}} \ind_{\{\lfloor n/4 \rfloor < i \leq \lfloor n/2 \rfloor \} } =:  S_{i,n}^{\leq} + S_{i,n,}^{\geq}.
    \end{equation*}

    For the term $ S_{i,n}^{\leq}$, note that we have the expansion 
\begin{equation}\label{eq:expansion-S-two-orders}
S_{i,n}^{\leq} = 1 + \frac{i (1-2i)}{n} + O\left(\frac{i^3}{n^{2}}\right)
\end{equation}  
    which follows by considering the exponential of the identity (for $i\ge 1$)
    \begin{equation*}
        \log \frac{n! }{(n-2i)!n^{2i}} =  \log  \prod_{j=0}^{2i-1} \left(1-\frac{j}{n}\right) = \sum_{j =0}^{2i-1}\log\left(1-\frac{j}{n}\right) = \sum_{j =0}^{2i-1}\left( \frac{-j}{n} +O \left( \frac{j^2}{n^2} \right) \right) = \frac{i(1-2i)}{n} + O\left( \frac{i^3}{n^2}\right)
    \end{equation*}
 (in deriving the above, we used the trivial  bound $j\leq 2i-1 \leq 2 \lfloor n/4 \rfloor -1 \leq n/2$  to expand  $\log\left(1-\frac{j}{n}\right)= -\frac{j}{n} + O\left(\frac{j^2}{n^2}\right)$).  For the term $ S_{i,n,}^{\geq}$, note that it satisfies $0\leq S_{i,n,}^{\geq}\leq 1$ and $S_{i,n,}^{\geq} \neq 0$ only for $\lfloor n/4 \rfloor < i \leq \lfloor n/2 \rfloor $. In particular, we can calculate:
    \begin{equation}\label{eq:estiamte-second-term}
      \left|   \sum_{i=0}^{\infty } \frac{1}{(i!)^2} \left( \frac{\alpha}{2}\right)^{2i}S_{i,n}^\geq \right| \leq n^{-2} \sum_{i=0}^{\infty } n^2 \frac{1}{(i!)^2} \left( \frac{|\alpha|}{2}\right)^{2i}S_{i,n}^\geq \leq  n^{-2} \sum_{i=0}^{\infty }  \frac{(4i)^2}{(i!)^2} \left( \frac{|\alpha|}{2}\right)^{2i}S_{i,n}^\geq \leq n^{-2} 4 |\alpha|^2 I_0(|\alpha|)
    \end{equation}
(thus, the part of the sum in \eqref{Sum in terms of alpha} corresponding to $S_{i,n,}^{\geq}$  is of lower order compared to that corresponding to $ S_{i,n}^{\leq}$). 
    
    Thus, combining  \eqref{eq:expansion-S-two-orders}  and  \eqref{eq:estiamte-second-term},  we obtain that 
\begin{align*}
\sum_{i=0}^{\infty } \frac{1}{(i!)^2} \left( \frac{\alpha}{2}\right)^{2i} \frac{n! }{(n-2i)!n^{2i}} 1_{\{ i \leq \lfloor n/2 \rfloor \} } & = \sum_{i=0}^{\infty } \frac{1}{(i!)^2} \left( \frac{\alpha}{2}\right)^{2i} S_{i,n}^{\leq}  +O(n^{-2}) \\ & = \sum_{i=0}^{\infty } \frac{1}{(i!)^2} \left( \frac{\alpha}{2}\right)^{2i}  + \frac{1}{n} \sum_{i=0}^{\infty } \frac{i(1-2i)}{(i!)^2} \left( \frac{\alpha}{2}\right)^{2i} + O(n^{-2})\\ & = I_0(\alpha) + \frac{1}{2n}  \left(\alpha  I_1(\alpha )-\alpha ^2 I_0(\alpha )\right) + O(n^{-2}).
\end{align*}
 Moreover, $ A^n = (1+\frac{\alpha}{n})^n = e^\alpha ( 1 - \frac{\alpha^2}{2n} + O(n^{-2}) )$  and $\sqrt A = 1 + \frac{\alpha}{2n} + O(n^{-2})$ as $n\to \infty$ and thus,
\begin{equation*}
    \frac{1}{2^{n-1} \sqrt \pi  n!} \int_{0}^{\infty} e^{- A  x^2} H^2_{n}(x) dx = e^{-\alpha} I_0(\alpha) -\frac{e^{-\alpha } \alpha  ( I_0(\alpha )-I_1(\alpha ))}{2 n} + O(n^{-2})
\end{equation*}
as $n\to\infty$. 

The more general result \eqref{eq:weighted-projections} follows by differentiating \eqref{Simple weighted-projections} with respect to $\alpha$.
\end{proof}

\begin{lemma}\label{lem:Difference spectral projections estimate}
Let $N, \lambda, L$ be as in the statement of \cref{prop:The 3 times 3 system}. The spectral projection terms $\mathbb I_{\minusone, \minusone,0,0}$ and $\mathbb I_{1,1,0,0}$ defined by \eqref{Pure spectral projections} satisfy (in the regime $N\gg 1$):
\begin{equation}\label{Estimate difference spectral projections radial}
 \mathbb I_{\minusone,\minusone,0,0} -  \mathbb I_{1,1,0,0}   =\mathcal D (\beta) \frac{1}{\lambda}+ O(\lambda^{-2}),
\end{equation}
where $\mathcal D(\beta) = -4.311320\dots $ is given by
\begin{equation*}
\mathcal D(\beta) =   -64\frac{\beta^2-1}{\sqrt \pi}  e^{-2 \beta ^2} \beta ^2 \left(\left(64 \beta ^4-40 \beta ^2-1\right) \beta ^2 I_0\left(2 \beta ^2\right)+\left(-64 \beta ^6+24 \beta ^4+5 \beta ^2+1\right) I_1\left(2 \beta ^2\right)\right).
\end{equation*}
\end{lemma}

\begin{proof}
We compute: 
\begin{align*}
\mathbb I_{\minusone,\minusone,0,0} & = \int_0^\infty \lambda^{\frac 12} e_{n_\minusone}(\lambda^{\frac 12} x) e_{n_0}^2(x) x^6 dx \\
& = \lambda^{\frac 12}\frac{4}{\sqrt \pi } c_{n_\minusone}^2  \int_0^\infty  H^2_{2N-1}(\lambda^{\frac 12} x) x^8 e^{-(1+\lambda) x^2 }dx \\
& = \lambda^{-4}\frac{4}{\sqrt \pi } c_{n_\minusone}^2  \int_0^\infty  H^2_{2N-1}( x) x^8 e^{-\frac{1+\lambda}{\lambda} x^2 }dx .
\end{align*}
We now use \cref{lem:weighted-hermite-functions-integral} for $n = 2N-1$ and $\alpha = \alpha_{\minusone} \doteq   \frac{{2N-1}}{\lambda}$ to obtain
\begin{equation*}
    \mathbb I_{\minusone,\minusone,0,0} = \frac{4}{\sqrt \pi} \alpha_{\minusone}^4 \left(L_{8,0}(\alpha_{\minusone} ) + L_{8,1}(\alpha_{\minusone} )(2N-1)^{-1}  + O(N^{-2})\right).
\end{equation*}
Similarly, setting  $n= 2N+3$ and $\alpha = \alpha_1 \doteq \frac{2N+3}{\lambda +2}$  in \cref{lem:weighted-hermite-functions-integral} we obtain 
\begin{equation*}
    \mathbb I_{1,1,0,0} = \frac{4}{\sqrt \pi}\alpha_1^4 \left(L_{8,0}(\alpha_1 ) + L_{8,1}(\alpha_1 )(2N+3)^{-1}  + O(N^{-2})\right).
\end{equation*}
Using the fact that $\alpha_\minusone -\alpha_1 = 4( \beta^2 - 1 )\lambda^{-1} + O(N^{-2})$ and $\alpha_{-1} = 2 \beta^{2} + O(N^{-1})$, we obtain: 
\begin{align*}
    \lambda (  \mathbb I_{\minusone,\minusone,0,0} -  \mathbb I_{1,1,0,0} )
& = \lambda \frac{4}{\sqrt \pi } \left[ \alpha_{\minusone}^4 L_{8,0}(\alpha_\minusone) - \alpha_{1}^4 L_{8,0}(\alpha_1) + O(N^{-2})\right] \\
 & = \frac{16}{\sqrt \pi} (\beta^2-1) \frac{d}{d\alpha} \big(\alpha^4 L_{8,0}(\alpha)\big)\Big|_{\alpha = 2\beta^{2}}   + O(N^{-1}) \\
    & =-64\frac{\beta^2-1}{\sqrt \pi}  e^{-2 \beta ^2} \beta ^2 \big[ \left(64 \beta ^4-40 \beta ^2-1\right) \beta ^2 I_0\left(2 \beta ^2\right)\\
    & \qquad +\left(-64 \beta ^6+24 \beta ^4+5 \beta ^2+1\right) I_1\left(2 \beta ^2\right)\big]  + O(N^{-1}).
\end{align*}
\end{proof}

\begin{corollary}\label{cor:Radial estimates spectral coefficients}
Let $N, \lambda, L$ be as in the statement of \cref{prop:The 3 times 3 system}. Provided $N$ is larger than a (fixed) absolute constant, the spectral projection coefficients $\mathbb I_{\minusone,1,0,0}$,  $\mathbb I_{\minusone,\minusone,0,0}$ and $\mathbb I_{1,1,0,0}$ satisfy the bound 
\begin{equation}\label{Spectral projection ratio}
0.96 < \lambda \f{ \big|\mathbb I_{\minusone,\minusone,0,0} -  \mathbb I_{1,1,0,0}\big|}{2\big|\mathbb I_{\minusone,1,0,0}\big|} < 0.97. 
\end{equation}
\end{corollary}

\begin{proof}
Combining \eqref{Estimate mixed spectral projection radial} and \eqref{Estimate difference spectral projections radial}, we obtain
\begin{align*}
\f{\lambda \big|\mathbb I_{\minusone,\minusone,0,0} -  \mathbb I_{1,1,0,0}\big|}{2\big|\mathbb I_{\minusone,1,0,0}\big|} 
& = \f{64\frac{\beta^2-1}{\sqrt \pi}  e^{-2 \beta ^2} \beta ^2 \left[\left(64 \beta ^4-40 \beta ^2-1\right) \beta ^2 I_0\left(2 \beta ^2\right)\! +\! \left(-64 \beta ^6+24 \beta ^4+5 \beta ^2+1\right) I_1\left(2 \beta ^2\right)\right]}{16 \frac{  \beta^8}{\sqrt\pi} e^{-2\beta^2}(14 - 8 \beta^2 + \beta^4)} \\
& \hphantom{=}+O(N^{-1})   = 0.964377\dots + O(N^{-1}).
\end{align*}
\end{proof}

\subsection{The slowly oscillating approximation}
\label{sec:slowly-oscillating-approximation}
In this section, we will study the effective dynamics of solutions to \eqref{eq:full-system-for-tildeak 2} which oscillate ``slowly'' with time (i.e.~at a timescale large compared to $L^{-1}$). For such solutions, we expect that their dynamics are not affected by the terms in the equations \eqref{eq:full-system-for-tildeak 2} with coefficients that either decay with $L$ as $L\rightarrow +\infty$ or oscillate with frequency $\gtrsim L^{-1}$; the latter ones are precisely the oscillating terms $e^{i L^{2s+2} ( \tilde{\omega}_k - \tilde{\omega}_{j_1}+\tilde{\omega}_{j_2}-\tilde{\omega}_{j_3})}$ with $| \tilde{\omega}_k - \tilde{\omega}_{j_1}+\tilde{\omega}_{j_2}-\tilde{\omega}_{j_3} | \gtrsim 1$ (note that the only combinations of $k, j_i$ for which the last expression is $0$ have to yield an expression which is a multiple of $\tilde{\omega}_{-1}-2\tilde{\omega}_0 +\tilde{\omega}_1=0$). In particular, if we \emph{formally} drop all such terms from \eqref{eq:full-system-for-tildeak 2} (together with all the terms for which the coefficient $\llangle \bar E_k E_{j_1} \bar E_{j_2} E_{j_3}\rrangle$ vanishes identically), we obtain the following $1^{st}$ order system of ODEs (which we will call the \emph{slowly oscillating approximation} of \eqref{eq:full-system-for-tildeak 2}):
\begin{equation}\label{Slowly oscillating system}
\begin{cases}
     - 2 i {\tilde {\omega}}_{\minusone} \frac{d}{d\tilde t} b^{\mathrm{sl}}_\minusone =  \Big( \displaystyle\sum_{j=-1}^1\llangle |E_{\minusone}|^2 |E_{j}|^2    \rrangle \left( \tilde{\omega}_{\minusone} \tilde {\omega}_j - \tilde {\omega}_\minusone^2 \right) |b^{\mathrm{sl}}_{j}|^2\Big)  b^{\mathrm{sl}}_{\minusone}
+ \Big(\! \llangle E_{-1} \bar E_0^2  E_{1}   \rrangle \left( \tilde {\omega}_{0} \tilde  {\omega}_{1} - \tilde {\omega}_{0}^2 \right) (b^{\mathrm{sl}}_{0})^2\Big)\overline{ b^{\mathrm{sl}}_{1}}, \\[5pt]
      - 2 i {\tilde {\omega}}_{0} \frac{d}{d\tilde t} b^{\mathrm{sl}}_0 =  \Big( \displaystyle\sum_{j=-1}^1 \! \! \llangle |E_0|^2 |E_{j}|^2    \rrangle \! \left(  \tilde{\omega}_0 \tilde {\omega}_j - \tilde {\omega}_0^2 \right) |b^{\mathrm{sl}}_{j}|^2\Big)  b^{\mathrm{sl}}_{0}\! 
+ \! \Big(\! \llangle E_{-1} \bar E_0^2  E_{1}   \rrangle \left( (\tilde {\omega}_{1}+\tilde{\omega}_{\minusone}) \tilde  {\omega}_{0} - \tilde {\omega}_{\minusone}^2-{\omega}_1^2 \right) b^{\mathrm{sl}}_{\minusone} b^{\mathrm{sl}}_{1} \Big)\overline{ b^{\mathrm{sl}}_{0}}, \\[5pt]
      - 2 i {\tilde {\omega}}_{1} \frac{d}{d\tilde t} b^{\mathrm{sl}}_1 =  \Big( \displaystyle\sum_{j=-1}^1\llangle |E_1|^2 |E_{j}|^2    \rrangle \left( \tilde{\omega}_1 \tilde {\omega}_j - \tilde {\omega}_1^2 \right) |b^{\mathrm{sl}}_{j}|^2\Big)  b_{1}^{\mathrm{sl}}
+ \Big(\! \llangle E_{-1} \bar E_0^2  E_{1}   \rrangle \left( \tilde {\omega}_{0} \tilde  {\omega}_{\minusone} - \tilde {\omega}_{0}^2 \right) (b^{\mathrm{sl}}_{0})^2\Big)\overline{ b_{\minusone}^{\mathrm{sl}}}.
\end{cases}
\end{equation}
We will study the solutions to the above system arising from the initial data set
\begin{equation} \label{eq:initial-data-for-ak 3}
        b_j^{\mathrm{sl}}(0) = \begin{cases}
            \epsilon c_\minusone \lambda^{-s} & \text{ if } j=\minusone, \\
              \epsilon  & \text{ if } j=0, \\  
              \epsilon c_1 \lambda^{-s} & \text{ if } j=1
        \end{cases} 
\end{equation}
(i.e.~the same as \eqref{eq:initial-data-for-ak 2}). Note that, in view of the way we performed the renormalization \eqref{Renormalization}, the coefficients of \eqref{Slowly oscillating system} and the initial data set \eqref{eq:initial-data-for-ak 3} satisfy bounds which are \textbf{independent} of $L$ (but depend on the parameters $\epsilon, \lambda, N, s$).

We will establish the following approximation result:

\begin{proposition}\label{prop:Slowly oscillating approximation}
 Let $\tilde T>0$ be such that the initial value problem  \eqref{Slowly oscillating system}--\eqref{eq:initial-data-for-ak 3} has a smooth solution $\big\{ b_{\minusone}^{\mathrm{sl}}(\tilde t),  b_{0}^{\mathrm{sl}}(\tilde t), b_{1}^{\mathrm{sl}}(\tilde t)\big\}$ on the interval $\tilde t \in [0,\tilde T]$. Then, provided $L$ is large enough in terms of the parameters $\tilde T, \lambda, N, s$ and $\| b_j^{\mathrm{sl}}\|_{L^\infty([0,\tilde T])}$, the solution  $\big\{ \tilde b_{\minusone}(\tilde t),  \tilde b_{0}(\tilde t), \tilde b_{1}(\tilde t)\big\}$ of the original system \eqref{eq:full-system-for-tildeak 2} with initial data
\begin{equation}\label{Initial data tilde a k}
\big( \tilde b_j(0), \frac{d}{d\tilde t} \tilde b_j (0)\big) = \big( b_j^{\mathrm{sl}}(0), \frac{d}{d\tilde t} b_j^{\mathrm{sl}}(0) \big) \quad \text{for} \quad j\in \{\minusone, 0, 1\}
\end{equation}
exists on the interval $\tilde t\in [0, \tilde T]$ and satisfies 
\begin{equation}\label{Difference C 1 norm approximation}
\sum_{j\in \{\minusone, 0, 1\}} \big\| \tilde b_j - b^{\mathrm{sl}}_j\|_{L^\infty([0,\tilde T])} \le \f{C}{L^{2s+2}}, \quad \sum_{j\in \{\minusone, 0, 1\}} \big\| \frac{d}{d\tilde t}\big( \tilde b_j -  b^{\mathrm{sl}}_j\big) \big\|_{L^\infty([0,\tilde T])} \le C
\end{equation}
and, for any integer $q \ge 2$,
\begin{equation}\label{Difference higher order approximation}
\sum_{j\in \{\minusone, 0, 1\}} \big\| \frac{d^q}{d\tilde t^q}\big(\tilde b_j - b^{\mathrm{sl}}_j\big)\|_{L^\infty([0,\tilde T])} \le C_q L^{(2s+2)(q-1)}
\end{equation}
for some constants $C, C_q>0$ depending on $\tilde T, \lambda, N, s$, $\max_{j\in \{\minusone, 0, 1\}}\| b_j^{\mathrm{sl}}\|_{L^\infty([0,\tilde T])}$ and $q \in \mathbb N_{\ge 2}$ (but \textbf{not} on $L$).
\end{proposition}

\begin{proof}
Let us adopt the shorthand notation
\[
\tilde B \doteq 
\begin{bmatrix}
\tilde b_\minusone\\
\tilde b_0 \\
\tilde b_1
\end{bmatrix}
\quad \text{and} \quad
B^{\mathrm{sl}} \doteq 
\begin{bmatrix}
b^{\mathrm{sl}}_\minusone\\
b^{\mathrm{sl}}_0 \\
b^{\mathrm{sl}}_1
\end{bmatrix}.
\]
We will also introduce the small parameter
\[
h \doteq L^{-2-2s}.
\]
Note that the slowly oscillating system \eqref{Slowly oscillating system} takes the schematic form
\begin{equation}\label{Slowly oscillating matrix}
-2i \Omega \cdot \frac{d  B^{\mathrm{sl}}}{d\tilde t} = \mathcal G(B^{\mathrm{sl}}),
\end{equation}
where 
\[
\Omega \doteq 
\begin{bmatrix}
\tilde {\omega}_\minusone & 0 & 0\\
0 & \tilde {\omega}_0 & 0\\
0 & 0 & \tilde {\omega}_1
\end{bmatrix}
\]
and
\[
 \mathcal G(B^{\mathrm{sl}}) \doteq 
\begin{bmatrix}
  \Big( \displaystyle\sum_{j=-1}^1\llangle |E_{\minusone}|^2 |E_{j}|^2    \rrangle \left( \tilde{\omega}_{\minusone} \tilde {\omega}_j - \tilde {\omega}_\minusone^2 \right) |b^{\mathrm{sl}}_{j}|^2\Big)  b_{\minusone}^{\mathrm{sl}}
+ \Big(\llangle E_{-1} \bar E_0^2  E_{1}   \rrangle \left( \tilde {\omega}_{0} \tilde  {\omega}_{1} - \tilde {\omega}_{0}^2 \right) (b^{\mathrm{sl}}_{0})^2\Big)\overline{ b_{1}^{\mathrm{sl}}} \\
     \Big( \displaystyle\sum_{j=-1}^1\!\!\! \llangle |E_0|^2 |E_{j}|^2    \rrangle \left( \tilde{\omega}_0 \tilde {\omega}_j - \tilde {\omega}_0^2 \right) |b^{\mathrm{sl}}_{j}|^2\Big)  b_{0}^{\mathrm{sl}}\! 
+ \! \Big(\llangle E_{-1} \bar E_0^2  E_{1}   \rrangle \left( (\tilde {\omega}_{1}+\tilde{\omega}_{\minusone}) \tilde  {\omega}_{0} - \tilde {\omega}_{\minusone}^2-{\omega}_1^2 \right) b^{\mathrm{sl}}_{\minusone} b_{1}^{\mathrm{sl}} \Big)\overline{ b_{0}^{\mathrm{sl}}}, \\
    \Big( \displaystyle\sum_{j=-1}^1\llangle |E_1|^2 |E_{j}|^2    \rrangle \left( \tilde{\omega}_1 \tilde {\omega}_j - \tilde {\omega}_1^2 \right) |b^{\mathrm{sl}}_{j}|^2\Big)  b_{1}^{\mathrm{sl}}
+ \Big(\llangle E_{-1} \bar E_0^2  E_{1}   \rrangle \left( \tilde {\omega}_{0} \tilde  {\omega}_{\minusone} - \tilde {\omega}_{0}^2 \right) (b^{\mathrm{sl}}_{0})^2\Big)\overline{ b_{\minusone}^{\mathrm{sl}}},
\end{bmatrix}
\]
while the full system \eqref{eq:full-system-for-tildeak 2} takes the schematic form
\begin{equation}\label{Renormalized system matrix}
h\frac{d^2  \tilde B}{d\tilde t^2} - 2i \Omega \cdot \frac{d  \tilde B}{d\tilde t} = \mathcal G(\tilde B) + \mathcal O (\tilde t, \tilde B) + h \mathcal G^{(1)}(\tilde t, \tilde B, \frac{d \tilde B}{d\tilde t}) + h^2 \mathcal G^{(2)}(\tilde t, \tilde B, \frac{d \tilde B}{d\tilde t}, \frac{d^2 \tilde B}{d\tilde t^2}),
\end{equation}
where the oscillatory terms are defined by the formula
\[
\mathcal O (\tilde t, \tilde B) 
\doteq
\begin{bmatrix} 
\displaystyle\sum_{\substack{j_1,j_2,j_3 \in \{\minusone, 0, 1 \}:\\ \tilde {\omega}_\minusone - \tilde {\omega}_{j_1} + \tilde {\omega}_{j_2} - \tilde {\omega}_{j_3} \neq 0}} e^{i h^{-1} ( \tilde {\omega}_\minusone - \tilde {\omega}_{j_1} + \tilde {\omega}_{j_2} - \tilde {\omega}_{j_3} )\tilde t} \llangle \bar E_\minusone E_{j_1} \bar E_{j_2} E_{j_3}   \rrangle \left( \tilde {\omega}_{j_1}\tilde  {\omega}_{j_2} - \tilde {\omega}_{j_3}^2 \right)\tilde b_{j_1} \overline{ \tilde b_{j_2}} \tilde b_{j_3} \\[10pt]
\displaystyle\sum_{\substack{j_1,j_2,j_3 \in \{\minusone, 0, 1 \}:\\ \tilde {\omega}_0 - \tilde {\omega}_{j_1} + \tilde {\omega}_{j_2} - \tilde {\omega}_{j_3} \neq 0}} e^{i h^{-1} ( \tilde {\omega}_0 - \tilde {\omega}_{j_1} + \tilde {\omega}_{j_2} - \tilde {\omega}_{j_3} )\tilde t} \llangle \bar E_0 E_{j_1} \bar E_{j_2} E_{j_3}   \rrangle \left( \tilde {\omega}_{j_1}\tilde  {\omega}_{j_2} - \tilde {\omega}_{j_3}^2 \right)\tilde b_{j_1} \overline{ \tilde b_{j_2}} \tilde b_{j_3} \\[10pt]
\displaystyle\sum_{\substack{j_1,j_2,j_3 \in \{\minusone, 0, 1 \}:\\ \tilde {\omega}_1 - \tilde {\omega}_{j_1} + \tilde {\omega}_{j_2} - \tilde {\omega}_{j_3} \neq 0}} e^{i h^{-1} ( \tilde {\omega}_1 - \tilde {\omega}_{j_1} + \tilde {\omega}_{j_2} - \tilde {\omega}_{j_3} )\tilde t} \llangle \bar E_1 E_{j_1} \bar E_{j_2} E_{j_3}   \rrangle \left( \tilde {\omega}_{j_1}\tilde  {\omega}_{j_2} - \tilde {\omega}_{j_3}^2 \right)\tilde b_{j_1} \overline{ \tilde b_{j_2}} \tilde b_{j_3} 
\end{bmatrix},
\]
while the entries of the lower order terms $\mathcal G^{(1)}= \big[ \mathcal (G^{(1)})_k\big]_{k\in \{-1,0,1\}}$ and $\mathcal G^{(2)} = \big[ \mathcal (G^{(2)})_k\big]_{k\in \{-1,0,1\}}$ are given by
\begin{align*}
\big(\mathcal G^{(1)}(\tilde t, \tilde B, \frac{d \tilde B}{d\tilde t})\big)_k \!\!\!\!\!\!
\doteq
\sum_{j_1,j_2,j_3 \in \{\minusone, 0, 1 \}}\!\!\!\!\!\!  & e^{i h^{-1} ( \tilde {\omega}_k - \tilde {\omega}_{j_1} + \tilde {\omega}_{j_2} -  \tilde {\omega}_{j_3} )\tilde t} \llangle \bar E_k E_{j_1} \bar E_{j_2} E_{j_3}   \rrangle \\
&\hphantom{\sum\sum\sum}
 \times \left( - i \tilde  {\omega}_{j_1} \tilde  b_{j_1} \frac{d}{d\tilde t}{\overline{\tilde b_{j_2}}}  \tilde b_{j_3} +  i  \tilde  {\omega}_{j_2}   \frac{d}{d\tilde t}{\tilde b_{j_1} }{\overline{\tilde b_{j_2}}} \tilde  b_{j_3}   - 2i  \tilde  {\omega}_{j_3} b_{j_1} {\overline{ \tilde b_{j_2}}} \frac{d}{d\tilde t}{\tilde{b_{j_3}}}  \right)
\end{align*}
and
\[
\big(\mathcal G^{(2)}(\tilde t, \tilde B, \frac{d \tilde B}{d\tilde t}, \frac{d^2 \tilde B}{d\tilde t^2})\big)_k
\doteq \!\!\! \!\!\!
\sum_{j_1,j_2,j_3 \in \{\minusone, 0, 1 \}}\!\!\!\!\!\! e^{i h^{-1} (  \tilde {\omega}_k -  \tilde {\omega}_{j_1} +  \tilde {\omega}_{j_2} -  \tilde {\omega}_{j_3} )\tilde t} \llangle \bar E_k E_{j_1} \bar E_{j_2} E_{j_3}   \rrangle \left( \frac{d}{d\tilde t}{\tilde b_{j_1}}  \frac{d}{d\tilde t}{\overline{\tilde b_{j_2}}} \tilde b_{j_3}  + \tilde b_{j_1} \overline{\tilde b_{j_2}} \frac{d^2}{d\tilde t^2}{\tilde b_{j_3}}  \right).
\]
If we set
\[
Z \doteq \tilde B - B^{\mathrm{sl}}, 
\]
then $Z$ formally solves the initial value problem
\begin{equation}\label{Renormalized system difference}
\begin{cases}
h\frac{d^2  Z}{d\tilde t^2} - 2i \Omega \cdot \frac{d  Z}{d\tilde t} =- h\frac{d^2  \tilde B^{\mathrm{sl}}}{d\tilde t^2} + \mathcal O(\tilde t ,B^{\mathrm{sl}}+Z) + \mathcal G'(B^{\mathrm{sl}}+Z, B^{\mathrm{sl}}) \cdot Z  \\[5pt]
\hphantom{h\frac{d^2  \tilde Z}{d\tilde t^2} - 2i \Omega \cdot \frac{d  \tilde Z}{d\tilde t} =}
 + h \mathcal G^{(1)}(\tilde t, B^{\mathrm{sl}}+Z, \frac{d (B^{\mathrm{sl}}+Z)}{d\tilde t}) + h^2 \mathcal G^{(2)}(\tilde t, B^{\mathrm{sl}}+Z, \frac{d (B^{\mathrm{sl}}+Z)}{d\tilde t}, \frac{d^2 (B^{\mathrm{sl}}+Z)}{d\tilde t^2}),\\[5pt]
\big( Z(0), \frac{d Z}{d\tilde t}(0)\big) = \big(0,0\big),
\end{cases}
\end{equation}
where
\[
\mathcal G'(A, B) \doteq \int_0^1 [D\mathcal G](s A + (1-s) B) \, ds
\]
(with $[D\mathcal G]$ being the matrix differential of the map $\mathcal G:\mathbb C^3 \rightarrow \mathbb C^3$). We will also make use of the following $L^\infty$-type norm for $Z$:
\begin{equation}
\label{Norm Z}
\mathscr E[Z](\tilde t) \doteq \sup_{\tau \in [0,\tilde \tau]} \Big( \f{1}{h}\big| Z(\tau)\big| + \big| \frac{d Z}{d\tilde t}(\tau)\big| + h \big| \frac{d^2 Z}{d\tilde t^2}(\tau)\big| \Big).
\end{equation}
\begin{remark}
    Note that, formally at least, $\mathscr E[Z](\tilde t)$ is increasing in $\tilde t$.
\end{remark}

The existence of the solution $\tilde B$ on the whole interval $\tilde t \in [0, \tilde T]$ and the bounds \eqref{Difference C 1 norm approximation} and   \eqref{Difference higher order approximation} for $q=2$ will follow simultaneously via the continuity method:
Let $C_{\mathrm{boot}}$ be a (bootstrap) constant that will be fixed later in terms of $\tilde T, \lambda, s$ and $\|B^{\mathrm{sl}}\|_{L^\infty([0,\tilde T])}$ (in particular, $C_{\mathrm{boot}}$ will be independent of $h$) and let $\tilde T_* \in (0, \tilde T]$ be the largest time such that for  $Z$   the following bound holds:
\begin{equation}\label{Bootstrap approximation}
\mathscr E[Z](\tilde T_*) \le C_{\mathrm{boot}}. 
\end{equation}
We will show that, provided $h$ is small enough in terms of $C_{\mathrm{boot}}$, $\tilde T, \lambda, s$ and $\|B^{\mathrm{sl}}\|_{L^\infty([0,\tilde T])}$, the following \emph{improvement} of the bound \eqref{Bootstrap approximation} actually holds:
\begin{equation}\label{Bootstrap approximation improvement}
\mathscr E[Z](\tilde T_*) \le \f12 C_{\mathrm{boot}}.
\end{equation}

\begin{remark}
For the rest of this proof, the constants implicit in the $\lesssim$ notation will be allowed to depend on $\tilde T, \lambda, s$ and $\|B^{\mathrm{sl}}\|_{L^\infty([0,\tilde T])}$, but \textbf{not} on $C_{\mathrm{boot}}$ and $h$. 
    \end{remark}

We first consider the following inhomogeneous linear initial value problem
\begin{equation}\label{Inhomogeneous equation Z}
\begin{cases}
h \frac{d^2 X}{d\tilde t^2} - 2i \Omega \cdot \frac{d X}{d\tilde t} = F,\\[5pt]
\big(X(0), \frac{d X}{d\tilde t}(0) \big) = \big( 0, 0\big).
\end{cases}
\end{equation}
The solution to \eqref{Inhomogeneous equation Z} is given as
\[
X(\tilde t) = -\f{i}{2}\int_0^{\tilde t}\big(e^{2i h^{-1}\Omega\cdot (\tilde t-s)} - \mathbb I\big)\cdot\Omega^{-1}\cdot  F(s) \, ds
\]
and
\[
\frac{ d X}{d\tilde t}(\tilde t) =\f{1}{h}\int_0^{\tilde t} e^{2i h^{-1}\Omega\cdot (\tilde t-s)} \cdot F(s) \, ds.
\]
Therefore, we have the following representation formulas for $Z$ and $\frac{d Z}{d\tilde t}$ for $\tilde t \in [0, \tilde T_*]$:
\begin{equation}\label{Representation formula Z}
Z(\tilde t) = -\f{i}{2} \Big(\mathrm{\text{IA}} + \mathrm{\text{IB}}+\mathrm{\text{IC}}\Big)
\end{equation}
and
\begin{equation}\label{Representation formula d Z}
\frac{dZ}{d\tilde t}(\tilde t) = \f1h \Big(\mathrm{\text{IIA}} + \mathrm{\text{IIB}}+\mathrm{\text{IIC}}\Big),
\end{equation}
where
\begin{align*}
\mathrm{\text{IA}} & \, \,\doteq \int_0^{\tilde t} \big( e^{2i h^{-1}\Omega\cdot (\tilde t-s)} - \mathbb I \big) \cdot \Omega^{-1} \cdot \big\{ - h\frac{d^2  \tilde B^{\mathrm{sl}}}{d\tilde t^2}\big\} \, ds
, \\
\mathrm{\text{IIA}} & \, \, \doteq \int_0^{\tilde t} e^{2i h^{-1}\Omega\cdot (\tilde t-s)} \cdot \big\{ - h\frac{d^2  \tilde B^{\mathrm{sl}}}{d\tilde t^2}\big\} \, ds,
\end{align*}

\begin{align*}
\mathrm{\text{IB}} & \, \, \doteq \int_0^{\tilde t} \big( e^{2i h^{-1}\Omega\cdot (\tilde t-s)} - \mathbb I \big) \cdot \Omega^{-1} \cdot \Big\{\mathcal G'(B^{\mathrm{sl}}+Z, B^{\mathrm{sl}}) \cdot Z + h \mathcal G^{(1)}(s, B^{\mathrm{sl}}+Z, \frac{d (B^{\mathrm{sl}}+Z)}{d\tilde t}) \\
&\hphantom{ \, \, \doteq \int_0^{\tilde t} \big( e^{2i h^{-1}\Omega\cdot (\tilde t-s)} - \mathbb I \big) \cdot \Omega^{-1} \cdot \Big\{\mathcal G}
+ h^2 \mathcal G^{(2)}(s, B^{\mathrm{sl}}+Z,  \frac{d (B^{\mathrm{sl}}+Z)}{d\tilde t}, \frac{d^2 (B^{\mathrm{sl}}+Z)}{d\tilde t^2})   \Big\} \, ds
,\\
\mathrm{\text{IIB}}  & \, \, \doteq \int_0^{\tilde t} e^{2i h^{-1}\Omega\cdot (\tilde t-s)} \cdot  \Big\{ \mathcal G'(B^{\mathrm{sl}}+Z, B^{\mathrm{sl}}) \cdot Z + h \mathcal G^{(1)}(s, B^{\mathrm{sl}}+Z, \frac{d (B^{\mathrm{sl}}+Z)}{d\tilde t})\\ 
&\hphantom{ \, \, \doteq \int_0^{\tilde t} e^{2i h^{-1}\Omega\cdot (\tilde t-s)} \cdot  \Big\{ \mathcal G}
+ h^2 \mathcal G^{(2)}(s, B^{\mathrm{sl}}+Z, \frac{d (B^{\mathrm{sl}}+Z)}{d\tilde t},  \frac{d^2 (B^{\mathrm{sl}}+Z)}{d\tilde t^2})   \Big\} \, ds
\end{align*}
and

\begin{align*}
\mathrm{\text{IC}} & \, \, \doteq  \int_0^{\tilde t} \big( e^{2i h^{-1}\Omega\cdot (\tilde t-s)} - \mathbb I \big) \cdot \Omega^{-1} \cdot  \mathcal O(s,B^{\mathrm{sl}}+Z)   \, ds
,\\
\mathrm{\text{IIC}} & \, \, \doteq  \int_0^{\tilde t} e^{2i h^{-1}\Omega\cdot (\tilde t-s)} \cdot  \mathcal O (s,B^{\mathrm{sl}}+Z)    \, ds.
\end{align*}

We will estimate the terms above as follows:
\begin{enumerate}
\item
Using equation \eqref{Slowly oscillating matrix} to calculate derivatives of $B^{\mathrm{sl}}$ in terms of lower order terms, we can readily show inductively that, for any $q\in \mathbb N$, 
\begin{equation}\label{Bound A sl}
\sup_{\tilde t \in [0, \tilde T]} \big| \frac{d^q B^{\mathrm{sl}}}{d\tilde t^q} (\tilde t) \big| \lesssim_q 1.
\end{equation}
Therefore, we can immediately estimate
\[
| \mathrm{\text{IA}}| + |\mathrm{\text{IIA}}| \lesssim \int_0^{\tilde t} h \Big| \frac{d^2 B^{\mathrm{sl}}}{d\tilde t^2} (s) \Big| \, ds \lesssim h. 
\]

\item
Using \eqref{Bound A sl} and the bootstrap assumption \eqref{Bootstrap approximation} we can estimate 
\[
|B^{\mathrm{sl}}|+|Z| \lesssim 1+C_{\mathrm{boot}} h.
\]
Now, notice that the term $\mathcal G^{(1)}(s, A,B)$ is linear in its last argument and can be expressed as
\[
\mathcal G^{(1)}(s, A,B) = \mathcal G_\flat^{(1)}(s, A) \cdot B,
\]
 while the term $\mathcal G^{(2)}(s, A,B,C)$ is linear in $C$ and quadratic in $B$, taking the form
\begin{equation}\label{Decomposition G 2 term}
\mathcal G^{(2)}(s, A,B,C) = B^\top \cdot \mathcal G^{(2)}_\flat (s, A)\cdot B + \mathcal G^{(2)}_\sharp(s, A) \cdot C,
\end{equation}
with $|\mathcal G_\flat^{(1)}(s, A)| \lesssim |A|^2$, $|\mathcal G_\flat^{(2)}(s, A)| \lesssim |A|$ and $|\mathcal G_\sharp^{(2)}(s, A)| \lesssim |A|^2$.
Therefore, we can readily bound:
\begin{align*}
| \mathrm{\text{IB}}| & + |\mathrm{\text{IIB}}|  \lesssim \int_0^{\tilde t} \Bigg|   \mathcal G'(B^{\mathrm{sl}}+Z, B^{\mathrm{sl}}) \cdot Z  + h \mathcal G^{(1)}(s, B^{\mathrm{sl}}+Z, \frac{d (B^{\mathrm{sl}}+Z)}{d\tilde t}) \\
&\hphantom{\lesssim \int_0^{\tilde t}
\Bigg( } + h^2 \mathcal G^{(2)}(s, B^{\mathrm{sl}}+Z, \frac{d (B^{\mathrm{sl}}+Z)}{d\tilde t}, \frac{d^2 (B^{\mathrm{sl}}+Z)}{d\tilde t^2}) \Bigg| \, ds \\
& \lesssim \int_0^{\tilde t}
\Bigg( \Big|G'(B^{\mathrm{sl}}+Z, B^{\mathrm{sl}})\Big| \cdot \Big| Z \Big| + h \Big| \mathcal G^{(1)}_\flat(s, B^{\mathrm{sl}}+Z) \Big| \cdot \Big|\frac{d (B^{\mathrm{sl}}+Z)}{d\tilde t}\Big|\\
&\hphantom{\lesssim \int_0^{\tilde t}
\Bigg( }
 + h^2  \Big| \mathcal G^{(2)}_\flat(s, B^{\mathrm{sl}}+Z) \Big| \cdot \Big|\frac{d (B^{\mathrm{sl}}+Z)}{d\tilde t}\Big|^2 
+ h^2 \Big| \mathcal G^{(2)}_\sharp(s, B^{\mathrm{sl}}+Z) \Big| \cdot \Big|\frac{d^2 (B^{\mathrm{sl}}+Z)}{d\tilde t^2}\Big|
\Bigg) \, ds
\\
& \lesssim \int_0^{\tilde t}
\Bigg( h (1+C_{\mathrm{boot}} h)^2  \cdot \f{1}{h} |Z(s)| + h (1+C_{\mathrm{boot}} h)^2 \cdot \big(\big| \frac{dZ}{d\tilde t} (s)\big| + 1 \big) \\
&\hphantom{\lesssim \int_0^{\tilde t}
\Bigg( }
+ h^2 (1+C_{\mathrm{boot}} h)^2 C_{\mathrm{boot}} \cdot  \big(\big| \frac{dZ}{d\tilde t} (s)\big| + 1 \big) + h (1+C_{\mathrm{boot}} h)^2  \cdot \big(h\big| \frac{d^2 Z}{d\tilde t^2} (s)\big| + h \big) 
\Bigg) \, ds\\
& \lesssim h (1+C_{\mathrm{boot}} h)^3 + h (1+C_{\mathrm{boot}} h)^2  \int_0^{\tilde t}\mathscr E[Z](s) \, ds\Bigg). 
\end{align*}

\item For the oscillating terms $\mathrm{\text{IC}}$ and $\mathrm{\text{IIC}}$, we can decompose them as
\begin{align}\label{Formulas oscillating terms}
\mathrm{\text{IC}} & = \Omega^{-1} \cdot \Bigg( e^{2ih^{-1} \Omega \tilde t}\int_0^{\tilde t} \mathcal O_1\big(s, B^{\mathrm{sl}}(s)+Z(s)\big) \, ds - \int_0^{\tilde t} \mathcal O_2\big(s, B^{\mathrm{sl}}(s)+Z(s)\big) \, ds\Bigg), \\
\mathrm{\text{IIC}}  & = e^{2ih^{-1} \Omega \tilde t}\int_0^{\tilde t} \mathcal O_1\big(s,B^{\mathrm{sl}}(s)+Z(s)\big) \, ds,   \nonumber 
\end{align}
where, for any $B = [b_{\minusone}, b_0, b_1]:\mathbb R \rightarrow \mathbb C^3$,
\begin{align*}
& \mathcal O_1(s, B) \doteq e^{-2ih^{-1} \Omega s} \mathcal O(s, B)\\ &=
\begin{bmatrix} 
\displaystyle\sum_{\substack{j_1,j_2,j_3 \in \{\minusone, 0, 1 \}:\\ \tilde {\omega}_\minusone -  \tilde {\omega}_{j_1} +  \tilde {\omega}_{j_2} -  \tilde {\omega}_{j_3} \neq 0}} e^{i h^{-1} ( -\tilde {\omega}_\minusone -  \tilde {\omega}_{j_1} +  \tilde {\omega}_{j_2} -  \tilde {\omega}_{j_3} )s } \llangle \bar E_\minusone E_{j_1} \bar E_{j_2} E_{j_3}   \rrangle \left(  \tilde {\omega}_{j_1}\tilde  {\omega}_{j_2} - \tilde {\omega}_{j_3}^2 \right) b_{j_1} \overline{ b_{j_2}} b_{j_3} \\[10pt]
\displaystyle\sum_{\substack{j_1,j_2,j_3 \in \{\minusone, 0, 1 \}:\\ \tilde {\omega}_0 -  \tilde {\omega}_{j_1} +  \tilde {\omega}_{j_2} -  \tilde {\omega}_{j_3} \neq 0}} e^{i h^{-1} ( -\tilde {\omega}_0 - \tilde {\omega}_{j_1} + \tilde {\omega}_{j_2} - \tilde {\omega}_{j_3} )s} \llangle \bar E_0 E_{j_1} \bar E_{j_2} E_{j_3}   \rrangle \left( \tilde {\omega}_{j_1}\tilde  {\omega}_{j_2} - \tilde {\omega}_{j_3}^2 \right) b_{j_1} \overline{ b_{j_2}} b_{j_3} \\[10pt]
\displaystyle\sum_{\substack{j_1,j_2,j_3 \in \{\minusone, 0, 1 \}:\\ \tilde {\omega}_1 - \tilde {\omega}_{j_1} + \tilde {\omega}_{j_2} - \tilde {\omega}_{j_3} \neq 0}} e^{i h^{-1} ( -\tilde {\omega}_1 - \tilde {\omega}_{j_1} + \tilde {\omega}_{j_2} - \tilde {\omega}_{j_3} )s} \llangle \bar E_1 E_{j_1} \bar E_{j_2} E_{j_3}   \rrangle \left( \tilde {\omega}_{j_1}\tilde  {\omega}_{j_2} - \tilde {\omega}_{j_3}^2 \right)b_{j_1} \overline{ b_{j_2}} b_{j_3} 
\end{bmatrix}
\end{align*}
and
\begin{align*} & \mathcal O_2(s,B) \doteq \mathcal O(s, B)  \\ &=
\begin{bmatrix} 
\displaystyle\sum_{\substack{j_1,j_2,j_3 \in \{\minusone, 0, 1 \}:\\ \tilde {\omega}_\minusone - \tilde {\omega}_{j_1} + \tilde {\omega}_{j_2} - \tilde {\omega}_{j_3} \neq 0}} e^{i h^{-1} ( \tilde {\omega}_\minusone - \tilde {\omega}_{j_1} + \tilde {\omega}_{j_2} - \tilde {\omega}_{j_3} )s } \llangle \bar E_\minusone E_{j_1} \bar E_{j_2} E_{j_3}   \rrangle \left( \tilde {\omega}_{j_1}\tilde  {\omega}_{j_2} - \tilde {\omega}_{j_3}^2 \right) b_{j_1} \overline{ b_{j_2}} b_{j_3} \\[10pt]
\displaystyle\sum_{\substack{j_1,j_2,j_3 \in \{\minusone, 0, 1 \}:\\ \tilde {\omega}_0 - \tilde {\omega}_{j_1} + \tilde {\omega}_{j_2} - \tilde {\omega}_{j_3} \neq 0}} e^{i h^{-1} ( \tilde {\omega}_0 - \tilde {\omega}_{j_1} + \tilde {\omega}_{j_2} - \tilde {\omega}_{j_3} )s} \llangle \bar E_0 E_{j_1} \bar E_{j_2} E_{j_3}   \rrangle \left( \tilde {\omega}_{j_1}\tilde  {\omega}_{j_2} - \tilde {\omega}_{j_3}^2 \right) b_{j_1} \overline{ b_{j_2}} b_{j_3} \\[10pt]
\displaystyle\sum_{\substack{j_1,j_2,j_3 \in \{\minusone, 0, 1 \}:\\ \tilde {\omega}_1 - \tilde {\omega}_{j_1} + \tilde {\omega}_{j_2} - \tilde {\omega}_{j_3} \neq 0}} e^{i h^{-1} ( \tilde {\omega}_1 - \tilde {\omega}_{j_1} + \tilde {\omega}_{j_2} - \tilde {\omega}_{j_3} )s} \llangle \bar E_1 E_{j_1} \bar E_{j_2} E_{j_3}   \rrangle \left( \tilde {\omega}_{j_1}\tilde  {\omega}_{j_2} - \tilde {\omega}_{j_3}^2 \right)b_{j_1} \overline{ b_{j_2}} b_{j_3} 
\end{bmatrix}.
\end{align*}

\begin{remark}
    Note that the phase of the exponential terms in the $k$-th component of $\mathcal O_1(B)$ is $i h^{-1} ( -\tilde {\omega}_k - \tilde {\omega}_{j_1} + \tilde {\omega}_{j_2} - \tilde {\omega}_{j_3} )s $, while the phase in the corresponding term for $\mathcal O_2(B)$ is $i h^{-1} ( \tilde {\omega}_k - \tilde {\omega}_{j_1} + \tilde {\omega}_{j_2} - \tilde {\omega}_{j_3} )s $. 
    \end{remark}

In order to proceed, we will make use of the following observation: For any $k, j_1, j_2, j_3 \in \{\minusone, 0, 1\}$ we have 
\[
-\tilde{\omega}_k - \tilde {\omega}_{j_1} + \tilde {\omega}_{j_2} - \tilde {\omega}_{j_3} \neq 0.
\]
As a result, all the frequencies in the exponential factors appearing in the expressions above for $\mathcal O_1(B)(s) $ and $\mathcal O_2(B)(s) $ are non-zero, and, in fact, we can write
\[
\mathcal O_1(s,B) = \sum_{\mu\in I} e^{i h^{-1} \sigma^{(\mu)}_1 s} \mathcal O^{(\mu)}_1 (B)
\]
and
\[
\mathcal O_2(s,B) = \sum_{\mu\in I} e^{i h^{-1} \sigma^{(\mu)}_2 s} \mathcal O^{(\mu)}_2 (B),
\]
where the index set $I$ parametrizes tetrads $\Big\{ k, j_1, j_2, j_3 \in \{\minusone, 0, 1\}: \, \tilde{\omega}_k - \tilde {\omega}_{j_1} + \tilde {\omega}_{j_2} - \tilde {\omega}_{j_3}\neq 0 \Big\}$, the frequencies $\sigma_j^{(\mu)}$ satisfy $|\sigma_j^{(\mu)}|\gtrsim 1$ and the functions $O^{(\mu)}_j (B)$ are cubic polynomials of $B$ with constant coefficients defined in terms of $\llangle E_k E_{j_1} E_{j_2} E_{j_3} \rrangle$ and the $\tilde {\omega}_j$'s. 

We are now in a position to apply a (non-)stationary phase argument: Writing
\[
e^{i h^{-1} \sigma^{(\mu)}_j s} = - i \f{h}{\sigma^{(\mu)}_j } \frac{d}{ds}\big( e^{i h^{-1} \sigma^{(\mu)}_j s} \big)
\]
in the expressions for $\mathcal O_j(s, B^{\mathrm{sl}}+Z)$ in formulas \eqref{Formulas oscillating terms} for $\mathrm{\text{IC}}$--$\mathrm{\text{IIC}}$ and then integrating by parts, we can readily infer that
\begin{align*}
|\mathrm{\text{IC}}| + |\mathrm{\text{IIC}}| & \lesssim h \sup_{\tau \in [0, \tilde t]} \big|B^{\mathrm{sl}}(\tau)+Z(\tau)\big|^3 + h \int_0^{\tilde t}  \big|B^{\mathrm{sl}}+Z\big|^2 |\f{d B^{\mathrm{sl}}}{d\tilde t}+\f{d Z}{d\tilde t}\big|\, ds \\
& \lesssim h (1+ C_{\mathrm{boot}} h)^3 +h (1+ C_{\mathrm{boot}} h)^2\int_0^{\tilde t} \mathscr E[Z](s) \, ds.
\end{align*}
\end{enumerate}

Combining the above bounds for the terms $\mathrm{\text{IA}}$, $\mathrm{\text{IIA}}$, $\mathrm{\text{IB}}$, $\mathrm{\text{IIB}}$, $\mathrm{\text{IC}}$ and $\mathrm{\text{IIC}}$ and returning to \eqref{Representation formula Z}--\eqref{Representation formula d Z}, we infer that
\begin{equation}\label{First bound Z and d Z}
\f1h \big| Z(\tilde t) \big| + \big| \frac{d Z}{d\tilde t} (\tilde t) \big| \lesssim  (1+ C_{\mathrm{boot}} h)^3 + (1+ C_{\mathrm{boot}} h)^2\int_0^{\tilde t} \mathscr E[Z](s) \, ds.
\end{equation}

In order to estimate the second order term $\frac{d^2 Z}{d\tilde t^2}$, we will use equation \eqref{Renormalized system difference} directly: Recall that the last term in the right-hand side of \eqref{Renormalized system difference} is linear in $\frac{d^2(B^{\mathrm{sl}}+Z)}{d\tilde t^2}$, in view of the decomposition \eqref{Decomposition G 2 term}  of the $\mathcal G^{(2)}$ term. Therefore, assuming that $h$ is small enough in terms of $C_{\mathrm{boot}}$, $\|A\|_{L^\infty([0,\tilde T])}$, $\lambda, N$ and $s$ so that the matrix $\mathbb I - h \mathcal G^{(2)}_\sharp (\tilde t, B^{\mathrm{sl}}+Z)$ is invertible, we can rearrange \eqref{Renormalized system difference} as follows:
\begin{align}\label{Second derivative Z expression}
h \frac{d^2 Z}{d\tilde t^2} = \big( \mathbb I - h \mathcal G^{(2)}_\sharp (\tilde t, B^{\mathrm{sl}}+Z) \big)^{-1} \cdot \Bigg( &
2i \Omega \cdot \frac{d  Z}{d\tilde t} - h\frac{d^2  \tilde B^{\mathrm{sl}}}{d\tilde t^2} + \mathcal O(\tilde t ,B^{\mathrm{sl}}+Z) + \mathcal G'(B^{\mathrm{sl}}+Z, B^{\mathrm{sl}}) \cdot Z  \\[5pt]
& 
 + h \mathcal G^{(1)}(\tilde t, B^{\mathrm{sl}}+Z, \frac{d (B^{\mathrm{sl}}+Z)}{d\tilde t})
+h^2 \mathcal G^{(2)}_\sharp (\tilde t, B^{\mathrm{sl}}+Z) \frac{d^2 B^{\mathrm{sl}}}{d\tilde t^2}
  \nonumber\\[5pt]
&
+h^2 \Big(\frac{d(B^{\mathrm{sl}}+Z)}{d\tilde t} \Big)^\top\cdot  \mathcal G^{(2)}_\flat (\tilde t, B^{\mathrm{sl}}+Z) \cdot \Big(\frac{d(B^{\mathrm{sl}}+Z)}{d\tilde t} \Big)   \Bigg)  \nonumber
\end{align}
Using the bounds \eqref{Bootstrap approximation} and \eqref{First bound Z and d Z} for $Z$ and $\frac{d Z}{d\tilde t}$, as well as the bound \eqref{Bound A sl} for $B^{\mathrm{sl}}$, we readily obtain from \eqref{Second derivative Z expression}:
\[
h \big|\frac{d^2 Z}{d\tilde t^2}(\tilde t)\big|  \lesssim  \f{(1+ C_{\mathrm{boot}} h)^3}{1-C_{\mathrm{boot}}h} + \f{(1+ C_{\mathrm{boot}} h)^2}{1-C_{\mathrm{boot}}h}\int_0^{\tilde t} \mathscr E[Z](s) \, ds.
\]
Combining the above bound with \eqref{First bound Z and d Z}, we obtain for any $\tilde t \in [0, \tilde T_*]$:
\begin{equation}\label{Gronwall for closing bootstrap Z}
\mathscr E[Z](\tilde t) \lesssim  \f{(1+ C_{\mathrm{boot}} h)^3}{1-C_{\mathrm{boot}}h} + \f{(1+ C_{\mathrm{boot}} h)^2}{1-C_{\mathrm{boot}}h}\int_0^{\tilde t} \mathscr E[Z](s) \, ds.
\end{equation}
Assuming, now that $h$ is small enough in terms of $C_{\mathrm{boot}}$ so that $C_{\mathrm{boot}} h \ll 1$, an application of Gr\"onwall's lemma yields:
\[
\mathscr E[Z](\tilde T_*) \lesssim 1.
\]
Note that the constant implicit in the $\lesssim$ notation above depends on $\tilde T, \lambda, N, s, \|B^{\mathrm{sl}}\|_{L^\infty([0,\tilde T])}$, but \textbf{not} on $C_{\mathrm{boot}}$, $h$. Therefore, provided  $C_{\mathrm{boot}}$ was chosen large enough in terms of $\tilde T, \lambda, N, s, \|B^{\mathrm{sl}}\|_{L^\infty([0,\tilde T])}$, we infer the improved inequality \eqref{Bootstrap approximation improvement}.

Since we showed that  \eqref{Bootstrap approximation} implies \eqref{Bootstrap approximation improvement}, it follows from a standard continuity argument that $\tilde T_* = \tilde T$ and that \eqref{Bootstrap approximation} holds on the whole of the interval $\tilde t \in [0, \tilde T]$. Thus, we have established the existence of the solution $\tilde B$ for $\tilde t \in [0, \tilde T]$, together with the bounds \eqref{Difference C 1 norm approximation} and   \eqref{Difference higher order approximation} for $q=2$. The higher order bound \eqref{Difference higher order approximation} for $q\ge 3$ follows inductively, by assuming the bounds \eqref{Difference C 1 norm approximation} and   \eqref{Difference higher order approximation} for $\bar q\le q-1$ and then differentiating the expression \eqref{Second derivative Z expression} in order to obtain an expression for $\frac{d^q Z}{d\tilde t^q}$ in terms of the lower order terms $\big\{Z, \ldots, \frac{d^{q-1} Z}{d\tilde t^{q-1}} \big\}$, which can be estimated using the inductive bounds. We will leave the details of this process to the reader.

\end{proof}

\subsection{Analysis of the approximating system}
\label{sec:Analysis of the approximating system}
In this section, we will show that the solution $\big\{b^{\mathrm{sl}}_\minusone, b^{\mathrm{sl}}_0, b^{\mathrm{sl}}_1\big\}$ of the initial value problem \eqref{Slowly oscillating system}--\eqref{eq:initial-data-for-ak 3} satisfies the properties required by \cref{prop:The 3 times 3 system}  (provided $c_{\pm 1}$ are suitably fixed in \eqref{eq:initial-data-for-ak 3}). In particular, we will establish the following result:

\begin{proposition}\label{prop:Growth slowly oscillating system}
Let $C_{\mathrm{amp}}>1$ and $\epsilon\in (0,1]$ be given constants\footnote{Here, we consider $C_{\mathrm{amp}}$ and $\epsilon$ as independently chosen parameters, however we will later only be interested in the case when $C_{\mathrm{amp}}$ is a function of $\epsilon$; see the remark below \cref{prop:The 3 times 3 system}.} and assume that the parameters $N, \lambda, L$ are as in the statement of \cref{prop:The 3 times 3 system} (with $\lambda$, in particular, being large in terms of $\epsilon, s, C_{\mathrm{amp}}$ and with $L$ being large in terms of all the other parameters). Then, for some appropriately chosen $c_{\pm 1}$ satisfying 
\[
|c_\minusone|^2+|c_1|^2=1
\]
(see \eqref{eq:initial-data-for-ak 3}), there exists a $\tilde T = \tilde T(\epsilon, C_{\mathrm{amp}}, \lambda, s) >0$ such that the maximal solution $\{ b^{\mathrm{sl}}_k(\tilde t)\}_{k\in \{\minusone, 0, 1\}}$ of the initial value problem  \eqref{Slowly oscillating system}--\eqref{eq:initial-data-for-ak 3}  is defined on the whole time interval $\tilde t \in [0, \tilde T]$ and satisfies
\begin{equation}\label{Upper bound amplification}
\frac{\sup_{\tilde t \in [0,\tilde T]}\Big( \displaystyle \sum_{j\in \{\minusone, 0, 1\}} \tilde{\omega}_j^{2s} |b^{\mathrm{sl}}_j (\tilde t)|^2 \Big)}{\displaystyle \sum_{j\in \{\minusone, 0, 1\}} \tilde{\omega}_j^{2s} |b^{\mathrm{sl}}_j (0)|^2 } \le 8 C_{\mathrm{amp}}
\end{equation}
and 
\begin{equation}\label{Amplification final time}
\frac{ \displaystyle \sum_{j\in \{\minusone, 0, 1\}} \tilde{\omega}_j^{2s} |b^{\mathrm{sl}}_j (\tilde T)|^2}{\displaystyle \sum_{j\in \{\minusone, 0, 1\}} \tilde{\omega}_j^{2s} |b^{\mathrm{sl}}_j (0)|^2 } \ge \f32 C_{\mathrm{amp}}.
\end{equation}
\end{proposition}

\begin{proof}
First of all, let us note that the choice \eqref{eq:initial-data-for-ak 3} for the initial data $\{b^{\mathrm{sl}}_k(0)\}_{k\in \{\minusone, 0, 1\}}$ and the fact that 
\begin{equation}\label{Useful expressions omega tilde}
\tilde {\omega}_{\minusone} = -\lambda + O\big(\f{N}{L}\big), \quad \tilde {\omega}_0 = 1 + O\big(\f{1}{L}\big), \quad \tilde  {\omega}_1 = \lambda+2+ O\big(\f{N}{L}\big)
\end{equation}
imply that 
\begin{equation}\label{Size initial data slowly oscillating}
|b^{\mathrm{sl}}_\minusone (0)|^2 = \lambda^{-2s}|c_\minusone|^2 \epsilon^2, \quad |b^{\mathrm{sl}}_0(0)|^2 = \epsilon^2, \quad   |b^{\mathrm{sl}}_1 (0)|^2 = \lambda^{-2s} |c_1|^2 \epsilon^2
\end{equation}
and
\[
\sum_{j\in \{\minusone, 0, 1\}} \tilde{\omega}_j^{2s} |b^{\mathrm{sl}}_j (0)|^2 = 3\big( 1+ O\big(\f{N}{L}+\f{1}{\lambda}\big)\big) \epsilon^2
\]
 (recall that $\f{N}{L}+\f{1}{\lambda}\ll 1$ since $L$ is assumed large in terms of the rest of the parameters and $\lambda \gg 1$).

We will show that, assuming that $\lambda$ is large in terms of $C_{\mathrm{amp}}$ and $s$, there exists a time 
\[
\tilde T_* = \tilde T_*(\epsilon, C_{\mathrm{amp}}, \lambda, s) >0
\]
 such that the solution of \eqref{Slowly oscillating system}--\eqref{eq:initial-data-for-ak 3} exists for $\tilde t \in [0, \tilde T_*]$ and (provided $c_\minusone, c_1$ are suitably fixed):
\begin{equation}\label{Growth bound for tilde T star}
\sup_{\tilde t\in [0, \tilde T]} |b^{\mathrm{sl}}_0(\tilde t)|^2 \le 2 \epsilon^2, \qquad
\frac{\sup_{\tilde t\in [0, \tilde T_*]} \big(|b^{\mathrm{sl}}_\minusone(\tilde t)|^2 + |b^{\mathrm{sl}}_1(\tilde t)|^2\big)}{|b^{\mathrm{sl}}_\minusone(0)|^2 + |b^{\mathrm{sl}}_1(0)|^2} \le 7C_{\mathrm{amp}}.
\end{equation}
and
\begin{equation}\label{Amplification bound for tilde T star}
\frac{|b^{\mathrm{sl}}_\minusone(\tilde T_*)|^2 + |b^{\mathrm{sl}}_1(\tilde T_*)|^2}{|b^{\mathrm{sl}}_\minusone(0)|^2 + |b^{\mathrm{sl}}_1(0)|^2} \ge 4 C_{\mathrm{amp}}
\end{equation}
Note that, in view of the expressions \eqref{Useful expressions omega tilde} for the $\tilde\omega_j$'s and the relations \eqref{Size initial data slowly oscillating} for the $b^{\mathrm{sl}}_j(0)$'s,  the bounds \eqref{Growth bound for tilde T star} and \eqref{Amplification bound for tilde T star} imply that
\[ 
\frac{\sup_{\tilde t \in [0,\tilde T_*]}\Big( \displaystyle \sum_{j\in \{\minusone, 0, 1\}} \tilde{\omega}_j^{2s} |b^{\mathrm{sl}}_j (\tilde t)|^2 \Big)}{\displaystyle \sum_{j\in \{\minusone, 0, 1\}} \tilde{\omega}_j^{2s} |b^{\mathrm{sl}}_j (0)|^2 } \le 7C_{\mathrm{amp}} + O(\lambda^{-1})
\]
and
\[
\frac{ \displaystyle \sum_{j\in \{\minusone, 0, 1\}} \tilde{\omega}_j^{2s} |b^{\mathrm{sl}}_j (\tilde T_*)|^2 }{\displaystyle \sum_{j\in \{\minusone, 0, 1\}} \tilde{\omega}_j^{2s} |b^{\mathrm{sl}}_j (0)|^2 } \ge 
2 C_{\mathrm{amp}}+ O(\lambda^{-1}).
\]
Therefore, setting $\tilde T = \tilde T_*$, we infer that \eqref{Upper bound amplification} and \eqref{Amplification final time} hold provided $\lambda$ is large enough in terms of $C_{\mathrm{amp}}$.

\begin{remark} For the rest of the proof, we will treat $\lambda$ as a \emph{large} parameter. We will assume that the constants implicit in the $\lesssim$, $\ll$ and $O(\cdot)$ notations are allowed to depend on $C_{\mathrm{amp}}$ and $s$ but \textbf{not} on $\lambda, L$ and $\epsilon \in (0, 1]$.
\end{remark}

Note that, since $\lambda \gg 1$, the initial data \eqref{Size initial data slowly oscillating} satisfy
\begin{equation}\label{Smallness bound initial data slowly oscillating}
\f{|b^{\mathrm{sl}}_\minusone (0)|}{|b^{\mathrm{sl}}_0(0)|}, \, \f{|b^{\mathrm{sl}}_1 (0)|}{|b^{\mathrm{sl}}_0(0)|} \ll 1,
\end{equation}
while, if $\tilde T_*$ indeed exists such that \eqref{Growth bound for tilde T star} holds, then \eqref{Growth bound for tilde T star} would imply that
\begin{equation}\label{Smallness bound remains slowly oscillating}
\f{|b^{\mathrm{sl}}_\minusone (\tilde t)|}{|b^{\mathrm{sl}}_0(\tilde t)|}, \, \f{|b^{\mathrm{sl}}_1 (\tilde t)|}{|b^{\mathrm{sl}}_0(\tilde t)|} \ll 1 \quad \text{for all } \tilde t \in [0,\tilde T_*].
\end{equation}
It is easy to verify that the solution to the system  \eqref{Slowly oscillating system} arising from the initial data set 
\[
\big\{b^{\mathrm{sl}}_\minusone(0), b^{\mathrm{sl}}_0 (0), b^{\mathrm{sl}}_1(0)  \big\} = \{ 0, B_0, 0\}
\]
for any $B_0\in \mathbb C$ is simply the constant solution
\[
\big\{b^{\mathrm{sl}}_\minusone(\tilde t), b^{\mathrm{sl}}_0 (\tilde t), b^{\mathrm{sl}}_1(\tilde t)  \big\} = \{ 0, B_0, 0\}.
\]
Therefore, the bounds \eqref{Smallness bound initial data slowly oscillating}--\eqref{Smallness bound remains slowly oscillating} suggest that we should analyze the behavior of the \textbf{linearization} of the system \eqref{Slowly oscillating system} around the constant solution $\big\{b^{\mathrm{sl}}_\minusone(\tilde t), b^{\mathrm{sl}}_0 (\tilde t), b^{\mathrm{sl}}_1(\tilde t)  \big\} = \{ 0, B_0, 0\}$ with $B_0=b^{\mathrm{sl}}_0(0)= \epsilon$; this is the system:
\begin{align}\label{eq:Linearized slowly oscillating system}
     - 2 i\frac{d}{d\tilde t}  \ob^{\mathrm{sl}}_\minusone &=  \llangle |E_{\minusone}|^2 |E_0|^2   \rrangle (  {\tilde {\omega}}_0 - {\tilde {\omega}}_\minusone )  |B_0|^2 \ob_\minusone^{\mathrm{sl}} + \llangle E_{\minusone} \bar E_{0}^2 E_1  \rrangle \frac{ {\tilde {\omega}}_0 {\tilde {\omega}}_1  - {\tilde {\omega}}_0^2 }{ {\tilde {\omega}}_{\minusone} } 
  (B_0)^2 \overline{ \ob_1^{\mathrm{sl}}} \\ \label{eq:ode-0}
     - 2 i\frac{d}{d\tilde t}   \ob^{\mathrm{sl}}_0 &=  0\\\label{eq:ode1}
     - 2 i\frac{d}{d\tilde t}   \ob^{\mathrm{sl}}_1 &=  \llangle |E_{1}|^2 |E_0|^2  \rrangle (  {\tilde {\omega}}_0 - {\tilde {\omega}}_1 )  |B_0|^2 \ob_1^{\mathrm{sl}} + \llangle E_{\minusone} \bar E_{0}^2 E_1 \rrangle \frac{ {\tilde {\omega}}_0 {\tilde {\omega}}_\minusone  - {\tilde {\omega}}_0^2 }{ {\tilde {\omega}}_{1} }  (B_0)^2  \overline{ \ob_\minusone^{\mathrm{sl}}}.
\end{align}
We can also express the above system as a $6\times 6$ \textbf{real} linear ODE in matrix form (in terms of the variables $\{ \Re \ob^{\mathrm{sl}}_k, \Im \ob^{\mathrm{sl}}_k\}_{k\in \{\minusone, 0, 1\}}$) as follows: Setting
\begin{equation}\label{Vector for linearized system}
\ob^{\mathrm{sl}} \doteq
\begin{bmatrix}
\Re \big(\ob^{\mathrm{sl}}_\minusone\big) \\ \Im\big(\ob^{\mathrm{sl}}_\minusone\big) \\ \Re \big(\ob^{\mathrm{sl}}_0\big) \\ \Im\big(\ob^{\mathrm{sl}}_0\big) \\ \Re \big(\ob^{\mathrm{sl}}_1\big) \\ \Im\big(\ob^{\mathrm{sl}}_1 \big)
\end{bmatrix} \in \mathbb R^6,
\end{equation}
the system \eqref{eq:Linearized slowly oscillating system} reads
\begin{equation}\label{eq:Linearized slowly oscillating system matrix}
\frac{d }{d\tilde t} \ob^{\mathrm{sl}} = \f{\epsilon^2}2 \mathbb M \cdot \ob^{\mathrm{sl}},
\end{equation}
where (in view of the fact that $B_0 = \epsilon$)
\begin{equation}\label{eq:matrix-M}
\mathbb M \doteq
\left[
\begin{array}{c|c|c}
\mathbb M_{-1, -1} & \bigzero & \mathbb M_{-1, 1} \\
\hline
\stackrel{\hphantom{\circ}}{\bigzero} & \bigzero & \bigzero \\
\hline
\mathbb M_{1, -1} & \stackrel{\hphantom{\circ}}{\bigzero} & \mathbb M_{1,1}
\end{array}
\right]
\end{equation}
with each of the $2\times 2$ blocks $\mathbb M_{i,j}$ above given by the expressions 
\[
\mathbb M_{-1, -1} \doteq
\llangle |E_\minusone|^2 |E_0|^2 \rrangle
\begin{bmatrix}
0 & - (\tilde{\omega}_0-\tilde{\omega}_\minusone)\\[10pt]
\tilde{\omega}_0-\tilde{\omega}_\minusone & 0
\end{bmatrix},
\]

\medskip
\[
\mathbb M_{-1, 1} \doteq
\llangle E_\minusone \bar E_0^2 E_1 \rrangle 
\begin{bmatrix}
0 & 
 \f{\tilde{\omega}_0\tilde{\omega}_1 - \tilde{\omega}_0^2}{\tilde{\omega}_\minusone} 
\\[10pt]
 \f{\tilde{\omega}_0\tilde{\omega}_1 - \tilde{\omega}_0^2}{\tilde{\omega}_\minusone}  & 
  0
\end{bmatrix},
\]

\medskip
\[
\mathbb M_{1, -1} \doteq
\llangle E_\minusone \bar E_0^2 E_1 \rrangle 
\begin{bmatrix}
 0 & 
 \f{\tilde{\omega}_0\tilde{\omega}_\minusone - \tilde{\omega}_0^2}{\tilde{\omega}_1}
\\[10pt]
 \f{\tilde{\omega}_0\tilde{\omega}_\minusone - \tilde{\omega}_0^2}{\tilde{\omega}_1} & 
  0
\end{bmatrix}
\]
\medskip
and
\[
\mathbb M_{1, 1} \doteq
\llangle |E_1|^2 |E_0|^2 \rrangle
\begin{bmatrix}
0 & - (\tilde{\omega}_0-\tilde{\omega}_1) \\[10pt]
\tilde{\omega}_0-\tilde{\omega}_1 & 0
\end{bmatrix}.
\]

Notice that if $\{\ob^{\mathrm{sl}}_{\minusone}, \ob^{al}_0, \ob^{\mathrm{sl}}_1\}$ is a solution to \eqref{eq:Linearized slowly oscillating system} and we set
\[
\ob^{\mathrm{sl}}_\sharp \doteq
 \begin{bmatrix}
\Re \big(\ob^{\mathrm{sl}}_\minusone\big) \\ \Im\big(\ob^{\mathrm{sl}}_\minusone\big) \\ \Re \big(\ob^{\mathrm{sl}}_1\big) \\ \Im\big(\ob^{\mathrm{sl}}_1 \big)
\end{bmatrix}
\]
 (i.e.~$\ob^{\mathrm{sl}}_\sharp$ contains the $\ob^{\mathrm{sl}}_{\pm1}$ components of the vector $\ob^{\mathrm{sl}}$ defined by \eqref{Vector for linearized system}), then we have
\begin{equation}\label{Matrix solution of linearized system}
\ob^{\mathrm{sl}}_0 = \text{const} \quad \text{and} \quad 
\ob^{\mathrm{sl}}_\sharp(\tilde t) = e^{\f{\epsilon^2}2\widetilde{\mathbb M} \tilde t}  \cdot\ob^{\mathrm{sl}}_\sharp(0),
\end{equation}
where
\begin{equation}\label{The reduced M matrix}
\widetilde{\mathbb M} \doteq \left[
\begin{array}{c|c}
\mathbb M_{-1, -1} &  \mathbb M_{-1, 1} \\
\hline
\mathbb M_{1, -1}  & \mathbb M_{1,1}
\end{array}
\right].
\end{equation}

Let us now return to the nonlinear initial value problem \eqref{Slowly oscillating system}--\eqref{eq:initial-data-for-ak 3} for 
\[
b^{\mathrm{sl}} =
\begin{bmatrix}
\Re \big(b^{\mathrm{sl}}_\minusone\big) \\ \Im\big(b^{\mathrm{sl}}_\minusone\big) \\ \Re \big(b^{\mathrm{sl}}_0\big) \\ \Im\big(b^{\mathrm{sl}}_0\big) \\ \Re \big(b^{\mathrm{sl}}_1\big) \\ \Im\big(b^{\mathrm{sl}}_1 \big)
\end{bmatrix}
\]
and let us use the following ansatz for $b^{\mathrm{sl}}(\tilde t)$:
\begin{equation}\label{Ansatz a sl}
b^{\mathrm{sl}}(\tilde t) = \begin{bmatrix}
0 \\ 0 \\ \epsilon  \\ 0 \\ 0\\0
\end{bmatrix} +\f{1}{\lambda^s}\ob^{\mathrm{sl}}(\tilde t) + \f{1}{\lambda^{2s}} b^{\mathrm{er}}(\tilde t),
\end{equation}
where $\ob^{\mathrm{sl}}(\tilde t)$ solves the linearized initial value problem:
\begin{equation}\label{Linear initial value problem ring a}
\begin{cases}
\frac{d }{d\tilde t} \ob^{\mathrm{sl}} = \f{\epsilon^2}2 \mathbb M \cdot \ob^{\mathrm{sl}}, \\[5pt]
\ob^{\mathrm{sl}}(0) =
\begin{bmatrix}
\epsilon \Re(c_\minusone) & \epsilon \Im(c_\minusone) & 0 & 0 & \epsilon \Re(c_1) & \epsilon \Im(c_1)
\end{bmatrix}^\top.
\end{cases}
\end{equation}
Note that \eqref{Slowly oscillating system}--\eqref{eq:initial-data-for-ak 3} then imply that the ``error'' term $b^{\mathrm{er}}$ solves the inhomogeneous, nonlinear initial value problem
\begin{equation}\label{Initial value problem b}
\begin{cases}
\frac{d }{d\tilde t} b^{\mathrm{er}} = \f{\epsilon^2}2 \mathbb M \cdot b^{\mathrm{er}} + \mathcal Q(\ob^{\mathrm{sl}},\f{1}{\lambda^s}b^{\mathrm{er}}),
\\[5pt]
b^{\mathrm{er}}(0)=0,
\end{cases}
\end{equation}
where the nonlinear term $\mathcal Q(\ob^{\mathrm{sl}},\f{1}{\lambda^s}b^{\mathrm{er}})$ satisfies the bound
\begin{equation}\label{Bound Q}
|\mathcal Q(A,B)| \lesssim \Big(\max_{j_k\in \{-1,0,1\}} \f{|\tilde{\omega}_{j_1} \tilde{\omega}_{j_2}|}{|\tilde{\omega}_{j_3}|}\big|\llangle \bar E_{j_4} E_{j_5} \bar E_{j_6} E_{j_7} \rrangle \big|\Big) \Big( \epsilon+|A|+|B|\Big)^3 \lesssim \lambda^2 \Big( \epsilon+|A|+|B|\Big)^3.
\end{equation}
For any $\tau>0$ such that the solution of \eqref{Slowly oscillating system}--\eqref{eq:initial-data-for-ak 3} exists on $0\le \tilde t\le \tau$, we can immediately estimate from \eqref{Initial value problem b} and \eqref{Bound Q}:
\begin{equation}\label{A priori estimate b}
\sup_{0\le \tilde t \le \tau} |b^{\mathrm{er}}(\tilde t)| \lesssim \exp\Big( \epsilon^2 \max_{\sigma_j \in \mathrm{\text{Spec}} \mathbb M} |\Re (\sigma_j) | \tau\Big)\cdot \lambda^2 \int_0^\tau \Big(\epsilon+|\ob^{\mathrm{sl}}(s)|+\f{1}{\lambda^s}|b^{\mathrm{er}}(s)| \Big)^3 \, ds.
\end{equation}

In view of \cref{lem:The matrix spectrum} below, there exists a $\sigma \in \mathrm{\text{Spec}} \widetilde{\mathbb M}$ with $\Re(\sigma)\gtrsim 1$. Let us chose $c_{\pm 1}$ so that
\[
c = \begin{bmatrix}
\Re(c_\minusone) & \Im(c_\minusone) & \Re(c_1) &\Im(c_1)
\end{bmatrix}^\top
\]
is the corresponding eigenvector with $|c|^2=1$.
Then,  in view of the explicit expression \eqref{Matrix solution of linearized system} for the solution to the linearized initial value problem \eqref{Linear initial value problem ring a}, we infer that there exists a 
\begin{equation}\label{T dagger}
T_\dagger \sim \log(C_{\mathrm{amp}})+1
\end{equation}
with the constant implicit in the $\sim$ notation here depending \emph{only} on the lower bound for $\{|\Re(\sigma)|\}_{\sigma \in \mathrm{\text{Spec}} \widetilde{\mathbb M}}$ in \cref{lem:The matrix spectrum} (and thus is independent of $C_{\mathrm{amp}}$, $\lambda$, $L$), such that, for 
\begin{equation}\label{Definition T tilde star}
\tilde T_* = \frac{1}{\epsilon^2}T_\dagger,
\end{equation}
we have
\begin{equation}\label{Growth bound for tilde T star linearized}
\ob^{\mathrm{sl}}_0 (\tilde t) = 0 \, \text{ for all }\tilde t\in [0, \tilde T_*], \qquad
 \frac{\sup_{\tilde t\in [0, \tilde T_*]} |\ob^{\mathrm{sl}}_\sharp(\tilde t)|^2}{|\ob^{\mathrm{sl}}_\sharp(0)|^2} \le 6 C_{\mathrm{amp}}
\end{equation}
and
\begin{equation}\label{Amplification bound for tilde T star linearized}
 \frac{ |\ob^{\mathrm{sl}}_\sharp(\tilde T_*)|^2}{|\ob^{\mathrm{sl}}_\sharp(0)|^2} \ge 5 C_{\mathrm{amp}}.
\end{equation}

We will now show that the error term $b^{\mathrm{er}}$ remains under control for $\tilde t\in [0, \tilde T_*]$. Note that \cref{lem:The matrix spectrum} below implies that
\begin{equation}\label{Spectrum upper bound M}
\max_{\sigma_j \in \mathrm{\text{Spec}} \mathbb M} |\Re (\sigma_j) | \lesssim 1.
\end{equation}
Therefore, the a~priori estimate \eqref{A priori estimate b} for $b^{\mathrm{er}}$, the definition \eqref{Definition T tilde star} of $\tilde T_*$ and the bound \eqref{Growth bound for tilde T star linearized} for $\ob^{\mathrm{sl}}$ (which implies that $|\ob^{\mathrm{sl}}|\lesssim C_{\mathrm{amp}}^{\f12}\epsilon$) yield, for any $\tau \in [0, \tilde T_*]$ such that  the solution of \eqref{Slowly oscillating system}--\eqref{eq:initial-data-for-ak 3} exists on $0\le \tilde t\le \tau$:
\begin{equation}\label{A priori estimate b again}
\sup_{0\le \tilde t \le \tau} |b^{\mathrm{er}}(\tilde t)| \lesssim \exp\Big( C (1+\log(C_{\mathrm{amp}}))\Big)\cdot \lambda^2 \int_0^\tau \Big(C_{\mathrm{amp}}^{\f32} \epsilon^3+\f{1}{\lambda^{3s}}|b^{\mathrm{er}}(s)|^3 \Big) \, ds
\end{equation}
for some constant $C>0$ depending only on the implicit constant in the $\sim$ notation in \eqref{T dagger} (which, as we commented before, is independent of $C_{\mathrm{amp}}$, $\lambda$, $L$). A straightforward continuity argument now implies (in view of the fact that $\lambda^{2-s} \ll 1$ since $s\ge 4$) that the solution $b^{\mathrm{er}}(\tilde t)$ of \eqref{Initial value problem b} exists on the whole interval $\tilde t\in [0, \tilde T_*]$ and satisfies:
\begin{equation}\label{Upper bound b}
\sup_{\tilde t\in [0, \tilde T_*]} |b^{\mathrm{er}}(\tilde t)|\lesssim \exp\big((C+\f32)(\log(C_{\mathrm{amp}})+1)\big)\lambda^2 \epsilon^3 \tilde T_* \ll \lambda^{5/2}
\end{equation}
(provided $\lambda$ is large enough in terms of $C_{\mathrm{amp}}$). Combining \eqref{Growth bound for tilde T star linearized}, \eqref{Amplification bound for tilde T star linearized} and \eqref{Upper bound b} and returning to the expression \eqref{Ansatz a sl} for $b^{\mathrm{sl}}$, we obtain that
\[
|b^{\mathrm{sl}}_0(\tilde t)|^2 = \epsilon^2+O(\f{1}{\lambda^{s-5/2}})\quad \text{for all }\tilde t\in [0, \tilde T_*], \qquad
 \frac{\sup_{\tilde t\in [0, \tilde T_*]} \big(|b^{\mathrm{sl}}_\minusone(\tilde t)|^2 + |b^{\mathrm{sl}}_1(\tilde t)|^2\big)}{|b^{\mathrm{sl}}_\minusone(0)|^2 + |b^{\mathrm{sl}}_1(0)|^2} \le 6C_{\mathrm{amp}}+O(\f{1}{\lambda^{s-5/2}})
\]
and
\[
 \frac{ \big(|b^{\mathrm{sl}}_\minusone(\tilde T_*)|^2 + |b^{\mathrm{sl}}_1(\tilde T_*)|^2\big)}{|b^{\mathrm{sl}}_\minusone(0)|^2 + |b^{\mathrm{sl}}_1(0)|^2} \ge 5 C_{\mathrm{amp}} +O(\f{1}{\lambda^{s-5/2}}),
\]
which imply \eqref{Growth bound for tilde T star} and \eqref{Amplification bound for tilde T star} if $\lambda$ is assumed to be large enough in terms of $C_{\mathrm{amp}}$ and $s$ (recall that $s\ge 4$). Thus, the proof of \cref{prop:Growth slowly oscillating system} is complete.

\end{proof}

\begin{lemma}\label{lem:The matrix spectrum}
Let $\widetilde{\mathbb M}$ be the $4\times 4$ real matrix \eqref{The reduced M matrix}. The spectrum of  $\widetilde{\mathbb M}$ consists of the two pairs of eigenvalues $\pm \sigma_1 $ and $\pm \sigma_2 $, with non-zero real and imaginary parts satisfying
\begin{align}\label{Expression four eigenvalues}
\Re(\sigma_1) = \Re(\sigma_2) & \sim 1  \\
\Im(\sigma_1) = -\Im(\sigma_2) & \sim \lambda,  \nonumber  
\end{align}
where the constants implicit in the notation $\sim$ above are independent of $N, \lambda, L$.
\end{lemma}

\begin{proof}
In this proof, we will adopt the convention that constants implicit in the notations $\lesssim$, $O(\cdot)$ etc. might depend on $s$ but are \textbf{independent} of $\lambda, N, L$. 

The spectrum of $\widetilde{\mathbb M}$ consists of the roots of the characteristic equation:
\begin{align}\nonumber
0 = \det \big( \widetilde{\mathbb M} - \sigma \mathbb I_{4\times 4} \big) 
& = \det \left(
\begin{array}{c|c}
\mathbb M_{-1, -1} - \sigma \mathbb I_{2\times 2} &  \mathbb M_{-1, 1} \\
\hline
\mathbb M_{1, -1}  & \mathbb M_{1,1}- \sigma \mathbb I_{2\times 2}
\end{array}
\right)
\\ & = \det \mathbb M_{1, -1} \cdot \det\Big(\mathbb M_{-1, 1} - \big(\mathbb M_{-1, -1} - \sigma \mathbb I_{2\times 2}\big) \mathbb M_{1, -1}^{-1} \big(\mathbb M_{1,1}- \sigma \mathbb I_{2\times 2}\big)  \Big).\label{Characteristic equation}
\end{align}
Defining the real scalars
\begin{align*}
\mathfrak A & = \llangle |E_\minusone|^2 |E_0|^2 \rrangle (\tilde{\omega}_0-\tilde{\omega}_\minusone), \\
\mathfrak B & = \llangle E_\minusone \bar E_0^2 E_1 \rrangle \f{\tilde{\omega}_0\tilde{\omega}_1 - \tilde{\omega}_0^2}{\tilde{\omega}_\minusone}, \\
\mathfrak C & =  \llangle E_\minusone \bar E_0^2 E_1 \rrangle \f{\tilde{\omega}_0\tilde{\omega}_\minusone - \tilde{\omega}_0^2}{\tilde{\omega}_1},\\
\mathfrak D & = \llangle |E_1|^2 |E_0|^2 \rrangle (\tilde{\omega}_0-\tilde{\omega}_1),
\end{align*}
we compute that
\[
\mathbb M_{-1, 1} - \big(\mathbb M_{-1, -1} - \sigma \mathbb I_{2\times 2}\big) \mathbb M_{1, -1}^{-1} \big(\mathbb M_{1,1}- \sigma \mathbb I_{2\times 2}\big) 
= 
\f{1}{\mathfrak C}
\begin{bmatrix}
- (\mathfrak A -\mathfrak D) \sigma &
-\sigma^2+ \mathfrak C \mathfrak B - \mathfrak A \mathfrak D\\[10pt]
-\sigma^2+ \mathfrak C \mathfrak B - \mathfrak A \mathfrak D &
  (\mathfrak A -\mathfrak D) \sigma
\end{bmatrix}
\]
and, therefore, the characteristic equation \eqref{Characteristic equation} becomes
\[
\sigma^4 + \big(\mathfrak A^2  + \mathfrak D^2  - 2\mathfrak B \mathfrak C\big)\sigma^2 + (\mathfrak A \mathfrak D - \mathfrak B \mathfrak C)^2  = 0.
\]
with roots
\begin{equation}\label{Roots characteristic polynomial}
\sigma^2 = - \frac 12 \big(\mathfrak A^2  + \mathfrak D^2  - 2\mathfrak B \mathfrak C\big) \pm \f12 (\mathfrak A - \mathfrak D) \sqrt{(\mathfrak A+\mathfrak D)^2 -4\mathfrak B \mathfrak C}.
\end{equation}
Since
\[
\tilde{\omega}_\minusone = -\lambda +O(\f{N}{L}), \quad\tilde{\omega}_0 = 1 + O(\f{1}{L}), \quad \tilde{\omega}_1 = \lambda + 2 + O(\f{N}{L})
\]
and
\[
\llangle E_{j_1} E_{j_2} E_{j_3} E_{j_4} \rrangle \lesssim 1 \quad \text{ for }\,  j_1, j_2, j_3, j_4 \in \{\minusone, 0, 1\},
\]
we can readily compute (assuming that $L$ is large in terms of $\lambda, N, s$) that
\begin{align*}
\mathfrak A & = \llangle |E_\minusone|^2 |E_0|^2 \rrangle (\lambda+1) + O(\f{N}{L}), \\
\mathfrak B & = - \llangle E_\minusone \bar E_0^2 E_1 \rrangle \f{\lambda +1}{\lambda} + O(\f{N}{L}), \\
\mathfrak C & =  - \llangle E_\minusone \bar E_0^2 E_1 \rrangle \f{\lambda +1}{\lambda+2}+ O(\f{N}{L}),\\
\mathfrak D & = - \llangle |E_1|^2 |E_0|^2 \rrangle (\lambda+1)+ O(\f{N}{L}),
\end{align*}
and, therefore,
\[
\mathfrak A^2  + \mathfrak D^2  - 2\mathfrak B \mathfrak C = \big( \llangle |E_\minusone|^2 |E_0|^2\rrangle^2 + \llangle |E_1|^2 |E_0|^2\rrangle^2 \big)(\lambda+1)^2 +O(1),
\]
\[
\mathfrak A - \mathfrak D = \big( \llangle |E_\minusone|^2 |E_0|^2\rrangle + \llangle |E_1|^2 |E_0|^2\rrangle\big) \lambda +O(1)
\]
and
\begin{equation}\label{Discriminant expression}
(\mathfrak A+\mathfrak D)^2 -4\mathfrak B \mathfrak C = \Big((\lambda+1) \big(\llangle |E_\minusone|^2 |E_0|^2\rrangle - \llangle |E_1|^2 |E_0|^2\rrangle \big) \Big)^2 - 4 \llangle E_\minusone \bar E_0^2 E_1 \rrangle^2 + O(\f{1}{\lambda}).
\end{equation}

Let us set
\[
\mathfrak E \doteq \llangle |E_\minusone|^2 |E_0|^2 \rrangle \sim 1
\]
and applying   \cref{prop:Estimates for spectral projections} (in particular, the estimate \eqref{Crucial spectral estimate}),\footnote{Recall that   \cref{prop:Estimates for spectral projections} also implies that $\llangle |E_1|^2 |E_0|^2 \rrangle = \mathfrak E +O(\lambda^{-1})$ and $\llangle E_\minusone \bar E_0^2 E_1 \rrangle = \mathfrak E+O(\lambda^{-1})$.} we readily obtain that
\[
- 0.32\mathfrak E^2 \le (\mathfrak A+\mathfrak D)^2 -4\mathfrak B \mathfrak C \le -0.23 \mathfrak E^2,
\]
i.e.~the discriminant \eqref{Discriminant expression} is strictly \emph{negative} (and hence the imaginary part of $\sigma^2$ is non-zero). In particular, returning to the expression \eqref{Roots characteristic polynomial} for $\sigma^2$, we obtain (provided $\lambda\gg 1$) that
\[
\Re\{\sigma^2\}  = -\lambda^2 \mathfrak E^2 + O(\lambda)
\]
and
\[
0.48 \lambda \mathfrak E^2 \le |\Im\{\sigma^2\}| \le  0.57\lambda \mathfrak E^2,
\]
from which \eqref{Expression four eigenvalues}  readily follows.
\end{proof}

We are now in the position of giving the proof of \cref{prop:The 3 times 3 system}.

\begin{proof}[Proof of \cref{prop:The 3 times 3 system}]
 The statement of \cref{prop:The 3 times 3 system} follows as a corollary of \cref{prop:Slowly oscillating approximation,prop:Growth slowly oscillating system}:
Let $C_{\mathrm{amp}}$ and $\epsilon$ be as in the statement of   \cref{prop:The 3 times 3 system} and let $c_{\pm 1} \in \mathbb C$ and $\tilde T$ be as in the statement of \cref{prop:Growth slowly oscillating system}. Let  $\{ b^{\mathrm{sl}}_k(\tilde t)\}_{k\in \{\minusone, 0, 1\}}$ be the solution of the initial value problem \eqref{Slowly oscillating system}--\eqref{eq:initial-data-for-ak 3} for $\tilde t\in [0, \tilde T]$ (it exists, according to \cref{prop:Growth slowly oscillating system}). Then, provided $L$ is large enough in terms of $C_{\mathrm{amp}}, \epsilon, \lambda, N, s$ (recall that $\tilde T$ and $\|b^{\mathrm{sl}}\|_{L^\infty([0, \tilde T])}$ also depend on those parameters, according to the statement of \cref{prop:Growth slowly oscillating system}), \cref{prop:Slowly oscillating approximation} implies that the solution $\{ \tilde b_k(\tilde t)\}_{k\in \{\minusone, 0, 1\}}$ to \eqref{eq:full-system-for-tildeak 2} with initial data 
\begin{equation}\label{Initial data for dominant modes from slowly oscillating}
\big( \tilde b_j(0), \frac{d}{d\tilde t} \tilde b_j (0)\big) = \big( b_j^{\mathrm{sl}}(0), \frac{d}{d\tilde t} b_j^{\mathrm{sl}}(0) \big) \quad \text{for} \quad j\in \{\minusone, 0, 1\}
\end{equation}
exists on the whole interval $\tilde t \in [0, \tilde T]$ and satisfies \eqref{Difference C 1 norm approximation}--\eqref{Difference higher order approximation}. 

In view of the fact that the $b^{\mathrm{sl}}_k$'s satisfy the system  \eqref{Slowly oscillating system}, we can readily estimate at any time $\tilde t\in [0, \tilde T]$:
\begin{equation}\label{Uniform bound a sl low}
\sum_{j\in \{\minusone, 0, 1\}}\Big|\frac{d}{d\tilde t} b^{\mathrm{sl}}_j(\tilde t) \Big| \lesssim \sum_{j\in \{\minusone, 0, 1\}} | b^{\mathrm{sl}}_j (\tilde t) |^3,
\end{equation}
where the constant implicit in the $\lesssim$ notation above depends on $\lambda, N, s$ but \textbf{not} on $C_{\mathrm{amp}}, \epsilon, L$. Thus, differentiating \eqref{Slowly oscillating system} and arguing inductively  on $q\in \mathbb N$, we obtain:
\begin{equation}\label{Uniform bound a sl higher order}
\sum_{j\in \{\minusone, 0, 1\}}\Big|\frac{d^q}{d\tilde t^q} b^{\mathrm{sl}}_j(\tilde t) \Big| \le C_q,
\end{equation}
where the constant $C_q$ depends on $q, \|b^{\mathrm{sl}}\|_{L^\infty([0, \tilde T])}\lambda, N, s$ (and, thus, $C_q = C_q(q,C_{\mathrm{amp}}, \epsilon, \lambda, N, s)$).

The initial bound \eqref{Initial derivative tilde a bound} for $\Big|\frac{d}{d\tilde t} \tilde b_j(0) \Big| = \Big|\frac{d}{d\tilde t} b^{\mathrm{sl}}_j(0) \Big|$ now follows from \eqref{Uniform bound a sl low} applied for $\tilde t=0$. The bounds \eqref{Bound C 1 norm tilde a}--\eqref{Bound C q norm tilde a} follow by combining the approximation estimates \eqref{Difference C 1 norm approximation}--\eqref{Difference higher order approximation} with \eqref{Uniform bound a sl higher order}. Finally, the amplification bound \eqref{Amplification lower bound renormalized system} follows from \eqref{Difference C 1 norm approximation} and  \eqref{Amplification final time}.  
\end{proof}

\subsection{Proof of \texorpdfstring{\cref{cor:The 3 times 3 system}}{Corollary 5.4}}\label{sec:Proof of corollary dominant modes}
In this section, we will establish \cref{cor:The 3 times 3 system} using the estimates provided by \cref{prop:The 3 times 3 system}.

\begin{proof}[Proof of \cref{cor:The 3 times 3 system}] We will assume that the constants implicit in the $\lesssim$, $\sim$ and $O(\cdot)$ notation are allowed to depend on $M_0$ and $s$, but are independent of the parameters $C_{\mathrm{amp}}, \epsilon, N, \lambda$ and $L$.

It can be readily verified that the solutions $\{\tilde\phi_j\}_{j\in \mathcal K_{\mathrm{D}}}$ of the boundary value problem  \eqref{Boundary value problem tilde phi} with initial data \eqref{Initial data dominant modes final} take the form
\begin{equation}\label{Ansatz again dominant modes}
r(y) \tilde\phi_j(t,y,\theta, \varphi) =  b_j(t) e^{-i \varepsilon_j \omega_j t} E_j(y,\theta,\varphi)
\end{equation}
with
\[
b_j(t) \doteq L^{-s} \tilde b_j\big(L^{-1-2s} t\big),
\]
where the renormalized amplitudes $\{\tilde b_j(\tilde t)\}_{j\in \mathcal K_{\mathrm{D}}}$ are the ones provided by \cref{prop:The 3 times 3 system} (and which satisfy \eqref{Initial data b tilde dominant}--\eqref{Amplification lower bound renormalized system}). 

We will show that the spatial eigenfunctions $\{E_j\}_{j\in \mathcal K_{\mathrm{D}}}$ satisfy for any $\bar s, p\in \mathbb N$ with $\bar s, p\le s$:
\begin{equation}\label{H s norm dominant eigenfunctions}
\Big\|\big|\f{1}{\sin^p \theta}\chi\cdot \partial_\varphi^{\bar s} E_j\big| -| \omega_j^{\bar s } E_j| \Big\|_{L^2(\sin\theta dy d\theta d\varphi)} \lesssim \omega_j^{\bar s -1+2\delta_0}
\end{equation}
and, for any $s_y, s_\theta, s_\varphi, p\in \mathbb N$ with $\bar s \doteq s_y+s_\theta+s_\varphi\le s$, $p\le s$ and $s_y+s_\theta\ge 1$:
\begin{equation}\label{H s norm subdominant eigenfunctions}
\Big\|\f{1}{\sin^p \theta} \partial_y^{s_y} \partial_{\theta}^{s_\theta} \partial_\varphi^{s_\varphi} E_j  \Big\|_{L^2(\sin\theta dy d\theta d\varphi)} \lesssim \omega_j^{\bar s -\f12+2\delta_0}.
\end{equation}
Assuming, for a moment, that \eqref{H s norm dominant eigenfunctions}--\eqref{H s norm subdominant eigenfunctions} hold, it then readily follows (using the fact that the $E_j$'s are orthonormal in $L^2(\sin\theta dy d\theta d\varphi)$ and $\omega_j \gg 1$) that, for any coefficients $\beta_j \in \mathbb C$, we have:
\[
\Big\| \sum_{j\in \mathcal K_{\mathrm{D}}} \sum_{s_2+s_3=0}^s \beta_j \partial_y^{s_2} \nabla_{\mathbb S^2}^{s_3} E_j \Big\|_{L^2(\sin\theta dy d\theta d\varphi)} \sim \Big(\sum_{j\in \mathcal K_{\mathrm{D}}} \beta_j^2  \omega_j^{2s}\Big)^{\f12}.
\]
Therefore, in view of the expression \eqref{Ansatz again dominant modes} for $r\tilde \phi_j$ and the fact that $\omega_j = \ell_j +O(N)$ for $j\in \mathcal K_{\mathrm{D}}$ (which is a consequence of  \eqref{Approximation estimate frequency}), the bound \eqref{Smallness initial norm dominant} follows immediately from the bounds \eqref{eq:initial-data-for-ak 2}--\eqref{Initial derivative tilde a bound} for the coefficients $\big(\tilde B^{(0)}_j, \tilde B^{(1)}_j\big)$ and the estimates \eqref{Bound C 1 norm tilde a}--\eqref{Bound C q norm tilde a} for higher order derivatives of the slowly oscillating coefficients $\tilde b_j(\tilde t)$, while the bound \eqref{Largeness final norm dominant} follows readily from the estimate \eqref{Amplification lower bound renormalized system} for $\tilde b_j(\tilde T)$ and the higher order estimates  \eqref{Bound C 1 norm tilde a}--\eqref{Bound C q norm tilde a}.

In order to complete the proof of \cref{cor:The 3 times 3 system}, it remains to establish \eqref{H s norm dominant eigenfunctions} and \eqref{H s norm subdominant eigenfunctions}. Recall that, for $j = (n_j, \ell_j, m_j) \in \mathcal K_{\mathrm{D}}$, the spatial eigenfunction $E_j$ takes the form:
\[
E_j(y, \theta, \varphi) = R_{n_j,\ell_j}(y) Y_{\ell_j, \varepsilon_j \ell_j}(\theta, \varphi),
\]
where $\varepsilon_j \in \{-1,1\}$ and the parameters $n_j$ and $\ell_j$ are fixed in terms of $N,\lambda, L$ by \cref{def:Dominant mode parameters}. Recall also that the spherical harmonic $Y_{\ell_j,\pm\ell_j}$ takes the form
\begin{equation}\label{Spherical harmonic m equals l}
Y_{\ell_j,\pm\ell_j}(\theta, \varphi)= \f{\gamma_{\ell_j}}{\sqrt{2\pi}} (\sin\theta)^{\ell_j}  e^{\pm i\ell_j \varphi},
\end{equation}
 where
    \[
    \gamma_\ell \doteq \left( \frac{ \Gamma(  \ell+\frac 32)}{\sqrt\pi \Gamma(\ell + 1)})\right)^{\frac 12 }.
    \]
  In particular, we have for any $\bar s\in \mathbb N$:
  \begin{equation}\label{D phi derivative E j}
 \partial^{\bar s}_{\varphi} E_j = (i \varepsilon_j \ell_j)^{\bar s} E_j.
  \end{equation}  
Moreover, the following estimates hold:
\begin{itemize}
\item As a consequence of \cref{lem:eigenfunctions-estimate-m>0} for the functions $R_{n_j,\ell_j}$ and \cref{lem: WKB asymptotics Airy} of the Appendix for the asymptotics of the normalized Hermite functions $e_n$ (and using the fact that $R_{n_j,\ell_j}$ satisfies the ODE  \eqref{Radial boundary value problem} in order to calculate higher order derivatives of $R_{n_j,\ell_j}$ in terms of lower order terms), we can readily estimate for any $j\in \mathcal K_{\mathrm{D}}$ and any $s_y \in \mathbb N$ with $s_y \le s$: 
\begin{equation}\label{Dominant radial eigenfunction L infty again}
\Big|\f{d^{s_y}}{dy^{s_y}} R_{n_j, \ell_j}\Big|\lesssim \ell_j^{\f14+\f{s_y}2} n_j^{\f{s_y}2}
\end{equation}
and
\begin{equation}\label{Decay estimate dominant R}
\sup_{y\ge \sqrt{\f{4n_j-1}{\ell_j}}} \Big|\ell_j^{-\f14-\f{s_y}2}n_j^{-\f{s_y}2} e^{\f12 n_j^{-\f14} \ell_j^{\f34}(y-\sqrt{\f{4n_j-1}{\ell_k}})^{\f32}}\f{d^{s_y}}{dy^{s_y}} R_{n_j, \ell_j}\Big|\lesssim 1
\end{equation}
(the latter bound interpreted as the statement that the support of $R_{n_j,\ell_j}$ is concentrated in the region $\{y\lesssim n_j^{\f12} \ell_j^{-\f12}\}$).

\item  Using the explicit expression \eqref{Spherical harmonic m equals l} for $Y_{\ell_j,\pm\ell_j}$, we can estimate for any $s_\theta, s_\varphi \in \mathbb N$ with $s_\theta+s_\varphi\le s$ and any $0\le p\le s$ (provided $L$ is large enough so that $p+s_\varphi < \ell_j$):
 \begin{equation}\label{Dominant angular eigenfunction L infty again}
 \Big|\ell_j^{-\f14-\f{s_\theta}2 - s_\varphi} \f{\partial^{s_\theta}}{\partial \theta^{s_\theta}} \f{\partial^{s_\varphi}}{\partial \varphi^{s_\varphi}} Y_{\ell_j, \pm\ell_j} \Big| \lesssim 1
 \end{equation}
 and
 \begin{equation}\label{Dominant angular eigenfunction L infty again decay}
\sup_{|\f\pi2-\theta|\gtrsim \ell_j^{-\f12}} \Big|\ell_j^{-\f14-\f{s_\theta}2-s_\varphi} e^{\f12 \ell_j |\f\pi2-\theta|}\f1{\sin^p\theta}\f{\partial^{s_\theta}}{\partial \theta^{s_\theta}} \f{\partial^{s_\varphi}}{\partial \varphi^{s_\varphi}} Y_{\ell_j, \pm\ell_j} \Big|\lesssim 1
 \end{equation}
(the latter bound interpreted as the statement that the support of $Y_{\ell_j, \pm\ell_j} $ is concentrated in the region $\{|\f\pi2-\theta| \lesssim \ell_j^{-\f12}\}$).
\end{itemize}
As a result, \eqref{H s norm dominant eigenfunctions} follows directly from the expression \eqref{D phi derivative E j} for $\partial_\varphi^{\bar s} E_j$, the decay estimates \eqref{Decay estimate dominant R} and \eqref{Dominant angular eigenfunction L infty again decay} and the fact that $\omega_j = \ell_j +O(N)= \ell_j +O(\ell_j^{\delta_0})$ (which is a consequence of \eqref{Approximation estimate frequency}). Similarly, the bound \eqref{H s norm subdominant eigenfunctions} follows by combining the bounds \eqref{Dominant radial eigenfunction L infty again}--\eqref{Dominant angular eigenfunction L infty again decay}. Thus, the proof of \cref{cor:The 3 times 3 system} is complete.
\end{proof}

\section{The non-dominant modes}\label{sec:Non dominant modes}
In this section, we will analyze the long time dynamics of the system of non-dominant modes $\tilde\phi_k$, $k\in \mathcal K \setminus \mathcal K_{\mathrm{D}}$; for this analysis, the non-resonant condition \eqref{Non resonant condition time frequencies} will play a crucial role.

Recall that the functions $\{\tilde\phi_k\}_{k\in \mathcal K \setminus \mathcal K_{\mathrm{D}}}$ solve the (decoupled) system of inhomogeneous boundary value problems \eqref{Boundary value problem tilde phi} sourced by terms involving only $\{\tilde \phi_k\}_{k\in \mathcal K_{\mathrm{D}}}$:
\begin{equation}\label{Boundary value problem tilde phi again}
\begin{cases}
\square_g \tilde\phi_k + 2\tilde\phi_k = \f{1}{r\big(1-\f{2M}r+r^2\big)}\sum_{k_1, k_2, k_3\in \mathcal K_{\mathrm{D}}} \mathbb P_k \Big(r\big(1-\f{2M}r+r^2\big) \mathcal N[\tilde\phi_{k_1}, \tilde\phi_{k_2}, \tilde\phi_{k_3}]\Big), \\
\tilde\phi_k|_{y=y_\mathrm{mirror}}=0, \quad r\tilde\phi_k|_{y=0}=0.
\end{cases}
\end{equation} 
We will fix the functions $\{\tilde\phi_k\}_{k \in \mathcal K \setminus \mathcal K_{\mathrm{D}}}$ by choosing for them  \textbf{vanishing initial data}:
\[
\big( \tilde\phi_k(0), \partial_t\tilde\phi_k(0)\big) = \big(0, 0\big) \quad \text{for all} \quad k \in \mathcal K \setminus \mathcal K_{\mathrm{D}}.
\]
In particular, using the slowly oscillating $b_k$ variables defined by \eqref{B k variables} for the dominant modes $\{\tilde\phi_k\}_{k\in \mathcal K_{\mathrm{D}}}$, we infer that the non-dominant amplitudes $\{a_k(t)\}_{k\in \mathcal K\setminus \mathcal K_{\mathrm{D}}}$ solve the linear inhomogeneous initial value problem:
\begin{align} \nonumber
     \f{d^2 a_k}{dt^2} +{\omega}_k^2 a_k = \sum_{j_1,j_2,j_3 \in \mathcal K_{\mathrm{D}} } & e^{i ( - \varepsilon_{j_1}\omega_{j_1} +  \varepsilon_{j_2}\omega_{j_2} -  \varepsilon_{j_3}\omega_{j_3})t  } \langle \bar E_k E_{j_1} \bar E_{j_2} E_{j_3} \rangle \\  \nonumber
     \times &\big[ \left(  \varepsilon_{j_1} \varepsilon_{j_2}\omega_{j_1} \omega_{j_2} - \omega_{j_3}^2 \right)b_{j_1} \overline{ b_{j_2}} b_{j_3}   \\  
\nonumber & - i  \varepsilon_{j_1}\omega_{j_1}  b_{j_1} \f{d \overline{b_{j_2}}}{dt}  b_{j_3}   +  i  \varepsilon_{j_2} \omega_{j_2}   \f{d b_{j_1}}{dt} {\overline{b_{j_2}}}  b_{j_3}   - 2i  \varepsilon_{j_3}\omega_{j_3} b_{j_1} {\overline{b_{j_2}}} \f{d b_{j_3}}{dt} \\
 & +    \f{d b_{j_1}}{dt}  \f{d {\overline{b_{j_2}}}}{dt} b_{j_3} + b_{j_1}\overline{b_{j_2}} \f{d^2 b_{j_3}}{dt^2}  \big] \label{ODE system non dominant modes}
\end{align}
with
\begin{equation}\label{Initial data a k}
\big(a_k(0), \frac{d a_k}{dt}(0) \big) = \big( 0,0 \big).
\end{equation}
Recall also that the averaging operator $\langle \cdot \rangle$ is defined by \eqref{Definition average}. Let us also set for any $k\in \mathcal K$:
\begin{equation}\label{Non dominant spectral projection}
\mathcal P_k \doteq \max_{j_1, j_2, j_3 \in \mathcal K_{\mathrm{D}}} L^2 \big| \langle \bar E_k E_{j_1} \bar E_{j_2} E_{j_3} \rangle \big|.
\end{equation}

The main result of this section will be the following:

\begin{proposition}\label{prop:Total bound non-dominant modes}
Let $M_0>0$, $C_{\mathrm{amp}}$, $N, \lambda, L$ and $M$ be as in the statement of   \cref{prop:The 3 times 3 system}. Let also $T_1>0$ be defined as 
\[
T_1 \doteq L^{2s+1} \tilde T,
\]
where $\tilde T = \tilde T(C_{\mathrm{amp}}, \epsilon,\lambda, N,  s, M_0)$ is the parameter appearing in the statement of   \cref{prop:The 3 times 3 system}. Then, we have for any integer $p\ge 0$:
\begin{equation}\label{Estimate sum non dominant modes}
\sup_{t\in [0,T_1]}\sum_{k\in \mathcal K \setminus \mathcal K_{\mathrm{D}}}\Bigg( \sum_{ p_1+p_2+p_3=0}^p  \| \partial_t^{p_1} \partial_y^{p_2} \nabla_{\mathbb S^2}^{p_3} (r\tilde\phi_k)\|_{L^2(dy \dvol_{\mathbb S^2})} \Bigg) \le C_p L^{p+\f12-3s}, 
\end{equation}
where the constant $C_{p}>0$ might depend only on the parameters $C_{\mathrm{amp}}, N, \lambda, \epsilon, s, M_0$ and the integer $p$ (in particular, it is independent of $L$). Moreover, for any $k=(n_k, \ell_k, m_k)\in \mathcal K$ with $\ell_k \notin [L, 3(\lambda+2) L]$, we have
\[
\tilde\phi_k \equiv 0.
\]
\end{proposition}

The proof of \cref{prop:Total bound non-dominant modes} can be found at the end of  \cref{sec:Non dominant amplitudes}.

\subsection{Bounds for the spectral projection coefficients}
The following estimate for the spectral coefficients appearing in the right-hand side  of \eqref{ODE system non dominant modes} will be useful for estimating the modes with frequency parameters differing substantially from those of the dominant modes:
\begin{lemma}\label{lem:Bounds P k}
For any $k=(n_k, \ell_k, m_k)\in \mathcal K$, the quantity $\mathcal P_k$ defined by \eqref{Non dominant spectral projection} satisfies 
\begin{equation}\label{Bound P k}
\mathcal P_k \lesssim \ind_{[L, 3(\lambda+2) L]}(\ell_k) \cdot \ind_{\mathrm{res}}(m_k) \cdot \f{1}{n_k^5}
\end{equation}
where 
\[
\ind_{[L, 3(\lambda+2) L]}(\ell_k) =
\begin{cases}
1, \quad \ell_k \in [L, 3(\lambda+2) L],\\
0, \quad \ell_k \notin [L, 3(\lambda+2) L],
\end{cases} 
\]
\begin{equation}\label{Indicator set m}
\ind_{\mathrm{res}}(m_k) = 
\begin{cases}
1, \quad    m_k \in \big\{ m_{j_1}-m_{j_2}+m_{j_3}: \, j_1, j_2, j_3 \in \{-1, 0, +1\} \big\} ,\\
0, \quad \text{otherwise}
\end{cases}
\end{equation}
and the constants implicit in the $\lesssim$ notation in \eqref{Bound P k} might depend on $M, N, \lambda$ but are \textbf{independent} of $L$ and $k$.
\end{lemma}

\begin{proof}
For this proof, we will assume that the constants implicit in the $\lesssim$ and $O(\cdot)$ notation  are independent of $L$ (but might depend on $M, N, \lambda$).

In view of the decomposition \eqref{Spatial eigenfunction} of the functions $E_k$, we have for any $k, j_1, j_2, j_3 \in \mathcal K$:
\begin{align}\nonumber
L^{2} \langle \bar E_{k} E_{j_1} \bar E_{j_2} E_{j_3} \rangle = L^2 & \Bigg(\int_0^{y_\mathrm{mirror}} \! \! \! \!   R_{n_{k},\ell_{k}}(y) R_{n_{j_1},\ell_{j_1}}(y) R_{n_{j_2},\ell_{j_2}}(y) R_{n_{j_3},\ell_{j_3}}(y) \Big(1+\f{1}{r(y)^2} -\f{2M}{r(y)^3}\Big) \big(r(y)\big)^{-6}\, dy \Bigg)\\
& \times   \Bigg(\int_0^{2\pi}\int_0^\pi \bar Y_{\ell_{k},m_{k}}(\theta, \varphi) Y_{\ell_{j_1},m_{j_1}}(\theta, \varphi) \bar Y_{\ell_{j_2},m_{j_2}}(\theta, \varphi) Y_{\ell_{j_3},m_{j_3}}(\theta, \varphi) \, \sin\theta d\theta d\varphi \Bigg). \label{Expression spectral coefficient again}
\end{align}

We will first focus on the second factor of the right-hand side of \eqref{Expression spectral coefficient again}. In view of the fact that $Y_{\ell,m}$ takes the form
\[
Y_{\ell,m}(\theta, \varphi) = c_{\ell,m} P^m_\ell(\cos\theta) e^{im\varphi},
\]
it can be readily verified that 
\begin{equation}\label{Resonance in m for projection}
\int_0^{2\pi}\int_0^\pi \bar Y_{\ell_{k},m_{k}}(\theta, \varphi) Y_{\ell_{j_1},m_{j_1}}(\theta, \varphi) \bar Y_{\ell_{j_2},m_{j_2}}(\theta, \varphi) Y_{\ell_{j_3},m_{j_3}}(\theta, \varphi) \, \sin\theta d\theta d\varphi =0 \quad 
\text{if} 
\quad
-m_k+m_{j_1}-m_{j_2}+m_{j_3} \neq 0.
\end{equation}
In particular, in view of the fact
\[
\min_{j_1, j_2, j_3 \in \mathcal K_{\mathrm{D}}}\Big| m_{j_1}-m_{j_2}+m_{j_3}\Big|=L
\]
and $\ell\ge |m|$, we infer that
\begin{equation}\label{Lower bound angular support projection}
\mathcal P_k=0 \quad \text{if}\quad \ell_k<L.
\end{equation}

Recall that the product of two spherical harmonics $Y_{\ell_1,m_1}$ and $Y_{\ell_2, m_2}$ can be decomposed as a sum of spherical harmonics of order at most $\ell_1+\ell_2$, i.e.
\[
Y_{\ell_1,m_1}(\theta,\varphi) \cdot Y_{\ell_2,m_2}(\theta, \varphi) = \sum_{\ell=0}^{\ell_1+\ell_2}\sum_{m=-\ell}^\ell a_{\ell, m } Y_{\ell, m}(\theta, \varphi)
\]
(where $a_{\ell, m }$ are known as the Clebsch--Gordan coefficients). Using the above property (and the fact that $\bar Y_{\ell, m} = (-1)^m Y_{\ell, -m}$), we infer that
\begin{equation}\label{Vanishing spectral coefficient angular component}
\int_0^{2\pi}\int_0^\pi \bar Y_{\ell_{k},m_{k}}(\theta, \varphi)  Y_{\ell_{j_1},m_{j_1}}(\theta, \varphi) \bar Y_{\ell_{j_2},m_{j_2}}(\theta, \varphi) Y_{\ell_{j_3},m_{j_3}}(\theta, \varphi) \, \sin\theta d\theta d\varphi  = 0 \quad \text{if} \quad \ell_k > \ell_{j_1}+\ell_{j_2}+\ell_{j_3}.
\end{equation}
Therefore,
\begin{equation}\label{Vanishing spectral coefficient}
\mathcal P_k =0 \quad \text{if} \quad \ell_k > \max_{j_1, j_2, j_3 \in \mathcal K_{\mathrm{D}}} \big( \ell_{j_1}+\ell_{j_2}+\ell_{j_3} \big) = 3(\lambda+2) L.
\end{equation}
Combining \eqref{Lower bound angular support projection} with \eqref{Vanishing spectral coefficient}, we infer that
\begin{equation}\label{Bound angular support projection}
\mathcal P_k =0 \quad \text{if} \quad \ell_k\notin [L,  3(\lambda+2) L].
\end{equation}

Note also that, in view of the fact that $\|Y_{\ell,\pm\ell}\|_{L^\infty(\mathbb S^2)} \le C \ell^{\f14}$ for some absolute constant $C>0$ and $\|Y_{\ell_k, m_k}\|_{L^2(\mathbb S^2)}=1$, we have for any $j_1, j_2, j_3 \in \mathcal K_{\mathrm{D}}$ and any $k\in \mathcal K$:
\begin{equation}\label{Bound product spherical harmonics again}
L^{-\f12} \int_0^{2\pi}\int_0^\pi \bar Y_{\ell_{k},m_{k}}(\theta, \varphi) \Big| Y_{\ell_{j_1},m_{j_1}}(\theta, \varphi) \bar Y_{\ell_{j_2},m_{j_2}}(\theta, \varphi)  Y_{\ell_{j_3},m_{j_3}}(\theta, \varphi) \Big| \, \sin\theta d\theta d\varphi \lesssim 1.
\end{equation}

Let $k\in \mathcal K$ satisfy $L\le \ell_k \le 3(\lambda+2) L$. We will consider two regimes for the parameter $n_k$ (recalling that the definition of the set of parameters $\mathcal K$ implies that $n_k \lesssim \ell^{1-\delta_0}_k$):
\begin{itemize}
\item In the case when 
\[
n_k \lesssim 1,
\]
 we can use the approximation \cref{lem:eigenfunctions-estimate-m>0} for the radial eigenfunctions $R_{n, \ell}$ and the fact that
\[
\|e_n\|_{L^\infty([0,+\infty))} \le C n^{\f16-\f14} \le C 
\]
for some absolute constant $C>0$ (following readily for instance from the Hermite asymptotics provided by \cref{lem: WKB asymptotics Airy}), together with the normalization condition \eqref{Normalization radial eigenfunctions} for $R_{n,\ell}$, in order to obtain for any $j_1, j_2, j_3 \in \mathcal K_{\mathrm{D}}$:
\begin{equation}\label{Bound product radial eigenfunctions again}
L^{\f52} \Bigg| \int_0^{y_\mathrm{mirror}} R_{n_{k},\ell_{k}}(y) R_{n_{j_1},\ell_{j_1}}(y) R_{n_{j_2},\ell_{j_2}}(y) R_{n_{j_3},\ell_{j_3}}(y) \Big(1+\f{1}{r(y)^2} -\f{2M}{r(y)^3}\Big) \big(r(y)\big)^{-6}\, dy\Bigg| \lesssim \f{\ell_k^{\f14}}{L^{\f14}} \lesssim 1.
\end{equation}
\item In the case when
\[
n_k \gg 1,
\]
applying \cref{lem:Crude non stationary phase} with $V(y) = 1+\f1{r(y)^2}-\f{2M}{r(y)^3}$, $\lambda_-=\f{1}{4\lambda}$, $\lambda_+=4\lambda$ and $(n,\ell)=(n_k,\ell_k)$, we immediately infer that
\begin{equation}\label{Bound product radial eigenfunctions again once more}
L^{\f52} \Bigg|\int_0^{y_\mathrm{mirror}} R_{n_{k},\ell_{k}}(y) R_{n_{j_1},\ell_{j_1}}(y) R_{n_{j_2},\ell_{j_2}}(y) R_{n_{j_3},\ell_{j_3}}(y) \Big(1+\f{1}{r(y)^2} -\f{2M}{r(y)^3}\Big) \big(r(y)\big)^{-6}\, dy\Bigg| \lesssim \f{1}{n_k^5}.
\end{equation}
\end{itemize}

Combining \eqref{Resonance in m for projection}, \eqref{Bound angular support projection} and the bounds \eqref{Bound product spherical harmonics again}, \eqref{Bound product radial eigenfunctions again} and \eqref{Bound product radial eigenfunctions again once more} for the expression \eqref{Expression spectral coefficient again} for $\langle \bar E_k E_{j_1} \bar E_{j_2} E_{j_3}\rangle$ (when $L\le \ell_k \le 3(\lambda+2)L$) and substituting the above bounds for the integrals in the right-hand side of \eqref{Expression spectral coefficient again}, we infer \eqref{Bound P k}.

\end{proof}

\subsection{Bounds for the non-dominant mode amplitudes}\label{sec:Non dominant amplitudes}
The following estimate for the non-dominant amplitudes $\{a_k\}_{k\in \mathcal K \setminus \mathcal K_{\mathrm{D}}}$:
\begin{lemma}\label{lem:The non dominant modes}
Let $M_0>0$, $C_{\mathrm{amp}}$, $N, \lambda, L$ and $M$ be as in the statement of  \cref{prop:The 3 times 3 system} and $T_1>0$ be as in the statement of  \cref{prop:Total bound non-dominant modes}.
Then, for any $k = (n_k, \ell_k, m_k)\in \mathcal K \setminus \mathcal K_{\mathrm{D}}$, the amplitude $a_k$ of the function $\tilde \phi_k$ satisfies the following bound on the time interval $t\in [0, T_1]$ for any integer $p\ge 0$:
\begin{equation}\label{Estimate non dominant modes}
\sup_{t\in [0,T_1]} \Big | \f{d^p a_k}{dt^p} (t) \Big|  \le C_p \mathcal P_k \omega_k^{p-1} L^{\f12-3s}.
\end{equation}
In the above, the constant $C_{p}>0$ depends only on the parameters $N, \lambda, \epsilon, s$ and the integer $p$ (in particular, it is independent of $L$ and the index $k$).
\end{lemma}

\begin{proof}
For this proof, we will assume that the constants implicit in the $\lesssim$ and $O(\cdot)$ notation  are independent of $L$ (but might depend on $M, N, \lambda, \epsilon, s$). 

The initial value problem for the non-dominant mode amplitudes \eqref{ODE system non dominant modes} takes the form
\begin{equation}\label{ODE non dominant modes again}
\begin{cases}
\f{d^2 a_k}{dt^2} + \omega_k^2 a_k = \sum_{j_1, j_2, j_3 \in \mathcal K_{\mathrm{D}}} e^{i(-\varepsilon_{j_1} \omega_{j_1}+\varepsilon_{j_2}\omega_{j_2}-\varepsilon_{j_3}\omega_{j_3})t} \langle \bar E_k E_{j_1} \bar E_{j_2} E_{j_3} \rangle \mathcal B_{j_1,j_2,j_3},\\[7pt]
\big(a_k(0), \f{d a_k}{dt}(0)\big) = (0,0),
\end{cases}
\end{equation}
where
\begin{align*}
\mathcal B_{j_1,j_2,j_3} \doteq  \, & 
      \left(  \varepsilon_{j_1} \varepsilon_{j_2}\omega_{j_1} \omega_{j_2} - \omega_{j_3}^2 \right)b_{j_1} \overline{ b_{j_2}} b_{j_3}   \\  
 & - i  \varepsilon_{j_1}\omega_{j_1}  b_{j_1} \f{d \overline{b_{j_2}}}{dt}  b_{j_3}   +  i  \varepsilon_{j_2} \omega_{j_2}   \f{d b_{j_1}}{dt} {\overline{b_{j_2}}}  b_{j_3}   - 2i  \varepsilon_{j_3}\omega_{j_3} b_{j_1} {\overline{b_{j_2}}} \f{d b_{j_3}}{dt} \\
 & +    \f{d b_{j_1}}{dt}  \f{d {\overline{b_{j_2}}}}{dt} b_{j_3} + b_{j_1}\overline{b_{j_2}} \f{d^2 b_{j_3}}{dt^2}.
\end{align*}
Therefore, applying Duhamel's principle, we obtain the following representation formula for $a_k$ for each $k\in \mathcal K \setminus \mathcal K_{\mathrm{D}}$ and each $t\in [0,T_1]$:
\begin{align} \label{Representation a k0}
a_k(t) = & \sum_{j_1, j_2, j_3 \in \mathcal K_{\mathrm{D}}}\langle \bar E_k E_{j_1} \bar E_{j_2} E_{j_3} \rangle  \int_0^t \f{e^{i\omega_k (t-s)} - e^{-i\omega_k (t-s)}}{2i \omega_k} e^{i(-\varepsilon_{j_1} \omega_{j_1}+\varepsilon_{j_2}\omega_{j_2}-\varepsilon_{j_3}\omega_{j_3})s} \mathcal B_{j_1,j_2,j_3}(s) \, ds\\ \label{Representation a k1}
& = \sum_{j_1, j_2, j_3 \in \mathcal K_{\mathrm{D}}}\f{\langle \bar E_k E_{j_1} \bar E_{j_2} E_{j_3} \rangle}{2i\omega_k}  \Bigg( e^{i\omega_k t}\int_0^t e^{i(-\omega_k-\varepsilon_{j_1} \omega_{j_1}+\varepsilon_{j_2}\omega_{j_2}-\varepsilon_{j_3}\omega_{j_3})s} \mathcal B_{j_1,j_2,j_3}(s) \, ds \\\label{Representation a k2}
&\hphantom{ = \sum_{j_1, j_2, j_3 \in \mathcal K_{\mathrm{D}}}\f{\langle \bar E_k E_{j_1} \bar E_{j_2} E_{j_3} \rangle}{2i\omega_k}  \Bigg(}
+ e^{-i\omega_k t}\int_0^t e^{i(\omega_k-\varepsilon_{j_1} \omega_{j_1}+\varepsilon_{j_2}\omega_{j_2}-\varepsilon_{j_3}\omega_{j_3})s} \mathcal B_{j_1,j_2,j_3}(s) \, ds \Bigg).
\end{align}

We only have to consider those values of $k\in \mathcal K \setminus \mathcal K_{\mathrm{D}}$ for which there exists $j_1, j_2, j_3\in \mathcal K_{\mathrm{D}}$ such that
\begin{equation}\label{Resonant condition in ms}
  m_k = m_{j_1}-m_{j_2}+m_{j_3}.
\end{equation}
Indeed, note that for  any $k \in \mathcal K \setminus \mathcal K_{\mathrm{D}}$ not satisfying the above condition, the right-hand side of \eqref{ODE non dominant modes again} vanishes (in view of the bound \eqref{Bound P k} for $\mathcal P_k$) and, thus, $a_k\equiv 0$. On the other hand,  in view of  the non-resonant condition \eqref{Non resonant condition time frequencies},  for any $k\in \mathcal K \setminus \mathcal K_{\mathrm{D}}$ satisfying \eqref{Resonant condition in ms} we must have:
\begin{equation}\label{Non resonant condition time frequencies again}
\min_{j_1, j_2, j_3 \in \mathcal K_{\mathrm{D}}} \Big| \pm{\omega}_k - \varepsilon_{j_1}{\omega}_{j_1} + \varepsilon_{j_2}{\omega}_{j_2} - \varepsilon_{j_3}{\omega}_{j_3} \Big| \gtrsim L^{-\f12}.
\end{equation}
Therefore, by writing
\[
e^{i(\pm \omega_k-\varepsilon_{j_1} \omega_{j_1}+\varepsilon_{j_2}\omega_{j_2}-\varepsilon_{j_3}\omega_{j_3})s} = \f{1}{i(\pm \omega_k-\varepsilon_{j_1} \omega_{j_1}+\varepsilon_{j_2}\omega_{j_2}-\varepsilon_{j_3}\omega_{j_3})} \f{d}{ds}\big( e^{i(\pm \omega_k-\varepsilon_{j_1} \omega_{j_1}+\varepsilon_{j_2}\omega_{j_2}-\varepsilon_{j_3}\omega_{j_3})s}\big)
\]
and integrating by parts under the integrals in \eqref{Representation a k1} and \eqref{Representation a k2}, we obtain:
\begin{equation}\label{Almost there bound a k}
|a_k(t)| \lesssim \f{L^{\f12}}{\omega_k} \max_{j_1, j_2, j_3 \in \mathcal K_{\mathrm{D}}}\Big\{|\langle \bar E_k E_{j_1} \bar E_{j_2} E_{j_3}\rangle | \Big( \sup_{s\in [0,t]} |\mathcal B_{j_1,j_2, j_3}(s)| + \int_0^t  \Big| \f{d \mathcal B_{j_1,j_2, j_3}(s)}{ds}\Big| \, ds \Big)\Big\}.
\end{equation}
Using the bounds \eqref{Bound C 1 norm tilde a}--\eqref{Bound C q norm tilde a} from  \cref{prop:The 3 times 3 system} for the slowly oscillating amplitudes  $\{b_j\}_{j\in \mathcal K_{\mathrm{D}}}$ (and recalling that in the statement of  \cref{prop:The 3 times 3 system}, we have $\tilde b_j = L^s b_j$ and $\tilde t = L^{-2s-1}t$), we obtain that
\[
\max_{j_1, j_2, j_3 \in \mathcal K_{\mathrm{D}}} \sup_{s\in [0,t]} |\mathcal B_{j_1,j_2, j_3}(s)| \lesssim L^{-3s+2} 
\]
and (recall that $T_1 \sim L^{2s+1}$)
\begin{equation*}
\max_{j_1, j_2, j_3 \in \mathcal K_{\mathrm{D}}}\int_0^{T_1}  \Big| \f{d \mathcal B_{j_1,j_2, j_3}(s)}{ds}\Big| \, ds \lesssim T_1 L^{-5s+1} \lesssim L^{-3s+2}.
\end{equation*}
Thus, returning to \eqref{Almost there bound a k} and using the fact that $\mathcal P_k = L^{2} \max_{j_1, j_2, j_3\in \mathcal K_{\mathrm{D}}}|\langle \bar E_k E_{j_1} \bar E_{j_2} E_{j_3}\rangle |$, we obtain:
\[
|a_k(t)| \lesssim \f{L^{\f12-3s}}{\omega_k} \mathcal P_k.
\]
Differentiating \eqref{Representation a k0} with respect to $t$ and repeating the same process, we finally obtain \eqref{Estimate non dominant modes}.

\end{proof}

\begin{proof}[Proof of \cref{prop:Total bound non-dominant modes}]
For the rest of this proof, we will assume that the constants implicit in the $\lesssim$ notation  might depend on  $C_{\mathrm{amp}}, N, \lambda,\epsilon, s M_0$, but are independent of the mode parameter $k$ and the frequency parameter $L$.

Recall that the mode $\tilde\phi_k$ has the form
\[
r \tilde\phi_k(t, y, \theta, \varphi) = a_k(t) E_k(y, \theta, \varphi).
\]
Moreover, let us collect the following facts about $E_k(y,\theta, \varphi) = R_{n_k,\ell_k}(y) Y_{\ell_k,m_k}(\theta, \varphi)$:
\begin{itemize}
\item 
For any $(n_k, \ell_k)\in (\mathbb N^*)^2$, in view of the fact that $\|R_{n_k,\ell_k}\|_{L^2\big([0,y_\mathrm{mirror}]\big)} = 1$ and satisfies \[\f{d^2}{dy^2} R_{n_k,\ell_k} = \big(V_{\ell_k}-\omega_{n_k,\ell_k}^2) R_{n_k,\ell_k}= O(\ell_k^2+n_k^2) R_{n_k,\ell_k}\] (the fact that $\omega_{n_k,\ell_k}\lesssim \ell_k+n_k$ following from \cref{lem:Crude Weyl law}), we can readily estimate for any $j\in \mathbb N$:  
\[
\big\|(\ell_k+n_k)^{-j} \f{d^j}{dy^j} R_{n_k,\ell_k}\big\|_{L^2([0,y_\mathrm{mirror}])} \lesssim_j 1.
\]

\item The spherical harmonics $Y_{\ell_k,m_k}$ (given by \eqref{Expression spherical harmonic}) satisfy
\[
\Big\| \ell_k^{-j} \nabla_{\mathbb S^2}^j Y_{\ell_k,m_k}\Big\|_{L^2(\mathbb S^2)} \lesssim 1.
\]
\end{itemize}
Therefore, we can readily estimate for any $k\in \mathcal K$ and any $p\in \mathbb N$ (using the fact that $n_k \lesssim \ell_k^{1-\delta_0}$):
\begin{equation}\label{Bound non dominant sum almost there}
\sup_{t\in[0,T_1]}\sum_{0\leq p_1+p_2+p_3\leq p} \| \partial_t^{p_1} \partial_y^{p_2}\nabla_{\mathbb S^2}^{p_3} (r\tilde\phi_k)\|_{L^2(dy dvol_{\mathbb S^2})} \lesssim_p \sup_{t\in[0,T_1]}\sum_{p_1=0}^{p} \ell_k^{p-p_1} |\frac{d^{p_1}}{d t^{p_1}} a_k|,
\end{equation}
where the constant implicit in the notation $\lesssim_p$ above is allowed to depend on $p$.  

The upper bound for $\mathcal P_k$ provided by \cref{lem:Bounds P k} implies that  $\ell_k \sim L$ for any $k=(n_k, \ell_k, m_k) \in \mathcal K$ with $\mathcal P_k \neq 0$. Thus, using also the fact that $\omega_k-\ell_k \sim n_k$ (as a consequence of   \cref{lem:Crude Weyl law}), we obtain by combining \eqref{Bound non dominant sum almost there} with the bound \eqref{Estimate non dominant modes} for $\{a_k\}_{k\in \mathcal K \setminus \mathcal K_{\mathrm{D}}}$ and the bound for $\mathcal P_k$ from   \cref{lem:Bounds P k}:
\[
\sup_{t\in[0,T_1]}\sum_{0\leq p_1+p_2+p_3\leq p} \!\!  \!\!  \!\!  \!\! \| \partial_t^{p_1} \partial_y^{p_2}\nabla_{\mathbb S^2}^{p_3} (r\tilde\phi_k)\|_{L^2(dy dvol_{\mathbb S^2})} 
\lesssim_p \f{1}{n_k^5} \ind_{[L,3(\lambda+2)L]}(\ell_k) \cdot \ind_{\mathrm{res}}(m_k) L^{p-\f12-3s}   \text{ for any }  k\in \mathcal K \setminus \mathcal K_{\mathrm{D}}.
\]
Summing the above expression over $\Big\{(n_k, \ell_k, m_k) \in \mathcal K\setminus \mathcal K_{\mathrm{D}}: \, \ind_{[L,3(\lambda+2)L]}(\ell_k) \cdot \ind_{\mathrm{res}}(m_k) \neq 0\Big\}$ (noting that this sum contains $\lesssim L$ values of $(m_k, \ell_k)$), we obtain \eqref{Estimate sum non dominant modes}.

\end{proof}

\section{Estimates for the error term}\label{sec:Estimates error term}

In this section, we will estimate the size of the error term $\psi$ in the decomposition \eqref{The ansatz} for $\phi$. Recall that $\psi$ solves the initial-boundary value problem on $\mathcal M_\mathrm{ext}$, i.e.: 
\begin{equation}\label{IVP Psi}
\begin{cases}
\square_g \psi + 2 \psi -r^{-6}|\chi\tilde\phi+\psi|^2 \partial_{t^*}^2\psi -\chi^2 \mathcal N^{(1)}[\tilde \phi;\psi]   - \chi \mathcal N^{(2)}[\tilde\phi; \psi] -\mathcal N^{(3)}[\psi] = - \mathcal F[\tilde \phi],\\[5pt]
(\psi, \partial_{t^*} \psi)|_{t^*=0} = (0,0), \\
r\psi|_{r=\infty} =0.
\end{cases}
\end{equation}
where
\[
\mathcal F[\tilde \phi] \doteq   \square_g (\chi \tilde\phi) + 2 \chi\tilde\phi - \mathcal N[\chi\tilde\phi, \chi\tilde\phi, \chi\tilde\phi]
\]
and $\mathcal N^{(1)}[\tilde\phi; \psi]$, $\mathcal N^{(2)}[\tilde\phi; \psi]$ and $\mathcal N^{(3)}[\psi]$ denote the lower-order linear, quadratic and cubic terms in $\psi$, respectively, i.e.
\begin{align*}
\mathcal N^{(1)}[\tilde\phi; \psi] \doteq & \,  r^{-6} \Big( 2\partial_{t^*}^2 \tilde\phi \, \Re\{\tilde\phi \bar \psi\}  +  |\partial_{t^*} \tilde\phi|^2 \psi + 2 \tilde\phi \Re\{\partial_{t^*} \tilde\phi \partial_{t^*} \bar\psi\}\Big),\\[5pt]
\mathcal N^{(2)}[\tilde\phi; \psi] \doteq &\,  r^{-6} \Big(\partial_{t^*}^2 \tilde\phi |\psi|^2 + \tilde\phi |\partial_{t^*} \psi|^2 + 2\Re\{\partial_{t^*} \tilde\phi \partial_{t^*} \bar \psi\} \psi   \Big)\\[5pt]
\mathcal N^{(3)}[\psi] \doteq& \, r^{-6} \psi |\partial_{t^*} \psi|^2.
\end{align*}

We will establish the following result:
\begin{proposition}\label{prop: Estimates error term}
Let $M_0>0$, $C_{\mathrm{amp}}$, $N, \lambda, L$, $s\ge 4$ and $M$ be as in the statement of  \cref{prop:The 3 times 3 system}. Let $T_1>0$ be defined as 
\begin{equation*}
T_1 \doteq L^{2s+1} \tilde T,
\end{equation*}
where $\tilde T $ is the parameter appearing in the statement of  \cref{prop:The 3 times 3 system} (and which is independent of $L$). Let $\psi$ be the solution of the initial value problem \eqref{IVP Psi}. Then, provided $L$ is large enough in terms of the parameters $\lambda, N, s$ and the geometry of $(\mathcal M_\mathrm{ext},g_M)$, the solution $\psi$ exists on the domain $\{0\le t^*\le T_1\}$ and satisfies
\begin{equation}\label{Bound psi}
\sum_{\bar j=0}^{s} \sup_{\tau\in [0,T_1]} \Big( L^{-\bar j}  \sum_{j_1+j_2+j_3=\bar j}  \big\|\partial_{t^*}^{j_1} (r^2\partial_r)^{j_2}\nabla_{\mathbb S^2}^{j_3} (r \psi)\big|_{\Sigma^*_\tau} \big\|_{L^2(r^{-2}\sin\theta dr d\theta d\varphi)} \Big) \le L^{-\f52-s} .
\end{equation}

\end{proposition}
The proof of \cref{prop: Estimates error term} will be established via a bootstrap argument; see \cref{sec:The bootstrap argument}. In \cref{sec:Commutation notations}--\cref{subsec:Energy estimates linear}, we will set up the commutation scheme and the norms that will be involved in the bootstrap argument; the main technical tool for the proof of \cref{prop: Estimates error term} is the a priori energy estimate established in \cref{sec:A priori energy bounds}.

\begin{remark} Let us note that \eqref{Bound psi} cannot be inferred by only relying on energy estimates using $\partial_{t^*}$ as a commutator and multiplier; this can be seen by noting that in the linearized version of equation \eqref{IVP Psi}, namely
\[
\square_g \psi + 2 \psi -r^{-6} \chi^2|\tilde\phi|^2 \partial_{t^*}^2\psi -\chi^2 \mathcal N^{(1)}[\tilde \phi;\psi]   = - \mathcal F[\tilde \phi],
\]
the oscillations in time of the coefficients involving $\tilde\phi$ are in principle consistent with exponential growth at a timescale much smaller than $T_1$. The structure that we need to exploit is that $\tilde\phi$ behaves like a wave-packet localized in phase space around null geodesics at the conformal boundary at $r=\infty$. In particular, the helical Killing vector field $V$ defined by \eqref{Helical vector field} (and which is  timelike away from $\mathcal H^+$ and degenerates to an almost null vector field on the conformal boundary at $r=\infty$)
commutes well with $\tilde\phi$, in the sense that near $y=0$, we have schematically
\[
\Big|V^i (L^{-1}\partial)^{j_1}(L^{-\f12} \bar\partial)^{j_2}\tilde\phi \Big| \sim \big|\tilde\phi \big| \quad \text{for } \, i\in \{0,1\}, j_1, j_2 \in \mathbb N, 
\]
where the constants implicit in the $\sim$ notation are independent of the frequency  parameter $L$ (note, however, that any additional differentiation with respect to $V$ will result in a quantity satisfying an estimate which is worse by an additional factor of $L$). We will make use of this observation while setting up a commutation scheme that will allow us to close energy estimates for \eqref{IVP Psi}.
\end{remark}

\begin{remark} For the rest of this section, we will assume that the constants implicit in the $\lesssim$ and $O(\cdot)$ notation might depend on $M_0, M, \tilde T, \epsilon, s, \lambda, N$ (where $\tilde T$ is the timescale appearing in the statement of \cref{prop:The 3 times 3 system}) but will be \textbf{independent} of $L$.
\end{remark}

\subsection{Notations on commutations with \texorpdfstring{$Y$, $\partial_{t^*}$}{Y, partial t*}} \label{sec:Commutation notations}
In order to establish non-degenerate energy estimates for solutions to the conformally coupled wave equation 
\begin{equation}\label{Conformally coupled wave}
\square_g f + 2f =F
\end{equation}
 we will need to commute \eqref{Conformally coupled wave} with both the Killing vector field $\partial_{t^*}$ and the red-shift vector field $Y$. When considering solutions with frequency support concentrated at scales $\sim L$ (such as $\tilde \phi$ or $\psi$), each such derivative will increase the ``size'' of the solution by $\sim L$. For this reason, we will adopt the following shorthand notation:
\begin{definition}\label{def:Commutations}
For any sufficiently regular function $f:\mathcal M_\mathrm{ext} \rightarrow \mathbb C$ and any $j_1, j_2 \in \mathbb N$, we will set
\[
f^{(j_1; j_2)} \doteq \f1{L^{j_1+j_2}} Y^{j_1} \partial_{t^*}^{j_2} f.
\]
We denote with $f^{(j)}$ any linear combination of terms of the form $f^{(j_1;j_2)}$ with $j_1+j_2=j$ and with $f^{(\le j)}$ any linear combination of terms of the form $f^{(\bar j)}$ with $\bar j\le j$.
\end{definition}

\begin{remark} 
In the region $\{r\ge r_++\delta_2\}$ (where $\delta_2(M)>0$ is fixed in \cref{sec:The red shift vector field}), we have $Y=\partial_{t^*}$ and, therefore, $f^{(j_1; j_2)}|_{\{r\ge r_++\delta_2\}} = f^{(0;j_1+j_2)}|_{\{r\ge r_++\delta_2\}} = f^{(j_1+j_2;0)}|_{\{r\ge r_++\delta_2\}}$.
\end{remark}

\subsection{The bootstrap norms}\label{subsec: Bootstrap norms}
In this section, we will introduce the norms that will be used in the bootstrap argument in the proof of \cref{prop: Estimates error term}. To this end, let us also introduce the modified distance function
\begin{equation}\label{Distance function}
w(y,\theta) \doteq \min\Big\{\sqrt{\f1L+y^2+\left|\theta-\frac{\pi}{2}\right|^2}, \, 1\Big\}.
\end{equation}
Note that, for fixed $t^*$, $w(y, \theta)$ measures the coordinate distance from the equatorial circle $\{y=0\}\cap\{\theta=\frac{\pi}2\}$; the support of the approximate solution $\tilde\phi$ is concentrated in the region where $w\sim L^{-\f12}$.

For the following definition, we will make use of the shorthand notation \eqref{Shorthand square derivatives} and \eqref{Shorthand square derivatives near horizon} for coordinate derivatives of $f$ in the far away and the near horizon regions (i.e.\ the regions $\{y\le y_\mathrm{mirror}\}=\{r\ge r_\mathrm{mirror}\}$ and $\{r\le r_\mathrm{mirror}\}$, respectively).

\begin{definition}\label{def:Norms}
Let $f:\mathcal M_\mathrm{ext}\rightarrow \mathbb C$ be a sufficiently regular function. For any $\tau\in \mathbb R$, we will define 
\begin{align}\nonumber
\mathcal E[f](\tau) \doteq 
&  \int_{\{t^*=\tau\}\cap\{y\le y_\mathrm{mirror}\}} \Bigg[|\bar\partial^2(rf)  |^2+\big(w(y, \theta)\big)^2|\partial\bar\partial (rf)|^2+\big( w(y, \theta)\big)^4|\partial^2(rf)|^2  \\
& \hphantom{ \Bigg[|\bar\partial^2(rf)  |^2+\big(w+++}\qquad
+  |\partial V(rf)|^2+ |\partial (rf)|^2 + \frac{1}{y^2}|rf|^2\Bigg]\, \sin\theta dy d\theta d\varphi   \nonumber \\
& + \int_{\{t^*=\tau\}\cap\{r\le r_\mathrm{mirror}\}} \Big[
 |\partial (V f)|^2+  |\partial f|^2 + |f|^2 \Big] \,  \, r^2\sin\theta dr d\theta d\varphi \nonumber \\
& + \big(\log L\big)^{-1} \sup_{j\in \mathbb N} \Bigg(\int_{\{t^*=\tau\}\cap\{y\le y_\mathrm{mirror}\}\cap\{ w(y, \theta) \le 2^{-j}\}}  
\f{2^j}{w(y,\theta)}|\bar\partial (rf)|^2\, \sin\theta dy d\theta d\varphi      \Bigg)   \nonumber \\
& + \big(\log L\big)^{-1} \int_0^{2\pi} \Bigg[\sup_{(y,\theta)\in(0,y_\mathrm{mirror})\times(0,\pi)}\frac{|rf|^2(\tau,y,\theta, \varphi)}{\big(w(y,\theta)\big)^2}\Bigg] \, d\varphi. \label{Basic norm}
\end{align}
 For any integer $j\ge 1$, we will also set
\begin{equation}\label{Higher order norm}
\mathcal E^{(j)}[f](\tau) \doteq \sum_{\bar j=0}^k L^{-2\bar j} \mathcal E[r^{-1}\partial^{\bar j} (rf)](\tau).
\end{equation}
\begin{remark}
    We note here the derivative $r^{-1}\partial^{\bar j}(rf)$ appearing in the previous definition which is adapted to the conformal boundary at $r=\infty$. These derivatives are the natural ones for measuring regularity with respect to the conformal geometry at infinity and coincide with the twisted derivatives introduced in \cite{W13,HW14}.
\end{remark}
 We will also define for any $A\ge 0$:
\begin{align} \nonumber
\mathscr F^{(j)}_A[f]&(\tau) \doteq  
\sum_{\bar j =0}^j \sum_{i=0}^1\sum_{j_1+j_2=0}^2\sum_{J=0}^1 \Big\| L^{J-\bar j} \big( w(y,\theta) \big)^{A + 2j_1+ j_2+2J}\partial^{j_1+\bar j} \bar\partial^{j_2} V^i (rf)|_{t^*=\tau}  \Big\|^2_{L^\infty_{y, \theta, \varphi}(\{y\le y_\mathrm{mirror}\})}\\
& + \sum_{\bar j=0}^j \sum_{i=0}^1\sum_{j_0=0}^2 \Big\| L^{1-\bar j} \partial^{j_0+\bar j}  V^i (rf)|_{t^*=\tau}  \Big\|^2_{L^\infty_{r, \theta, \varphi}(\{ r\le r_\mathrm{mirror}\})}\nonumber \\
& 
+\sum_{\bar j =0}^j \sum_{i=0}^1\sum_{j_1+j_2=0}^2  \sum_{J=0}^1 \Big\|  L^{\f12+J-\bar j} \big( w(y,\theta) \big)^{A+2j_1+ j_2+2J}\partial^{j_1+\bar j} \bar\partial^{j_2} V^i (rf)|_{t^*=\tau}  \Big\|^2_{L^\infty_{\varphi}L^2_{y,\theta}(\{y\le y_\mathrm{mirror}\}, \sin\theta dy d\theta)}      \nonumber\\
& + \sum_{\bar j=0}^j \sum_{i=0}^1\sum_{j_0=0}^2 \Big\| L^{\f32-\bar j} \partial^{j_0+\bar j}  V^i (rf)|_{t^*=\tau}  \Big\|^2_{L^\infty_{\varphi}L^2_{r,\theta}(\{r\le r_\mathrm{mirror}\}, r^{-2}\sin\theta dr d\theta)}
. \label{Driving norm}
\end{align}
Finally, we will set for any $A\ge 0$:
\begin{equation}\label{Weaker driving norm}
\mathscr G^{(j)}_A[f](\tau) \doteq \,  \sum_{\bar j=0}^j  \Big( \big(w(y,\theta)\big)^A L^{-\bar j} \partial^{\bar j} (rf)  \big\|^2_{L^\infty_{y,\theta, \varphi}(\{y\le y_\mathrm{mirror}\})}+ \big\|  L^{-\bar j} \partial^{\bar j} (rf)  \big\|^2_{L^\infty_{r,\theta, \varphi}(\{r\le r_\mathrm{mirror}\})} \Big)
\end{equation}
\end{definition}

\begin{remark}
Note that the last two terms in \eqref{Basic norm} come with a degenerating  $\big(\log L\big)^{-1}$ factor; this is necessitated by the Sobolev inequalities implemented in our bootstrap argument and the logarithmic loss in the functional inequalities of \cref{lem:Hardy type inequality 2d} in the Appendix. In order to close the bootstrap estimates appearing in the proof of \cref{prop: Estimates error term}, it is crucial that the exponent of $(\log L)^{-1}$ factor appearing above is strictly less than $2$. 
\end{remark}

\begin{remark}
The quantities $\mathcal E[f](\tau)$  and $\mathcal E^{(j)}[f](\tau)$ emerge naturally while deriving an energy identity for the conformal wave equation $\square_g f+2f=0$ using $Y$ as a multiplier and $\big\{V, Y, \partial_{t^*} \big\}$ as commutators (see also \cref{lem:Coercivity energy}). 
\end{remark}
\begin{remark}
The norm \eqref{Driving norm} is adjusted to the behavior of wave packets of frequency $\sim L$ travelling along the null geodesics in the direction of $\partial_{t^*}+\partial_\varphi$ on the conformal boundary $y=0$. In particular, functions $f$ which are ``well controlled'' by $\mathscr F^{(j)}_0[f](\tau)$ are expected to be localized in the region $\{y^2+|\theta-\frac{\pi}2|^2\} \lesssim L^{-1}$ (where $w(y,\theta)\sim L^{-\f12}$) and their derivatives with respect to $V$, $L^{-1}\partial_{t^*}$ and $L^{-\f12}\bar\partial $ to be well-behaved. The approximate solution $\tilde\phi$ is a prototypical example of such a function; see \cref{lem: Bound F norm phi tilde} below. The weaker norm \eqref{Weaker driving norm} will be used to control lower order terms in $\psi$ in the context of our bootstrap argument; see \eqref{Estimate C term via Holder}. We will mainly use  \eqref{Driving norm} -- \eqref{Weaker driving norm}  for $A=3$ and $A=6$  (in view of the $r^{-6}$ weight in front of the nonlinear terms in \eqref{IVP Psi}; note that $r^{-A} \lesssim w^A$).
\end{remark}
\begin{remark}
Strictly speaking, it is only the functionals $f\rightarrow \sqrt{\mathcal E^{(j)}[f](\tau)}$, $f\rightarrow \sqrt{\mathscr F^{(j)}_A[f](\tau)}$ and $f\rightarrow \sqrt{\mathscr G^{(j)}_A[f](\tau)}$  that are norms (i.e.~satisfy the triangle inequality). Nevertheless, through a minor abuse of terminology, we will also refer to $\mathcal E^{(j)}[\cdot](\tau)$, $\mathscr F^{(j)}_A[\cdot](\tau)$ and $\mathscr G^{(j)}_A[\cdot](\tau)$ as norms.
\end{remark}

\bigskip

The following result shows that the norms $\mathcal E^{(j)}[\cdot]$ and $\mathscr F^{(j)}[\cdot]$ are well-adapted to the structure of the approximate solution $\tilde \phi$:

\begin{lemma}\label{lem: Bound F norm phi tilde}
Let $k=(n_k, \ell_k, m_k) \in \mathcal K$. Then, for any $j\in \mathbb N$, $A\ge 0$ and any $\tau\in [0,T_1]$, the following bounds hold:
\begin{itemize}
\item If $k\in \mathcal K\setminus \mathcal K_{\mathrm{D}}$, we have
\begin{equation}\label{Source norm non dominant modes}
 \mathscr F^{(j)}_A[\chi \tilde\phi_k](\tau) \lesssim_{j,A}  \f{L^{9-6s}}{n_k^{10}} \cdot \ind_{[L,3(\lambda+2)L]}(\ell_k) \cdot \ind_{\mathrm{res}}(m_k),
\end{equation}
and
\begin{equation}\label{Energy norm non dominant modes}
 \mathcal E^{(j)}[\chi \tilde\phi_k](\tau) \lesssim_j  \f{L^{10-6s}}{n_k^{10}} \cdot \ind_{[L,3(\lambda+2)L]}(\ell_k) \cdot \ind_{\mathrm{res}}(m_k),
\end{equation}
where $\ind_{\mathrm{res}}(m_k) \in \{0,1\}$ is defined by \eqref{Indicator set m}.
\item If $k\in \mathcal K_{\mathrm{D}}$, we have
\begin{equation}\label{Source norm bound dominant modes}
 \mathscr F^{(j)}_A[\chi\tilde\phi_k](\tau) \sim_{j,A} L^{1-A}  |b_k(\tau)|^2 +O(L^{-6s-1-A})
\end{equation}
and
\begin{equation}\label{Energy norm bound dominant modes}
 \mathcal E^{(j)}[\chi\tilde\phi_k](\tau) \sim_j L^2  |b_k(\tau)|^2 +O(L^{-6s}),
\end{equation}
where $b_k(t)$ is the slowly oscillating amplitude defined by \eqref{B k variables}.
\end{itemize}
In particular, the approximate solution $\tilde \phi$  satisfies provided $A<4s-11$:
\begin{equation}\label{Estimate scri F model solution}
 \mathscr F^{(j)}_A[\chi\tilde\phi](\tau) \sim_{j,A} L^{1-A}  \sum_{k\in \mathcal K_{\mathrm{D}}} |b_k(\tau)|^2 \sim_{j,A} L^{1-2s-A}
\end{equation}
and
\begin{equation}\label{Estimate energy E model solution}
\mathcal E^{(j)}[\chi\tilde\phi](\tau) \sim_j L^2  \sum_{k\in \mathcal K_{\mathrm{D}}} |b_k(\tau)|^2 \sim_{j} L^{2-2s}.
\end{equation}
In  all the   above, the constants implicit in the $\lesssim$ (and $\lesssim_{j,A}$) notation are assumed to depend on $N, \lambda, M$ (and $j,A$), but are \textbf{independent} of $k$, $\tau$ and $L$.
\end{lemma}

\medskip
\begin{proof}
In this proof, we will assume that the constants implicit in the $\lesssim$ notation might depend on $N, \lambda, M$, but will be independent of $\tau$,  $k$ and the frequency scale $L$. Recall that, according to \cref{prop:Total bound non-dominant modes}, we have $\tilde\phi_k \equiv 0$ unless 
\begin{equation}\label{Similar l}
  L\le  \ell_k \le 3(\lambda+2) L.
\end{equation}
Thus, for the rest of the proof, we will assume that $k$ is such that \eqref{Similar l} holds.

For any $k\in \mathcal K$, recall that the function $\tilde\phi_k:\{r\ge r_\mathrm{mirror}\}\rightarrow \mathbb C$ takes the following form (noting also that $t=t^*$ in the region $\{r\ge r_\mathrm{mirror}\}$):
\[
\tilde\phi_k(t,y, \theta, \varphi) = \f{a_k(t)}{r} R_{n_k,\ell_k}(y) Y_{\ell_k, m_k}(\theta, \varphi).
\]
Let us also note the following facts:
\begin{itemize}
\item 
For any $(n_k, \ell_k)\in( \mathbb N^*)^2$, in view of the fact that $\|R_{n_k,\ell_k}\|_{L^2\big([0,y_\mathrm{mirror}]\big)} = 1$ and satisfies \[\f{d^2}{dy^2} R_{n_k,\ell_k} = \big(V_{\ell_k}-\omega_{n_k,\ell_k}^2) R_{n_k,\ell_k}= O(\ell_k^2+n_k^2) R_{n_k,\ell_k}\] (the fact that $\omega_{n_k,\ell_k}\lesssim \ell_k+n_k$ following from \cref{lem:Crude Weyl law}), we can readily estimate for any $j\in \mathbb N$:  
\[
\big\|(\ell_k+n_k)^{-\f12-j} \f{d^j}{dy^j} R_{n_k,\ell_k}\big\|_{L^\infty([0,y_\mathrm{mirror}]} \lesssim_j 1.
\]

\item The spherical harmonics $Y_{\ell,m}$ (given by \eqref{Expression spherical harmonic}) satisfy the pointwise bounds 
\[
\Big\| \ell^{-j-\f12} \nabla_{\mathbb S^2}^j Y_{\ell,m}\Big\|_{L^\infty(\mathbb S^2)} \lesssim 1.
\]
\end{itemize}
Therefore, we can readily estimate for any $k\in \mathcal K$ satisfying \eqref{Similar l}:
\[
\big\| \partial^j \tilde\phi_k \big\|_{L^\infty_{y,\theta,\varphi}} \lesssim \sum_{\bar j=0}^j L^{1+\bar j}\Big|\f{d^{j-\bar j} a_k}{dt^{j-\bar j}}(t)\Big|,
\]
from which we immediately infer the (non-optimal) estimate for any $A\ge 0$:
\[
\mathscr F^{(j)}_A[\chi\tilde\phi_k](\tau) \lesssim L^{10} \sum_{\bar j=0}^{j+3} L^{-2\bar j}\Big|\f{d^{\bar j} a_k}{dt^{\bar j}}(t)\Big|^2.
\]
In the case of when $k$ belongs to the non-dominant set of parameters $\mathcal K \setminus \mathcal K_{\mathrm{D}}$, \cref{lem:The non dominant modes} then implies that
\[
\sup_{\tau \in [0,T_1]} \mathscr F^{(j)}_A[\chi\tilde\phi_k](\tau) \lesssim_j L^{11-6s} \mathcal P_k^2 \sum_{\bar j=0}^{j+3} L^{-2\bar j}\omega_k^{2\bar j-2}.
\]
Using  \cref{lem:eigenfunctions-estimate-m>0,lem:Crude Weyl law} to estimate $\omega_k-\ell_k\sim n_k = O(\ell_k^{1-\delta_0})$ (the last relation following from our definition of the set $\mathcal K$) and \cref{lem:Bounds P k} to estimate $\mathcal P_k$, we deduce \eqref{Source norm non dominant modes} for any $k\in \mathcal K\setminus \mathcal K_{\mathrm{D}}$ and any $j\in \mathbb N$. The bound \eqref{Energy norm non dominant modes} follows similarly.

In the case of the dominant modes $\{\tilde{\phi}_k\}_{k\in \mathcal K_{\mathrm{D}}}$, we will utilize the structure of the eigenfunctions $\{E_k\}_{k\in \mathcal K_{\mathrm{D}}}$ to obtain more precise estimates. In particular, recall that the dominant modes take the form
\[
\tilde\phi_k(t,y, \theta, \varphi) = \f{b_k(t)}{r} e^{-i\varepsilon_k \omega_k t} R_{n_k,\ell_k}(y) Y_{\ell_k, \varepsilon_k \ell_k}(\theta, \varphi).
\]
where $\varepsilon_k \in \{-1,1\}$, $\ell_k\sim L$, $n_k \lesssim 1$ and the slowly oscillating amplitudes $b_k$ satisfy the bounds provided by \cref{prop:The 3 times 3 system}. Note also the following facts:
\begin{itemize}
\item As a consequence of \cref{lem:eigenfunctions-estimate-m>0} for the functions $R_{n_k,\ell_k}$ and \cref{lem: WKB asymptotics Airy} of the Appendix for the asymptotics of the normalized Hermite functions $e_n$ (and using the fact that $R_{n_k,\ell_k}$ satisfies the ODE  \eqref{Radial boundary value problem} in order to calculate higher order derivatives of $R_{n_k,\ell_k}$ in terms of lower order terms), we can readily estimate for any $k\in \mathcal K_{\mathrm{D}}$ and any $j\in \mathbb N$: 
\begin{equation}\label{Dominant radial eigenfunction L infty}
\Big|\f{d^j}{dy^j} R_{n_k, \ell_k}\Big|\lesssim_j L^{\f14+\f12j}
\end{equation}
and
\begin{equation}\label{Dominant radial eigenfunction decay L infty}
\sup_{y\ge \sqrt{\f{4n_k-1}{\ell_k}}} \Big|L^{-\f14-\f12j} e^{\f12 N^{-\f14} L^{\f34}(y-\sqrt{\f{4n_k-1}{\ell_k}})^{\f32}}\f{d^j}{dy^j} R_{n_k, \ell_k}\Big|\lesssim_j 1
\end{equation}
(the latter bound can be interpreted as the statement that the support of $R_{n_k,\ell_k}$ is concentrated in the region $\{y\lesssim L^{-\f12}\}$).

\item The spherical harmonic $Y_{\ell,\pm\ell}$ takes the form
\[
Y_{\ell,\pm\ell}(\theta, \varphi)= \f{\gamma_\ell}{\sqrt{2\pi}} (\sin\theta)^\ell  e^{\pm i\ell \varphi},
\]
 where
    \[
    \gamma_\ell \doteq \left( \frac{ \Gamma(  \ell+\frac 32)}{\sqrt\pi \Gamma(\ell + 1)}\right)^{\frac 12 }.
    \]
 Therefore, we can estimate for any $j_1, j_2 \in \mathbb N$:
 \begin{equation}\label{Dominant angular eigenfunction L infty}
 \Big|L^{-\f14-\f12j_1 - j_2} \f{\partial^{j_1}}{\partial \theta^{j_1}} \f{\partial^{j_2}}{\partial \varphi^{j_2}} Y_{\ell_k, \pm\ell_k} \Big| \lesssim_{j_1, j_2} 1
 \end{equation}
 and
 \begin{equation}\label{Dominant angular eigenfunction decay L infty}
\sup_{|\f\pi2-\theta|\gtrsim L^{-\f12}} \Big|L^{-\f14-\f12j_1-j_2} e^{\f12 L |\f\pi2-\theta|}\f{\partial^{j_1}}{\partial \theta^{j_1}} \f{\partial^{j_2}}{\partial \varphi^{j_2}} Y_{\ell_k, \pm\ell_k} \Big|\lesssim_{j_1, j_2} 1
 \end{equation}
(the latter bound can be interpreted as the statement that the support of $Y_{\ell_k, \pm\ell_k} $ is concentrated in the region $\{|\f\pi2-\theta| \lesssim L^{-\f12}\}$).

\item The weight function $w(y,\theta)$ satisfies 
\[
w\lesssim L^{-\f12} \quad \text{on} \quad \{y\lesssim L^{-\f12}\} \cap \{|\f\pi2-\theta| \lesssim L^{-\f12}\}.
\]

\item The helical vector field $V$ almost commutes with $e^{-i \varepsilon_k \omega_k t}E_k(y, \theta, \varphi)$, in the sense that
\[
V \big( e^{-i \varepsilon_k \omega_k t}E_k \big) = O(1) \cdot   e^{-i \varepsilon_k \omega_k t}E_k,
\]
\end{itemize}
where the constant implicit in $O(1)$ is independent of $L$.  
Combining the above facts (in view also of the inclusion $\mathrm{\text{supp}}(\chi \tilde\phi_k \subset \{y\le y_\mathrm{mirror}\}$ and the fact that $\bar\partial \in \{\partial_\theta, \partial_y\}$ in the region $\{y\ll1\}\cap \{|\f\pi2-\theta|\ll 1\}$), we can readily calculate that, for any $k\in \mathcal K_{\mathrm{D}}$:

\begin{align}
 \mathscr F^{(j)}_A[\chi\tilde\phi_k](\tau)  = \, &
\sum_{\bar j =0}^j \sum_{i=0}^1\sum_{j_1+j_2=0}^2\sum_{J=0}^1
 \Big\|
   L^{J-\bar j}
   w(y,\theta)^{A+2j_1+ j_2+2J}
   \partial^{j_1+\bar j} \bar\partial^{j_2}
   V^i (\chi r \tilde\phi_k)\big|_{t^*=\tau}
 \Big\|^2_{L^\infty_{y, \theta, \varphi}(\{y\le y_\mathrm{mirror}\})}
     \nonumber\\
&
+\sum_{\bar j =0}^j \sum_{i=0}^1\sum_{j_1+j_2=0}^2  \sum_{J=0}^1
 \Big\|
   L^{\f12+J-\bar j}
   w(y,\theta)^{A+2j_1+ j_2+2J}
   \partial^{j_1+\bar j} \bar\partial^{j_2}
   V^i (\chi r \tilde\phi_k)\big|_{t^*=\tau}
 \Big\|^2_{\substack{L^\infty_{\varphi}L^2_{y,\theta}(\{y\le y_\mathrm{mirror}\},\\
                    \sin\theta \, dy \, d\theta)}}
     \nonumber   \\
\sim\, 
&  \sum_{\bar j=0}^{j+2}\sum_{i=0}^1
 \Big|
   L^{\f12-\f{A}2}
   \Big(L^{-1}\f{d}{dt}\Big)^{\bar j}
   \f{d^i}{dt^i} b_k(\tau)
 \Big|^2.     
\end{align}

Using the bound $\Big| \f{d^{\bar j} b_k}{dt^{\bar j}} \Big| \lesssim_{\bar j} L^{-3s+\bar j-2}$ for any $k\in \mathcal K_{\mathrm{D}}$ and  $\bar j\ge 1$ (as a consequence of \cref{prop:The 3 times 3 system}), we finally infer \eqref{Source norm bound dominant modes}. The bound \eqref{Energy norm bound dominant modes} follows similarly, by noting that

\begin{align*}
 \mathcal E^{(j)}[\chi\tilde\phi_k](\tau)  \sim_j \, &
\sum_{\bar j =0}^j
 \int_{\{t^*=\tau\}\cap\{y\le y_\mathrm{mirror}\}}
 \Big[
   |L^{-\bar j} \bar\partial^2\partial^{\bar j}(r\tilde\phi_k)|^2
   + w(y, \theta)^2
     |L^{-\bar j}\bar\partial \partial^{1+\bar j}(r \tilde\phi_k)|^2
\\
&\hphantom{
\sum_{\bar j =0}^j
 \int_{\{t^*=\tau\}\cap\{y\le y_\mathrm{mirror}\}} \Big[}
   + w(y, \theta)^4
     |L^{-\bar j}\partial^{2+\bar j}(r \tilde\phi_k)|^2
   + |L^{-\bar j} \partial^{1+\bar j} V (r \tilde\phi_k)|^2
\\
&\hphantom{
\sum_{\bar j =0}^j
 \int_{\{t^*=\tau\}\cap\{y\le y_\mathrm{mirror}\}} \Big[}
   + |L^{-\bar j} \partial^{1+\bar j}(r \tilde\phi_k)|^2
   + \frac{1}{y^2}|r L^{-\bar j} \partial^{\bar j} \tilde\phi_k|^2
 \Big]\, \sin\theta \, dy \, d\theta \, d\varphi
\\
&
+ \big(\log L\big)^{-1}
  \sum_{\bar j =0}^j   \sup_{p \in \mathbb N}
 \Bigg(
   \int_{\substack{\{t^*=\tau\}\cap\{y\le y_\mathrm{mirror}\}\\
                   \cap\{ w(y, \theta) \le 2^{-p}\}}}
   \frac{2^p}{w(y,\theta)}
   |L^{-\bar j} \bar\partial  \partial^{\bar j} (r \tilde\phi_k)|^2\,
   \sin\theta \, dy \, d\theta \, d\varphi
 \Bigg)
\\
&
+ \big(\log L\big)^{-1}
  \sum_{\bar j =0}^j   \int_0^{2\pi}
 \Bigg[
   \sup_{\substack{(y,\theta)\in(0,y_\mathrm{mirror})\\\times(0,\pi)}}
   \frac{| L^{-\bar j} \partial^{\bar j} (r \tilde\phi_k)|^2
          (\tau,y,\theta, \varphi)}
        {w(y,\theta)^2}
 \Bigg] \, d\varphi
\\
\sim\, 
&  \sum_{\bar j=0}^{j+2}
 \Big|
   L \Big(L^{-1}\f{d}{dt}\Big)^{\bar j}  b_k(\tau)
 \Big|^2.
\end{align*}

Noting that the functional $f\rightarrow \sqrt{\mathscr F^{(j)}[f](\tau)}$ is a norm, hence satisfies the triangle inequality, we can estimate
\[
\sum_{k\in \mathcal K \setminus \mathcal K_{\mathrm{D}}} \mathscr F^{(j)}_A[\chi \tilde\phi_k](\tau) \lesssim
\Big(\sum_{k\in \mathcal K \setminus \mathcal K_{\mathrm{D}}} \sqrt{\mathscr F^{(j)}_A[\chi \tilde\phi_k](\tau)}\Big)^2
 \stackrel{\eqref{Source norm non dominant modes}}{\lesssim_{j,A}} L^{12-6s}.
\]
Thus,  the bound \eqref{Estimate scri F model solution} follows readily combining the above with \eqref{Source norm bound dominant modes}, provided $A<4s-11$. Similarly for the energy bound \eqref{Estimate energy E model solution}

\end{proof}

\subsection{Sobolev-type estimates for the energy norm \texorpdfstring{$\mathcal E^{(j)}[\cdot]$}{Ej}}

In this section, we will collect a number of Sobolev-type estimates providing bounds  for the $L^\infty$-type norms $\mathscr F^{(j)}_A$ and $\mathscr G^{(j)}_A$ in terms of the energy norms $\mathcal E^{(j')}$; these estimates will then be used to control the nonlinear terms in the equation \eqref{IVP Psi} for $\psi$ in the context of the bootstrap argument of \cref{sec:The bootstrap argument}.

In particular, the following holds:
\begin{lemma}\label{lem:Sobolev estimates driving norms}
For any sufficiently regular function $f:\{0\le t^* \le T_1\}\rightarrow \mathbb C$, any $\tau \in [0,T_1]$, any $j\in \mathbb N$ and any $A\ge 0$, we can estimate (provided $L\gg_{j,A} 1$):
\begin{equation}\label{Sobolev estimate G and G V norm}
\mathscr G^{(j)}_A[f](\tau) + \mathscr G^{(j)}_A[Vf](\tau) \le C_{j, A}  L^2 \cdot \mathcal E^{(j+1)}[f](\tau)
\end{equation}
and
\begin{equation}\label{Sobolev estimate F norm}
\mathscr F^{(j)}_A [f](\tau) \le C_{j,A} L^{9} \cdot \mathcal E^{(j+2)}[f](\tau)
\end{equation}
In the above, the constant $C_{A,j}$ depends only on $A\ge0$, $j$ and the geometry of $(\mathcal M_\mathrm{ext}, g)$ (in particular, they are \textbf{independent} of $L$).
\end{lemma}

\begin{proof}
In this proof, the constants implicit in the $\lesssim$ (and $\lesssim_j$) notation will only depend on the geometry of $(\mathcal M_\mathrm{ext}, g)$ (and, in addition, $j$, respectively). In view of the fact that, for any $A\ge 0$, we have $(w(y,\theta))^{A}\lesssim_A 1$, it trivially follows from the expressions \eqref{Driving norm}--\eqref{Weaker driving norm} for $\mathscr F^{(j)}_A[\cdot]$ and $\mathscr G^{(j)}_A[\cdot]$ that, for any smooth function $f$,
\[
\mathscr F^{(j)}_A[f]\lesssim_A \mathscr F^{(j)}_0[f] \quad \text{and} \quad \mathscr G^{(j)}_A[f]\lesssim_A \mathscr G^{(j)}_0[f].
\]
Therefore, it suffices to establish \eqref{Sobolev estimate G and G V norm}--\eqref{Sobolev estimate F norm} for $A=0$. Moreover, it suffices to establish \eqref{Sobolev estimate G and G V norm} for $j=0$ (since the higher order bound follows by applying the $j=0$ estimate to terms of the form  $r^{-1}L^{-\bar j} \partial^{\bar j} (rf)$ for $\bar j=0, \ldots, j$). In particular, it suffices to establish that (provided $L\gg 1$):
\begin{equation}\label{Sobolev estimate f and V f energy}
\big\| rf \big\|^2_{L^{\infty}_{\Sigma^*_\tau}}+ \big\| V(rf) \big\|^2_{L^{\infty}_{\Sigma^*_\tau}}\lesssim \mathcal E[r^{-1}\partial (rf)](\tau) +  \mathcal E[f](\tau).
\end{equation}
The above follows directly from the standard Sobolev embedding: Since the energy norm \eqref{Basic norm} controls all  first order derivatives of $f$ and $Vf$ without degeneracy, namely
\[
\sum_{\bar j=0}^1\sum_{i=0}^1 \Big(\big\|  \partial^{\bar j} V^i (rf)|_{\{y\le y_\mathrm{mirror}\}} \big\|^2_{L^2_{y, \theta, \varphi}(\sin\theta dy d\theta d\varphi)}  +
\big\|  \partial^{\bar j} V^i (rf)|_{\{r\le r_\mathrm{mirror}\}} \big\|^2_{L^2_{r, \theta, \varphi}(\sin\theta dr d\theta d\varphi)}\Big)
 \lesssim \mathcal E[f],
\]
we can readily bound:
\begin{align*}
\big\| V(rf) \big\|^2_{L^{\infty}_{\Sigma^*_\tau}}+ \big\| rf \big\|^2_{L^{\infty}_{\Sigma^*_\tau}} 
& \lesssim 
\sum_{\bar j=0}^2\sum_{i=0}^1 \Big(\big\|  \partial^{\bar j} V^i (rf)|_{\{y\le y_\mathrm{mirror}\}} \big\|^2_{L^2_{y, \theta, \varphi}(\sin\theta dy d\theta d\varphi)}  \\ & \hphantom{\lesssim}+
\big\|  \partial^{\bar j} V^i (rf)|_{\{r\le r_\mathrm{mirror}\}} \big\|^2_{L^2_{r, \theta, \varphi}(\sin\theta dr d\theta d\varphi)}\Big)
\\
& \lesssim \mathcal E[r^{-1} \partial(rf)](\tau)+ \mathcal E[f](\tau),
\end{align*}
i.e.~\eqref{Sobolev estimate f and V f energy} holds.

From \eqref{Driving norm} by keeping the $V$ derivatives separate we obtain 
\[   \mathscr F^{(j)}[f](\tau)\lesssim L^7 ( \mathscr G^{(j+2)} [f] (\tau) + \mathscr G^{(j+2)} [Vf] (\tau) ).\]
Then applying the bound \eqref{Sobolev estimate G and G V norm} to each term on the right-hand side gives 
\[ \mathscr F^{(j)}[f](\tau)\lesssim L^9 \cdot \mathcal E^{(j+3)}[f](\tau),\]
which then gives the desired bound \eqref{Sobolev estimate F norm}.
\end{proof}

\subsection{The commuted equations for \texorpdfstring{$\psi$}{psi}}
Assume that $\psi$ is a sufficiently regular solution of the initial-boundary value problem \eqref{IVP Psi} on a region of the form $\{0\le t^* < T^*\}$. Commuting \eqref{IVP Psi} with $V$, we can readily calculate that $V\psi$ solves the boundary value problem:
\begin{equation}\label{IVP V Psi}
\begin{cases}
\square_g (V \psi) + 2 V \psi -r^{-6}|\chi\tilde\phi+\psi|^2 \partial_{t^*}^2(V \psi) -  2\chi^2 r^{-6} \partial_{t^*}^2 \tilde\phi \, \Re\{\tilde\phi\, V\bar \psi\} - \chi^2  r^{-6} |\partial_{t^*} \tilde\phi|^2 V\psi\\
\hphantom{\square_g (V \psi) + 2 V \psi -|\chi\tilde\phi+\psi|^2 \partial_{t^*}V\psi}
 -\chi^2 \mathcal N^{(1)}_V[\tilde \phi;\psi]   - \chi \mathcal N^{(2)}_V[\tilde\phi; \psi] -\mathcal N^{(3)}_V[\psi] = - V(\mathcal F[\tilde \phi]),\\[5pt]
rV\psi|_{r=\infty} =0,
\end{cases}
\end{equation}
where
\begin{align*}
\mathcal N_{(1)}^{[V]} [\tilde\phi; \psi] \doteq & \, r^{-6} \Big( 2\Re\{\bar{\tilde\phi} V(\tilde\phi) \}\partial_{t^*}^2\psi  + 2\partial_{t^*}^2 (V \tilde\phi)  \Re\{\tilde\phi \bar \psi\} +  2\partial_{t^*}^2 \tilde\phi \, \Re\{V\tilde\phi \bar \psi\}  \\
&  
+  2\Re\{\partial_{t^*} \bar{\tilde\phi}  \partial_{t^*} (V\tilde\phi)\} \psi  
+ 2 V\tilde \phi  \Re\{\partial_{t^*} \tilde\phi \partial_{t^*} \bar\psi\} + 2 \tilde \phi \Re\{\partial_{t^*} (V\tilde\phi) \partial_{t^*} \bar\psi\}+ 2 \tilde\phi \Re\{\partial_{t^*} \tilde\phi  \partial_{t^*} (V\bar\psi)\}\Big),\\[5pt]
\mathcal N_{(2)}^{[V]} [\tilde\phi; \psi] \doteq &\,  r^{-6} \Big(\partial_{t^*}^2 (V\tilde\phi) \, |\psi|^2  +
2\partial_{t^*}^2 \tilde\phi \, \Re\{\psi \, V\bar\psi\} \\
&\quad 
+ V\tilde\phi\, |\partial_{t^*} \psi|^2 + 2\tilde\phi\, \Re\{ \partial_{t^*} \psi \, \partial_{t^*} (V\psi) \} \\
&\quad + \Re\{\partial_{t^*} (V\tilde\phi) \partial_{t^*} \bar \psi\} \psi + \Re\{\partial_{t^*} \tilde\phi \partial_{t^*} (V\bar \psi)\} \psi + \Re\{\partial_{t^*} \tilde\phi \partial_{t^*} \bar \psi\} V\psi \\
&\quad + \partial_{t^*} (V\tilde\phi)\, \psi \partial_{t^*} \bar\psi + \partial_{t^*} \tilde\phi\, V\psi \partial_{t^*} \bar\psi + \partial_{t^*} \tilde\phi\, \psi \partial_{t^*} (V\bar\psi) + 2\Re\{\tilde\phi \, V\bar\psi\} \partial_{t^*}^2 \psi + 2\Re\{(V\tilde\phi) \, \bar\psi\} \partial_{t^*}^2 \psi\Big)\\[5pt]
\mathcal N_{(3)}^{[V]}[\psi] \doteq & \, r^{-6} \Big(V\psi |\partial_{t^*} \psi|^2 + 2\psi \Re\{ \partial_{t^*} \psi \partial_{t^*} (V\bar\psi) \} + 2\Re\{\psi V\bar\psi\} \partial_{t^*}^2 \psi \Big).
\end{align*}
Note that, while writing down \eqref{IVP V Psi}, we have separated the linear terms $|\chi\tilde\phi+\psi|^2 \partial_{t^*}^2(V \psi)$,  $2\chi^2 \partial_{t^*}^2 \tilde\phi \, \Re\{\tilde\phi\, V\bar \psi\}$ and $\chi^2  |\partial_{t^*} \tilde\phi|^2 V\psi$ from the rest of the terms in $\mathcal N_{(1)}^{[V]}$; this is because these terms will require some special treatment when deriving an energy inequality for \eqref{IVP V Psi}.

Similarly, commuting \eqref{IVP Psi} $j_1$ times  with $L^{-1} Y$ and $j_2$ times with $L^{-1} \partial_{t^*}$, we infer that $\psi^{(j_1;j_2)}$ solves:
\begin{equation}\label{IVP Psi k}
\begin{cases}
\square_g  \psi^{(j_1;j_2)} + 2 \psi^{(j_1;j_2)}-L^{-j_1}\big[\square_g, Y^{j_1}\big]\psi^{(0;j_2)} -r^{-6}|\chi\tilde\phi+\psi|^2 \partial_{t^*}^2\psi^{(j_1;j_2)}\\
\hphantom{\square_g  \psi^{(k)} +}
 -\chi^2 \mathcal N_{(1,j_1, j_2)}[\tilde \phi;\psi]   - \chi \mathcal N_{(2,j_1, j_2)}[\tilde\phi; \psi] -\mathcal N_{(3,j_1, j_2)}[\psi] = - (\mathcal F[\tilde \phi])^{(j_1;j_2)},\\[5pt]
r\psi^{(j_1;j_2)}|_{r=\infty} =0,
\end{cases}
\end{equation}
where, schematically (setting $j\doteq j_1+j_2$):
\begin{align*}
\mathcal N_{(1,j_1, j_2)}[\tilde \phi;\psi] =
& \,   r^{-6}\Big( \Phi^{(\le j)} \partial \Phi^{(\le j)} \cdot  \partial \Psi^{(\le j)}
 + \Big(\Phi^{(\le j)} \partial^2 \Phi^{(\le j)}+\partial \Phi^{(\le j)}  \partial \Phi^{(\le j)}\Big)\cdot  \Psi^{(\le j)}\Big)
,\\[5pt]
\mathcal N_{(2,j_1, j_2)}[\tilde \phi;\psi] =
& \, 
 r^{-6}\Big[\Big(\Phi^{(\le j)}  \partial \Psi^{(\le \f j2 +1)} + \partial \Phi^{(\le j)}  \Psi^{(\le \f j2 +1)}\Big) \cdot \partial \Psi^{(\le j)} \\
&\hphantom{\, \sum_{\le k=0}^k \Bigg\{\big(\Phi}
+ \Big(\partial^2\Phi^{(\le j)}  \Psi^{(\le \f j2+1)}+\partial\Phi^{(\le j)} \cdot \partial \Psi^{(\le \f j2+1)}+\Phi^{(\le j)}  \partial^2\Psi^{(\le \f j2+1)} \Big) \cdot \Psi^{(\le j)}
\Big],\\[5pt]
\mathcal N_{(3,j_1, j_2)}[\psi] =
& \,  
r^{-6} \Big( \Psi^{(\le \f j2 +1)}  \partial \Psi^{(\le \f j2 +1)} \cdot \partial \Psi^{(\le j)} 
+ \partial \Psi^{(\le \f j2+1)}  \partial \Psi^{(\le \f j2+1)} \cdot \Psi^{(\le j)}\Big).
\end{align*}

  \begin{remark} In the above, we use the shorthand notation that $\Phi$ denotes any linear combination of $\Re\{\tilde\phi\}$ and $\Im\{\tilde\phi\}$ with coefficients which are bounded functions with bounded derivatives of any order; similarly, $\Psi$
denotes any linear combination of $\Re\{\psi\}$ and $\Im\{\psi\}$ with coefficients as before. 
  \end{remark}

Finally, commuting \eqref{IVP Psi k} with $V$, we calculate:
\begin{equation}\label{IVP V Psi k}
\begin{cases}
\square_g  (V \psi^{(j_1;j_2)}) + 2 V \psi^{(j_1;j_2)} -L^{-j_1}\big[\square_g, Y^{j_1}\big] V\psi^{(0;j_2)} -r^{-6}|\chi\tilde\phi+\psi|^2 \partial_{t^*}^2(V\psi^{(j_1;j_2)}) \\
\hphantom{\square_g (V \psi) }
-  \chi^2 r^{-6} \big(\partial_{t^*}^2 \tilde\phi \, \bar{\tilde\phi}+|\partial_{t^*} \tilde\phi|^2 \big)^{(j_1;j_2)}  \cdot V\psi- \chi^2 r^{-6} \big(\partial_{t^*}^2 \tilde\phi \, \tilde\phi \big)^{(j_1;j_2)}  \cdot V\bar\psi \\
\hphantom{\square_g (V \psi) }
 -\chi^2 \mathcal N_{(1,j_1,j_2)}^{[V]}[\tilde \phi;\psi]   - \chi \mathcal N_{(2,j_1,j_2)}^{[V]}[\tilde\phi; \psi] -\mathcal N_{(3,j_1,j_2)}^{[V]}[\psi] = - V(\mathcal F[\tilde \phi])^{(j_1;j_2)},\\[5pt]
r V\psi^{(j_1, j_2)}|_{r=\infty} =0,
\end{cases}
\end{equation}
where, setting $j\doteq j_1+j_2$ as before:
\begin{align*}
\mathcal N_{(1,j_1,j_2)}^{[V]}[\tilde \phi;\psi] = 
& \, r^{-6} \Bigg( \sum_{\bar j=1}^j \Bigg\{  \Phi^{(\le j)} \partial \Phi^{(\le j)}  \cdot  \partial V \Psi^{(\bar j)}
+ \big( V\Phi^{(\le j)} \partial \Phi^{(\le j)} +\Phi^{(\le j)} \partial V \Phi^{(\le j)} \big) \cdot  \partial  \Psi^{(\bar j)}\\
& \hphantom{r^{-6} (\sum_{\bar j=1}^j \Bigg\{\Phi}
 + \Big(\Phi^{(\le j)} \partial^2\Phi^{(\le j)}+\partial \Phi^{(\le j)}  \partial \Phi^{(\le j)}\Big)\cdot  V\Psi^{(\bar j)}
\Bigg\}\\
&
 \!\! +  \Phi^{(\le j)} V\Phi^{(\le j)} \cdot  \partial^2 \Psi^{(\le j)} 
+  \Big(\Phi^{(\le j)} \partial^2 V\Phi^{(\le j)}+ V\Phi^{(\le j)} \partial^2 \Phi^{(\le j)}+\partial \Phi^{(\le j)}  \partial V\Phi^{(\le j)}\Big)\cdot  \Psi^{(\le j)}\Bigg)
\end{align*}
\begin{align*}
\mathcal N_{(2,j_1,j_2)}^{[V]}[\tilde \phi;\psi] = 
& \, 
r^{-6} \Bigg[\Big( V\Phi^{(\le j)} \Psi^{(\le \f j2 +1)} + \Phi^{(\le j)} V\Psi^{(\le \f j2 +1)} \Big) \cdot \partial^2 \Psi^{(\le j)}\\
&\hphantom{\, \sum_{\le j=0}^k \Bigg\{\big(\Phi}
+ \Big(\Phi^{(\le j)}  \partial \Psi^{(\le \f j2 +1)} + \partial \Phi^{(\le j)}  \Psi^{(\le \f j2 +1)}\Big) \cdot \partial V \Psi^{(\le j)} \\
&\hphantom{\, \sum_{\le k=0}^k \Bigg\{\big(\Phi}
+\Big(V\Phi^{(\le j)}  \partial \Psi^{(\le \f j2 +1)}+\Phi^{(\le j)}  \partial V\Psi^{(\le \f j2 +1)} \\
&\hphantom{\, \sum_{\le k=0}^k \Bigg\{\big(\Phi+\Big(V}
+ \partial V\Phi^{(\le j)}  \Psi^{(\le \f j2 +1)}+ \partial \Phi^{(\le j)}  V\Psi^{(\le \f j2 +1)}\Big) \cdot \partial \Psi^{(\le j)}\\
&\hphantom{\, \sum_{\le k=0}^k \Bigg\{\big(\Phi}
+ \Big(\partial^2\Phi^{(\le j)}  \Psi^{(\le \f j2+1)}+\partial \Phi^{(\le j)} \cdot \partial \Psi^{(\le \f j2+1)}+\Phi^{(\le j)}  \partial^2\Psi^{(\le \f j2+1)} \Big) \cdot V\Psi^{(\le j)}\\
&\hphantom{\, \sum_{\bar k=0}^k \Bigg\{\big(\Phi}
+ \Big(\partial^2\Phi^{(\le j)}  \Psi^{(\le \f j2+1)}+\partial \Phi^{(\le j)} \cdot \partial \Psi^{(\le \f j2+1)}+\Phi^{(\le j)}  \partial^2\Psi^{(\le \f j2+1)} \Big) \cdot V\Psi^{(\le j)}\\
&\hphantom{\, \sum_{\bar k=0}^k \Bigg\{\big(\Phi}
+ \Big(\partial^2V\Phi^{(\le j)}  \Psi^{(\le \f j2+1)}+\partial^2\Phi^{(\le j)}  V\Psi^{(\le \f j2+1)}
+\partial V\Phi^{(\le j)} \cdot \partial \Psi^{(\le \f j2+1)}\\
&\hphantom{\, \sum_{\bar k=0}^k \Bigg\{\big(\Phi}
+\partial \Phi^{(\le j)} \cdot \partial V\Psi^{(\le \f j2+1)}+V\Phi^{(\le j)}  \partial^2\Psi^{(\le \f j2+1)}
+\Phi^{(\le j)}  \partial^2V\Psi^{(\le \f j2+1)} \Big) \cdot \Psi^{(\le j)}\Bigg],
\end{align*}
\begin{align*}
\mathcal N_{(3,j_1,j_2)}^{[V]} [\psi] = 
& \,
r^{-6} \Bigg[V \Psi^{(\le \f j2 +1)}  \Psi^{(\le \f j2 +1)} \cdot \partial^2 \Psi^{(\le j)}
+ \Psi^{(\le \f j2 +1)}  \partial\Psi^{(\le \f j2 +1)} \cdot \partial V\Psi^{(\le j)} \\
&\hphantom{ \, \sum_{\le k=0}^k \Bigg\{ \Phi}
+\Big(V\Psi^{(\le \f j2 +1)}  \partial \Psi^{(\le \f j2 +1)}+ \Psi^{(\le \f j2 +1)}  \partial V\Psi^{(\le \f j2 +1)}\Big) \cdot \partial \Psi^{(\le j)}\\
&\hphantom{ \, \sum_{\le k=0}^k \Bigg\{ \Phi} 
+ \sum_{\bar j=1}^j \partial \Psi^{(\le \f j2+1)}  \partial \Psi^{(\le \f j2+1)} \cdot V\Psi^{(\bar j)}
+ \sum_{\bar j=1}^j \partial \Psi^{(\le \f j2+1)}  \partial V\Psi^{(\le \f j2+1)} \cdot \Psi^{(\bar j)}\Bigg]
\end{align*}
(see the remark above \eqref{IVP V Psi k} for the $\Phi, \Psi$ notation). Note that we have separated in  \eqref{IVP V Psi k} the linear terms in $V\psi$ containing no  derivatives of $V\psi$ from $\mathcal N_{(1,j_1,j_2)}^{[V]}[\tilde\phi;\psi]$.

\subsection{Energy estimates for the linear inhomogeneous wave equation}\label{subsec:Energy estimates linear}
Let $f:\mathcal M_\mathrm{ext}\rightarrow \mathbb C$ be a smooth function satisfying 
\begin{equation}\label{Inhomogeneous wave equation}
\begin{cases}
\square_g f + 2f = F,\\
rf|_{r=\infty} =0.
\end{cases}
\end{equation}
In this section, we will derive the degenerate energy identity for $f$ associated to the Killing vector field $\partial_{t^*}$, as well as the non-degenerate energy inequality associated to the red-shift vector field $Y$ defined by \eqref{Red shift vector field}.

\subsubsection{Coercivity properties of the energy fluxes}
We will now study the coercivity properties of the energy fluxes associated to the vector fields $\partial_{t^*}$ and $Y$ (see  \cref{sec:Energy momentum} for the definition of the notion of the energy flux associated to a vector field).

 In the case of the vector field  $\partial_{t^*}$, the associated energy flux is positive-definite but degenerates at $r=r_+$; in particular, the following lower bound holds (see \cite{HS13}):

\begin{lemma}\label{lem:Degenerate energy trivial bound}
Let $f:\mathcal M_\mathrm{ext}\rightarrow \mathbb C$ be a smooth and sufficiently decaying function. Then, for any $\tau \ge 0$:
\begin{align}\label{Trivial dt energy coercivity}
\int_{t^*=\tau} \frac12\Bigg(
& g_{rr} |\partial_{t^*}f|^2+  \f1{18}(-g_{t^* t^*})|\partial_r f|^2\\
&  \hphantom{\Sigma\Sigma\Sigma\Sigma\Sigma}
+ r^{-2}|\nabla_{\mathbb S^2} f|^2_{g_{\mathbb S^2}}   + \f18 |f|^2\Bigg) \,  \, r^2\sin\theta dr d\theta d\varphi
\le \int_{t^*=\tau} J_\mu^{(\partial_{t^*})}[f]  \, \hat n^\mu_{\Sigma^*} \, \dvol_{\Sigma^*}, \nonumber
\end{align}
where $g_{rr}$ and $g_{t^* t^*}$ denote the corresponding components of $g$ in the $(t^*, r, \theta, \varphi)$ coordinate system.
\end{lemma}

\begin{remark} Recall that, near $r=r_+$, we have $g_{rr}\sim 1$ and $g_{t^* t^*} \sim -(r-r_+)$, while as $r\rightarrow +\infty$ we have $g_{rr}\sim r^{-2}$, $g_{t^* t^*} \sim -r^2$. 
\end{remark}

\begin{proof}
We can explicitly calculate:
\begin{align}\label{First expression flux}
\int_{t^*=\tau} J_\mu^{(\partial_{t^*})}[f] & \, \hat n^\mu_{\Sigma^*} \, \dvol_{\Sigma^*}\\
 = &  \int_{t^*=\tau} \Big( \Re\big\{\hat n_{\Sigma^*} f \cdot \partial_{t^*} \bar f \big\} - \f12 \big(\partial^\alpha f \partial_\alpha \bar f - 2|f|^2\big)g(\hat n_{\Sigma^*}, \partial_{t^*})\Big) \, \dvol_{\Sigma^*} \nonumber\\
= & \int_{t^*=\tau} \frac{1}{2\sqrt{-g^{t^* t^*}}}\Re\Bigg\{
(-g^{t^*t^*})|\partial_{t^*}f|^2+  g^{rr} |\partial_r f|^2 + r^{-2}|\nabla_{\mathbb S^2} f|^2_{g_{\mathbb S^2}} - 2 |f|^2\Bigg\} \,  \,\sqrt{g_{rr}}r^2\sin\theta dr d\theta d\varphi   \nonumber\\
= & \int_{t^*=\tau} \frac12\Bigg(
g_{rr}|\partial_{t^*}f|^2+  (-g_{t^*t^*}) |\partial_r f|^2 + r^{-2}|\nabla_{\mathbb S^2} f|^2_{g_{\mathbb S^2}} - 2 |f|^2\Bigg) \,  \, r^2\sin\theta dr d\theta d\varphi,\nonumber 
\end{align}
where, in passing to the last line above, we made use of the identities \eqref{Inverse metric components} for the components $g^{\mu\nu}$ of the Schwarzschild--AdS metric in the $(t^*, r, \theta, \varphi)$ coordinate system.

Using the Hardy inequality \eqref{Hardy SchAdS} provided by \cref{lem:Hardy inequality global} from the Appendix to estimate
\[
\int_{t^*=\tau} 2(1+\f1{18})r^2 |f|^2 \, \sin\theta drd\theta d\varphi \le \int_{t^*=\tau} (1-\f1{18}) r^2 (-g_{t^* t^*}) |\partial_r f|^2 \, \sin\theta drd\theta d\varphi,
\]
we immediately deduce \eqref{Trivial dt energy coercivity} from \eqref{First expression flux}.
\end{proof}

In the case of the red-shift vector field \eqref{Red shift vector field}, the corresponding energy flux does not degenerate near $r=r_+$:
\begin{lemma}\label{lem:Non degenerate energy trivial bound}
Let us assume that the parameter $\delta_2>0$ appearing in the definition \eqref{Red shift vector field} of the red-shift vector field $Y$ has been fixed small enough depending only on the geometry of $(\mathcal M_\mathrm{ext}, g_M)$.
Then there exists a $C>0$ such that for any smooth function $f:\mathcal M_\mathrm{ext}\rightarrow \mathbb C$ and any $\tau \ge 0$, we have:
\begin{equation}\label{Trivial Y energy coercivity}
\int_{t^*=\tau} \Bigg( r^{-2} |\partial_{t^*}f|^2+  r^2 |\partial_r f|^2
+ r^{-2}|\nabla_{\mathbb S^2} f|^2_{g_{\mathbb S^2}}   +  |f|^2\Bigg) \,  \, r^2\sin\theta dr d\theta d\varphi
\le C \int_{t^*=\tau} J_\mu^{(Y)}[f]  \, \hat n^\mu_{\Sigma^*} \, \dvol_{\Sigma^*}.
\end{equation}
\end{lemma}

\begin{proof}

Using the formula \eqref{Red shift vector field} for the vector field $Y$, we can readily calculate that
\begin{align}\nonumber
 \int_{t^*=\tau} J_\mu^{(Y)}[f]  \, \hat n^\mu_{\Sigma^*} \, \dvol_{\Sigma^*}
= & \int_{t^*=\tau}\Big(\zeta(r) J_\mu^{(\partial_{t*}-\lambda' \partial_r)}[f] +  J_\mu^{(\partial_{t*})}[f]\Big)  \, \hat n^\mu_{\Sigma^*} \, \dvol_{\Sigma^*} \\
= &  \int_{t^*=\tau}\!\! \!\! \zeta(r) T_{\mu\nu}[f] (\partial_{t^*}-\lambda' \partial_r)^\mu \hat n^\nu_{\Sigma^*} \, \dvol_{\Sigma^*} 
+  \int_{t^*=\tau} J_\mu^{(\partial_{t*})}[f]  \, \hat n^\mu_{\Sigma^*} \, \dvol_{\Sigma^*} .  \label{First inequality non degenerate flux}
\end{align}
In view of the fact that $\eta\in (0,1]$ was chosen in \eqref{Red shift vector field} so that the vector field $\partial_{t*}-\eta \partial_r$ is uniformly timelike on $\mathrm{\text{supp}}\zeta \subset \{r_+ \le r \le r_+ +\delta_2\}$, we can readily estimate that
\[
 T_{\mu\nu}[f] (\partial_{t^*}-\eta \partial_r)^\mu \hat n^\nu_{\Sigma^*} 
\ge c_{\delta_2, \eta} |\partial f|^2 - C_* |f|^2 \quad \text{for} \quad r\in [r_+, r_++\delta_2],
\]
where $c_{\delta_2, \eta}>0$ depends on the precise choice of the parameters $\eta, \delta_2$ appearing in the definition \eqref{Red shift vector field} of $Y$ while $C_*>0$ depends only on the geometry of $(\mathcal M_\mathrm{ext}, g_M)$ in the near horizon region. Therefore, in view of the degenerate lower bound \eqref{Trivial dt energy coercivity} for $\int_{t^*=\tau} J_\mu^{(\partial_{t*})}[f]  \, \hat n^\mu_{\Sigma^*} \, \dvol_{\Sigma^*} $, we can readily estimate from \eqref{First inequality non degenerate flux}:
\begin{align} \nonumber
 \int_{t^*=\tau} J_\mu^{(Y)}[f]  \, \hat n^\mu_{\Sigma^*} \, \dvol_{\Sigma^*}
& \ge 
\f{c_{\delta_2, \eta}}{100}  \int_{t^*=\tau} \Bigg( r^{-2} |\partial_{t^*}f|^2+  r^2 |\partial_r f|^2
+ r^{-2}|\nabla_{\mathbb S^2} f|^2_{g_{\mathbb S^2}}   +  |f|^2\Bigg) \,  \, r^2\sin\theta dr d\theta d\varphi
\\ & +  \int_{\{t^*=\tau\}\cap\{r\le r_++2\delta_2\}}\!\!\! \Big(\f{(r-r_+)}{100}|\partial_r f|^2+ (\f{1}{100}- C_* \zeta(r)) |f|^2 \Big) \sin\theta dr d\theta d\varphi.   \label{Second inequality non degenerate flux}
\end{align}
Using the fact that
\[
0\le \zeta \le 2 \quad \text{and} \quad \mathrm{\text{supp}}\zeta \subseteq \{r_+\le r \le r_++\delta_2\}
\]
together with the trivial Hardy-type inequality
\[
\int_{r_+}^{r_++\delta_2} \f{1}{(r-r_+)^\f12} |f|^2 \,  dr \le C \int_{r_+}^{r_++2\delta_2} \Big((r-r_+)|\partial_r f|^2+ |f|^2 \Big) \, dr,
\]
we obtain \eqref{Trivial Y energy coercivity} from \eqref{Second inequality non degenerate flux} provided $\delta_2$ has been chosen small enough in terms of the constant $C_*$ (which only depends on the geometry  of $(\mathcal M_\mathrm{ext}, g_M)$).
\end{proof}

Using the expression for the wave equation (together with appropriate elliptic estimates), we can show that the $Y$-energy flux of $Vf$ and $f$ controls the energy-type norm \eqref{Basic norm}:

\begin{lemma}\label{lem:Coercivity energy}
Let $f:\{0\le t^*\le T\}\rightarrow \mathbb C$ be a smooth function satisfying
\begin{equation}\label{Inhomogeneous wave equation 2}
\begin{cases}
\square_g f + 2f = F,\\
rf|_{r=\infty}=0.
\end{cases}
\end{equation}
Then, for any $\tau\in [0,T]$, we can estimate
\begin{equation}\label{Coercivity}
\mathcal E[f](\tau) \lesssim \int_{t^*=\tau} \Big(  J_\mu^{(Y)}[f] +J_\mu^{(Y)}[Vf]  \Big)\, \hat n^\mu_{\Sigma^*} \, \dvol_{\Sigma^*}  + \int_{\{t^*=\tau\}\cap\{y\le y(2M)\}} |r^3F|^2\, \sin\theta dy d\theta d\varphi , 
\end{equation}
where $\mathcal E[f](\tau)$ was defined by \eqref{Basic norm} and  the constant implicit in the $\lesssim$ notation above depends only on the geometry of $(\mathcal M_\mathrm{ext}, g)$ and the precise choice of the vector field $Y$.
\end{lemma}

\begin{proof} 
Using the bound \eqref{Trivial Y energy coercivity} for $Vf$ and $f$, we infer:
\begin{align}\label{Second expression flux}
 \int_{t^*=\tau} \Bigg(
r^{-2} & |\partial_{t^*}Vf|^2+ r^2 |\partial_r Vf|^2 + r^{-2}|\nabla_{\mathbb S^2} Vf|^2_{g_{\mathbb S^2}}+|Vf|^2 \\
&\quad + r^{-2}|\partial_{t^*}f|^2+ r^2 |\partial_r f|^2 + r^{-2}|\nabla_{\mathbb S^2} f|^2_{g_{\mathbb S^2}} + |f|^2 \Bigg) \,  \, r^2\sin\theta dr d\theta d\varphi \nonumber\\
 & \quad \quad \quad \quad \quad \quad \hphantom{\Sigma\Sigma}
\lesssim 
 \int_{t^*=\tau}\Big( J_\mu^{(Y)}[Vf]   \, \hat n^\mu_{\Sigma^*} + J_\mu^{(Y)}[f] \, \hat n^\mu_{\Sigma^*} \Big) \, \dvol_{\Sigma^*}   . \nonumber 
\end{align}

Let us now focus on the region $\{y\lesssim 1\}$ and switch to $(t^*, y, \theta, \varphi)$ coordinates. We will use equation \eqref{Inhomogeneous wave equation} (in particular, the relation \eqref{Wave operator V}), together with standard elliptic estimates, to show that the left-hand side of \eqref{Second expression flux} in fact controls every second derivative of $rf$ with appropriate $w$-weights as $y\rightarrow 0$. More precisely, the relation \eqref{Wave operator V} implies that
\begin{align}\label{Substitution from equation}
\int_{\{t^*=\tau\}\cap\{y\le y(2M)\}} & \Big(|\mathfrak L(rf) |^2 +|rf|^2\Big)\, \sin\theta dy d\theta d\varphi \\
  \stackrel{\hphantom{\eqref{Second expression flux}}}{\lesssim} &
 \int_{\{t^*=\tau\}\cap\{y\le y(2M)\}}\Big(|\partial_{t^*} V(rf) |^2 +|\partial_{\varphi} V(rf)|^2 + |rf|^2 + |r^3F|^2\Big)\, \sin\theta dy d\theta d\varphi,     \nonumber \\
 \stackrel{\eqref{Second expression flux}}{\lesssim} &
 \int_{t^*=\tau} J_\mu^{(Y)}[Vf]  \, \hat n^\mu_{\Sigma^*} \, \dvol_{\Sigma^*} + \int_{t^*=\tau} J_\mu^{(Y)}[f]  \, \hat n^\mu_{\Sigma^*} \, \dvol_{\Sigma^*} \nonumber
\\ + &   \int_{\{t^*=\tau\}\cap\{y\ge y(2M)\}} |r^3F|^2\, \sin\theta dy d\theta d\varphi,     \nonumber 
\end{align}
where $\mathfrak L$ is the operator \eqref{Elliptic operator} which is elliptic in the region $\{r>2M\}= \{y< y(2M)\}$. Using standard elliptic estimates (essentially expanding the square $|\mathfrak L(rf) |^2$ and integrating by parts; see also the remark below \eqref{Elliptic operator} about the behavior of the coefficients of $\mathfrak L$ as $y\rightarrow 0$), together with the Dirichlet boundary condition $rf|_{y=0}=0$, we can readily show that
\begin{align*}
\int_{\{t^*=\tau\}\cap\{y\le y_\mathrm{mirror}\}}\Big( & |\partial_y^2(rf)  |^2 +  |\partial_y \partial_\theta(rf)|^2 +|\partial_\theta^2(rf)|^2+\big(w(y,\theta)\big)^4\|\nabla^2_{\mathbb S^2}(rf)\|^2_{g_{\mathbb S^2}} \\
& + \f{\big(w(y,\theta)\big)^2}{\sin^2\theta} \big(|\partial_y \partial_\varphi (rf)|^2 +  |(\nabla^2_{\mathbb S^2})_{\theta\phi} (rf)|^2 \big) +|rf|^2\Big)\, \sin\theta dy d\theta d\varphi \nonumber  \\
& \hphantom{+ \f{\big(w(y,\theta)\big)^2}{\sin\theta} \big(|\partial_y \partial_\varphi}
 \lesssim \int_{\{t^*=\tau\}\cap\{y\le y(2M)\}}\Big(|\mathfrak L(rf) |^2 +|rf|^2\Big)\, \sin\theta dy d\theta d\varphi,
\end{align*}
where $w(y,\theta)$ is the modified distance function from $(y,\theta)=(0, \f{\pi}2)$ defined by \eqref{Distance function} and the constants implicit in the notation $\lesssim$ above depend on the geometry of $(\mathcal M_\mathrm{ext}, g)$. Thus, combining the above inequality with \eqref{Substitution from equation} \eqref{Second expression flux} and using the fact that
\[
\partial_{t^*}(rf) = \big(1+L^{-1}\big)^{-1} \Big( V(rf)-\partial_\varphi (rf)\Big)  \text{ and }  \partial_{t^*}^2(rf) = \big(1+L^{-1}\big)^{-2} \Big( (1+L^{-1})\partial_{t^*}V(rf) -  \partial_{\varphi} V(rf) + \partial_\varphi^2(rf) \Big),
\]
we infer that:
\begin{align}\label{Top terms control V energy}
\int_{\{t^*=\tau\}\cap\{y\le y_\mathrm{mirror}\}}& \Big[|\bar\partial^2(rf)  |^2+\big(w(y, \theta)\big)^2|\partial\bar\partial (rf)|^2+\big( w(y, \theta)\big)^4|\partial^2(rf)|^2 \\
& \hphantom{ \Big[|\bar\partial^2(rf)  |^2+\big(w}
+  |\partial V(rf)|^2+ |\partial (rf)|^2 + \frac{1}{y^2}|rf|^2\Big]\, \sin\theta dy d\theta d\varphi   \nonumber \\
\lesssim
 \int_{t^*=\tau} &  \Big(J_\mu^{(Y)}[Vf]  +J_\mu^{(Y)}[f]  \Big)\, \hat n^\mu_{\Sigma^*} \, \dvol_{\Sigma^*}    
+  \int_{\{t^*=\tau\}\cap\{y\le y(2M)\}} |r^3F|^2\, \sin\theta dy d\theta d\varphi,     \nonumber 
\end{align}
where the shorthand notation regarding $|\partial h|^2$, $|\bar\partial h|^2$ etc.\ was introduced in \eqref{Shorthand square derivatives}.

Applying the functional inequalities of   \cref{lem:Hardy type inequality 2d} for the function $rf$ with respect to the  $(y, \theta)$ variables (with the center chosen to be at  $(y,\theta)=(0, \f{\pi}2))$), combined with the simpler Sobolev inequality
\[
\sup_{(y,\theta)\in(0,y_\mathrm{mirror})\times(0,\pi)}\big(|rf|^2(\tau, y, \theta, \varphi)\big) \lesssim \int_0^\pi \int_0^{y_\mathrm{mirror}} \Big(|\bar\partial^2 (rf)|^2  +|rf|^2 \Big)(\tau, y, \theta, \varphi) \, \sin\theta dy d\theta,
\]
and the observation that 
\[
w(y, \theta)\sim L^{-\f12}\quad \text{for} \quad |y|^2+|\theta-\f\pi2|^2\lesssim L^{-1},
\]
 we readily deduce that
\begin{multline}
\sup_{k\in \mathbb N} \Bigg(\int_{\{t^*=\tau\}\cap\{y\le y_\mathrm{mirror}\}\cap\{ w(y, \theta) \le 2^{-k}\}}   
\f{2^k}{w(y,\theta)}|\bar\partial (rf)|^2\, \sin\theta dy d\theta d\varphi      \Bigg)  \\  
  + 
\int_0^{2\pi} \Bigg[\sup_{(y,\theta)\in(0,y_\mathrm{mirror})\times(0,\pi)}\frac{|rf|^2(\tau,y,\theta, \varphi)}{\big(w(y,\theta)\big)^2}\Bigg] \, d\varphi \\
 \lesssim \big(\log L\big)\int_{\{t^*=\tau\}\cap\{y\le y_\mathrm{mirror}\}}\Big( |\bar\partial^2 (rf)|^2 +|rf|^2\Big) \,\sin\theta dy d\theta d\varphi.   \label{Linfty bound final}
\end{multline}
Combining \eqref{Second expression flux}, \eqref{Top terms control V energy} and \eqref{Linfty bound final}, we finally obtain \eqref{Coercivity}.

\end{proof}

The following result is a higher order analog of \cref{lem:Coercivity energy}:

\begin{lemma}\label{lem:Coercivity energy higher order}
Let $f:\{t^*\ge 0\}\rightarrow \mathbb C$ be a smooth function satisfying
\begin{equation}\label{Wave equation system}
\begin{cases}
\square_g f + 2f = F,\\
rf|_{r=\infty}=0.
\end{cases}
\end{equation}
Then, for any integer $j \ge 0$ and any $\tau \ge 0$, we can estimate
\begin{equation}\label{Coercivity higher order}
 \mathcal E^{(j)}[f](\tau)
 \lesssim_j  \sum_{\bar j=0}^j \Bigg(\int_{t^*=\tau} \Big(  J_\mu^{(Y)}[f^{(\bar j;0)}] +J_\mu^{(Y)}[Vf^{(\bar j; 0)}]  \Big)\, \hat n^\mu_{\Sigma^*} \, \dvol_{\Sigma^*} + \int_{t^*=\tau} |r^3F^{(\bar j;0)}|^2  
 \, \sin\theta \f{1}{r^2}dr d\theta d\varphi  \Bigg), 
\end{equation}
where the higher order norm $\mathcal E^{(j)}[f](\tau)$ was defined by \eqref{Higher order norm} and  the constant implicit in the $\lesssim_j$ notation above depends only on $j$ and the geometry of $(\mathcal M_\mathrm{ext}, g)$.
\end{lemma}

\begin{proof}
We will establish \eqref{Coercivity higher order} by arguing inductively on $j$. For $j=0$, \eqref{Coercivity higher order} reduces to  \eqref{Coercivity}; let us assume, then, that $j\ge 1$ and that  \eqref{Coercivity higher order} holds for $\mathcal E^{(l)}[f](\tau)$ for $0\le l\le j-1$.

Let us consider the quantity $f^{(j;0)}=L^{-j} Y^j f$. Commuting \eqref{Wave equation system} with $L^{-j}Y^{j}$, we infer that $f^{(j;0)}$ satisfies
\[
\begin{cases}
\square_g f^{(j;0)} + 2f^{(j;0)} = F^{(j;0)} + L^{-j}[\square_g, Y^{j}] f,\\
r f^{(j;0)}|_{r=\infty} =0.
\end{cases}
\]
 Applying   \cref{lem:Coercivity energy} for $f^{(j;0)}$, we obtain 
\begin{align} \nonumber
\mathcal E[f^{(j;0)}](\tau) \lesssim  \int_{t^*=\tau} \Big( & J_\mu^{(Y)}[f^{(j;0)}] +J_\mu^{(Y)}[Vf^{(j;0)}]  \Big)\, \hat n^\mu_{\Sigma^*} \, \dvol_{\Sigma^*}\\
& 
+ \int_{\{t^*=\tau\}\cap\{y\le y(2M)\}} \Big(|r^3F^{(j;0)}|^2 + L^{-2j}|r^3 [\square_g, Y^{j}] f|^2\Big)\, \sin\theta dy d\theta d\varphi  .  \label{Coercivity higher order first}
\end{align}
Recall that, since $Y=\partial_{t^*}$ in the region $\{r\ge r_++\delta_2\}$, the term $[\square_g, Y^j] f$  is supported in the near-horizon region $\{r_+\le r\le r_++\delta_2\}$. In particular, $[\square_g, Y^j] f=0$ in the region $\{y\le y(2M)\}$; therefore,  \eqref{Coercivity higher order first} trivially reduces to:
\begin{align} \nonumber 
\mathcal E[f^{(j;0)}](\tau) \lesssim  \int_{t^*=\tau} \Big( & J_\mu^{(Y)}[f^{(j;0)}] +J_\mu^{(Y)}[Vf^{(j;0)}]  \Big)\, \hat n^\mu_{\Sigma^*} \, \dvol_{\Sigma^*}\\
& 
+ \int_{\{t^*=\tau\}\cap\{y\le y(2M)\}} |r^3F^{(j;0)}|^2\, \sin\theta dy d\theta d\varphi. \label{Coercivity higher order first 2}
\end{align}

In view of the expression \eqref{Two expression wave elliptic}, we can schematically write for any smooth function $h:\{t^*\ge 0\}\rightarrow \mathbb C$:
\begin{equation}\label{Two expression wave elliptic again}
r^3\big(\square_g h + 2h\big) = \mathcal L_* (rh) + \partial Y (r h) + \partial (r h) +O(r^{-2}) h,
\end{equation}
where $\mathcal L_* $ is a second-order elliptic operator in the spatial variables which is \emph{uniformly} elliptic up to $r=r_+$ and satisfies
\[
\mathcal L_* \sim \big(1+O(y^2)\big) \partial_y^2 +  \Delta_{\mathbb S^2} \quad \text{as} \quad r\rightarrow +\infty
\]
and
\[
\mathcal L_* \sim \big(1+O(r-r_+)\big) \partial_r^2 + r^{-2} \Delta_{\mathbb S^2}  \quad \text{as} \quad r\rightarrow  r_+.
\]
Therefore, we obtain via standard elliptic estimates (using also the functional inequalities of   \cref{lem:Hardy type inequality 2d} to obtain the analog of the bound \eqref{Linfty bound final} for $L^-{j}r^{-1}\partial^j(rf)$ in place of $f$):
\begin{align} \nonumber
\mathcal E[L^{-j}r^{-1} & \partial^j(r f)](\tau)  \lesssim \mathcal E[f^{(j;0)}](\tau)+\mathcal E^{(j-1)}[f](\tau) + \sum_{\bar j=0}^{j-1} \int_{\{t^*=\tau\}} \Big(|r^3(\square_g f^{(\bar j;0)}+ 2 f^{(\bar j;0)})|^2 \Big)\, \f1{r^2} \sin\theta dr d\theta d\varphi \\
& \lesssim
 \mathcal E[f^{(\bar j;0)}](\tau)+\mathcal E^{(j-1)}[f](\tau) + \sum_{\bar j=0}^{j-1} \int_{\{t^*=\tau\}} \Big(|r^3 F^{(\bar j;0)}|^2+ |r^3 L^{-\bar j}[\square_g,Y^{\bar j}]f|^2 \Big)\, \f1{r^2} \sin\theta dr d\theta d\varphi \nonumber \\
& \stackrel{\eqref{Coercivity higher order first 2}}{\lesssim}
\int_{t^*=\tau} \Big( J_\mu^{(Y)}[f^{(j;0)}] +J_\mu^{(Y)}[Vf^{(j;0)}]  \Big)\, \hat n^\mu_{\Sigma^*} \, \dvol_{\Sigma^*}   \nonumber \\
& \hphantom{\stackrel{\eqref{Coercivity higher order first 2}}{\lesssim}}
+\mathcal E^{(j-1)}[f](\tau) +  \int_{\{t^*=\tau\}} \Big(\sum_{\bar j=0}^j |r^3 F^{(\bar j;0)}|^2+ \sum_{\bar j=0}^{j-1} |r^3 L^{-\bar j}[\square_g,Y^{\bar j}]f|^2 \Big)\, \f1{r^2} \sin\theta dr d\theta d\varphi\label{After uniform elliptic estimate energy}
\end{align}
Note that the term $[\square_g, Y^{\bar j}] f$ is of the form $\partial^{\le \bar j+1} f$ and is supported in the near-horizon region $\{r_+\le r\le r_++\delta_2\}$; thus, we can estimate
\begin{align*}
 \sum_{\bar j=0}^{j-1} \int_{\{t^*=\tau\}} |r^3 L^{-\bar j}[\square_g,Y^{\bar j}]f|^2 \, \f1{r^2} \sin\theta dr d\theta d\varphi  \lesssim 
& \sum_{\bar j=0}^{j-1} \int_{\{t^*=\tau\}\cap \{r\le r_++\delta_2} | L^{-\bar j}\partial^{\bar j+1} f|^2 \, \sin\theta dr d\theta d\varphi  \\
\lesssim & \mathcal E^{(j-1)}[f](\tau).
\end{align*}
Therefore, using the above estimate to control the last term in the right-hand side of \eqref{After uniform elliptic estimate energy}, we obtain
\begin{align}\label{Coercivity higher order second}
\mathcal E[L^{-j}r^{-1}\partial^j (rf)](\tau) \lesssim & 
\int_{t^*=\tau} \Big( J_\mu^{(Y)}[f^{(j;0)}] +J_\mu^{(Y)}[Vf^{(j;0)}]  \Big)\, \hat n^\mu_{\Sigma^*} \, \dvol_{\Sigma^*}  \\
& 
+\mathcal E^{(j-1)}[f](\tau) + \sum_{\bar j=0}^j  \int_{\{t^*=\tau\}}|r^3 F^{(\bar j;0)}|^2\, \f1{r^2} \sin\theta dr d\theta d\varphi. \nonumber
\end{align}
Using the trivial bound
\[
\mathcal E^{(j)}[f](\tau)\lesssim \mathcal E[L^{-j}r^{-1}\partial^j(r f)](\tau)  + \mathcal E^{(j-1)}[f](\tau),
\]
(following directly from the definition \eqref{Higher order norm} of $\mathcal E^{(j)}[\cdot]$), we obtain from \eqref{Coercivity higher order second}:
\begin{align}\label{Coercivity higher order almost there}
\mathcal E^{(j)}[f](\tau) \lesssim & 
\int_{t^*=\tau} \Big( J_\mu^{(Y)}[f^{(j;0)}] +J_\mu^{(Y)}[Vf^{(j;0)}]  \Big)\, \hat n^\mu_{\Sigma^*} \, \dvol_{\Sigma^*}  \\
& 
+\mathcal E^{(j-1)}[f](\tau) + \sum_{\bar j=0}^k  \int_{\{t^*=\tau\}}|r^3 F^{(\bar j;0)}|^2\, \f1{r^2} \sin\theta dr d\theta d\varphi. \nonumber
\end{align}
Using our inductive assumption that \eqref{Coercivity higher order} holds for $\mathcal E^{(l)}[f](\tau)$ for $l\le j-1$ to estimate the second term in the right-hand side of \eqref{Coercivity higher order almost there}, we deduce  \eqref{Coercivity higher order}.

\end{proof}

\subsubsection{The \texorpdfstring{$\partial_{t^*}$}{partial t}- and \texorpdfstring{$Y$}{Y}-energy estimates}
In this section, we will make use of the divergence identity \eqref{Basic energy identity div} to obtain a number of a priori energy estimates for solutions to \eqref{Inhomogeneous wave equation}.

The following lemma is a straightforward consequence of the conservation of the (degenerate) energy flux associated to the Killing vector field $\partial_{t^*}$:

\begin{lemma}\label{lem:Energy inequality basic}
Let $f:\mathcal M_\mathrm{ext}\rightarrow \mathbb C$ satisfy \eqref{Inhomogeneous wave equation}. Then $f$ satisfies the following estimate for any $\tau\ge 0$:
\begin{align}\label{Energy inequality}
\int_{t^*=\tau} J_\mu^{(\partial_{t^*})}[f] \hat n^\mu_{\Sigma^*} \, \dvol_{\Sigma^*} +& \int_{\mathcal H^+\cap\{0\le t^*\le\tau\}} J_\mu^{(\partial_{t^*})}[f] \hat n^\mu_{\mathcal H^+} \, \dvol_{\mathcal H^+}\\
&  = \int_{t^*=0} J_\mu^{(\partial_{t^*})} \hat[f] n^\mu_{\Sigma^*} \, \dvol_{\Sigma^*} - \int_{0\le t^*\le \tau}\Re\big\{ F \cdot \partial_{t^*} \bar f\big\} \, \dvol_{g},    \nonumber
\end{align}
where, along the future event horizon $\mathcal H^+ = \{r=r_+\}$ in the $(t^*,r,\theta,\varphi)$ coordinate system we have $n_{\mathcal H^+} =\partial_{t^*}$, $ \dvol_{\mathcal H^+} = r^2 \sin\theta dr d\theta d\varphi$.\footnote{Note that, unlike the case of a spacelike or timelike hypersurface, the choices of a normal $n_{\mathcal H^+}$ and associated volume form $\dvol_{\mathcal H^+}$ for a null hypersurface like $\mathcal H^+$ are not canonical and, in particular, they are not unique; they can always be replaced by $(n_{\mathcal H^+},\dvol_{\mathcal H^+})  \rightarrow (\lambda' \cdot n_{\mathcal H^+},(\lambda')^{-1} \cdot\dvol_{\mathcal H^+}) $ for any smooth function $\lambda':\mathcal H^+ \rightarrow (0,+\infty)$. The corresponding flux term across $\mathcal H^+$, however, remains invariant under these changes.}
\end{lemma}

\begin{remark} Note that $J_\mu^{(\partial_{t^*})}[f] \hat n^\mu_{\mathcal H^+} = |\partial_{t^*} f|^2 \ge 0$.
\end{remark}

\begin{proof}
Integrating \eqref{Basic energy identity div} for $X=\partial_{t^*}$ over the domain $\{0\le t^* \le \tau \}$ and using the Dirichlet boundary condition $rf|_{r=\infty}=0$ (and the fact that $\partial_{t^*}$ is a Killing vector field), we obtain \eqref{Energy inequality}.

\end{proof}

The energy flux $J^{(\partial_{t^*})}_\mu[f] \hat n^\mu_{\Sigma^*}$ appearing in the left-hand side of \eqref{Energy inequality} degenerates at $r=r_+$, where $\partial_{t^*}$ becomes null. However, using the red-shift vector field \eqref{Red shift vector field} as a multiplier allows us to obtain the following non-degenerate estimate \cite{DR09}:

\begin{lemma}\label{lem:Non degenerate energy estimate}
There exists a constant $C>0$ (depending on the precise choice of the vector field \eqref{Red shift vector field}) such that for any function $f:\mathcal M_\mathrm{ext}\rightarrow \mathbb C$ satisfying \eqref{Inhomogeneous wave equation}, we can estimate for any $\tau\ge 0$:
\begin{align}\label{Non degenerate energy inequality}
\int_{t^*=\tau} J_\mu^{(Y)}[f] \hat n^\mu_{\Sigma^*} \, \dvol_{\Sigma^*} &
+ \int_{\mathcal H^+ \cap \{0\le t^*\le\tau\} } J_\mu^{(\partial_{t^*})}[f] \hat n^\mu_{\mathcal H^+} \, \dvol_{\mathcal H^+} \\
\le C \Bigg\{ &  \int_{t^*=0} J_\mu^{(Y)}[f] \hat n^\mu_{\Sigma^*} \, \dvol_{\Sigma^*} + \sup_{0\le \bar\tau \le \tau} \Big| \int_{\{0\le t^* \le \bar\tau\}} \Re\big\{ F \cdot \partial_{t^*} \bar f\big\} \, \dvol_{g}    \Big| \nonumber\\
& + \sup_{\substack{0\le \tau_1 \le \tau_2 \le \tau,\\ \tau_2-\tau_1 \le 1}}  \Big|  \int_{\{\tau_1\le t^*\le \tau_2\}}(t^*-\tau_1)\Re\big\{ F \cdot \partial_{t^*} \bar f\big\} \, \dvol_{g} \Big|  \nonumber \\
  & + \sup_{\substack{0\le \tau_1 \le \tau_2 \le \tau,\\ \tau_2-\tau_1 \le 1}}  \Big| \int_{\{\tau_1\le t^*\le \tau_2\}}\Re\big\{ F \cdot Y \bar f\big\} \, \dvol_{g} \Big| \Bigg\},  \nonumber
\end{align}
where $Y$ is the vector field \eqref{Red shift vector field} and, along the future event horizon $\mathcal H^+ = \{r=r_+\}$ in the $(t^*,r,\theta,\varphi)$ coordinate system we have $n_{\mathcal H^+} =\partial_{t^*}$, $ \dvol_{\mathcal H^+} = r^2 \sin\theta dr d\theta d\varphi$.
\end{lemma}

\begin{proof}
Integrating \eqref{Basic energy identity div} for $X=Y$ over the domain $\{\tau_1\le t^* \le \tau_2 \}$ for any $0\le \tau_1\le \tau_2 $ and using the Dirichlet boundary condition $rf|_{r=\infty}=0$, we obtain:
\begin{multline}
\int_{t^*=\tau_2} J_\mu^{(Y)}[f]  \hat n^\mu_{\Sigma^*} \, \dvol_{\Sigma^*} 
+ \int_{\mathcal H^+ \cap \{0\le t^*\le\tau\} } J_\mu^{(Y)}[f] \hat n^\mu_{\mathcal H^+} \,  \dvol_{\mathcal H^+}
+ \int_{\{\tau_1\le t^* \le \tau_2\}} T_{\mu\nu}[f] (\pi^{(Y)})^{\mu\nu} \, \dvol_g \\
  = \int_{t^*=\tau_1} J_\mu^{(Y)} \hat[f] n^\mu_{\Sigma^*} \, \dvol_{\Sigma^*} - \int_{\tau_1\le t^*\le \tau_2}\Re\big\{ F \cdot Y \bar f\big\} \, \dvol_{g}. \label{First div identity Y first}
\end{multline}
In view of the fact that 
\[
J_\mu^{(Y)}[f] \hat n^\mu_{\mathcal H^+} \ge c \big(|\partial_{t^*} f|^2 + |\nabla_{\mathbb S^2} f|^2 \big) - C_* |f|^2
\]
for some constants $c,C_*>0$ (following from the fact that $Y|_{\mathcal H^+}$ is future directed timelike and $\partial_{t^*}|_{\mathcal H^+}$ is future directed null), we infer that
\begin{multline}
\int_{t^*=\tau_2} J_\mu^{(Y)}[f]  \hat n^\mu_{\Sigma^*} \, \dvol_{\Sigma^*} 
+ \int_{\{\tau_1\le t^* \le \tau_2\}} T_{\mu\nu}[f] (\pi^{(Y)})^{\mu\nu} \, \dvol_g \\
  \le \int_{t^*=\tau_1} J_\mu^{(Y)} \hat[f] n^\mu_{\Sigma^*} \, \dvol_{\Sigma^*}
+C_* \int_{\mathcal H^+ \cap \{\tau_1\le t^* \le \tau_2\}}|f|^2 \, \dvol_{\mathcal H^+}
 - \int_{\tau_1\le t^*\le \tau_2}\Re\big\{ F \cdot Y \bar f\big\} \, \dvol_{g}. \label{First div identity Y}
\end{multline}
Recall that, in view of the way we defined $Y$, we have that $\pi^{(Y)} =0$  on $\{r\ge r_++\delta_2\}$ (where $Y = \partial_{t^*}$) and there exists a $\delta_3=\delta_3(M) \in (0, \delta_2)$ independent of $f$ such that
\begin{equation}\label{Red shift positivity again}
T^{\mu\nu}[f] \pi^{(Y)}_{\mu\nu} \ge C_2 |\partial f|^2 - C_3 |f|^2 \quad \text{on} \quad \{r_+\le r \le r_++\delta_3\}.
\end{equation}
Thus, we can estimate:
\begin{align}
\int_{\tau_1}^{\tau_2}
 \Bigg( \int_{t^*=\sigma} J_\mu^{(Y)}[f]\,
        \hat n^\mu_{\Sigma^*} \, \dvol_{\Sigma^*}\Bigg)\, d\sigma
& \stackrel{\hphantom{\eqref{Trivial Y energy coercivity}}}{\le}
  C \int_{\tau_1}^{\tau_2}
    \Bigg( \int_{t^*=\sigma}
      \big( |\partial(rf)|^2 +|rf|^2  \big)\, r^{-2}
      \sin\theta \, dr \, d\theta \, d\varphi \Bigg)\, d\sigma
  \nonumber \\
& \stackrel{\eqref{Trivial dt energy coercivity}}{\le}
  C' \int_{\{\tau_1\le t^* \le \tau_2\}}
      T_{\mu\nu}[f]\, (\pi^{(Y)})^{\mu\nu}\, \dvol_g
  \nonumber\\
&\qquad
  + C''\int_{\tau_1}^{\tau_2}
    \Bigg( \int_{t^*=\sigma}
      J_\mu^{(\partial_{t^*})}[f]\, \hat n^\mu_{\Sigma^*}\,
      \dvol_{\Sigma^*}\Bigg)\, d\sigma
  \label{Estimate deformation tensor}
\end{align}

for some constants $C,C',C''>0$ independent of $f, \tau_1, \tau_2$. Thus, if we set 
\[
E_Y[f](\sigma) \doteq \int_{t^*=\sigma} J_\mu^{(Y)}[f]  \hat n^\mu_{\Sigma^*} \, \dvol_{\Sigma^*}, \quad \text{and} \quad E_{\partial_{t^*}}[f](\sigma) \doteq \int_{t^*=\sigma} J_\mu^{(\partial_{t^*})}[f]  \hat n^\mu_{\Sigma^*} \, \dvol_{\Sigma^*},
\]
 we obtain by combining \eqref{First div identity Y} and \eqref{Estimate deformation tensor} and using the trivial trace inequality
\begin{align*}
\int_{\mathcal H^+ \cap \{\tau_1\le t^* \le \tau_2\}}|f|^2 \, \dvol_{\mathcal H^+} \lesssim_{\delta}&  \Big( \int_{\{\tau_1\le t^* \le \tau_2\}\cap \{r\le r_++\delta\}}  |\partial f|^2 \, \dvol_g \Big)^{\f12}\Big( \int_{\{\tau_1\le t^* \le \tau_2\}\cap \{r\le r_++\delta\}} |f|^2 \, \dvol_g \Big)^{\f12}\\
& +  \int_{\{\tau_1\le t^* \le \tau_2\}\cap \{r_++\f12\delta r\le r_++\delta\}} |f|^2 \, \dvol_g
\end{align*}
that, for any $0\le \tau_1 \le \tau_2$:
\begin{align}
E_Y[f](\tau_2) - &E_Y[f](\tau_1)  + \int_{\tau_1}^{\tau_2}  E_Y[f](\sigma) \, d\sigma  \nonumber \\
& \stackrel{\hphantom{\eqref{Energy inequality}}}{\le }  
C \Bigg( \int_{\tau_1}^{\tau_2} E_{\partial_{t^*}}[f](\sigma)\, d\sigma +\Big| \int_{\{\tau_1\le t^*\le \tau_2\}}\Re\big\{ F \cdot Y \bar f\big\} \, \dvol_{g} \Big| \Bigg)   \nonumber\\
& \stackrel{\eqref{Energy inequality}}{\le}
C' \Bigg( (\tau_2 -\tau_1) \sup_{0\le \bar\tau \le \tau_2}E_{\partial_{t^*}}[f](\bar \tau) +\Big| \int_{\tau_1}^{\tau_2} \Bigg(\int_{\{\tau_1\le t^*\le \bar\tau\}}\Re\big\{ F \cdot \partial_{t^*} \bar f\big\} \, \dvol_{g}  \Bigg) \, d\bar\tau\Big| \nonumber  \\
& \hphantom{ \stackrel{\eqref{Energy inequality}}{\lesssim} C' \Bigg( ( +}
  +  \Big| \int_{\{\tau_1\le t^*\le \tau_2\}}\Re\big\{ F \cdot Y \bar f\big\} \, \dvol_{g} \Big| \Bigg),     \label{Almost there for non degenerate inequality}
\end{align}
where the constants $C, C'>0$ are independent of $f$, $\tau_1$, $\tau_2$ and are not necessarily the same as before. Let us set, for convenience, 
\begin{align*}
\mathfrak A[\tau] \doteq
E_Y[f](0) &  + \sup_{0\le \bar\tau \le \tau} E_{\partial_t^*}[f](\bar\tau) 
+  \sup_{\substack{0\le \tau_1 \le \tau_2 \le \tau \\ \tau_2-\tau_1 \le 1}} \Big|\int_{\{\tau_1 \le t^* \le \tau_2 \}} (t^* -\tau_1) \Re \Big\{ F \cdot \partial_{t^*} \bar f \Big\} \, \dvol_g  \Big|  \\
& +  \sup_{\substack{0\le \tau_1 \le \tau_2 \le \tau \\ \tau_2-\tau_1 \le 1}}\Big| \int_{\{\tau_1\le t^*\le \tau_2\}}\Re\big\{ F \cdot Y \bar f\big\} \, \dvol_{g} \Big|
\end{align*}
From \eqref{Almost there for non degenerate inequality}, we can infer that, for all $\tau \ge 0$:
\begin{equation}\label{Contradiction non degenerate energy}
\sup_{0\le \bar\tau \le \tau} E_Y[f](\bar\tau) \le 4C' \mathfrak A[\tau].
\end{equation}
Indeed, assume, for the sake of contradiction, that there exists some $\tau\ge0$ for which \eqref{Contradiction non degenerate energy} does not hold. Let $\tau_* \in [0,\tau]$ be the time for which $E_Y[f](\bar\tau)$ achieves its maximum in $[0,\tau]$, i.e.
\begin{equation}\label{Contradiction 1}
E_Y[f](\tau_*)  = \max_{\bar\tau \in [0,\tau]} E_Y[f](\bar\tau)>4C' \mathfrak A[\tau].
\end{equation}
In particular,
\begin{equation}\label{Contradiction 2}
E_Y[f](\tau_*) \ge E_Y[f](\bar\tau) \quad \text{for all} \quad \bar\tau \in [0,\tau].
\end{equation}
 Without loss of generality, we can assume that $\tau, \tau_* \ge 1$ (since \eqref{Contradiction non degenerate energy} follows trivially on $\{0\le t^*\le 1\}$ through an application of Gr\"onwall's inequality on \eqref{Almost there for non degenerate inequality}). Then, applying \eqref{Almost there for non degenerate inequality} for $\tau_2=\tau_*$ and $\tau_1 \in[\tau_*-1, \tau_*]$, we obtain
\begin{align}\label{Almost there for contradiction}  
E_Y[f](\tau_*) - E_Y[f](\tau_1) & + \int_{\tau_1}^{\tau_*}  E_Y[f](\sigma) \, d\sigma \\
& \le 
C' \Bigg( (\tau_* -\tau_1) \sup_{0\le \bar\tau \le \tau_2}E_{\partial_{t^*}}[f](\bar \tau) +\Big| \int_{\tau_1}^{\tau_*} \Bigg(\int_{\{\tau_1\le t^*\le \bar\tau\}}\Re\big\{ F \cdot \partial_{t^*} \bar f\big\} \, \dvol_{g}  \Bigg) \, d\bar\tau\Big|  \nonumber \\
& \hphantom{ \stackrel{\eqref{Energy inequality}}{\lesssim} C' \Bigg( ( +}
  +  \Big| \int_{\{\tau_1\le t^*\le \tau_*\}}\Re\big\{ F \cdot Y \bar f\big\} \, \dvol_{g} \Big| \Bigg)  \nonumber    \\
& \le 
C' \mathfrak A[\tau].
\nonumber
\end{align}
In view of our assumption \eqref{Contradiction 2} for $E_Y[f](\tau_*)$, the above estimate yields for any $\tau_1 \in [\tau_*-1, \tau_*]$:
\begin{equation}\label{Pair of estimates contradiction}
\Big|E_Y[f](\tau_*) - E_Y[f](\tau_1)  \Big| \le C' \mathfrak A[\tau] \quad \text{and} \quad \int_{\tau_1}^{\tau_*}  E_Y[f](\sigma) \, d\sigma \le C' \mathfrak A[\tau].
\end{equation}
Since we assumed that $E_Y[f](\tau_*)$ satisfies \eqref{Contradiction 1}, we infer from the first bound in \eqref{Pair of estimates contradiction} that
\[
\inf_{\tau_1 \in [\tau_*-1, \tau_*] } E_Y[f](\tau_1) \ge 3 C' \mathfrak A[\tau]
\]
and, therefore,
\begin{equation}\label{Lower bound integral contradiction}
 \int_{\tau_*-1}^{\tau_*}  E_Y[f](\sigma) \, d\sigma \ge 3 C' \mathfrak A[\tau]
\end{equation}
which is in contradiction with the second bound in \eqref{Pair of estimates contradiction}. Therefore, \eqref{Contradiction non degenerate energy} holds.

Using \eqref{Energy inequality} for the term $\sup_{\bar\tau \in [0, \tau]} E_{\partial_{t^*}}[f](\bar \tau)$, we obtain from \eqref{Contradiction non degenerate energy} that, for any $\tau\ge 0$:
\begin{align}
\int_{t^*=\tau} J_\mu^{(Y)}[f] \hat n^\mu_{\Sigma^*} \, \dvol_{\Sigma^*} 
\le C \Bigg\{ &  \int_{t^*=0} J_\mu^{(Y)}[f] \hat n^\mu_{\Sigma^*} \, \dvol_{\Sigma^*} + \sup_{0\le \bar\tau \le \tau} \Big| \int_{\{0\le t^* \le \bar\tau\}} \Re\big\{ F \cdot \partial_{t^*} \bar f\big\} \, \dvol_{g}    \Big| \\
& + \sup_{\substack{0\le \tau_1 \le \tau_2 \le \tau,\\ \tau_2-\tau_1 \le 1}}  \Big|  \int_{\{\tau_1\le t^*\le \tau_2\}}(t^*-\tau_1)\Re\big\{ F \cdot \partial_{t^*} \bar f\big\} \, \dvol_{g} \Big|  \nonumber \\
  & + \sup_{\substack{0\le \tau_1 \le \tau_2 \le \tau,\\ \tau_2-\tau_1 \le 1}}  \Big| \int_{\{\tau_1\le t^*\le \tau_2\}}\Re\big\{ F \cdot Y \bar f\big\} \, \dvol_{g} \Big| \Bigg\}.  \label{Non degenerate energy inequality almost}
\end{align}
Adding a multiple of the $\partial_{t^*}$ identity \eqref{Energy inequality} to \eqref{Non degenerate energy inequality almost}, we finally obtain \eqref{Non degenerate energy inequality}.
\end{proof}

\subsection{A priori energy bounds for \texorpdfstring{$\psi$}{psi}}\label{sec:A priori energy bounds}
The following energy bound combines the estimates that were established earlier in this section and  lies at the heart of the proof of \cref{prop: Estimates error term}:

\begin{lemma}\label{lem:Energy estimates Psi}
Let $\psi$ be a solution of the initial-boundary value problem \eqref{IVP Psi} on the domain $\{0\le t^*\le T^*\}$ for some $ 1 \le T^* \le T_1$. Let us also set for any integer $j \ge 0$ and any $\tau \in [0,T^*]$:
\[
\mathscr E^{(j)}[\psi](\tau) \doteq \sup_{t^*\in [0,\tau]} \mathcal E^{(j)}[\psi](t^*)
\]
and:
\[
\mathscr H^{(j)}[\psi](\tau) \doteq \sum_{\bar j=0}^j \int_{\mathcal H^+ \cap \{0\le t^* \le \tau\}} \big(|\partial_{t^*} V \psi^{(\bar j;0)}|^2 + |\partial_{t^*} \psi^{(\bar j;0)}|^2 \big) \,r^2\sin\theta dt^* d\theta d\varphi.
\] 
Then, we can estimate for any $\tau \in [0,T^*]$:
\begin{align}\label{Energy estimate Psi fundamental}
\mathscr E^{(j)}[\psi](\tau) &  +\mathscr H^{(j)}[\psi](\tau) \\
\lesssim_j & \,
\mathscr E^{(j)}[\psi](0) \nonumber\\
& +
 L  \sup_{\bar \tau \in [0,\tau]} \Bigg\{ \Big(\mathscr F_3^{(j+1)}[\chi\tilde\phi](\bar\tau) + \mathscr G_3^{(\lceil \f{j}{2}\rceil+2)}[\psi](\bar\tau) + \big( \mathscr G_3^{(\lceil \f{j}{2}\rceil+2)}[V \psi](\bar\tau)\big)^{\f12}\big(\mathscr G_3^{(\lceil \f{j}{2}\rceil+2)}[\psi](\bar\tau)\big)^{\f12} \Big) 
 \nonumber \\
 & \hphantom{+
 L  \sup_{\bar \tau \in [0,\tau]} \Bigg\{ \Big(\mathscr F_3^{(j+1)}[\chi\tilde\phi](\bar\tau) }
 \times  \Big( 1+ L^3 \big( \mathscr G_3^{(j)}[\chi\tilde\phi](\bar\tau)+\mathscr G_3^{(\lceil \f{j}{2}\rceil+2)}[\psi](\bar\tau) \big) \Big)\Bigg\} 
 \nonumber\\
 & \hphantom{ L  \sup_{\bar \tau \in [0,\tau]} \Bigg\{ \Big(\mathscr F_3^{(j+1)}[\chi\tilde\phi](\bar\tau) + \mathscr F_3^{(\lceil \f{j}{2}\rceil+2)}[\psi](\bar\tau) \Big)\cdot }
 \times  (\log L)^{\f12} \int_0^\tau \mathscr E^{(j)}[\psi](t^*) \, dt^* \nonumber \\[5pt]
& + 
 \sup_{\bar \tau \in [0,\tau]}  \Big(\big(\mathscr F_3^{(j+2)}[\chi\tilde\phi](\bar\tau)  +\mathscr F_3^{(\lceil \f{j}{2}\rceil+2)}[\psi](\bar\tau)\big) \big(1+\mathscr F_3^{(j+2)}[\chi\tilde\phi](\bar\tau)+\mathscr F_3^{(\lceil \f{j}{2}\rceil+2)}[\psi](\bar\tau) \big) \Big) \nonumber \\
&\hphantom{+ 
 \sup_{\bar \tau \in [0,\tau]}  \Big(\big(\mathscr F_3^{(j+2)}[\chi\tilde\phi](\bar\tau) +}
\times  \big(\mathscr E^{(j)}[\psi](\tau)+\mathscr H^{(j)}[\psi](\tau)\big)   \nonumber \\
& +
\Bigg\{1+L^4  \sup_{\bar \tau \in [0,\tau]}  \Big(\mathscr F_3^{(j+1)}[\chi\tilde\phi](\bar\tau) \Big)^2\Bigg\}  \cdot \Bigg( \int_0^\tau \big\| r^3 V^{\le 1}\big( \mathcal F[\tilde \phi]\big)^{(\le j+1)} \Big|_{\Sigma^*_{t^*}}\big\|_{L^2_{r,\theta,\varphi}(r^{-2}\sin\theta dr d\theta d\varphi)} \, dt^*\Bigg)^2  \nonumber \\
& +
\sup_{\bar\tau \in[0,\tau]} \Big\|\Big(r^3 \mathcal F[\tilde \phi]
\Big)^{(\le j)}\big|_{\Sigma^*_{\bar\tau}} \Big\|^2 _
 {L^2(\sin\theta \f{1}{r^2}dr d\theta d\varphi)},   \nonumber 
\end{align}
where the constant implicit in the $\lesssim_j$ notation depends only on $j$ and the geometry of the spacetime $(\mathcal M_\mathrm{ext}, g_M)$, while $\mathscr F^{(j)}_3[\cdot]$ and $\mathscr G^{(j)}_3[\cdot]$ denote the $\|\cdot\|_{L^\infty_{r,\theta,\varphi}}^2$-type quantities defined by \eqref{Driving norm}--\eqref{Weaker driving norm} for $A=3$. 
\end{lemma}

Before proceeding to the proof of \cref{lem:Energy estimates Psi}, we will need the following auxiliary result regarding the structure of the model solution $\tilde \phi$:

\begin{lemma}\label{lem:Structure of tilde Phi}
There exist smooth functions $ \Phi_i^\sharp, \Phi_i^\star: \{0\le t^*\le T_1\} \rightarrow \mathbb C$, $i=1,2$, supported in $\{r \ge r_\mathrm{mirror}\}$, such that
\begin{align}\label{Decomposition Phi tilde}
\chi^2(r) \Big(\partial_{t^*}^2 \tilde \phi \cdot \bar{\tilde \phi} + |\partial_{t^*} \tilde \phi|^2\Big) 
&= \f1L \partial_{t^*}\Phi^\sharp_1 + \f1L \Phi^\star_1,\\
\chi^2(r) \Big(\partial_{t^*}^2 \tilde \phi \cdot \tilde \phi \Big) 
&= \f1L \partial_{t^*} \Phi^\sharp_2 + \f1L \Phi^\star_2   \nonumber
\end{align}
and satisfying the following bounds for any integer $j\ge 1$ for some constants $C_j>0$ depending only on  $j$, the geometry of $(\mathcal M_\mathrm{ext}, g_M)$ and the parameters $\epsilon, C_{\mathrm{amp}}, s,\lambda, N$ (being, in particular, \emph{independent} of $L$):
\begin{itemize}
\item The ``time-integral'' terms $\Phi^\sharp_i$ satisfy
\begin{equation}\label{Bound Phi sharp i}
\sup_{\tau \in [0,T_1]}\mathscr F^{(j)}_6[r\Phi^\sharp_i](\tau) \le C_j  L^4 \Big(\sup_{\tau \in [0,T_1]}\mathscr F^{(j)}_3[\chi \tilde \phi](\tau)\Big)^2
\end{equation}
\item The ``error'' terms $\Phi^\star_i$ satisfy
\begin{equation}\label{Bound Phi star i}
\sup_{\tau \in [0,T_1]}\mathscr F^{(j)}_6[r\Phi^\star_i](\tau) \le C_j L^{9-4s}\Big(  \sup_{\tau \in [0,T_1]}\mathscr F^{(j+1)}_3[\chi \tilde \phi](\tau)\Big)^2.
\end{equation}
\end{itemize}
In the above, $\mathscr F^{(j)}_\cdot[\cdot]$ denotes the norm defined by \eqref{Driving norm}. 
\end{lemma}

\begin{remark} In the decomposition \eqref{Decomposition Phi tilde}, the term $ \f1L \partial_{t^*}\Phi^\sharp_i$ contains terms of time frequency $\sim \f1L$ (and which can therefore be expressed as time derivatives of functions satisfying similar bounds as the original expressions).
In view of the fact that $s>\f54$, the right-hand side of \eqref{Bound Phi star i} is much smaller than that of \eqref{Bound Phi sharp i}. In particular, we can think of $\f1L \partial_{t^*}\Phi_i^\sharp$ as the dominant terms in  \eqref{Decomposition Phi tilde}.
\end{remark}

\begin{proof}
The construction of $\Phi_i^\sharp$ and $\Phi_i^\star$ will be done by expanding $\tilde\phi = \sum_{k \in \mathcal K} \tilde\phi_k$  in the expressions $\partial_{t^*}^2 \tilde \phi \cdot \bar{\tilde \phi} + |\partial_{t^*} \tilde \phi|^2 $ and $\partial_{t^*}^2 \tilde \phi \cdot \tilde \phi $ into products of modes $\tilde\phi_k$. 

More precisely, recall that $t=t^*$ in the region $\{r\ge r_\mathrm{mirror}\}$ and that we have
\[
\tilde\phi_k(t, y, \theta, \varphi) = \f{a_k(t)}{r}E_k(y, \theta, \varphi).
\]
Moreover, for $k\in \mathcal K_{\mathrm{D}}$,
\[
a_k(t) = b_k(t) e^{-i \varepsilon_k \omega_k t}.
\]
Recall also that
\[
|\varepsilon_{k_1}\omega_{k_1} \pm \varepsilon_{k_2}\omega_{k_2}| \gtrsim L \quad \text{for} \quad k_1\neq k_2 \in \mathcal K_{\mathrm{D}}.
\]
We will set
\begin{align*}
&\Phi_1^\flat(t,y,\theta,\varphi) \doteq\, 
 \chi^2(r(y)) \sum_{k \in \mathcal K_{\mathrm{D}}} \Big[\Big(i\varepsilon_k \omega_k \big(\f{d \bar b_k}{dt} b_k-\f{d b_k}{dt} \bar b_k) + \f{d^2 b_k}{dt^2}\bar b_k + \big|\f{d b_k}{dt} \big|^2 \Big) \f{|E_k|^2(y,\theta,\varphi)}{r(y)^2}   \Big]\\
& + \chi^2(r(y)) \sum_{\substack{k_1, k_2 \in \mathcal K_{\mathrm{D}},\\ k_1\neq k_2}} \Big[\Big(-i\varepsilon_{k_1} \omega_{k_1} \f{d b_{k_1}}{dt} \bar b_{k_2} + \f{d^2 b_{k_1}}{dt^2}\bar b_{k_2} -i\varepsilon_{k_1} \omega_{k_1} b_{k_1} \f{d\bar b_{k_2}}{dt} + i \varepsilon_{k_2} \omega_{k_2} \f{d b_{k_1}}{dt} \bar b_{k_2}+ \f{d b_{k_1}}{dt} \f{d \bar b_{k_2}}{dt} \Big) \\
& \hphantom{+ \chi^2(r(y)) \sum_{\substack{k_1, k_2 \in \mathcal K_{\mathrm{D}},\\ k_1\neq k_2}} \Big[}
\times e^{-i(\varepsilon_{k_1}\omega_{k_1}-\varepsilon_{k_2}\omega_{k_2})t} \f{E_{k_1}(y,\theta,\varphi) \overline{E_{k_2}}(y,\theta,\varphi)}{r(y)^2}   \Big]\\
& - 
\chi^2(r(y)) \sum_{\substack{k_1, k_2 \in \mathcal K_{\mathrm{D}},\\ k_1\neq k_2}} \Big[\Big(-\omega^2_{k_1}   +\varepsilon_{k_1}\varepsilon_{k_2} \omega_{k_1}\omega_{k_2} \Big) 
\f{e^{-i(\varepsilon_{k_1}\omega_{k_1}-\varepsilon_{k_2}\omega_{k_2})t}}{-i(\varepsilon_{k_1}\omega_{k_1}-\varepsilon_{k_2}\omega_{k_2})} \big( \f{ d b_{k_1}}{dt} \bar b_{k_2} + b_{k_1} \f{d \bar b_{k_2}}{dt}\big)\\
&\hphantom{- 
\chi^2(r(y)) \sum_{\substack{k_1, k_2 \in \mathcal K_{\mathrm{D}},\\ k_1\neq k_2}} \Big[}
\times \f{E_{k_1}(y,\theta,\varphi) \overline{E_{k_2}}(y,\theta,\varphi)}{r(y)^2} \Big],
\end{align*}

\begin{align*}
\Phi_1^\sharp (t,y,\theta,\varphi) \doteq \, & 
\chi^2(r(y)) \sum_{\substack{k_1, k_2 \in \mathcal K_{\mathrm{D}},\\ k_1\neq k_2}}
\Big[\Big(-\omega^2_{k_1} + \varepsilon_{k_1}\varepsilon_{k_2} \omega_{k_1}\omega_{k_2} \Big)
\f{e^{-i(\varepsilon_{k_1}\omega_{k_1}-\varepsilon_{k_2}\omega_{k_2})t} \, L}
  {-i(\varepsilon_{k_1}\omega_{k_1}-\varepsilon_{k_2}\omega_{k_2})}
\, b_{k_1}(t) \bar b_{k_2}(t)
\\
&\hphantom{\chi^2(r(y)) \sum_{\substack{k_1, k_2 \in \mathcal K_{\mathrm{D}},\\ k_1\neq k_2}}
\Big[}
\times \f{E_{k_1}(y,\theta,\varphi) \overline{E_{k_2}}(y,\theta,\varphi)}{r(y)^2}
\Big],
\end{align*}

\begin{align*}
\Phi_1^\natural (t,y,\theta,\varphi)  \doteq \, & 
L  \cdot \chi^2(r(y)) \sum_{\substack{k_1, k_2 \in \mathcal K,\\ k_1\notin \mathcal K_{\mathrm{D}} \text{\textbf{ or }} k_2 \notin \mathcal K_{\mathrm{D}}}} 
\Big[ \Big( \f{d^2 a_{k_1}}{dt^2} \bar a_{k_1} + \f{d a_{k_1}}{dt}\f{d \bar a_{k_2}}{dt}  
\Big)\f{E_{k_1}(y,\theta,\varphi) \overline{E_{k_2}}(y,\theta,\varphi)}{r(y)^2}   \Big], 
\end{align*}
and
\begin{align*}
\Phi_2^\flat \doteq \, & 
\chi^2(r(y)) \sum_{k_1, k_2 \in \mathcal K_{\mathrm{D}}} \Big[
\Big(-2i \varepsilon_{k_1}\omega_{k_1} \f{d b_{k_1}}{dt} b_{k_2} + \f{d^2 b_{k_1}}{dt^2} b_{k_2} \Big)
\f{E_{k_1}(y,\theta,\varphi) E_{k_2}(y,\theta,\varphi)}{r(y)^2}   \Big]
\\
& - \chi^2(r(y)) \sum_{k_1, k_2 \in \mathcal K_{\mathrm{D}}} \Big[
(-\omega^2_{k_1}) 
\f{e^{-i(\varepsilon_{k_1}\omega_{k_1}+\varepsilon_{k_2}\omega_{k_2})t}}{-i(\varepsilon_{k_1}\omega_{k_1}+\varepsilon_{k_2}\omega_{k_2})} \Big( \f{d b_{k_1}}{dt}  b_{k_2} + b_{k_1} \f{d b_{k_2}}{dt} \Big)\f{E_{k_1}(y,\theta,\varphi) E_{k_2}(y,\theta,\varphi)}{r(y)^2}   \Big],
\\[10pt]
\Phi_2^\sharp \doteq \, & 
\chi^2(r(y)) \sum_{k_1, k_2 \in \mathcal K_{\mathrm{D}}} \Big[
(-\omega^2_{k_1}) 
\f{e^{-i(\varepsilon_{k_1}\omega_{k_1}+\varepsilon_{k_2}\omega_{k_2})t} L}{-i(\varepsilon_{k_1}\omega_{k_1}+\varepsilon_{k_2}\omega_{k_2})} b_{k_1}(t)  b_{k_2}(t)\f{E_{k_1}(y,\theta,\varphi) E_{k_2}(y,\theta,\varphi)}{r(y)^2}   \Big],
\\[10pt]
\Phi_2^\natural \doteq \, & 
L \cdot \chi^2(r(y)) \sum_{\substack{k_1, k_2 \in \mathcal K\\ k_1\notin \mathcal K_{\mathrm{D}} \text{\textbf{ or }} k_2 \notin \mathcal K_{\mathrm{D}}}} \Big[
 \f{d^2 a_{k_1}}{dt}(t) a_{k_2}(t) \f{E_{k_1}(y,\theta,\varphi) E_{k_2}(y,\theta,\varphi)}{r(y)^2}   \Big].
\end{align*}

Note that if we split $\tilde\phi$ into the dominant and non-dominant parts
\[
\tilde\phi_{\mathrm{D}} \doteq \sum_{k\in \mathcal K_{\mathrm{D}}} \tilde\phi_k \quad \text{and} \quad \tilde\phi_{\mathrm{ND}} \doteq \sum_{k\in \mathcal K \setminus \mathcal K_{\mathrm{D}}} \tilde\phi_k,
\]
then 
\begin{align*}
\Phi^\flat_1+\f1L \partial_t \Phi_1^\sharp & = \chi^2(r) \Big(\partial_{t^*}^2 \tilde \phi_{\mathrm{D}} \cdot \bar{\tilde \phi}_{\mathrm{D}} + |\partial_{t^*} \tilde \phi_{\mathrm{D}}|^2\Big),\\
\f1L \Phi_1^\natural & = \chi^2(r) \Big(\partial_{t^*}^2 \tilde \phi \cdot \bar{\tilde \phi}_{\mathrm{ND}}+\partial_{t^*}^2 \tilde \phi_{\mathrm{ND}} \cdot \bar{\tilde \phi} + 2\Re\{\partial_{t^*} \tilde \phi \partial_{t^*} \tilde\phi_{\mathrm{ND}}\}\Big)
\end{align*}
and
\begin{align*}
\Phi^\flat_2+\f1L \partial_t \Phi_2^\sharp & = \chi^2(r) \Big(\partial_{t^*}^2 \tilde \phi_{\mathrm{D}} \cdot \tilde \phi_{\mathrm{D}} \Big),\\
\f1L \Phi_2^\natural & = \chi^2(r) \Big(\partial_{t^*}^2 \tilde \phi \cdot \tilde \phi_{\mathrm{ND}}+\partial_{t^*}^2 \tilde \phi_{\mathrm{ND}} \cdot \tilde \phi\Big).
\end{align*}
Note also that the quantities $\Phi_i^\flat$ contain only products of the slowly oscillating amplitudes $b_k$'s where at least one factor has a time derivative applied to it. The terms $\Phi_i^\star$ will then be defined as
\[
\Phi_i^\star \doteq L \Phi_i^\flat + \Phi^\natural_i, \quad i=1,2.
\]

The following pointwise bounds  for the spatial eigenfunctions $E_k(y, \theta, \varphi)$ were derived in the proof of \cref{lem: Bound F norm phi tilde}:

\begin{itemize}
\item 
For any $(n_k, \ell_k)\in( \mathbb N^*)^2$ and any $j\in \mathbb N$, we can estimate:  
\[
\big\|(\ell_k+n_k)^{-\f12-j} \f{d^j}{dy^j} R_{n_k,\ell_k}\big\|_{L^\infty([0,y_\mathrm{mirror}]} \lesssim_j 1.
\]

\item The spherical harmonics $Y_{\ell,m}$ (given by \eqref{Expression spherical harmonic}) satisfy
\[
\Big\| \ell^{-j-\f12} \nabla_{\mathbb S^2}^j Y_{\ell,m}\Big\|_{L^\infty(\mathbb S^2)} \lesssim 1.
\]

\item In the case when $k\in \mathcal K_{\mathrm{D}}$, we have for any $j\in \mathbb N$: 
\[
\Big|\f{d^j}{dy^j} R_{n_k, \ell_k}\Big|\lesssim_j L^{\f14+\f12j}
\]
and
\[
\sup_{y\ge \sqrt{\f{4n_k-1}{\ell_k}}} \Big|L^{-\f14-\f12j} e^{\f12 N^{-\f14} L^{\f34}(y-\sqrt{\f{4n_k-1}{\ell_k}})^{\f32}}\f{d^j}{dy^j} R_{n_k, \ell_k}\Big|\lesssim_j 1
\]
(the latter bound interpreted as the statement that the support of $R_{n_k,\ell_k}$ is concentrated in the region $\{y\lesssim L^{-\f12}\}$).

\item The spherical harmonic $Y_{\ell,\pm\ell}$ satisfies
 \[
 \Big|L^{-\f14-\f12j_1 - j_2} \f{\partial^{j_1}}{\partial \theta^{j_1}} \f{\partial^{j_2}}{\partial \varphi^{j_2}} Y_{\ell_k, \pm\ell_k} \Big| \lesssim_{j_1, j_2} 1
 \]
 and
 \[
\sup_{|\f\pi2-\theta|\gtrsim L^{-\f12}} \Big|L^{-\f14-\f12j_1-j_2} e^{\f12 L |\f\pi2-\theta|}\f{\partial^{j_1}}{\partial \theta^{j_1}} \f{\partial^{j_2}}{\partial \varphi^{j_2}} Y_{\ell_k, \pm\ell_k} \Big|\lesssim_{j_1, j_2} 1
 \]
(the latter bound interpreted as the statement that the support of $Y_{\ell_k, \pm\ell_k} $ is concentrated in the region $\{|\f\pi2-\theta| \lesssim L^{-\f12}\}$ for $k\in \mathcal K_{\mathrm{D}}$).

\item The weight function $w(y,\theta)$ satisfies 
\[
w\lesssim L^{-\f12} \quad \text{on} \quad \{y\lesssim L^{-\f12}\} \cap \{|\f\pi2-\theta| \lesssim L^{-\f12}\}.
\]

\item The helical vector field $V$ almost commutes with $e^{-i \varepsilon_k \omega_k t}E_k(y, \theta, \varphi)$ when $k\in \mathcal K_{\mathrm{D}}$, in the sense that
\[
V \big( e^{-i \varepsilon_k \omega_k t}E_k \big) = \f{-i \varepsilon_k \omega_k}{L} \cdot e^{-i \varepsilon_k \omega_k t}E_k = O(1) \cdot e^{-i \varepsilon_k \omega_k t}E_k.
\]
\end{itemize}

Using the above observations for the eigenfunctions $E_k$, the expression \eqref{Driving norm} for the norm $\mathscr F^{(j)}_A[\cdot]$ and the expressions defining the functions $\Phi_i^\flat$, $\Phi_i^\sharp$ and $\Phi_i^\natural$ (together with the estimate $\omega_k-\ell_k \sim n_k = O(\ell_k^{1-\delta_0})$, following from   \cref{lem:Crude Weyl law}), we can readily calculate that, for $i=1,2$:
\[
\mathscr F^{(j)}_6[r \Phi^\sharp_i] \lesssim_j 
\sum_{k_1, k_2\in \mathcal K_{\mathrm{D}}} \sum_{j_1+j_2=0}^{j+2}\sum_{j_3+j_4=0}^1 \Bigg(  L^{-j_1-j_2} \Big| \f{d^{j_1+j_3} b_{k_1}}{dt^{j_1+j_3}} \f{ d^{j_2+j_4} b_{k_2}}{dt^{j_2+j_4}} \Big| \Bigg)^2,
\]
\begin{multline*}
\mathscr F^{(j)}_6[ r\Phi_i^\flat] \lesssim_j  \sum_{k_1, k_2\in \mathcal K_{\mathrm{D}}} \sum_{j_1+j_2=0}^{j+2}\sum_{j_3+j_4=0}^1 \Bigg( L^{-2} \cdot  \Big(   L^{1-j_1-j_2} \Big| \f{d^{1+j_1+j_3} b_{k_1}}{dt^{1+j_1+j_3}} \f{ d^{j_2+j_4} b_{k_2}}{dt^{j_2+j_4}} \Big| \\
  +   L^{-j_1-j_2} \Big| \f{d^{2+j_1+j_3} b_{k_1}}{dt^{2+j_1+j_3}} \f{ d^{j_2+j_4} b_{k_2}}{dt^{j_2+j_4}} \Big|  +   L^{-j_1-j_2} \Big| \f{d^{1+j_1+j_3} b_{k_1}}{dt^{1+j_1+j_3}} \f{ d^{1+j_2+j_4} b_{k_2}}{dt^{1+j_2+j_4}} \Big| \Big) \Bigg)^2
\end{multline*}
(note that the right-hand side above does not contain a product of zeroth order terms of the form $b_{k_1} \cdot b_{k_2}$)
and
\[
\mathscr F^{(j)}_6[r \Phi_i^\natural] \lesssim_j
\Bigg(\!\!\!\! \sum_{\substack{k_1, k_2 \in \mathcal K,\\ k_1\notin \mathcal K_{\mathrm{D}} \text{\textbf{ or }} k_2 \notin \mathcal K_{\mathrm{D}}}} \!\!\! \! \sum_{j_1+j_2=0}^{j+2}\sum_{j_3+j_4=0}^1  
L^{-j_1-j_2}\cdot  \Big( \Big| \f{d^{2+j_1+j_3} a_{k_1}}{dt^{2+j_1+j_3}} \f{ d^{j_2+j_4} a_{k_2}}{dt^{j_2+j_4}} \Big| + \Big| \f{d^{1+j_1+j_3} a_{k_1}}{dt^{1+j_1+j_3}} \f{ d^{1+j_2+j_4} a_{k_2}}{dt^{1+j_2+j_4}} \Big| \Big)  \Bigg)^2
\]

 Using the bounds \eqref{Bound C 1 norm tilde a}--\eqref{Bound C q norm tilde a} for the slowly oscillating amplitudes $\{b_k\}_{k\in \mathcal K_{\mathrm{D}}}$ (recalling that $b_k = L^{-s} \tilde b_k$ and $\f{d}{d\tilde t}= L^{2s+1} \f{d}{d t}$) and the bounds \eqref{Estimate non dominant modes} for the non-dominant amplitudes $\{a_k\}_{k\in \mathcal K \setminus \mathcal K_{\mathrm{D}}}$ (together with the bound \eqref{Bound P k} for the spectral projection coefficients $\mathcal P_k$ in \eqref{Estimate non dominant modes}), we obtain for any $\tau \in [0, T_1]$:
 \[
  \mathscr F^{(j)}_6[r \Phi_i^\sharp](\tau) \lesssim_j   L^{-4s},
 \]
 \[
 \mathscr F^{(j)}_6[r \Phi_i^\flat](\tau) \lesssim_j L^{-4-8s}
 \]
 and
  \[
  \mathscr F^{(j)}_6[r \Phi_i^\natural](\tau) \lesssim_j   L^{5-8s}.
 \]
 In view of our definition $\Phi_i^\star = L \Phi_i^\flat+\Phi_i^\natural$, we infer that
 \[
  \mathscr F^{(j)}[r \Phi_i^\star](\tau) \lesssim_j   L^{5-8s}
 \]
 The estimates \eqref{Bound Phi sharp i}--\eqref{Bound Phi star i} now follow from the above, together with the fact that 
 \[
 \sup_{\tau}\mathscr F^{(j)}_3[\chi \tilde \phi](\tau) \sim_j L^{-2} \cdot \sup_{\tau}\sum_{k\in \mathcal K_{\mathrm{D}}}|b_k(\tau)|^2 \sim_j L^{-2-2s}
 \]
  (see \eqref{Estimate scri F model solution}).

\end{proof}

\begin{proof}[Proof of \cref{lem:Energy estimates Psi}]

Using the higher-order coercivity estimate \eqref{Coercivity higher order} and the fact that $\psi$ solves \eqref{IVP Psi}, we can control $\mathscr E^{(j)}[\psi](\tau)$ by the higher-order $Y$-energy flux of $\psi$ as follows: 
\begin{align}\label{Coercivity higher order again}
 \mathscr E^{(j)}[\psi](\tau)
 \lesssim_j  \sup_{\bar\tau \in[0,\tau]} \sum_{\bar j=0}^j \Bigg(
& \int_{t^*=\bar\tau} \Big(  J_\mu^{(Y)}[\psi^{(\bar j;0)}] +J_\mu^{(Y)}[V\psi^{(\bar j; 0)}]  \Big)\, \hat n^\mu_{\Sigma^*} \, \dvol_{\Sigma^*} \\
& + \int_{t^*=\bar\tau} \Big|r^3 \Big(
r^{-6}|\chi\tilde\phi+\psi|^2 \partial_{t^*}^2\psi +\chi^2 \mathcal N^{(1)}[\tilde \phi;\psi]   
\nonumber\\
& \hphantom{+ \int_{t^*=\bar\tau} \Big|r^3 \Big(} 
+ \chi \mathcal N^{(2)}[\tilde\phi; \psi]    
 +\mathcal N^{(3)}[\psi] - \mathcal F[\tilde \phi]
\Big)^{(\bar k;0)}\Big|^2  
 \, \sin\theta \f{1}{r^2}dr d\theta d\varphi  \Bigg)     \nonumber
\end{align}
(see below equation \eqref{IVP Psi} for the definition of the terms $\mathcal N^{(i)}[\tilde \phi;\psi] $ and $ \mathcal F[\tilde \phi]$). Expanding the terms in the last integral above and using H\"older-type inequalities of the form
\begin{align*}
\Big\| r^{-3}\prod_{i=1}^3 z_i^{(j_i;0)} \Big\|^2_{L^2(r^{-2} \sin\theta drd\theta d\varphi)}
\lesssim \Big\|   & r^{-6} \big(w(y(r),\theta)\big)^{-A} (rz_1^{(j_1;0)}) (r z_2^{(j_2;0)}) \Big\|^2_{L^\infty_{r,\theta,\varphi}} \\
& \times \Big\| \big(w(y(r),\theta)\big)^{A} r z_3^{(j_3;0)} \Big\|^2_{L^2(r^{-2}\sin\theta dr d\theta d\varphi)}
\end{align*}
where $z_i  \in \{\partial^{\le 2}\tilde\phi, \partial^{\le 2}\psi\}$ (making sure to always include the terms involving the highest number of derivatives on $\psi$ in the $L^2$-norm of the right-hand side above), we can immediately estimate (recall also that $\f1{r(y)}\lesssim w(y,\theta)$):
\begin{align*}
\sum_{\bar j=0}^j    \int_{t^*=\bar\tau} \Big|r^3 \Big(
r^{-6}|\chi\tilde\phi+\psi|^2 \partial_{t^*}^2\psi +\chi^2 \mathcal N^{(1)}[\tilde \phi;\psi]  & + \chi \mathcal N^{(2)}[\tilde\phi; \psi]   
 +\mathcal N^{(3)}[\psi]
\Big)^{(\bar j;0)}\Big|^2  
 \, \sin\theta \f{1}{r^2}dr d\theta d\varphi  \\
& \lesssim_j \big( \mathscr F_3^{(j+2)}[\tilde\phi](\bar\tau) + \mathscr F_3^{(\lceil \f j2 \rceil+2)}[\psi](\bar\tau)  \big)^2 \cdot \mathcal E^{(j)}[\psi](\bar\tau).
\end{align*}
Thus, returning to \eqref{Coercivity higher order again}, we have
\begin{align} \nonumber
 \mathscr E^{(j)}[\psi](\tau)
 \lesssim_j &  \sup_{\bar\tau \in[0,\tau]} \sum_{\bar j=0}^j \Bigg(
 \int_{t^*=\bar\tau} \Big(  J_\mu^{(Y)}[\psi^{(\bar j;0)}] +J_\mu^{(Y)}[V\psi^{(\bar j; 0)}]  \Big)\, \hat n^\mu_{\Sigma^*} \, \dvol_{\Sigma^*} \\
&  + \Big[ \sup_{\bar\tau \in [0,\tau]}\big( \mathscr F^{(j+2)}_3[\chi \tilde\phi](\bar\tau) + \mathscr F_3^{(\lceil \f j2 \rceil+2)}[\psi](\bar\tau)  \big)^2 \Big] \mathscr E^{(j)}[\psi](\tau)  \nonumber \\
& 
 + \sup_{\bar\tau \in[0,\tau]} \Big\|\Big(r^3 \mathcal F[\tilde \phi]
\Big)^{(\le j)}\big|_{\Sigma^*_{\bar\tau}} \Big\|^2 _
 {L^2(\sin\theta \f{1}{r^2}dr d\theta d\varphi)}.    \label{Coercivity higher order final estimate}
\end{align}

We will now proceed to estimate the first term in the right-hand side of \eqref{Coercivity higher order final estimate} using the energy estimate provided by \cref{lem:Non degenerate energy estimate}. More precisely, 
applying \eqref{Non degenerate energy inequality} for $f=\psi^{(\bar j;0)}$ and $f=V\psi^{(\bar j;0)}$, $0\le \bar j\le j$
and using the fact that $\psi^{(\bar j;0)}$ and $V\psi^{(\bar j;0)}$ satisfy \eqref{IVP Psi k} and \eqref{IVP V Psi k}, we obtain for any $\bar\tau \in [0,\tau]$:
\begin{align}
\int_{t^*=\bar\tau} \Big(  J_\mu^{(Y)} & [\psi^{(\bar j;0)}] +J_\mu^{(Y)}[V\psi^{(\bar j; 0)}]  \Big)\, \hat n^\mu_{\Sigma^*} \, \dvol_{\Sigma^*} + \mathscr H^{(k)}[\psi](\tau) \\
&\lesssim_j  \int_{t^*=0} \Big(  J_\mu^{(Y)}[\psi^{(\bar j;0)}] +J_\mu^{(Y)}[V\psi^{(\bar j; 0)}]  \Big)\, \hat n^\mu_{\Sigma^*} \, \dvol_{\Sigma^*}  
+ \mathcal A[\bar\tau]+\mathcal B[\bar\tau]+\mathcal C[\bar\tau]+\mathcal F_*[\bar\tau],    \nonumber
\end{align}
where, denoting by $Z$ any derivative among $\{\partial_{t^*},Y\}$:
\begin{align*}
\mathcal A[\bar\tau] \doteq  \sum_{\bar j=0}^j 
\sup_{\substack{[\tau_1,\tau_2]\subseteq [0,\bar\tau],\\
h:[\tau_1,\tau_2] \rightarrow \mathbb R,\\ \|h\|_{C^1} \le 1 }} \Bigg( & 
\Bigg|    
\int_{\{\tau_1\le t^* \le \tau_2\}} h(t^*) \Re\Big\{ \Big(r^{-6}|\chi\tilde\phi+\psi|^2 \partial_{t^*}^2(V \psi^{(\bar j;0)})\Big)  \cdot Z V  \psi^{(\bar j;0)}  \Big\} \, \dvol_g
\Bigg| \\
& + \Bigg|    
\int_{\{\tau_1\le t^* \le \tau_2\}} h(t^*) \Re\Big\{ \Big(r^{-6}|\chi\tilde\phi+\psi|^2 \partial_{t^*}^2 \psi^{(\bar j;0)}\Big)  \cdot Z \psi^{(\bar j;0)}  \Big\} \, \dvol_g
\Bigg|
\Bigg),
\end{align*}

\begin{align*}
\mathcal B[\bar\tau] \doteq  \sum_{\bar j=0}^j 
\sup_{\substack{[\tau_1,\tau_2]\subseteq [0,\bar\tau],\\
h:[\tau_1,\tau_2] \rightarrow \mathbb R, \\ \|h\|_{C^1} \le 1 }} \Bigg(  
\Bigg|    
\int_{\{\tau_1\le t^* \le \tau_2\}} h(t^*) \Re\Big\{
 r^{-6} &\Big( \chi^2  \big(\partial_{t^*}^2 \tilde\phi \,  \bar{\tilde\phi}+|\partial_{t^*} \tilde\phi|^2 \big)^{(\bar j;0)}  \cdot V\psi \\
& + \chi^2 \big(\partial_{t^*}^2 \tilde\phi \, \tilde\phi \big)^{(\bar j;0)}  \cdot V\bar\psi \Big)\cdot Z (V\bar \psi^{(\bar j;0)})
 \Big\} \, \dvol_g
\Bigg| \Bigg),
\end{align*}

\begin{align*}
\mathcal C[\bar\tau] \doteq  \sum_{\bar j=0}^j 
\sup_{\substack{[\tau_1,\tau_2]\subseteq [0,\bar\tau],\\
h:[\tau_1,\tau_2] \rightarrow \mathbb R,\\ \|h\|_{L^\infty} \le 1 }} \!\!\Bigg( & 
\Bigg|    
\int_{\{\tau_1\le t^* \le \tau_2\}} \!\!\!\!\!\! \!\!\!\!\!\! h(t^*) \Re\Big\{ 
\Big( \chi^2 \mathcal N_{(1,\bar j,0)}^{[V]}[\tilde \phi;\psi]   +\chi \mathcal N_{(2,\bar j,0)}^{[V]}[\tilde\phi; \psi] +\mathcal N_{(3,\bar j,0)}^{[V]}[\psi] \Big)  \cdot 
Z V  \psi^{(\bar j;0)}  
\Big\} \, \dvol_g
\Bigg| \\
+ &  \Bigg|    
\int_{\{\tau_1\le t^* \le \tau_2\}}  \!\!\!\!\!\!  \!\!\!\!\!\! h(t^*) \Re\Big\{ 
\Big( \chi^2 \mathcal N_{(1,\bar j,0)}[\tilde \phi;\psi]   +\chi \mathcal N_{(2,\bar j,0)}[\tilde\phi; \psi] +\mathcal N_{(3,\bar j,0)}[\psi]  \Big) \cdot
 Z \psi^{(\bar j;0)}  \Big\} \, \dvol_g
\Bigg|
\Bigg),
\end{align*}
and
\begin{align*}
\mathcal F_*[\bar \tau] \doteq  \sum_{\bar j=0}^j 
\sup_{\substack{[\tau_1,\tau_2]\subseteq [0,\bar\tau],\\
h:[\tau_1,\tau_2] \rightarrow \mathbb R, \\ \|h\|_{L^\infty} \le 1 }} \Bigg(  
 \Bigg|    
\int_{\{\tau_1\le t^* \le \tau_2\}}  \!\!\!\!\!\! \!\!\!\!\!\! h(t^*) \Re\Big\{ 
 V(\mathcal F[\tilde \phi])^{(\bar j;0)} \Big)\cdot Z (V\bar \psi^{(\bar j;0)}) 
+
\Big(  (\mathcal F[\tilde \phi])^{(\bar j;0)}\Big)\cdot Z (\bar \psi^{(\bar j;0)})
\Big\} \, \dvol_g
\Bigg|
\Bigg).
\end{align*}
Returning to \eqref{Coercivity higher order final estimate}, we therefore obtain:
\begin{align}\label{Coercivity higher order final estimate again}
 \mathscr E^{(j)}[\psi](\tau) + \mathscr H^{(j)}[\psi](\tau)
 \lesssim_j &  \, \mathscr E^{(j)}[\psi](0) \\
& + \mathcal A[\tau]+\mathcal B[\tau]+\mathcal C[\tau] \nonumber \\
&   + \Big[ \sup_{\bar\tau \in [0,\tau]}\big( \mathscr F_3^{(j+2)}[\chi \tilde\phi](\bar\tau) + \mathscr F_3^{(\lceil \f j2 \rceil+2)}[\psi](\bar\tau)  \big)^2 \Big] \mathscr E^{(j)}[\psi](\tau)  \nonumber \\
& 
 +\mathcal F_*[\tau]+ \sup_{\bar\tau \in[0,\tau]} \Big\|\Big(r^3 \mathcal F[\tilde \phi]
\Big)^{(\le j)}\big|_{\Sigma^*_{\bar\tau}} \Big\|^2 _
 {L^2(\sin\theta \f{1}{r^2}dr d\theta d\varphi)}.    \nonumber
\end{align}

We will now estimate the terms $A[\tau]$, $\mathcal B[\tau]$, $\mathcal C[\tau]$ appearing in the right-hand side of \eqref{Coercivity higher order final estimate again}; among those terms, the ``trickiest'' one is $\mathcal B[\tau]$, which requires implementing some slightly non-standard integration by parts scheme.
\begin{enumerate}
\item The term $A[\tau]$ contains derivatives of higher order than those appearing in the left-hand side (namely, the terms of the form $\partial_{t^*}^2(V \psi^{(\bar j;0)})$) and, hence, cannot be directly controlled by  $\mathscr E^{(j)}[\psi](\tau)$ using simply a H\"older-type inequality (this is to be expected, since these are precisely the error terms in the energy estimate associated to the quasilinear terms in our equation for $\psi$). However, this can be achieved after performing a straightforward integrations-by-parts scheme, namely writing
\[
\Re 
\{\partial_{t^*}^2(V \psi^{(\bar j;0)}) \cdot Z V\bar\psi^{(\bar j;0)}\} = \partial_{t^*}\Big(\Re 
\{\partial_{t^*}(V \psi^{(\bar j;0)}) \cdot Z V\bar\psi^{(\bar j;0)}\}\Big) - \f12 Z \Big(\big| \partial_{t^*}(V \psi^{(\bar j;0)}) \big|^2 \Big)
\]
and then integrating by parts in the $\partial_{t^*}$,$Z$-derivatives above 
(note that, for this process, we have to make use of the fact that the lower order coefficient $|\chi \tilde\phi +\psi|^2$ appearing inside $\Re\big\{ \cdot \big\}$ in the expression for $\mathcal A[\tau]$ is real valued); similarly for the term
\[
\Re 
\{\partial_{t^*}^2 \psi^{(\bar j;0)} \cdot Z \bar\psi^{(\bar j;0)}\} = \partial_{t^*}\Big(\Re 
\{\partial_{t^*} \psi^{(\bar j;0)} \cdot Z \bar\psi^{(\bar j;0)}\}\Big) - \f12 Z \Big(\big| \partial_{t^*} \psi^{(\bar j;0)} \big|^2 \Big).
\]
After following through with this process (and noticing that only integrations by parts with respect to $Y$ give rise to boundary terms on $\mathcal H^+$), we end up with the following schematic expression for $\mathcal A[\tau]$:
\begin{align*}
\mathcal A[\tau]  \le & \sum_{\bar j=0}^j \Bigg(
2\sup_{\bar\tau\in [0,\tau]}\Bigg[ \Bigg| \int_{t^*=\bar\tau}r^{-6}|\chi \tilde\phi +\psi|^2\Re\{\partial V \psi^{(\bar j;0)} \cdot \partial V \bar\psi^{(\bar j;0)} \}   \,r^2\sin\theta dr d\theta d\varphi \Bigg| \\
&\hphantom{ = \sum_{\bar j=0}^j \Bigg(
2\sup_{\bar\tau\in [0,\tau]}\Bigg[}
+  \Bigg| \int_{t^*=\bar\tau}r^{-6}|\chi \tilde\phi +\psi|^2\Re\{\partial \psi^{(\bar j;0)} \cdot \partial \bar\psi^{(\bar j;0)} \}  \,r^2\sin\theta dr d\theta d\varphi \Bigg|
\Bigg] \\
&\hphantom{ = \sum_{\bar j=0}^j \Bigg(}
+c_* \int_{\mathcal H^+ \cap \{0\le t^* \le \tau\}}r^{-6}|\chi \tilde\phi +\psi|^2\Big( |\partial_{t^*} V\psi^{(\bar j;0)}|^2 + |\partial_{t^*} \psi^{(\bar j;0)}|^2  \Big) \, r^2 \sin\theta dr d\theta d\varphi
\Bigg)\\
&  + \mathcal C'[\tau] \\
 \lesssim &\Big(\sup_{\bar \tau \in [0,\tau]} \big( \mathscr F_3^{(0)}[\chi\tilde\phi](\bar\tau)+\mathscr F_3^{(0)}[\psi](\bar\tau)  \big) \Big) \cdot \big(\mathscr E^{(j)}[\psi](\tau)+\mathscr H^{(j)}[\psi](\tau)\big) + \mathcal C'[\tau],
\end{align*}
where $c_*>0$ is a constant that can be explicitly computed in terms of the vector field $Y$, while $\mathcal C'[\tau]$ contains only bulk terms of similar form as those of $\mathcal C[\tau]$ and, thus, can be estimated in exactly the same way (see point 3.~below).\footnote{The fact that all the boundary terms that appeared on $\mathcal H^+$ through the above process are controlled by our energy fluxes is a consequence of the structure of our quasilinear terms in equation \eqref{Initial Boundary Value Problem Phi}. In fact, it is crucial for this argument that $\mathcal H^+$, which is a null hypersurface with respect to the background metric $g_M$ is also achronal with respect to the metric associated to the principal symbol of our equation, which is of the form $g[\phi] = g_M + |\phi|^2 \big(O(1) dr^2 + O(1) dt^* dr \big)$. In the case of a more general quasilinear term, we would be forced to perform our energy estimates in a domain going slightly inside the event horizon and having a strictly spacelike future boundary with respect to $g_M$.}

\item The term $\mathcal B[\tau]$ does not contain higher order derivatives compared to the term $\mathscr E^{(j)}[\psi](\tau)$ on the left-hand side; however, it cannot be estimated directly via H\"older type estimates by terms of the form $\mathscr F_3^{(j+c_1)}[\chi\tilde\phi] \cdot \mathscr E^{(j)}[\psi]$, in view of the fact that $\mathscr E^{(j)}[\psi](\tau)$ does \textbf{not} control $\|\f{1}{w(y,\theta)}V(r\psi)\|^2_{L^2_\varphi L^\infty_{y,\theta}}$ or any other norm of $V\psi$ with the same scaling properties near $(y,\theta)=(0,\f\pi2)$ (even when allowing for a $(\log L)^{-1}$ degenerating factor). For this reason, we will implement an integrations-by-parts scheme that makes use of the decomposition \eqref{Decomposition Phi tilde} for the quadratic terms in $\tilde \phi$. 

Let us first notice that the integrand appearing in the term  $\mathcal B[\tau]$ contains a factor of $\chi^2$ and is, thus, supported in $\{r\ge r_\mathrm{mirror}\}$. In this region, $Y=\partial_{t^*}$. Therefore,  $\mathcal B[\tau]$ can be expressed as

\begin{multline*}
\mathcal B[\tau] =  \sum_{\bar j=0}^j 
\sup_{\substack{[\tau_1,\tau_2]\subseteq [0,\tau],\\
h:[\tau_1,\tau_2] \rightarrow \mathbb R, \\ \|h\|_{C^1} \le 1 }} \Bigg(  
\Bigg|    
\int_{\{\tau_1\le t^* \le \tau_2\}} h(t^*) r^{-6} \Re\Big\{
 \Big( \chi^2  \big(\f1L \partial_{t^*}  \big)^{\bar j}\big(\partial_{t^*}^2 \tilde\phi \,  \bar{\tilde\phi}+|\partial_{t^*} \tilde\phi|^2 \big)  \cdot V\psi \\
 + \chi^2 \big(\f1L \partial_{t^*}\big)^{\bar j} \big(\partial_{t^*}^2 \tilde\phi \, \tilde\phi \big)  \cdot V\bar\psi \Big)\cdot \partial_{t^*} (V\bar \psi^{(0,\bar j)})
 \Big\} \, \dvol_g
\Bigg| \Bigg).
\end{multline*}
Using the decomposition \eqref{Decomposition Phi tilde}, we therefore have
\begin{equation}\label{Decomposition B}
\mathcal B[\tau] \le  \sum_{\bar j=0}^j \Big( \mathcal B_1^{(\bar j)}[\tau] + \mathcal B_2^{(\bar j)}[\tau]  \Big),
\end{equation}
where
\begin{equation}\label{def B k 1}
\mathcal B_1^{(\bar j)}[\tau] \doteq 
\sup_{\substack{[\tau_1,\tau_2]\subseteq [0,\tau],\\
h:[\tau_1,\tau_2] \rightarrow \mathbb R, \\ \|h\|_{C^1} \le 1 }}  
\Bigg|    
\int_{\{\tau_1\le t^* \le \tau_2\}} h(t^*)r^{-6}  \Re\Big\{
    \big(
\f1L \partial_{t^*}\Phi^\sharp_1 + \f1L \Phi^\star_1
 \big)^{(0;\bar j)}  \cdot V\psi \cdot \partial_{t^*} (V\bar \psi^{(0,\bar j)})
 \Big\} \, \dvol_g
\Bigg| 
\end{equation}
and
\begin{equation}\label{def B k 2}
\mathcal B_2^{(\bar j)}[\tau] \doteq 
\sup_{\substack{[\tau_1,\tau_2]\subseteq [0,\tau],\\
h:[\tau_1,\tau_2] \rightarrow \mathbb R, \\ \|h\|_{C^1} \le 1 }}  
\Bigg|    
\int_{\{\tau_1\le t^* \le \tau_2\}} h(t^*) r^{-6} \Re\Big\{
    \big(
 \f1L \partial_{t^*}\Phi^\sharp_2 + \f1L \Phi^\star_2
 \big)^{(0;\bar j)}  \cdot V\bar \psi \cdot \partial_{t^*} (V\bar \psi^{(0,\bar j)})
 \Big\} \, \dvol_g
\Bigg|.
\end{equation}

We will show how to estimate  $\mathcal B_1^{(\bar j)}[\tau]$; the analogous estimate for $\mathcal B_2^{(\bar j)}[\tau]$ will then follow in exactly the same way. 

We start by noting that
\begin{equation}\label{Decomposition B 1}
\mathcal B_1^{(\bar j)}[\tau] \le \mathcal  \mathcal B_1^{(\bar j,\sharp)}[\tau] + \mathcal B_1^{(\bar j,\star)}[\tau],
\end{equation}
where
\[
\mathcal B_1^{(\bar j,\sharp)}[\tau] \doteq 
\sup_{\substack{[\tau_1,\tau_2]\subseteq [0,\tau],\\
h:[\tau_1,\tau_2] \rightarrow \mathbb R, \\ \|h\|_{C^1} \le 1 }}  
\Bigg|    
\int_{\{\tau_1\le t^* \le \tau_2\}} h(t^*) r^{-6} \Re\Big\{
    \f1L \partial_{t^*} \big(
\Phi^\sharp_1 
 \big)^{(0;\bar j)}  \cdot V\psi \cdot \partial_{t^*} (V\bar \psi^{(0,\bar j)})
 \Big\} \, \dvol_g
\Bigg| 
\]
and
\[
\mathcal B_1^{(\bar j,\star)}[\tau] \doteq 
\sup_{\substack{[\tau_1,\tau_2]\subseteq [0,\tau],\\
h:[\tau_1,\tau_2] \rightarrow \mathbb R, \\ \|h\|_{C^1} \le 1 }}  
\Bigg|    
\int_{\{\tau_1\le t^* \le \tau_2\}} h(t^*) r^{-6} \Re\Big\{
   \f1L \big(
\Phi^\star_1
 \big)^{(0;\bar j)}  \cdot V\psi \cdot \partial_{t^*} (V\bar \psi^{(0,\bar j)})
 \Big\} \, \dvol_g
\Bigg|.
\]

\begin{itemize}
\item For the term $\mathcal B_1^{(\bar j,\sharp)}[\tau]$, we argue as follows: Let $\zeta:\mathcal M_\mathrm{ext}\rightarrow [0,1]$ be a smooth cut-off function which is $t^*,\varphi$-invariant and which satisfies
\[
\mathrm{\text{supp}}\zeta \subset \{w(y,\theta) <1 \}, \quad \zeta \equiv 1 \text{ on } \{w(y,\theta) \le \f12\}.
\]
We will split
\begin{multline*}
\mathcal B_1^{(\bar j,\sharp)}[\tau] 
=  
\sup_{\substack{[\tau_1,\tau_2]\subseteq [0,\tau],\\
h:[\tau_1,\tau_2] \rightarrow \mathbb R, \\ \|h\|_{C^1} \le 1 }}  
\Bigg|    
\int_{\{\tau_1\le t^* \le \tau_2\}} h(t^*)r^{-6} \Re\Big\{
    \f1L \partial_{t^*} \big(
\Phi^\sharp_1 
 \big)^{(0;\bar j)}  \cdot V\psi \cdot \partial_{t^*} (V\bar \psi^{(0,\bar j)})
 \Big\} \, \dvol_g
\Bigg| \\
\le  
\sup_{\substack{[\tau_1,\tau_2]\subseteq [0,\tau],\\
h:[\tau_1,\tau_2] \rightarrow \mathbb R, \\ \|h\|_{C^1} \le 1 }}  
\Bigg|    
\int_{\{\tau_1\le t^* \le \tau_2\}} \zeta(y,\theta) h(t^*) r^{-6} \Re\Big\{
    \f1L \partial_{t^*} \big(
\Phi^\sharp_1 
 \big)^{(0;\bar j)}  \cdot V\psi \cdot \partial_{t^*} (V\bar \psi^{(0,\bar j)})
 \Big\} \, \dvol_g
\Bigg| \\
 + 
\sup_{\substack{[\tau_1,\tau_2]\subseteq [0,\tau],\\
h:[\tau_1,\tau_2] \rightarrow \mathbb R, \\ \|h\|_{C^1} \le 1 }}  
\Bigg|    
\int_{\{\tau_1\le t^* \le \tau_2\}} \big(1-\zeta(y,\theta)\big) h(t^*) r^{-6} \Re\Big\{
    \f1L \partial_{t^*} \big(
\Phi^\sharp_1 
 \big)^{(0;\bar j)}  \cdot V\psi \cdot \partial_{t^*} (V\bar \psi^{(0,\bar j)})
 \Big\} \, \dvol_g
\Bigg| \\
 \doteq \,  
\, \mathcal B_{1,\zeta}^{(\bar j,\sharp)}[\tau] + \mathcal B_{1,1-\zeta}^{(\bar j,\sharp)}[\tau].
\end{multline*}
Note that $\mathcal B_{1,1-\zeta}^{(\bar j,\sharp)}[\tau]$ can be immediately estimated using H\"older-type estimates, in view of the fact that the norms $\mathscr E^{(j)}$ and $\mathscr F_\cdot^{(j)}$ provide improved control in the region where $w(y,\theta) \gtrsim 1$, namely where $1-\zeta$ is supported:
\begin{align}
\mathcal B_{1,1-\zeta}^{(\bar j,\sharp)} & [\tau] \nonumber   \\
& = \sup_{\substack{[\tau_1,\tau_2]\subseteq [0,\tau],\\
h:[\tau_1,\tau_2] \rightarrow \mathbb R, \\ \|h\|_{C^1} \le 1 }}  
\Bigg|      
\int_{\{\tau_1\le t^* \le \tau_2\}}  \!\!\!\!\!\! \!\!\!\!\!\! \big(1-\zeta(y,\theta)\big) h(t^*) r^{-6} \Re\Big\{
    \f1L \partial_{t^*} \big(
\Phi^\sharp_1 
 \big)^{(0;\bar j)}  \cdot V\psi \cdot \partial_{t^*} (V\bar \psi^{(0,\bar j)})
 \Big\} \, \dvol_g
\Bigg|  \nonumber   \\[5pt]
&\quad \stackrel{\hphantom{w\ge 1/2 \text{ in }\mathrm{\text{supp}}(1-\zeta)}}{\lesssim_j} 
\int_0^\tau   \big\| (1-\zeta) r^{-5} (\Phi^\sharp_1)^{(0,\bar j+1)}\big\|_{L^\infty_{r,\theta,\varphi}}  \nonumber    \\
&\hphantom{\quad \stackrel{w\ge 1/2 \text{ in }\mathrm{\text{supp}}(1-\zeta)}{\lesssim_j} \int_0^\tau   \big\|\f1L }
\times \big\| rV\psi   \big\|_{L^2_{r,\theta,\varphi}(\sin\theta dr d\theta d\varphi)}
\big\| \partial V\psi^{(0,\bar j)}   \big\|_{L^2_{r,\theta,\varphi}(\sin\theta dr d\theta d\varphi)}
   \, dt^*  \nonumber    \\
&\quad \stackrel{w\ge 1/2 \text{ in }\mathrm{\text{supp}}(1-\zeta)}{\lesssim_j} 
\int_0^\tau   \big\|\f1L \cdot(L w^2)\cdot r^{-5} (\Phi^\sharp_1)^{(0,\bar j+1)}\big\|_{L^\infty_{r,\theta,\varphi}}  \nonumber    \\
&\hphantom{\quad \stackrel{w\ge 1/2 \text{ in }\mathrm{\text{supp}}(1-\zeta)}{\lesssim_j} \int_0^\tau   \big\|\f1L }
\times \big\| rV\psi   \big\|_{L^2_{r,\theta,\varphi}(\sin\theta dr d\theta d\varphi)}
\big\| \partial V\psi^{(0,\bar j)}   \big\|_{L^2_{r,\theta,\varphi}(\sin\theta dr d\theta d\varphi)}
   \, dt^*  \nonumber    \\
&\quad \stackrel{\hphantom{w\ge 1/2 \text{ in }\mathrm{\text{supp}}(1-\zeta)}}{\lesssim_j}
 L \sup_{\bar \tau \in [0,\tau]} \mathscr F_3^{(j+1)}[\chi\tilde\phi](\bar\tau) \cdot \int_0^\tau \mathscr E^{(j)}[\psi](t^*) \, dt^*,  \label{Bound B 1 sharp 1 minus zeta}
\end{align}
where, in the last step above, we made use of the bound for $\Phi_1^\sharp$ provided by \cref{lem:Structure of tilde Phi}.

In order to estimate the term $\mathcal B_{1,\zeta}^{(\bar j,\sharp)}[\tau]$,  we begin by integrating in the $\partial_{t^*}$ derivative appearing in front of $\Phi^\sharp_1$:

\begin{align}
\mathcal B_{1,\zeta}^{(\bar j,\sharp)}[\tau] = &
\sup_{\substack{[\tau_1,\tau_2]\subseteq [0,\tau],\\
h:[\tau_1,\tau_2] \rightarrow \mathbb R, \\ \|h\|_{C^1} \le 1 }}  
\Bigg|    
\int_{\{\tau_1\le t^* \le \tau_2\}} \zeta(y,\theta)h(t^*) r^{-6} \Re\Big\{
    \f1L \partial_{t^*} \big(
\Phi^\sharp_1 
 \big)^{(0;\bar j)}  \cdot V\psi \cdot \partial_{t^*} (V\bar \psi^{(0,\bar j)})
 \Big\} \, \dvol_g
\Bigg|  \nonumber \\
\le & 
\sup_{\substack{[\tau_1,\tau_2]\subseteq [0,\tau],\\
h:[\tau_1,\tau_2] \rightarrow \mathbb R, \\ \|h\|_{C^1} \le 1 }}  
\Bigg|    
\int_{\{\tau_1\le t^* \le \tau_2\}} \!\!\!\!\zeta(y,\theta) h(t^*) r^{-6} \Re\Big\{
    \f1L  \big(
\Phi^\sharp_1 
 \big)^{(0;\bar j)}  \cdot (\partial_{t^*} V\psi) \cdot \partial_{t^*} (V\bar \psi^{(0,\bar j)})
 \Big\} \, \dvol_g
\Bigg|     \nonumber  \\
& + 
\sup_{\substack{[\tau_1,\tau_2]\subseteq [0,\tau],\\
h:[\tau_1,\tau_2] \rightarrow \mathbb R, \\ \|h\|_{C^1} \le 1 }}  
\Bigg|    
\int_{\{\tau_1\le t^* \le \tau_2\}} \zeta(y,\theta) h(t^*) r^{-6} \Re\Big\{
    \f1L  \big(
\Phi^\sharp_1 
 \big)^{(0;\bar j)}  \cdot  V\psi \cdot \partial_{t^*}^2 (V\bar \psi^{(0,\bar j)})
 \Big\} \, \dvol_g
\Bigg|     \nonumber\\
& + 
\sup_{\substack{[\tau_1,\tau_2]\subseteq [0,\tau],\\
h:[\tau_1,\tau_2] \rightarrow \mathbb R, \\ \|h\|_{C^1} \le 1 }}  
\Bigg|    
\int_{\{\tau_1\le t^* \le \tau_2\}} \zeta(y,\theta) h'(t^*) r^{-6} \Re\Big\{
    \f1L  \big(
\Phi^\sharp_1 
 \big)^{(0;\bar j)}  \cdot  V\psi \cdot \partial_{t^*} (V\bar \psi^{(0,\bar j)})
 \Big\} \, \dvol_g
\Bigg|  \nonumber
\\
& + \Big|\mathrm{\text{Boundary terms}}\Big|  \nonumber
\\
\doteq  \, & \mathrm{\text{I}} + \mathrm{\text{II}} + \mathrm{\text{III}} +  \Big|\mathrm{\text{Boundary terms}}\Big|,   \label{Bound B 1 zeta sharp}
\end{align}
where the boundary terms  can be estimated using the bounds for $\Phi_1^\sharp$ from \cref{lem:Structure of tilde Phi} as follows:
\begin{equation}\label{Boundary terms estimate}
\Big| \mathrm{\text{Boundary terms}}\Big| \lesssim_j \sup_{\bar \tau \in [0,\tau]}  \mathscr F^{(j)}_3[\chi\tilde\phi](\bar\tau)  \cdot\mathscr E^{(j)}[\psi](\tau).
\end{equation}

The terms $\mathrm{\text{I}}$  and  $\mathrm{\text{III}}$ can be estimated as follows (recall again that $r^{-1}\lesssim w$):
\begin{align}\label{Bound I term sharp high}
\mathrm{\text{I}} + \mathrm{\text{III}} \, 
 \stackrel{\hphantom{\eqref{Bound Phi sharp i}}}{\lesssim}
  & \f1L \int_0^\tau \big\| r^{-6} \cdot r^2 (\Phi_1^\sharp)^{(0;\bar j)} \big\|_{L^{\infty}_{r,\theta,\varphi}} \big\| \partial V \psi \big\|_{L^2_{r,\theta,\varphi}(\sin\theta dr d\theta d\varphi)} \big\| \partial V \psi^{(0,\bar j)} \big\|_{L^2_{r,\theta,\varphi}(\sin\theta dr d\theta d\varphi)} \, dt^* \\
& +   \f1L \int_0^\tau \big\| r^{-6} \cdot r^2 (\Phi_1^\sharp)^{(0;\bar j)} \big\|_{L^{\infty}_{r,\theta,\varphi}} \big\| V \psi \big\|_{L^2_{r,\theta,\varphi}(\sin\theta dr d\theta d\varphi)} \big\| \partial V \psi^{(0,\bar j)} \big\|_{L^2_{r,\theta,\varphi}(\sin\theta dr d\theta d\varphi)} \, dt^*  \nonumber  \\
\stackrel{\eqref{Bound Phi sharp i}}{\lesssim_j} & 
\f 1L \int_0^\tau \Big( L^2 \mathscr F_3^{(j+1)}[\chi\tilde\phi](t^*)  \cdot \mathscr E^{(j)}[\psi](t^*)\Big) \, dt^*
   \nonumber \\
\stackrel{\hphantom{\eqref{Bound Phi sharp i}}}{\lesssim_j} &
L  \sup_{\bar \tau \in [0,\tau]}  \mathscr F_3^{(j+1)}[\chi\tilde\phi](\bar\tau)  \cdot \int_0^\tau \mathscr E^{(j)}[\psi](t^*) \, dt^*.     \nonumber
\end{align}

For the term $\mathrm{\text{II}}$, we will need to use the equation satisfied by $\psi^{(0,\bar j)}$ and perform some additional integrations by parts. More precisely, in the support of $\zeta$ (which lies in the region $r\gtrsim 1, \min\{\theta, \pi-\theta\} \gtrsim 1$), we can schematically express using the identity \eqref{Wave operator V} of the wave operator (using also the fact that $\partial_\varphi = V - (1+L^{-1})\partial_{t^*}$ to express all $\partial_\varphi$ derivatives in terms of $\partial_{t^*}$ and $\partial_{\varphi}$) and equation \eqref{IVP Psi k}:
\begin{align}\label{Wave operator V again}
 r \partial_{t^*}V\psi^{(0;\bar j)} 
=  \big(1+ & O(w(y,\theta))\big)\bar\partial^2(r\psi^{(0;\bar j)}) + \big(1+O(w(y,\theta))\big) V^2(r\psi^{(0;\bar j)}) + O(1) \psi^{(0;\bar j)} \\
& +O(r^3)\Big( r^{-6} |\chi\tilde\phi+\psi|^2 \partial_{t^*}^2\psi^{(0;\bar j))}
 +\chi^2 \mathcal N_{(1,0,\bar j)}[\tilde \phi;\psi]  + \chi \mathcal N_{(2, 0,\bar j)}[\tilde\phi; \psi] \nonumber \\
& \hphantom{++O(r^3)\Big(}
+\mathcal N_{(3,0,\bar j)}[\psi]  - (\mathcal F[\tilde \phi])^{(0;\bar j)}
\Big),   \nonumber
\end{align}
where the terms $O(w)$, $O(1)$ and $O(r)$ are all $t,\varphi$-independent (and hence their $V$ derivative vanishes) and the $O(w)$ terms satisfy
\[
|\bar\partial\big(O(w)\big)|, \quad |\bar\partial^2\big(O(w)\big)| \lesssim 1.
\]

\begin{remark} From this point and until \eqref{Final estimate IIA}, we will use the notation $O(w)$, $O(1)$ etc. to denote possibly different functions satisfying the same properties as above.
\end{remark}
Substituting the expression \eqref{Wave operator V again} for $\partial_{t^*}V\psi^{(0;\bar j)}$ in  $\partial_{t^*}^2 V\bar\psi^{(0;\bar j)} = \partial_{t^*}\Big(\partial_{t^*} V\bar\psi^{(0;\bar j)} \Big)$ inside the term $\mathrm{\text{II}}$, we have
\begin{align}
\mathrm{\text{II}} \doteq \, & 
\sup_{\substack{[\tau_1,\tau_2]\subseteq [0,\tau],\\
h:[\tau_1,\tau_2] \rightarrow \mathbb R, \\ \|h\|_{C^1} \le 1 }}  
\Bigg|    
\int_{\{\tau_1\le t^* \le \tau_2\}} \!\!\!\!\!\! \!\!\!\!\!\! \!\!\!\!\!\! \zeta(y,\theta) h(t^*) r^{-6} \Re\Big\{
    \f1L  \big(
\Phi^\sharp_1 
 \big)^{(0;\bar j)}  \cdot  V\psi \cdot \partial_{t^*}^2 (V\bar \psi^{(0,\bar j)})
 \Big\} \, \dvol_g
\Bigg|
\nonumber \\
\le & 
\sup_{\substack{[\tau_1,\tau_2]\subseteq [0,\tau],\\
h:[\tau_1,\tau_2] \rightarrow \mathbb R, \\ \|h\|_{C^1} \le 1 }}  
\Bigg|    
\int_{\{\tau_1\le t^* \le \tau_2\}}  \!\!\!\!\!\!\!\!\!\!\!\!  \!\!\!\!\!\!\zeta(y,\theta) h(t^*) r^{-6} \Re\Big\{
    \f1L  \big(
\Phi^\sharp_1 
 \big)^{(0;\bar j)}  \cdot  V\psi \cdot \f{1+ O(w(y,\theta))}r \partial_{t^*}\bar\partial^2(r\bar\psi^{(0;\bar j)})
 \Big\} \, \dvol_g
\Bigg|
\nonumber \\
& + 
\sup_{\substack{[\tau_1,\tau_2]\subseteq [0,\tau],\\
h:[\tau_1,\tau_2] \rightarrow \mathbb R, \\ \|h\|_{C^1} \le 1 }}  
\Bigg|    
\int_{\{\tau_1\le t^* \le \tau_2\}} \!\!\!\!\!\! \!\!\!\!\!\! \!\!\!\!\!\! \zeta(y,\theta) h(t^*) r^{-6} \Re\Big\{
    \f1L  \big(
\Phi^\sharp_1 
 \big)^{(0;\bar j)}  \cdot  V\psi \cdot \partial_{t^*}V^2\big((1+O(w)) \bar \psi^{(0;\bar j)} \big)
 \Big\} \, \dvol_g
\Bigg|
\nonumber \\
& + 
\sup_{\substack{[\tau_1,\tau_2]\subseteq [0,\tau],\\
h:[\tau_1,\tau_2] \rightarrow \mathbb R, \\ \|h\|_{C^1} \le 1 }}  
\Bigg|    
\int_{\{\tau_1\le t^* \le \tau_2\}}\!\!\!\!\!\! \!\!\!\!\!\! \!\!\!\!\!\! \zeta(y,\theta) h(t^*) r^{-6} \Re\Big\{
    \f1L  \big(
\Phi^\sharp_1 
 \big)^{(0;\bar j)}  \cdot  V\psi 
\nonumber \\
& \hphantom{+ 
\sup_{\substack{[\tau_1,\tau_2]\subseteq [0,\tau],\\
h:[\tau_1,\tau_2] \rightarrow \mathbb R, \\ \|h\|_{C^1} \le 1 }}  
\Bigg|    
\int_{\{\tau_1\le t^* \le \tau_2\}}}
\times \partial_{t^*}\big( O(r^{-1}) \bar \psi^{(0;\bar j)} + O(r^{-4})|\chi\tilde\phi+\psi|^2 \partial_{t^*}^2 \bar \psi^{(0;\bar j)} \big)
 \Big\} \, \dvol_g
\Bigg|
\nonumber \\
& + 
\sup_{\substack{[\tau_1,\tau_2]\subseteq [0,\tau],\\
h:[\tau_1,\tau_2] \rightarrow \mathbb R, \\ \|h\|_{C^1} \le 1 }}  
\Bigg|    
\int_{\{\tau_1\le t^* \le \tau_2\}} \!\!\!\!\!\! \!\!\!\!\!\! \!\!\!\!\!\!  \zeta(y,\theta) h(t^*) r^{-6} \Re\Big\{
    \f1L  \big(
\Phi^\sharp_1 
 \big)^{(0;\bar j)}  \cdot  V\psi 
\nonumber \\
& \hphantom{+ 
\sup_{\substack{[\tau_1,\tau_2]\subseteq [0,\tau],\\
h:[\tau_1,\tau_2] \rightarrow \mathbb R, \\ \|h\|_{C^1} \le 1 }}  
\Bigg|    
\int_{\{\tau_1\le t^* \le \tau_2\}}}
\times O(r^2)\partial_{t^*}\big(\chi^2 \mathcal N_{(1,0,\bar j)}[\bar{\tilde \phi}; \bar\psi] + \chi \mathcal N_{(2,0,\bar j)}[\bar{\tilde \phi}; \bar\psi] + \mathcal N_{(3,0,\bar j)}[\bar\psi] \big)
 \Big\} \, \dvol_g
\Bigg|
\nonumber \\
& + 
\sup_{\substack{[\tau_1,\tau_2]\subseteq [0,\tau],\\
h:[\tau_1,\tau_2] \rightarrow \mathbb R, \\ \|h\|_{C^1} \le 1 }}  
\Bigg|    
\int_{\{\tau_1\le t^* \le \tau_2\}} \!\!\!\!\!\! \!\!\!\!\!\! \!\!\!\!\!\!  \zeta(y,\theta) h(t^*) r^{-6} \Re\Big\{
    \f1L  \big(
\Phi^\sharp_1 
 \big)^{(0;\bar j)}  \cdot  V\psi \cdot O(r^2)\partial_{t^*}\big( \mathcal F[\tilde \phi]\big)^{(0,\bar j)}
 \Big\} \, \dvol_g
\Bigg|
\nonumber \\
\doteq \, & 
\mathrm{\text{IIA}}+\mathrm{\text{IIB}}+\mathrm{\text{IIC}}+\mathrm{\text{IID}}+\mathrm{\text{IIE}}.\label{Expansion II term}
\end{align}
For the term $\mathrm{\text{IIA}}$, we first integrate by parts with respect to both of the $\bar\partial$ derivatives on the third factor (note that this process does not generate boundary terms in view of the support of $\zeta$) and then we integrate by parts with respect to the $V$ derivative on the second factor; the result of this process is as follows (recall also that $\dvol_g  = \big(1+\f1{r^2}-\f{2M}{r^3}\big) r^4  \sin\theta dt^* dy d\theta d\varphi$ and that $r(y) \sim y$):

\begin{align*}
\mathrm{\text{IIA}} 
\stackrel{\hphantom{\text{Int. by parts}}}{\doteq} \, & 
\sup_{\substack{[\tau_1,\tau_2]\subseteq [0,\tau],\\
h:[\tau_1,\tau_2] \rightarrow \mathbb R, \\ \|h\|_{C^1} \le 1 }}  
\Bigg|
\int_{\{\tau_1\le t^* \le \tau_2\}} \zeta(y,\theta) h(t^*) r^{-6}
 \Re\Big\{
    \f1L (\Phi^\sharp_1)^{(0;\bar j)} \cdot V\psi
\\
&\hphantom{
\sup_{\substack{[\tau_1,\tau_2]\subseteq [0,\tau],\\
h:[\tau_1,\tau_2] \rightarrow \mathbb R, \\ \|h\|_{C^1} \le 1 }}  
\Bigg|
\int_{\{\tau_1\le t^* \le \tau_2\}} }
\times \f{1+ O(w(y,\theta))}{r}\,
 \partial_{t^*}\bar\partial^2(r\bar\psi^{(0;\bar j)})
 \Big\} \, \dvol_g
\Bigg|
\\[0.5em]
\stackrel{\hphantom{\text{Int. by parts}}}{=} \, & 
\sup_{\substack{[\tau_1,\tau_2]\subseteq [0,\tau],\\
h:[\tau_1,\tau_2] \rightarrow \mathbb R, \\ \|h\|_{C^1} \le 1 }}  
\Bigg|
\int_{\{\tau_1\le t^* \le \tau_2\}} \zeta(y,\theta) h(t^*) r(y)^{-6}
 \Re\Big\{
    \f1L (r^2 \Phi^\sharp_1)^{(0;\bar j)}  (rV\psi)
\\
&\hphantom{
\sup_{\substack{[\tau_1,\tau_2]\subseteq [0,\tau],\\
h:[\tau_1,\tau_2] \rightarrow \mathbb R, \\ \|h\|_{C^1} \le 1 }}  
\Bigg|
\int_{\{\tau_1\le t^* \le \tau_2\}} }
\times \big(1+ O(w(y,\theta))\big)\,
 \partial_{t^*}\bar\partial^2(r\bar\psi^{(0;\bar j)})
 \Big\} \,
 \sin\theta \, dy \, d\theta \, d\varphi \, dt^*
\Bigg|
\\[0.5em]
\stackrel{\text{Int. by parts}}{\lesssim} \, &
\sum_{A=0}^2 \sum_{B=0}^A 
\int_{\{0\le t^* \le \tau\}\cap \mathrm{\text{supp}}\zeta}
 \f1L r^{-6+A-B}
 \big|\bar\partial^{B} V^{\le 1}(r^2 \Phi^\sharp_1)^{(0;\bar j)}\big|
  \big| \bar\partial^{2-A} (r \psi) \big|
\\
&\hphantom{
\sum_{A=0}^2 \sum_{B=0}^A 
\int_{\{0\le t^* \le \tau\}\cap \mathrm{\text{supp}}\zeta}
 \f1L r^{-6+A-B}}
\times \big| \partial_{t^*}(r\psi^{(0,\bar j)}) \big|
 \, \sin\theta \, dy \, d\theta \, d\varphi \, dt^*
\\
& \quad +
\sum_{A=0}^2 \sum_{B=0}^A 
\int_{\{0\le t^* \le \tau\}\cap \mathrm{\text{supp}}\zeta}
 \f1L r^{-6+A-B}
 \big|\bar\partial^{B} (r^2 \Phi^\sharp_1)^{(0;\bar j)}\big|
  \big| \bar\partial^{2-A} (r \psi) \big|
\\
&\hphantom{\quad +
\sum_{A=0}^2 \sum_{B=0}^A 
\int_{\{0\le t^* \le \tau\}\cap \mathrm{\text{supp}}\zeta}
 \f1L r^{-6+A-B}}
\times \big| \partial_{t^*}V(r\psi^{(0,\bar j)}) \big|
 \, \sin\theta \, dy \, d\theta \, d\varphi \, dt^*
\\
& \quad + \big| \mathrm{\text{Boundary terms}}  \big|,
\end{align*}
where the term $\big| \mathrm{\text{Boundary terms}}  \big|$ can be estimated as in \eqref{Boundary terms estimate}.
Using the standard $L^\infty-L^2-L^2$ H\"older-type estimate to control the above bulk integral when the summation index $A$ satisfies $A=0$ and the estimates
\[
 \Bigg[\sup_{j\in \mathbb N} \Bigg(\int_{\{t^*=\tau\}\cap\{y\le y_\mathrm{mirror}\}\cap\{ w(y, \theta) \le 2^{-j}\}}  
\f{2^j}{w(y,\theta)}|\bar\partial (r\psi)|^2\, \sin\theta dy d\theta d\varphi      \Bigg)  \Bigg]^{\f12} \le (\log L)^{\f12} \big(\mathcal E[\psi](\tau)\big)^{\f12}
\]
and
\[
\Bigg(\int_0^{2\pi} \Bigg[\sup_{(y,\theta)\in(0,y_\mathrm{mirror})\times(0,\pi)}\frac{|r\psi|^2(\tau,y,\theta, \varphi)}{\big(w(y,\theta)\big)^2}\Bigg] \, d\varphi \Bigg)^{\f12}  \le (\log L)^{\f12} \big(\mathcal E[\psi](\tau)\big)^{\f12}
\]
(obtained directly from the definition \eqref{Basic norm} of the norm $\mathcal E[\cdot]$) to control the second factors in the bulk terms when $A=1,2$ (i.e.~$\bar\partial (r \psi)$ and $r \psi$, respectively), we can readily estimate (noting that, for $0\le B\le A\le 2$, we have $w^A r^{-6+A-B}\lesssim w^{6+B}$):
\begin{align} 
\mathrm{\text{IIA}} & 
\lesssim_j 
 \,  \f1L \sum_{A=0}^2\sum_{B=0}^A \sup_{\bar \tau \in [0,\tau]}  \Big\|  (w(y,\theta))^A r^{-6+A-B} \bar\partial^{B}\big(r^2 \Phi^\sharp_1\big)^{(0;\bar j)} \Big\|_{L^\infty_{r,\theta,\varphi}(\{t^*=\bar \tau\})} \!\!\!\!\!\! \!\!\!\!\!\! \!\!\!\!\!\!\!   \cdot (\log L)^{\f12} \int_0^\tau \!\!  \mathscr E^{(j)}[\psi](t^*) \, dt^*
\nonumber \\
& \hphantom{\lesssim_j \, \sum_{i=0}^2 \sup_{\bar \tau \in [0,\tau]}  \Big\| \f1L r^{-6+i} \bar\partial^{2-i}\big(r^2 \Phi^\sharp_1\big)}
+\sup_{\bar \tau \in [0,\tau]}  \mathscr F_3^{(j)}[\chi\tilde\phi](\bar\tau)  \cdot \mathscr E^{(j)}[\psi](\tau)
\nonumber \\
&
\lesssim_j 
 \, \f1L \sup_{\bar \tau \in [0,\tau]}  \mathscr F^{(j)}_6[r \Phi^\sharp_1](\bar\tau)  \cdot (\log L)^{\f12} \int_0^\tau \mathscr E^{(j)}[\psi](t^*) \, dt^*
+\sup_{\bar \tau \in [0,\tau]}  \mathscr F_3^{(j)}[\chi\tilde\phi](\bar\tau)  \cdot \mathscr E^{(j)}[\psi](\tau)
\nonumber \\
& 
\lesssim_j 
 \, L  \sup_{\bar \tau \in [0,\tau]}  \mathscr F_3^{(j+1)}[\chi\tilde\phi](\bar\tau)   (\log L)^{\f12} \int_0^\tau \mathscr E^{(j)}[\psi](t^*) \, dt^*  +\sup_{\bar \tau \in [0,\tau]}  \mathscr F_3^{(j)}[\chi\tilde\phi](\bar\tau)  \cdot \mathscr E^{(j)}[\psi](\tau)  \label{Final estimate IIA}
\end{align}
where, in the last step, we used the bound \eqref{Bound Phi sharp i} for $\mathscr F^{(j)}[r\Phi^\sharp_1]$.

For the term $\mathrm{IIB}$, we obtain after integrating by parts with respect to one of the $V$ derivatives in front of the third factor (recall again that $\dvol_g  = \big(1+\f1{r^2}-\f{2M}{r^3}\big) r^4  \sin\theta dt^* dy d\theta d\varphi$):
\begin{align*}
\mathrm{\text{IIB}} \doteq \, &
\!\!\!\!
\sup_{\substack{[\tau_1,\tau_2]\subseteq [0,\tau],\\
h:[\tau_1,\tau_2] \rightarrow \mathbb R, \\ \|h\|_{C^1} \le 1 }}  
\Bigg|    
\int_{\{\tau_1\le t^* \le \tau_2\}} \!\! \!\!\!\!\!\!\zeta(y,\theta) h(t^*) r^{-6} \Re\Big\{
    \f1L  \big(
\Phi^\sharp_1 
 \big)^{(0;\bar j)}  \cdot  V\psi \cdot \partial_{t^*}V^2 \big((1+O(w)) \bar \psi^{(0;\bar j)}\big)
 \Big\} \, \dvol_g
\Bigg|\\
\lesssim & 
\sum_{A=0}^1 \int_{\{0\le t^* \le \tau\}\cap \mathrm{\text{supp}}\zeta}
\f1Lr^{-6}   \big| V^{\le A} (r^2 \Phi^\sharp_1)^{(0;\bar j)} \big| \cdot \big| V^{2-A} (r\psi) \big| \cdot \big|\partial_{t^*}V(r\psi)^{(0;\bar j)} \big| \, \sin\theta dy d\theta d\varphi dt^*
\\
& + \big|\mathrm{\text{Boundary terms}}\big|,
\end{align*}
where the terms $\big|\mathrm{\text{Boundary terms}}\big|$ can be estimated as in \eqref{Boundary terms estimate}. Therefore,
\begin{align*}
\mathrm{\text{IIB}} \lesssim 
& \int_0^\tau\!\! \f1L\big\| r^{-4} V^{\le 1} (\Phi^\sharp_1)^{(0;\le \bar j+1)}\big\|_{L^\infty_{y,\theta,\varphi}}\!\!\!\! \big\| V^{\le 2} (r\psi) \big\|_{L^2_{y,\theta,\varphi}(\sin\theta dy d\theta d\varphi)}\big\| \partial_{t^*}V (r\psi) \big\|_{L^2_{y,\theta,\varphi}(\sin\theta dy d\theta d\varphi)} \, dt^*\\
& + \big|\mathrm{\text{Boundary terms}}\big| \\
& \lesssim_j 
L  \sup_{\bar \tau \in [0,\tau]}  \mathscr F_3^{(j+1)}[\chi\tilde\phi](\bar\tau)  \cdot \int_0^\tau \mathscr E^{(j)}[\psi](t^*) \, dt^*  +\sup_{\bar \tau \in [0,\tau]}  \mathscr F_3^{(j)}[\chi\tilde\phi](\bar\tau)  \cdot \mathscr E^{(j)}[\psi](\tau),
\end{align*}
where, in passing to the last line above, we made use of the bound \eqref{Bound Phi sharp i} for $\Phi_1^\sharp$ from \cref{lem:Structure of tilde Phi}.

For the term $\mathrm{IIC}$, we obtain after integrating by parts with respect to the $\partial_{t^*}$ derivative in front of the third factor:
\begin{align}
\mathrm{\text{IIC}} \doteq \, &
\, 
\sup_{\substack{[\tau_1,\tau_2]\subseteq [0,\tau],\\
h:[\tau_1,\tau_2] \rightarrow \mathbb R, \\ \|h\|_{C^1} \le 1 }}  
\Bigg|    
\int_{\{\tau_1\le t^* \le \tau_2\}} \zeta(y,\theta) h(t^*) r^{-6}  \Re\Big\{
    \f1L  \big(
\Phi^\sharp_1 
 \big)^{(0;\bar j)}  \cdot  V\psi 
 \nonumber\\
& \hphantom{+ 
\sup_{\substack{[\tau_1,\tau_2]\subseteq [0,\tau],\\
h:[\tau_1,\tau_2] \rightarrow \mathbb R, \\ \|h\|_{C^1} \le 1 }}  
\Bigg|    
\int_{\{\tau_1\le t^* \le \tau_2\}}}\qquad 
\times \partial_{t^*}\big[O(r^{-1}) \bar \psi^{(0;\bar j)} + O(r^{-4})|\chi\tilde\phi+\psi|^2 \partial_{t^*}^2 \bar \psi^{(0;\bar j)} \big]
 \Big\} \, \dvol_g
\Bigg|
\nonumber\\
\lesssim & 
\sum_{A=0}^1 \int_{\{0\le t^* \le \tau\}\cap \mathrm{\text{supp}}\zeta}
\big|r^{-6}  \big(\f1L \partial_{t^*}\big)^{\le A} (r^2 \Phi^\sharp_1)^{(0;\bar j)} \big| \cdot \big| \big(\f1L \partial_{t^*}\big)^{1- A} V(r\psi) \big| 
\nonumber\\
& \hphantom{\sum_{A=0}^1 \int_{\{0\le t^* \le \tau\}\cap \mathrm{\text{supp}}\zeta}
\big| \big(\f1L }
\times \Big(\big|\psi^{(0;\bar j)}\big| + r^{-3} |\chi\tilde\phi+\psi|^2 \cdot \big| \partial_{t^*}^2 \psi ^{(0;\bar j)}\big|\Big) \, \sin\theta dy d\theta d\varphi dt^*
\nonumber
\\
& + \big|\mathrm{\text{Boundary terms}}\big| + \big|\mathrm{\text{Boundary terms}}'\big| \label{First bound IIC},
\end{align}
where the terms $\big|\mathrm{\text{Boundary terms}}\big|$ can be estimated  as in \eqref{Boundary terms estimate}, while $\big|\mathrm{\text{Boundary terms}}'\big|$ (which contain the boundary terms associated to the summand $r^{-6}  |\chi\tilde\phi+\psi|^2 \partial_{t^*}^2 \bar \psi^{(0;\bar j)}$ in the third factor of the integrand) can be estimated by
\begin{equation}\label{Boundary terms estimate 2}
\Big| \mathrm{\text{Boundary terms}}'\Big| \lesssim_j \sup_{\bar \tau \in [0,\tau]}  \Big(\mathscr F_3^{(j)}[\chi\tilde\phi](\bar\tau) \big(\mathscr F_3^{(0)}[\chi\tilde\phi](\bar\tau)+\mathscr F_3^{(0)}[\psi](\bar\tau) \big) \Big) \cdot\mathscr E^{(j)}[\psi](\tau).
\end{equation}
Using  the estimate
\begin{multline*}
\Big\|\f{1}{r^3(y) w^2(y,\theta)} |\chi\tilde\phi+\psi|^2 \cdot \partial_{t^*}^2 \psi ^{(0;\bar j)} \Big\|_{L^2_{y,\theta,\varphi}(\sin\theta dy d\theta d\varphi)} 
\\  \lesssim \Big\| \f{1}{w^2}L^2 r^{-4}|\chi\tilde\phi+\psi|^2 \Big\|_{L^\infty_{y,\theta,\varphi}} \Big\| \f1{L^2} \partial_{t^*}^2(r \psi ^{(0;\bar j)})\big| \Big|_{L^2_{y,\theta,\varphi}(\sin\theta dy d\theta d\varphi)} \\
 \lesssim \Big\| L^3 r^{-4}|\chi\tilde\phi+\psi|^2 \Big\|_{L^\infty_{y,\theta,\varphi}} \Big\| w^4 \partial^2(r \psi ^{(0;\bar j)})\big| \Big|_{L^2_{y,\theta,\varphi}(\sin\theta dy d\theta d\varphi)} \\
\lesssim L^3 \cdot \Big(\mathscr G_3^{(0)}[\chi\tilde\phi] + \mathscr G_3^{(0)}[\psi] \Big) \cdot \big( \mathscr E^{(\bar j)}[\psi]\big)^{\f12}
\end{multline*}
to control the term $r^{-4}|\chi\tilde\phi+\psi|^2 \big| \partial_{t^*}^2(r \psi ^{(0;\bar j)})\big|$ in the third factor (recall the definition \eqref{Weaker driving norm} of the $L^\infty$ type norm $\mathscr G_3^{(j)}[\cdot]$), we obtain from \eqref{First bound IIC} (using again  the bounds for $\Phi_1^\sharp$ from \cref{lem:Structure of tilde Phi}, as well as the fact that $\f1r \lesssim w$) that

\begin{align}
\mathrm{\text{IIC}}
\lesssim_j
\, &
 \int_0^\tau \Bigg(
   \Big\| L^{-1} \cdot L \cdot w^2 \cdot r^{-6}
          (L^{-1}\partial)^{\le 1}
          (r^2\Phi_1^\sharp)^{(0,\bar j)}
          \big|_{\mathrm{\text{supp}}\zeta}
   \Big\|_{L^\infty_\varphi L^2_{y,\theta}(\sin\theta \, dy \, d\theta)}
\nonumber\\
&  \hphantom{\int_0^\tau \Bigg(}
   \times  \Big\|(L^{-1}\partial )^{\le 1} V(r\psi)
            \big|_{\mathrm{\text{supp}}\zeta}\Big\|_{L^2_{y,\theta,\varphi}
            (\sin\theta \, dy \, d\theta \, d\varphi)}
   \cdot  \Big\|\f1{r w^2}\,|r\psi^{(0;\bar j)}|
          \Big\|_{L^2_\varphi L^\infty_{y,\theta}}
 \Bigg) \, dt^*
\nonumber \\ 
& +  \int_0^\tau \Bigg(
   \Big\| L^{-1} \cdot L \cdot w^2 \cdot r^{-6}
          (L^{-1}\partial)^{\le 1}
          (r^2\Phi_1^\sharp)^{(0,\bar j)}
          \big|_{\mathrm{\text{supp}}\zeta}
   \Big\|_{L^\infty_{y,\theta,\varphi}}
\nonumber\\
&  \hphantom{+  \int_0^\tau \Bigg(}
   \times  \Big\|(L^{-1}\partial )^{\le 1} V(r\psi)
            \big|_{\mathrm{\text{supp}}\zeta}\Big\|_{L^2_{y,\theta,\varphi}
            (\sin\theta \, dy \, d\theta \, d\varphi)}
\nonumber\\
&  \hphantom{+  \int_0^\tau \Bigg(}
   \times  \Big\| \f{1}{r^3 w^2} |\chi\tilde\phi+\psi|^2
                  \big| \partial_{t^*}^2 \psi ^{(0;\bar j)}\big|
                  \Big|_{\mathrm{\text{supp}}\zeta}
           \Big\|_{L^2_{y,\theta,\varphi}
           (\sin\theta \, dy \, d\theta \, d\varphi)}
 \Bigg) \, dt^* 
 \nonumber \\
& + \big|\mathrm{\text{Boundary terms}}\big|
  + \big|\mathrm{\text{Boundary terms}}'\big|
\nonumber \\
\lesssim_j
\, &
 \int_0^\tau \Bigg(
   L^{-1} \big(\mathscr F_6^{(j+1)}[r \Phi^\sharp_1]\big)^{\f12}
   \cdot \Big\| w^2 \cdot (L^{-1}\partial)^{\le 1}
              (r^2\Phi_1^\sharp)^{(0,\bar j)}
              \big|_{\mathrm{\text{supp}}\zeta}
        \Big\|_{L^\infty_\varphi L^2_{y,\theta}
        (\sin\theta \, dy \, d\theta)}
\nonumber\\
&  \hphantom{\int_0^\tau \Bigg(}
   \times  \Big\|(L^{-1}\partial )^{\le 1} V(r\psi)
            \big|_{\mathrm{\text{supp}}\zeta}\Big\|_{L^2_{y,\theta,\varphi}
            (\sin\theta \, dy \, d\theta \, d\varphi)}
   \cdot  \Big\|\f1{w}\,|r\psi^{(0;\bar j)}|
          \Big\|_{L^2_\varphi L^\infty_{y,\theta}}
 \Bigg) \, dt^*
\nonumber \\ 
& +  \int_0^\tau \Bigg(
   L^{-1} \big(\mathscr F_6^{(j+1)}[r \Phi^\sharp_1]\big)^{\f12}
   \cdot  \Big\|(L^{-1}\partial )^{\le 1} V(r\psi)
            \big|_{\mathrm{\text{supp}}\zeta}\Big\|_{L^2_{y,\theta,\varphi}
            (\sin\theta \, dy \, d\theta \, d\varphi)}
\nonumber\\
&  \hphantom{+  \int_0^\tau \Bigg(}
   \times  \Big\| \f{1}{r^3w^2} |\chi\tilde\phi+\psi|^2
                  \big| \partial_{t^*}^2 \psi ^{(0;\bar j)}\big|
                  \Big|_{\mathrm{\text{supp}}\zeta}
           \Big\|_{L^2_{y,\theta,\varphi}
           (\sin\theta \, dy \, d\theta \, d\varphi)}
 \Bigg) \, dt^* 
 \nonumber \\
& + \big|\mathrm{\text{Boundary terms}}\big|
  + \big|\mathrm{\text{Boundary terms}}'\big|
\nonumber \\
\lesssim_j 
\,   &
L  \sup_{\bar \tau \in [0,\tau]}
   \Big\{ \mathscr F_3^{(j+1)}[\chi\tilde\phi](\bar\tau)\Big\}
   \cdot \big(\log L\big)^{\f12}
   \int_0^\tau \mathscr E^{(j)}[\psi](t^*) \, dt^*
\nonumber \\
&
+ L^4  \sup_{\bar \tau \in [0,\tau]}
   \Bigg\{
     \mathscr F_3^{(j+1)}[\chi\tilde\phi](\bar\tau)
     \cdot \big( \mathscr G_3^{(0)}[\chi\tilde\phi](\bar\tau)
                +\mathscr G_3^{(0)}[\psi](\bar\tau) \big)
   \Bigg\}
\nonumber\\
&\hphantom{+}
   \times \int_0^\tau \mathscr E^{(j)}[\psi](t^*) \, dt^*
\nonumber \\
& +  \sup_{\bar \tau \in [0,\tau]}
 \Big(
   \mathscr F_3^{(j)}[\chi\tilde\phi](\bar\tau)
   \big(1+\mathscr F_3^{(0)}[\chi\tilde\phi](\bar\tau)
        +\mathscr F_3^{(0)}[\psi](\bar\tau) \big)
 \Big)
 \cdot \mathscr E^{(j)}[\psi](\tau)
\end{align}

where, in the last step above, we used the bound for $\Phi_1^\sharp$  from  \cref{lem:Structure of tilde Phi}. For the terms $\mathrm{\text{IID}}$ and $\mathrm{\text{IIE}}$, we can estimate directly using H\"older-type inequalities (i.e.~without the need to perform any integrations by parts) using the estimates for $\Phi_1^\sharp$  from  \cref{lem:Structure of tilde Phi} and the quantities controlled by $\mathcal E^{(j)}[\cdot]$ (see \eqref{Basic norm}):
\begin{align*}
\mathrm{\text{IID}} \doteq \, & \, 
\sup_{\substack{[\tau_1,\tau_2]\subseteq [0,\tau],\\
h:[\tau_1,\tau_2] \rightarrow \mathbb R, \\ \|h\|_{C^1} \le 1 }}  
\Bigg|    
\int_{\{\tau_1\le t^* \le \tau_2\}} \zeta(y,\theta) h(t^*) r^{-6} \Re\Big\{
    \f1L  \big(
\Phi^\sharp_1 
 \big)^{(0;\bar j)}  \cdot  V\psi \\
& \hphantom{+ 
\sup_{\substack{[\tau_1,\tau_2]\subseteq [0,\tau],\\
h:[\tau_1,\tau_2] \rightarrow \mathbb R, \\ \|h\|_{C^1} \le 1 }}  
\Bigg|    
 \int}
\times O(r^2)\partial_{t^*}\big(\chi^2 \mathcal N_{(1,0,\bar j)}[\bar{\tilde \phi}; \bar\psi] + \chi \mathcal N_{(2,0,\bar j)}[\bar{\tilde \phi}; \bar\psi] + \mathcal N_{(3,0,\bar j)}[\bar\psi] \big)
 \Big\} \, \dvol_g
\Bigg|\\
\lesssim & \int_0^\tau \f1L \big\| r^{-6} \cdot  r^2\big(\Phi^\sharp_1\big)^{(0;\bar j)}\|_{L^\infty_{y,\theta,\varphi}} \big\| V(r\psi) \big\|_{L^2_{y,\theta,\varphi}(\sin\theta dy d\theta d\varphi)} \\
& \hphantom{\lesssim \int_0^\tau \f1L \big\|  r^2}
\times \big\| r^3 \partial_{t^*}\big(\chi^2 \mathcal N_{(1,0,\bar j)}[\bar{\tilde \phi}; \bar\psi] + \chi \mathcal N_{(2,0,\bar j)}[\bar{\tilde \phi}; \bar\psi] + \mathcal N_{(3,0,\bar j)}[\bar\psi] \big) \big\|_{L^2_{y,\theta,\varphi}(\sin\theta dy d\theta d\varphi)} \, dt^* \\
\lesssim_j & 
L^2  \sup_{\bar \tau \in [0,\tau]} \Bigg\{ \mathscr F_3^{(j+1)}[\chi\tilde\phi](\bar\tau)\Big( \mathscr F_3^{(j)}[\chi\tilde\phi](\bar\tau)+\mathscr F_3^{(\lceil \f{j}{2}\rceil+2)}[\psi](\bar\tau) \Big)\Bigg\}  \cdot (\log L)^{\f12} \int_0^\tau \mathscr E^{(j)}[\psi](t^*) \, dt^*
\end{align*}
and
\begin{align*}
\mathrm{\text{IIE}} \doteq \, & \,
\sup_{\substack{[\tau_1,\tau_2]\subseteq [0,\tau],\\
h:[\tau_1,\tau_2] \rightarrow \mathbb R, \\ \|h\|_{C^1} \le 1 }}  
\Bigg|    
\int_{\{\tau_1\le t^* \le \tau_2\}} \!\!\!\!\!\!\!\!\!\zeta(y,\theta) h(t^*) r^{-6} \Re\Big\{
    \f1L  \big(
\Phi^\sharp_1 
 \big)^{(0;\bar j)}  \cdot  V\psi \cdot O(r^2) L \cdot \f1L \partial_{t^*}\big( \mathcal F[\tilde \phi]\big)^{(0,\bar j)}
 \Big\} \, \dvol_g
\Bigg|\\
\lesssim & \int_0^\tau\!\!  \big\| \f1L r^{-6} \cdot r^2\big(\Phi^\sharp_1\big)^{(0;\bar j)}\|_{L^\infty_{y,\theta,\varphi}} \!\big\| V(r\psi) \big\|_{L^2_{y,\theta,\varphi}\!(\sin\theta dy d\theta d\varphi)} 
 \big\| L r^3 \big( \mathcal F[\tilde \phi]\big)^{(0,\bar j+1)} \big\|_{L^2_{y,\theta,\varphi}\!(\sin\theta dy d\theta d\varphi)} \, dt^* \\
\lesssim_j &
L^2  \sup_{\bar \tau \in [0,\tau]}  \mathscr F_3^{(j+1)}[\chi\tilde\phi](\bar\tau)\cdot \big(\mathscr E^{(j)}[\psi](\tau)\big)^{\f12}  \cdot \int_0^\tau \big\| r^3 \big( \mathcal F[\tilde \phi]\big)^{(0,\bar j+1)} \big\|_{L^2_{y,\theta,\varphi}(\sin\theta dy d\theta d\varphi)} \, dt^*.
\end{align*}
Collecting the above bounds for $\mathrm{\text{IIA}}$, $\mathrm{\text{IIB}}$, $\mathrm{\text{IIC}}$, $\mathrm{\text{IID}}$ and $\mathrm{\text{IIE}}$ and returning to \eqref{Expansion II term}, we infer that
\begin{align*}
\mathrm{\text{II}} \lesssim_j & \,
 L  \sup_{\bar \tau \in [0,\tau]} \Bigg\{ \mathscr F_3^{(j+1)}[\chi\tilde\phi](\bar\tau)\Big( 1+ L^3 \big( \mathscr G_3^{(j)}[\chi\tilde\phi](\bar\tau)+\mathscr G_3^{(\lceil \f{j}{2}\rceil+2)}[\psi](\bar\tau) \big) \Big)\Bigg\}   (\log L)^{\f12} \int_0^\tau \mathscr E^{(j)}[\psi](t^*) \, dt^* \\
& + 
 \sup_{\bar \tau \in [0,\tau]}  \Big(\mathscr F_3^{(j)}[\chi\tilde\phi](\bar\tau) \big(1+\mathscr F_3^{(j)}[\chi\tilde\phi](\bar\tau)+\mathscr F_3^{(\lceil \f{j}{2}\rceil+2)}[\psi](\bar\tau) \big) \Big) \cdot\mathscr E^{(j)}[\psi](\tau)\\
& +
L^2  \sup_{\bar \tau \in [0,\tau]}  \mathscr F_3^{(j+1)}[\chi\tilde\phi](\bar\tau)\cdot \big(\mathscr E^{(j)}[\psi](\tau)\big)^{\f12}  \cdot \int_0^\tau \big\| r^3 \big( \mathcal F[\tilde \phi]\big)^{(0,\bar j+1)} \big\|_{L^2_{y,\theta,\varphi}(r^{-2}\sin\theta dr d\theta d\varphi)} \, dt^*.
\end{align*}
Therefore, returning to \eqref{Bound B 1 zeta sharp}, we have:
\begin{align*}
\mathcal B_{1,\zeta}^{(\bar j,\sharp)}[\tau] 
\lesssim_j & \,
 L  \sup_{\bar \tau \in [0,\tau]} \Bigg\{ \mathscr F_3^{(j+1)}[\chi\tilde\phi](\bar\tau)\Big( 1+ L \big( \mathscr F_3^{(j)}[\chi\tilde\phi](\bar\tau)+\mathscr F_3^{(\lceil \f{j}{2}\rceil+2)}[\psi](\bar\tau) \big) \Big)\Bigg\}  \\ & \hphantom{ L  \sup_{\bar \tau \in [0,\tau]}}\times  (\log L)^{\f12} \int_0^\tau \mathscr E^{(j)}[\psi](t^*) \, dt^* \\
& + 
 \sup_{\bar \tau \in [0,\tau]}  \Big(\mathscr F_3^{(j)}[\chi\tilde\phi](\bar\tau) \big(1+\mathscr F_3^{(j)}[\chi\tilde\phi](\bar\tau)+\mathscr F_3^{(\lceil \f{j}{2}\rceil+2)}[\psi](\bar\tau) \big) \Big) \cdot\mathscr E^{(j)}[\psi](\tau)\\
& +
L^2  \sup_{\bar \tau \in [0,\tau]}  \mathscr F_3^{(j+1)}[\chi\tilde\phi](\bar\tau)\cdot \big(\mathscr E^{(j)}[\psi](\tau)\big)^{\f12}   \int_0^\tau \big\| r^3 \big( \mathcal F[\tilde \phi]\big)^{(0,\bar j+1)} \big\|_{L^2_{r,\theta,\varphi}(r^{-2}\sin\theta dr d\theta d\varphi)} \, dt^*.
\end{align*}
Combining the above bound with the estimate \eqref{Bound B 1 sharp 1 minus zeta} for $\mathcal B_{1,1-\zeta}^{(\bar j,\sharp)}[\tau]$ and summing over $\bar j \in \{0,\ldots,j\}$, we finally obtain:
\begin{align}\nonumber
\sum_{\bar j=0}^j \mathcal B_{1}^{(\bar j,\sharp)}[\tau] &
\lesssim_j  \,
 L  \sup_{\bar \tau \in [0,\tau]} \Bigg\{ \mathscr F_3^{(j+1)}[\chi\tilde\phi](\bar\tau)\Big( 1+ L^3 \big( \mathscr G_3^{(j)}[\chi\tilde\phi](\bar\tau)+\mathscr G_3^{(\lceil \f{j}{2}\rceil+2)}[\psi](\bar\tau) \big) \Big)\Bigg\} \\
 & \hphantom{L  \sup_{\bar \tau \in [0,\tau]}} \times   (\log L)^{\f12} \int_0^\tau \mathscr E^{(j)}[\psi](t^*) \, dt^* \\
& + 
 \sup_{\bar \tau \in [0,\tau]}  \Big(\mathscr F_3^{(j)}[\chi\tilde\phi](\bar\tau) \big(1+\mathscr F_3^{(j)}[\chi\tilde\phi](\bar\tau)+\mathscr F_3^{(\lceil \f{j}{2}\rceil+2)}[\psi](\bar\tau) \big) \Big) \cdot\mathscr E^{(j)}[\psi](\tau)   \nonumber \\
& +
L^2  \sup_{\bar \tau \in [0,\tau]}  \mathscr F_3^{(j+1)}[\chi\tilde\phi](\bar\tau) \big(\mathscr E^{(j)}[\psi](\tau)\big)^{\f12}   \int_0^\tau \big\| r^3 \big( \mathcal F[\tilde \phi]\big)^{(\le j+1)} \big\|_{L^2_{r,\theta,\varphi}(r^{-2}\sin\theta dr d\theta d\varphi)} \, dt^*.   \label{Bound B 1 sharp}
\end{align}

\item For the term $B_1^{(\bar j,\star)}[\tau]$, we can directly estimate:
\begin{align*}
\mathcal B_1^{(\bar j,\star)}[\tau] \doteq \, & \, 
\sup_{\substack{[\tau_1,\tau_2]\subseteq [0,\tau],\\
h:[\tau_1,\tau_2] \rightarrow \mathbb R, \\ \|h\|_{C^1} \le 1 }}  
\Bigg|    
\int_{\{\tau_1\le t^* \le \tau_2\}} h(t^*) r^{-6}\Re\Big\{
   \f1L \big(
\Phi^\star_1
 \big)^{(0;\bar j)}  \cdot V\psi \cdot \partial_{t^*} (V\bar \psi^{(0,\bar j)})
 \Big\} \, \dvol_g
\Bigg| \\
\lesssim & 
\int_0^\tau \f1L \big\|r^{-6} \cdot  r^2 \big(\Phi^\star_1 \big)^{(0;\bar j)} \big\|_{L^\infty_{r,\theta,\varphi}} \big\| V\psi \big\|_{L^2_{r,\theta,\varphi}(\sin\theta  dr d\theta d\varphi)} \big\| \partial_{t^*} V\psi^{(0;\bar j)} \big\|_{L^2_{r,\theta,\varphi}(\sin\theta  dr d\theta d\varphi)} \, dt^* \\
\lesssim_j &
L  \sup_{\bar \tau \in [0,\tau]} \mathscr F_3^{(j+1)}[\chi\tilde\phi](\bar\tau) \int_0^\tau \mathscr E^{(j)}[\psi](t^*) \, dt^*
\end{align*}
(where, in the last step above, we made use of the bound \eqref{Bound Phi star i} for $\Phi_1^\star$) and, therefore, after summing over $\bar j\in \{0,j\}$:
\begin{equation}\label{Bound B 1 star}
\sum_{\bar j=0}^j \mathcal B_1^{(\bar j,\star)}[\tau] \lesssim_j 
L  \sup_{\bar \tau \in [0,\tau]} \mathscr F_3^{(j+1)}[\chi\tilde\phi](\bar\tau) \int_0^\tau \mathscr E^{(j)}[\psi](t^*) \, dt^*.
\end{equation}
\end{itemize}
Returning to \eqref{Decomposition B 1} and combining the bounds  \eqref{Bound B 1 sharp} and \eqref{Bound B 1 star}, we therefore obtain the required bound for $\mathcal B_1^{(\bar j)}[\tau]$:
\begin{align} \nonumber
\sum_{\bar j=0}^j \mathcal B_{1}^{(\bar j)}[\tau] 
\lesssim_j & \,
 L \!\!\! \sup_{\bar \tau \in [0,\tau]}\!\! \!\Bigg\{ \! \mathscr F_3^{(j+1)}[\chi\tilde\phi](\bar\tau)\Big( 1+ L^3 \big( \mathscr G_3^{(j)}[\chi\tilde\phi](\bar\tau)+\mathscr G_3^{(\lceil \f{j}{2}\rceil+2)}[\psi](\bar\tau) \big) \Big)\Bigg\}    (\log L)^{\f12} \!\int_0^\tau \!\!\mathscr E^{(j)}[\psi](t^*) \, dt^* \\
& + 
 \sup_{\bar \tau \in [0,\tau]}   \Big(\mathscr F_3^{(j)}[\chi\tilde\phi](\bar\tau) \big(1+\mathscr F_3^{(j)}[\chi\tilde\phi](\bar\tau)+\mathscr F_3^{(\lceil \f{j}{2}\rceil+2)}[\psi](\bar\tau) \big) \Big) \cdot\mathscr E^{(j)}[\psi](\tau)   \nonumber \\
& +
L^2  \sup_{\bar \tau \in [0,\tau]}  \mathscr F_3^{(j+1)}[\chi\tilde\phi](\bar\tau)\cdot \big(\mathscr E^{(j)}[\psi](\tau)\big)^{\f12}  \cdot \int_0^\tau \big\| r^3 \big( \mathcal F[\tilde \phi]\big)^{(\le j+1)} \big\|_{L^2_{r,\theta,\varphi}(r^{-2}\sin\theta dr d\theta d\varphi)} \, dt^*.  \label{Bound B 1}
\end{align}
In the case of $\mathcal B_2^{(\bar j)}[\tau]$ (see \eqref{def B k 2}), repeating exactly the same arguments as for the proof of \eqref{Bound B 1}, we similarly get:
\begin{align} \nonumber
\sum_{\bar j=0}^j \mathcal B_{2}^{(\bar j)}[\tau] 
\lesssim_j & \,
 L  \!\! \sup_{\bar \tau \in [0,\tau]}\!\!\! \Bigg\{ \!\mathscr F_3^{(j+1)}[\chi\tilde\phi](\bar\tau)\Big( 1+ L^3 \big( \mathscr G_3^{(j)}[\chi\tilde\phi](\bar\tau)+\mathscr G_3^{(\lceil \f{j}{2}\rceil+2)}[\psi](\bar\tau) \big) \Big)\Bigg\} (\log L)^{\f12} \!\int_0^\tau\!\!\mathscr E^{(j)}[\psi](t^*) \, dt^* \\
& + 
 \sup_{\bar \tau \in [0,\tau]}  \Big(\mathscr F_3^{(j)}[\chi\tilde\phi](\bar\tau) \big(1+\mathscr F_3^{(j)}[\chi\tilde\phi](\bar\tau)+\mathscr F_3^{(\lceil \f{j}{2}\rceil+2)}[\psi](\bar\tau) \big) \Big) \cdot\mathscr E^{(j)}[\psi](\tau)   \nonumber \\
& +
L^2  \sup_{\bar \tau \in [0,\tau]}  \mathscr F_3^{(j+1)}[\chi\tilde\phi](\bar\tau)\cdot \big(\mathscr E^{(j)}[\psi](\tau)\big)^{\f12}  \cdot \int_0^\tau \big\| r^3 \big( \mathcal F[\tilde \phi]\big)^{(\le j+1)} \big\|_{L^2_{r,\theta,\varphi}(r^{-2}\sin\theta dr d\theta d\varphi)} \, dt^*.   \label{Bound B 2}
\end{align}
Returning to \eqref{Decomposition B} and combining \eqref{Bound B 1} and \eqref{Bound B 2}, we finally obtain:
\begin{align}\nonumber
\mathcal B [\tau] 
\lesssim_j & \,
 L  \sup_{\bar \tau \in [0,\tau]} \Bigg\{ \mathscr F_3^{(j+1)}[\chi\tilde\phi](\bar\tau)\Big( 1+ L^3 \big( \mathscr G_3^{(j)}[\chi\tilde\phi](\bar\tau)+\mathscr G_3^{(\lceil \f{j}{2}\rceil+2)}[\psi](\bar\tau) \big) \Big)\Bigg\}  \cdot (\log L)^{\f12} \int_0^\tau \mathscr E^{(j)}[\psi](t^*) \, dt^* \\
& + 
 \sup_{\bar \tau \in [0,\tau]}  \Big(\mathscr F_3^{(j)}[\chi\tilde\phi](\bar\tau) \big(1+\mathscr F_3^{(j)}[\chi\tilde\phi](\bar\tau)+\mathscr F_3^{(\lceil \f{j}{2}\rceil+2)}[\psi](\bar\tau) \big) \Big) \cdot\mathscr E^{(j)}[\psi](\tau)   \nonumber \\
& +
L^2  \sup_{\bar \tau \in [0,\tau]}  \mathscr F_3^{(j+1)}[\chi\tilde\phi](\bar\tau)\cdot \big(\mathscr E^{(j)}[\psi](\tau)\big)^{\f12}  \cdot \int_0^\tau \big\| r^3 \big( \mathcal F[\tilde \phi]\big)^{(\le j+1)} \big\|_{L^2_{r,\theta,\varphi}(r^{-2}\sin\theta dr d\theta d\varphi)} \, dt^*.   \label{Bound B}
\end{align}

\item For the term $\mathcal C[\tau]$, we can estimate directly using H\"older-type inequalities:
\begin{align} \nonumber
\mathcal C[\tau] \doteq  &  
  \sum_{\bar j=0}^j 
\sup_{\substack{[\tau_1,\tau_2]\subseteq [0,\tau],\\
h:[\tau_1,\tau_2] \rightarrow \mathbb R,\\ \|h\|_{L^\infty} \le 1 }} \Bigg( 
\Bigg|    
\int_{\{\tau_1\le t^* \le \tau_2\}}\!\!\!\!\!\!\!\!\!\! \!\!\!\!\! \!\!\!\!\! \!\!\!  h(t^*)  \Re\Big\{ 
\Big( \chi^2 \mathcal N_{(1,\bar j,0)}^{[V]}[\tilde \phi;\psi]   +\chi \mathcal N_{(2,\bar j,0)}^{[V]}[\tilde\phi; \psi] +\mathcal N_{(3,\bar j,0)}^{[V]}[\psi] \Big)   
Z V  \psi^{(\bar j;0)}  
\Big\} \, \dvol_g
\Bigg| \\
& + \Bigg|    
\int_{\{\tau_1\le t^* \le \tau_2\}} h(t^*) \Re\Big\{ 
\Big( \chi^2 \mathcal N_{(1,\bar j,0)}[\tilde \phi;\psi]   +\chi \mathcal N_{(2,\bar j,0)}[\tilde\phi; \psi] +\mathcal N_{(3,\bar j,0)}[\psi]  \Big) \cdot
 Z \psi^{(\bar j;0)}  \Big\} \, \dvol_g
\Bigg|
\Bigg)
\nonumber\\[5pt]
\lesssim_j &
\int_0^{\tau} \Big( 
\big\|r^3\chi^2 \mathcal N_{(1,\bar j,0)}^{[V]}[\tilde \phi;\psi] \big\|_{L^2_{r,\theta,\varphi}(r^{-2}\sin\theta dr d\theta d\varphi)}  +\big\|r^3 \chi \mathcal N_{(2,\bar j,0)}^{[V]}[\tilde\phi; \psi]\big\|_{L^2_{r,\theta,\varphi}(r^{-2}\sin\theta dr d\theta d\varphi)} \nonumber\\
 & \hphantom{\int_0^{\tau} \Big( 
\big\|\chi^2 \mathcal N}
 +\big\|r^3 \mathcal N_{(3,\bar j,0)}^{[V]}[\psi]\big\|_{L^2_{r,\theta,\varphi}(r^{-2}\sin\theta dr d\theta d\varphi)}
 \Big) 
 \times  \big\| \partial V (r\psi^{(\bar j;0)})\big\|_{L^2_{r,\theta,\varphi}(r^{-2}\sin\theta dr d\theta d\varphi)}  \, dt^* 
\nonumber \\[5pt]
& + 
\int_0^{\tau} \Big( 
\big\|r^3 \chi^2 \mathcal N_{(1,\bar j,0)}[\tilde \phi;\psi] \big\|_{L^2_{r,\theta,\varphi}(r^{-2}\sin\theta dr d\theta d\varphi)}  +\big\|r^3 \chi \mathcal N_{(2,\bar j,0)}[\tilde\phi; \psi]\big\|_{L^2_{r,\theta,\varphi}(r^{-2}\sin\theta dr d\theta d\varphi)} \nonumber\\
 & \hphantom{\int_0^{\tau} \Big( 
\big\|\chi^2 \mathcal N}
  +\big\|r^3 \mathcal N_{(3,\bar j,0)}[\psi]\big\|_{L^2_{r,\theta,\varphi}(r^{-2}\sin\theta dr d\theta d\varphi)}
 \Big) 
 \times \big\| \partial (r\psi^{(\bar j;0)})\big\|_{L^2_{r,\theta,\varphi}(r^{-2}\sin\theta dr d\theta d\varphi)}  \, dt^* 
\nonumber\\
\lesssim_j & 
L  \sup_{\bar \tau \in [0,\tau]} \Bigg\{ \mathscr F_3^{(j+1)}[\chi\tilde\phi](\bar\tau) + \mathscr G_3^{(\lceil \f{j}{2}\rceil+2)}[\psi](\bar\tau) + \Big(\mathscr G_3^{(\lceil \f{j}{2}\rceil+2)}[\psi](\bar\tau)\Big)^{\f12}\Big(\mathscr G_3^{(\lceil \f{j}{2}\rceil+2)}[V\psi](\bar\tau)\Big)^{\f12}\Bigg\} \nonumber \\
  & \hphantom{\lesssim_j} \times \big(\log L\big)^{\f12} \int_0^\tau \mathscr E^{(j)}[\psi](t^*) \, dt^*.
\label{Estimate C term via Holder}
\end{align}
In the above, we estimated the terms $r^3 \chi^2 \mathcal N_{(1,\bar j,0)}^{[V]}[\tilde \phi;\psi]  $, $r^3 \chi \mathcal N_{(2,\bar j,0)}^{[V]}[\tilde\phi; \psi]$,  $r^3 \mathcal N_{(3,\bar j,0)}^{[V]}[\psi]$, $r^3 \chi^2 \mathcal N_{(1,\bar j,0)}[\tilde \phi;\psi] $, $r^3 \chi \mathcal N_{(2,\bar j,0)}[\tilde\phi; \psi]$ and $r^3 \mathcal N_{(3,\bar j,0)}[\psi]$
simply using an $L^\infty-L^\infty-L^2$ H\"older type estimate; in the case of $ \chi^2 \mathcal N_{(1,\bar j,0)}^{[V]}[\tilde \phi;\psi]  $, for instance, which can be written schematically as
\begin{align*}
\mathcal N_{(1,\bar j,0)}^{[V]}[\tilde \phi;\psi] = &
 \, \sum_{j_1=1}^{\bar j} r^{-6} \Bigg\{  \tilde \phi^{(\le \bar j)} \partial \tilde \phi^{(\le \bar j)}  \cdot  \partial V \psi^{(j_1)}
+ \big( V\tilde \phi^{(\le \bar  j)} \partial \tilde \phi^{(\le \bar j)} +\tilde \phi^{(\le \bar  j)} \partial V \tilde \phi^{(\le\bar  j)} \big) \cdot  \partial  \psi^{(j_1)}\\
& \hphantom{\sum_{j_1=1}^j \Bigg\{\tilde \phi r^{-6} }
 + \Big(\tilde \phi^{(\le \bar j)} \partial^2\tilde \phi^{(\le \bar j)}+\partial \tilde \phi^{(\le \bar j)}  \partial \tilde \phi^{(\le \bar j)}\Big)\cdot  V\psi^{(j_1)}
\Bigg\}+  r^{-6}  \tilde \phi^{(\le\bar  j)} V\tilde \phi^{(\le\bar  j)} \cdot  \partial^2 \psi^{(\le \bar j)} \\
& + r^{-6}  \Big(\tilde \phi^{(\le \bar j)} \partial^2 V\tilde \phi^{(\le\bar  j)}+ V\tilde \phi^{(\le\bar  j)} \partial^2 \tilde \phi^{(\le\bar  j)}+\partial \tilde \phi^{(\le \bar j)}  \partial V\tilde \phi^{(\le\bar  j)}\Big)\cdot  \psi^{(\le\bar  j)}
\\
= & \, \sum_{j_2=0}^{\bar j-1} r^{-6}  \Bigg\{  \tilde \phi^{(\le \bar j)} \partial \tilde \phi^{(\le \bar j)}  \cdot  \partial V \psi^{(j_2+1)}
+ \big( V\tilde \phi^{(\le \bar  j)} \partial \tilde \phi^{(\le \bar j)} +\tilde \phi^{(\le \bar  j)} \partial V \tilde \phi^{(\le\bar  j)} \big) \cdot  \partial  \psi^{(j_2+1)}\\
& \hphantom{\sum_{j_1=1}^j \Bigg\{\tilde \phi}
 + \Big(\tilde \phi^{(\le \bar j)} \partial^2\tilde \phi^{(\le \bar j)}+\partial \tilde \phi^{(\le \bar j)}  \partial \tilde \phi^{(\le \bar j)}\Big)\cdot  \f1L \partial V\psi^{(j_2)}
\Bigg\}  +  r^{-6}  \tilde \phi^{(\le\bar  j)} V\tilde \phi^{(\le\bar  j)} \cdot  \partial^2 \psi^{(\le \bar j)}  \\
&+  r^{-6} \Big(\tilde \phi^{(\le \bar j)} \partial^2 V\tilde \phi^{(\le\bar  j)}+ V\tilde \phi^{(\le\bar  j)} \partial^2 \tilde \phi^{(\le\bar  j)}+\partial \tilde \phi^{(\le \bar j)}  \partial V\tilde \phi^{(\le\bar  j)}\Big)\cdot  \psi^{(\le\bar  j)}
\end{align*}
(with all $\partial$ derivatives above being of the form $\partial_{t^*}$ or $Y$), one estimates:
\begin{align*}
\big\| r^3 \chi^2 & \mathcal N_{(1,\bar j,0)}^{[V]}[\tilde \phi;\psi]  \big\|_{L^2_{r,\theta,\varphi}(r^{-2}\sin\theta dr d\theta d\varphi)} \\
\lesssim_j \, & 
  \sum_{j_2=0}^{\bar j-1} \Bigg\{ L \big\|r^{-3}\chi r \tilde \phi^{(\le \bar j)} \big\|_{L^\infty_{r,\theta,\varphi}} \big\|r^{-3} \f1L \chi \partial (r \tilde \phi^{(\le \bar j)})\big\|_{L^\infty_{r,\theta,\varphi}}   \big\| \partial V (r \psi^{(j_2+1)})\big\|_{L^2_{r,\theta,\varphi}(r^{-2}\sin\theta dr d\theta d\varphi)}  \\
  &  
+ L \cdot  \big( \big\|r^{-3} \chi V\tilde (r \phi^{(\le \bar  j)})\big\|_{L^\infty_{r,\theta,\varphi}} \big\| r^{-3} \f1L \chi \partial ( r\tilde \phi^{(\le \bar j)})\big\|_{L^\infty_{r,\theta,\varphi}} 
 \\ &  
 + \big\|r^{-3} \chi \tilde ( r\phi^{(\le \bar  j)})\big\|_{L^\infty_{r,\theta,\varphi}} \big\|r^{-3} \f1L  \chi \partial V (r \tilde \phi^{(\le\bar  j)}) \big\|_{L^\infty_{r,\theta,\varphi}}  \big)   \big\|  \partial  (r \psi^{(j_2+1)})\big\|_{L^2_{r,\theta,\varphi}(r^{-2}\sin\theta dr d\theta d\varphi)} \\
&  
 +  L\cdot  \Big( \big\|r^{-3}\chi r \tilde \phi^{(\le \bar j)}\big\|_{L^\infty_{r,\theta,\varphi}} \big\| r^{-3}\f1{L^2}  \chi  \partial^2(r \tilde \phi^{(\le \bar j)})\big\|_{L^\infty_{r,\theta,\varphi}}\\
 &  +\big\|r^{-3}\f1L \chi \partial (r \tilde \phi^{(\le \bar j)} )\big\|_{L^\infty_{r,\theta,\varphi}} \big\|r^{-3} \f1L\chi  \partial (r \tilde \phi^{(\le \bar j)})\big\|_{L^\infty_{r,\theta,\varphi}} \Big) 
   \big\| \partial V( r\psi^{(j_2)})\big\|_{L^2_{r,\theta,\varphi}(r^{-2}\sin\theta dr d\theta d\varphi)} 
\Bigg\},\\
&\, + L \big\|r^{-3}\chi r \tilde \phi^{(\le\bar  j)}\big\|_{L^\infty_{r,\theta,\varphi}}\big\| r^{-3}\chi V( r\tilde \phi^{(\le\bar  j)}) \big\|_{L^\infty_{r,\theta,\varphi}} \cdot \big\| \f1L \partial^2 ( r\psi^{(\le \bar j)}) \big\|_{L^2_{r,\theta,\varphi}(r^{-2}\sin\theta dr d\theta d\varphi)} \\
&+  L \Big(\big\|r^{-3}\chi r \tilde \phi^{(\le \bar j)}\big\|_{L^\infty_{r,\theta,\varphi}} \big\| r^{-3} \f1L \chi w \partial^2 V(r \tilde \phi^{(\le\bar  j)})\big\|_{L^\infty_{\varphi}L^2_{r,\theta}(r^{-2}\sin\theta dr d\theta )} \\
& + \big\|r^{-3}\chi V( r\tilde \phi^{(\le\bar  j)}) \big\|_{L^\infty_{r,\theta,\varphi}}\big\|r^{-3}\f1L  \chi w \partial^2 (r \tilde \phi^{(\le\bar  j)})\big\|_{L^\infty_{\varphi}L^2_{r,\theta}(r^{-2}\sin\theta dr d\theta)} \\
&  +\big\|r^{-3} L^{-\f12} \chi w \partial (r \tilde \phi^{(\le \bar j)})\big\|_{L^\infty_{r,\theta,\varphi}} \big\| r^{-3} L^{-\f12} \chi  \partial V(r \tilde \phi^{(\le\bar  j)})\big\|_{L^\infty_\varphi L^2_{r,\theta,}(r^{-2}\sin\theta dr d\theta)} \Big)
   \big\| \f1{w} r \psi^{(\le\bar  j)}\big\|_{L^2_\varphi L^\infty_{r,\theta}} 
 \\
 \lesssim_j \, &
 \mathscr F_3^{(\bar j+1)}[\chi\tilde\phi] \cdot \big(\log L \big)^{\f12} \big(\mathcal E^{(\bar j)}[\psi] \big)^{\f12} .
\end{align*}

Similarly,  in the case of $ \mathcal N_{(3,\bar j,0)}^{[V]}[\psi]$, which can be written schematically as
\begin{align*}
\mathcal N_{(3,\bar j,0)}^{[V]} [\psi] = 
& \, 
r^{-3} \Bigg[ V \psi^{(\le \f{\bar j}2 +1)}  \psi^{(\le \f{\bar j}2 +1)} \cdot \partial^2 \psi^{(\le j)}
+ \psi^{(\le \f {\bar j}2 +1)}  \partial\psi^{(\le \f {\bar j}2 +1)} \cdot \partial V\psi^{(\le j)} \\
&\hphantom{ \, \sum_{\le k=0}^k \Bigg\{ \Phi}
+\Big(V\psi^{(\le \f {\bar j}2 +1)}  \partial \psi^{(\le \f {\bar j}2 +1)}+ \psi^{(\le \f {\bar j}2 +1)}  \partial V\psi^{(\le \f {\bar j}2 +1)}\Big) \cdot \partial \psi^{(\le j)} \\
&\hphantom{ \, \sum_{\le k=0}^k \Bigg\{ \Phi} 
+ \sum_{j_1=1}^{\bar j} \Big(\partial \psi^{(\le \f {\bar j}2+1)}  \partial \psi^{(\le \f {\bar j}2+1)} \cdot V\psi^{(j_1)}
+ \partial \psi^{(\le \f {\bar j}2+1)}  \partial V\psi^{(\le \f {\bar j}2+1)} \cdot \psi^{(j_1)}\Big)\Bigg]
\\
=
& \, 
r^{-3}V \psi^{(\le \f{\bar j}2 +1)}  \psi^{(\le \f{\bar j}2 +1)} \cdot \partial^2 \psi^{(\le j)}
+ r^{-3}\psi^{(\le \f {\bar j}2 +1)}  \partial\psi^{(\le \f {\bar j}2 +1)} \cdot \partial V\psi^{(\le j)} \\
&\hphantom{ \, \sum_{\le k=0}^k \Bigg\{ \Phi}
+r^{-3}\Big(V\psi^{(\le \f {\bar j}2 +1)}  \partial \psi^{(\le \f {\bar j}2 +1)}+ \psi^{(\le \f {\bar j}2 +1)}  \partial V\psi^{(\le \f {\bar j}2 +1)}\Big) \cdot \partial \psi^{(\le j)}\\
&\hphantom{ \, \sum_{\le k=0}^k \Bigg\{ \Phi} 
+r^{-3} \sum_{j_2=0}^{\bar j-1} \Big(\partial \psi^{(\le \f {\bar j}2+1)}  \partial \psi^{(\le \f {\bar j}2+1)} \cdot \f1L \partial V\psi^{(j_2)}
+ \partial \psi^{(\le \f {\bar j}2+1)}  \partial V\psi^{(\le \f {\bar j}2+1)} \cdot \f1L \partial \psi^{(j_2)}\Big)
\end{align*}
(and with all $\partial$ derivatives above as before being of the form $\partial_{t^*}$ or $Y$), one similarly estimates
\begin{align*}
\big\|r^3 &\mathcal N_{(3,\bar j,0)}^{[V]}[\psi]  \big\|_{L^2_{r,\theta,\varphi}(r^{-2} \sin\theta dr d\theta d\varphi)}\\
\lesssim_j \, & 
L \big\|r^{-3} V (r\psi^{(\le \f{\bar j}2 +1)})\big\|_{L^\infty_{r,\theta,\varphi}} \big\|r^{-3} \cdot  r \psi^{(\le \f{\bar j}2 +1)}\big\|_{L^\infty_{r,\theta,\varphi}} \big\| \f1L\partial^2(r \psi^{(\le j)})\big\|_{L^2_{r,\theta,\varphi}(r^{-2} \sin\theta dr d\theta d\varphi)}
\\ + L &  \big\|r^{-3} \cdot  r \psi^{(\le \f {\bar j}2 +1)}\big\|_{L^\infty_{r,\theta,\varphi}}  \big\|r^{-3}\f1L \partial( r \psi^{(\le \f {\bar j}2 +1)})\big\|_{L^\infty_{r,\theta,\varphi}} \big\| \partial V(r \psi^{(\le j)})\big\|_{L^2_{r,\theta,\varphi}(r^{-2} \sin\theta dr d\theta d\varphi)} \\
 +L &\Big( \big\|r^{-3}V(r \psi^{(\le \f {\bar j}2 +1)})\big\|_{L^\infty_{r,\theta,\varphi}} \big\| r^{-3}\f{1}{L} \partial ( r \psi^{(\le \f {\bar j}2 +1)})\big\|_{L^\infty_{r,\theta,\varphi}}\\
 &   + \big\| r^{-3} \cdot r \psi^{(\le \f {\bar j}2 +1)} \big\|_{L^\infty_{r,\theta,\varphi}} \big\| r^{-3} \f1L \partial V( r\psi^{(\le \f {\bar j}2 +1)})\big\|_{L^\infty_{r,\theta,\varphi}}\Big)
 \big\| \partial (r \psi^{(\le j)})\big\|_{L^2_{r,\theta,\varphi}(r^{-2} \sin\theta dr d\theta d\varphi)}\\
+&  \sum_{j_2=0}^{\bar j-1} L   \Big(\big\|r^{-3}\f1L \partial(r \psi^{(\le \f {\bar j}2+1)})\big\|_{L^\infty_{r,\theta,\varphi}}  \big\| r^{-3}\f1L \partial (r \psi^{(\le \f {\bar j}2+1)} )\big\|_{L^\infty_{r,\theta,\varphi}} \big\| \partial V(r \psi^{(j_2)})\big\|_{L^2_{r,\theta,\varphi}(r^{-2} \sin\theta dr d\theta d\varphi)} \\
& \hphantom{\sum}+ \big\|r^{-3}\f1L \partial (r \psi^{(\le \f {\bar j}2+1)})\big\|_{L^\infty_{r,\theta,\varphi}}  \big\| r^{-3} \f1L \partial V( r\psi^{(\le \f {\bar j}2+1)})\big\|_{L^\infty_{r,\theta,\varphi}}  \big\| \partial(r \psi^{(j_2)})\big\|_{L^2_{r,\theta,\varphi}(r^{-2} \sin\theta dr d\theta d\varphi)} \Big)\\
\lesssim_j \, &
L \Big(\big(\mathscr G_3^{(\lceil \f{\bar j}2\rceil+2)}[\psi]\big)^{\f12} \big(\mathscr G_3^{(\lceil \f{\bar j}2\rceil+2)}[V\psi]\big)^{\f12} + \mathscr G_3^{(\lceil \f{\bar j}2\rceil+2)}[\psi]\Big) \cdot \big(\mathcal E^{(\bar j)}[\psi] \big)^{\f12} .
\end{align*}
The rest of the terms  $r^3 \chi \mathcal N_{(2,\bar j,0)}^{[V]}[\tilde\phi; \psi]$,  $r^3 \chi^2 \mathcal N_{(1,\bar j,0)}[\tilde \phi;\psi] $, $r^3 \chi \mathcal N_{(2,\bar j,0)}[\tilde\phi; \psi]$ and $r^3 \mathcal N_{(3,\bar j,0)}[\psi]$ are estimated similarly.

\item For the term $\mathcal F_*[\tau]$, we can estimate directly:
\begin{align*}
\mathcal F_*[\tau] \doteq \, & \,
  \sum_{\bar j=0}^j 
\sup_{\substack{[\tau_1,\tau_2]\subseteq [0,\tau],\\
h:[\tau_1,\tau_2] \rightarrow \mathbb R, \\ \|h\|_{L^\infty} \le 1 }} \Bigg(  
 \Bigg|    
\int_{\{\tau_1\le t^* \le \tau_2\}} \!\!\!\! \!\!\!\! \!\!\!\! h(t^*) \Re\Big\{ 
 V(\mathcal F[\tilde \phi])^{(\bar j;0)} \Big) Z (V\bar \psi^{(\bar j;0)}) 
+
\Big(  (\mathcal F[\tilde \phi])^{(\bar j;0)}\Big) Z (\bar \psi^{(\bar j;0)})
\Big\} \, \dvol_g
\Bigg|
\Bigg) \\
\lesssim & 
\sum_{\bar j=0}^j \int_0^\tau \big\| r^3 \big( V^{\le 1}\mathcal F[\tilde \phi]\big)^{(\bar j;0)} \big\|_{L^2_{r,\theta,\varphi}(r^{-2}\sin\theta dr d\theta d\varphi)} \big\| \partial V^{\le 1} (r \psi^{(\bar j;0)}) \big\|_{L^2_{r,\theta,\varphi}(r^{-2} \sin\theta dr d\theta d\varphi)}\, dt^* \\
\lesssim_j  & 
\big(\mathscr E^{(j)}[\psi](\tau)\big)^{\f12} \int_0^\tau \big\| r^3 \big( V^{\le 1}\mathcal F[\tilde \phi]\big)^{(\le j)} \big\|_{L^2_{r,\theta,\varphi}(r^{-2}\sin\theta dr d\theta d\varphi)} \, dt^*.
\end{align*}

\end{enumerate}
Combining the above bounds for $\mathcal A[\tau]$, $\mathcal B[\tau]$, $\mathcal C[\tau]$ and $\mathcal F_*[\tau]$ and returning to \eqref{Coercivity higher order final estimate again}, we finally obtain the sought after energy estimate \eqref{Energy estimate Psi fundamental}.
\end{proof}

\subsection{Estimates for the \texorpdfstring{$\tilde\phi$ terms}{tilde phi terms} in the commuted equations for \texorpdfstring{$\psi$}{psi}}
In this section, we will establish a number of estimates for the terms involving the approximate solution $\tilde\phi = \sum_{k\in \mathcal K} \tilde\phi_k$ appearing in equations \eqref{IVP Psi k}--\eqref{IVP V Psi k}. These estimates will be used in combination with the a priori energy bounds of \cref{sec:A priori energy bounds} in the context of our bootstrap argument in the next section. To this end, we will make use of all the bounds established for the modes $\{\tilde\phi_k\}_{k\in \mathcal K}$ in \cref{sec:Dominant modes,sec:Non dominant modes}.

In particular, we will show the following:

\begin{lemma}\label{lem:Bounds for F tilde phi combined}
For any integer $j \ge 1$, the model solution $\tilde\phi$ satisfies for any $\tau\in[0,T_1]$:
\begin{equation}\label{Bound infty norm tilde phi}
\mathscr F_3^{(j)}[\chi \tilde\phi](\tau) + \f{1}{L^4}\mathcal E^{(j)}[\chi\tilde\phi](\tau) \lesssim_j \f1{T_1 \cdot L}
\end{equation}
(recall the definitions \eqref{Higher order norm} and \eqref{Driving norm} of the norms appearing above) and
\begin{equation}\label{Bound inhomogeneous term best}
\big\|  r^3 \big(\mathcal F[\tilde \phi]  \big)^{(\le j)}  \big|_{\Sigma^*_{\tau}}\big\|_{L^2_{r,\theta,\varphi}(r^{-2}\sin\theta drd\theta d\varphi)} + \big\|  r^3 V \big(\mathcal F[\tilde \phi]  \big)^{(\le j)}  \big|_{\Sigma^*_{\tau}}\big\|_{L^2_{r,\theta,\varphi}(r^{-2}\sin\theta drd\theta d\varphi)} \lesssim_{j} \f1{T_1 \cdot L^{s+3-\f92\delta_0}},
\end{equation}
where the constants implicit in the $\lesssim_j$ notation above depend only on $j$, the parameters $\lambda, N$ and the geometry of the spacetime $(\mathcal M_\mathrm{ext}, g_M))$ (they are, in particular, independent of $L$). 
\end{lemma}

\begin{proof}
The bound \eqref{Bound infty norm tilde phi} is a direct consequence of \cref{lem: Bound F norm phi tilde}. In order to establish \eqref{Bound inhomogeneous term best}, we will use the definition $\tilde\phi = \sum_{k\in \mathcal K} \tilde\phi_k$ and the fact that the $\tilde\phi_k$'s satisfy equation \eqref{Boundary value problem tilde phi} to  expand $\mathcal F[\tilde\phi]$ as follows:
\begin{align*}
\mathcal F[\tilde \phi] & \, \doteq \square_g (\chi \tilde\phi) + 2\chi \tilde\phi - \mathcal N[\chi \tilde\phi, \chi \tilde\phi,\chi \tilde\phi] \\
& = \partial \chi \cdot \partial \tilde\phi + \partial^2 \chi \cdot \tilde \phi + \chi \cdot \Big(\square_g \tilde\phi + 2\tilde\phi - \chi^2 \sum_{k_1, k_2, k_3\in \mathcal K} \mathcal N[\tilde \phi_{k_1}, \tilde \phi_{k_2}, \tilde \phi_{k_3}] \Big)\\
& = \partial \chi \cdot \partial \tilde\phi + \partial^2 \chi \cdot \tilde \phi + \chi \cdot (1-\chi^2) \sum_{k_1, k_2, k_3\in \mathcal K} \mathcal N[\tilde \phi_{k_1}, \tilde \phi_{k_2}, \tilde \phi_{k_3}] \\
& \hphantom{=}
+ \chi \cdot \Big(\sum_{k\in \mathcal K} (\square_g \tilde\phi_k + 2\tilde\phi_k) -  \sum_{k_1, k_2, k_3\in \mathcal K} \mathcal N[\tilde \phi_{k_1}, \tilde \phi_{k_2}, \tilde \phi_{k_3}] \Big),
\end{align*}
which can be reexpressed as
\begin{align} \nonumber
\mathcal F[\tilde \phi] 
& = \partial \chi \cdot \partial \tilde\phi + \partial^2 \chi \cdot \tilde \phi + \chi \cdot (1-\chi^2) \sum_{k_1, k_2, k_3\in \mathcal K} \mathcal N[\tilde \phi_{k_1}, \tilde \phi_{k_2}, \tilde \phi_{k_3}] \\
& \hphantom{=}
+ \chi \cdot \Bigg(\f{1}{r\big(1-\f{2M}r+r^2\big)}\sum_{k\in \mathcal K} \sum_{j_1, j_2, j_3\in \mathcal K_{\mathrm{D}}}\!\!\!\! \mathbb P_k \Big(r\big(1-\f{2M}r+r^2\big) \mathcal N[\tilde\phi_{j_1}, \tilde\phi_{j_2}, \tilde\phi_{j_3}]\Big) - \!\!\!\! \!\!\!\!\sum_{j_1, j_2, j_3\in \mathcal K_{\mathrm{D}}}\!\!\!\! \mathcal N[\tilde \phi_{j_1}, \tilde \phi_{j_2}, \tilde \phi_{j_3}] \Bigg)
\nonumber \\
&\hphantom{=}
+ \chi \cdot \sum_{\substack{k_1, k_2, k_3\in \mathcal K :\\ k_1\notin \mathcal K_{\mathrm{D}} \text{ \textbf{or} } k_2\notin \mathcal K_{\mathrm{D}} \text{ \textbf{or} } k_3 \notin \mathcal K_{\mathrm{D}} }} \mathcal N[\tilde \phi_{k_1}, \tilde \phi_{k_2}, \tilde \phi_{k_3}]    \nonumber  \\[5pt]
& = \partial \chi \cdot \partial \tilde\phi + \partial^2 \chi \cdot \tilde \phi + \chi \cdot (1-\chi^2) \sum_{k_1, k_2, k_3\in \mathcal K} \mathcal N[\tilde \phi_{k_1}, \tilde \phi_{k_2}, \tilde \phi_{k_3}] \nonumber \\
 &\hphantom{=}
 +\f{1}{r\big(1-\f{2M}r+r^2\big)}\sum_{k \notin \mathcal K} \, \sum_{j_1, j_2, j_3\in \mathcal K_{\mathrm{D}}} \mathbb P_k \Big(r\big(1-\f{2M}r+r^2\big) \mathcal N[\tilde\phi_{j_1}, \tilde\phi_{j_2}, \tilde\phi_{j_3}]\Big)
\nonumber \\
&\hphantom{=}
+ \chi \cdot \sum_{\substack{k_1, k_2, k_3\in \mathcal K :\\ k_1\notin \mathcal K_{\mathrm{D}} \text{ \textbf{or} } k_2\notin \mathcal K_{\mathrm{D}} \text{ \textbf{or} } k_3 \notin \mathcal K_{\mathrm{D}} }} \mathcal N[\tilde \phi_{k_1}, \tilde \phi_{k_2}, \tilde \phi_{k_3}],   \label{Expansion source term}
\end{align}
where the notation $k\notin \mathcal K$ means that $k=(n_k, \ell_k, m_k)\in \Big\{ \mathbb N^* \times \mathbb N_{\ge|m|} \times \mathbb Z : \quad n_k > \ell_k^{1-\delta_0}\Big\}$ (recall that $\mathcal K = \big\{ (n_k, \ell_k, m_k)\in \mathbb N^* \times \mathbb N_{\ge|m|} \times \mathbb Z: \, n_k \le \ell_k^{1-\delta_0}  \big\}$).

\begin{remark} In view of the bounds established below, we expect that the dominant term in the expansion \eqref{Expansion source term} for $\mathcal F[\tilde \phi]$ is the one occupying the second line in the right-hand side, namely 
\[
\f{1}{r\big(1-\f{2M}r+r^2\big)}\sum_{k \notin \mathcal K} \, \sum_{j_1, j_2, j_3\in \mathcal K_{\mathrm{D}}} \mathbb P_k \Big(r\big(1-\f{2M}r+r^2\big) \mathcal N[\tilde\phi_{j_1}, \tilde\phi_{j_2}, \tilde\phi_{j_3}]\Big).
\]
The fact that the above term satisfies bounds which are small in terms of $L$ merely at a fixed (i.e.~independent of $s$) polynomial rate and no better is due to the fact that the high frequency projection estimates for radial eigenfunctions provided by \cref{lem:Crude non stationary phase} are merely polynomially small in terms of the high frequency overtone $n$ (see also the remark below the statement of that lemma).
\end{remark}

As a consequence of \cref{prop:Total bound non-dominant modes}, the modes $\tilde\phi_k$ that do not vanish identically satisfy
\[
\ell_k \sim L.
\]
Moreover, note that the functions $\partial\chi$, $\partial^2\chi$ and $\chi \cdot (1-\chi^2)$ are supported in the region $\f12 y_\mathrm{mirror} \le y \le y_\mathrm{mirror}$. In this region, the modes $\{\tilde\phi_k\}_{k\in \mathcal K}$ with $\ell_k\sim L$ satisfy the following smallness bound, as a consequence of \cref{lem:eigenfunctions-estimate-m>0}, the trivial bound $\|\nabla_{\mathbb S^2}^j Y_{\ell_k, m_k}\|_{L^\infty} \lesssim_j \ell_k^{\f14}$ and the bounds \eqref{Bound C 1 norm tilde a}, \eqref{Bound C q norm tilde a} and \eqref{Estimate non dominant modes} for the coefficients $a_k(t)$:
\[
\sup_{\tau\in [0, T_1]} \sum_{j=0}^2 \| \partial^j \tilde\phi_k\|_{L^\infty(\{\f12 y_\mathrm{mirror}\le y \le y_\mathrm{mirror}\})}\lesssim e^{-L^{\f12}},
\]
where the constant implicit in the $\lesssim_k$ notation is independent of $L$ and $k$. Therefore, after summing over $k\in \mathcal K$, we obtain:
\[
\sup_{\tau\in [0, T_1]} \sum_{j=0}^2 \| \partial^j \tilde\phi \|_{L^\infty(\{\f12 y_\mathrm{mirror}\le y \le y_\mathrm{mirror}\})}\lesssim e^{-L^{\f14}}
\]
and, as a result, the first line in the expansion \eqref{Expansion source term} can be estimated as follows:
\begin{equation}\label{Bound cut off term}
\sup_{\tau\in [0, T_1]} \Big\|r^3 \Big(\partial \chi \cdot \partial \tilde\phi + \partial^2 \chi \cdot \tilde \phi + \chi \cdot (1-\chi^2) \sum_{k_1, k_2, k_3\in \mathcal K} \mathcal N[\tilde \phi_{k_1}, \tilde \phi_{k_2}, \tilde \phi_{k_3}]\Big)\Big|_{\Sigma^*_\tau} \Big\|_{L^2(r^{-2} \sin\theta dr d\theta d\varphi)}  \lesssim e^{-L^{\f14}}.
\end{equation}

The last line in the expansion \eqref{Expansion source term} involves products of modes $\tilde\phi_k$ for which at least one factor is a non-dominant mode; therefore, using \cref{prop:Total bound non-dominant modes} for the non-dominant modes and the $L^\infty$ bound $\|r^{-3}\partial^j (r\tilde\phi_k)\|_{L^\infty}\lesssim_j L^{j-1-s}$ for the dominant ones (following directly from \cref{prop:The 3 times 3 system} for the dominant amplitudes $b_k$ and the bounds \eqref{Dominant radial eigenfunction L infty}--\eqref{Dominant angular eigenfunction decay  L infty} for the dominant eigenfunctions $E_k$), we can readily estimate:
\begin{equation}\label{Bound non dominant part}
\sup_{\tau \in [0, T_1]}\Big\|  r^3  \chi \cdot \sum_{\substack{k_1, k_2, k_3\in \mathcal K :\\ k_1\notin \mathcal K_{\mathrm{D}} \text{ \textbf{or} } k_2\notin \mathcal K_{\mathrm{D}} \text{ \textbf{or} } k_3 \notin \mathcal K_{\mathrm{D}} }} \mathcal N[\tilde \phi_{k_1}, \tilde \phi_{k_2}, \tilde \phi_{k_3}]   \Big|_{\Sigma^*_\tau} \Big\|_{L^2(r^{-2} \sin\theta dr d\theta d\varphi)} \lesssim L^{\f12-5s}.
\end{equation}

Finally, we will estimate the second line in the expansion  \eqref{Expansion source term} as follows: We can rewrite for any dominant parameters $j_1, j_2, j_3 \in \mathcal K_{\mathrm{D}}$
\begin{multline}  
 \f{1}{r\big(1-\f{2M}r+r^2\big)}\sum_{k \notin \mathcal K} \mathbb P_k \Big(r\big(1-\f{2M}r+r^2\big) \mathcal N[\tilde\phi_{j_1},   \tilde\phi_{j_2}, \tilde\phi_{j_3}]\Big)
 \\
=   \f{1}{r\big(1-\f{2M}r+r^2\big)} \Big( \f{d a_{j_1}}{dt} \f{d \overline{a_{j_1}}}{dt} a_{j_3} + a_{j_1} \overline{a_{j_2}} \f{d^2 a_{j_3}}{dt^2} \Big)  
 \\
 \times \sum_{k\notin \mathcal K} \Big( \int_0^{y_\mathrm{mirror}} \int_{\mathbb S^2}\f1{r^6} \big(1+\f1{r^2}-\f{2M}{r^3}\big) \overline{E_k} E_{j_1} \overline{E_{j_2}} E_{j_3}  \, \sin\theta d\theta d\varphi dy\Big)\cdot E_k. 
 \label{One more expression for projection term high frequencies}
\end{multline}
Note that we can decompose for any $k=(n_k, \ell_k, m_k)\in \mathbb N^* \times \mathbb N_{\ge|m|} \times 
 \mathbb Z$:
\begin{align}\nonumber
 \int_0^{y_\mathrm{mirror}} \int_{\mathbb S^2} & \f1{r^6}  \big(1+\f1{r^2}-\f{2M}{r^3}\big) \overline{E_k} E_{j_1} \overline{E_{j_2}} E_{j_3}  \, \sin\theta d\theta d\varphi dy \\
 = & \Big(\int_0^{y_\mathrm{mirror}} \f1{r^6}  \big(1+\f1{r^2}-\f{2M}{r^3}\big) R_{n_k, \ell_k} R_{n_{j_1}, \ell_{j_1}} R_{n_{j_2}, \ell_{j_2}} R_{n_{j_3}, \ell_{j_3}} dy \Big)
 \nonumber \\
 & \quad \times \Big( \int_{\mathbb S^2} \overline{Y}_{\ell_k,m_k} Y_{\ell_{j_1}, m_{j_1}} \overline{Y}_{\ell_{j_2}, m_{j_2}}Y_{\ell_{j_3}, m_{j_3}} \, \sin\theta d\theta d\varphi \Big).  \label{High frequency total projection operator}
\end{align}
In view of the relations \eqref{Resonance in m for projection} and \eqref{Vanishing spectral coefficient angular component}, we have
\[
\int_{\mathbb S^2} \overline{Y}_{\ell_k,m_k} Y_{\ell_{j_1}, m_{j_1}} \overline{Y}_{\ell_{j_2}, m_{j_2}}Y_{\ell_{j_3}, m_{j_3}} \, \sin\theta d\theta d\varphi =0 \quad \text{ if } \ell_k \notin [L, 3(\lambda+2)L] \, \text{ or } \, m_k \neq m_{j_1}-m_{j_2}+m_{j_3}.
\]
Moreover, using the fact that $\|Y_{\ell,m}\|_{L^2(\mathbb S^2)} =1$ and $\|Y_{\ell, \pm\ell}\|_{L^{\infty}(\mathbb S^2)} \lesssim \ell^{\f14}$ (and the fact that $m_j =\pm \ell_j$ for $j\in \mathcal K_{\mathrm{D}}$), we can bound
\[
\Big|\int_{\mathbb S^2} \overline{Y}_{\ell_k,m_k} Y_{\ell_{j_1}, m_{j_1}} \overline{Y}_{\ell_{j_2}, m_{j_2}}Y_{\ell_{j_3}, m_{j_3}} \, \sin\theta d\theta d\varphi\Big| \lesssim L^{\f12} \quad \text{when} \quad \ell_k \in [L, 3(\lambda+2)L] \, \text{ and } \, m_k \neq m_{j_1}-m_{j_2}+m_{j_3}.
\]
As a consequence of \cref{lem:Crude non stationary phase} (applied with $(n,\ell)=(n_k, \ell_k)$, $(n_i,\ell_i)=(n_{j_i}, \ell_{j_i})$, $\lambda_-=\f1{(3\lambda+2)}$, $\lambda_+=(3\lambda+2)$ and $V(y) = \f{1}{y^6 \cdot r(y)^6}\big(1+\f1{r(y)^2}-\f{2M}{r(y)^3}\big)$), we then obtain that, for $\ell_k \in [L, 3(\lambda+2)L]$:
\[
\Big|\int_0^{y_\mathrm{mirror}} \f1{r^6} \big(1+\f1{r^2}-\f{2M}{r^3}\big) R_{n_k, \ell_k} R_{n_{j_1}, \ell_{j_1}} R_{n_{j_2}, \ell_{j_2}} R_{n_{j_3}, \ell_{j_3}} dy\Big| \lesssim \f{L^{-\f{5}2}}{n_k^5}.
\]
Combining the above bounds and returning to \eqref{High frequency total projection operator}, we infer that, for any $k\in \mathbb N^* \times \mathbb N_{\ge|m|} \mathbb Z$:
\begin{equation}\label{Polynomial bound projection radial modes}
\Big| \int_0^{y_\mathrm{mirror}} \int_{\mathbb S^2} \f1{r^6} \big(1+\f1{r^2}-\f{2M}{r^3}\big) \overline{E_k} E_{j_1} \overline{E_{j_2}} E_{j_3}  \, \sin\theta d\theta d\varphi dy\Big| \lesssim \f{1}{L^2 n_k^5} \ind_{[L, 3(\lambda+2)L]}(\ell_k) \ind_{\{m_{j_1}-m_{j_2}+m_{j_3}\}}(m_k).
\end{equation}
In particular, using the fact that $k\notin \mathcal K$ if and only if $n_k > \ell_k^{1-\delta_0}$, we have 
\[
\Bigg(\sum_{k\notin \mathcal K} \Big(\int_0^{y_\mathrm{mirror}} \int_{\mathbb S^2}\f1{r^6} \big(1+\f1{r^2}-\f{2M}{r^3}\big) \overline{E_k} E_{j_1} \overline{E_{j_2}} E_{j_3}  \, \sin\theta d\theta d\varphi dy\Big)^2  \Bigg)^{\f12} \lesssim L^{-6+\f92\delta_0}.
\]
Thus, in view of the fact that the $E_k$'s form an orthonormal sequence in $L^2(\sin\theta dy d\theta d\varphi)$ in $\{r\ge r_\mathrm{mirror}\}$, as well as the fact that in the region where $\chi\neq 1$ we have the exponentially small bound 
\[
\|E_k\|_{L^\infty(\mathrm{\text{supp}}(1-\chi))} \lesssim e^{-L^\f12}
\]
 for $k\in \mathcal K$ (following from \cref{lem:eigenfunctions-estimate-m>0} and the trivial bound $\|\nabla_{\mathbb S^2}^j Y_{\ell_k, m_k}\|_{L^\infty} \lesssim_j \ell_k^{\f14}$), we obtain from \eqref{One more expression for projection term high frequencies} that, for any $j_1, j_2, j_3\in \mathcal K_{\mathrm{D}}$:
\begin{align*}
\Big\|  \chi \cdot \f{1}{\big(1+\f1{r^2}-\f{2M}{r^3}\big)}\sum_{k\notin \mathcal K} \mathbb P_k \Big(r\big(1-\f{2M}r+r^2\big) &  \mathcal N[\tilde\phi_{j_1}, \tilde\phi_{j_2}, \tilde\phi_{j_3}]\Big) \Bigg|_{\Sigma^*_\tau}   \Big\|_{L^2(\sin\theta dy d\theta d\varphi)}\\
&  \lesssim L^{-6+\f92 \delta_0} \Big| \f{d a_{j_1}}{dt} \f{d \overline{a_{j_1}}}{dt} a_{j_3}(\tau) + a_{j_1} \overline{a_{j_2}} \f{d^2 a_{j_3}}{dt^2}(\tau) \Big|.
\end{align*}
Summing over $j_1, j_2, j_3\in \mathcal K_{\mathrm{D}}$ and using the bounds \eqref{Bound C 1 norm tilde a}--\eqref{Bound C q norm tilde a} for the dominant modes $a_{j_i} = b_{j_i} e^{-i \varepsilon_{j_i} \omega_{j_i} t}$, we finally infer that
\begin{equation}\label{Bound for projection term in high frequencies}
\sup_{\tau \in [0, T_1]} \Bigg\|    \f{r^3 \chi}{r\big(1-\f{2M}{r}+r^2\big)}\sum_{k\notin \mathcal K} \sum_{j_1, j_2, j_3\in \mathcal K_{\mathrm{D}}} \mathbb P_k \Big(r\big(1-\f{2M}r+r^2\big) \mathcal N[\tilde\phi_{j_1}, \tilde\phi_{j_2}, \tilde\phi_{j_3}]\Big) \Bigg|_{\Sigma^*_\tau}      \Bigg\|_{L^2(r^{-2} \sin\theta dr d\theta d\varphi)}\!\!\!\!\!\!\!\!\!\!\!\!\!\!\!\!\lesssim L^{-4+\f92 \delta_0-3s}.
\end{equation}

Combining \eqref{Bound cut off term}, \eqref{Bound non dominant part} and \eqref{Bound for projection term in high frequencies}, we obtain from \eqref{Expansion source term} (using also the fact that $s\ge 4$):
\[
\big\|  r^3 \mathcal F[\tilde \phi] \big|_{\Sigma^*_\tau} \big\|_{L^2(r^{-2} \sin\theta dr d\theta d\varphi)} \lesssim L^{-4+\f92 \delta_0-3s}.
\]
Repeating the same steps after commuting $j$ times with $L^{-1}\partial_{t^*}$ and once $V$ (noting that, for any dominant mode $\tilde\phi_k$, $k \in \mathcal K_{\mathrm{D}}$, we have for any $j_0, j_1\in \mathbb N$ and $i\in \{0,1\}$ that $\|\partial^{j_0} \partial_{t^*}^{j_1} V^i (r\tilde\phi_k)\|_{L^2(\sin\theta dy d\theta d\varphi )}$ and $\|\partial^{j_0} \partial_{t^*}^{j_1} V^i (r\tilde\phi_k)\|_{L^\infty(\sin\theta dy d\theta d\varphi )}$ satisfy similar bounds as $\|r\tilde\phi_k\|_{L^2(\sin\theta dy d\theta d\varphi )} $ and   $\|r\tilde\phi_k\|_{L^\infty(\sin\theta dy d\theta d\varphi )} $, respectively, as a consequence of \cref{prop:The 3 times 3 system} and the fact that $V\big(e^{i\varepsilon_k \omega_k t}E_k(y, \theta, \varphi)\big) = O(1) \cdot e^{i\varepsilon_k \omega_k t}E_k(y, \theta, \varphi)$), we obtain 
\[
\big\|  r^3 \big(\mathcal F[\tilde \phi]  \big)^{(\le j)}  \big|_{\Sigma^*_{\tau}}\big\|_{L^2_{r,\theta,\varphi}(r^{-2}\sin\theta drd\theta d\varphi)} + \big\|  r^3 V \big(\mathcal F[\tilde \phi]  \big)^{(\le j)}  \big|_{\Sigma^*_{\tau}}\big\|_{L^2_{r,\theta,\varphi}(r^{-2}\sin\theta drd\theta d\varphi)} \lesssim_{j} L^{-4+\f92 \delta_0-3s}.
\]
The bound \eqref{Bound inhomogeneous term best} now readily follows in view of the fact that $T_1 \sim L^{2s+1}$.
\end{proof}

\subsection{Proof of \texorpdfstring{\cref{prop: Estimates error term}}{Proposition~7.1}: The bootstrap argument}\label{sec:The bootstrap argument}
In this section, we will establish   \cref{prop: Estimates error term}; the proof will proceed via a standard continuity argument, the main ingredients of which are going to be the a priori energy estimate \eqref{Energy estimate Psi fundamental} from   \cref{lem:Energy estimates Psi} and the bounds for the model solution $\tilde \phi$ provided by   \cref{lem:Bounds for F tilde phi combined}

Let us set
\[
j \doteq \max\Big\{ s+4, \, 20   \Big\}.
\]
Let also $C_{\mathrm{boot}} = C_{\mathrm{boot}}(\epsilon, s,\tilde T, \lambda, N)$ be a (large) bootstrap constant that will be fixed later in terms only of the parameters $\epsilon, s,\lambda, N, \tilde T$ (where $\tilde T$ is the timescale appearing in the statement of  \cref{prop:The 3 times 3 system}) and the geometry of $(\mathcal M_\mathrm{ext},g)$; hence, $C_{\mathrm{boot}}$ is chosen independently of $L$, which will be assumed to be large enough compared to the other parameters. We will employ the same convention that we have adopted so far in this section regarding the dependence of the constants implicit in the $\lesssim$, $\sim$ notations on the various parameters, namely that they will be allowed to depend on all the parameters \textbf{except} for $L$ and $C_{\mathrm{boot}}$.

We will define $T_{\mathrm{boot}}$ to be the largest time in the interval $(0, T_1]$ satisfying both of the following properties:
\begin{enumerate}
\item \textbf{Existence:}  The solution $\psi$ of the initial-boundary value problem \eqref{IVP Psi} exists and remains smooth for $0 \le t^* \le T_{\mathrm{boot}}$.
\item \textbf{The bootstrap assumption:} The function $\psi$ satisfies on $\{0\le t^* \le T_{\mathrm{boot}}\}$:
\begin{equation}\label{Bootstrap assumption}
\sup_{\tau \in [0,T_{\mathrm{boot}}]}\mathcal E^{(j)}[\psi](\tau) \le C_{\mathrm{boot}} L^{-8+10\delta_0} \mathcal E [\chi \tilde\phi](0).
\end{equation}
\end{enumerate}
\begin{remark} From now on, we will explicitly state the dependence of any constant on $C_{\mathrm{boot}}$. In particular, it will be assumed that the implicit constant associated to the notation $\lesssim$ does not depend on $C_{\mathrm{boot}}$ and $L$ (but is allowed to depend on the spacetime geometry and the parameters $\lambda, N, s$).
\end{remark}

Note that, due to the well-posedness of the initial boundary value problem \eqref{IVP Psi} in the smooth category and the fact that
\begin{equation}\label{Initial higher order bound}
\mathcal E^{(j)}[\psi](0) \lesssim \mathcal E^{(j-2)}[r^{-4} (1+|\chi \tilde \phi|^2)^{-1} \mathcal F[\tilde \phi]](0) \stackrel{\cref{lem:Bounds for F tilde phi combined}}{\lesssim} L^{-6s-2} \stackrel{\cref{lem: Bound F norm phi tilde}}{\ll} L^{-8} \mathcal E[\chi \tilde\phi](0)   
\end{equation}
 (obtained using the equation \eqref{IVP Psi} along the initial slice $t^*=0$ to solve for $\partial_{t^*}^2 \psi|_{t^*=0}$ via the identity $(g_{rr}+r^{-6}|\chi \tilde\phi|^2)\partial_{t^*}^2 \psi \big|_{t^*=0} = \mathcal F[\tilde \phi] \big|_{t^*=0}$), there certainly exists some $T_{\mathrm{boot}}>0$ such that the above two conditions are satisfied on $\{0\le t^* \le T_{\mathrm{boot}}\}$ provided $C_{\mathrm{boot}}$ is chosen large enough in terms of $s,\lambda, N$ and the geometry of $(\mathcal M_\mathrm{ext},g)$, while $L$ is chosen large enough in terms of $C_{\mathrm{boot}}, s, \lambda, N$. 

Following the standard line of arguments employed in proofs using the continuity method, the proof of  \cref{prop: Estimates error term}  (in particular, the estimate \eqref{Bound psi}) will immediately follow once we show that, if $T_{\mathrm{boot}} < T_1$, then the following improvement of the bootstrap assumption \eqref{Bootstrap assumption} holds on $\{0\le t^* \le T_{\mathrm{boot}}\}$ (provided $L$ is chosen large enough depending only on $C_{\mathrm{boot}}, s, \lambda, N, \tilde T$ and the geometry of $(\mathcal M_\mathrm{ext},g)$):
\begin{equation}\label{Bootstrap improvement}
\sup_{\tau \in [0,T_{\mathrm{boot}}]}\mathcal E^{(j)}[\psi](\tau) \le \f12 C_{\mathrm{boot}} L^{-8+10\delta_0} \mathcal E[\chi \tilde\phi](0).
\end{equation}

We will apply \cref{lem:Energy estimates Psi} for $T^* = T_{\mathrm{boot}}$: Using the notation $\mathscr E^{(j)}[\psi]$ and $\mathscr H^{(j)}[\psi]$ appearing in the  statement of \cref{lem:Energy estimates Psi}, we obtain the following energy estimate for $\psi$ for any $\tau \in [0, T_{\mathrm{boot}}]$:
\begin{align}\nonumber
\mathscr E^{(j)}[\psi](\tau) &  +\mathscr H^{(j)}[\psi](\tau) \\
\lesssim & \,
\mathscr E^{(j)}[\psi](0) \nonumber\\
& +
 L  \sup_{\bar \tau \in [0,\tau]} \Bigg\{ \Big(\mathscr F_3^{(j+1)}[\chi\tilde\phi](\bar\tau) + \mathscr G_3^{(\lceil \f{j}{2}\rceil+2)}[\psi](\bar\tau) + \big( \mathscr G_3^{(\lceil \f{j}{2}\rceil+2)}[V \psi](\bar\tau)\big)^{\f12}\big(\mathscr G_3^{(\lceil \f{j}{2}\rceil+2)}[\psi](\bar\tau)\big)^{\f12} \Big) 
 \nonumber \\
 & \hphantom{+
 L  \sup_{\bar \tau \in [0,\tau]} \Bigg\{ \Big(\mathscr F_3^{(j+1)}[\chi\tilde\phi](\bar\tau) }
 \times  \Big( 1+ L^3 \big( \mathscr G_3^{(j)}[\chi\tilde\phi](\bar\tau)+\mathscr G_3^{(\lceil \f{j}{2}\rceil+2)}[\psi](\bar\tau) \big) \Big)\Bigg\} 
 \nonumber\\
 & \hphantom{ L  \sup_{\bar \tau \in [0,\tau]} \Bigg\{ \Big(\mathscr F_3^{(j+1)}[\chi\tilde\phi](\bar\tau) + \mathscr F_3^{(\lceil \f{j}{2}\rceil+2)}[\psi](\bar\tau) \Big)\cdot }
 \times  (\log L)^{\f12} \int_0^\tau \mathscr E^{(j)}[\psi](t^*) \, dt^* \nonumber \\[5pt]
& + 
 \sup_{\bar \tau \in [0,\tau]}  \Big(\big(\mathscr F_3^{(j+2)}[\chi\tilde\phi](\bar\tau) + \mathscr F_3^{(\lceil \f{j}{2}\rceil+2)}[\psi](\bar\tau)\big)\big(1+\mathscr F_3^{(j+2)}[\chi\tilde\phi](\bar\tau)+\mathscr F_3^{(\lceil \f{j}{2}\rceil+2)}[\psi](\bar\tau) \big) \Big) \nonumber\\
 & \hphantom{+\sup}\times  \big(\mathscr E^{(j)}[\psi](\tau)+\mathscr H^{(j)}[\psi](\tau)\big)   \nonumber \\
& +
\Bigg\{1+L^4  \sup_{\bar \tau \in [0,\tau]}  \Big(\mathscr F_3^{(j+1)}[\chi\tilde\phi](\bar\tau) \Big)^2\Bigg\}  \cdot \Bigg( \int_0^\tau \big\| r^3 V^{\le 1}\big( \mathcal F[\tilde \phi]\big)^{(\le j+1)} \Big|_{\Sigma^*_{t^*}}\big\|_{L^2_{r,\theta,\varphi}(r^{-2}\sin\theta dr d\theta d\varphi)} \, dt^*\Bigg)^2  \nonumber \\
& +
\sup_{\bar\tau \in[0,\tau]} \Big\|\Big(r^3 \mathcal F[\tilde \phi]
\Big)^{(\le j)}\big|_{\Sigma^*_{\bar\tau}} \Big\|^2 _
 {L^2(\sin\theta \f{1}{r^2}dr d\theta d\varphi)}. \label{Energy-estimate-for-Bootstrap}
\end{align}
We will estimate the coefficients involving the $L^\infty$-type norms of $\psi$ in the right-hand side above using the Sobolev embedding estimates of \cref{lem:Sobolev estimates driving norms} and our bootstrap assumption \eqref{Bootstrap assumption} (and the fact that $j\ge 20$): For any $\bar \tau \in [0,T_{\mathrm{boot}}]$, we have 
\[
\mathscr G_3^{(\lceil \f{j}{2}\rceil+2)}[\psi](\bar\tau)+\mathscr G_3^{(\lceil \f{j}{2}\rceil+2)}[V\psi](\bar\tau)\stackrel{\cref{lem:Sobolev estimates driving norms}}{\lesssim} L^{2} \mathscr E^{(j)}[\psi](\tau) \stackrel{\eqref{Bootstrap assumption}}{\lesssim}C_{\mathrm{boot}} L^{-6+10\delta_0} \mathcal E[\chi\tilde\phi](0), 
\]
and
\[
\mathscr F_3^{(\lceil \f{j}{2}\rceil+2)}[\psi](\bar\tau)\stackrel{\cref{lem:Sobolev estimates driving norms}}{\lesssim} L^{10} \mathscr E^{(j)}[\psi](\tau) \stackrel{\eqref{Bootstrap assumption}}{\lesssim}C_{\mathrm{boot}} L^3 \mathcal E[\chi\tilde\phi](0).
\]
In particular, combining the above with the estimate \eqref{Bound infty norm tilde phi} for $\tilde\phi$ (noting also that $\mathscr G_3^{(j)}[\chi\tilde\phi]\lesssim \mathscr F_3^{(j)}[\chi\tilde\phi]$ just in view of the definition of the quantities $\mathscr G_A[\cdot], \mathscr F_A[\cdot]$) and  using the fact that $s\ge 4$ and $T_1 \sim L^{2s+1}$, we infer that, provided $L$ is sufficiently large in terms of $C_{\mathrm{boot}}$:
\begin{align*}
 \sup_{\bar \tau \in [0,\tau]}  \Bigg\{ \Big(\mathscr F_3^{(j+1)}[\chi\tilde\phi](\bar\tau) + &  \mathscr G_3^{(\lceil \f{j}{2}\rceil+2)}[\psi](\bar\tau) + \big( \mathscr G_3^{(\lceil \f{j}{2}\rceil+2)}[V \psi](\bar\tau)\big)^{\f12}\big(\mathscr G_3^{(\lceil \f{j}{2}\rceil+2)}[\psi](\bar\tau)\big)^{\f12} \Big) \\ & 
\hphantom{\lesssim}\times  \Big( 1+ L^3 \big( \mathscr G_3^{(j)}[\chi\tilde\phi](\bar\tau)+\mathscr G_3^{(\lceil \f{j}{2}\rceil+2)}[\psi](\bar\tau) \big) \Big)\Bigg\} 
\\
 & \lesssim \Big(\f{1}{L\cdot T_1} + C_{\mathrm{boot}} \f1{L^{3-10\delta_0}T_1}\Big)\cdot \Big(1+L^3 \big(\f{1}{L T_1}+ C_{\mathrm{boot}}\f{1}{L^{3-10\delta_0}T_1}\big)\Big)
 \\
 & \lesssim \f{1}{L\cdot T_1},
 \end{align*}
 
\begin{multline*}
\sup_{\bar \tau \in [0,\tau]}  \Big(\big(\mathscr F_3^{(j+2)}[\chi\tilde\phi](\bar\tau) + \mathscr F_3^{(\lceil \f{j}{2}\rceil+2)}[\psi](\bar\tau)\big)\big(1+\mathscr F_3^{(j+2)}[\chi\tilde\phi](\bar\tau)+\mathscr F_3^{(\lceil \f{j}{2}\rceil+2)}[\psi](\bar\tau) \big) \Big) 
\\  \lesssim \big(\f1{L \cdot T_1}+C_{\mathrm{boot}} \f{L^6}{T_1}\big)\big(1+\f{1}{L\cdot T_1} + C_{\mathrm{boot}}\f{ L^6}{T_1}\big) 
 \lesssim C_{\mathrm{boot}} L^{5-2s} \ll 1
\end{multline*}
and
\[
L^4  \sup_{\bar \tau \in [0,\tau]}  \Big(\mathscr F_3^{(j+1)}[\chi\tilde\phi](\bar\tau) \Big)^2 \lesssim \f{L^4}{L^2 T_1^2} \ll 1.
\]
 Returning to \eqref{Energy-estimate-for-Bootstrap}, we therefore obtain:
 \begin{align*}
\mathscr E^{(j)}[\psi](\tau)   +\mathscr H^{(j)}[\psi](\tau)  
\lesssim & \,
\mathscr E^{(j)}[\psi](0)  +
 \f1{T_1} (\log L)^{\f12} \int_0^\tau \mathscr E^{(j)}[\psi](t^*) \, dt^* \nonumber \\
& + 
 C_{\mathrm{boot}}L^{5-2s} \big(\mathscr E^{(j)}[\psi](\tau)+\mathscr H^{(j)}[\psi](\tau)\big)   \nonumber \\
& +
  \Bigg( \int_0^\tau \big\| r^3 V^{\le 1}\big( \mathcal F[\tilde \phi]\big)^{(\le j+1)} \Big|_{\Sigma^*_{t^*}}\big\|_{L^2_{r,\theta,\varphi}(r^{-2}\sin\theta dr d\theta d\varphi)} \, dt^*\Bigg)^2 \nonumber \\
  & +
\sup_{\bar\tau \in[0,\tau]} \Big\|\Big(r^3 \mathcal F[\tilde \phi]
\Big)^{(\le j)}\big|_{\Sigma^*_{\bar\tau}} \Big\|^2 _
 {L^2(\sin\theta \f{1}{r^2}dr d\theta d\varphi)},   \nonumber 
\end{align*}
and, after absorbing the third term on the right-hand side into the left-hand side:
 \begin{equation}\label{Energy estimate for Bootstrap again}
\mathscr E^{(j)}[\psi](\tau) 
\lesssim 
\mathscr E^{(j)}[\psi](0)  +
 \f1{T_1} (\log L)^{\f12} \int_0^\tau \mathscr E^{(j)}[\psi](t^*) \, dt^* +\mathcal F_1,   \nonumber 
\end{equation}
where
\[
\mathcal F_1 \doteq   \Bigg( \!\! \int_0^\tau \big\| r^3 V^{\le 1}\big( \mathcal F[\tilde \phi]\big)^{(\le j+1)} \Big|_{\Sigma^*_{t^*}}\big\|_{L^2_{r,\theta,\varphi}(r^{-2}\sin\theta dr d\theta d\varphi)} \, dt^*\Bigg)^2 \!+\!
\sup_{\bar\tau \in[0,\tau]} \Big\|\Big(r^3 \mathcal F[\tilde \phi]
\Big)^{(\le j)}\big|_{\Sigma^*_{\bar\tau}} \Big\|^2 _
 {L^2(\sin\theta \f{1}{r^2}dr d\theta d\varphi)}.
\]

Applying Gr\"onwall's lemma for the bound \eqref{Energy-estimate-for-Bootstrap}, we infer that, for a constant $C>0$ depending only on the parameters $\lambda, N, s$ and the geometry of $(\mathcal M_\mathrm{ext}, g)$ (and not on $C_{\mathrm{boot}}, L$), we have for any $\tau \in [0,T_{\mathrm{boot}}]$:
\begin{equation}\label{From Gronwall bootstrap}
\mathscr E^{(j)}[\psi](\tau) \le C e^{C (\log L)^{\f12} \f{T_{\mathrm{boot}}}{T_1}} \Big(\mathscr E^{(j)}[\psi](0) + \mathcal F_1\Big).
\end{equation}
As a consequence of the bound \eqref{Bound inhomogeneous term best} for $\mathcal F[\tilde\phi]$, we have
\[
\mathcal F_1 \lesssim L^{-2s-6+9\delta_0} \lesssim L^{-8+9\delta_0} \mathcal E[\chi\tilde\phi](0).
\]
Combining the above with the initial bound \eqref{Initial higher order bound} for $\mathscr E^{(j)}[\psi](0)$ and the fact that $T_{\mathrm{boot}}\le T_1$, we can then estimate from \eqref{From Gronwall bootstrap}:
\begin{equation}\label{From Gronwall bootstrap again}
\sup_{\tau \in [0,T_{\mathrm{boot}}]} \mathscr E^{(j)}[\psi](\tau) \le C^2 \exp \big( C^2 (\log L)^{\f12}\big) L^{-8+9\delta_0} \mathcal E [\chi \tilde \phi](0) 
\end{equation}
from which the improved bootstrap bound \eqref{Bootstrap improvement} immediately follows provided, $C_{\mathrm{boot}}$ was originally chosen large enough compared to the constant $C$ above and $L$ is sufficiently large in terms of $C_{\mathrm{boot}}$ (so that $C^2 \exp \big( C^2 (\log L)^{\f12}\big)<\f12 C_{\mathrm{boot}} L^{\delta_0}$).
\qed

\section{Putting everything together: Proof of \texorpdfstring{\cref{thm:Main theorem}}{Theorem 1}}\label{sec:The proof of the main theorem}

In this section, we will combine the results established so far in this paper in order to obtain the proof of \cref{thm:Main theorem}. 
Let us note that, as a trivial consequence of the well-posedness of the initial-boundary value problem \eqref{Initial Boundary Value Problem Phi} in the class of smooth initial data (see e.g.\ \cite{HKM76}) and the fact that the coefficients of \eqref{Initial Boundary Value Problem Phi} depend smoothly on the Schwarzschild--AdS mass parameter $M$, the set of mass parameters $M\in (0,+\infty)$ for which a solution $\phi$ of \eqref{Initial Boundary Value Problem Phi} exists satisfying the conditions of \cref{thm:Main theorem} is an \textbf{open} subset of $(0, +\infty)$. More precisely, the following holds: 

\bigskip
\noindent \textbf{Cauchy stability with respect to the mass parameter:} Suppose that, for some $\epsilon>0$, $s\ge 4$, $T_1>0$ and some value $M_0\in (0,+\infty)$ of the mass parameter, there exists a smooth initial data set $f_0, f_1:(r_+(M_0),+\infty)\times \mathbb S^2\rightarrow \mathbb C$ satisfying
\begin{equation}\label{Support of initial data openness}
\overline{\mathrm{\text{supp}}f_0}, \, \overline{\mathrm{\text{supp}}f_1} \subset \{r>3M_0\}
\end{equation}
and
\begin{equation}\label{Smallness initial data openness}
\|f_0\|_{H^s}+\|f_1\|_{H^{s-1}} <\epsilon
\end{equation}
and such that the solution $\phi=\phi^{(M_0)}$ of the boundary value problem \eqref{Initial Boundary Value Problem Phi} on $(\mathcal M_\mathrm{ext}^{(M_0)}, g_{M_0})$ with initial data $(f_0, f_1)$ satisfies
\begin{equation}\label{Largeness development openness}
\| \phi^{(M_0)}|_{t^*=T_1}\|_{H^s} > \f1\epsilon.
\end{equation}
Then, there exists a $\delta = \delta(\epsilon, s, M_0, T_1, f_0, f_1)>0$ sufficiently small such that, for any $M\in (M_0-\delta, M_0+\delta)$, the same initial data set $(f_0, f_1)$ (which can be thought of as a pair of functions on $(r_+(M), +\infty)\times \mathbb S^2$, since $f_0, f_1=0$ in a neighborhood of $r=r_+(M_0)$) satisfies \eqref{Support of initial data openness}--\eqref{Smallness initial data openness} with $M$ in place of $M_0$, and the corresponding solution $\phi^{(M)}$ of \eqref{Initial Boundary Value Problem Phi} on $(\mathcal M_\mathrm{ext}^{(M)}, g_{M})$ satisfies \eqref{Largeness development openness}.

\bigskip

As a result, in order to establish \cref{thm:Main theorem}, it suffices to show that the set of mass parameters $\mathcal J_{\epsilon, s}$ (see the statement of  \cref{thm:Main theorem}) is merely \textbf{dense}, since openness would then follow automatically from the discussion above. In particular, \cref{thm:Main theorem} can be restated as follows:

\begin{manualtheorem}{1}[Second version]\label{thm:Main theorem again}
\emph{For any $\epsilon, \epsilon'>0$, any $M'\in (0, +\infty)$ and any integer $s\ge 4$, there exists a Schwarzschild--AdS mass parameter $M\in (M'-\epsilon', M'+\epsilon')$ such that the following statement holds: There exists a smooth initial data set $(f_0, f_1)$ for the initial--boundary value problem \eqref{Initial Boundary Value Problem Phi} on $(\mathcal M_\mathrm{ext}, g_M)$ with
\[
\| f_0\|^2_{H^{s}} + \|f_1\|^2_{H^{s-1}} \le \epsilon^2 \quad \text{and} \quad \overline{\supp f_0 }, \overline{\supp f_1} \subset \{r> 3M\}
\]
and a time $T_1>0$ such that a smooth solution for \eqref{Initial Boundary Value Problem Phi} exists for $t^*\le T_1$ and satisfies}
\begin{equation}\label{Norm inflation again}
\|\phi |_{t^*=T_1}\|^2_{H^{s}} > \f1{\epsilon^2}.
\end{equation}
\end{manualtheorem}

\begin{proof}
Let $\epsilon$, $\epsilon'$, $M'$ and $s\ge 4$ be as in the statement of the theorem. Without loss of generality, we can assume that $\epsilon$ and $\epsilon'$ are small constants in terms of $M'$ and $s$ (otherwise, we can replace them by smaller constants without affecting the statement of the theorem). We will also set 
\[
M_0 \doteq \f32 M'
\]
(so that $M' \in (\f12 M_0, M_0)$) and
\[
C_{\mathrm{amp}} = \f{4}{\epsilon^4}.
\]
Let also $\delta_0 \in (0,1)$ be a constant which is small in terms of $\epsilon$, $\epsilon'$, $s$ and $M_0$.

Let $N, L \in \mathbb N^*$ be such that $N$ is large in terms of $\epsilon$, $\epsilon'$, $s$ and $M_0$,  while $L$ is large in terms of $N$, $\epsilon$, $\epsilon'$, $s$ and $M_0$. In particular, we have
\[
L^{-4\delta_0} < \f12 \epsilon'.
\]
Let $M\in [M'-L^{-4\delta_0}, M'+L^{-4\delta_0}]$ and $\lambda \in \f1{L} \mathbb N^*$ be defined in terms of $N$, $L$ and $M'$ as in \cref{prop:Resonance conditions}. Note, in particular, that
\[
|M-M'| < \epsilon'
\]
and (as a consequence of \eqref{eq:ansatz-lambda}):
\[
\lambda = \beta^{-2} N +O(1).
\]
We also have $M\in \mathcal C_{N, L;M_0}$ (according to \cref{def:Dense set of mass parameters}).

\begin{remark} For the rest of this section, we will assume that the constants implicit in the $\lesssim$, $O(\cdot)$ etc.~notation are allowed to depend on $\epsilon$, $\epsilon'$, $M_0$ and $s$ but are \textbf{independent} of $N$, $\lambda$, $L$.
\end{remark}

On the Schwarzschild--AdS exterior spacetime $(\mathcal M^{(M)}_\mathrm{ext}, g_M)$ with mass parameter $M$, let us consider the ansatz \eqref{The ansatz} for a (formal) solution $\phi$ of \eqref{Initial Boundary Value Problem Phi}, i.e. 
\begin{equation}\label{The ansatz again}
\phi(t^*, r, \theta,\varphi) = \chi(r) \tilde\phi(t^*, r, \theta, \varphi) + \psi(t^*, r, \theta, \varphi),
\end{equation}
where 
\[
\tilde\phi = \sum_{k \in \mathcal K} \tilde\phi_k,
\]
with the mode parameter set $\mathcal K$ defined (in terms of $N, \lambda, L$ as in   \cref{sec:Hierarchy of parameters}),  the functions $\tilde\phi_k:\{r\ge r_\mathrm{mirror}(M,M_0)\}\rightarrow \mathbb C$ being of the form
\[
\tilde\phi_k(t^*,y,\theta, \varphi) = \f{a_k(t^*)}{r(y)} E_k(y, \theta, \varphi)
\]
(with $E_k$ the spatial eigenfunction defined by \eqref{Spatial eigenfunction}) and satisfying the boundary value problem \eqref{Boundary value problem tilde phi}, 
while the error term $\psi$ satisfies the initial-boundary value problem \eqref{IVP Psi again} (see \cref{sec:The mode based ansatz} for more details on the ansatz \eqref{The ansatz again}). In the case of the dominant modes $\big\{\tilde \phi_j\big\}_{j\in \mathcal K_{\mathrm{D}}}$, we will also use the representation \eqref{B k variables}, namely
\begin{equation}\label{B representation again dominant}
\tilde\phi_j (t^*, y, \theta, \varphi) = b_j(t^*) e^{-i \varepsilon_j \omega_j t^*} \f{1}{r(y)} E_j(y, \theta, \varphi).
\end{equation}

Let us define $\tilde T= \tilde T(C_{\mathrm{amp}}, \epsilon, \lambda, N, s, M_0)$ as in the statement of   \cref{prop:The 3 times 3 system} and let us set
\begin{equation}\label{Fixing T1}
T_1 \doteq L^{2s+1} \tilde T.
\end{equation}
The results established in   \cref{sec:Dominant modes,sec:Estimates error term} now imply the following facts:
\begin{itemize}
\item \emph{For the dominant modes $\{\tilde\phi_j\}_{j\in \mathcal K_{\mathrm{D}}}$}: Let us consider the following initial data set for $\{\tilde\phi_j\}_{j\in \mathcal K_{\mathrm{D}}}$:
\begin{align}\label{Initial data dominant modes final again again}
 \tilde\phi_j(0,y, \theta, \varphi) = & L^{-s} \tilde B^{(j)}_0 \cdot \f{1}{r(y)}E_j(y, \theta, \varphi) ,
 \\
  \partial_{t^*} \tilde\phi_j(0, y, \theta, \varphi)  = &  L^{-s} \big(-i\varepsilon_j \omega_j \tilde B^{(j)}_0 + L^{-2s-1}\tilde B^{(j)}_1\big) \cdot \f{1}{r(y)}E_j (y, \theta, \varphi) ,\nonumber 
\end{align}
where the pair $\big\{ \big( \tilde B^{(j)}_0, \tilde B^{(j)}_1\big)\big\}_{j\in \mathcal K_{\mathrm{D}}}$ is as in the statement of   \cref{prop:The 3 times 3 system}.
Then,   \cref{cor:The 3 times 3 system} implies that the corresponding solution $\{\tilde\phi_j\}_{j\in \mathcal K_{\mathrm{D}}}$ of \eqref{Boundary value problem tilde phi} with initial data as above exists on the whole domain $\{0\le t^* \le T_1\}\cap\{r\ge r_\mathrm{mirror}\}$ and satisfies
\begin{equation}\label{Final final estimate initial data}
 \big(\| \chi \cdot \sum_{j\in \mathcal K_{\mathrm{D}}} \tilde\phi_j |_{t^*=0}\|^2_{H^s}+ \| \chi \cdot \sum_{j\in \mathcal K_{\mathrm{D}}} \partial_{t^*} \tilde\phi_j |_{t^*=0}\|^2_{H^{s-1}}\big) < C \epsilon^2
\end{equation}
for some constant $C>0$ \textbf{independent} of $\epsilon, \epsilon',  N, \lambda, L$ (but which might depend on $s$ and $M_0$) and:
\begin{equation}\label{Final bound amplification}
\| \chi \cdot \sum_{j\in \mathcal K_{\mathrm{D}}} \tilde\phi_j |_{t^*=T_1}\|^2_{H^s}\ge  \f1{C} C_{\mathrm{amp}} \epsilon^2 >  \f{2}{C \epsilon^2}.
\end{equation}
Recall that, in the above, the Sobolev space $H^s(\{t^*=\tau\})$ is defined according to \eqref{Integer Sobolev norm}.

\item \emph{For the non-dominant modes $\{\tilde\phi_k\}_{k\in \mathcal K \setminus \mathcal K_{\mathrm{D}}}$}: We will choose vanishing initial data for  $\{\tilde\phi_k\}_{k\in \mathcal K \setminus \mathcal K_{\mathrm{D}}}$, i.e.
\[
\big( \tilde\phi_k(0), \partial_t\tilde\phi_k(0)\big) = \big(0, 0\big) \quad \text{for all} \quad k \in \mathcal K \setminus \mathcal K_{\mathrm{D}}.
\]
Then,   \cref{prop:Total bound non-dominant modes} implies that, if $T_1$ is fixed by \eqref{Fixing T1} and the dominant modes $\{\tilde\phi_j\}_{j\in \mathcal K_{\mathrm{D}}}$ are constructed as above,  the corresponding solutions  $\{\tilde\phi_k\}_{k\in \mathcal K\setminus \mathcal K_{\mathrm{D}}}$ exist on the whole domain   $\{0\le t^* \le T_1\}\cap\{r\ge r_\mathrm{mirror}\}$ and satisfy  (provided $L$ is sufficiently large in terms of $N, \lambda, s, \epsilon, M',\delta_0$)
\begin{equation}\label{Final bound non dominant modes}
\sup_{\tau\in [0, T_1]} \sum_{k\in \mathcal K \setminus \mathcal K_{\mathrm{D}}} \big\|\chi \tilde\phi_k |_{t^*=\tau} \big\|_{H^s} \le  L^{\f12-2s}.
\end{equation}

\item \emph{For the error term $\psi$}: According to   \cref{prop: Estimates error term}, the solution $\psi$ to the initial-boundary value problem \eqref{IVP Psi again} (with $\tilde\phi = \sum_k\tilde\phi_k$ as constructed above on $\{0\le t^*\le T_1\}\cap\{r\ge r_\mathrm{mirror}\}$ and $T_1$ defined by \eqref{Fixing T1}) exists on the whole domain $\{0\le t^*\le T_1\}$ and satisfies (provided $L$ is sufficiently large in terms of $N, \lambda, s, \epsilon, M',\delta_0$)
\begin{equation}\label{Final bound error term}
\sup_{\tau \in [0, T_1]} \big\|\psi|_{t^*=\tau} \big\|_{H^s} \le L^{-\f52}.
\end{equation}
\end{itemize}

Therefore, the above results imply that if we choose as initial data for $\phi$ those induced by $\chi\tilde\phi+\psi$, namely
\[
\big( f_0, \, f_1 \big) = \big( \sum_{j\in \mathcal K_{\mathrm{D}}} \chi \tilde \phi_j|_{t^*=0}, \, \sum_{j\in \mathcal K_{\mathrm{D}}} \chi \partial_{t^*} \tilde \phi_j|_{t^*=0} \big),
\]
(where the functions $(\tilde \phi_j|_{t^*=0}, \partial_{t^*} \tilde \phi_j|_{t^*=0})$ are given by \eqref{Initial data dominant modes final}), then the corresponding solution $\phi = \chi\tilde\phi+\psi$ to the initial-boundary value problem \eqref{Initial Boundary Value Problem Phi} exists (and is smooth) on the whole domain $\{0\le t^* \le T_1\}$ and satisfies (as a consequence of \eqref{Final final estimate initial data})
\[
\|f_0\|^2_{H^s} + \|f_1\|^2_{H^{s-1}} < C \epsilon^2
\]
and (as a consequence of \eqref{Final bound amplification}, \eqref{Final bound non dominant modes} and \eqref{Final bound error term}):
\[
\big\| \phi|_{t^*=T_1} \big\|_{H^s}\! \ge  \|\!\sum_{j\in \mathcal K_{\mathrm{D}}} \chi \tilde\phi_j |_{t^*=T_1}\|_{H^s} - \Big(\!\sum_{k\in \mathcal K \setminus \mathcal K_{\mathrm{D}}} \| \chi \tilde\phi_k |_{t^*=T_1}\|_{H^s} + \big\|\psi|_{t^*=T_1} \big\|_{H^s} \Big) > \f{\sqrt2}{C^{\f12} \epsilon} - O(L^{-\f52}+ L^{\f12-2s})> \f{1}{C^{\f12}\epsilon}.
\]
Thus, $\phi$ satisfies the conditions of \cref{thm:Main theorem again} (with $C^{\f12} \epsilon$ in place of $\epsilon$; since $C$ only depends on $s$ and not on the rest of the parameters one could have eliminated this constant by repeating the same steps but with $\f{\epsilon}{C^{\f12}}$ in place of $\epsilon$). 
\end{proof}

\appendix
\section{Appendix}
\subsection{The Airy equation}\label{sec:The Airy equation}
The Airy differential equation takes the form
\begin{equation}\label{Airy differential equation}
\f{d^2 u}{dx^2} - x u =0.
\end{equation}
The \emph{Airy function} (of the first kind) $\Ai:\mathbb R \rightarrow \mathbb R$ is defined as the unique solution of \eqref{Airy differential equation} with the property that
\[
\Ai(x) \rightarrow 0 \quad \text{as} \quad x\rightarrow +\infty.
\]
Similarly, the Airy function of the second kind $\Bi:\mathbb R \rightarrow \mathbb R$ is defined as the unique solution of \eqref{Airy differential equation} satisfying
\[
\sqrt\pi \zeta^{-\f16} \Bi(-\zeta^{\f23}) = \cos(\f23\zeta+\f\pi4) +o(1) \quad \text{as} \quad \zeta \rightarrow +\infty.
\]
The functions $\Ai(\cdot)$ and $\Bi(\cdot)$ are linearly independent solutions of \eqref{Airy differential equation} and obey the following asymptotics \cite[Chapter~11]{Olver97}:
\begin{align}\nonumber 
\Ai(x) &= \f1{\sqrt\pi} |x|^{-\f14} \sin\big(\f23 |x|^{\f32} +\f\pi4\big)+O(|x|^{-\f74}) \quad \text{for } x \le - 1, \\ 
 \Ai(x) &= \f1{2\sqrt\pi} x^{-\f14} e^{-\f23 x^{\f32}} \cdot \Big(1 +O(x^{-\f32}) \Big) \quad \text{for } x \ge 1 \label{Airy asymptotics}
\end{align}
and
\begin{align}\nonumber
\Bi(x) &= \f1{\sqrt\pi} |x|^{-\f14} \cos\big(\f23 |x|^{\f32} +\f\pi4\big)+O(|x|^{-\f74}) \quad \text{for } x \le - 1, \\  
 \Bi(x) &=  \f1{\sqrt\pi} x^{-\f14} e^{\f23 x^{\f32}} \cdot \Big(1 +O(x^{-\f32}) \Big) \quad \text{for } x \ge 1. \label{Airy asymptotics 2nd kind}
\end{align}
Moreover, the functions $\Ai(\cdot)$ and $\Bi(\cdot)$  and their derivatives satisfy the following bounds:
\begin{equation}\label{Bounds Airy functions B}
\big|\Ai(y)\big| + (1+|y|)^{-\f12}\big|\Ai'(y)\big| \lesssim (1+|y|)^{-\f14} \mathscr E(y) \quad \text{and} \quad \big|\Bi(y)\big| + (1+|y|)^{-\f12}\big|\Bi'(y)\big| \lesssim  \f{(1+|y|)^{-\f14}}{\mathscr E(y)},
\end{equation}
where
\begin{equation}\label{B error bound function again}
\mathscr E (y) \doteq 
\begin{cases}
1, \quad y\in (-\infty, 0], \\[5pt]
e^{-\f23 y^{\f32}}, \quad y\in [0, +\infty).
\end{cases}
\end{equation}

The following representation formula allows us to capture more effectively the oscillating character of $\Ai(x)$ and $\Bi(x)$ in the region $x\le 0$: Using the notation of \cite[Chapter~11]{Olver97}, we have
\begin{equation}\label{Sine representation Airy function}
E(x)\Ai(x) = M(x) \sin\big(\theta(x)\big) \quad \text{and} \quad \f{1}{E(x)} \Bi(x) = M(x) \cos\big(\theta(x)\big),
\end{equation}
where the functions $E,M:\mathbb R \rightarrow (0,+\infty)$ (called, respectively, the \emph{weight} and \emph{modulus} functions) and $\theta:\mathbb R \rightarrow \mathbb R$ (called the \emph{phase} function) satisfy the following conditions (with $\rho_*\simeq -0.366$ defined as the first point on the left of $x=0$ where $\Ai(x)=\Bi(x)$):
\begin{itemize}
\item In the region $x\le \rho_*$, we have
\[
E(x) =1,
\]
the functions 
$M(x)$, $\theta(x)$ solve
\begin{equation}\label{ODE M}
M''(x) = x M(x) +\f1{\pi^2 M^3(x)}
\end{equation}
and
\[
\theta'(x) = -\f1{\pi M^2(x)}
\]
and, as $x\rightarrow -\infty$, we have
\[
M(x) = \f1{\sqrt\pi} (-x)^{-\f14}\cdot\Big(1 +O\big((-x)^{-2}\big)\Big), \quad \text{and} \quad M'(x) = \f14 \f{1}{\sqrt\pi} (-x)^{-\f54} \cdot\Big(1+O\big((-x)^{-2}\big)\Big)
\]
(note that the asymptotic expansions for $M(x)$ and $M'(x)$ as $x\rightarrow -\infty$ do not contain terms of the form $(-x)^{-\f54}$ and $(-x)^{-\f94}$, respectively, as can be verified by computing the coefficients $c_i$ in $M(x)= \f1{\sqrt\pi} (-x)^{-\f14} + c_1 (-x)^{-\f54}+c_2 (-x)^{-\f94}+\dots$ by formal substitution in \eqref{ODE M}).
The above conditions imply that $M'(x), M''(x)>0$ for all $x\le \rho_*$ and, as $x\rightarrow -\infty$,
\[
\theta(x) = \f23 (-x)^{\f32}\cdot\Big(1 +O\big( (-x)^{-2}\big)\Big).
\]

\item In the region $x\ge \rho_*$, we have
\[
\theta(x) = \f{\pi}{4}
\]
and, as $x\rightarrow +\infty$, we have
\[
E(x) = \sqrt 2 e^{\f23 x^{\f32}}\Big(1+O(x^{-\f32})\Big), \quad \text{and} \quad  M(x) = \f1{\sqrt\pi} x^{-\f14} +O\big(x^{-\f54}\big).
\]

\item The functions $E(x)$, $M(x)$ and $\theta(x)$ are smooth on $\mathbb R \setminus \{\rho_*\}$ and continuous at $x=\rho_*$.
\end{itemize}
An analogous representation also holds for $\Ai'$ and $\Bi'$.

\subsection{Asymptotics for integrals of the half-line Hermite polynomials}\label{sec:Integrals Hermite}

For $n\in \mathbb N$, the $n$-th Hermite polynomial can be expressed as
\begin{equation}\label{Rodrigues formula}
H_n (x) = (-1)^n e^{x^2} \Big( \frac{d^n}{dx^n} e^{-x^2}\Big).
\end{equation}
The above expression implies that the functions $H_n$ satisfy the recurrence relations
\begin{align}\label{Recurrence relations}
    H_{n}' (x) & = 2n H_{n-1}(x),\\
    H_{n+1}(x) & =2x H_n(x) - 2nH_{n-1}(x). \nonumber
\end{align}
Moreover, the functions $H_n(x)$ are odd or even  depending on whether $n$ is odd or even, respectively.

\begin{lemma}
Let $p,q\in \mathbb N$ and let us set
\begin{equation}\label{Product different parity}
\mathcal I_{p,q} \doteq \int_0^{\infty}  H_{2p}(x) H_{2q+1}(x) e^{-x^2}\,dx.
\end{equation}
Then:
\begin{equation}\label{Inner product half line}
  \mathcal I_{p,q} = \f{(-1)^{p+q+1}}{\big (2(p-q)-1\big)\pi} 2^{2(p+q)+1} \Gamma\left(p+\f12\right) \Gamma\left(q+\f32\right). 
\end{equation}
\end{lemma}

\begin{proof}
The functions $e^{-\f{x^2}2} H_{2p}(x)$ and $e^{-\f{x^2}2} H_{2q+1}(x)$ solve 
\begin{align*}
     \big(e^{-\f{x^2}2} H_{2p}\big)^{\prime\prime} +\big(4p+1-x^2\big) e^{-\f{x^2}2} H_{2p} =0\\
      \big(e^{-\f{x^2}2} H_{2q+1}\big)^{\prime\prime} +\big(4q+3-x^2\big) e^{-\f{x^2}2} H_{2q+1} =0.
\end{align*}
Therefore, the Wronskian
\[
W(x) \doteq \f{d}{dx}\big( e^{-\f{x^2}2} H_{2p}(x)\big) \cdot e^{-\f{x^2}2} H_{2q+1}(x) - e^{-\f{x^2}2} H_{2p}(x) \f{d}{dx}\big( e^{-\f{x^2}2} H_{2q+1}(x)\big)
\]
satisfies
\[
\f{1}{4(q-p)+2}\f{dW}{dx}(x) =  H_{2p}(x) H_{2q+1}(x) e^{-x^2}.
\]
Integrating the above relation over $[0,+\infty)$ and using the fact that $W(0)=-H^\prime_{2q+1}(0)H_{2p}(0)$ (since $H_{2q+1}(0)=0$), we infer
\begin{equation}\label{Almost product formula}
\int_0^{\infty}  H_{2p}(x) H_{2q+1}(x) e^{-x^2}\,dx
=\f{1}{4(q-p)+2} H^\prime_{2q+1}(0)H_{2p}(0).
\end{equation}
Using the recurrence relations \eqref{Recurrence relations} we can readily calculate that
\begin{align*}
H_{2p}(0) &= (-1)^p 2^p \Big(\prod_{k=0}^{p-1} (2p-2k-1)\Big) H_0(0)\\
& = (-1)^p 2^{2p} \Big(\prod_{k=0}^{p-1} (p-k-\f12)\Big) \\
& = (-1)^p 2^{2p} \f{\Gamma(p+\f12)}{\Gamma(\f12)} 
\end{align*}
and
\begin{align*}
H^\prime_{2q+1}(0) & = 2(2q+1) H_{2q}(0) \\
& = 2(2q+1) (-1)^q 2^{2q} \f{\Gamma(q+\f12)}{\Gamma(\f12)}\\
& = (-1)^q 2^{2q+2} \f{\Gamma(q+\f32)}{\Gamma(\f12)}.
\end{align*}
Thus, \eqref{Almost product formula} yields the formula \eqref{Inner product half line}.
\end{proof}

\begin{lemma}\label{lem:Asymptotic-functions-f-and-h}
Let us define the functions $f_3, h$ by the relations
\[
f_3(n',n) \doteq \int_0^\infty x^3 e_n(x)e_{n'}(x)  \,dx,
\]

\[
f_3(n)\doteq f_3(n,n) = \int_0^\infty x^3 (e_n(x)\big)^2  \,dx
\]
and
\begin{equation}\label{H definition again}
h(n) \doteq \sum_{n'\neq n } \frac{|f_3(n',n)|^2}{n'-n} = \sum_{1 \leq k \leq n-1} \frac{|f_3(n+k,n)|^2 - |f_3(n-k,n)|^2}{k} +  \sum_{n \leq k} \frac{|f_3(n+k,n)|^2}{k},
\end{equation}
where $e_n(x)$ is the normalized odd Hermite function given by \eqref{Normalized Hermite function}. Then, we have 
\begin{equation}
    f_3(1) = f_3(1,1) = \frac{4}{\sqrt \pi} 
\end{equation}
and 
\begin{equation}\label{Value h 1}
    \frac{123}{5\pi} \leq h(1) \leq \frac{123}{5\pi} + 10^{-1}.
\end{equation}
Moreover, as $n\rightarrow +\infty$, the following asymptotic formulas hold:
\begin{equation}\label{Asymptoics f3}
  f_3(n) = \f{32}{3\pi} n^{\f32}-\f{4}{\pi}n^{\f12}+\f{5}{12\pi}n^{-\f12}+O(n^{-\f32}) 
\end{equation}
and 
\begin{equation}\label{Asymptoics h}
  h(n) = \Big(30-\f{512}{3\pi^2}\Big) n^2 -  \Big(15-\f{256}{3\pi^2}\Big) n +O(1).
\end{equation}
\end{lemma}

\begin{proof}
Using the recurrence relation \eqref{Recurrence relations} successively, we can readily compute for any $n>2$:
\begin{align}\label{x3 Hn}
x^3 H_{2n-1} &= x^2 \big(\f12 H_{2n} +(2n-1) H_{2n-2}\big)\\
& = x \Big(\f14 H_{2n+1} + (2n-\f12) H_{2n-1} +(2n-1)(2n-2)H_{2n-3} \Big) \nonumber\\
&= \f18 H_{2n+2}+\f{3n}2 H_{2n} + \f32(2n-1)^2H_{2n-2} +(2n-1)(2n-2)(2n-3)H_{2n-4}. \nonumber
\end{align}
Therefore, we have
\begin{align*}
     f_3(n) &= c_{n}^2 \int_0^\infty x^3  \big( H_{2n-1}(x)\big)^2 e^{-x^2} \, dx \\
     &= c_{n}^2\int_0^\infty \Big(x^3  H_{2n-1}(x)\Big) H_{2n-1}(x) e^{-x^2} \, dx \\
    &= \f{c_{n}^2}{8} \Big( \mathcal I_{n+1,n-1} +12n \mathcal I_{n,n-1} +12(2n-1)^2 \mathcal I_{n-1,n-1}+8(2n-1)(2n-2)(2n-3)\mathcal I_{n-2,n-1} \Big)
\end{align*}
and, similarly:
\begin{align*}
     f_3(n',n) 
    &= \f{ c_{n} c_{n'}}{8} \Big( \mathcal I_{n+1,n'-1} +12n \mathcal I_{n,n'-1} +12(2n-1)^2 \mathcal I_{n-1,n'-1}+8(2n-1)(2n-2)(2n-3)\mathcal I_{n-2,n'-1} \Big),
\end{align*}
where $\mathcal I_{p,q}$ are defined by \eqref{Product different parity} and
\[
c_{n}^2 = \frac{ 1}{\sqrt \pi}  \cdot \frac{1}{2^{2n-2} \Gamma(2n)  }
\]
(see \eqref{Normalized Hermite function}). Thus, using the expression \eqref{Inner product half line} for $\mathcal I_{p,q}$, we obtain:
\begin{align}\nonumber
    f_3(n) & = \f1{8 \pi^{\f12}2^{2n-2} \Gamma(2n)} \Bigg(-\f{1}{3\pi} 2^{4n+1} \Gamma(n+\f32) \Gamma(n+\f12) \\ \nonumber
     &\hphantom{=\f1{8 \pi^{\f12}2^{2n-2} \Gamma(2n)}\Bigg(}\,
    +\f{12n}{ \pi} 2^{4n-1} \Gamma(n+\f12) \Gamma(n+\f12) \\ \nonumber
    &\hphantom{=\f1{8 \pi^{\f12}2^{2n-2} \Gamma(2n)}\Bigg(}\,
    +12\f{(2n-1)^2}{\pi} 2^{4n-3} \Gamma(n-\f12) \Gamma(n+\f12)\\\nonumber
    &\hphantom{=\f1{8 \pi^{\f12}2^{2n-2} \Gamma(2n)}\Bigg(}\,
    -\f{8(2n-1)(2n-2)(2n-3)}{3\pi} 2^{4n-5} \Gamma(n-\f32) \Gamma(n+\f12)\Bigg)\\
    & = \f4{3\pi^{\f32} } 2^{2n}( 4n -1) \frac{\big( \Gamma(n+\f12)\big)^2}{\Gamma(2n)} =\frac{8 (4n-1) \Gamma(n+\f12)}{3\pi \Gamma(n )} .  \label{eq:compu-of-f3}
\end{align}

We also directly compute 
\begin{equation}
    f_3(1) = f_3(1,1) = \frac{4}{\sqrt \pi}.
\end{equation}

Using Stirling's approximation
\begin{equation}\label{Stirling formula}
\Gamma(z+1) = \sqrt{2\pi z} \cdot \big( \f{z}{e}\big)^z \Big( 1 + \f{1}{12z} + \f{1}{288 z^2}+ O(z^{-3})\Big),
\end{equation}
we directly obtain from \eqref{eq:compu-of-f3}:
\begin{align*}
f_3(n) & = \frac{8 (4n-1) }{3\pi} \sqrt{\frac{n-\f 12}{n-1}}  \cdot \frac{ (\frac{n-\f 12 }{e})^{n-\f12}}{(\frac{n-1}{e})^{n-1}} \cdot \frac{1 +  \f{1}{12n}   + \f{13}{288n^2}+ O(n^{-3})}{1 + \f{1}{12n} + \f{25}{288 n^2}+ O(n^{-3})} \\
& = \f{32}{3\pi} n^{\f32}-\f{4}{\pi}n^{\f12}+\f{5}{12\pi}n^{-\f12}+O(n^{-\f32}),
\end{align*}
thus obtaining \eqref{Asymptoics f3}.

Arguing similarly as for the derivation of \eqref{eq:compu-of-f3} and using Legendre's duplication formula $\Gamma(2z) = \f{2^{2z-1}}{\sqrt\pi} \Gamma(z) \cdot \Gamma(z+\f12)$, we obtain
\begin{align}\label{Expression f3 n n}
 f_3(n',n) = f_3(n,n') & = (-1)^{n+n'}\f{3}{\pi^{\f32}} 2^{n+n'}\f{\big(2(n+n')-1\big)\big(4n'(n'-2)+3\big)}{\big(4(n-n')^2-9\big)\big(4(n-n')^2-1\big)} \f{\Gamma(n+\f12)\Gamma(n'-\f32)}{\sqrt{\Gamma(2n)\Gamma(2n')}}\\[5pt]
&= (-1)^{n+n'}\f{24}{\pi}  \frac{ (2(n'+n)-1)}{\big(4(n'-n)^2-9 \big) \big(4(n'-n)^2-1\big)} \sqrt{\f{\Gamma(n'+\f12)\Gamma(n+\f12)}{\Gamma(n')\Gamma(n)}}\nonumber
\end{align}

In order to compute $h(n)$, let us first define for $1 \leq k \leq n-1$:
\[
\Delta(n,k)\doteq   |  f_3(n+k,n)|^2 -  |  f_3(n-k,n)|^2.
\]
We can then compute from the expression \eqref{Expression f3 n n} for $f_3(n+p,n)$ for $p\in \mathbb Z$ with $p \ge 1-n$:
\begin{equation}\label{Expression f3 squared}
|f_3(n+p,n)|^2 = \frac{576}{\pi^2  } \frac{(4n+2p -1)^2}{(9-40p^2 + 16p^4)^2 }  \f{ \Gamma(n+\frac 12)}{\Gamma(n)}   \frac{\Gamma(n+p+\frac 12)}{\Gamma(n+p)}.
\end{equation}
Therefore,
\begin{align}\label{Formula Delta n k}
 \Delta(n,k)  = \frac{576}{\pi^2}\f{ \Gamma(n+\frac 12)}{\Gamma(n)} \frac{\left( (4n+2k -1)^2  \frac{\Gamma(n+k+\frac 12)}{\Gamma(n+k)}-(4n-2k -1)^2 \frac{\Gamma(n-k+\frac 12)}{\Gamma(n-k)}\right)}{(16k^4-40k^2 + 9 )^2 }   . 
\end{align}

In view of the definition \eqref{H definition again} of $h(n)$,  we can express in terms of $\Delta(\cdot, \cdot)$ as
\begin{equation}\label{Alternative expression h n}
    h(n) = \sum_{1\leq k \leq n-1} \frac{\Delta(n,k)}{k} + \sum_{k\geq n} \frac{|f_3(n+k,n)|^2}{k} .
\end{equation}
Since the function $x\mapsto \frac{\Gamma(x+\frac 12)}{\Gamma(x)}$ is increasing, the formula \eqref{Formula Delta n k} immediately implies that
\[
\Delta (n,k)\geq 0 \quad \text{for} \quad 1\leq k \leq n-1.
\]
Using the trivial bound $\sqrt{x-\tfrac 12} \leq \frac{\Gamma(x+\tfrac 12)}{\Gamma(x)} \leq \sqrt{x+\tfrac 12}$ and the fact that, for any $0\le B \le A$,
\[
\big( \sqrt{A+B} -\sqrt{A-B}\big) \le \sqrt{2} \f{B}{\sqrt{A}},
\]
 we can also estimate directly from the expression \eqref{Formula Delta n k} for any $1\le k \le n-1$:
\begin{align}\label{Upper bound Delta k}
    \frac{\Delta(n,k)}{k} \le \f{576}{\pi^2}\big(n+\f12\big)^{\f12} \f{\left((4n+2k-1)^2 (n+k+\tfrac 12)^{\frac 12} - (4n-2k-1)^2 (n-k-\tfrac 12)^{\frac 12} \right)}{k (16k^4-40k^2 + 9 )^2} \le 
   C \frac{n^2}{k^8} 
\end{align}
for some absolute constant $C>0$. Similarly, for $k\geq n$ we, can estimate directly from the expression \eqref{Expression f3 squared} for $|f_3(n+k,n)|^2$:
\begin{equation}\label{Upper bound f 3 k}
    \frac{|f_3(n+k,n)|^2}{k} \lesssim  \frac{k^2}{k^9}\frac{\Gamma(n+\tfrac12 )}{\Gamma(n)} \frac{\Gamma(n+k+\tfrac 12)}{\Gamma(n+k)} \lesssim \frac{1}{k^6}
\end{equation}
(where the constants implicit in the $\lesssim$ notation are independent of $n,k$). Using the bounds \eqref{Upper bound Delta k}--\eqref{Upper bound f 3 k} to control the right-hand side of the expression \eqref{Alternative expression h n} for $h(n)$, we therefore obtain for any $n\ge 1$ and for any cut-off $1\le k_0\le n-1$:
\begin{equation}\label{Expression h finite sum}
 h(n) = \sum_{k=1}^{k_0} \frac{\Delta(n,k)}{k} + O\Big(\f{n^2}{k^7_0}+\f1{k_0^5}\Big),
\end{equation}
where the constant implicit in the $O(\cdot)$ notation is independent of $k_0, n$.

In the case when $k< n^{\f12}$, Stirling's formula \eqref{Stirling formula} yields:
\[
\f{\Gamma(n+\f12)}{\Gamma(n)} = n^{\f12}-\f18 n^{-\f12}+\f1{128}n^{-\f32}+O(n^{-\f52})
\]
and, therefore,
\begin{align*}
\f{\Gamma(n+k+\f12)}{\Gamma(n+k)} & = n^{\f12}+ (\f12 k -\f18)n^{-\f12}+(-\f18k^2+\f1{16}k+\f1{128})n^{-\f32} +O(n^{-\f52}k^3), \\
 \f{\Gamma(n-k+\f12)}{\Gamma(n-k)} & =  n^{\f12}+ (-\f12 k -\f18)n^{-\f12}+(-\f18k^2-\f1{16}k+\f1{128})n^{-\f32} +O(n^{-\f52}k^3),
\end{align*}
where the constants implicit in the $O(\cdot)$ notation are independent of $k, n$.
Therefore, substituting the above in the expression \eqref{Formula Delta n k} for $\Delta(n,k)$, we compute for any $1\le k < n^{\f12}$:
\begin{align*}
\f{\Delta(n,k)}{k}
& =
\Bigg(\f{9}{256 \pi^2} \f{(1536n^3-192n^2-20n+3)(128n^2-16n+1)(4n-1)}{n^4}\Bigg) \f{1}{(16k^4-40k^2+9)^2} \\
& \hphantom{===\sum}
- \Bigg( \f{27}{256 \pi^2}\f{(128n^2-16n+1)(32n-5)}{n^5} \Bigg) \f{k^2}{(16k^4-40k^2+9)^2}
+O(k^{-5})\\
& = \Big(\f{9}{256 \pi^2} \big(786432 n^2 -393216 n +O(1)\big)\Big) \f{1}{(16k^4-40k^2+9)^2} 
+O(k^{-5})
\end{align*}
(where, as before, the constant implicit in the $O(\cdot)$ notation is independent of $k, n$). Using the fact that
\[
\sum_{k=1}^{k_0} \f1{(16k^4-40k^2+9)^2} = \f{45 \pi^2-256}{41472}+O(k_0^{-7}),
\]
we obtain from \eqref{Expression h finite sum} for $k_0 \sim n^{\f12}$:
\begin{align*}
h(n) & =
\Big(\f{9}{256 \pi^2} \big(786432 n^2 -393216 n +O(1)\big)\Big) \sum_{k=1}^{k_0} \f{1}{(16k^4-40k^2+9)^2} \\
& \hphantom{==\sum}
+\sum_{k=1}^{k_0} O(k^{-5})+ O(\f{n^2}{k^7_0}+\f1{k_0^5}\Big)
\\
& = 
\f{9}{256 \pi^2}\Big(786432 n^2 -393216 n\Big) \f{45 \pi^2-256}{41472} + O(1) \\
\\
& = \Big(30-\f{512}{3\pi^2}\Big) n^2 -  \Big(15-\f{256}{3\pi^2}\Big) n +O(1),
\end{align*}
where the constant implicit in the $O(\cdot)$ term above is independent of $n$. In particular, we infer \eqref{Asymptoics h}.

Finally,  from the expression \eqref{Expression f3 squared} for $|f_3(1+k,1)|^2$, which yields
 \begin{equation*}
|f_3(1+k,1)|^2 = \frac{288}{\pi^{\f32}  } \frac{1}{(2k-3)^2(4k^2-1)^2 }   \frac{\Gamma(k+\frac 32)}{\Gamma(k+1)}
\end{equation*}
we can readily  compute 
\begin{align}\label{Computation h 1}
    h(1) = \sum_{k\geq 1}\frac{|f_3(1+k,1)|^2}{k}  =  \frac{123}{5\pi} +  \sum_{k\geq 3}\frac{|f_3(1+k,1)|^2}{k} .
\end{align}
Note that, for $k\geq 3$, we can estimate
 \begin{equation}
  \frac{\Gamma(k+\tfrac 32)}{\Gamma(k+1)}\leq \sqrt{k+\tfrac 32}\leq k 
 \end{equation}
 and, thus,
 \begin{equation}
     \sum_{k\geq 3} \frac{|f_3(1+k,1)|^2}{k}\leq \f{288}{\pi^{\f32}}  \sum_{k\geq 3}\frac{1 }{(2k-3)^2(4k^2-1)^2} = \frac{675\pi^2 - 6656}{200 \pi^{\frac 32}} =0.0053\dots \leq 10^{-1}.
 \end{equation}
As a result, \eqref{Value h 1} follows from \eqref{Computation h 1}.
\end{proof}

\subsection{Recurrence relations for the Hermite polynomials}
The following result is a simple consequence of the well-known recurrence relations satisfied by Hermite polynomials (see  \eqref{Recurrence relations}):

\begin{lemma} \label{lem:iteration-hermite}
The Hermite polynomials satisfy for any $n\ge 8$:
    \begin{align}
    x^8 H_n(x) = \sum_{j=-8}^{8} p_n^j H_{n+j}(x),
\end{align}
where the coefficients $p^j_n$ vanish for odd $j$, while for even $j$ they are given by the following expressions:
\begin{equation}\label{Coefficients in final recurrence}
 \begin{split}
& p_n^{-8} = \frac{n!}{(n-8)!},\\
& p_n^{-4}  = \frac{7}{2} \frac{ n!}{ (n-4)! }  (2 (n-3) n+7),\\
& p_n^0  = \frac{35}{16} \left(2 n (n+1) \left(n^2+n+4\right)+3\right),\\
&  p_n^4 = \frac{7}{32} (2 n (n+5)+15),\\
&  p_n^8  = \frac{1}{256}.
\end{split}
\qquad
\begin{split}
& p_n^{-6} = 2(2n-5) \frac{n!}{(n-6)!}, \\
&  p_n^{-2}  = \frac{7}{2} \frac{n!}{(n-2)!} (2 n-1) ((n-1) n+3), \\
&  p_n^2  = \frac{7}{8} (2 n+3) (n (n+3)+5),\\
&    p_n^6  = \frac{1}{32} (2 n+7),\\
& \hphantom{~}
\end{split} 
\end{equation}
\end{lemma}

\begin{proof}
Using the second recurrence relation  in \eqref{Recurrence relations}, we readily calculate
\begin{align*}
    x^2 H_{n}  =& n  x H_{n-1} + \frac 12 x H_{n+1} \\ 
=& n (n-1) H_{n-2}  +  \frac{n}{2} H_n + \frac 12 (n+1) H_n + \frac 14 H_{n+2}\\
    =& n(n-1) H_{n-2} + (n+\frac 12) H_n + \frac 14 H_{n+2}. 
\end{align*}
Iterating the above formula yields  
\begin{align*}
    x^4 H_{n}    = &  n(n-1)(n-2)(n-3) H_{n-4}  + n(n-1)(2n-1) H_{n-2} \\ 
& \, + \frac{3}{4} (2n^2 + 2n +1)  H_{n} + \frac 14 (2n+3) H_{n+2} + \frac{1}{16} H_{n+4} 
\end{align*}
and, after one more iteration,
\begin{equation*}
    x^8 H_n = \sum_{j=-8}^{8} p_n^j H_{n+j},
\end{equation*}
where $p_n^j$ is zero if $j$ is odd and otherwise given as in the statement of \cref{lem:iteration-hermite}.
\end{proof}
\subsection{WKB asymptotics for the Hermite functions}\label{sec:WKB asymptotics}

 In this section, we will state an asymptotic formula for the Hermite functions $\psi_n(x) = H_n(x) e^{-\f{x^2}2}$ ($n\equiv 1 \mod 2$) as $n\rightarrow +\infty$, with an error term of relative size $O(\f1n)$ uniformly on domains containing the turning point at $x=\sqrt{2n+1}$.

\begin{lemma}\label{lem: WKB asymptotics Airy}
For any odd integer $n\in \mathbb N$ and $x\in[0,+\infty)$, let u set
\[
\nu \doteq \sqrt{2n+1} \quad \text{and} \quad z\doteq \f{x}{\nu}.
\]
Then, the Hermite function $\psi_n$ satisfies:
\begin{equation}\label{WKB asymptotics Airy}
\psi_n(\nu z) = (2\pi)^{\f12} e^{-\f{\nu^2}4} \nu^{\f{3\nu^2-1}6} \Big( \f{\Phi(z)}{z^2-1}\Big)^{\f14} \cdot \Big( \Ai\big( \nu^{\f43} \Phi(z)\big) + \mathcal E_{\nu}(z)\big)
\end{equation}
where, in the formula above:
\begin{itemize}
\item $\Ai(\cdot)$ is the Airy function of the first kind (see \cref{sec:The Airy equation}; in particular, see \eqref{Bounds Airy functions B} for bounds on $\Ai$ and $\Ai'$),
\item The function $\Phi:[0,+\infty)\rightarrow \mathbb R$ is the smooth function given by the expression
\begin{equation}\label{Definition Phi 0}
\Phi(z)  \doteq 
\begin{cases}
 - \Big[ \f34 \Big(\cos^{-1} z - z \sqrt{1 - z^2} \Big) \Big]^{\f23}, \quad 0\le z\le 1,\\
\Big[ \f34 \Big(z \sqrt{z^2-1} - \cosh^{-1} z\Big) \Big]^{\f23}, \quad z\ge 1
\end{cases}
\end{equation}
(note that $\Phi(1)=0$ and $\inf_{z\in [0,+\infty)} \Phi^\prime(z)>0$; in particular, $\f{\Phi(z)}{z^2-1}$ is smooth and strictly positive on $z\in[0,+\infty)$),
\item The error term $\mathcal E_{\nu}(z)$ satisfies
\[
\big| E_{\nu}(z) \big| + \f1{\nu^{\f43}(1+\nu^{\f23}|\Phi(z)|^{\f12})|\Phi'(z)|}\Big|\f{d E_{\nu}}{dz}(z) \Big|\le
\f{C_*}{\nu^2} \f{1}{(\nu^{\f43}|\Phi(z)|+1)^{\f14}}\mathscr E\big[ \nu^{\f43} \Phi(z)\big], 
\]
where $\mathscr E[\cdot]$ is the function given by \eqref{B error bound function again} and $C_*>0$  is an absolute constant (independent, in particular, of $\nu$).
\end{itemize}
\end{lemma}

\noindent For the proof of \cref{lem: WKB asymptotics Airy}, see Chapter 11 (in particular, pages 392--400 and 403) of \cite{Olver97}.

\begin{remark} The above is in some sense the most fundamental asymptotic formula for Hermite functions and its derivation is via a simple application of the WKB method; the bounds of the error term $E_{\nu}(z)$ are of order $O(n^{-1})$ (i.e.~$O(\nu^{-2})$) relative to the size of the dominant term involving $\Ai(\cdot)$ in the expansion. Utilizing a ``higher order'' WKB analysis (as described, for instance in \cite{Olver97}), one can in fact derive more precise asymptotic formulas (i.e.~with error terms of smaller relative size); however, the above result will suffice for the purposes of this paper (namely estimating the size of the Radial eigenfunctions $R_{n,\ell}$ defined in \cref{sec:Linear wave equation mirror}).
\end{remark}

\subsection{Functional inequalities}
In this section, we will establish the functional inequalities that are used in the proofs of \cref{sec:Estimates error term}.
The following result establishes two critical Hardy- and Sobolev-type inequalities on $\mathbb R^2$ with optimal logarithmic loss:

\begin{lemma}\label{lem:Hardy type inequality 2d}
There exists a constant $C>0$ such that the following statement holds: For any $0< R_0 < R$ and any smooth function $f:D^+_R\rightarrow \mathbb R$, where $D^+_z$ is the closed half disc 
\[
D^+_z = \big\{  x=(x^1, x^2)\in \mathbb R^2: \, \|x\|\le z \text{ and } x^2\ge 0      \big\},
\]
we can estimate:
\begin{equation}\label{Hardy inequality 2d}
\sup_{R_0<\rho<R}\Big( \f{1}{\rho}\int_{D^+_\rho}\f{f^2}{|x|} \, dx\Big) \le C \log\big(\f{R}{R_0}+1\big) \Big( \int_{D^+_R}|\nabla f|^2 \, dx + R^{-2}\int_{D^+_R}f^2 \, dx\Big),
\end{equation}
where $|\nabla f|^2= |\partial_{x^1} f|^2 + |\partial_{x^2} f|^2$. Moreover, if $f$ satisfies the Dirichlet-type boundary condition
\begin{equation}\label{Dirchlet condition lemma}
f(x^1, 0)=0 \quad \text{for } x^1\in [-R,R],
\end{equation}
then the following bound holds:
\begin{equation}\label{Sobolev estimate 2d}
\sup_{x\in D^+_R \setminus D^+_{R_0}}\Big||x|^{-1} f(x) \Big|^2  \le C\log\big(\f{R}{R_0}+1\big) \Big( \int_{D^+_R}|\nabla^2 f|^2 \, dx + R^{-2} \int_{D^+_R}|\nabla f|^2 \, dx \Big),
\end{equation}
where  $|\nabla^2 f|^2 \doteq |\partial_{x^1}^2 f|^2 + 2|\partial_{x^1}\partial_{x^2} f|^2 + |\partial_{x^2}^2 f|^2$.
\end{lemma}
\begin{remark}
The $\log(\f{R}{R_0})$ loss in the right-hand sides of \eqref{Hardy inequality 2d} and \eqref{Sobolev estimate 2d} is optimal in terms of polynomials factors of  $\log(\f{R}{R_0})$ (as can be seen, for instance, by plugging in $f(x)=\big|\log(|x|)\big|^{\f12-\delta}$ in \eqref{Hardy inequality 2d}).
\end{remark}

\begin{proof}
	In this proof, we will mainly work in the standard polar coordinates $(\rho, \vartheta)$  on $D^+_R \subset \mathbb R^2$. From the fundamental theorem of calculus we have \[|f(\rho, \vartheta)| \leq \int_\rho^R |\partial_\rho f(\rho', \vartheta)| \, d\rho' + |f(R,\vartheta)|\leq\left( \int_\rho^R \frac{1}{\rho'} d \rho' \int_\rho^R \rho' |\partial_\rho f(\rho', \vartheta)|^2 \, d\rho'\right)^{\frac 12 } + |f(R,\vartheta)|\]
	and, therefore, we can estimate
	\[ |f(\rho,\vartheta)|^2 \leq 2\log\left(\frac{R}{\rho}\right) \int_0^R |\partial_{\rho'} f(\rho', \vartheta)|^2 \rho' d \rho'   + 2 |f(R,\vartheta)|^2\] 
	which upon integrating in $\rho$ and in $\vartheta$ yields
	\[ \int_{D^+_\rho} \frac{|f|^2}{|x|} \, dx \leq 2\rho \left( 1+ \log\left(\frac{R}{\rho}\right) \right)\int_{D_R^+} |\nabla f|^2 \, dx + 2\rho \int_0^\pi |f(R,\vartheta)|^2 \, d\vartheta.\]
	Multiplying the above by $\f{1}{\rho}$ and taking the supremum over $\rho\in(R_0, R)$ as well as the standard trace inequality \begin{equation}
    \label{eq:trace-inequality}
    \int_0^\pi |f(R,\vartheta)|^2 d \vartheta \leq \frac{3}{R^2} \int_{D_R^+} | f|^2  \, dx + \int_{D_R^+} |\nabla f|^2 \, dx
  \end{equation}
	gives \eqref{Hardy inequality 2d}.

Let us now switch our attention to the case of functions satisfying the Dirichlet-type boundary condition \eqref{Dirchlet condition lemma}. Notice that, in the polar coordinate system, \eqref{Dirchlet condition lemma} translates to
\[
f|_{\vartheta=0,\pi} =0 
\]
implying, in particular, that
\[
\partial_\rho f|_{\vartheta=0,\pi}  =0.
\]
Fix \(\rho \in [R_0,R]\). Since \(f(\rho,0)=0\), we have
for any \(\vartheta \in [0,\pi]\):
\[
|f(\rho,\vartheta)|^2
=
\left|\int_0^\vartheta \partial_{\vartheta'}f(\rho,\vartheta')\,d\vartheta'\right|^2
\le \pi \int_0^\pi |\partial_\vartheta f(\rho,\vartheta')|^2\,d\vartheta'.
\]
Therefore,
\[
\sup_{\vartheta\in[0,\pi]}
\left|\frac{f(\rho,\vartheta)}{\rho}\right|^2
\le
\pi \int_0^\pi \left|\frac{1}{\rho}\partial_\vartheta f(\rho,\vartheta)\right|^2\,d\vartheta.
\]
Using the identity
\[
\frac{1}{\rho}\partial_\vartheta f
=
-\sin\vartheta\,\partial_{x_1}f+\cos\vartheta\,\partial_{x_2}f,
\]
we infer that
\[
\left|\frac{1}{\rho}\partial_\vartheta f\right|^2 \le |\nabla f|^2.
\]
Hence
\[
\sup_{\vartheta\in[0,\pi]}
\left|\frac{f(\rho,\vartheta)}{\rho}\right|^2
\le
\pi \int_0^\pi |\nabla f(\rho,\vartheta)|^2\,d\vartheta.
\]
We now apply the trace inequality \eqref{eq:trace-inequality} (with $R$ replaced by $\rho$) to
$g=\partial_{x_1}f$ and $g=\partial_{x_2}f$ to obtain
\begin{equation}
  \label{Final inequality for Sobolev}
\sup_{x\in D_R^+\setminus D_{R_0}^+}
\left|\frac{f(x)}{|x|}\right|^2
\le
C\left(
\int_{D_R^+}|\nabla^2 f|^2\,dx
+
\sup_{R_0<\rho<R}
\frac{1}{\rho}\int_{D_\rho^+}\frac{|\nabla f|^2}{|x|}\,dx
\right),
\end{equation}
where we used that $|x|\le \rho$ on $D_\rho^+$. 
The bound \eqref{Sobolev estimate 2d} now follows from the above estimate using \eqref{Hardy inequality 2d} with $\partial_{x^i} f$ in place of $f$ to estimate the last term in the right-hand side of \eqref{Final inequality for Sobolev}.

\end{proof}

The following lemma establishes a Hardy-type inequality on the hypersurfaces $\Sigma^*_\tau$ of the Schwarzschild--AdS exterior spacetime:

\begin{lemma}\label{lem:Hardy inequality global}
For any $\tau\in \mathbb R$, let $f:\Sigma^*_\tau\rightarrow \mathbb C$ be a sufficiently regular and decaying function (in the $(r,\theta, \varphi)$ coordinate system on the slice $\Sigma^*_\tau$ of $\mathcal M_\mathrm{ext}$). Then:
\begin{equation}\label{Hardy SchAdS}
\int_{\Sigma^*_\tau} 2r^2 |f|^2 \, \sin\theta drd\theta d\varphi \le \int_{\Sigma^*_\tau} \f89 r^2 \Big(1-\frac{2M}{r}+r^2\Big) |\partial_r f|^2 \, \sin\theta drd\theta d\varphi.
\end{equation}
\end{lemma}

\begin{proof}
We can compute for any $(\theta, \varphi)\in \mathbb S^2$:
\begin{align*}
\int_{r_{\mathcal H^+}}^{+\infty} 2r^2 |f(\tau, r, \theta, \varphi)|^2 \, dr & =  
\int_{r_{\mathcal H^+}}^{+\infty} \frac23 \frac{d (r^3-r_{\mathcal H^+}^3)}{r} |f(\tau, r, \theta, \varphi)|^2 \, dr\\
& = \f23 (r^3-r_{\mathcal H^+}^3)|f|^2 \Big|_{r=+\infty} - \f23 (r^3-r_{\mathcal H^+}^3)|f|^2 \Big|_{r=r_{\mathcal H^+}} \\
& \hphantom{ \f23 (r^3-r_{\mathcal H^+}^3)|f|^2 \Big|_{r=+\infty}}
- \int_{r_{\mathcal H^+}}^{+\infty} \frac23  (r^3-r_{\mathcal H^+}^3) \partial_r |f(\tau, r, \theta, \varphi)|^2 \, dr \\
& = 0 - 0 - \int_{r_{\mathcal H^+}}^{+\infty} \frac43  (r^3-r_{\mathcal H^+}^3)  \Re\{ f(\tau, r, \theta, \varphi)\partial_r \bar f(\tau, r, \theta, \varphi)\}\, dr \\
& \le \Big(\int_{r_{\mathcal H^+}}^{+\infty} 2r^2 |f(\tau, r, \theta, \varphi)|^2 \, dr  \Big)^{\f12} 
\Big(\int_{r_{\mathcal H^+}}^{+\infty} \f89 \f{(r^3-r_{\mathcal H^+}^3)^2}{r^2} |\partial_r f(\tau, r, \theta, \varphi)|^2 \, dr  \Big)^{\f12},
\end{align*}
from which we conclude that
\begin{equation}\label{Hardy}
\int_{r_{\mathcal H^+}}^{+\infty} 2r^2 |f|^2 \, dr \le \int_{r_{\mathcal H^+}}^{+\infty} \f89 \f{(r^3-r_{\mathcal H^+}^3)^2}{r^2} |\partial_r f|^2 \, dr.
\end{equation}
Let us consider the auxiliary  function
\[
W(r) \doteq r^4 \Big(1-\frac{2M}{r}+r^2\Big)-(r^3-r_+^3)^2 = r^4+2(r_+^3-M)r^3 - (2M-r_+)^2 = r^4+2(M-r_+)r^3 - r_+^6
\]
(in the above, we used the fact that  $r_+$ solves $1-\frac{2M}{r_+}+r_+^2=0$). Note that
\[
W(r_+)=0
\]
and
\[
W'(r) = 4r^3+6(M-r_+)r^2 = r^2(4r-6r_++6M)\ge 0 \quad \text{for } r\ge r_+.
\]
Therefore, $W(r)\ge 0$ for $r\ge r_+$, which is equivalent to
\[
\f{(r^3-r_{\mathcal H^+}^3)^2}{r^2} \le r^2 \Big(1-\frac{2M}{r}+r^2\Big).
\]
Substituting the above bound in \eqref{Hardy}, we obtain
\begin{equation}\label{Hardy 2}
\int_{r_{\mathcal H^+}}^{+\infty} 2r^2 |f|^2 \, dr \le \int_{r_{\mathcal H^+}}^{+\infty} \f89 r^2 \Big(1-\frac{2M}{r}+r^2\Big) |\partial_r f|^2 \, dr.
\end{equation}
Integrating \eqref{Hardy 2} over $(\theta, \varphi)$ with volume form $\sin\theta d\theta d\varphi$, we obtain \eqref{Hardy SchAdS}.
\end{proof}

\printbibliography[heading=bibintoc]

\end{document}